\documentclass[hyperpdf,bindnopdf]{hepthesis}

\usepackage[T1]{fontenc}
\usepackage{charter}
\usepackage[expert]{mathdesign}

\usepackage{morefloats,subfig,afterpage}
\usepackage{mathrsfs} 
\usepackage{verbatim}
\usepackage{amsmath}
\usepackage{xcolor}

\usepackage[english]{babel}
\usepackage{cite}
\usepackage{bm}
\usepackage{capt-of}   
\usepackage{array}
\usepackage{booktabs}
\newcolumntype{C}[1]{>{\centering\let\newline\\\arraybackslash\hspace{0pt}}m{#1}}
\usepackage{enumerate}
\usepackage[shortlabels]{enumitem}

\usepackage{stackrel}

\definecolor{cerulean}{rgb}{0.0, 0.48, 0.65}

\definecolor{ao(english)}{rgb}{0.0, 0.5, 0.0}

\definecolor{goldenrod}{rgb}{0.85, 0.65, 0.13}

\definecolor{blue-violet}{rgb}{0.54, 0.17, 0.89}

\definecolor{fandango}{rgb}{0.71, 0.2, 0.54}

\usepackage{anyfontsize}
\newcommand{\myparagraph}[1]{\noindent \textbf{{\fontsize{13}{13}\selectfont #1}} --}

\usepackage[ddmmyyyy]{datetime}
\usepackage{braket}
\usepackage{slashed}
\usepackage{graphicx}
\usepackage[export]{adjustbox}

\usepackage[subfigure]{tocloft}
\addtocontents{toc}{\vspace{-4ex}} 
\addtocontents{preface}{\vspace{-4ex}} 
\setkomafont{subsubsection}{\normalfont}
\makeatletter
\renewcommand\subsubsection{\@startsection{subsubsection}{3}{\z@}%
                                     {-3.25ex\@plus -1ex \@minus -.2ex}%
                                     {-1.5ex \@plus -.2ex}
                                     {\normalfont\normalsize\bfseries}}
\makeatother

\usepackage{amssymb}
\newcommand{\gsim}{\raisebox{-0.13cm}{~\shortstack{$>$ \\[-0.07cm]
      $\sim$}}~}
\newcommand{\lsim}{\raisebox{-0.13cm}{~\shortstack{$<$ \\[-0.07cm]
$\sim$}}~}

\usepackage{bibentry}

\usepackage{booktabs}
\usepackage{multirow}
\usepackage{tabularx}

\usepackage{comment}

\usepackage{wrapfig}
\usepackage{makecell}

\usepackage{neuralnetwork}
\newcommand{\mycite}[1]{\underline{\hspace{-0.5pt}\cite{#1}}}

\DeclareOldFontCommand{\rm}{\normalfont\rmfamily}{\mathrm}
\DeclareOldFontCommand{\sf}{\normalfont\sffamily}{\mathsf}
\DeclareOldFontCommand{\tt}{\normalfont\ttfamily}{\mathtt}
\DeclareOldFontCommand{\bf}{\normalfont\bfseries}{\mathbf}
\DeclareOldFontCommand{\it}{\normalfont\itshape}{\mathit}
\DeclareOldFontCommand{\sl}{\normalfont\slshape}{\@nomath\sl}
\DeclareOldFontCommand{\sc}{\normalfont\scshape}{\@nomath\sc}
\newcommand{\be}{\begin{equation}}
\newcommand{\ee}{\end{equation}}
\newcommand{\bea}{\begin{eqnarray}}
\newcommand{\eea}{\end{eqnarray}}
\newcommand{\bi}{\begin{itemize}}
\newcommand{\ei}{\end{itemize}}
\newcommand{\ben}{\begin{enumerate}}
\newcommand{\een}{\end{enumerate}}

\newcommand{\lc}{\left[}
\newcommand{\rc}{\right]}
\newcommand{\lp}{\left(}
\newcommand{\rp}{\right)}

\usepackage{floatrow}


\usepackage{caption}
\usepackage{sidecap}

\makeatletter
\@ifpackageloaded{hyperref}{%
\hypersetup{%
  pdftitle = {Exploring the substructure of nucleons and nuclei with machine learning},
  pdfsubject = {Rabah Abdul Khalek's PhD thesis},
  pdfkeywords = {proton parton distribution functions, nuclear parton distribution functions, PDFs, nPDFs, fragmentation functions, FFs, heavy-ion collisions, LHC, QCD, machine learning, neural networks, Monte Carlo method},
  pdfauthor = {\textcopyright\ Rabah Abdul Khalek}
}}{}
\makeatother

\begin{document}

\begin{frontmatter}
\thispagestyle{empty}
\begin{center}
\Huge{\textbf{Exploring the substructure of nucleons and nuclei with machine learning}}\\
\vspace*{\fill}
\Large{Rabah Abdul Khalek}
\end{center}
\newpage
\thispagestyle{empty}

\vspace*{\fill}
\noindent This work is financed by the Netherlands Organization for Scientific Research (NWO) and carried out at Nikhef and Vrije Universiteit Amsterdam.
\begin{figure*}[!h]
  \begin{center}
  \includegraphics[width=0.35\textwidth]{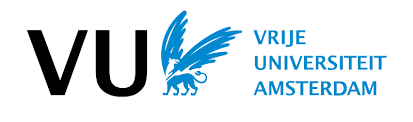}
  \includegraphics[width=0.25\textwidth]{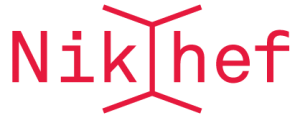}
  \includegraphics[width=0.38\textwidth]{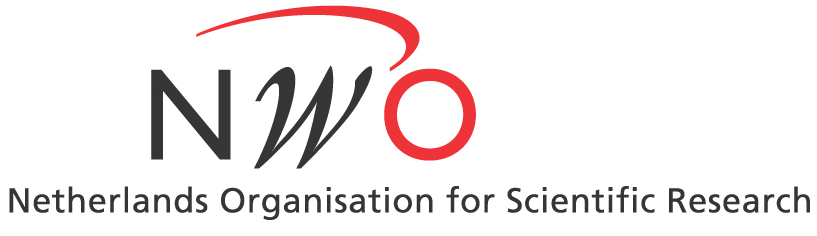}
  \end{center}
\end{figure*}

\newpage
\thispagestyle{empty}
\begin{center}
VRIJE UNIVERSITEIT
\\ \vspace{2cm}
\textbf{{\large E}XPLORING THE SUBSTRUCTURE OF NUCLEONS AND NUCLEI WITH} \\ \textbf{MACHINE LEARNING}
\\ \vspace{5cm}
ACADEMISCH PROEFSCHRIFT
\\ \vspace{1cm}
ter verkrijging van de graad Doctor of Philosophy aan\\
de Vrije Universiteit Amsterdam,\\
op gezag van de rector magnificus\\
prof.dr. C.M. van Praag,\\
in het openbaar te verdedigen\\
ten overstaan van de promotiecommissie\\
van de Faculteit der B\`{e}tawetenschappen\\
op maandag 27 september 2021 om 11.45 uur\\
in een bijeenkomst van de universiteit,\\
De Boelelaan 1105
\end{center}
\vspace*{\fill}
\begin{center}
door \\
Rabah Abdul Khalek \\
geboren te Baakline, Libanon
\end{center}

\newpage
\thispagestyle{empty}
\begin{table}
  \centering
  \begin{tabular}{ll} 
promotor: & prof.dr. J. Rojo  \\
& \\
& \\
copromotor: & dr. P.J.G. Mulders \\ 
& \\
& \\
promotiecommissie: &
prof.dr. H.G. Raven\\
&prof.dr. D. Boer\\
&prof.dr. E.L.M.P. Laenen\\
&dr. M. van Leeuwen\\
&dr. M. Verweij\\

\end{tabular}
\end{table}

\newpage
\begin{abstract}
  \thispagestyle{empty}
  Perturbative quantum chromodynamics (QCD) ceases to be applicable at low interaction energies due to the rapid increase of the strong coupling. In that limit, the non-perturbative regime determines the properties of quarks and gluons (partons) in terms of parton distribution functions (PDFs) or nuclear PDFs, based on whether they are confined within nucleons or nuclei respectively. Related non-perturbative dynamics describe the hadronisation of partons into hadrons and are encoded by the fragmentation functions (FFs). This thesis focuses on the detailed study of PDFs in protons and nuclei as well as the charged pions FFs by means of a statistical framework based on machine learning algorithms. The key ingredients are the Monte Carlo method for error propagation as well as artificial neural networks that act as universal unbiased interpolators. The main topics addressed are the inference of proton PDFs with theoretical uncertainties and the impact on the gluon PDF from dijet cross sections; a global determination of nuclear PDFs exploiting the constraints from proton-lead collisions at the LHC and using for the first time NNLO calculations; a new determination of FFs from single-inclusive annihilation and semi-inclusive deep-inelastic scattering data; and a quantitative assessment of the impact of future colliders such as the High-Luminosity LHC and the Electron Ion Collider on the proton and nuclear PDFs.
\end{abstract}

\newpage

\tableofcontents
\newpage
\markboth{}{}
\thispagestyle{plain}

\begin{preface}
\phantomsection
\addcontentsline{toc}{chapter}{Preface}

\noindent\textit{Beauty is truth, truth beauty -- that is all ye know on earth, and all ye need to know.} \\ \textcolor{white}{x} \hfill --- John Keats in ``Ode on a Grecian Urn''

The dissonance between \textit{Reductionism}, advocating that all properties of a phenomenon can be explained in terms of its constituents and their interactions, and \textit{Antireductionism} remains relevant in many open physics questions today. What are the most fundamental building blocks of matter? Is there only one or many irreducible entities at the heart of all building blocks? In fact, one of the most famous proponents of Antireductionism in the 20th century, Popper, classified phenomena into two types: "clock" phenomena with a mechanical underlying basis and "cloud" phenomena which are indivisible (or fundamental)~\cite{Popper1966-POPOCA}. This contrast highlights -- perhaps -- the human tendency to seek the simplest and most fundamental explanations possible to natural phenomena while constantly doubting their validity. It also hints -- perhaps -- to an inherently simple (and yet layered) nature -- or as Keats and Dirac~\cite{kragh2002paul} would prefer to call it \textit{beautiful} -- that we persistently seek to unfold and comprehend.

\myparagraph{Historical context}
One of the earliest signs of Reductionism in physics surfaced with Thales, around the 7th century BC, who hypothesized that the universe was made out of water, being the fundamental substance of which all others were composed~\cite{beresford2010medical}. Around a century later, Democritus proposed that all matter was composed of physically, but not geometrically, indivisible small particles called \textit{atoms}.

Democritus' atoms were revived by Dalton, around 1800, who performed many experiments studying the pressure of gases and defined atoms as tiny solid particles of matter~\cite{thackray1966origin, rocke2005search}. 
Then, the periodic table~\cite{mendelejew1869beziehungen} was published in 1869 by Mendeleev, listing the chemical elements in rows or columns in order of their relative atomic mass. 
In the nineteenth century, Prout, based on his observation that the elements' masses appeared to be integer multiples of the hydrogen atomic weight, proposed that all atoms are composed of hydrogen atoms~\cite{prout1815relation, prout1816correction}. Prout hypothesized that the hydrogen atom was the most fundamental object and that all other elements were actually groupings of many hydrogen atoms. Indeed in the early twentieth century, aligned with Prout's hypothesis, Rutherford discovered the proton and hinted for the neutron while proving that the hydrogen nucleus is present in other nuclei~\cite{rutherford2010collision}. That resulted from the well-known gold foil experiment in which Rutherford proved that some Helium particles emitted by a radioactive element, \textit{elastically} deflected from the foil instead of going through. The discovery of the neutron as a neutrally charged particle, distinct from the proton came later in 1932 with Chadwick~\cite{chadwick1932existence}. To explain their binding in the atomic nucleus, Hideki Yukawa proposed a model of interaction to describe the nuclear force between nucleons (proton and neutrons) mediated by pions~\cite{yukawa1935interaction,yukawa1937interaction,yukawa1938interaction}.

\myparagraph{Evolution of quantum chromodynamics} For decades after, the nucleons were considered as elementary (structureless) particles, until the discovery made by Stern and Frisch in 1933~\cite{frisch1933magnetische}, then Alvarez and Bloch in 1939~\cite{bacher1933note, alvarez1940quantitative}, proving that the anomalous magnetic moment of the proton and neutrons were different from that of point-particles, like electrons. During the 1950s, the first electron-nucleon elastic scattering experiments were performed by Hofstadter with electron beams having energy of the order of MeV~\cite{cowan1964electron}. These experiments were able to reveal the internal charge distribution of the nucleon supporting the conclusions drawn by the anomalous magnetic moment measurements and hinting strongly to a point-like substructure of the nucleon.

In 1964, Gell-Mann~\cite{gell1964schematic} and Zweig~\cite{zweig1964} proposed the quarks as elementary particles constituents of the variety of observed mesons (particles composed of an equal number of quarks and antiquarks) and baryons (particles containing at least three quarks). Only couple of years later the Stanford Linear Accelerator experiments~\cite{bloom1969high} measured \textit{deep-inelastic} electron-nucleon scattering~\cite{Blumlein:2012bf} covering electron energies ranging from 7 to 17 GeV. These experiments were of vital importance as they presented for the first time the proof of the nucleon compositeness. Moreover, the cross section measurements exhibited the feature of \textit{Bjorken scaling}~\cite{bjorken1969asymptotic}, which strongly suggested that the quarks within the nucleon behaved almost as collections of point-like constituents when probed at high energies. Furthermore, the data obeyed the \textit{Callan and Gross relation}~\cite{callan1969high}, which identified the quarks as fermions (spin-$1/2$ particles). These observations led Feynman to propose the \textit{parton model}~\cite{feynman1969very, feynman2018photon} of point-like fermions composing the nucleon that act as almost-free particles when probed at high energies in the deep-inelastic scattering process. 

In order to explain the quarks \textit{confinement} by gluon exchange inside hadrons (mesons and baryons), the state of the art quantum field theory of strong interactions, quantum chromodynamics or QCD, was developed over many years. In 1965, Nambu~\cite{nambu1966preludes} proposed a Yang-Mills theory based on \textit{colour}, a new three-valued charge degree of freedom quarks have. Two years later, the formalism needed for the quantisation of the theory was developed by Faddeev and Popov~\cite{faddeev1967feynman}. The renormalisation of massless Yang-Mills theories was proven by 't Hooft in 1971~\cite{hooft1971renormalization} and QCD as the theory of strong interactions was proposed by Fritzsch, Gell-Mann and Leutwyler~\cite{fritzsch2002current,fritzsch1973advantages}. In 1973, Gross, Wilczek~\cite{gross1973ultraviolet} and Politzer~\cite{politzer1973reliable} studied the running of the strong coupling constant of colour octet Yang-Mills theory with colour triplet quarks and found \textit{asymptotic freedom}.

\myparagraph{Aim of the thesis} 
The strong force, described by the theory of quantum chromodynamics (QCD), manifests itself through the constant exchange of \textit{gluons} between the quarks. Gluons, like photons, are elementary massless particles that define the nature of the force they carry. Gluons interact 137 stronger than photons at distances similar to the size of the nucleus ($\sim 10^{-15}$ m). The gluon interaction becomes even stronger for larger distances and weaker for smaller ones, a feature called asymptotic freedom. This property leads to the confinement of quarks and gluons (collectively called partons) within a nucleon and is also responsible for the fact that a parton cannot be measured or observed as a free particle. In fact whenever a parton gets knocked out of a nucleon, it instantly starts to fragment into other partons until it hadronises, forming a new hadron (a bound state of 2 or more quarks).

Knocking a parton out of a hadron requires a highly energetic kick, which is usually achieved by colliding it with another particle in experiments. An accelerated electron for instance, upon collision, exchanges a virtual photon with the hadron and one of two scenarios could occur. If the electrons acquire enough energy (typically more than the hadron's rest energy), then the emitted virtual photon would have a very short wavelength and thus the ability to probe smaller distances than the size of the hadron. This scenario is called deep-inelastic scattering (DIS), whereby the interaction happens between the virtual photon and a parton within the hadron. The struck parton acquires enough transferred energy resulting in a shattered hadron. The other scenario is the elastic scattering, wherein an electron's energy is not enough to probe the substructure of the hadron, it rather interacts with the whole hadron as a coherent-state particle that remains intact after the photon exchange.

DIS, among other processes, provides us with a channel to probe the internal structure of hadrons and understand the partonic dynamics within. However, the nature of the strong force makes it only possible to calculate QCD predictions at interaction energy scales (photon energy in our example) much higher than the characteristic scale of partons confinement called QCD scale ($\sim 200$ MeV). At such high energy scales, the partons are probed almost as if they were free and sharing the nucleon's momentum among them. That being said, the non-perturbative low energy scale dynamics, in particular the partons' momentum distribution, could be encoded in the so-called \textit{parton distribution functions} (PDFs) that are partly constrained by QCD and partly inferred from experimental data. This is facilitated by theorems and properties of QCD that allow the separation of non-perturbative dynamics from perturbative calculable ones as well as predicts their dependence on the energy scale. The partonic debris of DIS instantly fragments and turn into hadrons through a non-perturbative process called hadronisation that in turn is encoded by the so-called \textit{fragmentation functions} (FFs).

The partons distribution within nucleons bound in a nucleus is found to be significantly different than that of a free nucleon. This is ascribed to nuclear effects manifested by the interaction between the nucleons or their constituents, which are also considered to be non-perturbative effects. In order to incorporate these effects into the perturbative QCD framework, they were encoded in the so-called \textit{nuclear PDFs} (nPDFs) describing the PDFs of bound nucleons in nuclei, thus the modified counter-part of a free nucleon PDF. To this day, there is no rigorous theoretical explanation of the origin of these nuclear effects. Moreover, the fact that they significantly modify the partons momentum distribution even at energy scales ($\sim$ GeV) much higher than the nuclear binding effects scales ($\sim$ MeV) remains counter-intuitive.

In this thesis, I mainly study these non-perturbative dynamics and I will present my work on analysing the fragmentation of partons into hadrons as well as their distribution inside nucleons (free proton and neutron) and nuclei (bound states of protons and neutrons). In fact, the latter (distribution of partons in nuclei) is the main focus of my work. However, as I will try to show, these parton dynamics, be it fragmentation or distribution, inside nucleons or nuclei, are tightly connected and their analysis is a daunting task requiring cutting edge technologies and statistical methods. In this thesis, the principal and common tool used is the \textit{Artificial Neural Networks} (ANNs or NNs) which is perhaps one of the most famous machine learning algorithms. What makes NNs a suitable tool in describing non-perturbative dynamics of QCD is their remarkable feature of being able to approximate any continuous function for inputs within a specific range. This feature is a direct implication of the \textit{Universal Approximation Theorem}~\cite{csaji2001approximation} that is relied upon to decipher the theoretically indeterminable substructure of hadrons (nucleons and nuclei).

\myparagraph{Outline of the thesis} Chapter~\ref{chap:preQCD} is introductory and mainly based on Refs.~\cite{peskin2018introduction,schwartz2014quantum}. I introduce the elastic and inelastic scattering processes in the context of the theory quantum electrodynamics (QED). I discuss first the disagreement between the elastic scattering cross sections and the high-energy scattering measurements at leading-order. I then introduce next-to-leading corrections and renormalisation emphasizing on the nucleon's anomalous magnetic moment, the running of the QED coupling constant and the renormalisation group equation. Afterwards, I discuss the deep-inelastic scattering (DIS) in the context of the parton model and highlight its main features like the Bjorken-scaling and the Callan-Gross relation. With this, I set up the basics to introduce the theory of QCD.

Chapter~\ref{chap:QCD} contains most of the theoretical elements upon which my work is based on, that is the theory of QCD. I start by discussing the origin of the theory, the QCD Lagrangian and the strong coupling constant. Then, I supplement the parton model with QCD radiative corrections and discuss the DGLAP evolution equations, factorisation and the definition of parton distribution functions (PDFs). I reconsider then the factorised expressions of the DIS cross section and the treatment of heavy quarks effects. Subsequently, on one hand, I introduce the main theoretical aspects behind parton fragmentation into hadrons that is encoded in fragmentation functions (FFs). On the other hand, I review the nuclear effects on the distribution of partons in a nucleon bound within a nuclei, which is encoded in nuclear PDFs (nPDFs). Finally, I summarise the physical constraints on these non-perturbative objects as well as the kinematics and the cross section factorised expressions of all scattering processes that are considered in my work.

Chapter~\ref{chap:Stat} contains all the statistical elements upon which my work is based on and a review of my neural network analytical derivatives \texttt{C++} library~\mycite{AbdulKhalek:2020uza}. I start by discussing how the data uncertainties are provided by experimentalists in high-energy physics. Then, I introduce the main elements of Bayesian inference by examining the likelihood function, the gaussian assumption, the Monte Carlo and Hessian methods. I finally discuss neural networks and minimisation algorithms together with their main features and pitfalls.

Chapter~\ref{chap:PDF} is an overview of all my contributions to the \texttt{NNPDF} framework, dedicated for global determinations of proton PDFs parameterised in terms of a neural network. That covers my work on including missing higher order calculations as theoretical uncertainties in the NNPDF global analysis~\mycite{AbdulKhalek:2019ihb,AbdulKhalek:2019bux} as well as the systematic study of single inclusive jet and dijet measurements in the NNPDF global analysis~\mycite{AbdulKhalek:2020jut}.

Chapter~\ref{chap:nNNPDF10}, \ref{chap:nNNPDF20} and \ref{chap:nNNPDF30} are the summary of all my work on a framework that I lead its development: \texttt{nNNPDF}, dedicated for global determinations of nuclear PDFs parameterised in terms of a neural network. That covers the first \texttt{nNNPDF1.0} release of nPDF sets (Chapter~\ref{chap:nNNPDF10}) based solely on lepton-nucleus neutral current DIS scattering and the framework proof-of-concept~\mycite{Khalek:2018bbv,AbdulKhalek:2019mzd}, as well as the second \texttt{nNNPDF2.0} release (Chapter~\ref{chap:nNNPDF20}) based on neutral and charged current lepton-nucleus DIS as well as gauge boson production from proton-lead collisions from the LHC~\mycite{AbdulKhalek:2020yuc}. Finally, I discuss my results achieved to date related to the third \texttt{nNNPDF3.0} release (Chapter~\ref{chap:nNNPDF30}) that will be based on data that mostly constrains the gluon nPDF and performed at next-to-next-to-leading order (NNLO) accuracy in perturbative QCD.

Chapter~\ref{chap:FF} is the result of my work within the \texttt{MAPFF} framework, on the recent determination of unpolarised
charged-pion fragmentation functions (FFs) from a set of
single-inclusive electron-positron annihilation and lepton-nucleon
semi-inclusive deep-inelastic-scattering data. FFs are
parametrised in terms of a neural network and fitted to data
exploiting the knowledge of the analytic derivative of the NN itself
w.r.t.~its free parameters. This analysis is performed at
next-to-leading order (NLO) accuracy in perturbative QCD~\mycite{Khalek:2021gxf}.

Chapter~\ref{chap:Impact} is an overview of all the impact studies I contributed to. This includes the impact of future colliders like the High Luminosity Large Hadron Collider~\mycite{Khalek:2018mdn,Cepeda:2019klc,Azzi:2019yne} and Large Hadron electron Collider~\mycite{AbdulKhalek:2019mps} on proton PDFs. As well as, the impact of isolated photon and inclusive hadron production at the LHC in heavy-ion collisions~\mycite{AbdulKhalek:2020yuc} on nuclear PDFs. Finally, I dedicate a detailed section to discuss the impact of the Electron Ion Collider on both proton and nuclear PDFs in a fully consistent manner~\mycite{AbdulKhalek:2021gbh,Khalek:2021ulf,AbdulKhalek:2021xxx2}.


\end{preface}

\newpage
\phantomsection
\addcontentsline{toc}{chapter}{Publications}
\pagestyle{plain}
%
\begin{table}[!h]
\centering
\begin{adjustwidth}{-1.15cm}{}
\vspace{-2.2cm}
\resizebox{1.1\textwidth}{!}{%
\begin{tabular}{p{0.03\textwidth}|p{1.1\textwidth}|}
    \multicolumn{2}{c}{ {\Large \textbf{Publications}}}\\
    \cline{2-2}
    \parbox[c]{2mm}{\multirow{2}{*}{\rotatebox[origin=c]{90}{{\large \textbf{Ongoing}}\textcolor{white}{xx}}}}& $\triangleright$ R. Abdul Khalek,  T. Giani, E. R. Nocera, and J. Rojo, \textbf{nNNPDF3.0: A global analysis of nuclear parton distributions at NNLO} \\
    & \\[-8pt]
    & $\triangleright$ R. Abdul Khalek, V. Bertone, D. Pitonyak, A. Prokudin, and N. Sato, \textbf{New ideas for impact studies at the EIC}\\
    \cline{2-2}
    \multicolumn{2}{c}{ \vspace{-8pt} }\\
    \cline{2-2}
    \parbox[c]{2mm}{\multirow{2}{*}{\rotatebox[origin=c]{90}{{\large \textbf{arXiv~}}}}} & $\blacktriangleright$ R. Abdul Khalek and V. Bertone, \textbf{On the derivatives of feed-forward neural networks}, \href{http://arxiv.org/abs/arXiv:2005.07039}{arXiv:2005.07039}\\
    \cline{2-2}
    \multicolumn{2}{c}{ \vspace{-8pt} }\\
    \cline{2-2}
    &  $\blacktriangleright$ R. Abdul Khalek, V. Bertone, and E. R. Nocera, \textbf{A determination of unpolarised pion fragmentation functions using semi-inclusive deep-inelastic-scattering data: MAPFF1.0}, \textit{Phys. Rev. D} \textbf{104} (2021), no. 3 034007, 
    [\href{http://arxiv.org/abs/arXiv:2105.08725}{arXiv:2105.08725}] \\
    & \\[-8pt]
    & $\blacktriangleright$ R. Abdul Khalek, J. J. Ethier, E. R. Nocera, and J. Rojo, \textbf{Self-consistent determination of proton and nuclear PDFs at the Electron Ion Collider}, \textit{Phys. Rev. D} \textbf{103} (2021), no. 9 096005, [\href{http://arxiv.org/abs/arXiv:2102.00018}{arXiv:2102.00018}] \\
    & \\[-8pt]
    & $\blacktriangleright$ R. Abdul Khalek, J. J. Ethier, J. Rojo, and G. van Weelden, \textbf{nNNPDF2.0: quark flavor separation in nuclei from LHC data}, \textit{JHEP} \textbf{09} (2020) 183, [\href{http://arxiv.org/abs/arXiv:2006.14629}{arXiv:2006.14629}]\\
    & \\[-8pt]
    \parbox[c]{-2mm}{\multirow{10}{*}{\rotatebox[origin=c]{90}{{\large\textbf{Peer-reviewed}}}}} & $\blacktriangleright$ R. Abdul Khalek et al., \textbf{Phenomenology of NNLO jet production at the LHC and its impact on parton distributions}, \textit{Eur. Phys. J. C} \textbf{80} (2020), no. 8 797, [\href{http://arxiv.org/abs/arXiv:2005.11327}{arXiv:2005.11327}]\\
    & \\[-8pt]
    & $\blacktriangleright$ NNPDF Collaboration, R. Abdul Khalek et al., \textbf{Parton Distributions with Theory Uncertainties: General Formalism and First Phenomenological Studies}, \textit{Eur. Phys. J. C} \textbf{79} (2019), no. 11 931, [\href{http://arxiv.org/abs/arXiv:1906.10698}{arXiv:1906.10698}]\\
    & \\[-8pt]
    & $\blacktriangleright$ R. Abdul Khalek, S. Bailey, J. Gao, L. Harland-Lang, and J. Rojo, \textbf{Probing Proton Structure at the Large Hadron electron Collider}, \textit{SciPost Phys.} \textbf{7} (2019), no. 4 051, [\href{http://arxiv.org/abs/arXiv:1906.10127}{arXiv:1906.10127}]\\
    & \\[-8pt]
    & $\blacktriangleright$ NNPDF Collaboration, R. Abdul Khalek et al., \textbf{A first determination of parton distributions with theoretical uncertainties}, \textit{Eur. Phys. J.} \textbf{C} (2019) 79:838, [\href{http://arxiv.org/abs/arXiv:1905.04311}{arXiv:1905.04311}]\\
    & \\[-8pt]
    & $\blacktriangleright$ NNPDF Collaboration, R. Abdul Khalek, J. J. Ethier, and J. Rojo, \textbf{Nuclear parton distributions from lepton-nucleus scattering and the impact of an electron-ion collider}, \textit{Eur. Phys. J. C} \textbf{79} (2019), no. 6 471, [\href{http://arxiv.org/abs/arXiv:1904.00018}{arXiv:1904.00018}]\\
    & \\[-8pt]
    & $\blacktriangleright$ R. Abdul Khalek, S. Bailey, J. Gao, L. Harland-Lang, and J. Rojo, \textbf{Towards Ultimate Parton Distributions at the High-Luminosity LHC}, \textit{Eur. Phys. J. C} \textbf{78} (2018), no. 11962, [\href{http://arxiv.org/abs/arXiv:1810.03639}{arXiv:1810.03639}]\\
    \cline{2-2}
    \multicolumn{2}{c}{ \vspace{-8pt} }\\
    \cline{2-2}
    \parbox[c]{2mm}{\multirow{2}{*}{\rotatebox[origin=c]{90}{{\large\textbf{Proceedings \& Reports}}\textcolor{white}{xx}}}}& $\blacktriangleright$ R. Abdul Khalek et al., \textbf{Science Requirements and Detector Concepts for the Electron-Ion Collider: EIC Yellow Report}, \href{http://arxiv.org/abs/arXiv:2103.05419}{arXiv:2103.05419} \\
    & \\[-8pt]
    & $\blacktriangleright$ P. Azzi et al., \textbf{Report from Working Group 1: Standard Model Physics at the HL-LHC and HE-LHC}, \textit{CERN Yellow Rep. Monogr.} \textbf{7} (2019) 1-220, [\href{http://arxiv.org/abs/arXiv:1902.04070}{arXiv:1902.04070}] \\
    & \\[-8pt]
    & $\blacktriangleright$ M. Cepeda et al., \textbf{Report from Working Group 2: Higgs Physics at the HL-LHCand HE-LHC}, \textit{CERN Yellow Rep. Monogr.} \textbf{7} (2019) 221-584, [\href{http://arxiv.org/abs/arXiv:1902.00134}{arXiv:1902.00134}] \\
    & \\[-8pt]
    & $\blacktriangleright$ NNPDF Collaboration, R. Abdul Khalek, J. J. Ethier, and J. Rojo, \textbf{Nuclear Parton Distributions from Neural Networks}, \textit{Acta Phys. Polon. Supp.} \textbf{12} (2019), no. 4 927, [\href{http://arxiv.org/abs/arXiv:1811.05858}{arXiv:1811.05858}] \\
    \cline{2-2}
    \end{tabular}}
\end{adjustwidth}
\end{table}
\newpage




\clearpage
\pagestyle{headings}
\thispagestyle{empty}
\end{frontmatter}

\RedeclareSectionCommand[
  beforeskip=0pt,
  afterskip=2\baselineskip]{chapter}
\RedeclareSectionCommand[
  beforeskip=\baselineskip,
  afterskip=.5\baselineskip]{section}
\RedeclareSectionCommand[
  beforeskip=.75\baselineskip,
  afterskip=.5\baselineskip]{subsection}

\begin{mainmatter}
  \chapter{Probing hadrons}
\label{chap:preQCD}
\vspace{-1cm}
\begin{center}
\begin{minipage}{1.\textwidth}
    \begin{center}
        \textit{The derivations presented in this chapter are based on Refs.~\cite{peskin2018introduction,schwartz2014quantum}} 
    \end{center}
\end{minipage}
\end{center}

\myparagraph{Introduction} The electromagnetic visible spectrum lies at energies between 1.12 eV (near-infrared) and 4 eV (ultraviolet)~\cite{sliney2016light}. This amounts to a wavelength range of [310, 1100] nm, far too large to resolve even one of the heaviest nuclei such as lead ($^{208}$Pb) whose average radius is of the order of: 
\begin{equation} 
    \text{Radius($^{208}$Pb)}\simeq r_0A^{1/3} \simeq 7.4\times 10^{-15}\text{ m} \ll \text{Visible limit} \simeq 3.1 \times 10^{-11} \text{ m}\nonumber
\end{equation} 
with $r_0 \sim 1.25\times 10^{-15}$ m the average radius of a nucleon and $A$ the nuclei atomic mass.

This fm-level domain is governed by relativistic quantum mechanics, thus requires a set of sophisticated apparatus to be able to detect particle interactions and resolve their structure. 
For instance, a minimum photon energy of roughly 0.16 GeV is required to interact with the internal structure of a lead nucleus. One of the ways possible to attain such photonic energies and higher is by using highly energetic leptons targeted at a lead nucleus. A lepton scatters off the nucleus by emitting a highly energetic photon whose energy is measurable from that of the lepton before and after scattering. In general, photons are not the only probes that could be used, others include the rest of the electroweak bosons (Z-boson and W-boson) as well as partons in case of hadron-hadron collisions.

Some of the earliest particle detectors were the \textit{cloud chambers}~\cite{CHOPPIN2002192} that were used for visualising traces of energetic charged particles interacting with a certain gas mixture by ionising it. This apparatus helped discovering particles such as the positron~\cite{anderson1933positive} and the muon~\cite{street1937new}. Nowadays however, probe-oriented or collider experiments such as the Large Hadron Collider (see Chapter~\ref{chap:PDF}) are mostly used for studying the properties of the Standard Model and the hadronic substructure among other topics. Such experiments are able to achieve very high collision energies ($\sim$ TeV) and luminosities. Moreover, they rely on accurate and advanced electronic detector technologies such as cutting-edge tracking detectors, calorimeters and muon detectors, recording systematically enormous amount of events statistics to ensure best possible data precision (see Chapter~\ref{chap:Stat}).

\myparagraph{Outline} In this chapter, the focus is on the elastic and inelastic scattering process of nucleons and their historic contribution to the understanding of the hadronic structure. For that reason, I have divided the sections according to the gradual and historical understanding of this topic that happens to largely overlap with the logical steps to build the theory of quantum chromodynamics (QCD). 
In Sect.~\ref{s1:Elastic_scattering}, I start with the theoretical aspects of lepton-nucleon elastic scattering (order of MeV) and discuss its limitations in explaining the hadronic structure. 
In Sect.~\ref{s1:Inelastic_scattering}, I introduce the theory of lepton-nucleon deep-inelastic scattering (order of GeV) and the parton model by which I set up the introduction to the theory of the strong interactions, QCD.

\section{Elastic scattering off hadrons} \label{s1:Elastic_scattering}
In this section I focus on the elastic scattering process. In Sect.~\ref{s2:Elastic_scattering_LO}, I will summarise the results of the electron-nucleon elastic scattering at leading order (LO) in QED, discussing the cross section expressions (Mott's and Rutherford's) in different energy regimes.
%
In Sect.~\ref{s2:QED_Renormalisation}, I discuss the next-to-leading (NLO) order QED corrections that give rise to the anomalous magnetic moment through form factors and how this is considered the QED accuracy test. Then, I will briefly introduce renormalisation for the first time by focusing on the divergent NLO \textit{vacuum polarisation} contribution and deduce the renormalisation group equation. I will highlight the running-type of the QED coupling constant to draw later a comparison with that of QCD. I will end this section by discussing the NLO cross section at high energies given by the \textit{Rosenbluth} expression and its implications.


\subsection{Leading-order} \label{s2:Elastic_scattering_LO}
Elastic scattering is the dominant lepton-hadron scattering process if a lepton has an energy $E$ much lower than the mass $m_A$ of a hadron $A$ ($E\ll m_A$). From the lepton's ``perspective'', at low energies, the hadron is a solid and coherent object that is difficult to resolve. All the lepton can be sensitive to, is the hadron's macroscopic properties, such as its size, electric and magnetic moments.
Except for their masses, the use of any lepton is totally equivalent to derive the scattering cross section, I will therefore discuss only the scattering of electrons. At the lowest order in perturbation theory (leading order or simply LO), the amplitude of this process is given by the following t-channel Feynman diagram:
\begin{align*}
    i\mathcal{M}= \includegraphics[valign=c]{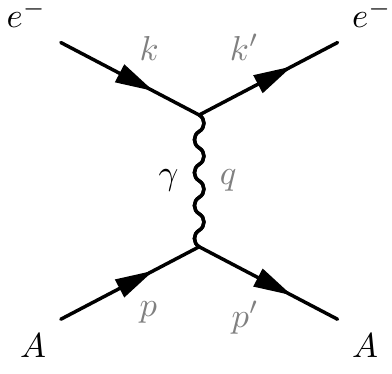} = (-ie)\bar{u}(k')\gamma^\mu u(k)\frac{-i g_{\mu\nu}}{q^2}(-ie)\bar{u}(p')\gamma^\nu u(p)
\end{align*} 
expressed in terms of the Dirac spinors $u(k)$ and $\bar{u}(k)$ describing the incoming and outgoing electron (similar for the hadron $A$) and based on the QED Feynman rules.
We can re-express the four-vector momenta using the following \textit{Mandelstam variables}:
\begin{equation}
    s=(k+p)^2, \qquad  t=(k-k')^2, \qquad u=(k-p')^2  \nonumber
\end{equation}
In order to calculate the cross section of this process, we first square the matrix-element $\mathcal{M}$. Then we assume that we do not know the polarisation of the initial states nor that we measure that of the final states, we therefore consider the case of repeating the measurement many times to eventually end up with the average over all spins or polarisations as follows:
\begin{equation}
|\tilde{\mathcal{M}}|=\frac{1}{4}\sum_{spins}|\mathcal{M}|^2= \frac{2e^4}{t^2} \left[u^2+s^2+4t(m_e^2+m_A^2)-2(m_e^2+m_A^2)^2\right]    \label{eq:ep_elastic_amplitude}
\end{equation}
The cross section in the centre-of-mass (CM) frame for two incoming and outgoing particles can be expressed generally as:
\begin{equation}
\left(\frac{d\sigma}{d\Omega}\right)_{\text{CM}}=\frac{1}{64\pi^2E_{cm}^2} \frac{|\bm{k'}|}{|\bm{k}|}|\tilde{\mathcal{M}}|^2 \quad \text{with: } d\Omega = \sin{\theta} d\theta d\phi \label{eq:ep_elastic_cross_section} 
\end{equation}
where $E_{\text{CM}}$ is the CM energy of the collision, $\theta$ is the scattering angle between the incoming and outgoing electron and $\phi$ is the azimuthal angle from the beam-axis which can be always made irrelevant by choosing a cylindrical symmetric reference frame.

To start, I consider the limit in which an electron has an energy $E \ll m_A \rightarrow \infty$ but with a finite mass $m_e$. The hadron is considered to have an infinite mass with no recoil. The four-momenta of the particles are therefore expressed in the CM frame as follows:
\begin{align}
        &\bigg\{k^\mu=(E,\bm{k}),\quad
        p^\mu=(m_A,0),\quad
        k'^\mu=(E,\bm{k'}),\quad
        p'^\mu=(m_A,0)\bigg\} \\
        k=|\bm{k}| &= |\bm{k'}|,\qquad \bm{k}\cdot\bm{k'}=k^2 \cos{\theta}, \qquad q^2=0 \text{ (on-shell)}, \qquad \Theta = \frac{\alpha^2}{(1-\cos{\theta})^2} \nonumber
\end{align}
Substituting these kinematics in Eq.~(\ref{eq:ep_elastic_amplitude}) and Eq.~(\ref{eq:ep_elastic_cross_section}) I distinguish between the following two ascending energy regimes: 
\begin{alignat}{2}
(k \ll E \sim m_e \ll m_A)&\stackrel[m_A\rightarrow\infty]{}{\longrightarrow}\quad \left(\frac{d\sigma}{d\Omega}\right)^{\text{Rutherford}}_{\text{CM}}&&=\Theta\cdot\frac{m_e^2}{k^4} \label{eq:Rutherford}\\
(E\ll m_A)&\stackrel[m_A\rightarrow\infty]{}{\longrightarrow}\quad \left(\frac{d\sigma}{d\Omega}\right)^{\text{Mott}}_{\text{CM}}&&=
\Theta\cdot\frac{E^2}{k^4}\left(1-\frac{k^2}{E^2}\sin^2\frac{\theta}{2}\right) \label{eq:Mott}
\end{alignat}
where the Mott expression Eq.~(\ref{eq:Mott}) includes relativistic corrections to Rutherford's expression Eq.~(\ref{eq:Rutherford}). Both of which describe the electron elastic scattering off a point-like hadron $A$ at low-energies.

Let us now consider the high energy regime ($E\gg m_e$) where the nucleon in the final state cannot be assumed to be at rest, i.e it has to recoil. The four-momenta of the particles in a lab frame in which the initial state nucleon are expressed as:
\begin{equation} \label{eq:labframe_ep}
    \bigg\{k^\mu=(E,\bm{k}),\quad
    p^\mu=(m_A,\bm{0}),\quad
    k'^\mu=(E',\bm{k'}),\quad
    p'^\mu=k^\mu - k'^\mu+p^\mu\bigg\} \\
    q^2 =-2k^\mu k'_\mu = -4E'E\sin^2\frac{\theta}{2} = -2m_A(E-E'), \qquad \Theta = \frac{\alpha^2}{(1-\cos{\theta})^2} \nonumber
\end{equation}
Similarly, with the right substitution of these kinematics in Eq.~(\ref{eq:ep_elastic_amplitude}) and Eq.~(\ref{eq:ep_elastic_cross_section}) the cross section reads:
\begin{align}
    (E \gg m_e)\longrightarrow \quad \left(\frac{d\sigma}{d\Omega}\right)^{\text{High energy}}_{\text{lab}}&=\Theta\cdot\frac{1}{E^2}\frac{E'}{E}\left(\cos^2{\frac{\theta}{2}} + \frac{E-E'}{m_A}\sin^2{\frac{\theta}{2}}\right)\\
    &=\Theta\cdot\frac{1}{E^2}\frac{E'}{E}\left(\cos^2{\frac{\theta}{2}} - \frac{q^2}{2m_A^2}\sin^2{\frac{\theta}{2}}\right) \label{eq:Rosenbluth_Qinf}
\end{align}

While all of the above point-like expressions were able to describe the scattering at low energies ($\sim 5$ MeV in Rutherford's gold foil experiment~\cite{rutherford2010collision}), they fail at higher energies.
This was first examined in the 1950s by a series of experiments initiated at Stanford by Hofstadter~\cite{Hofstadter:1953zjy,PhysRev.98.217,PhysRev.102.851}. One of the experiments~\cite{Hofstadter:1953zjy} focused on the study of $125$ MeV electron-nuclei scattering that showed clear deviation from the point-like Mott expression Eq.~(\ref{eq:Mott}). This did not came as a surprise as nuclei were expected to be composed by protons and neutrons long before the 1950s. However, another experiment~\cite{PhysRev.98.217,PhysRev.102.851} was based on $188$ MeV electron-proton scattering data and showed a similar incompatibility. This deviation provided a strong evidence that the proton is not a point-like particle, rather has a finite size, which first led to the conclusion that the proton must have a substructure.
Interestingly, at even higher energies than those considered by Hofstadter, the data seems to move towards an agreement with Eq.~(\ref{eq:Rosenbluth_Qinf}) hinting to the presence of point-like constituents within the nucleon, now known as quarks. This energy regime is the inelastic one, in which the incident electron's momentum is not conserved, transferring energy to the nucleon that end up shattering as a result. This process will be discussed in Sect.~\ref{s1:Inelastic_scattering}.

\subsection{Beyond leading-order and renormalisation} \label{s2:QED_Renormalisation}
In this section, I consider the correction of the photon-nucleon vertex without calculating it, only to draw some comparisons later when introducing the parton model, where actually with minimal assumptions, scattering inelastically off nucleon seems almost describable purely by QED. I discuss the divergences that come up the moment one considers the next-to-leading order (NLO) in the perturbative expansion of the scattering matrix. I finally introduce succinctly renormalisation in QED, then discuss the notion of a renormalisable theory, which are crucial ingredients to introduce QCD.

Let us start by considering the full photon-nucleon vertex NLO corrections to Eq.~(\ref{eq:Rosenbluth_Qinf}). In fact these will be valid to any fermion and therefore we will substitute the nucleon by a generic fermion $f$. Therefore the photon-fermion vertex correction reads:
\begin{align}\label{eq:fermion_vertex_correction}
    \includegraphics[width=0.15\textwidth,valign=c]{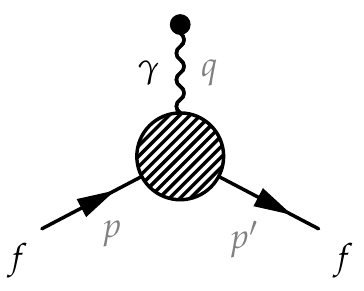}&= 
    \overbrace{
    \includegraphics[width=0.145\textwidth,valign=c]{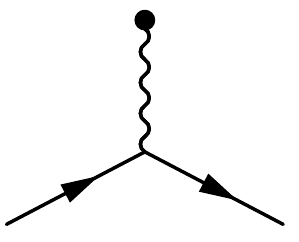}}^{\text{\text{(LO)}}} \hspace{-5pt}+\hspace{-5pt} 
    \overbrace{\underset{\text{vertex correction}}{\includegraphics[width=0.125\textwidth,valign=c]{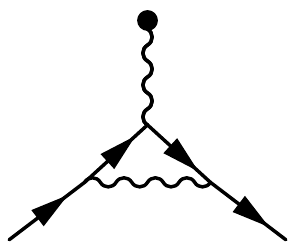}} \hspace{-5pt}+\hspace{-5pt}
    \underset{\text{self-energy}}{\includegraphics[width=0.125\textwidth,valign=c]{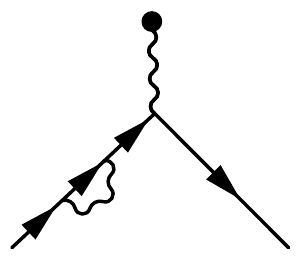} \hspace{-5pt}+\hspace{-5pt} 
    \includegraphics[width=0.125\textwidth,valign=c]{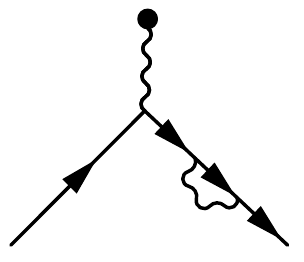}} \hspace{-5pt}+\hspace{-15pt} 
    \underset{\text{vacuum polarisation}}{\includegraphics[width=0.125\textwidth,valign=c]{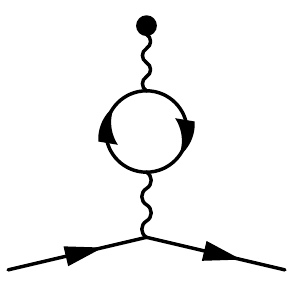}}}^{\text{\text{(NLO)}}}  \hspace{-15pt}+ 
   \mathcal{O}(e^5) \nonumber\\
   &=(-ie)\bar{u}(p')\Gamma^\mu u(p) 
\end{align}
where the hashed blob is designated by $\Gamma^\mu$, the sum of all spin-$1/2$ photon-fermion vertices and loop corrections at any order in perturbation theory (showed only the NLO order) and where photon is considered as a propagator (not incoming) between the fermion and a heavy target not shown in the diagram.

To start with, we note that $\Gamma^\mu= \gamma^\mu$ at leading-order in $\alpha$. Using Lorentz invariance
and the Ward identity ($q_\mu \Gamma^\mu =0$), we can write generally:
\begin{equation} \label{eq:Gamma_QED_vertex}
    \bar{u}(p')\Gamma^\mu u(p) = \bar{u}(p')\left[\gamma^\mu \mathcal{F}^f_1(q^2) + \frac{i\sigma^{\mu\nu}q_\nu}{2m_f}\mathcal{F}^f_2(q^2)\right]u(p) 
\end{equation}
in analogy with the Gordon decomposition identity:
\begin{equation}
    \bar{u}(p') \gamma^\mu u(p) = \bar{u}(p')\left[\frac{p'^\mu+p^\mu}{2m}+\frac{i\sigma^{\mu\nu}q_\nu}{2m}\right]u(p) 
\end{equation}
with $\sigma^{\mu\nu}=2S^{\mu\nu}$ the spinor generator of Lorentz transformations. $\mathcal{F}^f_1$ and $\mathcal{F}^f_2$ are the only unknown functions of $q^2$ allowed by relativistic invariance called \textit{form factors}, where at the lowest-order $\mathcal{F}^f_1=1$ and $\mathcal{F}^f_2=0$ leads back to the leading-order vertex. 

\myparagraph{QED accuracy test}
At NLO, the self-energy and vacuum polarisation diagrams in Eq.~(\ref{eq:fermion_vertex_correction}) involve a correction to the fermion and photon propagators respectively. These diagrams will result in divergent terms proportional to $\gamma^\mu$, therefore contributing only to $\mathcal{F}^f_1$. However, the vertex correction is convergent at all orders and contributes solely to $\mathcal{F}^f_2$ which in the limit $q^2\rightarrow 0$ provides the most accurate test of QED:
\begin{equation}
    \mathcal{F}^f_2(q^2\rightarrow 0) = |\bm{\mu_f}|-1 = g\frac{e}{2m_f}|\bm{S}_f|-1 \stackrel{\text{spin}-1/2}{=} \frac{\alpha}{2\pi}\frac{m_e}{m_f}
\end{equation} 
where $\bm{\mu_f}$ and $\bm{S}_f$ are the spin magnetic and angular moment of the fermion $f$ respectively and $g$ is their proportionality constant called g-factor. The prediction yielding the anomalous magnetic moment of an electron and equivalently the fine-structure constant agrees with the latest measurements within ten parts in a billion ($10^{-8}$). 
However, the measurement of the nucleons anomalous magnetic moment in 1933~\cite{frisch1933magnetische} and 1939~\cite{bacher1933note, alvarez1940quantitative} turned out to be significantly different from that of point-particles, like electrons. A further strong indication to the nucleons compositeness.

\myparagraph{Renormalisation of the electric charge}
Let us now consider vacuum polarisation, which is of particular importance as it will result in the renormalisation of the electric charge and explain the running of the $\alpha$ coupling and it reads:
\begin{align*}
    \includegraphics[valign=c]{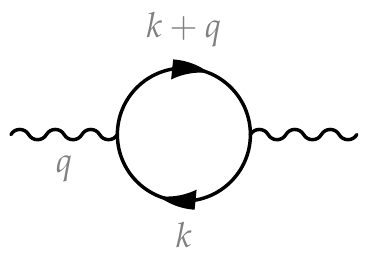}&= \frac{-i}{q^2}i\Pi_2^{\mu\nu}(q)\frac{-i}{q^2}\\
    i\Pi_2^{\mu\nu}(q) &= i(-q^2 g^{\mu\nu}+q^\mu q_\nu)e^2\Pi_2(q^2)
 \end{align*}
where $\Pi_2(q)$ is quadratically divergent and renders for instance the electron-proton scattering cross section or the measurable corrected Coulomb potential in the non-relativistic limit to diverge as well.
The way to solve this problem in renormalisation is to absorb this divergence in the non-observable electric charge that appears in the Lagrangian (called bare charge $e$), by following roughly the following procedure:
\begin{enumerate}
    \item \textbf{Regularisation}: introducing a new finite integral $\Pi_2(q^2,\Lambda)$ which depends on a cut-off scale $\Lambda$ in a way that:
    \begin{equation}
        \Pi_2(q^2,\Lambda) \stackrel{\Lambda \rightarrow \infty}{\longrightarrow} \Pi_2(q^2)
    \end{equation}
    which can be split into a divergent and a convergent part. the convergent part is known as radiative correction.
    \item \textbf{Renormalisation}: the divergent part can be combined with the leading-order contribution and treated as a modification to the bare quantities that become energy scale dependent. 
    \item \textbf{Cut-off scale elimination}: Taking the limit $\Lambda \rightarrow \infty$, the bare quantities become singular and now the renormalised quantities are finite. We call a theory renormalisable if all divergences can be removed by renormalisation of a finite number of couplings in the Lagrangian.
\end{enumerate}
Performing these steps on the vacuum polarisation diagram, the divergence is fixed by renormalising the electric charge or equivalently the coupling constant as follows:
\begin{align}
    \alpha_{\text{eff}}(-q^2) \stackrel{-q^2\rightarrow m_e^2}{\longrightarrow}& \alpha_R = \alpha - \alpha^2 \Pi_2(m_e^2) + \dddot{} = \frac{1}{137}\nonumber\\
    \alpha_{\text{eff}}(-q^2) \stackrel{-q^2 \gg m_e^2}{\longrightarrow}& \alpha_R - \alpha_R^2(\Pi_2(q^2)-\Pi_2(m_e^2)) + \dddot{} \nonumber
\end{align}
where $\alpha^{\text{NLO}}_{\text{eff}}(-q^2)$ is the effective coupling that runs with the scale $-q^2$, and $\alpha_{\text{eff}}(-m_e^2) = \alpha_R$ the renormalised coupling written in terms of $\alpha$ the bare coupling.

Finally, by \textit{resumming} all the higher-order corrections that purely
contain NLO loops into the perturbative series we can write: 
\begin{align}
    \alpha_{\text{eff}}(-q^2) &=\alpha_R\left[1+\frac{\alpha_R}{3\pi}\log\frac{-q^2}{m_e^2}+\left(\frac{\alpha_R}{3\pi}\log\frac{-q^2}{m_e^2}\right)^2+\dddot{}\right]\nonumber\\
    &= \frac{\alpha_R}{1-\frac{\alpha_R}{3\pi}\log{\left(\frac{-q^2}{m_e^2}\right)}} \label{eq:QED_running}
\end{align} 
where that last expression is called the \textit{running} of the coupling in the \textit{leading logarithmic resummation} at next-to-leading order (NLO) accuracy in the QED perturbative expansion. This expression implies that the electric coupling gets larger at shorter distances ($-q^2 \gg m_e^2$) which is consistent with what we know about QED interactions. This enlargement is slow as the coefficient in front of the logarithm is very small. However, in the limit where $\frac{\alpha_R}{3\pi}\log{\left(\frac{-q^2}{m_e^2}\right)} = 1$, the higher order corrections become as big as the LO contribution. This sets an upper limit on $-q^2$, which defines the \textit{Landau pole} of QED $\Lambda_{QED}$ where perturbation theory ceases to be applicable.  

\myparagraph{Renormalisation group equation} \label{s3:RGE} 
Instead of starting the running of the coupling constant from a reference value $\alpha_{\text{eff}}(-q^2=m_e^2)=\frac{1}{137}$, we can parameterise it at an arbitrary scale $\mu_R$ (renormalisation scale) whereby in the limit $-q^2 \gg m_e^2$ becomes:
\begin{equation} \label{eq:QED_alpha_eff}
    \alpha_{\text{eff}}(-q^2) &\stackrel{\mathcal{O}(\alpha_R)}{=} \frac{\alpha_R(\mu_R^2)}{1-\frac{\alpha_R(\mu_R^2)}{3\pi}\log{\left(\frac{-q^2}{\mu_R^2}\right)}}\\
    \frac{1}{e_{\text{eff}}^2(-q^2)}& \stackrel{\mathcal{O}(\alpha_R)}{=}\frac{1}{e_{\text{eff}}^2(\mu_R)}-\frac{1}{12\pi^2}\log{\frac{-q^2}{\mu_R^2}}
\end{equation}
Any observable, denoted $O$, should be independent on how we perform renormalisation, and that invariance could be expressed as ${\displaystyle \frac{d}{d\mu_R} O = 0 }$
where $O$ is an observable and $\mu_R$ the renormalisation scale.
Applying this condition on the coupling defines the \textit{Renormalisation group equation}:
\begin{align}
    \mu_R^2 \frac{d}{d\mu_R^2}(\alpha_{\text{eff}})&\equiv \frac{d}{d\log{\mu_R^2}}(\alpha_{\text{eff}}) \equiv \beta(\alpha_{\text{eff}}) \stackrel{\text{\text{(NLO)}}}{=} \frac{\alpha^2_{\text{eff}}}{3\pi} \label{eq:RGE_alpha} \\
    \mu_R \frac{d}{d\mu_R}(e_{\text{eff}})&\equiv \frac{d}{d\log{\mu_R}}(e_{\text{eff}}) \equiv \frac{1}{2}\beta(e_{\text{eff}}) \stackrel{\text{\text{(NLO)}}}{=} \frac{e^3_{\text{eff}}}{12\pi^2} \label{eq:RGE_e}
    \end{align}
The sign of $\beta(e_{\text{eff}})$ called also beta-function is very important, as it implies the nature of the interaction in a theory. As argued after Eq.~(\ref{eq:QED_running}), the QED beta-function being positive implies large coupling at short distances. 
A theory with a negative beta-function implies that the coupling constant tends to zero at a logarithmic rate as $-q^2$ increases. Such theory is called asymptotically free and oppositely to QED, becomes perturbative in the short-distance ($-q^2\rightarrow \infty$) limit.

\myparagraph{Elastic scattering at NLO}
Now we can rederive Eq.~(\ref{eq:Rosenbluth_Qinf}) with the vertex correction Eq.~(\ref{eq:Gamma_QED_vertex}). The resulting expression is called the \textit{Rosenbluth} formula and reads:
\begin{equation} \label{eq:Rosenbluth}
\left(\frac{d\sigma}{d\Omega}\right)_{\text{lab}}=\Theta\cdot \frac{1}{E^2}\frac{E'}{E}\left[\left((\mathcal{F}^p_1)^2-\frac{q^2}{4m_A^2}(\mathcal{F}^p_2)^2\right)\cos^2{\frac{\theta}{2}} - \frac{q^2}{2m_A^2}\left(\mathcal{F}^p_1+\mathcal{F}^p_2\right)^2 \sin^2{\frac{\theta}{2}}\right]
\end{equation}
While this expression agrees well with the measured cross section of lepton-lepton scattering $(f = e,\,\mu,\,\tau)$, it deviates for the lepton-nucleons at high energies. In the next Sect.~\ref{s1:Inelastic_scattering}, we discuss the dominant scattering process off nucleons at high energies, the deep-inelastic scattering, whereby the form factors in Eq.~(\ref{eq:Rosenbluth}) are replaced by structure functions encoding the confined substructure of the proton governed by the strong interactions.




\section{Deep-inelastic scattering off hadrons} \label{s1:Inelastic_scattering}
In this section, I discuss the dominant scattering process off hadrons at high energies, the deep-inelastic scattering.
In Sect.~\ref{s2:DIS}, I introduce the theory of unpolarised electron deep-inelastic scattering (DIS) in terms of the general cross section expression and kinematics.
In Sect.~\ref{s2:parton_model}, I discuss the parton model\footnote{By parton model we refer to a model prior to the development of QCD, agnostic to gluon radiation.}. I then deduce the origin of the \textit{Bjorken scaling}~\cite{bjorken1969asymptotic} and its implications. Furthermore, I emphasize on the \textit{Callan and Gross relation}~\cite{callan1969high} that identified quarks as fermions. Most importantly, I introduce the nucleon \textit{bare} \textbf{parton distribution functions} (PDFs), the main topic of this thesis, for the first time.

\subsection{Kinematics and cross section} \label{s2:DIS}
DIS was first measured in 1969 by the Stanford Linear Accelerator (SLAC) experiments~\cite{bloom1969high} covering electron energies ranging from 7 to 17 GeV.
At such high center-of-mass energies (Deep), a hadron A breaks apart (Inelastic) and the dominant scattering process becomes the so-called deep-inelastic scattering. That directly implies a different vertex correction derivation where instead we have to consider the photon-hadron-$X$ vertex, with $X$ is whatever the hadron can break apart into as in:
\begin{equation}
\begin{gathered}
    \underbrace{l^\mu = e\bar{u}(k')\gamma^\mu u(k)}\hspace{1.7cm}\\
   i\mathcal{M} = \includegraphics[valign=c]{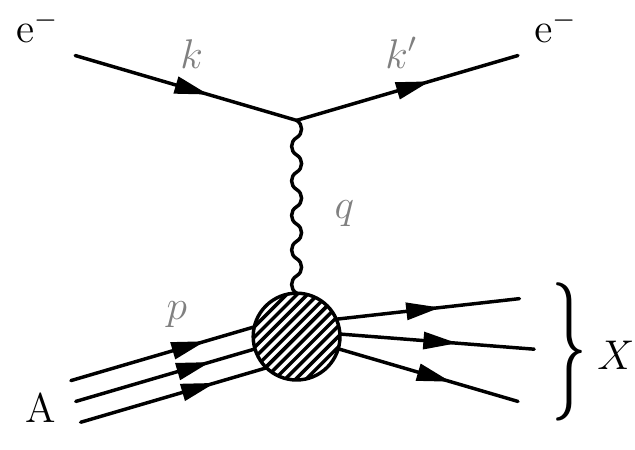} = 
   l^\mu \frac{-ig_{\mu\nu}}{q^2} h^\nu\\
    \overbrace{h^\nu = e\braket{X|j^\nu(0)|A}}\hspace{1.7cm}
   %
\end{gathered} 
\end{equation}
where $l^\mu$ is the electromagnetic current (in general, it could be any electroweak current) and $h^\nu$ is the electromagnetic current $j_\mu(0)$ sandwiched by the hadronic state $A$ and the final state $X$. In order to compute the amplitude of such process, we separate the leptonic and hadronic parts by defining a leptonic tensor $L^{\mu\nu}$ encoding the polarisation of the electron and a hadronic tensor $W_{\mu\nu}$ including the phase space of all final state particles $X$ as follows:
\begin{align}
    L_{\mu\nu} & = \frac{1}{2 e^2}\sum_{e^\pm \text{spins}} l_\mu l^*_\nu
    \equiv \frac{1}{2}\sum_{e^\pm \text{spins}} \bar{u}(k')\gamma^\mu u(k) \bar{u}(k)\gamma^\nu u(k') \nonumber\\[-5pt]
    & = 2(k'^\mu k^\nu + k'^\nu k^\mu - k \cdot k' g^{\mu\nu})\\ 
    W^{\mu\nu} &= \int d\Pi_X (2\pi)^4\delta^{(4)}(q+p-p_X) \frac{1}{e^2}  \sum_{X,\, \text{spins}} h^\mu h^{*\nu}  \nonumber\\[-7pt]
    & = \int d\Pi_X (2\pi)^4\delta^{(4)}(q+p-p_X) \sum_{X,\, \text{spins}} \braket{A|j_\nu(0)^\dagger|X}\braket{X|j_\mu(0)|A} \nonumber\\[-7pt]
    & = \int d^4y e^{iq\cdot y} \braket{A|j_\nu(y)^{\dagger}j_\mu(0)|A} \nonumber\\[-5pt]
    & = W_1(-g^{\mu\nu}+\frac{q^\mu q^\nu}{q^2})+W_2(p^\mu-\frac{p \cdot q}{q^2}q^\mu)(p^\nu - \frac{p \cdot q}{q^2}q^\nu) \label{eq:Hadronic_Tensor} 
\end{align}
where the phase space element $d\Pi_X$ is defined as:
\begin{equation}
d\Pi_X = \prod_{\text{final states } i} \frac{d^3p_{X,i}}{(2\pi)^3}\frac{1}{2E_{X,i}}
\end{equation}
and the final expression for $W^{\mu\nu}$ is the most general one, implied by the current conservation: 
\begin{equation}
\partial_\mu j^\mu = 0 \implies \begin{cases}
    q_\mu W^{\mu\nu}&=0 \\
    q_\nu W^{\mu\nu}&=0
\end{cases}
\end{equation}

The kinematics are taken to be in the lab frame as defined in Eq.~(\ref{eq:labframe_ep}). However, we will opt to use a set of Lorentz-invariant variables that are particularly suited for the description of the \textit{Bjorken scaling} feature that we will discuss in Sect.~\ref{s2:parton_model}. A feature that was predicted by the parton model but in fact violated in perturbative QCD. These variables read:
\begin{alignat}{2}
    \nu &= \frac{p\cdot q}{m} = E-E' \quad && \text{Lepton energy loss}\\
    Q^2 &= -q^2 > 0 \quad && \text{Energy scale squared}\\
    x &= \frac{Q^2}{2p\cdot q} = \frac{Q^2}{2m\nu} \quad && \text{Bjorken-$x$} \\
    y &= \frac{q\cdot p}{k\cdot p}=\frac{\nu}{E} \quad && \text{Inelasticity} \\
    W^2 &= (p+q)^2 = m_p^2 + Q^2\frac{1-x}{x} \quad && \text{Mass squared of the final state $X$}\\
    s &=(k+p)^2 = \frac{Q^2}{xy}+m_p^2 \quad && \text{Center-of-mass energy squared}
\end{alignat}
Taking into account the newly introduced Lorentz-invariant kinematic variables, we can finally write the double differential cross section as:
\begin{align} \label{eq:xsec_DIS_parton_model_v0}
    \left(\frac{d\sigma}{d\Omega dE'}\right)_{lab} & = \frac{\alpha^2}{4\pi m_p q^4}\frac{E'}{E}L_{\mu\nu}W^{\mu\nu} \\
    & = \frac{\alpha^2}{4\pi E^2 (1-\cos{\theta})^2}\left(\frac{m_p}{2}W_2(x,Q^2)\cos^2\frac{\theta}{2}+\frac{1}{m_p}W_1(x,Q^2)\sin^2\frac{\theta}{2}\right) \nonumber 
\end{align} 
We note that Eq.~(\ref{eq:xsec_DIS_parton_model_v0}) has the same form as the Rosenbluth formula Eq.~(\ref{eq:Rosenbluth}) for elastic scattering but with the quantities $W_1$ and $W_2$ encoding the hadronic structure instead of elastic form factors $\mathcal{F}^f_1$ and $\mathcal{F}^f_2$ that are purely electromagnetic. However, both of them are completely determined by measuring the energy and angular dependence of the outgoing electron. A feature that makes the DIS process one of the most experimentally accessible processes.

\subsection{Parton model} \label{s2:parton_model}
Now we return to a statement we mentioned in Sect.~\ref{s2:Elastic_scattering_LO}, at very high energy, the point-like scattering given by the expression Eq.~(\ref{eq:Rosenbluth_Qinf}) indicated the presence of point-like constituents within the hadron. Based on this observation, the parton model was developed by Feynman~\cite{feynman1969very, feynman2018photon}, assuming that some particles (possibly charged) called \textit{partons}\footnote{Partons: when we introduce QCD, partons will not only refer to quarks but also gluons.} within the hadron are very weakly interacting at high energy scales $Q^2$. Within this context, the electron scatters elastically off partons of mass $m_q$ inside a hadron A which will allow us to constrain the quantities $W_1$ and $W_2$. This is illustrated in the following diagram:
\begin{align*}
    \includegraphics[valign=c]{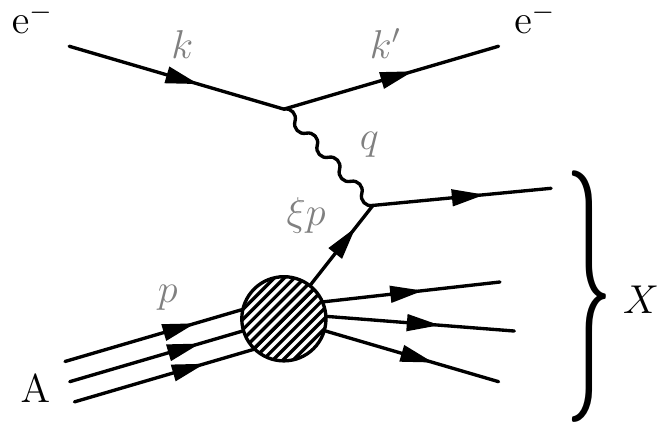}
\end{align*} 
We directly notice that the upper part of the diagram is equivalent to point-like elastic scattering that we've derived before in Sect.~\ref{s2:Elastic_scattering_LO}. To adapt it within the new context of the parton model, we call $p_q = \xi p$ the parton's fraction of the hadron's momentum $p$, which directly implies that:
\begin{equation} \label{eq:Bjorken_x}
    \begin{gathered}
     m_q^2 \stackrel{\text{parton model}}{=} (\xi p+ q)^2 = m_q^2 + 2\xi p \cdot q - Q^2  \implies \frac{Q^2}{2\xi p \cdot q}=1 \\
     {\displaystyle x\stackrel{\text{def}}{=}\frac{Q^2}{2 p \cdot q}\stackrel{\text{parton model}}{=}\xi}
    \end{gathered}
\end{equation}
therefore, measuring the Bjorken-$x$ would be equivalent to measuring the fraction of the hadron's momentum $\xi$ if the parton model is valid.

In the parton model, we assume that the DIS off hadron cross section of $e^- A \rightarrow e^-X$ is given by the off parton one $e^- q \rightarrow e^-X$. Therefore, the task of determining the unknown quantities $W_1$ and $W_2$ reduces to determining $f_{i}(\xi)d\xi$ the probability of the photon hitting parton type $i$ that has a fraction $\xi$ of the hadron's momentum $p$. We call $f_{i}(\xi)$ the bare parton distribution functions (PDFs).
This can be translated by the following:
\begin{equation} \label{eq:partonic_xsec}
    \sigma_{(e^-A\rightarrow e^-X)} \stackrel{\text{parton model}}{=} \sum_i\int_0^1d\xi f_{i}(\xi)\,\,\hat{\sigma}_{(e^- q\rightarrow e^-X)}
\end{equation}
where the hat refers to the partonic cross section of $e^-q\rightarrow e^-X$:
\begin{equation}
    \left(\frac{d\sigma}{d\Omega dE'}\right) \stackrel[\text{Eq.~(\ref{eq:xsec_DIS_parton_model_v0})}]{\text{parton model}}{=} 
    \sum_i f_{i}(x) \frac{\alpha^2e_{q_i}^2}{E^2(1-\cos{\theta})^2}\left(\frac{2m_p}{Q^2}x^2\cos^2\frac{\theta}{2}+\frac{1}{m_p}\sin^2\frac{\theta}{2}\right) \label{eq:xsec_DIS_parton_model_v1}\\
    \implies \begin{cases}
        W_1(x,Q) &= {\displaystyle 2\pi \sum_i e_{q_i}^2 f_{i}(x)}\\
        W_2(x,Q) &={\displaystyle 8\pi\frac{x^2}{Q^2}\sum_i e_{q_i}^2 f_{i}(x)}
    \end{cases}
    \stackrel{\text{or}}{\longleftrightarrow} 
    \begin{cases}
        F_1(x) &={\displaystyle \frac{1}{4\pi}W_1(x)}\\ 
        F_2(x) &= {\displaystyle \frac{Q^2}{8\pi x}W_2(x)}
    \end{cases}
\end{equation}

where $F_1$ and $F_2$ are the dimensionless form of $W_1$ and $W_2$. The fact that $F_1$ and $F_2$ are independent of the energy scale $Q$ is called Bjorken scaling that was the first success of the parton model as it was confirmed by the SLAC data~\cite{bloom1969high} up to a violation caused by QCD radiative corrections as we will see in Sect.~\ref{s2:DGLAP}. Another success of the parton model is the \textit{Callan-Gross} relation:
\begin{equation} \label{eq:Callan-Gross}
    W_1(x,Q^2) = \frac{Q^2}{4x^2}W_2(x,Q^2), \qquad Q \gg m_p
\end{equation}
This relation originates from the fact that when $x=1$ (elastic regime), $\frac{Q^2}{2m_q^2}=\frac{Q^2}{2x^2m_p^2}$ and the LO Rosenbluth Eq.~(\ref{eq:Rosenbluth_Qinf}) is retrieved which describes scattering of a spin-$1/2$ particle. Therefore this relation, verified by the same data, indicated that partons are actually spin-$1/2$ particles. 
  \chapter{Hadronic structure}
\label{chap:QCD}
\vspace{-1cm}
\begin{center}
    \begin{minipage}{1.\textwidth}
        \begin{center}
            \textit{The derivations presented in this chapter are partially based on Refs.~\cite{peskin2018introduction,schwartz2014quantum,Collins:1989gx}.} 
        \end{center}
    \end{minipage}
\end{center}

\myparagraph{Outline} In this chapter, I summarise the main features of QCD relevant to my thesis. In particular, the factorisation theorem that allows us to decouple the non-perturbative hadronic structure from perturbative QCD, the higher order corrections and the definition of the following unpolarised longitudinal (along the beam axis) non-perturbative objects:
\begin{itemize}
    \item \textbf{Parton distribution functions} (PDFs)
    \item \textbf{Fragmentation functions} (FFs) 
    \item \textbf{Nuclear parton distribution functions} (nPDFs).
\end{itemize}

In Sect.~\ref{s1:QCD}, I review some key aspects of QCD such as its development, running of the coupling constant, factorisation, evolution equations and definition of PDFs. 
In Sect.~\ref{s1:SF_heavyquarks} I emphasise on the DIS structure functions and the treatment of heavy quark mass effects as well as a summary of the mass schemes including the one I use in the results of the subsequent chapters.
In Sect.~\ref{s1:Beyond_PDF}, I introduce FFs and nPDFs, their interpretation, main features and challenges. I then discuss the main physical constraints on all non-perturbative objects and their commonalities. 
Finally, in Sect.~\ref{s1:QCD_summary}, I outline all the relevant scattering processes used in the subsequent results and clarify the procedure I follow to include NNLO QCD corrections in the cross section calculations.

\section{Quantum chromodynamics} \label{s1:QCD}
In order to explain the confinement of quarks inside hadrons, the state of the art quantum field theory of strong interactions QCD was developed over many years. 
We can deduce some of its main features by summarising the main deductions and limitations highlighted in the previous sections:
\begin{enumerate}
    \item \textbf{Beyond QED structure of the proton:} Assuming the nucleon as a point-like particle is a good approximation only at low center-of-mass energies. While the QED photon-fermion vertex corrections describes well the leptons, they deviate for the nucleon that has an anomalous magnetic moment significantly different from that of a point-like particle.
    \item \textbf{Weakly interacting partons at large-energies:} The parton model and Bjorken scaling (although approximate) were very successful in describing the first measured DIS data, which suggested that a proton is made of almost free partons of spin-$1/2$ which is also reinforced by the Callan-Gross relation when compared to the data.
    \item \textbf{Quarks:} Hadrons were sorted into groups having similar properties and masses using the eightfold way. Gell-Mann and Zweig proposed the \textit{up}, \textit{down} and \textit{strange} quarks to explain the structure of the groups. Heavier hadrons were discovered requiring the existence of \textit{charm}, \textit{bottom} and \textit{top} quarks. The quarks were assigned fractional charges to explain the charges of the observed hadrons. 
    \item \textbf{Fermi-statistics, ``colour'' and SU(3):} The spin-$3/2$ baryon $\ket{\Delta^{++}}=\ket{u_\uparrow u_\uparrow u_\uparrow}$, among others, suggested that the quarks wavefunction is overall symmetric in spin and flavours, contradicting the Fermi-statistics and the result of the observed Callan-Gross relation. This was remedied by introducing \textit{colour}, a new quantum number, under which the quarks wavefunction's anti-symmetry is restored. Colour was associated with the group SU(3), thereby $\ket{\Delta^{++}}$ can be written as $\epsilon^{ijk}\ket{u_{i\uparrow} u_{j\uparrow} u_{k\uparrow}}$. The fact that colour span SU($N_C$) with the colour factor $N_C=3$ was further confirmed by the measured ratio $R=\sigma(e^+e^-\rightarrow q\bar{q})/\sigma(e^+e^-\rightarrow \mu^+\mu^-)$ which turned out to be proportional to $N_C=3$. 
    \item \textbf{Asymptotic freedom and non-abelian theories}: A theory with negative beta-function seemed the best candidate to explain, on one hand, the success of the parton model at short-distances (asymptotic freedom) and, on the other hand, the strong binding at long-distances together with the fact that quarks are not observed as isolated particles. It was shown that \textit{non-abelian} (or non-commutative) gauge theories are the only asymptotically free field theories in four dimensions. Such theories can be constructed as generalisations of QED and this was the final piece that lead to QCD as the theory of quarks bound together by interacting vector bosons called gluons.
\end{enumerate}

In Sect.~\ref{s2:QCD_Lagrangian}, I discuss the non-abelian SU(3) Lagrangian of QCD and the running of the strong coupling constant, in analogy with QED, as well as the asymptotic freedom.
Then, in Sect.~\ref{s2:DGLAP}, I examine the higher order QCD corrections at the cause of the Bjorken scaling violation and how the Dokshitzer, Gribov, Lipatov, Altarelli and Parisi (DGLAP) evolution equations~\cite{Altarelli:1977zs} allows their systematic resummation. Subsequently, I establish a definition of PDFs, this time in QCD and beyond the parton model. 

\subsection{QCD Lagrangian} \label{s2:QCD_Lagrangian}
The Yang-Mills theory based on fermions having SU(3) colour as an additional degree of freedom was proposed by Nambu~\cite{nambu1966preludes} then quantised by Faddeev and Popov~\cite{faddeev1967feynman}. Its renormalisation was proven by Hooft in 1971~\cite{hooft1971renormalization}. Finally, the QCD as the theory of strong interactions was proposed by Fritzsch, Gell-Mann and Leutwyler~\cite{fritzsch2002current,fritzsch1973advantages}. The SU(3) QCD Lagrangian generalises the QED one and can be expressed as follows:
\begin{equation} \label{eq:L_QCD}
    L_{QCD}=\sum_i \bar{\psi}_i(i\slashed{D}-m_i)\psi_i - \frac{1}{4}F^a_{\mu\nu}F^{a,\mu\nu} 
\end{equation} 
where $\psi_i$ denotes a quark field with flavour index $i$ and mass $m_i$. The tensor $F^a_{\mu\nu}$ is the gluon field strength with colour index $a \in [1,\dddot{},8]$ and $\slashed{D}$ is the covariant derivative given by:
\begin{align}
    \slashed{D} & =\gamma^\mu D_\mu = \gamma^\mu (\partial_\mu - i g_s A^a_\mu t^a)\\
    F_{\mu\nu}^a &= \partial_\mu A_\nu^a - \partial_\nu A_\mu^a + g_s f^{abc}A_\mu^b A_\nu^c
\end{align}
where $g_s$ is the QCD equivalent of the QED electric charge, $f^{abc}$ are the structure constants of QCD, $t^a=\frac{1}{2}\lambda^a$ are the generators of SU(3) analogous of the Pauli matrices in SU(2), with $\lambda^a$ being the Gell-Mann matrices.
The QCD Lagrangian is invariant under the following SU(3) gauge transformations written in terms of an infinitesimal shift $\theta$ as follows:
\begin{align} \label{eq:QCD_gauge_transformations}
\psi_i &\rightarrow \psi'_i = e^{it^a\theta^a}\psi_i \\
A^a_\mu &\rightarrow A'^a_\mu = A_\mu^a + \frac{1}{g_s}\partial_\mu \theta^a + f^{abc} A_\mu^b \theta^c
\end{align}
which leads to the conservation of the SU(3) colour charge.

As discussed in Sect.~\ref{s3:RGE} the beta-function in QCD is negative which implies asymptotic freedom. The running of the strong coupling constant $\alpha_s = \frac{g_s^2}{4\pi}$ is given by the renormalisation group equation (RGE) analogously to Eq.~(\ref{eq:RGE_alpha}, \ref{eq:RGE_e}):
\begin{align} 
\mu^2 \frac{d}{d\mu^2} \alpha_s &= \beta(\alpha_s) = -(b_0\alpha_s^2 + b_1 \alpha_s^3 + b_2 \alpha_s^4 + 
\dddot{})\\
b_0 &= (\overbrace{11C_A}^{\text{gluon loops}} - \overbrace{4n_f T_R}^{\text{quark loops}})/(12\pi) \qquad \text{one-loop beta function}
\end{align}
where $C_A=3$ and $T_R=1/2$ are the colour-factors associated with the gluon emission from a gluon and the gluon split to a $q\bar{q}$ pair respectively. The value of the $\alpha_s$ is commonly specified at a reference scale $Q^2 = M^2_Z$ from which one can obtain it at any other perturbative scale $Q^2$. Therefore the \textit{running} of the strong coupling in the leading logarithmic resummation at next-to-leading order (NLO) accuracy in the QCD perturbative expansion is given by:
\begin{equation}
    \alpha_s(Q^2) =  \frac{\alpha_s(M_Z^2)}{1+b_0\alpha_s(M_Z^2)\log{\frac{Q^2}{M_Z^2}}+\mathcal{O}(\alpha_s^2)}
\end{equation}
which has a similar structure as the QED coupling in Eq.~(\ref{eq:QED_alpha_eff}) except for the opposite sign in the coefficient of the logarithm.


\subsection{DGLAP evolution equations} \label{s2:DGLAP}
In this section, we go beyond the parton model, to explain the violation of the Bjorken scaling and the origin of the logarithmic $Q^2$ dependence in the structure functions. We will continue to assume that the parton model holds but we will elevate it with elements of perturbative QCD.
We start by defining the partonic version $\hat{x}$ of the Bjorken-$x$:
\begin{equation} \label{eq:partonic_Bjorken_x}
    x = \frac{Q^2}{2p\cdot q} \longrightarrow \hat{x} = \frac{Q^2}{2p_q \cdot q} \implies x \stackrel{\text{parton model}}{=} \xi \hat{x}
\end{equation}
Similarly to Eq.~(\ref{eq:partonic_xsec}), let us consider the partonic version of the hadronic tensor where in the parton model it reads:
\begin{align} \label{eq:hadronic_tensor_def0}
W^{\mu\nu}(x,Q) \stackrel{\text{parton model}}{=} \sum_i \int_0^1 d\xi f_{i}(\xi) \hat{W}^{\mu\nu}(\xi,Q)
\end{align}
with $W^{\mu\nu}$ is associated to $\gamma^*A$ scattering and $\hat{W}^{\mu\nu}$ is associated to $\gamma^*q$ scattering.
Using the definition of the variable $\hat{x}$ in Eq.~(\ref{eq:partonic_Bjorken_x}) we can write Eq.~(\ref{eq:hadronic_tensor_def0}) as:
\begin{align} \label{eq:partonic_hadronic_tensor}
    W^{\mu\nu}(x,Q) &= \sum_i \int_0^1 d\hat{x} \int_0^1 d\xi f_{i}(\xi) \hat{W}^{\mu\nu}(\hat{x},Q) \delta(x-\xi \hat{x}) \nonumber\\
                    &= \sum_i \int_x^1 \frac{d\xi}{\xi}f_{i}(\xi)\hat{W}^{\mu\nu}(\frac{x}{\xi},Q)
\end{align}
where, analogously to Eq.~(\ref{eq:Hadronic_Tensor}), we can write the partonic hadronic tensor as:
\begin{equation} \label{eq:PartonicHadronicTensor}
    \hat{W}^{\mu\nu} = \hat{W}_1(-g^{\mu\nu}+\frac{q^\mu q^\nu}{q^2})+\hat{W}_2(p^\mu-\frac{p \cdot q}{q^2}q^\mu)(p^\nu - \frac{p \cdot q}{q^2}q^\nu)
    \end{equation}
At LO (order $\mathcal{O}(\alpha^0_s)$), the gluons are absent and the only contribution to $\hat{W}^{\text{(LO)}}_{\mu\nu}$ is from $(\gamma^*q\rightarrow q)$ diagrams which implies:
\begin{align} \label{eq:PartonicHadronicTensor_LO}
\hat{W}^{\text{(LO)}}_1 &= 2\pi e_{q_i}^2 \delta(1-\hat{x}) \\
\hat{W}^{\text{(LO)}}_2 &= \frac{4z}{Q^2} \delta(1-\hat{x})
\end{align}
with $e_{q_i}$ the electric charge of the incoming quark of type $i$. Plugging these expressions back in Eq.~(\ref{eq:partonic_hadronic_tensor}) we get back the parton model expression Eq.~(\ref{eq:xsec_DIS_parton_model_v1}).
Let us now focus solely on the unpolarised cross section for $\gamma^* A \rightarrow X$ given by the following projection of the hadronic tensor:\vspace{-5pt}
\begin{align}
W_0(x,Q) \equiv -g^{\mu\nu}W_{\mu\nu} \\[-30pt]\nonumber
\end{align}
where at order $\mathcal{O}(\alpha^0_s)$ and for $Q\gg m_p$:\vspace{-5pt}
\begin{align}
    \hat{W}^{\text{(LO)}}_0(x,Q) &\stackrel{\text{Eq.~(\ref{eq:PartonicHadronicTensor})}}{=}  4\pi e_{q_i}^2 \delta(1-\hat{x})\\
    W^{\text{(LO)}}_0(x,Q) &\stackrel{\text{Eq.~(\ref{eq:Hadronic_Tensor})}}{=} 4\pi \sum_i e_{q_i}^2 f_{i}(x) \label{eq:LO_W0} \\[-30pt]\nonumber
\end{align}
At NLO (order $\mathcal{O}(\alpha_s)$), summing all the leading logarithmic (LL) contributions from the virtual (V) vertex-corrected $(\gamma^*q\rightarrow q)$ and the real (R) gluon emission $(\gamma^*q\rightarrow qg)$ diagrams leads to the following:
\begin{align}
    \hat{W}^{\text{(NLO)}}_0(x,Q) &\stackrel{\text{(LL)}}{=} \hat{W}^{\text{(LO)}}_0 + \hat{W}^V_0 + \hat{W}_0^R\\
    W^{\text{(NLO)}}_0(x,Q) - W^{\text{(NLO)}}_0(x,Q_0) &\stackrel{\text{(LL)}}{=} 4\pi\sum_i e_{q_i}^2 \int^1_x \frac{d\xi}{\xi}f_{i}(\xi)\left[\frac{\alpha_s}{2\pi}P^{\text{(LO)}}_{qq}\left(\frac{x}{\xi}\right)\log{\frac{Q^2}{Q_0^2}}\right] \label{eq:NLO_W0}
\end{align}
where $Q_0$ is an arbitrary different energy scale from $Q$, $P^{\text{(LO)}}_{qq}(x)$ is the leading order \textit{DGLAP} splitting function for the quark-to-quark transition and $\frac{f(x)}{[1-x]_+}$ is the plus distribution, both of which are defined as:
\begin{align}
P^{\text{(LO)}}_{qq}(x) &= \frac{4}{3}\left(\frac{1+x^2}{[1-x]_+} + \frac{3}{2}\delta(1-x)\right) \\
\text{with: }\int_0^1 dx \frac{f(x)}{[1-x]_+} &\equiv \int_0^1 dx \frac{f(x)-f(1)}{1-x} \implies \int_0^1 P^{\text{(LO)}}_{qq}(x)dx = 0 \label{eq:integral_of_splittingF}
\end{align}
While the total DIS cross section at a given $Q$ integrated over $x$ is finite (mainly due to Eq.~(\ref{eq:integral_of_splittingF})), $W_0(x,Q)$ at a fixed $x$ is infrared divergent thus the need to take differences of cross sections to find the finite expression Eq.~(\ref{eq:NLO_W0}). The $\log$ term in the latter explains the violation of the Bjorken scaling. 

As it stands, Eq.~(\ref{eq:NLO_W0}) contains a singularity in the $\log$ term for $Q_0 \rightarrow 0$. This issue can be resolved by interpreting that the singularity is due to the breakdown of the perturbation theory, which is invalid at low energy scales where $\alpha_s$ is large. 
We therefore \textit{factorise} out the long distance behaviour of the structure function by replacing the bare quantities $f(x)$ with a physically accessible quantity measured at a factorisation scale $\mu_F$.
%
%
%
Although the divergent $\log$ terms has to be ``absorbed'' by the bare PDFs, choosing to include a regular finite part remains arbitrary and this choice defines a \textit{factorisation scheme}. In this thesis, we will be consistently using the modified minimal subtraction scheme ($\overline{\text{MS}}$) in which this regular term is chosen to be $\Delta_r = (\log4\pi -\gamma_E)$ with $\gamma_E$ being the Euler-Mascheroni constant. 
In this scheme, the renormalised PDF can be expressed in terms of an expansion in the bare PDFs as follows:
\begin{align}
f_{i}(x,\mu_F)&\stackrel{\overline{\text{MS}}}{=}f_{i}(x) +\frac{\alpha_s}{2\pi}\int^1_x \frac{d\xi}{\xi}f_{i}(\xi)\left[P^{\text{(LO)}}_{qq}\left(\frac{x}{\xi}\right)\log{\frac{\mu_F^2}{Q_0^2} + \Delta_r} \right] +\mathcal{O}(\alpha_s^2)
\end{align}
where $\mu_F$ is an arbitrary scale called factorisation scale similar in role to the renormalisation scale $\mu_R$. Subsequently, for any other perturbative scale $Q$ we can write:
\begin{align}
    f_{i}(x,Q)&\stackrel{\overline{\text{MS}}}{=}f_{i}(x,\mu_F) +\frac{\alpha_s}{2\pi}\int^1_x \frac{d\xi}{\xi}f_{i}(\xi,\mu_F)\left[P^{\text{(LO)}}_{qq}\left(\frac{x}{\xi}\right)\log{\frac{Q^2}{\mu_F^2} + \Delta_r} \right] +\mathcal{O}(\alpha_s^2)\label{eq:PDF_definition} 
\end{align}
Substituting the bare PDFs in Eq.~(\ref{eq:LO_W0}, \ref{eq:NLO_W0}) by the renormalised PDF in Eq.~(\ref{eq:PDF_definition}) leads to the following general expression:
\begin{align} 
W_0(x,Q) &\equiv 4\pi \sum_i e_{q_i}^2 f_{i}(x,Q)\label{eq:renormalised_partonic_hadronic_tensor}
\end{align}
where $Q$ is the physically accessible scale. As an observable, Eq.~(\ref{eq:renormalised_partonic_hadronic_tensor}) must be independent of the factorisation scheme and scale choices. The factorisation scale independence leads to the following RGE: 
\begin{equation}
    \mu_F \frac{d}{d\mu_F} W_0(x,Q^2)=0
\end{equation}
and consequently it leads to the PDFs DGLAP evolution equation expressed as follows:
\begin{equation} \label{eq:NS_DGLAP}
\mu_F \frac{d}{d\mu_F} f_{i}(x,\mu_F) = \frac{\alpha_s}{\pi} \int_x^1 \frac{d\xi}{\xi} f_{i}(\xi,\mu_F)P_{qq}\left(\frac{x}{\xi}\right) +\mathcal{O}(\alpha_s^2)
\end{equation}
where similarly to the RGE Eq.~(\ref{eq:RGE_alpha}), which resums contributions arising
from self-energy diagrams into the running of the strong coupling; the DGLAP equation Eq.~(\ref{eq:DGLAP}) resums scale logarithms arising
from collinear parton splittings into the structure functions.

At NLO, additionally to the contributions from the virtual $(\gamma^*q\rightarrow q)$ and the real gluon emission $(\gamma^*q\rightarrow qg)$ diagrams already considered, we have $(\gamma^*g\rightarrow q\bar{q})$ and $(\gamma^*q\rightarrow qg)$ which explains the reason why there's a probability of finding gluons and antiquarks in the proton. All of these PDFs mix under DGLAP evolution which is generally a coupled integro-differential equations of the form:
\begin{equation} \label{eq:DGLAP}
\mu_F \frac{d}{d\mu_F} \begin{pmatrix} f_{i}(x,\mu_F) \\ f_g(x,\mu_F)\end{pmatrix} = \sum_j \frac{\alpha_s}{\pi} \int_x^1 \frac{d\xi}{\xi} 
\begin{pmatrix} P_{q_i q_j} & P_{q_i g} \\ P_{gq_j} & P_{gg}\end{pmatrix}\begin{pmatrix} f_j(\xi,\mu_F) \\ f_g(\xi,\mu_F)\end{pmatrix}
\end{equation}
The splitting functions were known for some time at LO and NLO accuracy~\cite{Gross:1973ju,Georgi:1951sr,Floratos:1977au,Altarelli:1977zs,GonzalezArroyo:1979df,Floratos:1978ny,Furmanski:1980cm,Curci:1980uw,GonzalezArroyo:1979he,Floratos:1981hs,Hamberg:1991qt}, and more recently extended to NNLO accuracy ~\cite{Moch:2004pa,Vogt:2004mw}.
In order to solve Eq.~(\ref{eq:DGLAP}), we diagonalise as much as possible the evolution matrix, thus split the system of equations into two: the singlet and non-singlet sectors by introducing the \textit{evolution basis}:
\begin{alignat}{3} \label{eq:Evolution_basis} 
\Sigma &= \sum_i^{n_f}f_{i}^+ \text{ (Singlet)} \qquad && V &&= \sum_i^{n_f}f_{i}^- \text{ (Valence)}\\
T_{3} &= u^+ - d^+ \qquad && V_3 &&= u^- - d^-\nonumber \\
T_{8} &= u^+ + d^+ -2s^+ \qquad && V_8 &&= u^- + d^- - 2s^- \nonumber\\
T_{15} &= u^+ + d^+ + s^+ - 3c^+ \qquad && V_{15} &&= u^- + d^- + s^- - 3c^- \nonumber\\
T_{24} &= u^+ + d^+ + s^+ + c^+ - 4b^+ \qquad && V_{24} &&= u^- + d^- + s^- + c^- - 4b^- \nonumber\\
T_{35} &= u^+ + d^+ + s^+ + c^+ + b^+ - 5t^+ \qquad && V_{35} &&= u^- + d^- + s^- + c^- + b^- -5t^- \nonumber
\end{alignat}
where $f^+ = f_{i} + \bar{f}_{i}$ and $f^- = f_{i} - \bar{f}_{i}$. The non-singlet sector evolve according to Eq.~(\ref{eq:NS_DGLAP}) and is composed by the $T_i$, $V_i$ and $V$ as $f_{i}$. The coupled singlet sector evolve according to Eq.~(\ref{eq:DGLAP}) and is composed by the Singlet $\Sigma$ as $f_{i}$ and the gluon as $f_g$.

\section{Deep-inelastic scattering structure functions} \label{s1:SF_heavyquarks}
In the previous section, we derived the general renormalised expression of the $\gamma^*A \rightarrow X$ cross section $W_0=-g^{\mu\nu}W_{\mu\nu}$ in Eq.~(\ref{eq:renormalised_partonic_hadronic_tensor}) based on the definition of renormalised PDFs in the $\overline{\text{MS}}$ scheme.
I aim in Sect.~\ref{s2:factorisation}, to derive the other projections of the hadronic tensor $W^{\mu\nu}$, namely the dimensionless structure functions $F_1$ and $F_2$ in order to derive a factorised expression of the DIS cross section. 
In Sect.~\ref{s2:HQ}, I discuss the treatment of the heavy quark mass effects and the scheme I use consistently in the results of the subsequent chapters.

\subsection{Factorisation} \label{s2:factorisation}
We start by rewriting Eq.~(\ref{eq:renormalised_partonic_hadronic_tensor}) in its explicit form in $\overline{\text{MS}}$:\vspace{-0.1cm}
\begin{align}
    \frac{1}{4\pi} W_0(x,Q) &= \sum_i e_{q_i}^2 f_{i}(x,\mu_F) \nonumber\\[-0.8cm] 
    & + \left(\frac{\alpha_s}{2\pi}\right)\sum_{i}^{q\bar{q}}\sum_j^{q\bar{q}g} e_{q_i}^2 \int^1_x \frac{d\xi}{\xi}f_{j}(\xi,\mu_F)\overbrace{\left[P^{\text{(LO)}}_{ij}\left(\frac{x}{\xi}\right)\log{\frac{Q^2}{\mu_F^2}} + \Delta_r\right]}^{C_{0,ij}^{(1)}} \nonumber\\[-0.1cm]
    &+\mathcal{O}(\alpha_s^2) \nonumber\\[-0.1cm]
    &= \sum_j^{q\bar{q}g} \int_x^1 \frac{d\xi}{\xi} f_{j}(\xi,\mu_F) H_{0j}\left(\frac{x}{\xi},\frac{Q}{\mu_F},\alpha_s(\mu_R)\right) + \text{HT} \label{eq:factorisation_hadronic_tensor}
\end{align}
where $H$ is the hard scattering coefficient called \textit{coefficient functions} depending only on the parton type $j$ and expressed as a power series in $\alpha_s$ as follows:
\begin{equation}
H^{(n)}_{a,j} = \sum_n\frac{\alpha_s^n(\mu_R^2)}{(2\pi)^n} \sum_{i}^{q\bar{q}} e_{q_i}^2C_{a,ij}^{(n)}\left(\frac{x}{\xi},\frac{Q}{\mu_F}\right)    
\end{equation}
These coefficient functions $H^{(n)}$ evolve in $\mu_F$, similarly to PDFs but with a minus sign, by means of the DGLAP Eq.~(\ref{eq:DGLAP}). The ability to calculate $H$ is a great predictive feature of the factorisation theorem. 

In Eq.~(\ref{eq:factorisation_hadronic_tensor}), we distinguished between the renormalisation scale $\mu_R$ that addresses the ultraviolet (UV) divergences caused by the large momentum in the loops of the Feynman diagrams representing the amplitude; and the factorisation scale $\mu_F$ that addresses the infrared (IR) divergences caused by the long-range dynamics. This distinction will be useful when discussing the theory uncertainties in Sect.~\ref{s1:PDF_theory_uncertainties}.
We also introduced for the first time the \textit{Higher twist} effects (HT). 
So far we have only worked in the limit of $Q\gg m$ ($m$ being the mass of the hadron) and have not discussed contributions which are suppressed by powers of $1/Q^2$. Included in this category are:
\begin{itemize}
\item Kinematic corrections associated with the
mass $m$ of an initial state hadron, which are of order $m^2/Q^2$. In the case of a target nucleon, these target mass corrections (TMC) may be significant even when $Q^2$ is large enough so that perturbation theory in $\alpha_s(Q^2)$ is reliable. 
\item The other neglected effects that we refer to as HT involves the effect of the partons in the target being off-shell, coherence effects, etc. that are typically safer to neglect while keeping the TMC~\cite{Sheiman:1979ku, Ellis:1978ty, Schienbein:2007gr}.
\end{itemize}
TMC~\cite{Moffat:2019qll} actually compete with
higher-order QCD corrections. That is due to the fact that $\alpha_s(Q^2)/\pi \approx m^2/Q^2$ at $Q^2=6.5$ GeV$^2$, which makes TMC relevant in any attempt to include higher-order corrections at a moderate value of $Q^2$. For our purposes, we note that TMC is properly accounted for with the \texttt{APFEL} software~\cite{Bertone:2013vaa}, which we consistently use to compute the observables in this thesis. As for the other type of HT effects, we make sure to use proper kinematic cuts on the data in all of our global analyses to exclude the regions of phase-space where they become relevant. 

We can similarly write the dimensionless structure functions as follows (see Ref.~\cite{Collins:1989gx}):
\begin{align}
     F_1(x,Q) & = \frac{1}{4\pi} W_1(x) = \sum_i \int_x^1 \frac{d\xi}{\xi} f_{i}(\xi,\mu_F) H_{1i}\left(\frac{x}{\xi},\frac{Q}{\mu_F},\alpha_s(\mu_R)\right) + \text{HT}\\
    F_2(x,Q) &= \frac{Q^2}{8\pi x} W_2(x,Q) = x \sum_i \int_x^1 \frac{d\xi}{\xi} f_{i}(\xi,\mu_F) \frac{\xi}{x}H_{2i}\left(\frac{x}{\xi},\frac{Q}{\mu_F},\alpha_s(\mu_R)\right) + \text{HT} \label{eq:F2_collins}
\end{align}
and finally, without any loss of generality, we consider from now on the generic structure function $F(x,Q) \equiv \frac{1}{4\pi} W_0(x,Q)$ from which $F_{1,2}$ can be easily deduced, in this compact form:
\begin{align}
    F(x,Q) &=\sum_i f_{i} \otimes H^{\text{DIS}}_{i} \label{eq:compact_structure_function}
\end{align}
where we only considered the leading-twist contribution and where $F(x,Q)$ denotes a structure function associated with a hard scattering coefficient $H$ and a PDF $f$ coupled by the convolution as follows:
\begin{equation}
    f \otimes h = \int_x^1 \frac{d\xi}{\xi} f(\xi) h(\frac{x}{\xi}) \label{eq:convolution}
\end{equation}

\subsection{Treatment of heavy quarks} \label{s2:HQ}
This section is a summary based on a discussion in Ref.~\cite{Hartland:2014nha}.
So far, we derived DIS in the parton model improved by perturbative QCD and DGLAP evolution but while assuming that quarks are massless. An approximation that does not hold specially when $Q^2$ approaches the quarks' physical masses. 
Including the heavy quark mass effects in the calculation of the structure functions could be approached in different ways each with different regions of suitability. These are called \textit{heavy quark schemes}. The simplest of these are the fixed flavour number scheme (FFN) and the zero-mass variable flavour number scheme (ZM-VFN). Then comes the general-mass variable flavour number schemes (GM-VFN) that interpolates between the previous two and finally the FONLL approach that is implemented in \texttt{APFEL}~\cite{Bertone:2013vaa} and will be mainly adopted for the calculations in this thesis.

Let us start by considering a hard scale $Q^2 \lsim m_h^2$ of the order or smaller than some heavy quark mass $m_h$. Assuming that there's no intrinsic heavy quark component in the initial state hadron, the only remaining partons in the theory are $n_l$ light quark flavours and the gluon. We set the factorisation and renormalisation scales to be equal to a certain scale $\mu$ and rewrite Eq.~(\ref{eq:compact_structure_function}) as:
\begin{equation}
    F(n_l,Q,m_h) = \sum_i^{n_l} f_i(n_l,\mu) \otimes H_{i}^{DIS}(n_l,\frac{Q}{m_h},\frac{\mu}{m_h},\frac{Q}{\mu})
\end{equation}
where the sum is over light quark flavours only and where we omitted the $x$-dependence for clarity. We can separate the structure function into a light and heavy part where in the former only light flavours are present and in the latter only the heavy quark contributes such as:
\begin{equation}
F(n_l,Q,m_h) = F^L(n_l,Q) + F^H(n_l,Q,m_h) \qquad \text{FFN}
\end{equation}
where $F^L$ is derived from diagrams that do not contain heavy quarks, and $F^H$ from those that do through splitting of an initial state gluon into the heavy quarks at $\mathcal{O}(\alpha_s)$. This approach is known as the FFN scheme where the only quarks treated as partons are the $n_l$ light quarks. Although accurate in the quark mass threshold region and below, this scheme suffers from unresummed logarithms of the ratio $Q^2/m_h^2$ in the coefficient functions at larger scales.
This problem may be resolved by treating the heavy quark as a massless parton above its mass threshold with the introduction of an associated heavy quark PDF. That allows the resummation of the logarithmic contributions due to DGLAP. As it is identical to the FFN scheme but with an additional partonic flavour, this scheme is called the Zero-Mass Variable Flavour Number scheme or ZM-VFN where simply the structure functions read:
\begin{equation} \label{eq:ZM-VFN}
    F(n_l+1,Q) = \sum_i^{n_l+1} f_i(n_l+1,\mu) \otimes H_{i}^{DIS}(n_l+1,\frac{Q}{\mu}) \qquad \text{ZM-VFN}
\end{equation}
In this case, the heavy quark PDFs are set to zero below mass threshold and evolved as a massless parton according to DGLAP for scales greater than the heavy quark mass. Although this method solves the issue of logarithmic resummation at large scales, it still considers a heavy quark massless and suffers from inaccuracies in the region where $Q^2 \sim m_h^2$.

In order to address heavy quark effects at both near the quarks thresholds (where FFN is suitable) and large scale (where ZM-VFN is suitable), a general mass scheme has been developed. Generally this scheme reduces to the FFN regime at low scales and the ZM-VFN at high scales with the intermediate regime as some interpolation between the two. We start by requiring that the FFN scheme (heavy quarks only generated through gluon splitting) matches the ZM-VFN scheme (heavy quarks as massless but contributing to the cross section) at very large-scales $Q^2 \gg m_h^2$:
\begin{equation} 
F^L(n_l,Q) + \lim_{Q^2 \gg m_h^2} F^H(n_l,Q,m_h) = F(n_l+1,Q)
\end{equation}
where the dependence of the $F^H$ on the heavy quark mass can be neglected in that limit. This allows the PDFs in the two schemes to be related by a perturbative transformation as follows:
\begin{equation} \label{eq:GM_transformation}
    f_i(n_l+1,\mu) = \sum_j^{n_l} A_{ij}(n_l,\frac{\mu}{m_h}) \otimes f_j(n_l,\mu)
\end{equation}
where $A_{ij}$ is available up to NNLO in $\alpha_s$~\cite{Buza:1995ie,Buza:1996wv}.
We then impose continuity of the physical observables across the quark mass thresholds simply by setting:
\begin{align}
    F^{\text{GM}}(m_h) &= \sum_i^{n_l}H_{i}^{\text{GM}}(n_l,m_h)\otimes f_i(n_l)\\ 
    &= \sum_i^{n_l+1} H_{i}^{\text{GM}} (n_l+1,m_h)\otimes  f_i(n_l+1)\\
    &\stackrel{\text{Eq.~(\ref{eq:GM_transformation})}}{=} \sum_i^{n_l+1}\sum_j^{n_l} H_{i}^{\text{GM}}(n_l+1,m_h) \otimes A_{ij}(n_l,m_h) \otimes f_k(n_l)
\end{align}
which is called the \textit{matching condition}, where this last relation provides a basic description of the GM-VFN scheme:
\begin{equation}
    H_{j}^{\text{GM}}(n_l,m_h) = \sum_i^{n_l+1} H_{i}^{\text{GM}}(n_l+1,m_h) \otimes A_{ji}(n_l,m_h) \qquad \text{GM-VFN}
\end{equation}
We finally note that GM-VFN suffers from a degeneracy, as terms proportional to powers $m_h/Q$ may be interchanged between the coefficient functions of different flavours without changing the final value of the structure function therefore introducing a choice-dependence in the definition of the PDFs. 
However, the differences between these choices are always of higher order compared to the calculation at hand. Therefore the GM-VFN schemes still provides a considerable improvement over the simpler schemes when dealing with data sets spanning quark mass thresholds and large scales.

The last approach we introduce is called the FONLL (fixed-order calculations with next-to-leading log resummation)~\cite{Cacciari:1998it,Forte:2010ta}. The procedure relies on inverting the GM-VFN relation Eq.~(\ref{eq:GM_transformation}) and match it with the ZM-VFN relation Eq.~(\ref{eq:ZM-VFN}). The sum in the latter runs only over the $n_l$ light flavours in that case and the heavy flavour contribution (generated via DGLAP) is included in the modified hard part $\tilde{H}_{i}$ as follows:
\begin{align}
F(n_l,Q) &\stackrel{\text{Eq.~(\ref{eq:GM_transformation})}}{=} \sum_i^{n_l} B_{i}(n_l,\frac{Q}{m_h}) \otimes f_i(n_l+1,Q) \quad \text{GM-VFN} \label{eq:massive_calculation}\\
F(n_l+1,Q) &\stackrel{\text{Eq.~(\ref{eq:ZM-VFN})}}{=} \sum_i^{n_l} \tilde{H}_{i}^{DIS}(n_l+1,\frac{Q}{m_h}) \otimes f_i(n_l+1,Q)\quad \text{ZM-VFN} \label{eq:massless_calculation}
\end{align}
We note that $B_{i}$ contains a logarithmic dependence on $Q^2/m_h^2$ inherited from the FFN and DGLAP evolution and other terms suppressed by powers of $m_h/Q$ that we can be separated as such:
\begin{equation}
B_{i}(\frac{Q}{m_h})= \bar{B}_{i}(\log{\frac{Q^2}{m_h^2}})+\mathcal{O}(\frac{m_h}{Q})
\end{equation}
where in the limit $Q\gg m_h$ GM-VFN and ZM-VFN should match. Therefore, the massive structure in FONLL can expressed as the sum of the massive calculation in Eq.~(\ref{eq:massive_calculation}), and the massless calculation in Eq.~(\ref{eq:massless_calculation}) while matching the ZM-VFN in the asymptotic limit $Q\gg m_h$ by subtracting the duplicate DGLAP-originated terms as such:
\begin{equation}
F^{\text{FONLL}} = F(n_l,Q)+F(n_l+1,Q)-\bar{F}(n_l,Q)
\end{equation}
This procedure ensures the inclusion of the mass-suppressed terms present in the FFN calculation while also matching the ZM-VFN scheme at large scale.
We refer the reader to Refs.~\cite{Thorne:1997uu,Thorne:1997ga, Thorne:2008xf,Kramer:2000hn,Collins:1978wz,Collins:1998rz} for more details on the subject.


\section{Beyond parton distribution functions} \label{s1:Beyond_PDF}
I introduce in this section both fragmentation functions (FFs) that encode the production of hadrons in the final state and the nuclear PDFs (nPDFs) that encode the modified distribution of partons within nuclei due to nuclear effects.
In Sect.~\ref{s2:QCD_fragmentation}, I discuss the factorisation of FFs that describe the fragmentation of a parton into quarks and gluons until its hadronisation into a bound state hadron restoring its energetic stability and colour neutrality.
In Sect.~\ref{s2:Nuclear_modification}, I consider the nuclear effects on the distribution of quarks and gluons inside a nuclei. An effect that is yet to be fully understood theoretically but so far successfully incorporated into modified PDFs. These are also called nPDFs and are used similarly to PDFs, \textit{i.e.} conserving the whole theoretical framework (factorisation, DGLAP, heavy quark effects, etc.) but with minimal assumption related to the composition of nuclei in terms of nucleons. 
In Sect.~\ref{s2:Physical_constraints}, I summarise the main physical constraint on the FFs, PDFs and nPDFs like charge conjugation, isospin symmetry and sum rules. 

\subsection{Collinear fragmentation} \label{s2:QCD_fragmentation}
Due to the nature of the strong force and the increasing strong coupling at low-energies, quarks and gluons never appear as asymptotic states. It is rather energetically more favourable for partons to \textit{fragment} into \textit{jets} of quark-antiquark pairs until they end up \textit{hadronising}, \textit{i.e} transform into hadrons. \textit{Hadronisation} is therefore an intrinsically non-perturbative process~\cite{Webber:1999ui}, a mechanism by which quarks and gluons produced in hard processes transform into the hadrons that are observed in the final state. To date we have many theoretical models describing hadronisation, in this thesis however, we will be only concerned with \textit{fragmentation} referring to the inclusive production of hadron spectra without any assumptions concerning the underlying mechanism of hadron formation.

In inclusive deep-inelastic scattering, we only need to measure the incoming and outgoing lepton momenta to calculate the cross section, without the need to know anything about the final hadronic state X. However, whenever specific particles are identified in the final state, parton FFs appear as non-perturbative ingredient of QCD factorisation formulas~\cite{Metz:2016swz}. FFs are considered the counterpart of PDFs. While the latter are denoted by $f_{i}(x)$ and understood as probability densities for a struck parton of type $i$ to acquire a given momentum fraction $x$, inside a colour-neutral hadron, FFs are denoted by $D_{i}^h(z)$ and understood as probability densities for an unpolarised parton of type $i$ to fragment into an unpolarised hadron of type $h$, where the hadron carries the fraction $z$ of the parton momentum. In this thesis, we only consider the collinear longitudinal momentum of the hadron, that is, the component of the momentum along the direction of motion of the parton and therefore the transverse momentum of the parton integrated over thus the name \textit{collinear} FFs.

The main processes that are considered in the FFs determination discussed in Chapter~\ref{chap:FF} will be further summarised in Sect.~\ref{s2:Flavour_separation} and those include:
\begin{itemize}
    \item Single-inclusive hadron production in electron-positron annihilation (SIA) \vspace{-10pt}
    \begin{equation} e^+ + e^- \rightarrow h + X \nonumber \\[-40pt] \nonumber\end{equation}
    \item Semi-inclusive deep-inelastic lepton-nucleon scattering (SIDIS) \vspace{-10pt}
    \begin{equation} \ell + N \rightarrow \ell + h + X \nonumber \\[-40pt] \nonumber\end{equation}
\end{itemize}
The main result of the previous sections was to be able to define PDFs and how they enter the factorised hadronic tensor Eq.~(\ref{eq:factorisation_hadronic_tensor}) that encodes the process $\gamma^* A \rightarrow X$. The factorisation theorem in case of SIA and SIDIS is quite analogous~\cite{Collins:1989gx}. We start by considering SIA in the following and postpone the discussion of SIDIS to Sect.~\ref{s2:Flavour_separation} and \ref{s2:MAPFF10_theory}. The relevant tensor for SIA and specifically the process $\gamma^* \rightarrow h+X$ is: \vspace{-5pt}
\begin{equation}
    K^{\mu\nu}(x,Q) = \frac{1}{4\pi}\int d^4y e^{iq\cdot y} \sum_X \braket{0|j^\mu(y)|hX}\braket{hX|j^\nu(0)|0} \\[-40pt] \nonumber
\end{equation}
where $q^\mu$ is the photon's time-like\footnote{Time-like is associated in this context with SIA being an s-channel type diagram, while SIDIS for instance a space-like t-channel diagram.} momentum and $Q^2 = q^2$. We note that to make the transition from SIA to SIDIS one replaces the time-like photon by a space-like one, and the outgoing antiquark by a quark in the initial state. Defining the scaling variable $z=\frac{2p\cdot q}{Q^2}$ with $p^\mu$ being the momentum of $h$ and considering only the limit $Q \ll M_Z$, we can write the SIA cross section as:
\begin{align}
\frac{d\sigma^h}{dz}(z,Q) &= \frac{4\pi \alpha(Q)}{Q^2}F_2^h(z,Q) \nonumber \\
&= \frac{4\pi \alpha(Q)}{Q^2}\left(\sum_i D^h_i \otimes H^{\text{SIA}}_i + \text{HT}\right)
\end{align}
where $F_2^h$ is the fragmentation structure function defined in analogy with DIS and
where $H^{\mu\nu}_i$ is a perturbatively calculable hard function, $D^h_i$ the fragmentation function associated with a quark of type $i$.
The evolution equations for collinear FFs is identical in form to those of PDFs:
\begin{equation}
\frac{d}{d\log \mu^2} D^h_i(z,\mu^2) = \frac{\alpha_s(\mu^2)}{2\pi}\sum_j \int_z^1 \frac{d\zeta}{\zeta} P_{ji}(\zeta,\alpha_s(\mu^2))D_i^h(\frac{z}{\zeta},\mu^2)
\end{equation}
however, the matrix for the time-like splitting functions is $P_{ji}$ as opposed to $P_{ij}$ in case of PDFs. It is only at LO that the time-like splitting functions $P^{(0)}_{ji}$ agree with the LO space-like DGLAP splitting functions, as in the transpose of $P^{(0)}_{ij}$. We finally note that we will rely on the \textsc{APFEL++} software~\cite{Bertone:2013vaa, Bertone:2017gds} in this thesis to compute the SIA and SIDIS observables.

\subsection{Nuclear modification}\label{s2:Nuclear_modification}
Partons in nuclei have significantly different momentum distributions from those measured in free nucleons such as a proton. This was first discovered by the European Muon Collaboration (EMC) where in the original experiment~\cite{Aubert:1983xm} 120-280 GeV muons were scattered from an iron target and then compared to deuterium data.
The extracted structure function ratios (iron over deuterium) as a function of the per-nucleon Bjorken-$x$\footnote{ Within the leading twist approximation of QCD, the per-nucleon Bjorken-$x$ variable in DIS off nuclei is defined as $x=AQ^2/2(q\cdot p_A)$.} of the cross section ratio differed significantly from unity. Hence such a difference is referred to as the EMC effect~\cite{Malace:2014uea}. Even at higher energies where the nucleon is expected to be composed of weakly interacting quarks and gluons, effects of the nucleon being in the nucleus were still playing a role. This was at the least surprising as MeV-scale nuclear binding effects were expected to be negligible compared to the typical momentum transfers ($Q\gsim 1$ GeV) in hard-scattering reactions.

Many DIS off nuclei experiments have been performed after the EMC~\cite{Frankfurt:2012qs} and the first drawn conclusions were that the shape of the effect was relatively $Q^2$ independent, and that the nuclear effects increased logarithmically with the atomic mass A. 
In 2009, new high precision measurements of the EMC effect on light nuclei~\cite{PhysRevLett.103.202301} disagreed significantly with the logarithmic A dependence hypothesis (see also Sect.~\ref{s1:nNNPDF20_results}) putting once more the underlying cause of the EMC to question until today.

In perturbative QCD, as we learned from Sect.~\ref{s2:DGLAP}, the evolution of the parton distributions within a free nucleon is dictated by splitting functions that resum the contributions of quarks or antiquarks radiating into gluons and gluons converting into quark-antiquark pairs. However, this picture is ambiguous in the case of nuclei where in principle quarks and gluons can also \textit{fuse} together~\cite{Close:1989ca} to form a bound state different than nucleons. More clustering possibilities can in theory happen in a nucleus but these are currently inaccessible or rely heavily on models that are still poorly constrained by data.

\begin{figure}[!h]
    \begin{center}
    \makebox{\includegraphics[width=0.9\textwidth]{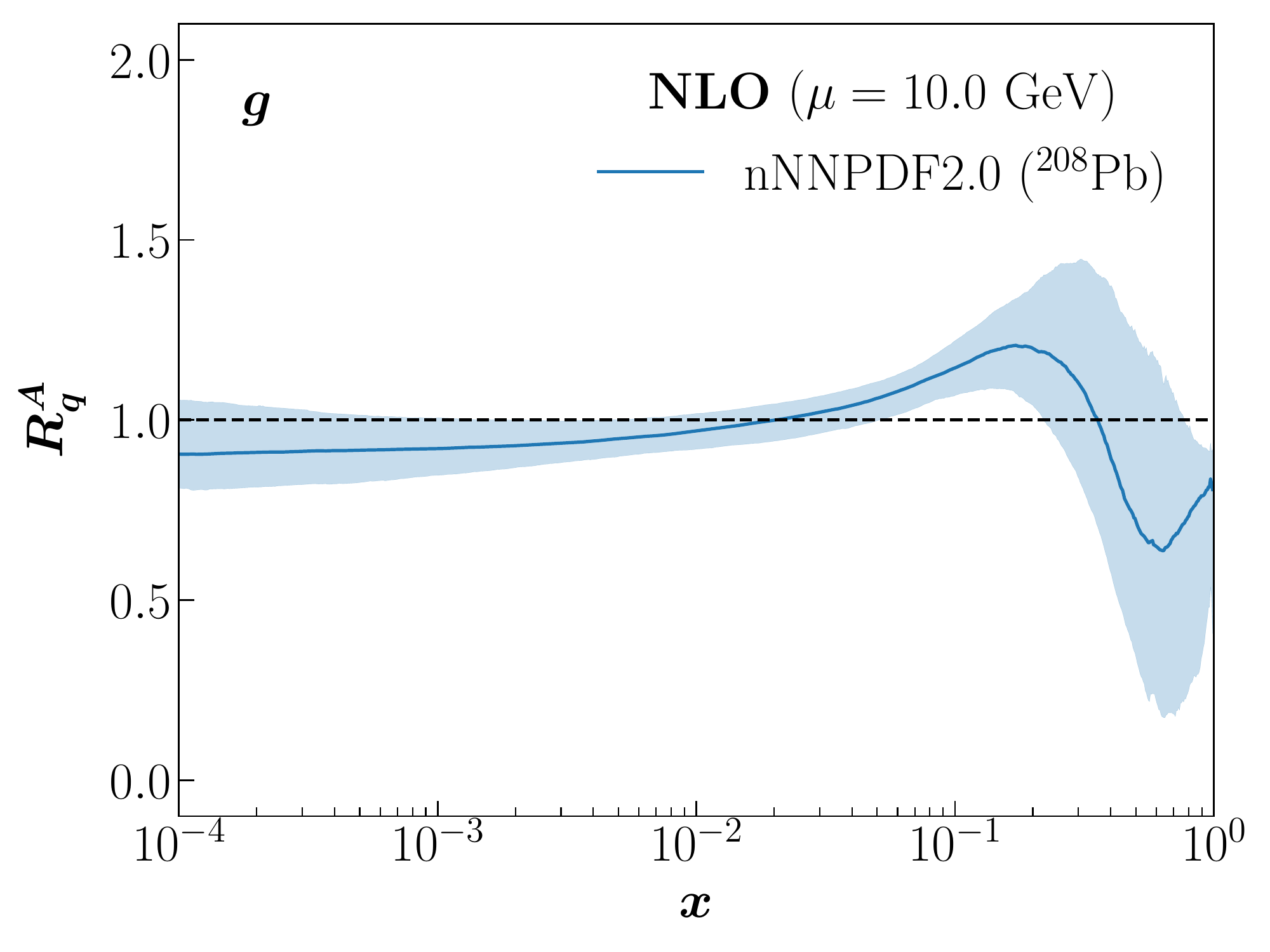}}
    \end{center}
    \caption{The nuclear modification ratio as defined in Eq.~(\ref{eq:nuclear_modification_ratio}) and its 1-$\sigma$ uncertainty band for the gluon of Lead ($A=208$) at energy scale $\mu=10$ GeV from the nNNPDF2.0~\mycite{AbdulKhalek:2020yuc} global analysis that will be discussed in Chapter~\ref{chap:nNNPDF20}.}
    \label{fig:nNNPDF20NuclearModification_Pb208}
    \end{figure}

Therefore, in this thesis we still adopt the most commonly accepted (and so far successful) assumption that the nucleons are the only degrees of freedom of the nuclear wave function. The nucleus structure function is therefore the additive sum of the constituent bound nucleons distributions. This assumption implies the following simplification. Let us consider a nucleus moving with a large momentum $p_A$ and a parton which carries a fraction $x_A$ of the nucleus' momentum such that $p_q = x_A p_A$. Each nucleon carries an average fraction $p_N = p_A/A$ of the nucleus' momentum up to very small Fermi motion corrections. This leads to the following:
\begin{equation}
    p_q = x_A p_A = x_A (Ap_N) = (Ax_A) p_N \implies x \equiv x^{N/A} = A x_A
\end{equation}
Hence a parton in a bound nucleon within a nucleus ($N/A$), carries a fraction $x=Ax_A$ varying between $[0,1]$ of the bound nucleon momentum and as a result we can define:
\begin{equation}\label{eq:nPDF_defintion}
f^{(N/A)}_i(x) = \frac{1}{A} \left( Z f^{(p/A)}_i(x) + (A-Z)f^{(n/A)}_i(x) \right)
\end{equation}
where $f^{(N/A)}_i$ is the PDF for the 
flavour $i$ associated to the average nucleon $N$ bound in
a nucleus with atomic number $Z$ and mass number $A$,
$f^{(p/A)}_i$ is the PDF for the 
flavour $i$ associated to a proton bound in the nucleus of mass number $A$. Similar for $f^{(n/A)}_i$ but instead for a neutron bound in the nucleus and linked to the proton by isospin symmetry (see Sect.~\ref{s2:Physical_constraints}).
Finally, we can asses the nuclear effects by means of calculating the ratio of a nuclear PDF (nPDF) Eq.~(\ref{eq:nPDF_defintion}) to the sum of free protons and neutrons as follows:
\begin{equation} \label{eq:nuclear_modification_ratio}
    R_i^A = \bigg( Z f^{(p/A)}_i(x) + (A-Z)f^{(n/A)}_i(x) \bigg)\bigg/ \bigg( Z f^{(p)}_i(x) + (A-Z)f^{(n)}_i(x) \bigg)
\end{equation}
This ratio is unity in absence of nuclear effects.
%
The dependence of the nuclear modification ratio Eq.~(\ref{eq:nuclear_modification_ratio}) on the per-nucleon Bjorken-$x$ is shown in Fig.~\ref{fig:nNNPDF20NuclearModification_Pb208} and can be summarised as follows:
\begin{itemize}
    \item \textbf{Shadowing}: In the small-$x$ region $x \lsim 0.1$, $R_i^A$ is suppressed, the suppression increasing with decreasing $x$.
    \item \textbf{Antishadowing}: In the intermediate-$x$ region $0.1 \lsim x \lsim 0.3$, $R_i^A$ is enhanced. 
    \item \textbf{EMC effect}: In the intermediate-$x$ region $0.3 \lsim x \lsim 0.8$, $R_i^A$ is suppressed. 
    \item \textbf{Fermi motion}: In the large-$x$ region $x \gsim 0.8$, $R_i^A$ increases dramatically above unity and this is ascribed to nucleon motion inside the nucleus. 
\end{itemize}

Various models attempt to describe the nuclear modification $R_i^A$ but there is no clear understanding of its behaviour in the entire $x$-dependence to date. Alongside endeavours to construct a satisfactory theoretical model, the experimental efforts to acquire data of various processes constraining nuclear PDFs is still ongoing and so far yields an impressive body of data as we will see in Chapters~\ref{chap:nNNPDF10}, \ref{chap:nNNPDF20} and \ref{chap:nNNPDF30}.

\subsection{Physical constraints} \label{s2:Physical_constraints}
The success of QED in describing elastic and deep-inelastic electron-proton scattering at very high energies discussed throughout Sects.~\ref{s1:Elastic_scattering} and \ref{s1:Inelastic_scattering} hinted strongly for a QCD theory as we know it.
The observed nature of the strong force is responsible for the non-perturbative interactions between quarks and gluons at low-energies, leading to their confinement within composite hadrons. 
Facilitated by the success of QCD in the high-energy regime, our ultimate aim in this thesis is to study the one-dimensional projection of the non-perturbative behaviour at low-energies. For that reason, we discuss in this section the properties of the \textit{nucleons}' PDFs and FFs as well the \textit{nuclei}' PDFs (nPDFs).
The latter distributions, although non-perturbative and not calculable from first principles, follow an important set of physical constraints that we lay down in this section. These constraints follow from boundary conditions and symmetries that, when released and tested phenomenologically, puts QCD to the test.

\myparagraph{Charge and isospin symmetry} \label{s3:Charge_Isospin_symmetry}
After the discovery of the neutron in 1932~\cite{chadwick1932existence}, it was suggested that the proton and neutron are two states of a single particle. 
This was motivated by the mass-energy equivalence and the fact that their masses are approximately equal which lead to the idea that there's an underlying approximate
SU(2) symmetry called isospin obeyed by strong interactions. This means that the proton and the neutron behave identically under the strong interactions and that their difference is solely due to their charge content. Historically, isospin was extended to an SU(3) isospin symmetry with the discovery of strange in an attempt to classify the hadrons using the Eightfold way. However a more fundamental and exact SU(3) colour symmetry was found to be a better fit, explaining consistently additional apparent contradictions that we discussed briefly in the introduction of Sect.~(\ref{s1:QCD})

In the same way spin is associated to a quantum number, isospin implied that the proton $p(I_3=\frac{1}{2})$ and neutron $n(I_3=-\frac{1}{2})$ formed a doublet of isospin $I=1/2$, while $\pi^+(I_3=1)$, $\pi^0(I_3=0)$ and $\pi^-(I_3=-1)$ a triplet of isospin $I=1$. Equivalently, in the parton model, the same previous claims are deduced by assuming that the up and down quarks form a doublet of isospin $I=1/2$.
Observing the violation of isospin symmetry required very precise data and dedicated analyses. For that reason, whenever a neutron PDF $f^n$ or an anti-pion FF $D^{\pi^-}$ are to be determined from a limited statistics, or imprecise data (moreover the case in nuclear collisions data), this approximate symmetry is still assumed.
Additionally to be able to write neutron PDFs in terms of the proton PDFs, and anti-pion in terms of pion FFs assuming isospin symmetry, charge conjugation is an another useful \textbf{exact} symmetry that connects particles to anti-particles thus reducing further their distribution degeneracies.

In Table.~\ref{tab:summary_CI_symmetry}, I summarise the implications of these two symmetries on the FFs and (n)PDFs by referring to SU(2) isospin symmetry as $\mathcal{I}$-symmetry and charge conjugation as $\mathcal{C}$-symmetry. Since no Kaon fragmentation are studied, we omit it and do not consider SU(3) isospin symmetry.
\begin{table}[!h] 
\centering
\begin{tabular}{c|ccc}  \toprule
    & \textbf{FFs} & \textbf{PDFs} & \textbf{nPDFs} \\ \cmidrule{1-4}
    $\mathcal{C}$ & $D^{h^+}_{q(\bar{q})} = D^{h^-}_{\bar{q}(q)}$  & N/R  & N/R  \\ \cmidrule{1-4}
    \multirow{2}{*}{$\mathcal{I}$ }& $D^{\pi^+}_{u(d)} = D^{\pi^-}_{d(u)}$  & $f^{(p)}_{u(d)} = f^{(n)}_{d(u)}$  & $f^{(N/A)}_{u(d)} = 2f^{(D)}_{d(u)} - f^{(N/A)}_{d(u)}$  \\
    & $D^{\pi^+}_{\bar{u}(\bar{d})} = D^{\pi^-}_{\bar{d}(\bar{u})}$  & $f^{(p)}_{\bar{u}(\bar{d})} = f^{(n)}_{\bar{d}(\bar{u})}$  & $f^{(N/A)}_{\bar{u}(\bar{d})} = 2f^{(D)}_{\bar{d}(\bar{u})} - f^{(N/A)}_{\bar{d}(\bar{u})}$  \\\cmidrule{1-4}
    $\mathcal{C}$ + $\mathcal{I}$  &  $D^{\pi^+}_{u(d)} = D^{\pi^-}_{d(u)} = D^{\pi^+}_{\bar{d}(\bar{u})}= D^{\pi^-}_{\bar{u}(\bar{d})}$   & N/R & N/R  \\ 
     \bottomrule
\end{tabular}
\caption{Charge conjugation ($\mathcal{C}$) and SU(2) isospin symmetry ($\mathcal{I}$) constraints on FFs and (n)PDFs taken to be valid for all values of $z$, $x$ and $Q^2$. N/R stands for "Not relevant". The superscript $p$ refers to proton, $n$ to neutron, $D$ to deuterium and $A$ to nucleus with $Z$ proton.}
\label{tab:summary_CI_symmetry}
\end{table}

\myparagraph{Sum Rules} \label{s3:Sum_Rules}
Following from the FFs and (n)PDFs bare definitions as probability distributions, they must satisfy a momentum and a number sum rules presented in the following Table~\ref{tab:summary_Sum_Rules}.
\begin{table}[!h] 
    \centering
    \begin{tabular}{c|ccc}  \toprule
        & \textbf{FFs} & \textbf{PDFs} & \textbf{nPDFs} \\ \cmidrule{1-4}
        Momentum  & \multirow{2}{*}{\(\displaystyle \sum_h \int_0^1 zD^{h}_idz = 1\)}  & \multirow{2}{*}{\(\displaystyle \sum_i \int_0^1 x f_{i}^{(p)}dx = 1 \)}   & \multirow{2}{*}{\(\displaystyle \sum_i \int_0^1 x f_{i}^{(p/A)}dx = 1 \)}  \\ 
        ($i=q,\bar{q},g)$ & & & \\\cmidrule{1-4}
        \multirow{2}{*}{Number}& \multirow{2}{*}{N/A}  &  \(\displaystyle \int_0^1 f_d^{-,(p)}dx = 1 \) & \(\displaystyle \int_0^1 f_d^{-,(p/A)}dx = 1 \) \\
        &   & \(\displaystyle \int_0^1 f_u^{-,(p)}dx = 2 \) & \(\displaystyle \int_0^1 f_u^{-,(p/A)}dx = 2 \) \\
        ($j=s,c,b,t$)&   & \(\displaystyle \int_0^1 f_j^{-,(p)}dx = 0 \) & \(\displaystyle \int_0^1 f_j^{-,(p/A)}dx = 0 \) \\
         \bottomrule
    \end{tabular}
    \caption{Sum rules for FFs and (n)PDFs taken to be valid for an arbitrary scale $Q$. N/A stands for "Not available". $p$ refers to proton and $A$ to nucleus with $Z$ proton. The combination $f^-$ is defined as $f^-=f_q - f_{\bar{q}}$.}
    \label{tab:summary_Sum_Rules}
    \end{table}
These rules stems from the fact that the bare (n)PDFs are probability densities for a struck parton of type $i$ to acquire a given momentum fraction $x$, inside a colour-neutral hadron $A$, and bare FFs are probability densities for an unpolarised parton of type $i$ to fragment into an unpolarised hadron of type $h$, where the hadron carries the fraction $z$ of the parton momentum. The DGLAP equations conserve these properties.
The momentum sum rule follows from momentum conservation and dictates that the sum of all partons momentum should be equal to the nucleon one. The number sum rules follows from quantum number conservation and dictates that the net number of quarks should reproduce the quantum number of the nucleon at all time.

%

\section{Beyond deep-inelastic scattering} \label{s1:QCD_summary}
In this section, I summarise the main outcomes of this chapter that will be of direct use in all the subsequent chapters.
In Sect.~\ref{s2:Flavour_separation}, I outline all the leading-order kinematics of the relevant processes as well as their cross section expressions factorised into non-perturbative objects and hard calculable part in perturbative QCD. 
In Sect.~\ref{s2:NNLO_corrections}, I explain how the NNLO QCD corrections are calculated and included in the results of the subsequent chapters.

\subsection{Relevant processes} \label{s2:Flavour_separation}
In this section, I summarise most of the processes used in the various analyses presented throughout this thesis by providing their cross section factorised expressions.

\myparagraph{Deep-Inelastic Scattering}
The high-energy lepton deep-inelastic scattering off a hadron A in both neutral-current NC (exchange of a photon or a Z-boson) and charged-current CC (exchange of a W-boson) is illustrated below:
\begin{alignat}{2}
&\text{neutral-current}\,\,(Z/\gamma^*):\,\, &&\ell^-+A \rightarrow \ell^{-}+X \nonumber\\
&\text{charged-current}\,\,(W^+):\,\, &&\nu+A \rightarrow \ell^{-}+X \nonumber\\
&\text{charged-current}\,\,(W^-):\,\, &&\nu+A \rightarrow \ell^{+}+X \nonumber
\end{alignat}
\vspace{-1cm}
\begin{align*}
    \includegraphics[valign=c]{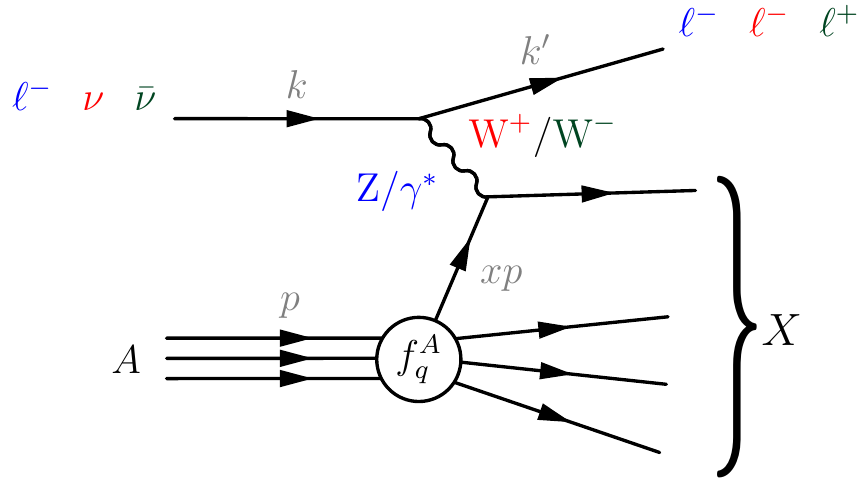} \qquad
    \begin{gathered}   
        \text{\textbf{DIS LO Kinematics}}\\
        x = {\displaystyle A\frac{Q^2}{2p\cdot q}}\\ 
        Q^2 = -q^2 \\
        y ={\displaystyle \frac{q\cdot p}{k \cdot p}}
    \end{gathered}
\end{align*}
The double differential cross section for DIS is given by:
\begin{equation} \label{eq:Master_DIS_xsec}
  \begin{gathered}
  \frac{d^2\sigma^{a}}{dxdQ^2}(x,Q^2,A)=\frac{2\pi\alpha^2}{xQ^4}\eta^a\left[Y_{+}F_2^a(x,Q^2,A) \stackrel[\ell^-\nu]{\ell^+\bar{\nu}}{\mp} Y_{-}xF_3^a(x,Q^2,A)-y^2F_L^a(x,Q^2,A)\right] \\
  \text{with: } a =
  \begin{cases}
    \text{NC} & \rightarrow \eta^{\text{NC}} = 1\\
    \text{CC} & \rightarrow \eta^{\text{CC}} = (1\stackrel[\nu]{\bar{\nu}}{\pm}\lambda)^2\eta_W
  \end{cases}
  \end{gathered}
\end{equation}
where $\alpha$ is the QED running coupling and $\eta_W$ denotes the 
squared ratio of the W-boson couplings and propagator
with respect to those of the photon, $\lambda$ 
is the helicity of the incoming lepton and $Y_{\pm} = 1 \pm (1-y)^2$.

We restrict our discussion of considering free proton as the hadron, however the neutron case is simply obtained by means of the approximate SU(2) isospin symmetry, \textit{i.e.} exchanging $u(\bar{u}) \leftrightarrow d(\bar{d})$.
The dimensionless NC DIS proton structure functions at leading twist can be written generally in terms of PDFs as:
\begin{equation}\label{eq:DIS_SF}
  F^{\text{NC},p}_i=\frac{1}{n_f}\left(\sum_j^{n_f}e_{q_j}^2\right) F_{i,\text{S}}^{\text{NC},p} + F_{i,\text{NS}}^{\text{NC},p},\quad (i=2,L)
\end{equation}
where the singlet $F_{i,\text{S}}^{\text{NC},p}$ and the non-singlet $F_{i,\text{NS}}^{\text{NC},p}$ components are given by:
\begin{align}
  F^{\text{NC},p}_{i,\text{S}} &= C_{i,\text{S}}^{\text{NC}}\otimes f^{(p)}_\Sigma + C_{i,g}^{\text{NC}} \otimes f^{(p)}_g \\
  F^{\text{NC},p}_{i,\text{NS}} &= C_{i,\text{NS}}^{\text{NC}} \otimes \sum_j^{n_f} e_{q_j} f^{(p)}_{j, \text{NS}}, \quad (i=2,\,L,\,3)
\end{align}
Therefore NC DIS provides a sensitivity on only 3 combinations of PDFs, the singlet ($\Sigma$), the sum of non-singlet ($V$, $V_i$ and $T_i$) and the gluon.

With charged-current DIS, the quark-mixing is described by unitary Cabibbo-Kobayashi-Maskawa (CKM) matrix elements:
\begin{equation}\label{eq:CKM_matrix}
    V_{ij} = 
\begin{pmatrix}
    V_{ud} & V_{us} & V_{ub} \\
    V_{cd} & V_{cs} & V_{cb} \\
    V_{td} & V_{ts} & V_{tb}
\end{pmatrix}
\end{equation}
which contains information on the strength of the flavour-changing weak interaction between the quarks. Due to this mixing, a different weight is associated to each PDF flavour such as:
\begin{align}
F_i^{\nu,p} = &C^{\text{CC}}_{i,q} \otimes  \bigg[\sum_{q=d,s,b}\sum_{j=u,c,t} |V_{jq}|^2 f^{(p)}_q + \sum_{q=u,c,t}\sum_{j=d,s,b} |V_{qj}|^2 f^{(p)}_{\bar{q}}\bigg] \nonumber\\
+ &C^{\text{CC}}_{i,g} \otimes 2\sum_{j=u,c,t}\sum_{k=d,s,b}|V_{jk}|^2  f^{(p)}_g, \quad i=(2,3, L)
\end{align}
therefore, CC DIS provides additional sensitivity to different PDF flavours with respect to NC DIS, and that is further the case when both $W^+(\nu)$ and $W^-(\bar{\nu})$ channels are considered with $F_i^{\bar{\nu},p}$ being obtained by exchanging $q \leftrightarrow \bar{q}$.
Finally, the nuclear structure function of a nucleus with mass number $A$ and atomic number $Z$ is given as:
\begin{equation}
    F_i^{a,A} = \frac{1}{A} \left(ZF_i^{a,p/A} + (A-Z)F_i^{a,n/A}\right)
\end{equation}
where $F_i^{a,p/A}$ and $F_i^{a,n/A}$ are both computed with bound proton and neutron PDFs within the nuclei A.

\myparagraph{Electroweak bosons production} 
The high-energy electroweak boson production process occurs in hadron-hadron scattering. The hadron could be a nucleus (with mass number $A$) or a nucleon ($A=1$). We denote the NC channel, the case where a quark of one hadron and an antiquark of another hadron annihilate, creating a virtual photon or Z-boson which then decays into a pair of oppositely-charged leptons. While we denote the CC channel, the case where the annihilation gives rise to a W-boson, which then decays into a charged lepton and a neutrino as illustrated below:
\begin{alignat}{2}
&\text{neutral-current}\,\,(Z/\gamma^*):\,\, &&A_1+A_2 \rightarrow \ell^{-}+\ell^+ \nonumber\\
&\text{charged-current}\,\,(W^+):\,\, && A_1+A_2 \rightarrow \nu+\ell^+ \nonumber\\
&\text{charged-current}\,\,(W^-):\,\, && A_1+A_2 \rightarrow \ell^{-}+\bar{\nu} \nonumber
\end{alignat}
\vspace{-1cm}
\begin{align*}
    \includegraphics[valign=c]{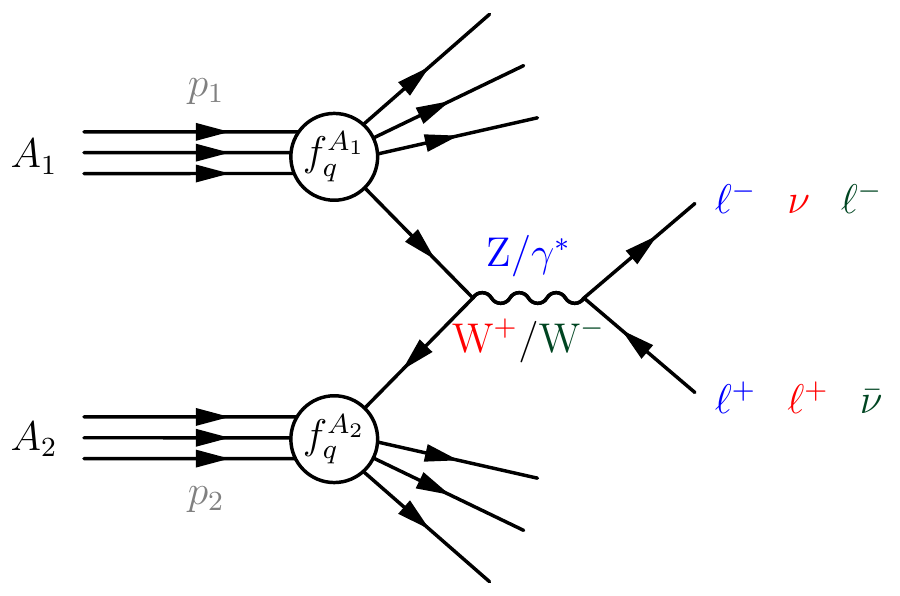} \qquad
    \begin{gathered}   
        \text{\textbf{DY LO Kinematics}}\\
        x_{1,2} = \frac{M}{\sqrt{s}}e^{\pm y}\\
        M^2 = x_1x_2s \\
        s = (p_1+p_2)^2 \\
        y_{\ell\ell} = \frac{1}{2}\ln{\frac{E_{\ell\ell}+p_{\ell\ell,z}}{E_{\ell\ell}-p_{\ell\ell,z}}}
    \end{gathered}
\end{align*} 
In general, the inclusive cross section for photon/Z (a=NC) and W (a=CC) production is given differential in the boson mass and the rapidity of the outgoing dilepton pair (in case of photon/Z) and outgoing lepton (in case of W) as follows:
\begin{align}
\frac{d^2\sigma^a}{dM^2dy} &= \sum_n \alpha_s^n(\mu_R^2) \sum_{ij}\int_{x_1}^1 d\xi_1 \int_{x_2}^1 d\xi_2 \nonumber\\ &f^{A_1}_i(\xi_1,\mu_F^2)f^{A_2}_j(\xi_2,\mu_F^2)\times \frac{d^2\hat{\sigma}_{ij\rightarrow W/Z + X}^a}{dM^2dy}(\frac{x_1}{\xi_1}, \frac{x_2}{\xi_2},Q,\mu_R^2,\mu_F^2)\\ 
& = \sum_{ij}f^{A_1}_i\otimes f^{A_2}_j \otimes \frac{d^2\hat{\sigma}_{ij\rightarrow W/Z + X}^a}{dM^2dy}
\end{align}
where $n$ is the perturbative order and $\hat{\sigma}$ is the partonic cross section of the process $q_i q_j \rightarrow W/Z + X$.
The NC DY provides sensitivity to different combinations of PDFs by means of the effective weak couplings:
\begin{equation*}
    a_q = (\bar{g}^q_V + \bar{g}^q_A), \quad\quad \bar{g}^f_V=(t^{(f)}_{3}-2Q_f \sin^2\theta_W), \quad\quad 
    \bar{g}^f_A=t^{(f)}_{3}
  \end{equation*}
with $t^{(f)}_{3}$ being the weak isospin of fermion $f$, $q_f$ the electric charge and $\theta_W$ is Weinberg's angle. While, similarly to CC DIS, CC DY provides sensitivity to additional flavours combinations through the CKM matrix.

\myparagraph{Jet production}
The inclusive jet cross section is the simplest hadron collider observable with a purely strongly interacting final state. For our purposes, we consider the two cases of detecting one (single jet) or two (dijet) jets in the final states as illustrated below (the diagrams are not meant to be an exhaustive list of all possibilities):
\begin{alignat}{2}
    &\text{single jet}:\,\, &&A_1+A_2 \rightarrow \text{jet}+X \nonumber\\
    &\text{dijet}:\,\, && A_1+A_2 \rightarrow 2\,\,\text{jets}+X\nonumber
    \end{alignat}
    \vspace{-1cm}
\begin{align*}
    \includegraphics[valign=c]{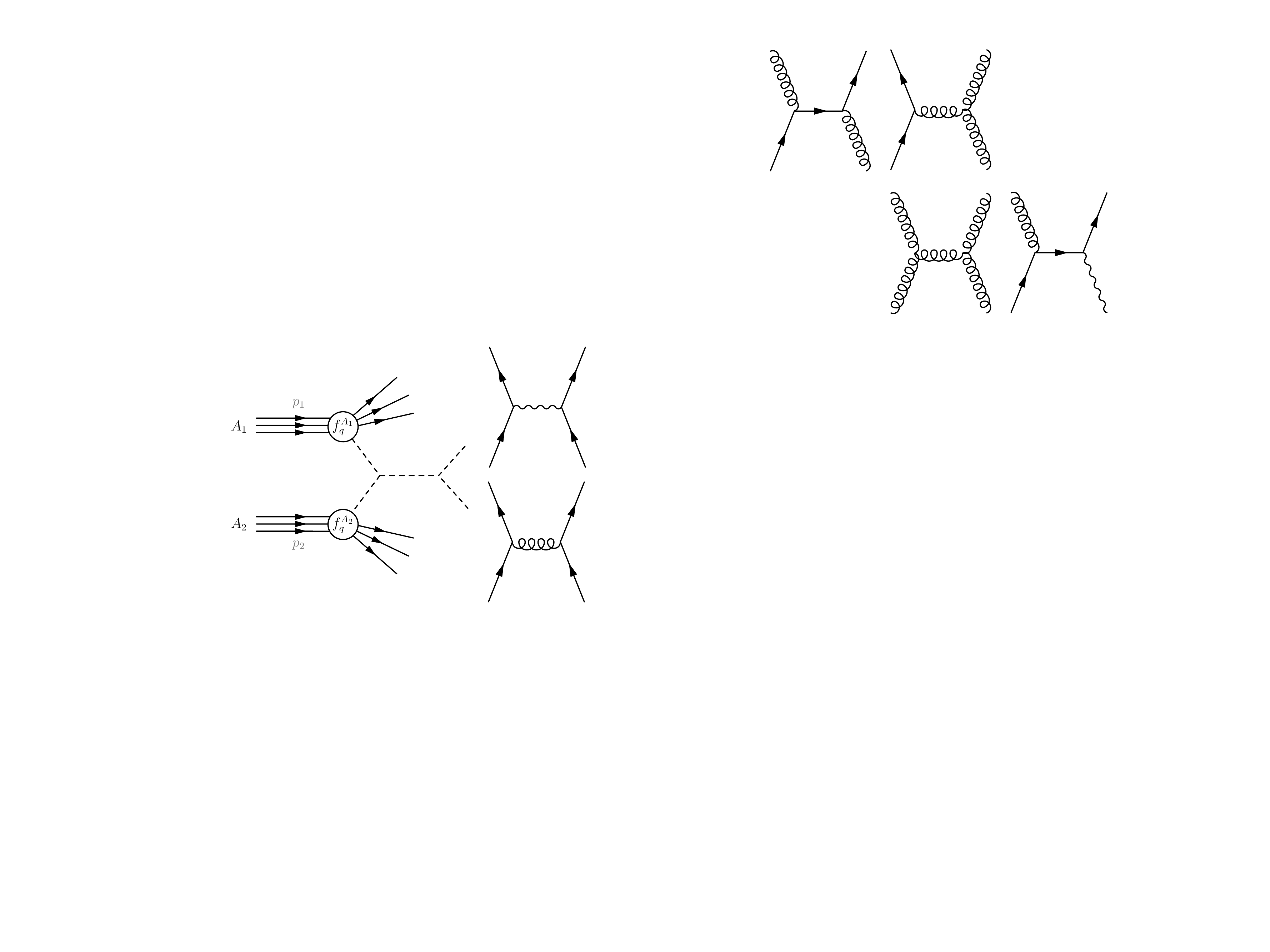} \quad
    \begin{gathered}   
        \text{\textbf{Jet production LO Kinematics}}\\
        x_{1,2} =\frac{p_T e^{\pm y}}{\sqrt{s}}\\
        p_T^2 = x_1x_2 s\\
        s = (p_1+p_2)^2 \\
        y = \frac{1}{2}\ln{\frac{E+p_{z}}{E-p_{z}}}
    \end{gathered}
\end{align*} 
In general, the inclusive jet cross section factorises similarly in form to the electroweak boson production one except that the partonic cross section can be further factorised into a hard matrix element and a jet function describing the chain of partons splitting (see Ref.~\cite{Catani:1996vz}). The hard scale of the process is the $p_T$ of the jets measured however the choice of a suitable factorisation scale is debatable and is discussed briefly in Sect.~\ref{s1:PDF_NNLO_jet}. For illustrative purposes, let us consider the double differential $n$-jet inclusive cross section in jet(s) transverse momentum ${\displaystyle p_T\equiv p_{T,\text{n-jets}}}$ and rapidity ${\displaystyle y\equiv y_{\text{n-jets}}}$:
\begin{align}
\frac{d^2\sigma^a}{dp_T^2dy} &= \sum_{ij}f_i\otimes f_j \otimes \frac{d^2\hat{\sigma}_{ij\rightarrow n\,\,\text{jets} + X}^a}{dp_T^2dy}
\end{align}
where $n$ is the number of jets in the final state and $\hat{\sigma}$ is the partonic cross section of the process $q_i q_j \rightarrow n\,\,\text{jets} + X$.
As we will observe in Sect.~\ref{s1:PDF_NNLO_jet}, the inclusive jet data on provides an important constraint on the gluon PDF.

\myparagraph{Single-inclusive annihilation}
The Single-inclusive annihilation (SIA) process is one of the easiest processes to derive. For our purposes, we only consider the specific case of a lepton and anti-lepton annihilation giving rise to a quark anti-quark pair that fragments with one $\pi^+$ or $\pi^-$ in the final states as illustrated below:
\begin{equation}\label{eq:SIAreaction}
    e^+(k_1)+e^-(k_2)\rightarrow \pi^{\pm}(p_h) + X\,.
  \end{equation}
  \vspace{-1cm}
\begin{align*}
    \includegraphics[width=0.35\textwidth,valign=c]{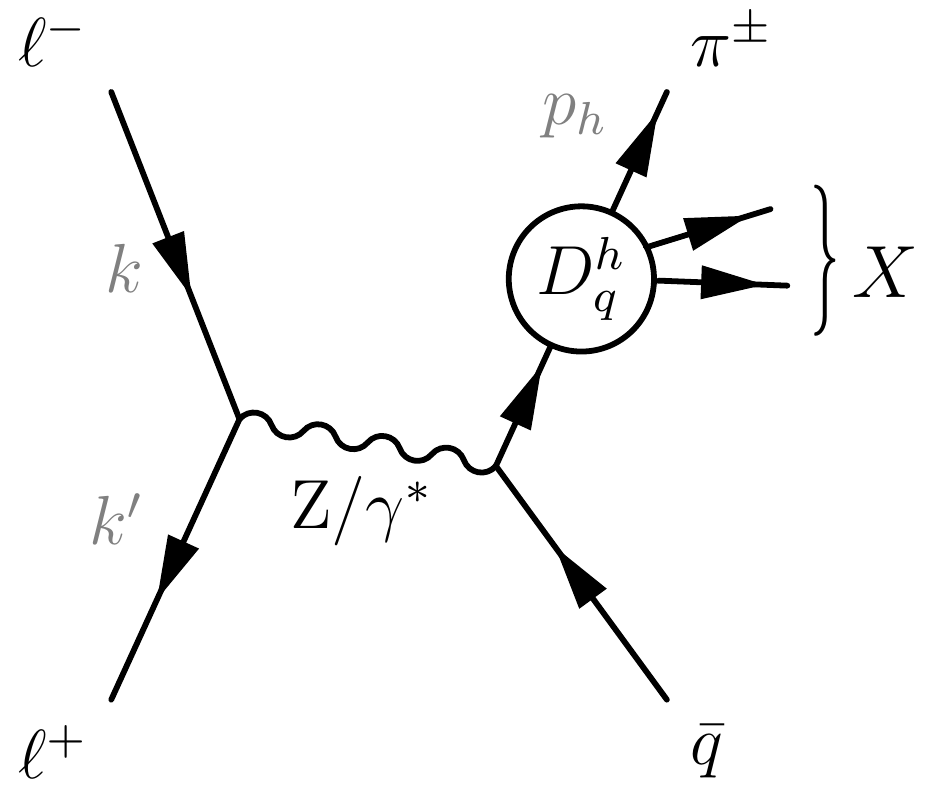} \qquad
    \begin{gathered}   
        \text{\textbf{SIA LO Kinematics}}\\
        Q^2 = q^2 \\
        z = \frac{2p_h\cdot q}{Q^2} \\
        s = Q^2
    \end{gathered}
\end{align*} 
The differential SIA cross section can be written as:
\begin{equation}
\frac{d\sigma^h}{dz}(z,Q)=\frac{4\pi\alpha(Q)}{Q^2} F_2^h(z,Q)
\end{equation}
where $\alpha$ is the QED running coupling and $F_2^h$ the dimensionless SIA fragmentation structure functions for $\pi^\pm$ at leading twist which can be written in terms of FFs as:
\begin{equation}
  F^{\text{NC},\pi^\pm}_2=\frac{1}{n_f}\left(\sum_j^{n_f}e_{q_j}^2\right) F_{2,\text{S}}^{\text{NC},\pi^\pm} + F_{2,\text{NS}}^{\text{NC},\pi^\pm}
\end{equation}
where the singlet $F_{2,S}^{\text{NC},\pi^\pm}$ and the non-singlet $F_{2,NS}^{\text{NC},\pi^\pm}$ components are given by:
\begin{align}
  F^{\text{SIA},\pi^\pm}_{2,\text{S}} = C_{2,\text{S}}^{\text{SIA}}\otimes D^h_\Sigma + C_{2,g}^{\text{SIA}} \otimes D^h_g, \qquad 
  F^{\text{SIA},\pi^\pm}_{2,\text{NS}} = C_{2,\text{NS}}^{\text{SIA}} \otimes \sum_j^{n_f} e_{q_j} D^{h}_{j, \text{NS}} \label{eq:SIA_SF}
\end{align}
Therefore SIA like NC DIS provides a sensitivity on only 3 combinations of FFs, the singlet ($\Sigma$), the sum of non-singlet ($V$, $V_i$ and $T_i$) and the gluon.

\myparagraph{Semi-Inclusive Deep-Inelastic Scattering}
The Semi-Inclusive Deep-Inelastic Scattering (SIDIS) process is similar to DIS except that a specific hadron $h$ has to be detected in the final states.
As in the case of SIA, we will only be concerned with the fragmentation into a charged pion $\pi^+$ or $\pi^-$ as illustrated below:
\begin{equation}\label{eq:SIDISreaction}
  \ell(k)+N(p)\rightarrow \ell(k')+ \pi^{\pm}(p_h) + X\,.
\end{equation}
\vspace{-1cm}
\begin{align*}
    \includegraphics[width=0.45\textwidth,valign=c]{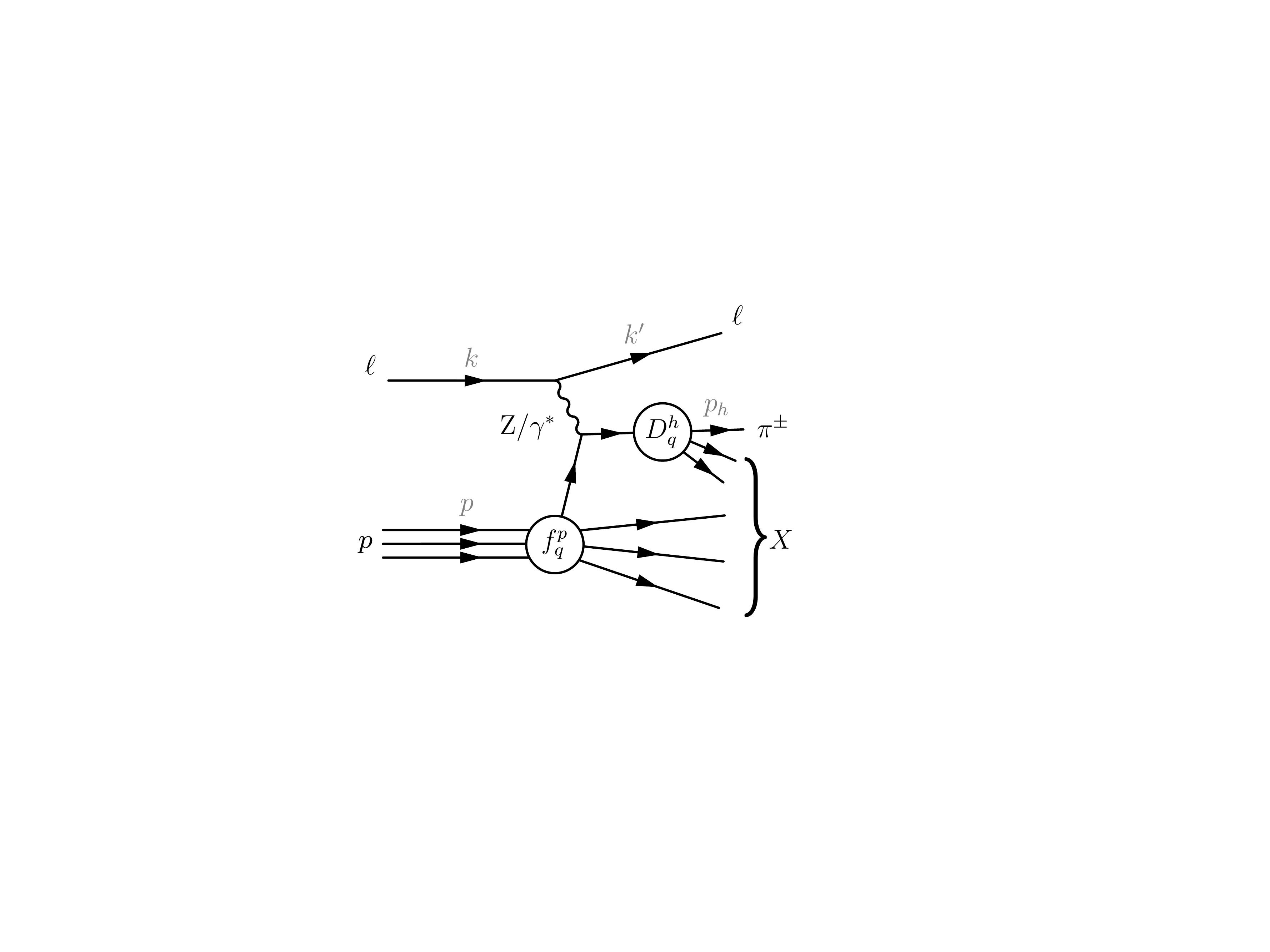}  \qquad
    \begin{gathered}   
        \text{\textbf{SIDIS LO Kinematics}}\\
        Q^2 = -q^2 \\
        x = \frac{Q^2}{2p\cdot q} \\
        z = \frac{p\cdot p_h}{p\cdot q} \\
        y = \frac{Q^2}{xs} \\
        s = (k^2 + p^2)
    \end{gathered}
\end{align*} 
Under the assumption $Q^2\ll M_Z^2$, where $M_Z$ is the mass of the
$Z$ boson, only the exchange of a virtual photon can be considered and the
triple-differential cross section for the reaction in
Eq.~(\ref{eq:SIDISreaction}) can be written as:
\begin{equation}\label{eq:sidis2}
  \frac{d^3\sigma}{dx dQ dz} = 
  \frac{4\pi\alpha^2(Q)}{xQ^3} 
  \left[\left(1+(1-y)^2\right) F_2(x,z,Q^2)
    -y^2 F_L(x,z,Q^2) \right]\,,
\end{equation}
where $\alpha$ is the QED running coupling, and $F_2$ and $F_L$ the dimensionless
fragmentation structure functions for $\pi^\pm$ at leading twist which can be written in terms of FFs as:
\begin{equation}\label{eq:f1sidis}
\begin{array}{rcl}
  \displaystyle  F_i(x,z,Q) &=& \displaystyle x\sum_{q\overline{q}} e_q^2 \bigg\{
                                    \left[C_{i,qq}(x,z,Q) \otimes f_q(x,Q) +   
                                    C_{i,qg}(x,z,Q) \otimes f_g(x,Q)\right]\otimes D^{\pi^\pm}_q(z,Q) \\
                                &+&\displaystyle   \left[C_{i,gq}(x,z,Q)
                                    \otimes f_q(x,Q)\right]\otimes
                                    D^{\pi^\pm}_g(z,Q) \bigg\}\,,\qquad i =2,L\,.
\end{array}
\end{equation}
The convolution symbol $\otimes$ acts equally on $x$ and $z$ and has
to be interpreted as follows:
\begin{equation}\label{eq:doubleconvolution}
C(x,z)\otimes f(x)\otimes D(z) =
\int_x^1\frac{dx'}{x'}\int_z^1\frac{dz'}{z'} C(x',z')f\left(\frac{x}{x'}\right) D\left(\frac{z}{z'}\right)\,.
\end{equation}
The sum in Eq.~(\ref{eq:f1sidis}) runs over \textit{both} quark and
antiquark flavours active at the scale $Q$, $e_q$ is the electric
charge of the quark flavour $q$, and $f_{q(g)}$ and
$D^{\pi^\pm}_{q(g)}$ denote the collinear quark (gluon) PDFs and FFs,
respectively. Since in this thesis (namely Chapter~\ref{chap:FF}), we are interested in determining the
FFs $D^{\pi^\pm}_{q(g)}$ using existing PDFs $f_{q(g)}$ as determined
in separate dedicated analyses, Eq.~(\ref{eq:f1sidis}) has been
arranged in a way that highlights the role of PDFs as effective
charges. Each quark FF
contributing to the cross section is weighted by a factor that, thanks
to PDFs, depends on the specific flavour or antiflavour. As a
consequence, SIDIS cross sections allow for FF quark-flavour separation,
a feature that is not present in SIA cross sections where quark and antiquark
FFs are always summed with equal weight (see {\it e.g.} Eq.~(3.1) in Ref.~\cite{Bertone:2017tyb}).

\subsection{NNLO QCD corrections} \label{s2:NNLO_corrections}
Generally in the results of the subsequent chapters, the QCD corrections to the DGLAP evolution equations and DIS coefficient functions are calculated \textit{analytically} up to NNLO using \texttt{APFEL}~\cite{Bertone:2013vaa} (see references therein).
However, for hadronic processes such as electroweak boson productions and Jet Production, the NLO and NNLO cross sections are calculated by means of Monte Carlo \textit{event-based} generators such as \texttt{MCFM6.8}~\cite{Campbell:2015qma,Boughezal:2016wmq} and \texttt{NLOjet}~\cite{Gehrmann:2018szu}.

Generally, it is very computationally expensive to compute a cross section including all the integrations over phase-space as well as solving the DGLAP equations during minimisation both for DIS and hadronic processes. For that reason, many softwares like \texttt{fastNLO}~\cite{Wobisch:2011ij} and \texttt{APPLgrid}~\cite{Carli:2010rw} were developed in order to concatenate the partonic cross section provided by Monte Carlo generators in form of interpolated weight tables. These tables convert the convolution $\otimes$ into a simple matrix multiplication without any reduction in numerical accuracy, speeding up the computation of observables. 
Moreover, the \texttt{APFELgrid}~\cite{Bertone:2016lga} software was developed as an interface between \texttt{APFEL}~\cite{Bertone:2013vaa} and the interpolated weight tables in order to combine the DGLAP evolution with the partonic cross section. We explain this procedure in Sect.~\ref{s2:fast_evolution}.

To date however, both \texttt{fastNLO} and \texttt{APPLgrid} are interfaced with versions of Monte Carlo generators that solely provide NLO and not NNLO calculations (the latter being provided by newer versions such as \texttt{MCFM9.0}~\cite{Campbell:2019dru} for instance). 
Therefore, due to this purely technical limitation, the $K$-factors approximation is used to compute NNLO prediction for hadronic processes as follows:
\begin{equation} \label{eq:NNLOxsec_kfac}
\sigma^{\text{(NNLO)}} \approx K_{\text{NNLO}}^{\text{QCD}} \times \sum_{ij}
\hat{\sigma}^{\text{NLO}}_{ij}\otimes \mathcal{L}^{\text{NNLO}}_{ij}
\end{equation}
where the sum runs over partonic subchannels,
$\hat{\sigma}_{ij}$ are  partonic cross sections, and 
$\mathcal{L}_{ij}$ the corresponding parton luminosities using a NNLO input PDF set and defined as follows (see Ref.~\cite{Campbell:2006wx}):
\begin{equation}
    \frac{\partial \mathcal{L}_{ij}}{\partial M^2_X}= \frac{1}{s}\int_\tau^1 \frac{dx}{x} f_{i}(x,M_X^2)f_j(\tau/x,M_X^2) \qquad \text{with: } \tau=\frac{M_X^2}{s}
\end{equation}
where $f_{i}(x,M_X^2)$ is the PDF of a parton of type-$i$ and $M_X$ is the invariant mass of the final state.
This quantity is in general a convenient measure of the PDF uncertainty impact on the production cross section at any collider energy and mass of the final state from a set of partons combinations $ij$. 
Finally, the $K$-factors are defined as follows: 
\begin{equation}
\label{eq:kfactor}
{\displaystyle
K_{\text{NNLO}}^{\text{QCD}}
\equiv 
\frac{\sum_{ij}\hat{\sigma}^{\text{NNLO}}_{ij} \otimes \mathcal{L}^{\text{NNLO}}_{ij}}{\sum_{ij}
\hat{\sigma}^{\text{NLO}}_{ij}\otimes \mathcal{L}^{\text{NNLO}}_{ij}}}
\end{equation}
When the approximated NNLO expression Eq.~(\ref{eq:NNLOxsec_kfac}) is included in the global analyses, the parameterised PDFs are forced to reproduce, as close as the minimisation procedure allows it, the NNLO prediction. To clarify further this point, we re-express Eq.~(\ref{eq:NNLOxsec_kfac}) as follows:
\begin{equation}
\sigma^{\text{(NNLO)}} \approx \frac{\sum_{ij}\hat{\sigma}^{\text{NNLO}}_{ij} \otimes \mathcal{L}^{\text{NNLO}}_{ij}}{\sum_{ij}
\hat{\sigma}^{\text{NLO}}_{ij}\otimes \mathcal{L}^{\text{NNLO}}_{ij}} \times \sum_{ij}
\hat{\sigma}^{\text{NLO}}_{ij}\otimes \mathcal{L}^{\text{parameterised}}_{ij}
\end{equation}
where $\mathcal{L}^{\text{parameterised}}_{ij}$ will converge to match $\mathcal{L}^{\text{NNLO}}_{ij}$ in order to contribute in the minimization procedure therefore incorporating the NNLO QCD corrections.

An additional technical limitation the K-factor method suffers from, is that most (if not all) of the Monte Carlo generators that provide NNLO QCD predictions do not accommodate the option of having two different input PDFs. This is particularly problematic for the calculation of any process containing different type of targets and/or projectiles, which is the case in proton-nucleus collisions relevant for nuclear PDFs analyses. In Chapter~\ref{chap:nNNPDF30} where this is relevant for \texttt{nNNPDF3.0}, we will opt for the approximation ${\displaystyle K_{\text{NNLO}}^{\text{QCD, proton-nucleus}} \approx K_{\text{NNLO}}^{\text{QCD, proton-proton}}}$ assuming that nuclear effects either cancel in the ratio Eq.~\ref{eq:kfactor} or can be neglected w.r.t. NNLO corrections. This assumption is not well justified however and we are currently looking into ways of computing K-factors for proton-nucleus collisions.
  \chapter{Statistical tools}
\label{chap:Stat}

\myparagraph{Introduction} Perturbative QCD breaks down at low-energy scales due to the rapid increase of the strong coupling $\alpha_s$. We established in Chapter~\ref{chap:QCD} that the low-energy contributions to a cross section could be incorporated by means of the QCD factorisation theorem. The latter allows us to separate the perturbative from non-perturbative contributions by absorbing the latter into bare quantities such as (n)PDFs or FFs that can be inferred from experimental measurements. 
Although these non-perturbative objects have to satisfy the physical constraints summarised in Sect.~\ref{s2:Physical_constraints}, their functional form cannot be deduced from perturbative QCD and thus must be assumed. Consequently, neural networks, have a major advantage as they can approximate any continuous function as dictated by the Universal Approximation Theorem~\cite{csaji2001approximation}. Nonetheless, many other choices of functional form are also adopted by different groups who infer these objects. Regardless of the choice adopted, it always suffices to parametrise these non-perturbative objects at a single initial scale $\mu$ since the perturbative DGLAP evolution equations, takes care of their evolution to any other perturbative energy scale.

To further illustrate what is meant by a PDF being inferred from experimental measurements, let us consider a PDF parameterised as $f(x,\mu;\bm{\theta})$ where $\bm{\theta}$ is a set of parameters and $f$ is a certain functional form.
To simplify the example, we consider only one data-point $\sigma\pm s$ representing a measured DIS structure function $F_2$ for a certain Bjorken-$x$ and energy transfer $Q$ with an associated additive uncertainty $s$. The corresponding theoretical prediction at ($x,Q$) is given by Eq.~(\ref{eq:F2_collins}) in DIS:
\begin{equation}
  t(x,Q;\bm{\theta}) = \sum_i H^{\text{hard}}_{i}(Q) \otimes \sum_j \Gamma_{ij}^{\text{DGLAP}}(Q,\mu) \otimes f_j(\mu;\bm{\theta}) \nonumber
\end{equation}
where $\Gamma$ is the DGLAP evolution kernel decoupled from the PDF, $H$ is the perturbative hard contribution and $f_j$ the functional form of the parton type-$j$ parameterised by $\bm{\theta}$ at initial scale $\mu$.
The goal of Bayesian inference is to deduce from $\sigma(x,Q)$ and its associated uncertainty $s$, the most likely probability distribution of $\bm{\theta}$ that gets $t(x,Q;\bm{\theta})$ within the range $[\sigma-s,\sigma+s]$.

There are different interpretations of probability in statistics, the Bayesian one~\cite{gelman2013bayesian,mcelreath2020statistical} interprets it as the degree of certainty or belief in a hypothesis to be true, while the frequentist~\cite{von1981probability} sees it as the limit of its relative frequency in many trials.
The result of a Bayesian approach is a probability distribution of the parameters given the data and their uncertainties. As for a frequentist approach, the result is either a "true or false" conclusion from a significance test.
In this thesis we adopt the Bayesian interpretation since we are interested in propagating the experimental uncertainties to the space of parameters and therefore determine the uncertainty associated with the non-perturbative objects. This uncertainty in fact, translates our degree of understanding of the hadronic structure and the limitations that are due to data, theoretical or methodological uncertainties~\cite{Gao:2017yyd, DelDebbio:2004xtd, Ball:2008by,Ball:2009qv,Ball:2014uwa, Ball:2017nwa,AbdulKhalek:2019ihb}.

\myparagraph{Outline} In Sect.~\ref{s1:Uncertainties_in_HEP}, I summarise how generally a particle production cross section is measured in high energy physics. Then, I discuss the different type of uncertainties that are associated with such measurement, their origin and how to treat them. Afterwards, I examine the gaussian assumption commonly used to define the data probability distribution. That is followed by the definition of the covariance matrix as well as the treatment of the D'Agostini's bias.
Then, in Sect.~\ref{s1:uncertainty_propagation}, I first define the likelihood function by means of Bayes' theorem and introduce the maximum log-likelihood as well as the Lagrange multiplier methods used during inferences. I then discuss two methods used to propagate uncertainties from the space of data to that of parameters, the Monte Carlo and Hessian methods.
Finally, in Sect.~\ref{s1:Machine_Learning}, I introduce the topic of machine learning focusing on feed forward neural networks, their structure, forward and backward modes of propagation, their universality, overfitting feature and how to over come it with cross-validation. I end this section by discussing the different minimisation algorithms and strategies I use for the different inferences in this thesis.

\section{Data uncertainties in high energy physics} \label{s1:Uncertainties_in_HEP} 
Since Rutherford gold foil experiment until today's high energy electron and hadron colliders, particle collisions remain the main type of experiments physicists use to confront their theories.
A cross section is the most commonly measured observable which together with the way detectors are built and segmented addresses questions related to the probability of a certain hypothesis to be true in a certain corner of phase space. Therefore generally, experiments in high energy physics are particles counting experiments coupled with novel detectors, tools and analyses to identify them and their properties.

\myparagraph{Cross section measurements}
Typically, the measurement of some particle production cross section from a certain collision relies on:
\begin{itemize} 
  \item $N_c$: Counting the number of candidate events in a sample of observed interactions.
  \item $N_b$: Estimating the number of ``background'' events in this sample from other processes.
  \item $\epsilon$: Estimating the acceptance of the apparatus including all selection requirements used to define the sample of events like the detector acceptance, trigger efficiency, reconstruction efficiency, selection efficiency and background rejection efficiency. The latter are usually benchmarked and inferred from a Monte Carlo simulation of the apparatus and collision events.
  \item $L_{int}$: The integrated luminosity that encapsulates the total number of collisions in a certain time defined as:
  \begin{equation} \label{eq:int_lumi}
  L_{int} = \int L dt = \int \frac{dN}{\sigma} \rightarrow \text{units of barn: } b^{-1} = 10^{28} m^{-2} 
  \end{equation}
  where $N$ is the number of events detected in a certain time $t$ and $\sigma$ the collision cross-section.
\end{itemize}
whereby the total production cross section of a certain particle is:
\begin{equation} \label{eq:master_formula}
\sigma = \frac{N_c - N_b}{\epsilon L_{int}}
\end{equation}
Analogously, one can measure a differential cross section in ``bins'' of phase space:
\begin{equation}
  \sigma_{i}= \frac{\Delta N_{i,\,c} - \Delta N_{i,\,b}}{\epsilon(\Delta k^{(1)}_i, \Delta k^{(2)}_i, \dddot{}) L_{int}}
\end{equation}
where $\Delta k^{(1)}_i$ and $\Delta k^{(2)}_i$ denotes well defined regions $i$ of phase space, $\Delta N_{i,\,c}$ and $\Delta N_{i,\,b}$ the number of candidate and background events in that region.

\myparagraph{Types of uncertainties}
Without exception, these counting measurements involve both \textit{statistical} and \textit{systematic} uncertainties. We refer to Ref.~\cite{Sinervo:2003wm} for the following definitions and additional case studies.

Statistical uncertainties are the result of random fluctuations originating from the fact that a measurement is based on a finite set of observations. Repeated measurements of the same phenomenon will therefore result in a set of observations that will differ, and the statistical uncertainty is a measure of the range of this variation. By definition, statistical variations between two identical measurements of the same phenomenon are uncorrelated. Examples of statistical uncertainties include the finite resolution of an instrument, the Poisson fluctuations associated with measurements involving finite sample sizes and random variations in the system one is examining.

Systematic uncertainties, on the other hand, arise from uncertainties associated with the nature of the measurement apparatus, assumptions made by the experimenter, or the model used to make inferences based on the observed data. Such uncertainties are generally correlated from one measurement to the next, as well as from one bin to the next which makes their estimation a daunting task. Common examples of systematic uncertainty include uncertainties that arise from the calibration of the measurement device, the fraction of signal events accessible by the detector due to its geometry (called the ``acceptance'' of the detector) and the fraction of signal events that pass the experimenter selection cuts (called the ``selection efficiency''). Conventionally, trigger and reconstruction efficiency are factored out of the latter. 

Experimentalists currently summarise their measurements and their associated uncertainties with the following indicators:
\begin{itemize}
  \item $\sigma_{i}$: the expectation value of all events (or observations) in a kinematic bin $i$.
  \item $s^{(a)}_{i,\text{unc}}$: uncorrelated additive uncertainties originating from a source $(a)$ and associated with a kinematic bin $i$.
  \item Correlated uncertainties:
  \begin{itemize}
    \item $s^{(a)}_{ij,\text{corr}}$: correlated \textit{additive} uncertainties originating from a source $(a)$ and associated with the kinematic bin $i$ and $j$.
    \item $\delta^{(a)}_{ij,\text{corr}}$: correlated \textit{multiplicative} uncertainties originating from a source $(a)$ and associated with the kinematic bin $i$ and $j$.
  \end{itemize}
  \item $\bm{C}$: In some cases, the breakdown of systematics is not provided, but instead the full combined experimental covariance matrix.
\end{itemize}
We also note that for some data sets, an asymmetric uncertainty is provided: $[s^{(a)+}_{ij},\,s^{(a)-}_{ij}]$: originating from a certain source $(a)$ which could be uncorrelated or correlated between kinematic bin $i$ and $j$.

\begin{figure}[!h]
  \begin{center}
    \makebox[\textwidth]{\includegraphics[width=\textwidth]{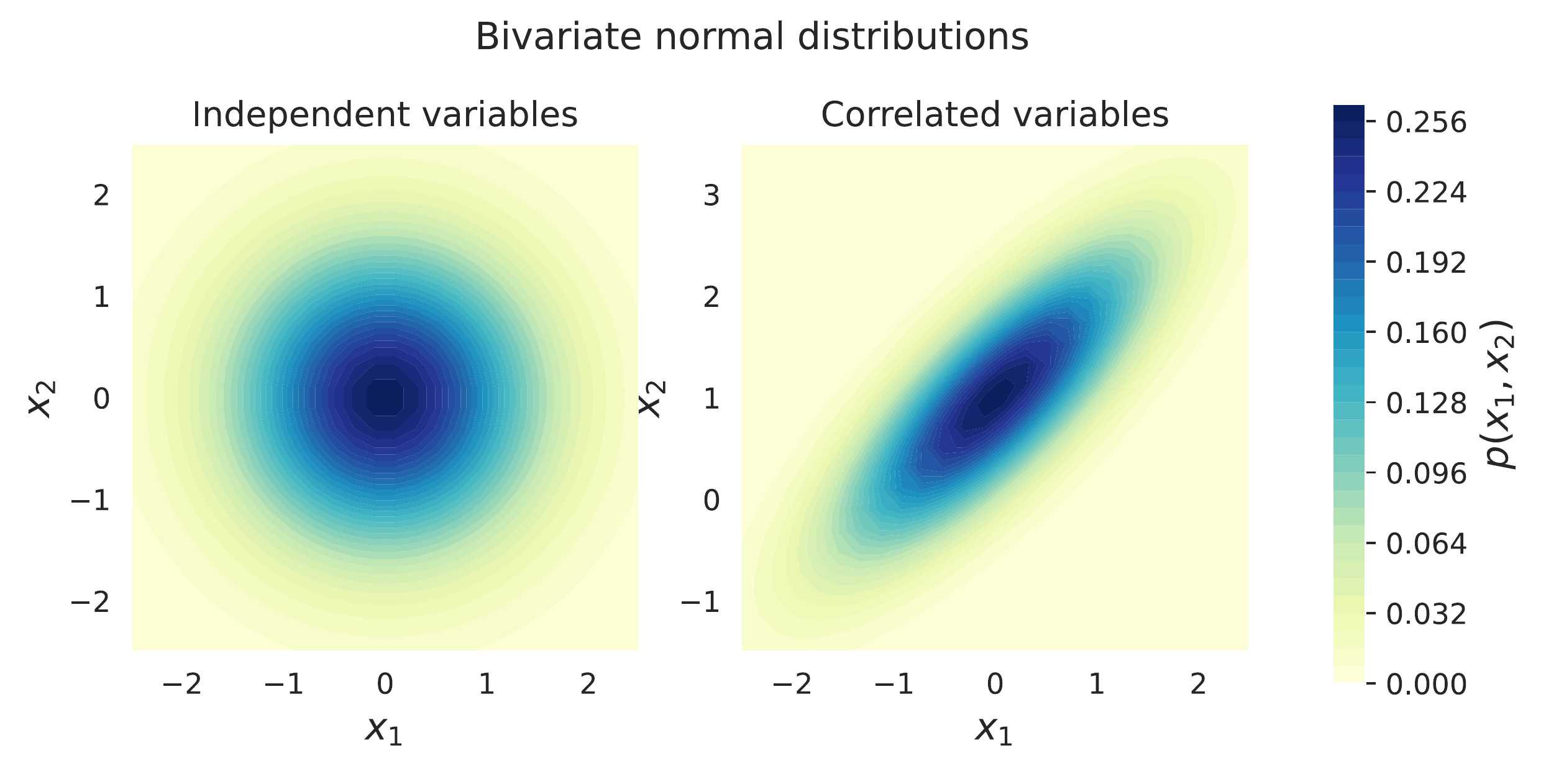}}
  \end{center}
\caption{\small A representative figure of two bivariate gaussian distributions. The figure on the left represents a distribution with uncorrelated variables $x_1$ and $x_2$ (zero off-diagonal elements in the covariance matrix). The figure on the right represents a distribution with correlated variables instead. The figure is generated by \href{https://github.com/peterroelants/peterroelants.github.io/blob/master/notebooks/misc/multivariate-normal-primer.ipynb}{peterroelants/peterroelants.github.io}.}
 \label{fig:gauss_dist}
\end{figure}

\myparagraph{Gaussian assumption}
Let us start by considering a total of $N_\text{rep}$ events observed leading to Eq.~(\ref{eq:master_formula}) which we can rewrite as:
\begin{equation}
  \sigma = \sum_k^{N_{\text{rep}}} \frac{N_c^{(k)}-N_b^{(k)}}{\epsilon L_{int}} = \frac{1}{N_{\text{rep}}} \sum_k^{N_{\text{rep}}} x^{(k)}= \sum_i^{N_{\text{dat}}} \sigma_{i}, \qquad \sigma_{i} =  \frac{1}{N_{\text{rep}}} \sum_k^{N_{\text{rep}}} x_i^{(k)}
\end{equation}
where $\epsilon$ is the acceptance of the apparatus including all selection requirements used to define the sample of events, $\bm{x}^{(k)} = \{x_1^{(k)}, x_2^{(k)}, \dddot{},
x^{(k)}_{N_{\text{dat}}}\}$ are equally probable observations,
$k=1,\dots,N_{\text{rep}}$, of a set of $N_{\text{dat}}$ measured quantities.
Then the indicators provided by the experimentalists could be seen as moments of the discrete data distribution $f(x_i,x_j,\dddot{})$ such that:
\begin{align}
\text{Expectation value: } \sigma_{i} &= \frac{1}{N_{\text{rep}}}\sum_k^{N_{\text{rep}}} x_i^{(k)} \\ 
\text{Variance: } s_{i} &= \frac{1}{N_{\text{rep}}}\sum_k^{N_{\text{rep}}} (x_i^{(k)}-\sigma_i)^2 \\
\text{Covariance: } C_{ij} &= \frac{1}{N_{\text{rep}}}\sum^n_{k}x^{(k)}_ix^{(k)}_j - \sigma_i \sigma_j \label{eq:def_covmat}
\end{align}
however, this set of indicators does not hold information about the distribution itself rather only its two first moments. For that reason, as illustrated in Fig.~\ref{fig:gauss_dist} the data is assumed to be sampled from a multivariate Gaussian distribution as defined below:
\begin{equation} \label{eq:gaussian}
  \mathcal{G}\left(\bm{x}^{(k)}\right) \propto
  \exp{\left[\left(\bm{x}^{(k)} - \bm{\sigma}\right)^T \cdot \bm{C}^{-1}\cdot \left(\bm{x}^{(k)} - \bm{\sigma}\right)\right]}
\end{equation}
One can already see that this assumption is not well suited for asymmetric uncertainties, however the approximation commonly adopted to symmetrise such uncertainties is by doing~\cite{DelDebbio:2004xtd}:
\begin{align}
\tilde{\sigma_i} &= \sigma_i + \frac{s_i^+-s_i^-}{2} \\
\tilde{s_i} &= \frac{s_i^+ + s_i^-}{2} 
\end{align}
where $\tilde{\sigma_i}$, $\tilde{s_i}$ are the new symmetrised mean and standard deviation. Ideally, one would opt to avoid such kind of assumptions on the data distribution and rather aim to include the original ``raw'' event distribution. However the latter are currently either not so commonly delivered or complicated to treat.

\myparagraph{Covariance matrix}
The gaussian assumption imposes a constraint on the covariance matrix, that has to be symmetric and positive-definite\footnote{$\bm{C}$ is set to be  positive-definite if the scalar $\bm{x}^T\cdot \bm{C} \cdot \bm{x}$ is strictly positive for every non-zero column vector $x$.}. Therefore we can express $\bm{C}$ following from its definition in Eq.~(\ref{eq:def_covmat}) as:
\begin{equation} \label{eq:covmat}
C_{ij} = \delta_{ij} \sum_s s^{(a)}_{i,\,\text{unc}}s^{(a)}_{j,\,\text{unc}} + \sum_{s} s^{(a)}_{i,\,\text{corr}}s^{(a)}_{j,\,\text{corr}} 
+ \sum_{s} \delta^{(a)}_{i,\,\text{corr}}\delta^{(a)}_{j,\,\text{corr}} \sigma_i\sigma_j
\end{equation}

\myparagraph{D'Agostini's bias}
The covariance matrix Eq.~(\ref{eq:covmat}), although valid to asses the quality of the data sets description \textit{after} a fit, it would lead to the so-called D'Agostini's bias in the result~\cite{DAgostini:1993arp} if used during minimisation.
To remedy this, we generally adopt the $t^0$-prescription presented in Ref.~\cite{Ball:2009qv} whereby we perform iterated fits using the
theory predictions central value instead of the experimental ones to render the multiplicative uncertainties into additive ones in the covariance matrix that becomes:
\begin{equation} \label{eq:t0_covmat}
  C^{t^0}_{ij} = \delta_{ij} \sum_s s^{(a)}_{i,\,\text{unc}}s^{(a)}_{j,\,\text{unc}} + \sum_{s} s^{(a)}_{i,\,\text{corr}}s^{(a)}_{j,\,\text{corr}} 
  + \sum_{s} \delta^{(a)}_{i,\,\text{corr}}\delta^{(a)}_{j,\,\text{corr}} t^0_i t^0_j
\end{equation}
where $t^0_i$ is the central prediction computed with a set of inferred parameters. 
These predictions are typically obtained from a previous fit which is then iterated until convergence is reached.
The use of the $t_0$ covariance matrix defined in Eq.~(\ref{eq:t0_covmat})
for the likelihood maximisation (to be discussed in Sect.~\ref{s2:likelihood})
avoids the bias associated with multiplicative uncertainties, which lead 
to a systematic underestimation of the best-fit values compared to their true 
values~\cite{DAgostini:1993arp}.

\section{Uncertainty propagation} \label{s1:uncertainty_propagation}

In Sect.~\ref{s2:likelihood}, I start by defining the likelihood function by means of Bayes' theorem, then I explain the difference between the marginal (mainly used by \texttt{NNPDF}, thus in Chapters~\ref{chap:PDF}, \ref{chap:nNNPDF10}, \ref{chap:nNNPDF20}, \ref{chap:nNNPDF30} and \ref{chap:Impact}) and conditional likelihoods (used by \texttt{MAPFF1.0} in Chapter~\ref{chap:FF}) when a subset of data is considered in the inference. Finally and most importantly, I discuss the minimisation of the chi-square function and how it originates from the maximum log-likelihood method within the gaussian assumption. This will be at the heart of all the minimisation procedures upon which the subsequent results will be based on. Furthermore, I discuss the case where the chi-square is subject to a generic equality constraint and how to deal with it by means of the Lagrange multiplier method.

The minimisation of the chi-square function result in a set of optimal parameters to describe the data.
However, since our main interest is to quantify the uncertainty associated to this optimal set, I discuss two methods that aim to propagate the uncertainties from the space of data to that of parameters. These are the Monte Carlo (Sect~\ref{s2:monte_carlo}) and Hessian (Sect.~\ref{s2:hessian_method}) methods. Starting from a certain parameters prior probability distribution, the application of either of these methods lead to a posterior\footnote{Posterior: refers to a distribution that was inferred from some data, \textit{i.e.} underwent the minimisation.}  probability distribution that encodes the uncertainties on the optimal parameters resulting from the minimisation.

\subsection{Likelihood function} \label{s2:likelihood}
In Bayesian statistics, the likelihood $\mathcal{L}(B|A)$ or equivalently the probability $\mathcal{P}(A|B)$ quantifies the degree with which the evidence $B$ supports a hypothesis $A$. To infer the probability of a hypothesis $A$ with a prior probability $\mathcal{P}(A)$ given the evidence $B$ with a prior probability $\mathcal{P}(B)$, one applies repetitively the Bayes' theorem:
\begin{equation}
  \mathcal{P}(A|B)=\frac{\mathcal{P}(B|A)\mathcal{P}(A)}{\mathcal{P}(B)}
\end{equation}
To infer $\mathcal{P}(A|B)$ while knowing/assuming $\mathcal{P}(A)$ and $\mathcal{P}(B)$, one tries to vary $A$ in a way to maximise $\mathcal{P}(B|A)$ and therefore maximise the probability that hypothesis $A$ is valid given the evidence $B$.
The likelihood function can be therefore defined as the probability of observing
a given sample of data for a given set of parameters $\bm{\theta}$. Based on the previous section, assuming that $\bm{D}$ the data is distributed gaussianly according to $\mathcal{G}(\bm{\sigma},\bm{C})$ whose expectation is $\bm{\sigma}$ and covariance matrix is $\bm{C}$, one can write the likelihood function as:
\begin{equation}
  \mathcal{L}(\bm{\theta}| \bm{\sigma}) \equiv \mathcal{P}(\bm{\sigma}|\bm{\theta}) =
  \frac{1}{\sqrt{(2\pi)^n|\bm{C}|}}\exp\left(-\frac{1}{2}[\bm{t(\theta)-\sigma}]^T\bm{C}^{-1}[\bm{t(\theta)-\sigma}]\right).
\end{equation}
where $\bm{t(\theta)}$ is the theoretical prediction to be compared with the data calculated with the parameters $\bm{\theta}$. When it comes to fitting a subset of the data (due to kinematic cuts for instance) one has to distinguish between using a marginal or conditional likelihoods. Let's consider $\bm{m_A}$ of dimension $n_A$ and $\bm{m_B}$ of dimension $n_B$ as subsets of $\bm{\sigma}$ such that $\bm{m_A, m_B} \subseteq \bm{\sigma}$ and $\bm{m_A \cup m_B} = \bm{\sigma}$ and the associated covariance to be: 
\begin{equation}
  \mathcal{G}\left(\bm{\sigma}=\begin{bmatrix}
    \bm{m_A}\\
    \bm{m_B}
    \end{bmatrix},
    \bm{C} = \begin{bmatrix}
      \bm{C_A}&\bm{C_C}\\
      \bm{C_C}&\bm{C_B}
      \end{bmatrix}\right)
\end{equation}

\begin{figure} 
  \floatbox[{\capbeside\thisfloatsetup{capbesideposition={right,center},capbesidewidth=0.3\textwidth}}]{figure}[\FBwidth]
  {\caption{Figure representing marginal distributions of a subset of random variables $x$ and $y$ from the original bivariate normal distribution. Each of the projected distributions represent a subset of the variables without reference to the other. The figure is generated by \href{https://github.com/peterroelants/peterroelants.github.io/blob/master/notebooks/misc/multivariate-normal-primer.ipynb}{peterroelants/peterroelants.github.io}.}\label{fig:gauss_marginal}}
  {\includegraphics[width=0.7\textwidth]{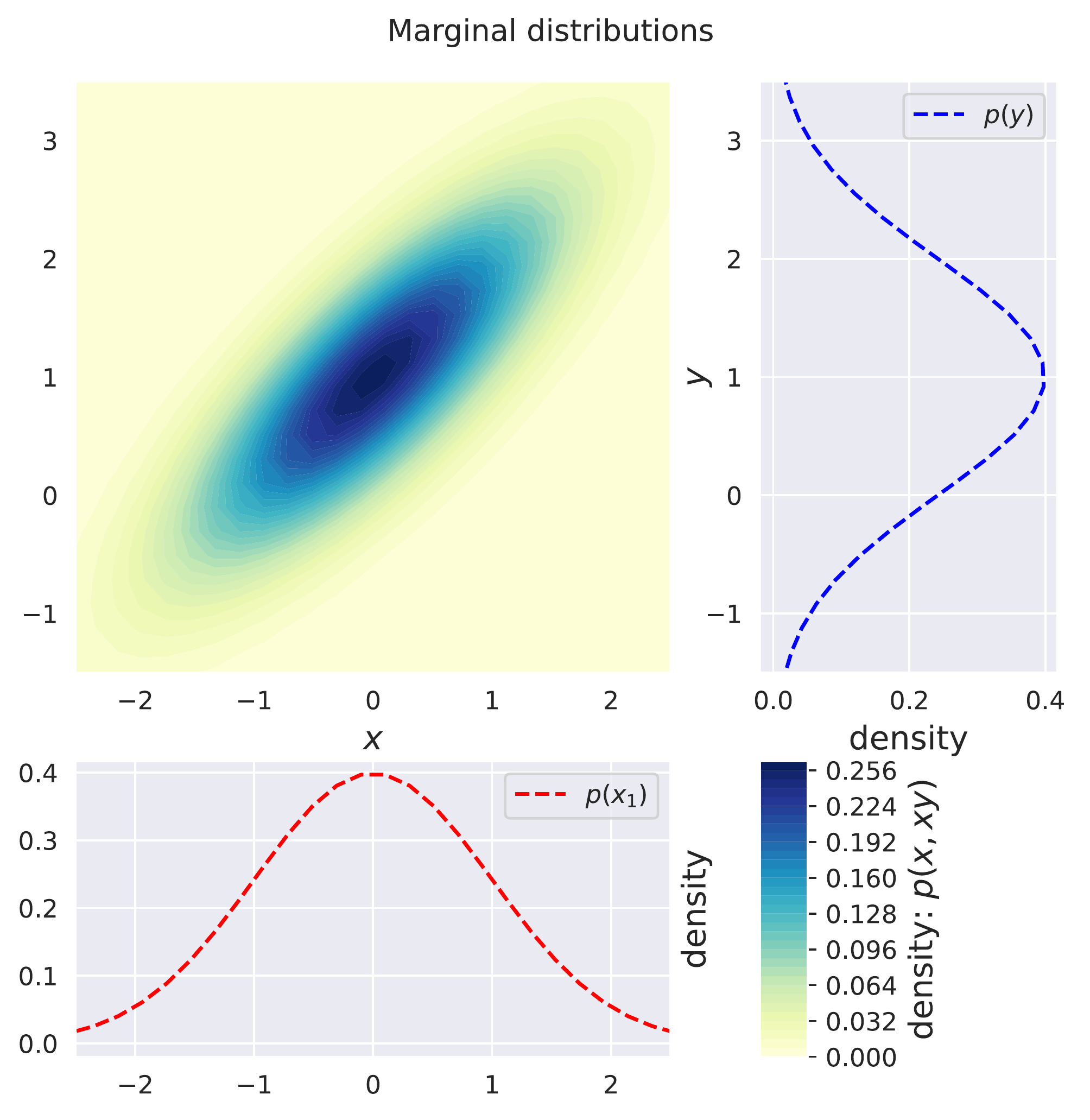}} 
  \end{figure}

\myparagraph{Marginal likelihood}
To obtain the marginal likelihood over $\bm{m_A}$, one only needs to drop the irrelevant values of $B$ (that one wants to marginalize out) from the mean vector and the covariance matrix. In which case one ends up with:
\begin{equation}
  \mathcal{L}(\bm{\theta_A}| \bm{m_A}) = 
  \frac{1}{\sqrt{(2\pi)^{n_A}|\bm{C_A}|}}\exp\left(-\frac{1}{2}[\bm{t(\theta_A)-m_A}]^T\bm{C_A}^{-1}[\bm{t(\theta_A)-m_A}]\right).
\end{equation}
where $\bm{\theta_A}$ are the parameters to be inferred from $\bm{m_A}$ without any influence or knowledge of $\bm{m_B}$ nor $\bm{C_B}$. This can be seen as inferring from one of the marginal projections $x$ (as A) or $y$ (as B) illustrated in Fig.~\ref{fig:gauss_marginal} without reference to each other.

\begin{figure} 
  \floatbox[{\capbeside\thisfloatsetup{capbesideposition={right,center},capbesidewidth=0.3\textwidth}}]{figure}[\FBwidth]
  {\caption{Figure representing conditional distributions of a subset of random variables $x$ and $y$ from the original bivariate normal distribution. Each of the projected distributions represent a subset of the variables accounting for the correlation of the other. The figure is generated by \href{https://github.com/peterroelants/peterroelants.github.io/blob/master/notebooks/misc/multivariate-normal-primer.ipynb}{peterroelants/peterroelants.github.io}.}\label{fig:gauss_conditional}}
  {\includegraphics[width=0.7\textwidth]{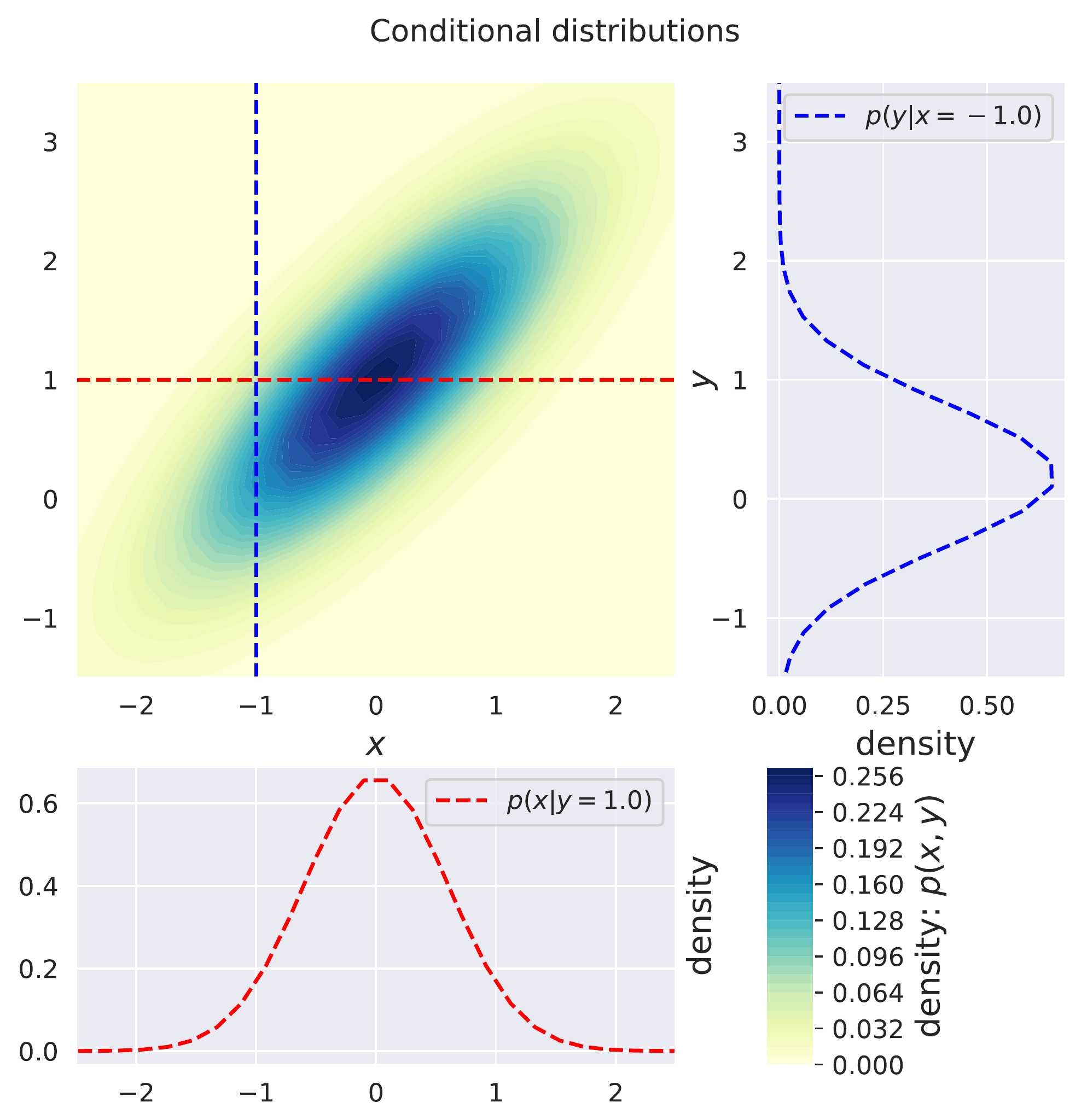}} 
  \end{figure}

\myparagraph{Conditional likelihood}
The computation of the conditional covariance matrix $\bm{C_{A|B}}$ can be seen as inverting the covariance matrix $\bm{C}$, dropping the rows and columns corresponding to the values of $\bm{m_B}$ that are being conditioned upon, and inverting back to get the conditional covariance matrix. In which case one ends up with the conditional likelihood defined as:
\begin{equation}
  \mathcal{L}(\bm{\theta_{A|B}}| \bm{\sigma_{A|B}}) = 
  \frac{1}{\sqrt{(2\pi)^{n_A}|\bm{C_{A|B}}|}}\exp\left(-\frac{1}{2}[\bm{t(\theta_{A|B})-\sigma_{A|B}}]^T\bm{C_{A|B}}^{-1}[\bm{t(\theta_{A|B})-\sigma_{A|B}}]\right).
\end{equation}
where $\bm{\sigma_{A|B}} = \bm{\sigma_{A}} + \bm{C_C C_B^{-1} (\bm{t(\theta_{A|B})}-\bm{m_B})}$, $\bm{C_{A|B}} = \bm{C_{A}} + \bm{C_C C_B^{-1} C_C^T}$ and the parameters $\bm{\theta_{A|B}}$ are to be inferred from $\bm{m_A}$ while accounting for correlations from $\bm{m_B}$ as illustrated in Fig.~\ref{fig:gauss_conditional}.

\myparagraph{Maximum log-likelihood}
The goal of a statistical inference is to find a particular set of
parameters $\bm{\theta}$ that maximises the likelihood
$\mathcal{L}(\bm{\theta}|\bm{\sigma})$, thus maximising the probability of
observing the given data. This procedure typically goes under the name
of maximum likelihood estimation (MLE). However, finding the maximum of the
likelihood could be computationally expensive and one can equivalently maximise the logarithm of the likelihood, that reads:
\begin{equation}
\ln\left(\mathcal{L}(\bm{\theta}|\bm{\sigma})\right) = -\frac{1}{2}\ln{\left([2\pi]^n|\bm{C}|\right)} -\frac{1}{2}[\bm{t(\theta)-\sigma}]^T\bm{C}^{-1}[\bm{t(\theta)-\sigma}].
\end{equation} 
Since the left-hand term is constant and the
right-term is negative, maximising the likelihood is equivalent to
minimising this last term, that is:
\begin{equation}\label{eq:MLE}
    \max_{\bm{\theta}}\ln\left(\mathcal{L}(\bm{\theta}|\bm{\sigma})\right) \rightarrow \min_{\bm{\theta}} \left(\frac{1}{2}[\bm{t(\theta)-\sigma}]^T\bm{C}^{-1}[\bm{t(\theta)-\sigma}]\right)
\equiv \frac{1}{2} \min_{\bm{\theta}} \chi^2(\bm{\theta})\,.
\end{equation}
This defines the chi-square $\chi^2$ as an effective loss function to be
minimised in a regression analysis. In the studies below, we will use
the $\chi^2$ as the function to minimise:
\begin{equation} \label{eq:chi2_general}
  \chi^2(\bm{\theta}) = [\bm{t(\theta)-\sigma}]^T\bm{C}^{-1}[\bm{t(\theta)-\sigma}]
\end{equation}
One can also reformulate this expression by breaking the covariance matrix into uncorrelated and correlated part~\cite{Gao:2017yyd} as follows:
\begin{equation}
  \chi^2(\bm{\theta}) = \sum_i^{N_{\text{dat}}} \frac{1}{s^{\text{tot}}_{i,\,\text{unc}}} \left( \sigma_i - t_i(\bm{\theta}) \sum_{s} s^{(a)}_{i,\,\text{corr}}\lambda^{(a)} \right)^2 + \sum_s \lambda^{(a)2}
\end{equation}
where each source of correlated systematic is described by a nuisance parameter $\lambda^{(a)}$. Thus the induced systematic shift to the experiment measurement is ${\displaystyle \sum_{s} s^{(a)}_{i,\,\text{corr}}\lambda^{(a)}}$. The ${\displaystyle \sum_s \lambda^{(a)2}}$ is called the penalty term. The gaussian assumption allows us to analytically determine the nuisance parameters to restore the definition of the $\chi^2$ as follows:
\begin{align}
  \lambda^{(a)} = \sum_i^{N_{\text{dat}}} \frac{\sigma_i-t_i(\bm{\theta})}{s^{\text{tot}}_{i,\,\text{unc}}} \sum_l \frac{1}{A^{(a)(l)}}\frac{s^{(l)}_i}{s^{\text{tot}}_{i,\,\text{unc}}}, \qquad A^{(a)(l)}  = \delta_{sl} + \sum_i^{N_{\text{dat}}} \frac{s^{(a)}_{i}s^{(l)}_{i}}{(s^{\text{tot}}_{i,\,\text{unc}})^2}\nonumber
\end{align}

\myparagraph{Lagrange multiplier} 
The various non-perturbative objects we are interested in, do not possess a functional form that can be deduced from perturbative QCD. Nevertheless, they have to satisfy the physical constraints summarised in Sect.~\ref{s2:Physical_constraints}. Among these constraints is the positivity of observables, which will be imposed by means of the Lagrange multiplier method~\cite{Ball:2013lla,Ball:2014uwa,AbdulKhalek:2020yuc,AbdulKhalek:2019mzd} in the subsequent results.

In order to find the maximum of the log-likelihood $\ln\mathcal{L}(\bm{\theta}|\bm{\sigma})$ subject to $N_{\text{EC}}$ equality constraints $\{(h_1(\bm{\theta})=0),\,\dddot{},(h_{N_{\text{EC}}}(\bm{\theta})=0)\}$, the method of Lagrange multipliers state that one should instead find the maximum of the Lagrange function defined as:
\begin{equation}
  \begin{gathered}
L(\bm{\theta},\bm{\sigma},\lambda) = \ln\mathcal{L}(\bm{\theta}|\bm{\sigma}) - \sum_i^{N_{\text{EC}}} \lambda_i h_i(\bm{\theta})\\
\max_\theta L(\bm{\theta},\bm{\sigma},\lambda) = \min_\theta \left[\chi^2(\bm{\theta}|\bm{\sigma}) + \sum_i^{N_{\text{EC}}} \lambda_i h_i(\bm{\theta})\right]
  \end{gathered}
\end{equation}
where $\lambda_i$ are called the lagrange multipliers that can be chosen in a way to determine the size of the constraint contribution during the minimisation. For instance, if $\lambda_i h_i(\bm{\theta}) \gg \chi^2(\bm{\theta}|\bm{\sigma})$, the minimisation algorithm will emphasize more on the satisfaction of this constraint contribution prior to minimising the likelihood (roughly speaking). Therefore, by maximising the Lagrange function, the optimisation algorithm seeks a point in parameter space where the gradient of the likelihood function points in the same direction as the gradients of its constraints, while also satisfying them.

\subsection{Monte Carlo method} \label{s2:monte_carlo}
The Monte Carlo (MC) method is nowadays widely used in QCD
analyses~\cite{Ball:2017nwa,
  Nocera:2014gqa,AbdulKhalek:2019mzd,AbdulKhalek:2020yuc,
  Bertone:2017tyb,Bertone:2018ecm
  ,Sato:2016tuz,Sato:2016wqj,Ethier:2017zbq,
  Barry:2018ort,Moutarde:2019tqa,Cuic:2020iwt}. It relies on estimating the parameters' posterior probability distribution by duplicating the inferences. Every inference is independent and performed on a replica of the data resulting in an optimal set starting from different initial conditions. The result of all inferences performed defines the parameters' posterior probability distribution which encapsulates both the optimal set (mean of this probability distribution) and its associated uncertainty (standard deviation of this probability distribution).
The majority of results~\mycite{Khalek:2021gxf,Khalek:2021ulf,AbdulKhalek:2020yuc,
AbdulKhalek:2020jut,
AbdulKhalek:2019ihb,
AbdulKhalek:2019bux,
AbdulKhalek:2019mzd,
Khalek:2018bbv} in the subsequent chapters are based on this method.

\begin{figure} 
  \floatbox[{\capbeside\thisfloatsetup{capbesideposition={right,center},capbesidewidth=0.3\textwidth}}]{figure}[\FBwidth]
  {\caption{A representative figure of the MC sampling Eq.~(\ref{eq:MCgen}) from a bivariate normal distribution. The figure is generated by \href{https://github.com/peterroelants/peterroelants.github.io/blob/master/notebooks/misc/multivariate-normal-primer.ipynb}{peterroelants/peterroelants.github.io}.}\label{fig:gauss_sampling}}
  {\includegraphics[width=0.7\textwidth]{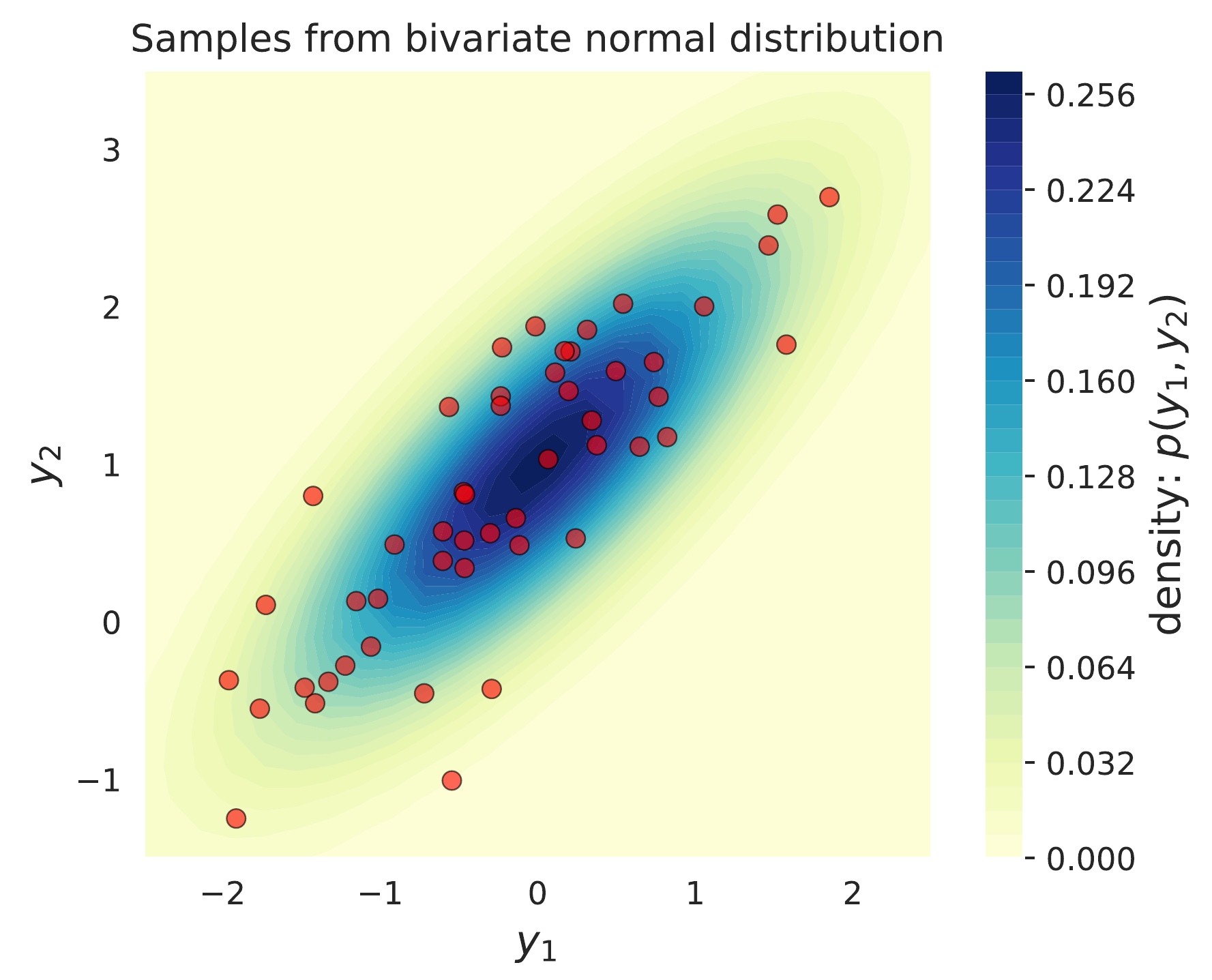}} 
  \end{figure}

\myparagraph{Inference} 
The MC method, relies on generating $N_{\text{rep}}$
replicas of the data, $\bm{x}^{(k)}$, using the Cholesky decomposition~\cite{press2007numerical}
$\bm{L}$ of the covariance matrix $\bm{C}$ (illustrated in Fig.~\ref{fig:gauss_sampling}) such that:
\begin{equation} \label{eq:MCgen}
    \bm{x}^{(k)} = \bm{\sigma} + \bm{L}\cdot\bm{r^{(k)}}\, ; \qquad \bm{C} = \bm{L}\cdot \bm{L}^T \, ,
\end{equation}
where $\bm{r}^{(k)}$ is an $N_{\text{dat}}$-dimensional normal random
vector such that the full set of replicas encodes properly the information on the data central values, variances and correlations provided by the experimental covariance matrix and therefore satisfying:
\begin{equation}
  \frac1{N_{\text{rep}}} {\displaystyle \sum^{N_{\text{rep}}}_k x_i^{(k)}} \simeq \sigma_i\, ,\qquad\frac1{N_{\text{rep}}} {\displaystyle \sum^{N_{\text{rep}}}_{k} x_i^{(k)}x_j^{(k)}} \simeq  \sigma_i\sigma_j + C_{ij}
  \label{eq:MCgeneration}
\end{equation}
in the limit of a sufficiently large number of replicas. Therefore, the MC method's accuracy in representing the underlying data probability distribution increases with larger $N_{\text{rep}}$.
%
%
Every replica $\bm{x}^{(k)}$ substitutes the data central value $\bm{\sigma}$ in Eq.~(\ref{eq:chi2_general}), which now reads:
\begin{equation} \label{eq:chi2_rep}
  \chi^{2(k)}(\bm{\theta}^{(k)}) = [\bm{t(\theta^{(k)})-x^{(k)}}]^T\cdot\bm{C}^{-1}\cdot[\bm{t(\theta^{(k)})-x^{(k)}}]
\end{equation}
Then, $N_{\text{rep}}$ independent inferences are carried out based on Eq.~(\ref{eq:chi2_rep}) starting from a flat parameters prior probability distribution.
The resulting set of parameters $\{\bm{\theta}^{(1)},\,\dddot{},\, \bm{\theta}^{(N_{\text{rep}})}\}$ defines their posterior probability distribution.
The average of this distribution is taken to be the best solution and the standard deviation its associated uncertainty.

\myparagraph{Reweighting}
Let us consider a new data set B and a set of parameters' posterior probability distribution $\{\bm{\theta}_A^{(1)},\,\dddot{},\, \bm{\theta}_A^{(N_{\text{rep}})}\}$ that were inferred from a data set A. 
This method~\cite{Ball:2011gg,Gao:2017yyd} relies on estimating the probability $\mathcal{P}(\bm{\theta}_A^{(k)}|\chi_B^{2(k)})$ of the posterior probability distribution given the chi-square function $\chi_B^{2(k)}$ calculated with the new data set B and defined as follows:
\begin{equation}  \label{eq:chi2_B}
  \chi_B^{2(k)}(\bm{\theta}_A^{(k)}) = [\bm{t(\theta_A^{(k)})-m_B^{(k)}}]^T\cdot\bm{C}^{-1}\cdot[\bm{t(\theta_A^{(k)})-m_B^{(k)}}]
\end{equation}
This probability is considered as a new weight $w_{A|B}^{(k)}$ to every posterior MC parameters in the set $\{\bm{\theta}_A^{(1)},\,\dddot{},\, \bm{\theta}_A^{(N_{\text{rep}})}\}$ accommodating the impact of the new data set B and defined as:
\begin{align}
  \displaystyle
  w_{A|B}^{(k)} \propto \mathcal{P}(\bm{\theta}_A^{(k)}|\chi_B^{2(k)}) &= \left(\sum_k^{N_{\text{rep}}} \Omega^{(k)} \right)^{-1} \Omega^{(k)} \\
  \text{with: } \Omega^{(k)}&= \left(\chi_B^{2(k)}\right)^{(n-1)/2}e^{-\chi_B^{2(k)}/2}.
\end{align}
If for a given replica, the agreement with the new data set is poor, it will result in a large $\chi_B^{2(k)}$ and thus the weight of this specific replica will be exponentially suppressed. Finally, the new parameters $\bm{\theta}_{A \bigcup B}^{(k)}$ accommodating the impact of the new data can be written as:
\begin{equation}
  \bm{\theta}_{A \bigcup B}^{(k)} = w_{A|B}^{(k)} \bm{\theta}_{A}^{(k)}
\end{equation}
hence the name ``reweighting''. This method however suffers from some limitations that are detailed in Refs.~\cite{Ball:2011gg,Gao:2017yyd}.


\subsection{Hessian method} \label{s2:hessian_method}
The Hessian method~\cite{Pumplin:2001ct,Paukkunen:2014zia, Carrazza:2015aoa,Schmidt:2018hvu} relies on estimating the uncertainty on the inferred optimal parameters from the behaviour of the $\chi^2$ close to the global minimum in the parameters space. I start by discussing the inference in the Hessian approach that will facilitate the introduction of the Hessian profiling upon which the impact studies~\mycite{Khalek:2018mdn,Azzi:2019yne,
Cepeda:2019klc,AbdulKhalek:2019mps} in Sect.~\ref{s2:PDF_HLLHC} and Sect.~\ref{s2:PDF_LHeC} are based on.

\myparagraph{Inference}
In the MC method, a set of $N_{\text{rep}}$ independent-inferences are considered of equal weights and defines the parameters' posterior probability distribution. Contrastingly, in the Hessian method, only one inference occurs from which the global minimum and a Hessian matrix encoding the region around the minimum are inferred. The eigenvectors of this matrix define the error set of parameters used to estimate the uncertainty on the best solution. I will rely on Ref.~\cite{Paukkunen:2014zia} in most of the following derivations.

Near the minimum, the $\chi^2$ can be approximated in terms of the following quadratic expansion:
\begin{align}
\Delta\chi^2 = \chi^2 - \chi^2_{\text{min}} \approx \sum_{ij}^{N_{\text{par}}} H_{ij}\Delta \theta_i \Delta \theta_j, \qquad \Delta \theta_i = \theta_i - \theta_{i, \text{min}} \label{eq:quadratic_expansion}
\end{align}
where $N_{\text{par}}$ is the total number of parameters and $\theta_{i, \text{min}}$ representing the best parameter $i$ leading to the global minimum of $\chi^2$.
The Hessian matrix $H_{ij}$ is defined as follows:
\begin{equation} \label{eq:Hessian_matrix}
H_{ij}(\bm{\theta}_{\text{min}})=\frac{1}{2}\frac{\partial^2\chi^2}{\partial\theta_i\partial\theta_j}\bigg|_{\bm{\theta}=\bm{\theta}_{\text{min}}}
\end{equation}
which contains all the information necessary to quantify the parameters' posterior probability distribution or in that case the \textit{Hessian error sets}. 
The Hessian matrix is symmetric and therefore has $N_{\text{eig}}$ orthonormal eigenvectors $\bm{v}^{(k)}$ and eigenvalues $\epsilon_k$ satisfying:
\begin{align} 
H_{ij}v_j^{(k)}= \epsilon_k v_i^{(k)},\qquad \sum_jv_j^{(k)}v_j^{(l)} = \sum_jv_k^{(j)}v_l^{(j)} = \delta_{kl}
\end{align}
We now can define a new set of parameters $z_k$ that diagonalises the Hessian matrix:
\begin{equation}
z_k = \sqrt{\epsilon_k}\sum_iv_i^{(k)}\Delta\theta_i, \Longleftrightarrow \Delta\theta_i = \sum_k\frac{1}{\sqrt{\epsilon_k}}v_i^{(k)}z_k 
\end{equation}
which if substituted in Eq.~(\ref{eq:quadratic_expansion}) leads to the following expression:
\begin{equation}
\Delta\chi^2 \stackrel{\text{Eq.~(\ref{eq:quadratic_expansion})}}{=} \sum_i z_i^2
\end{equation}
which can be generalised to any quantity $\mathcal{O}$ depending on the parameters as follows:
\begin{equation} \label{eq:Delta_O}
  (\Delta \mathcal{O})^2 = \Delta \chi^2 \sum_k \left(\frac{\partial \mathcal{O}}{\partial z_k}\right)^2
\end{equation}
We denote by $S_0$ the global minimum point in the space of parameters of the function $\chi^2$ and $S_k$ the different error sets issuing from the diagonalisation of the Hessian matrix as follows:
\begin{align}
\bm{z}(S_0) = (0,\,\dddot{},\,0),\qquad \bm{z_i}(S_k^\pm) = \pm T^\pm_k \bm{\delta_{ik}}
\end{align}
where $T^\pm_k$ is referred to as tolerance parameter chosen to delimit the neighbourhood of the the global minimum in which inferences are acceptable. In the following we consider the commonly used fixed tolerance $T^\pm_k=\sqrt{\Delta \chi^2}$.  Using these sets and the linear approximation of the derivatives in Eq.~(\ref{eq:Delta_O}) we can write:
\begin{equation} \label{eq:linear_approx}
\left(\frac{\partial \mathcal{O}}{\partial z_k}\right)\approx \frac{\mathcal{O}(S^+_k)-\mathcal{O}(S^-_k)}{2\sqrt{\Delta \chi^2}}
\end{equation}
which finally leads to the following master formulas in the case of asymmetric and symmetric error sets respectively: 
\begin{equation} 
  (\Delta \mathcal{O})^2_{\text{asym}} = \frac{1}{4}\sum_k(\mathcal{O}[S^+_k]- \mathcal{O}[S^-_k])^2,\qquad
  (\Delta \mathcal{O})^2_{\text{sym}} = \sum_k(\mathcal{O}[S^+_k]- \mathcal{O}[S_0])^2 \label{eq:hessian_errors}
\end{equation}

\myparagraph{Profiling}
Similar to the MC reweighting, the Hessian profiling method~\cite{Pumplin:2001ct,Schmidt:2018hvu, Paukkunen:2014zia} aims to gauge the impact and compatibility of a new data set B on the posterior parameters (inferred from a data set A) given the chi-square function $\chi_B^{2(k)}$ calculated with the new data set B and defined in Eq.~(\ref{eq:chi2_B}). For that purpose, we consider the combined $\chi^2_{A\bigcup B}$ defined as:
\begin{equation} \label{eq:new_chi2}
  \chi_{A\bigcup B}^{2} \equiv \chi_{A}^{2}+ \chi^2_B =  \chi_{0,A}^{2}+ \sum_k^{N_{\text{eig}}} z_k^2 + \chi^2_B
\end{equation}
Similarly to the linear approximation we used in Eq.~(\ref{eq:linear_approx}), we can write the theoretical prediction on the new data as follows:
\begin{equation} \label{eq:th_prediction_hessian}
  \begin{gathered}
t_i(\bm{\theta}) \approx t_i(S_0) + \sum_k^{N_\text{eig}} \frac{\partial t_i(S)}{\partial z_k}\bigg|_{S=S_0} z_k \approx t_i(S_0) + \sum_k^{N_\text{eig}} D_{ik}w_k \\
\text{with: }D_{ik} \equiv \frac{t_i(S_k^+)-t_i(S^-_k)}{2},\qquad w_k\equiv \frac{z_k}{(T^+_k - T^-_k)/2}
  \end{gathered}
\end{equation}
Substituting this expression in Eq.~(\ref{eq:new_chi2}), we deduce that $\chi_{A\bigcup B}^{2}$, within the considerations above, can be written as:
\begin{equation}
  \chi_{A\bigcup B}^{2} = \chi_{A\bigcup B}^{2}\bigg|_{\bm{w}=\bm{w}^{\text{min}}} + \sum_{ij}\Delta w_i G_{ij} \Delta w_j
\end{equation}
where $\Delta \bm{w} = \bm{w} - \bm{w}^{\text{min}}$, the $\bm{w^{\text{min}}}$ are the set of new parameters minimising $\chi_{A\bigcup B}^{2}$ and $\bm{G}$ the new Hessian matrix given by:
\begin{align}
\bm{w}^{\text{min}}&=-\bm{G}^{-1}\cdot \bm{a}\\
G_{kl} = \sum_{ij}D_{ik}C_{B,ij}^{-1}D_{jl} + \left(\frac{T_k^+-T_k^-}{2}\right)^2 &\delta_{kl},\qquad a_k= \sum_{ij}D_{ik}C^{-1}_{B,ij}(t_{B,j}(S_0)-\sigma_{B,j}) \nonumber
\end{align}
where the new Hessian matrix $\bm{G}$ and the vector $\bm{a}$ are computed solely from the experimental covariance matrix $C_B$ of the new data set B, the theoretical predictions $\bm{t}_B(\bm{\theta}_A)$ from the posterior parameters $\bm{\theta}_A$ and the tolerances $T^\pm_k$.

Using this new Hessian matrix $\bm{G}$, we can calculate the new error sets resulting from the impact of the data set B following the same steps highlighted in the Hessian method. 
In order to calculate any quantity $\mathcal{O}$ depending on the parameters with the new $\bm{w}^{\text{min}}$, it suffices to replace the theoretical prediction $t$ by $\mathcal{O}$ in Eq.~(\ref{eq:th_prediction_hessian}) or equivalently, use the new minimum parameters and error sets.
Finally, and in order to decide whether the new data set B is consistent with the original data set A, we define a criteria called the \textit{penalty term} as follows:
\begin{equation}
  P\equiv \sum_k^{N_\text{eig}} \left[\left(\frac{T^+_k+T^-_k}{2}\right)w_k^{\text{min}}\right]^2 \stackrel{T^\pm_k \rightarrow \sqrt{\Delta \chi_B^2}}{\longrightarrow} \Delta \chi_B^2 \sum_k^{N_\text{eig}} (w_k^{\text{min}})^2
\end{equation}
where having $P \ll \Delta \chi^2_A$ indicates that the new data B is not conflicting with A while having $P \gsim \Delta \chi^2_A$ indicates that B is in tension with A (See Ref.~\cite{Paukkunen:2014zia} for more details).

\section{Machine Learning for non-perturbative QCD} \label{s1:Machine_Learning}
Machine learning (ML) is a branch of Artificial intelligence (AI), the name given to the field developing technologies concerned with enabling a machine to simulate human behaviour. ML refers to the subfield of AI aimed at studying computer algorithms that are designed to operate and self-improve based on data and pattern recognition. There are 3 main ML algorithms categories:
\begin{itemize} 
  \item \textbf{Supervised learning}: the algorithm is presented with a data set containing a set of inputs and their associated outputs or labels. The task is to achieve an accurate mapping between the two while being predictive on new inputs. This category can be further subdivided into \textit{classification} and \textit{regression} algorithm types. In the former, the output is a discrete function or simply a category, while in the latter the output is a real continuous function of the inputs.
  \item \textbf{Unsupervised learning}: the algorithm is presented with an unlabelled data. The task is to describe its hidden structure and correlations. One example of this category is the \textit{clustering} algorithms that try to achieve a grouping of the data according to the patterns it can detect or is sensitive to.
  \item \textbf{Reinforcement learning}: the algorithm is provided by a desired performance within a specific context or constraints. The task is self-tuning based on trial and error trying to maximise the reward and minimise the penalty by approaching to the desired output.
\end{itemize}

As of the late 20th century, ML has been heavily applied to various problems in particle physics, such as simulations, real time analysis and triggering, object reconstruction, identification, and calibration, image recognition based on raw data in LHC analyses, approximating matrix elements, classifying the Standard Model events and finally determination of non-perturbative objects in QCD~\cite{Albertsson:2018maf,Schwartz:2021ftp}.

One of the most used ML algorithm in supervised learning, not only in particle physics, is the \textit{artificial neural networks} (ANN). This algorithm initially was inspired from the mode of operation biological neural networks has in the human brain. Particularly w.r.t. features like receiving inputs (Dendrites), structure (Axon), connectivity (Synapses) and processing of information (Soma). A more important similarity, is the ability for ANN to continuously update its parameters based on the data it \textit{trains} on and the patterns that it encounters~\cite{csaji2001approximation}.

In Sect.~\ref{s2:neural_networks}, I introduce the \textit{feed forward neural networks} (NNs) that will be the core functional form I use to parametrise all the non-perturbative objects discussed in the subsequent results. I start by discussing its forward and backward propagation modes, the motivation behind using them to infer QCD non-perturbative objects, their features and pitfalls.
Then, in Sect.~\ref{s2:minimisation}, I outline briefly the minimisation algorithms used to achieve the inferences in the subsequent chapters. 

\subsection{Neural networks} \label{s2:neural_networks}
There exists many models of neural networks~\cite{van2017neural} each suited to a particular problem. In this thesis and in the following I focus on feed forward neural networks (NNs).

\myparagraph{Forward propagation} A \textit{neuron} constitutes the basic unit of a neural network. In the following diagram\footnote{All neural network diagrams in this thesis are generated using the following public code: \href{http://github.com/battlesnake/neural}{github.com/battlesnake/neural}} we consider one neuron $N_1$ connected to 3 input neurons ${\pmb \xi}$.
\begin{center}
\begin{neuralnetwork}[height=3,nodesize=25pt] 
  \newcommand{\nodetextclear}[2]{}
  \newcommand{\nodetextx}[2]{$\xi_#2$}
  \newcommand{\nodetexty}[2]{$N_#2$}
  \inputlayer[count=3, bias=false, title=Inputs, text=\nodetextx]
  \outputlayer[count=1, title=Neuron, text=\nodetexty] \linklayers
\end{neuralnetwork}
\end{center}
Generally, every neuron is associated with one \textit{bias} (in this case $\theta_1$) and every link connecting the neuron corresponds to a set of \textit{weights} (in this case $w_{1i}$) that represent the parameters of the NN and are combined as follows:
\begin{equation} \label{eq:one_neuron}
N_1({\pmb \xi};\{w_{1i},\theta_1\}) = \phi\left(\sum^3_i w_{1i}\xi_i + \theta_1\right)
\end{equation}
where $\phi$ is the \textit{activation function} that defines the output of each neuron and has the property of being bounded. The most commonly used functions are:
\begin{itemize} 
  \item \textbf{Threshold function}: $\phi(x)=1$ if $x\geq 0$ and $\phi(x)=0$ if $x<0$. If used in the output layer, this function serves for discrete \textit{logistic regression} where the problem is classifying an input as belonging to one of two classes.
  \item \textbf{Sigmoid function}: $\phi(x)=\frac{1}{1+e^{-x}}$ that have an s-shape such that ${\displaystyle \lim_{x\rightarrow\infty}{\phi(x)}=1}$ and ${\displaystyle \lim_{x\rightarrow-\infty}{\phi(x)}=0}$. If used in the output layer, this function serves for continuous \textit{logistic regression} where the problem is classifying an input as belonging to one or more classes with a confidence level.
  \item \textbf{Hyperbolic tangent function}: $\phi(x)=\tanh{x}$ that have an s-shape such that ${\displaystyle \lim_{x\rightarrow\infty}\phi(x)=1}$ and ${\displaystyle \lim_{x\rightarrow-\infty}\phi(x)=-1}$. If used in the output layer, this function serves for continuous \textit{logistic regression}.
  \item \textbf{Linear function}: $\phi(x)=x$ that serves for \textit{regression} problems (inferring a continuous function). 
  \item \textbf{Quadratic linear function}: $\phi(x)=x^2$ that serves for \textit{regression} problems (inferring a positive-definite continuous function). 
\end{itemize}
Now, we can generalise Eq.~(\ref{eq:one_neuron}) by defining a general NN as a real-valued vector function $\bm{N}: \mathbb{R}^m \rightarrow \mathbb{R}^n$ that maps an input vector ${\pmb \xi}\in\mathbb{R}^m$ to an output vector $\bm{N}\in \mathbb{R}^n$. In the following diagram we consider a generic representative feed forward NN with $L$ total layers corresponding to 2 input neurons, 3 hidden layers of 3, 5 and 4 neurons each respectively, and 3 output neurons.
\begin{center}
\begin{neuralnetwork}[height=5,nodesize=25pt] 
  \newcommand{\nodetextclear}[2]{}
  \newcommand{\nodetextx}[2]{$\xi_#2$}
  \newcommand{\nodetexty}[2]{$N_#2$}
  \inputlayer[count=2, bias=false, title=Input\\layer (L-4), text=\nodetextx]
  \hiddenlayer[count=3, bias=false, title=Hidden\\layer (L-3), text=\nodetextclear]
  \linklayers 
  \hiddenlayer[count=5, bias=false, title=Hidden\\layer (L-2), text=\nodetextclear]\linklayers
  \hiddenlayer[count=4, bias=false, title=Hidden\\layer (L-1), text=\nodetextclear]\linklayers
  \outputlayer[count=3, title=Output\\layer (L), text=\nodetexty] \linklayers
\end{neuralnetwork}
\end{center}
We assume that all
the nodes belonging to the $\ell$-th layer have the same activation
function $\phi_\ell$. The $k$-th output of the NN can then be written
recursively as~\mycite{AbdulKhalek:2020uza}:
\begin{equation}\label{eq:nnexplicit}
\begin{array}{rcl}
  \displaystyle 
  N_k({\pmb \xi};\{\omega_{ij}^{(\ell)},\theta_{i}^{(\ell)}\}) &=&\displaystyle
    \phi_L\left(\sum_{j^{(1)}}^{N_{L-1}}\omega_{k
    j^{(1)}}^{(L)}y_{j^{(1)}}^{(L-1)}+\theta_{k}^{(L)}\right)\\
\\
&=&\displaystyle
    \phi_L\left(\sum_{j^{(1)}=1}^{N_{L-1}}\omega_{k
    j^{(1)}}^{(L)}\phi_{L-1}\left(\sum_{j^{(2)}=1}^{N_{L-2}}\omega_{j^{(1)}
    j^{(2)}}^{(L)}y_{j^{(2)}}^{(L-2)}+\theta_{j^{(1)}}^{(L-1)}\right)+\theta_{k}^{(L)}\right)\\
&=&\dots\,.
\end{array}
\end{equation}
The nesting in Eq.~(\ref{eq:nnexplicit}) continues until the input
layer $(L-4)$ is reached and is referred to as forward propagation.

\myparagraph{Backward propagation with analytical derivatives} The forward propagation is the mode in which a NN evaluates its outputs based on a given input by a successive evaluation of the layers. Now, I discuss the backward propagation, the mode in which the NN \textit{learn} by propagating back into the parameters the error found between its output and the data. Backward propagation relies on computing the gradient of a loss function (in our case $\chi^2$) w.r.t. the weights and biases of the NN by means of a repeated application of the chain-rule. This gradient is then used by a minimisation algorithm that adopts a gradient-descent strategy (see Sect.~\ref{s2:minimisation}).

In order to illustrate this mode of propagation~\mycite{AbdulKhalek:2020uza}, we consider for simplicity one single experimental point
measured at ${\pmb \xi}\in\mathbb{R}^m$ with central value
${\pmb \sigma}\in\mathbb{R}^n$ and standard deviation
$\pmb s\in\mathbb{R}^n$ that we want to fit with a NN
${\pmb N}:\mathbb{R}^m\rightarrow\mathbb{R}^n$ with $L$ layers and
parametrised by a set of weights and biases
$\{\omega_{ij}^{(\ell)},\theta_{i}^{(\ell)}\}$. The corresponding
$\chi^2$ reads:
\begin{equation}\label{eq:chi2def}
  \chi^2[\{\omega_{ij}^{(\ell)},\theta_{i}^{(\ell)}\}] =  \sum_{k=1}^n\left(\frac{ N_k({\pmb \xi};\{\omega_{ij}^{(\ell)},\theta_{i}^{(\ell)}\}) - \sigma_k}{s_k}  \right)^2\,,
\end{equation}
where $N_k$ is the $k$-th output of the NN.
The derivative of the $\chi^2$ in
Eq.~(\ref{eq:chi2def}) w.r.t. the weight
$\omega_{ij}^{(\ell)}$, relevant to the computation of the gradient,
takes the form:
\begin{equation}\label{eq:chi2derdef}
  \frac{\partial\chi^2}{\partial \omega_{ij}^{(\ell)}} =  2\sum_{k=1}^n\left(\frac{ N_k({\pmb \xi};\{\omega_{ij}^{(\ell)},\theta_{i}^{(\ell)}\}) - \sigma_k}{s_k^2}  \right) \frac{\partial N_k}{\partial \omega_{ij}^{(\ell)}}\,,
\end{equation}
and similarly for the derivative w.r.t. the bias
$\theta_{i}^{(\ell)}$.  Eq.~(\ref{eq:chi2derdef}) reduces the
computation of the derivatives of the $\chi^2$ in
Eq.~(\ref{eq:chi2def}) to the computation of the derivatives of
${\pmb N}$. In this respect, the feed-forward structure of the NN in
Eq.~(\ref{eq:nnexplicit}) is crucial to work out an explicit
expression for such derivatives. 

Starting with the following definitions:
\begin{equation}
\begin{array}{l}
  \displaystyle x_i^{(\ell)} =
  \sum_{j=1}^{N_{\ell-1}}\omega_{ij}^{(\ell)}y_{j}^{(\ell-1)}+\theta_{i}^{(\ell)},\qquad
  \displaystyle y_i^{(\ell)} =
  \phi_\ell\left(x_i^{(\ell)}\right),\qquad
  \displaystyle z_i^{(\ell)} = \phi'_\ell\left(x_i^{(\ell)}\right)
\end{array}
\end{equation}
we can derive the following expressions for the derivatives (see Ref.~\mycite{AbdulKhalek:2020uza} for more details):
\begin{equation}\label{eq:derivativesfinal}
\begin{array}{rcl}
  \displaystyle\frac{\partial N_k}{\partial \theta_{i}^{(\ell)}} =\displaystyle
                                                                     {\Sigma}_{ki}^{(\ell)}
                                                                     z_i^{(\ell)}, \qquad
  \displaystyle\frac{\partial N_k}{\partial \omega_{ij}^{(\ell)}} =\displaystyle{\Sigma}_{ki}^{(\ell)} z_i^{(\ell)} y_j^{(\ell-1)}
\end{array}
\end{equation}
where ${\Sigma}_{ki}^{(\ell)}$ is the $(k,i)$ entry of the matrix:
\begin{equation}
\mathbf{\Sigma}^{(\ell)} = \prod_{\alpha=L}^{\ell+1}\mathbf{S}^{(\alpha)}, \qquad
  S_{ij}^{(\ell)} = \frac{\partial y_i^{(\ell)}}{\partial
      y_j^{(\ell-1)}} = z_i^{(\ell)}\omega_{ij}^{(\ell)}
\end{equation}
The derivative expressions in Eq.~(\ref{eq:derivativesfinal}) can finally be used
to compute the gradient of the $\chi^2$ through
Eq.~(\ref{eq:chi2derdef}) (and its respective for the bias
$\theta_{i}^{(\ell)}$).  From the point of view of a numerical
implementation, it is crucial to notice that the matrix
$\mathbf{\Sigma}^{(\ell)}$ can be computed recursively moving
\textit{backwards} from the output layer as $\mathbf{\Sigma}^{(\ell-1)} = \mathbf{\Sigma}^{(\ell)} \cdot \mathbf{S}^{(\ell)}$, starting from the initial condition: $\mathbf{\Sigma}^{(L)} = \mathbf{I}$.
This feature allows one to compute the derivatives w.r.t. all free
parameters of a NN with a \textit{single} iteration of the chain-rule. We point out that the iterative nature of
Eq.~(\ref{eq:derivativesfinal}) is a direct consequence of the
structure of the object being derived, \textit{i.e.} a feed-forward
NN. Therefore, Eq.~(\ref{eq:derivativesfinal}) does \textit{not}
generally apply to any artificial NN.

\begin{figure}[t]
\centering
\includegraphics[width=0.7\textwidth]{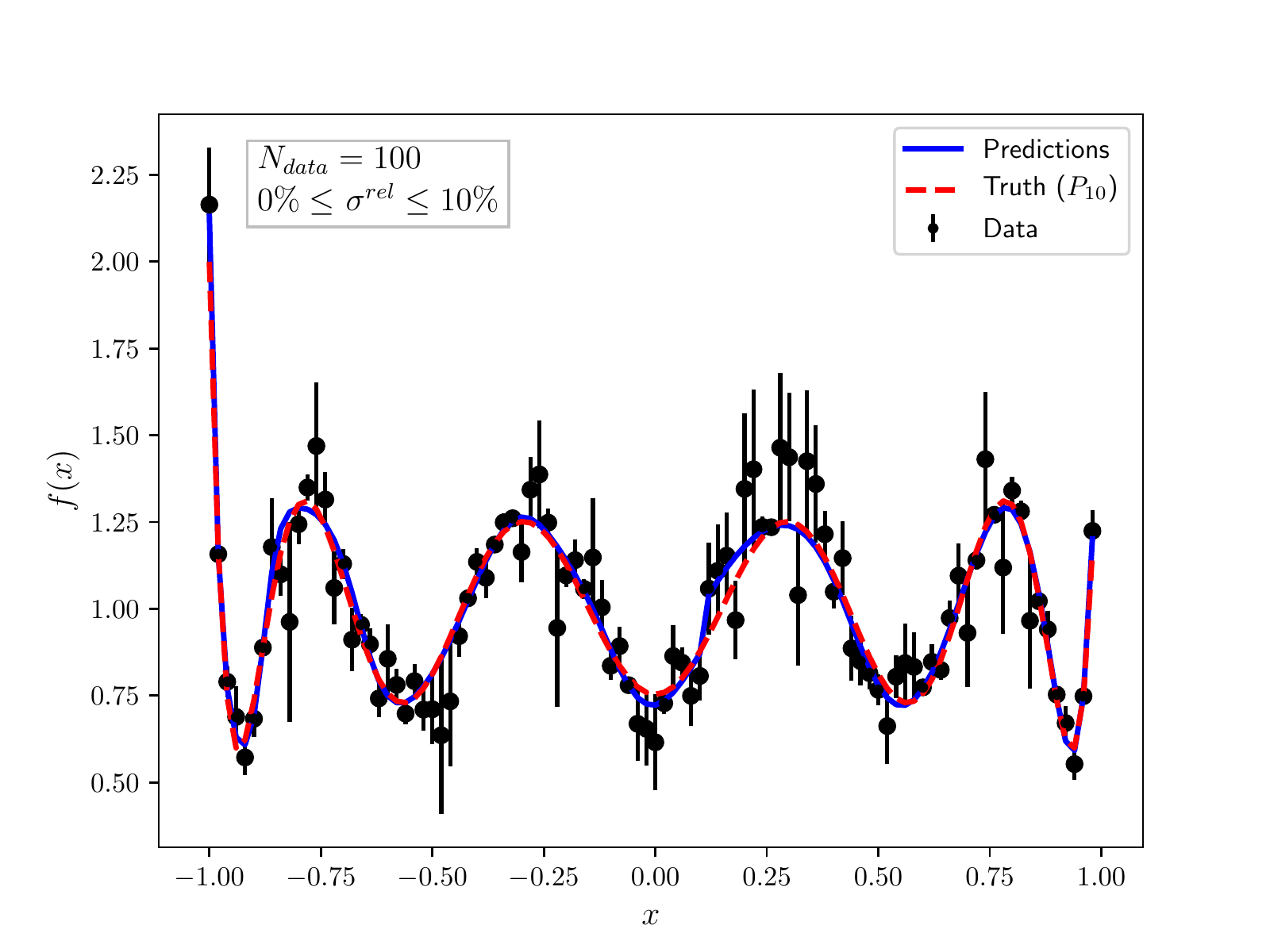}
\vspace{-0.3cm}
\caption{\label{fig:P10_fit} Fit of a NN with architecture $[1,25,1]$
  to 100 pseudodata points generated according to
  Eq.~(\ref{eq:legendrepoints}). The black points represent the
  pseudodata with the corresponding uncertainty, the red line is
  underlying law (truth), and the blue line is the result of a
  1000-iteration fit.}
\end{figure}

\myparagraph{Universality} 
One of the most important features of NNs is its compliance with the \textit{universal approximation theorem}~\cite{csaji2001approximation} that states that a feed forward NN with a single layer is sufficient to represent any function within the range of the given inputs. To highlight this feature, I take from Ref.~\mycite{AbdulKhalek:2020uza} the example of inferring a NN (as implemented in the \texttt{NNAD} library) from a pseudodata generated using an oscillating Legendre polynomial as an underlying
law.
To this purpose, $N_\text{dat}=100$ pseudodata points are generated over an equally-spaced grid $\{\xi_i\}$, with
$i=1,\dots,N_\text{data}$ and $\xi_i \in [-1,1]$. The corresponding
sets of central values $\{\sigma_i\}$ and uncertainties $\{s_i\}$ is
obtained as:
\begin{equation}\label{eq:legendrepoints}
\begin{array}{l}
    \sigma_i = [1 + P_{10}(\xi_i)]\times\mathcal{G}(1,0.1),\qquad s_i = [1+ P_{10}(\xi_i)]\times\mathcal{G}(0,0.1)
\end{array}
\end{equation}
where $P_{n}$ is the Legendre polynomial of degree $n$ and
$\mathcal{G}(\sigma,s)$ is the normal distribution with mean value
$\sigma$ and standard deviation $s$. The shift by $1$ in both
equations in Eq.~(\ref{eq:legendrepoints}) has the goal to make the
underlying law positive definite and facilitate thus the generation of
the pseudodata.
The model used to fit the data is a NN with one input node, one output
node, and a single hidden layer with 25 fully-connected nodes
(architecture $[1,25,1]$) for a total of 76 free parameters.
Fig.~\ref{fig:P10_fit} shows the result of a fit of 1000 iterations.
It is evident that the fitted NN reproduces the underlying law quite
accurately within the given pseudodata uncertainties. This provides a numerical and visual demonstration of the importance of the universal approximation theorem.

\begin{figure}[!h]
  \begin{center}
    \makebox[\textwidth]{\includegraphics[width=0.8\textwidth]{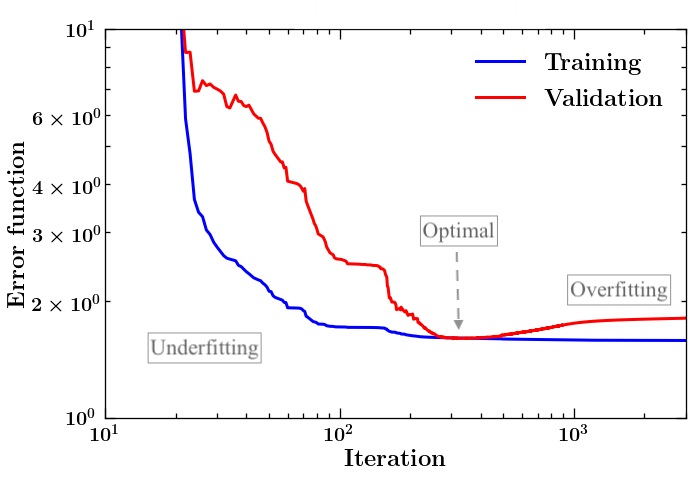}}
  \end{center}
  \vspace{-0.5cm}
\caption{\small A representative figure of the cross-validation method where the $\chi^2_{\text{tr}}$ and $\chi^2_{\text{val}}$ are displayed as a function of the minimisation iterations. The underfitting, optimal and overfitting regimes are also indicated.}
 \label{fig:CrossValidation}
\end{figure}

\myparagraph{Cross-validation and stopping criterion} Due to their convoluted structure and universality, NNs can be highly redundant for a given set of data. This can happen either due to a sparse data, or a large NN architecture w.r.t minimisation problem considered. In that case, the NN risk inferring the statistical fluctuations in the data rather the underlying functional form or physical law. One of the methods to prevent this feature, commonly called \textit{overfitting}, is the cross-validation. The basic idea behind this method is to first separate the input data sets into conditionally or marginally disjoint training (tr) and validation (val) data sets (see Sect.~\ref{s2:likelihood}).  Second, minimise only the $\chi^2_{\text{tr}}$ of the training data sets while evaluating the $\chi^2_{\text{val}}$ of the validation ones. The latter does not contribute in the backward propagation and the inference of parameters but serves as a measure of the NN's generalisation or predictive power on new data. If during minimisation, the $\chi^2_{\text{tr}}$ starts to continuously decreasing while the $\chi^2_{\text{val}}$ simultaneously increasing, this could be interpreted as a case of overfitting as shown in the representative Fig.~\ref{fig:CrossValidation}.

There are many strategies that could be adopted to deduce the set of optimised NN parameters that do not suffer from overfitting. One of these is the look-back stopping criterion presented for the first time in \texttt{NNPDF} fits in Ref.~\cite{Butterworth:2015oua}. This method dictates the inference to stop if the $\chi^2_{\text{val}}$ reaches some assigned threshold and subsequently ceases to decrease after a certain number of iterations $N_{\text{it}}^{\text{max}}$. The inference is then classified as successful and the set of parameters that minimise $\chi^2_{\text{val}}$ are selected as the best-fit parameters (look-back).

\myparagraph{Hyperparameters} 
Hyperparameters refer to the group of tunable parameters that are user-specified in the minimisation problem such as: NN architecture (number of layers, number of neurons in a layer), activation functions, initialisation of weights and biases, minimisation strategy, number of minimisation iterations, training/validation fraction, etc.
The choices of hyperparameters affect significantly the training process specially when large NN architectures and large data sets are considered in the minimisation problem. The procedure adopted to automate the tuning of these hyperparameters is called \textit{hyperoptimisation} (hyperparameter optimisation) that aims to find a set that yields an optimal minimisation model for some given data sets. For instance, in Ref.~\cite{Carrazza:2019mzf}, the authors present a new regression model for the determination of PDFs relying on a cross-validation mechanism through a hyperoptimisation procedure.

In this thesis however, we choose the hyperparameters by a trial-and-error (manual) approach instead of an automated hyperoptimisation algorithm. That is mainly due to the relative simplicity of NN models adopted as well as the moderate number and features of data included.

\subsection{Minimisation algorithms} \label{s2:minimisation}
In this section I address the optimisation problem Eq.~(\ref{eq:MLE}), \textit{i.e.} minimising the $\chi^2(\bm{\theta})$ function with respect to a set of parameters $\bm{\theta}$. It is generally cumbersome or even impossible to find the exact \textit{global minimum} of the $\chi^2$ analytically (direct methods). It is therefore vastly more efficient to solve a series of approximations to the original problem instead (iterative methods).

I will discuss the main three iterative minimisation algorithm used to achieve the results in the subsequent chapters:
\begin{enumerate} 
  \item \textbf{Genetic algorithms (GAs)} adopted for the \texttt{NNPDF} inference of proton PDFs (chapter~\ref{chap:PDF}).
  \item  \textbf{Stochastic gradient descent (SGD)} and in particular the Adaptive Moment Estimation algorithm (ADAM)~\cite{kingma2017adam} as provided by \texttt{TensorFlow}~\cite{tensorflow2015-whitepaper} and adopted for the \texttt{nNNPDF} inference of nuclear PDFs (chapters~\ref{chap:nNNPDF10}, \ref{chap:nNNPDF20} and \ref{chap:nNNPDF30}).
  \item \textbf{Trust-region} and in particular the Levenberg-Marquardt algorithm (LM)~\cite{levenberg1944method,marquardt1963algorithm} as provided by \texttt{ceres-solver}~\cite{ceres-solver} and  adopted for \texttt{MAPFF} inference of pion FFs (chapter~\ref{chap:FF}).
\end{enumerate}

The GA is a stochastic, randomness-based method of optimisation. Such algorithms do not benefit directly from the topology of the $\chi^2$ in space of parameters and therefore do not rely on gradients. Roughly speaking, the group of these methods rely on random-trial and error to scan for minimum of $\chi^2$. Some of the advantages of a GA are the ability of dealing with complex $\chi^2$ topologies being linear or nonlinear, continuous or discontinuous, or with random noise. 
However, GAs relies heavily on the hyperparameters choices 
and any inappropriate choice will make it difficult for the algorithm to converge~\cite{yang2020nature}.

All gradient based minimisation algorithms such as stochastic gradient descent and trust-region, considered as deterministic methods, depend the evaluation of the $\chi^2$ and its derivatives at arbitrary points in parameter space. Therefore, both the $\chi^2$ and its Jacobian (derivatives w.r.t. all parameters) are essential ingredients to solve the minimisation problem. There are three main approaches of calculating the $\chi^2$ Jacobian:
\begin{itemize}
  \item \textbf{Analytic}: in which the user provides the $\chi^2$ derivatives expression themselves as we did for the neural network in Sect.~\ref{s2:neural_networks} based on Ref.~\mycite{AbdulKhalek:2020uza}. 
  \item \textbf{Automatic}: in which the analytic derivatives are computed automatically by the decomposition of differentials by means of the chain-rule (more details in Ref.~\cite{ceres-solver}).
  \item \textbf{Numeric}: in which the derivatives are computed numerically using finite differences methods (more details in Ref.~\cite{ceres-solver}).
\end{itemize}
Some of the advantages of gradient based minimisation algorithms are their search efficiency and the guaranteed convergence to the global minimum for a convex\footnote{A convex function has one minimum.} $\chi^2$ and a local minimum for a non-convex one. However, these algorithms could be slow for complex topologies. The choice of optimal search-steps sizes or \textit{learning rate} is often difficult. Minimising highly non-convex $\chi^2$ could lead the algorithm to be stuck in a saddle point\footnote{Saddle point of a function: not its local minimum or maximum, rather a point where the derivatives in orthogonal directions are all zero.} or a local-minimum.

\begin{figure}[!h]
  \centering
  \includegraphics[width=0.8\textwidth]{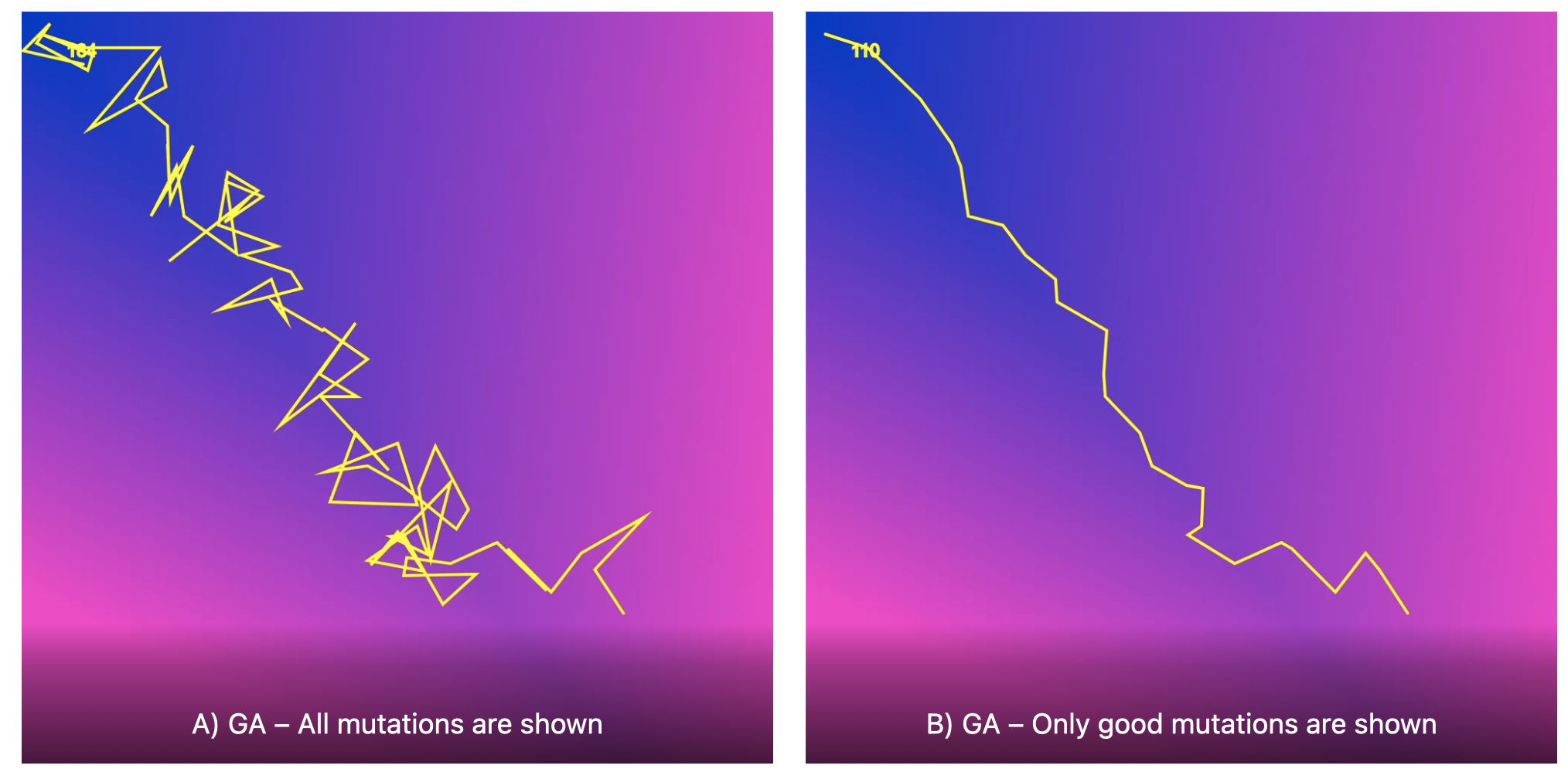}
  \caption{\label{fig:GA_example} A representative plot generated by \href{github.com/hunar4321/Genetic\_Algorithm}{github.com/hunar4321/Genetic\_Algorithm} showing the steps taken by a GA towards the global minimum (deep blue area).}
\end{figure}

\myparagraph{Genetic algorithm} 
The \texttt{NNPDF3.1}~\mycite{Khalek:2021ulf,AbdulKhalek:2020jut,AbdulKhalek:2019ihb} framework (Chapter~\ref{chap:PDF}) relies on a GA implemented as discussed in Ref.~\cite{Ball:2014uwa}.
GAs~\cite{whitley1994genetic,Ball:2014uwa} are a family of computational models inspired by the process of natural selection. The  initial iteration starts with a \textit{generation}\footnote{The population in each iteration called a generation.} of randomly generated \textit{population} of parameters $\{\bm{\theta}_1,\,\dddot{},\bm{\theta}_{N_\text{pop}}\}$. The $\chi^2(\bm{\theta}_i)$ is then evaluated for every individual $i$ in the population. The individuals with the least $\chi^2$ are stochastically selected from the population, and each of these individual's parameters are modified randomly to form a new generation. The new generation of candidate solutions is then used for the next iteration. The algorithm terminates when a satisfactory $\chi^2$ has been reached for the population as shown in the qualitative Fig.~\ref{fig:GA_example}.

\myparagraph{Stochastic gradient-descent} 
The \texttt{nNNPDF}~\mycite{AbdulKhalek:2019mzd,AbdulKhalek:2020yuc,AbdulKhalek:2021xxx1} framework (Chapters~\ref{chap:nNNPDF10}, \ref{chap:nNNPDF20} and \ref{chap:nNNPDF30}) relies on ADAM~\cite{kingma2017adam} from the SGD family provided by \texttt{TensorFlow}~\cite{tensorflow2015-whitepaper}. An open source \texttt{Python} ML library in which the gradients of the $\chi^2$ can be computed via automatic differentiation. I note that this algorithm was for the first time introduced in the NNPDF collaboration by the \texttt{nNNPDF} group in a separate framework to fit nPDFs. 

\begin{figure}[!h]
  \centering
  \includegraphics[width=0.8\textwidth]{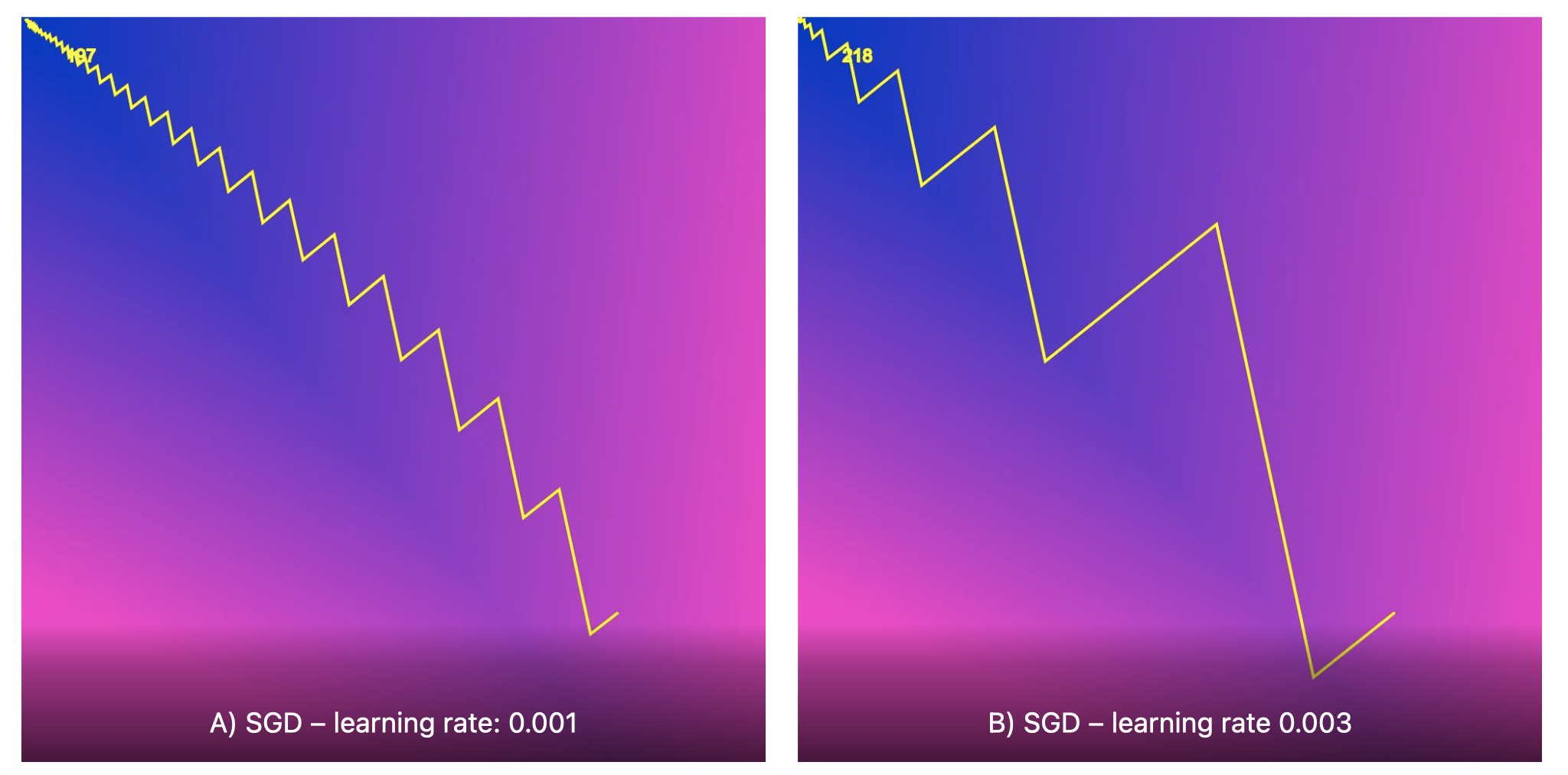}
  \caption{\label{fig:SGD_example} A representative plot generated by \href{github.com/hunar4321/Genetic\_Algorithm}{github.com/hunar4321/Genetic\_Algorithm} showing the steps taken by a SGD towards the global minimum (deep blue area) and its sensitivity to the learning rate.}
\end{figure}

The general idea of gradient descent (GD) is to take repeated steps in the opposite direction of the $\chi^2_\text{tot}$ Jacobian in parameter space. In a generic GD algorithm, the $\chi^2_\text{tot}$ is computed over all training data points available in the problem per iteration. The aim is that eventually these iterative steps will lead to the minimum of $\chi^2$. The main different feature SGDs have over GDs is that in case of the latter, the parameters-tuning happens based on the $\chi^2_\text{subset}$ Jacobian per iteration. The $\chi^2_\text{subset}$ is computed only on a subset of the training data points picked randomly every iteration (stochastic). Therefore, SGDs are considered to approximate the gradients, saving a lot of time compared to the case of full-data and
with a large enough number of iterations SGDs converge eventually to a minimum that describes well all the training data. The SGD mode of operation is shown in the qualitative Fig.~\ref{fig:SGD_example}.

One of the drawbacks of the GD approach, which is partially avoided by using GA-types of minimisers, is the risk of ending up trapped in local minima as well as the arbitrary choice of learning rate. To ensure that such situations are avoided as much as possible, the ADAM algorithm was used in \texttt{nNNPDF} to perform SGD. 
ADAM computes individual adaptive learning rates for different parameters from estimates of first and second moments of the gradients. By adjusting the learning rate of the parameters using averaged gradient information from previous iterations, local minima are more easily bypassed in the training procedure, which not only increases the likelihood of ending in a global minima but also significantly reduces the training time.

\myparagraph{Trust-region} 
The \texttt{MAPFF}~\mycite{Khalek:2021gxf} framework (Chapter~\ref{chap:FF}) relies on Levenberg-Marquardt (LM)~\cite{levenberg1944method,marquardt1963algorithm} from the GD family provided by \texttt{ceres-solver}~\cite{ceres-solver}. An open source \texttt{C++} library for modelling and solving large, complicated optimisation problems.

\begin{figure}[!h]
  \centering
  \includegraphics[width=0.8\textwidth]{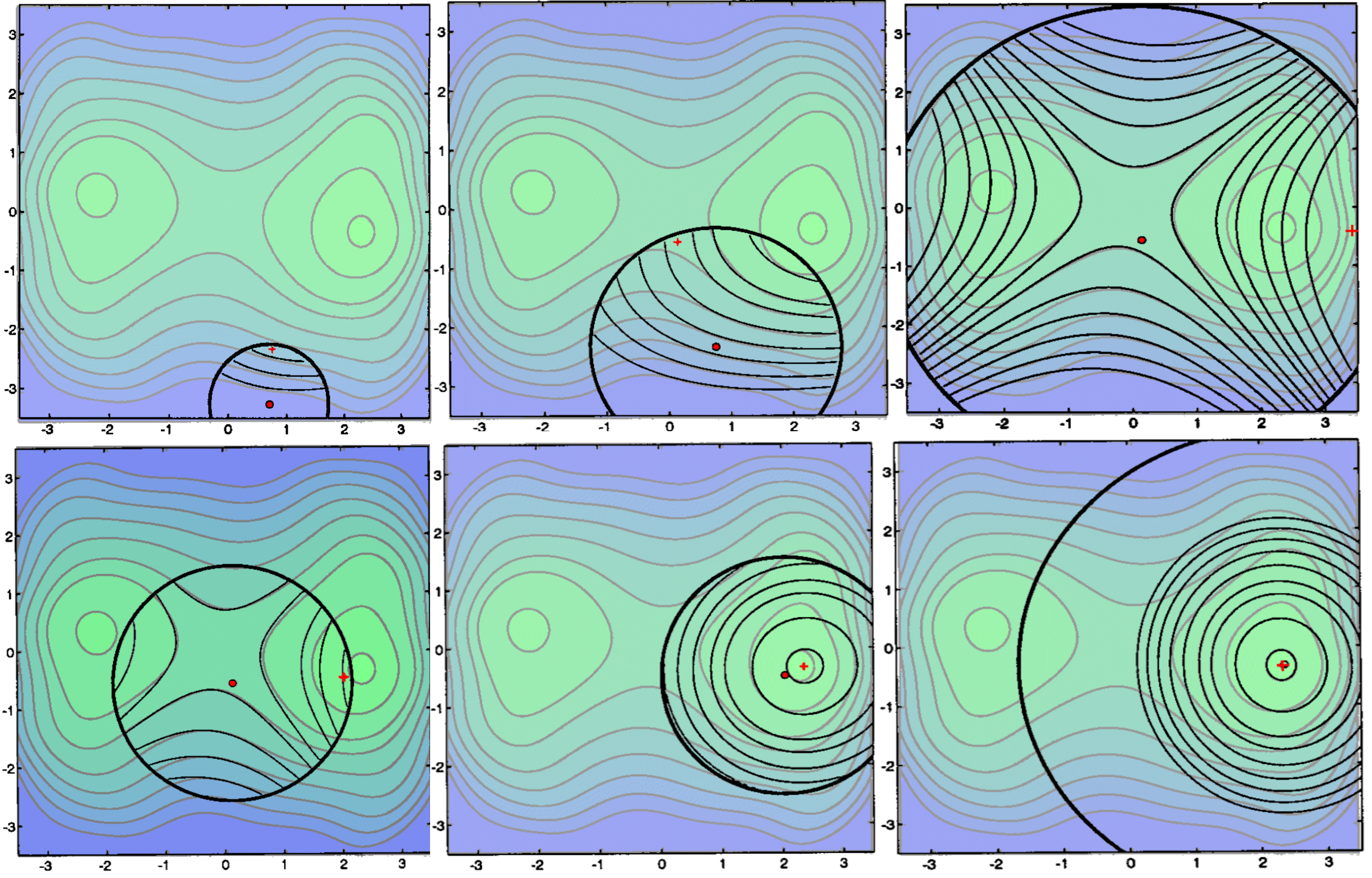}
  \caption{\label{fig:TrustRegion_example} A representative plot taken from Ref.~\cite{VANDENBERGHEN2005157} showing the steps taken by a Trust-region algorithm, towards the global minimum (on the right side of the figure). To be read from left to right and row by row.}
\end{figure}

The general idea of trust-region minimisation algorithms is to approximate the $\chi^2$ using a quadratic function over a subset of the search space known as the trust region. If a set of parameters leading to a small $\chi^2$ is found within the trust region, then the region is expanded, otherwise it is contracted and the minimisation problem is solved again. While the \textit{line-search} approach first finds a descent direction along which the $\chi^2$ will be reduced and then computes a step size, the trust-region approach first choose a step size (the size of the trust region) and then a step direction. The Trust-region algorithms mode of operation is shown in the qualitative Fig.~\ref{fig:TrustRegion_example}.

The LM~\cite{levenberg1944method,marquardt1963algorithm} is one of the most popular algorithm for solving non-linear least squares problems. It combines two minimisation algorithms: the GD method where the sum of the squared errors is reduced by updating the parameters in the steepest-descent direction, and the Gauss-Newton (GN) method where the sum of the squared errors is reduced by assuming the least squares function is locally quadratic in the parameters, and finding the minimum of this quadratic. Therefore the LM method acts more like a GD method when the parameters are far from their optimal value, and more like the GN method when the parameters are close to their optimal value~\cite{gavin2019levenberg}.
  \chapter{Proton parton distribution functions}
\label{chap:PDF}
\vspace{-1cm}
\begin{center}
  \begin{minipage}{1.\textwidth}
      \begin{center}
          \textit{This chapter is based on Refs.~\cite{Ball:2014uwa, Ball:2017nwa} and my results in Refs.~\mycite{AbdulKhalek:2020jut,AbdulKhalek:2019ihb,AbdulKhalek:2019bux}}.
      \end{center}
  \end{minipage}
  \end{center}

\myparagraph{Introduction} The continuous study of the composite nature of nucleons (protons and neutrons) over the past decades, lead to the formulation of QCD, the theory of the strong force that we discuss in Chapter~\ref{chap:QCD}. 
Our current understanding is that nucleons are bound states of confined partons (quarks and gluons) that collectively give its macro-properties that we measure, such as electric charge, spin and mass. 
%

%
%
At low energy scales, partons strongly interact and radiate into each other making them indescribable by perturbative QCD. The latter is only valid at high energy scales where the strong coupling constant $\alpha_s$ is small enough, thus in the case of weakly interacting partons. As at matter of fact, we saw in Sect.~\ref{s2:DGLAP}, that this is translated by an infrared divergence that emerges in the calculation of structure functions for a fixed-$x$ at a given $Q^2$. This divergence can be remedied by taking the finite difference of the divergent structure function between two different energy scales. This difference allows then the introduction of a factorisation scale that separate the perturbative from non-perturbative contributions that become ``absorbed'' by bare quantities, the parton distribution functions (PDFs). These renormalised non-perturbative objects can in turn be measured from experimental data and their running in $Q^2$ is governed by the perturbative DGLAP equations.
%


The nucleon substructure was first resolved by deep inelastic scattering at SLAC in 1966. However, due to the QCD collinear factorisation theorem, PDFs are universal and therefore can be determined from a variety of hard-scattering cross sections in lepton-proton and proton-proton collisions (such as at the LHC~\cite{Gao:2017yyd,Ethier:2020way,Forte:2020yip}), some of which are displayed in Sect.~\ref{s2:Flavour_separation}. 

\myparagraph{Motivation} In addition to improving our understanding of the nucleon substructure, there are various motivations behind the continuous efforts of studying PDFs. To start with, PDFs currently represent one of the dominant theoretical uncertainties for the determination of the Higgs boson couplings. This is particularly important in the
search for subtle deviations from the Standard Model (SM) predictions at
present~\cite{deFlorian:2016spz} and future~\mycite{Cepeda:2019klc}
high-energy colliders. 

Moreover, the production of new high-mass resonances, as predicted by many Beyond the Standard Model (BSM) scenarios~\cite{Beenakker:2015rna} is mainly sensitive to the region of large momentum fraction $x$ where data is generally very scarce. This is where measurements involving deuterium targets still play a significant role in the determination of proton PDFs~\cite{Ball:2018twp,Ball:2020xqw}, in particular to separate the up and down flavours.

In addition, PDFs also affect the precision measurement of the SM parameters at hadron colliders, such as the W-boson mass~\cite{Bozzi:2015hha,Bozzi:2015zja,Bozzi:2011ww,Aaboud:2017svj} or the strong coupling constant $\alpha_s$~\cite{Khachatryan:2014waa,Aaboud:2017fml, Chatrchyan:2013txa, Chatrchyan:2013haa,Ball:2018iqk,Forte:2020pyp}. These in turn can be also sensitive to BSM effects~\cite{Becciolini:2014lya, Dimopoulos:1981yj} and in many cases PDF uncertainties represent one of the limiting factors of their measurement.
Beyond the LHC, PDFs also affect the predictions for signal~\cite{CooperSarkar:2011pa} and background~\cite{Gauld:2015kvh,Garzelli:2016xmx,Zenaiev:2015rfa,Gauld:2016kpd} events at ultra-high energy neutrino telescopes.

Finally, PDFs play a major role at the moment in constraining nuclear PDFs (nPDFs), as we will see in Chapters~\ref{chap:nNNPDF10}, \ref{chap:nNNPDF20} and \ref{chap:nNNPDF30} where we discuss the nuclear PDF statistical framework \texttt{nNNPDF} and Sect.~\ref{s1:EIC} where we analyse the projection of the Electron-Ion collider. In our recently developed framework \texttt{nNNPDF} ~\mycite{AbdulKhalek:2019mzd,AbdulKhalek:2020yuc} we impose that nPDFs reduce to PDFs in the limit where the nucleus is a nucleon.  Other collaborations rely on determining the \textit{modification} of the PDFs due to nuclear effects rather than determining the nPDFs themselves, either by extending the parameterisation of PDFs~\cite{Kovarik:2015cma} or fitting a modification ratio~\cite{Eskola:2016oht}. In all these cases, PDFs are currently a crucial element in any nuclear PDF global analysis, where in particular it imposes heavy constraint on light-nuclei but also shapes our understanding of the lead PDF that is mainly considered in pA collisions at the LHC.

A number of collaborations provide regular updates of their PDF sets~\cite{Ball:2017nwa, Hou:2019efy, Bailey:2020ooq, Moffat:2021dji}. There are a range of differences between these analyses, arising from the selection of the input fitted dataset, the theoretical calculations of cross sections, methodological choices for the parametrisation of PDFs, the estimate and propagation of PDF uncertainties, and the treatment of experimental uncertainties. 
Despite these differences, it has been shown that, under some well-specified conditions, PDF sets can be statistically combined among them into a unified set. The most popular realisation of this paradigm are the \texttt{PDF4LHC15}  set~\cite{Butterworth:2015oua,Gao:2013bia,Carrazza:2015hva,Carrazza:2015aoa}, a combined
set of \texttt{CT14} ~\cite{Dulat:2015mca}, \texttt{MMHT2014}~\cite{Harland-Lang:2014zoa} and
\texttt{NNPDF3.0}~\cite{Ball:2014uwa}, using the Monte Carlo (MC) method (see Sect.~\ref{s2:monte_carlo}).

Now that we've established the broad importance of PDFs, let us discuss their source of uncertainties. We classify these into three classes: 
\begin{enumerate}[start=0,label={(\bfseries S\arabic*):}]
    \item The theory used to describe them including:
    \begin{itemize}
      \item Missing higher order uncertainty from fixed-order calculations.
      \item Uncertainty on the $\alpha_s$ determination.
      \item Heavy quark mass effects.
      \item Nuclear effects, mainly when considering data involving deuterium.
    \end{itemize} 
    \item The data uncertainties used in the global analyses.
    \item The methodology used for their statistical inference.
\end{enumerate}
PDFs are continuously confronted by new and more precise data mainly from hadron colliders like the LHC but also from the EIC in the near future.
We can asses this precision's order of magnitude by tracking the LHC integrated luminosity for example (defined in Eq.~\ref{eq:int_lumi}). Since the beginning of its operation in 2009 and until the end of Run 2 (2015-2018), the total integrated luminosity at the LHC was 189.3 $\text{fb}^{-1}$ for each of ATLAS and CMS, whereof 160 $\text{fb}^{-1}$ were accumulated during Run 2 alone~\cite{Steerenberg:2645638}. Run 3 is expected to start in 2021, achieving a total integrated luminosity of over 300 $\text{fb}^{-1}$ before the end of 2023. The HL-LHC will take over in 2026, with the goal of reaching 3 $\text{ab}^{-1}$ by 2037~\mycite{Azzi:2019yne, Cepeda:2019klc}. 

\myparagraph{Outline} In Sect.~\ref{s1:PDF_theory_uncertainties}, I discuss the uncertainty of type-\textbf{(S0)} where I present the first extraction of the proton PDFs that accounts for the missing higher order uncertainty (MHOU) in the fixed-order QCD calculations used in PDF determinations. This study is based on Refs.~\mycite{AbdulKhalek:2019bux, AbdulKhalek:2019ihb}.
%
In Sect.~\ref{s1:PDF_NNLO_jet}, I discuss the uncertainty of type-\textbf{(S1)} where I present a systematic investigation of single jet and dijet production at hadron colliders from a phenomenological point of view and their impact on PDFs uncertainties. This study is based on Ref.~\mycite{AbdulKhalek:2020jut}.
The results in the theory uncertainties Sect.~\ref{s1:PDF_theory_uncertainties}, single jet and dijet production Sect.~\ref{s1:PDF_NNLO_jet} are performed with the \texttt{NNPDF3.1}~\cite{Ball:2017nwa} framework that I discuss next. 

\section{\texttt{NNPDF3.1} framework}
\label{s1:NNPDF31_framework}
    The \texttt{NNPDF3.1} framework~\cite{Ball:2017nwa} relies on the Monte Carlo method discussed in Sect.~\ref{s2:monte_carlo} to infer/fit proton PDFs from experimental data. One of the main features of this framework is that the PDFs are parameterised with neural networks. The Universal Approximation Theorem discussed in Sect.~\ref{s2:neural_networks} makes neural networks very suitable objects to fit PDFs, as the latter's functional form is not theoretically motivated. The \texttt{NNPDF3.1} PDF sets are released at LO, NLO, and NNLO accuracy and validated by closure tests discussed in Refs.~\cite{DelDebbio:2004xtd, Ball:2014uwa}.

    In Sect.~\ref{s2:NNPDF31_data}, I will start by listing the various data sets by processes included in the \texttt{NNPDF3.1} inference as well as their kinematic coverage then in Sect.~\ref{s2:NNPDF31_parameterisation}, I present the parameterisation adopted in this framework including the imposition of the physical constraints defined in Sect.~\ref{s2:Physical_constraints}. Finally, In Sect.~\ref{s2:fast_evolution}, I discuss the fast evolution method adopted to transform the computation of the complicated convolutions in the cross section expressions into simple matrix multiplication.

    \subsection{Experimental data} \label{s2:NNPDF31_data}
    The data included in the \texttt{NNPDF3.1} global analysis is discussed in length in Sect.~2 of Ref.~\cite{Ball:2017nwa} and covers a broad type of processes and experiments:
    \begin{itemize}
      \item \textbf{Fixed-target neutral-current (NC) DIS}: NMC~\cite{Arneodo:1996kd,Arneodo:1996qe}, SLAC~\cite{Whitlow:1991uw} and BCDMS~\cite{Benvenuti:1989rh}.
      \item \textbf{Fixed-target charged-current (CC) DIS}: CHORUS~\cite{Onengut:2005kv} and NuTeV~\cite{Goncharov:2001qe,Mason:2006qa}.
      \item \textbf{Collider charged-current (NC and CC) DIS}: HERA~\cite{Abramowicz:2015mha,Abramowicz:1900rp,Aaron:2009af,Abramowicz:2014zub}.
      \item \textbf{Fixed-target Drell-Yan}: E866~\cite{Webb:2003ps,Webb:2003bj,Towell:2001nh} and E605~\cite{Moreno:1990sf}.
      \item \textbf{Collider Drell-Yan}: CDF~\cite{Aaltonen:2010zza} and D0~\cite{Abazov:2007jy,Abazov:2013rja,D0:2014kma}.
      \item \textbf{Drell-Yan inclusive gauge boson, and top-pair production}: ATLAS~\cite{Aad:2013iua,Aad:2014qja,Aad:2011dm,Aaboud:2016btc,Aad:2015auj,Aad:2014kva,Aaboud:2016pbd,Aad:2015mbv}, CMS~\cite{Chatrchyan:2012xt,
    Chatrchyan:2013mza,Chatrchyan:2013tia,Khachatryan:2016pev,Khachatryan:2015oaa,
    Khachatryan:2016mqs,Khachatryan:2015uqb,Khachatryan:2015oqa} 
    and LHCb~\cite{Aaij:2012vn,Aaij:2012mda,Aaij:2015gna,Aaij:2015zlq}.
    \item \textbf{Single-inclusive jet}: ATLAS~\cite{Aad:2011fc, Aad:2014vwa, Aad:2013lpa} and CMS~\cite{Chatrchyan:2012bja,Khachatryan:2015luy,Abulencia:2007ez}.
    \end{itemize}
    We note in particular that no nuclear corrections are applied to the deuteron structure function and neutrino charged-current cross section data taken on heavy nuclei, in particular NuTeV and CHORUS. We will return to this issue in Sect.~\ref{chap:nNNPDF20} where we promote these two last datasets from the proton PDFs determination into the nPDFs one.

\begin{figure}[!h]
  \floatbox[{\capbeside\thisfloatsetup{capbesideposition={right,top},capbesidewidth=0.35\textwidth}}]{figure}[\FBwidth]
  {\caption{\small The kinematic coverage of the \texttt{NNPDF3.1}~\cite{Ball:2017nwa} data sets surviving the kinematic cuts in the $(x,\,Q^2)$ plane. For hadronic data, leading-order kinematics (see Sect.~\ref{s2:Flavour_separation}) are assumed for illustrative purposes.}\label{fig:NNPDF31_kinematics}}
  {\includegraphics[width=0.65\textwidth]{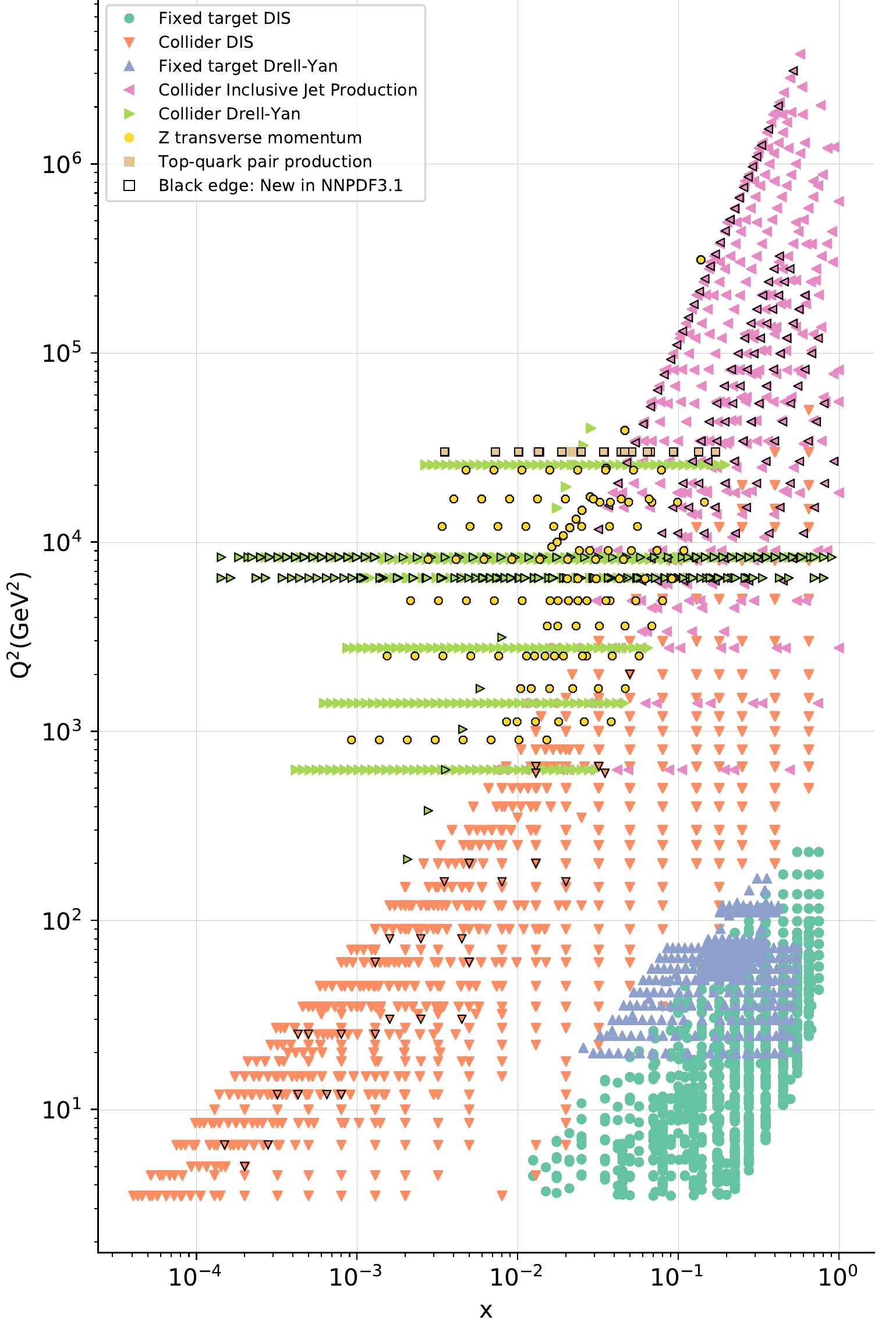}} 
  \end{figure}

    The kinematic cuts applied for DIS are $Q^2_{\text{min}}= 3.5$ GeV$^2$ and $W^2_\text{min}=12.5$ GeV$^2$, which delimits a region where higher twist effects might become relevant and thus the perturbative expansion less reliable. More details on the cuts can be found in Sect.~2 of Ref.~\cite{Ball:2017nwa}. Finally, I show in Fig.~\ref{fig:NNPDF31_kinematics} the kinematic coverage of the \texttt{NNPDF3.1} data sets in the $(x,\,Q^2)$ plane.

    \subsection{Parameterisation} \label{s2:NNPDF31_parameterisation}
    The flavours considered in \texttt{NNPDF3.1} are all the light quarks and anti-quarks, the gluon and optionally the total intrinsic charm ($\bm{c}^+ = \bm{c+\bar{c}}$) leading to a total of eight independent PDFs to fit. They are parameterised at a scale of $\mu_0=1$ GeV when charm is perturbatively generated and $\mu_0=1.65$ GeV when intrinsic charm is fitted. This ensures that the parameterisation scale is always above the charm mass when charm is independently parameterised, and below it when it is perturbatively generated. The FONLL GM-VFN has been adopted in order to include initial-state heavy quarks (see Sect.~\ref{s1:SF_heavyquarks}). This is accomplished using the formalism of Refs.~\cite{Ball:2015tna,Ball:2015dpa}.

\begin{figure}[!h]
    \begin{center}
    \makebox{\includegraphics[width=\textwidth]{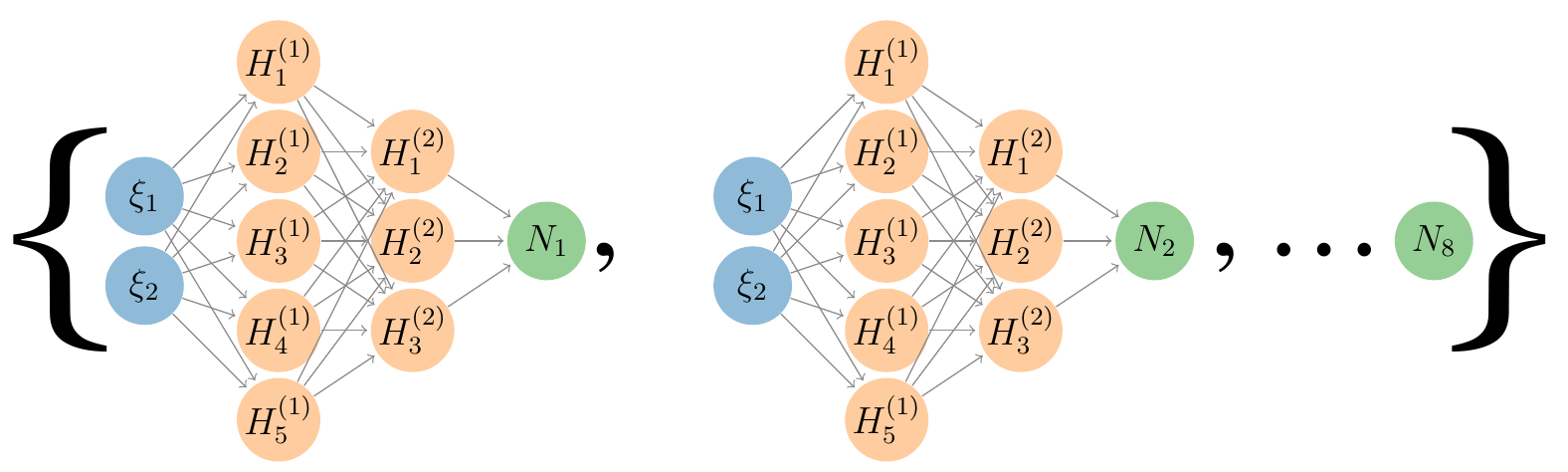}}
    \end{center}
  \caption{The NN architecture $\{2,\,5,\,3,\,1\}$ used per PDF flavour in \texttt{NNPDF3.1} having two input variables $x$ and $\log{1/x}$, two hidden layers with 5 and 3 nodes having a sigmoid activation function each and the output node with a linear activation function $N_i(x)$ of some PDF flavour $i$. There are as many independent NNs of that architecture as the number of flavours included in the fit.}
  \label{fig:NNPDF31_architecture}
  \end{figure}

    The parameterisation basis adopted in \texttt{NNPDF3.1} is defined in terms of the evolution basis detailed in Eq.~(\ref{eq:Evolution_basis}): 
    \begin{gather}
       \{\bm{g} \quad \bm{\Sigma} \quad \bm{T_3} \quad \bm{T_8} \quad \bm{V} \quad \bm{V_3} \quad \bm{V_8} \quad (\bm{c^+})\}
    \end{gather}
    which aims to speed-up the minimisation procedure. Each of the eight flavour above is parameterised in terms of an independent neural network (NN) as visualised in Fig.~(\ref{fig:NNPDF31_architecture}) whose inputs are the Bjorken-$x$ and $\log{1/x}$ and output node is multiplied by a preprocessing factor as follows:
    \begin{equation}
      f_i(x,\mu_0) = A_i \hat{f}_i(x,\mu_0), \quad \hat{f}_i(x,\mu_0)=x^{\alpha_i} (1-x)^{\beta_i} N_i(x)
    \end{equation}
where $A_i$ is an overall normalisation constant, $f_i$ and $\hat{f}_i$ denote the normalised and unnormalised PDF respectively. The preprocessing term $x^{\alpha_i}(1-x)^{\beta_i}$ is there to speed up the minimisation without biasing the fit and reflects the physical behaviour of the PDF at small- and large-$x$. The fitting of the parameters $\alpha_i$ and $\beta_i$ is explained in details in Sect.~3.2.2 of Ref.~\cite{Ball:2014uwa}.

Out of the eight normalisation constants $A_i$ three can be constrained by the valence sum rules (for up, down and strange quarks) and one by momentum sum rule as follows:
\begin{gather} \label{eq:NNPDF_sumrules}
A_g      = \frac{1-\int_0^1dx x \Sigma(x,\mu_0)}{\int_0^1 dx x \hat{g}(x,\mu_0)}, \quad A_\Sigma = A_{T_3} = A_{T_8} = 1 \nonumber\\
A_V     = \frac{3}{\int_0^1dx\hat{V}(x,\mu_0)}, \quad
A_{V_3}  = \frac{1}{\int_0^1 dx\hat{V}_3(x,\mu_0)}, \quad
A_{V_8} = \frac{3}{\int_0^1dx \hat{V}_8}
\end{gather}
which ensure that the PDFs satisfies the associated constraints as discussed in Sect.~\ref{s2:Physical_constraints} during the minimisation procedure.

The statistical framework adopted for the inference of the PDFs from experimental data relies on the Monte Carlo
sampling method explained in Sect.~\ref{s2:monte_carlo}. In 
\texttt{NNPDF3.1}, the minimisation the $\chi^2$ function defined in Eq.~(\ref{eq:chi2_general}) is performed by a Genetic Algorithm (Sect.~\ref{s2:minimisation} and Sect.~3.3.1 of Ref.~\cite{Ball:2014uwa}). Additionally, the treatment of the D'Agostini bias is achieved by means of the $t_0$-prescription of the covariance matrix as defined in Eq.~(\ref{eq:t0_covmat}). The Monte Carlo method aims at propagating the experimental
uncertainties into the space of parameters defined in our case by
 NNs. In order to do so, $N_{\text{rep}}$
replicas of the data are generated according to Eq.~(\ref{eq:MCgen}) that are then used independently for the inference.

The positivity of observables constraint discussed in Sect.~\ref{s1:positivity} is imposed by means of the Lagrange multiplier method as explained in Sect.~\ref{s2:minimisation}, which takes the following form:
\begin{equation}
\chi^2 + \sum_j^{N_{\text{pos}}} \lambda_{\text{pos}}^{(j)} \sum_i^{N_{\text{dat}}}\max{(0,-O_i^{(j)})}  
\end{equation}
where $j$ labels the positivity observable $O_i^{(j)}$ summarised in Table~\ref{tab:NNPDF31_positivity}. Every observable is evaluated at $N_{\text{dat}}$ pseudodata points that are chosen to cover an adequately large region of phase space relevant to various PDF combinations and $\lambda_{\text{pos}}^{(j)}$ is the associated Lagrange multiplier.

Finally we note that in \texttt{NNPDF3.1}, the theoretical predictions are performed using the FONLL-B general-mass variable flavour number scheme at NLO, the FONLL-C scheme at NNLO~\cite{Forte:2010ta}. All computations include target mass corrections.
The value of the strong coupling constant throughout the thesis is set to be $\alpha_s(m_Z)=0.118$, consistent
with the PDG average~\cite{Tanabashi:2018oca} and with recent high-precision
determinations~\cite{Ball:2018iqk,Verbytskyi:2019zhh,Bruno:2017gxd,Zafeiropoulos:2019flq}
(see~\cite{Pich:2018lmu} for an overview).

\subsection{Fast evolution} \label{s2:fast_evolution}
In the context of collinear QCD factorisation discussed in Sect.~\ref{s2:DGLAP}, the $F_2$ structure function
can be decomposed in terms of hard-scattering coefficient functions and PDFs as:
\begin{align} 
\label{eq:ev} 
F_2(x,Q^2) &= \sum_i^{n_f} C_i(x,Q^2) \otimes f_i(x,Q^2) \nonumber \\
&= \sum_{i,j}^{n_f} C_i(x,Q^2) \otimes \Gamma_{ij}(Q^2,\mu_0^2) \otimes f_j(x,\mu_0^2),
\end{align}
where $C_i(x,Q^2)$ are the process-dependent coefficient functions which
can be computed perturbatively as an expansion in the QCD and QED
couplings;  $\Gamma_{ij}(Q^2,\mu_0^2)$ is the evolution kernel, determined by the
solutions of the DGLAP equations, which evolves the nPDF from the initial
parameterisation scale $\mu_0^2$ into the hard-scattering scale $Q^2$,
$f_i(x,\mu_0^2)$ are the PDFs at the parameterisation scale, and
$\otimes$ denotes the convolution defined in Eq.~(\ref{eq:convolution}).
The sum over flavours $i,j$ runs over the $n_f$ active quarks and antiquarks flavours at a given
scale $Q$, as well as over the gluon.

The direct calculation of Eq.~(\ref{eq:ev}) during the PDF fit is not practical
since it requires first solving the DGLAP evolution equation for each new boundary
condition at $\mu_0$ and then convoluting with the coefficient
functions.
To evaluate Eq.~(\ref{eq:ev}) in a more computationally efficient way, it is far more efficient 
to precompute all the perturbative information, i.e. the coefficient functions $C_i$
and the evolution operators $\Gamma_{ij}$, with a suitable
interpolation basis.
Several of these approaches have been made available in the context of
PDF fits~\cite{DelDebbio:2013kxa,Carli:2010rw,Wobisch:2011ij}.
Here we use the \texttt{APFELgrid} tool~\cite{Bertone:2016lga} to precompute the perturbative
information of the DIS structure functions provided by the \texttt{APFEL} program~\cite{Bertone:2013vaa}.

Within this approach,
we can factorise the dependence on the PDFs at the input scale $\mu_0$ from
the rest of Eq.~(\ref{eq:ev}) as follows.
First, we introduce
an expansion over a set of interpolating functions $\{ I_{\beta}\}$ spanning both $Q^2$ and $x$ such that:
\begin{equation}
  f_i(x,Q^2) = \sum_{\beta} \sum_{\tau} f_{i,\beta \tau} I_{\beta}(x) I_{\tau}(Q^2)
\end{equation}
where the PDFs are now tabulated
in a grid in the $(x,Q^2)$ plane, $f_{i,\beta \tau}\equiv f_i(x_\beta,Q^2_{\tau})$.
We can express this result in terms of the PDFs at the input evolution scale
using the (interpolated) DGLAP evolution operators:
\begin{equation}
  f_{i,\beta \tau} = \sum_j \sum_{\alpha} \Gamma^{\tau}_{ij,\alpha \beta}\,f_j(x_{\alpha},\mu_0^2)
\end{equation}
so that the DIS structure function can be
evaluated as:
\begin{equation}
  F_2(x,Q^2) = \sum_i^{n_f} C_i(x,Q^2) \otimes \left(
  \sum_{\alpha,\beta,\tau} \sum_j \Gamma^{\tau}_{ij,\alpha \beta}\,f_j(x_{\alpha},\mu_0^2) I_{\beta}(x) I_{\tau}(Q^2)\right)
\end{equation}
which then can be rearranged to give:
\begin{align}
  \label{eq:ev_interp}
  F_2(x,Q^2) &= \sum_i^{n_f} \sum_{\alpha}^{n_x} \texttt{FK}_{i,\alpha}(x,x_{\alpha},Q^2,\mu_0^2) \, f_i(x_{\alpha},\mu_0^2) \end{align}
where all of the information about the partonic cross sections and the DGLAP
evolution operators is now encoded into the so-called Fast Kernel (FK) table, $\texttt{FK}_{i,\alpha}$.
Therefore, with the \texttt{APFELgrid} method we are able to
express the series of convolutions in Eq.(\ref{eq:ev}) by a matrix
multiplication in Eq.~(\ref{eq:ev_interp}), increasing the numerical 
calculation speed of the DIS structure functions by up to several orders
of magnitude.

Similarly, we can compute for any hadronic observable like Drell-Yan for instance the following expression:
\begin{align}
  \label{eq:ev_interp_hadronic}
  \frac{d\sigma^{DY}}{dQ^2dy}(y,Q^2) &= \sum_{ij}^{n_f} \sum_{\alpha\beta}^{n_x} \texttt{FK}_{ij,\alpha\beta}(x_1,x_2,x_{\alpha},x_{\beta},Q^2,\mu_0^2) \, f_i(x_{\alpha},\mu_0^2) \, f_j(x_{\beta},\mu_0^2) 
\end{align}
where throughout this thesis, the hard cross section in $\texttt{FK}_{ij,\alpha\beta}$ is pre-computed by means of fast interpolation grids at NLO accuracy in QCD either using \textsc{APPLgrid}~\cite{Carli:2010rw} generated by \textsc{MCFM}~\cite{Campbell:1999ah,Campbell:2011bn,Campbell:2015qma} for Z and W boson production or \textsc{NLOjet++}~\cite{Nagy:2001fj} interfaced to \textsc{FastNLO}~\cite{Wobisch:2011ij} for jet and dijet cross sections.

\section{Results}
In this section, I discuss my contributions to the NNPDF framework. In Sect.~\ref{s1:PDF_theory_uncertainties}, I discuss the first extraction of proton PDFs that accounts for the missing higher order uncertainty (MHOU) in fixed-order QCD calculations. In Sect.~\ref{s1:PDF_NNLO_jet}, I study the impact of inclusive jet and dijet production measurements from ATLAS and CMS at 7 and 8 TeV by including them in a global PDF determination at both NLO and NNLO QCD accuracy. 
\subsection{Theory uncertainties} 
\label{s1:PDF_theory_uncertainties}

Currently, PDF uncertainties only
account for the propagated statistical and
systematic errors on the measurements used in their
determination.
However, the same missing higher order uncertainty (MHOU)
that affects predictions at the LHC also affects predictions for the
various processes that enter the PDF determination.
These are
currently neglected, perhaps because they are believed to be
generally less important than experimental uncertainties.
However, as PDFs become more
precise, in particular thanks to ever tighter constraints
from LHC data~\cite{Rojo:2015acz}, eventually MHOUs in
PDF determinations will become significant.
Already in recent PDF sets  which make extensive
use of LHC data, such as \texttt{NNPDF3.1}~\cite{Ball:2017nwa},
the shift between PDFs at next-to-leading order (NLO) and the next
order (NNLO) is sometimes
larger than the PDF uncertainties from the experimental data.

In this section, I discuss the first PDF extraction that systematically accounts
for the MHOU in the QCD calculations used to extract them.
MHOUs are routinely estimated by varying the arbitrary renormalisation
$\mu_R$ and factorisation $\mu_F$ scales of perturbative
computations~\cite{deFlorian:2016spz}, though alternative methods
have also been proposed~\cite{Cacciari:2011ze,David:2013gaa,Bagnaschi:2014wea}.
Our inclusion of the MHOU in a PDF fit involves two steps:
first we establish how theoretical uncertainties can be included
in such a fit through a
covariance matrix~\cite{Ball:2018lag,Ball:2018twp}, and then we  find
a way of computing and validating the covariance matrix associated to the
MHOU using scale variations~\cite{Pearson:2018tim}.
By producing variants of \texttt{NNPDF3.1} which include the MHOU, we are
then able to finally address the long-standing question of their
impact on state-of-the-art PDF sets.
A detailed discussion of our results is presented in a companion
paper~\mycite{AbdulKhalek:2019ihb}, to which we refer for full
computational details, definitions, proofs and results.

Assuming that theory uncertainties can be modelled as Gaussian distributions (See Sect.~\ref{s1:Uncertainties_in_HEP}),
in the same way as experimental systematics, then the associated
theory covariance matrix $S_{ij}$ can be expressed in terms of nuisance parameters:
\begin{equation}
\label{eq:covth}
S_{ij} = \frac{1}{N}\sum_k \Delta_i^{(k)}\Delta_j^{(k)}
\end{equation}
where $\Delta_i^{(k)}=T^{(k)}_i-T_i^{(0)}$ is the expected shift with respect
to the central theory prediction for the $i$-th cross section,
$T_i^{(0)}$, due to the theory uncertainty, and $N$ is a normalisation
factor determined by the number of independent nuisance parameters.
Since theory uncertainties are independent of the
experimental ones, they can be
combined with them in quadrature:
the $\chi^2$ used to assess the agreement of theory and data is
given by:
\begin{equation}
\label{eq:th_chi2}
\chi^2=\sum_{i,j=1}^{N_{\text{dat}}}\left( D_i-T_i^{(0)}\right)
\left( S+C\right)^{-1}_{ij} \left(  D_j-T_j^{(0)}\right)
\end{equation}
with $D_i$ the central experimental value of the $i$-th data point,
and $C_{ij}$ the experimental covariance matrix.
More details of the implementation of the theory covariance matrix in PDF
fits may be found in Refs.~\cite{Ball:2018lag,Ball:2018twp}.

The choice of nuisance parameters $\Delta_i^{(k)}$ used in
Eq.~(\ref{eq:covth}) to estimate a particular theoretical uncertainty is
not unique, reflecting the fact that such estimates
always have some degree of arbitrariness.
Here we focus on the MHOU, and choose to use
scale variations to estimate $\Delta_i^{(k)}$.
A standard
procedure~\cite{deFlorian:2016spz}
is the so-called 7-point prescription, in which
the MHOU is estimated from the envelope of results obtained with
the following scales:
\begin{equation}
\nonumber
(k_F, k_R) \in \{ (1,1),(2,2),(\frac{1}{2},\frac{1}{2}),(2,1),(1,2),(\frac{1}{2},1),(1,\frac{1}{2}) \}
\end{equation}
where $k_R=\mu_R/\mu_R^{(0)}$ and $k_F=\mu_F/\mu_F^{(0)}$  are the ratios
of the renormalisation and factorisation scales to their central values.
Varying $\mu_R$ estimates the MHOU in the hard coefficient function of
the specific process, while the $\mu_F$ variation estimates the MHOU
in PDF evolution as can be seen in Eq.~(\ref{eq:factorisation_hadronic_tensor}).

In order to compute a covariance matrix, we must not only choose a set of
scale variations, but also make some assumptions about the way
they are correlated.
We do this by, first of all, classifying
the input datasets used in PDF fits
into processes as indicated in
Table~\ref{eq:expclassification}: charged-current (CC) and neutral-current (NC) deep-inelastic scattering (DIS), Drell-Yan (DY) production of gauge bosons (invariant mass, transverse momentum, and rapidity distributions), single-jet inclusive and top pair production cross sections.
Note that this step
requires making an educated guess as to which cross sections are likely to have
a similar structure of higher-order corrections.

\begin{table}[!h] 
  \centering
  \renewcommand*{\arraystretch}{1.3}
  \begin{tabular}{|c|c|}
    \hline
    Process Type  & Datasets \\
    \hline
    DIS NC  &   NMC, SLAC, BCDMS, HERA NC \\
    DIS CC  &   NuTeV, CHORUS, HERA CC \\
    DY  & CDF, D0, ATLAS, CMS, LHCb ($y$, $p_T$, $M_{ll}$) \\
    JET  & ATLAS, CMS inclusive jets \\
    TOP  & ATLAS, CMS total+differential cross sections \\
    \hline
  \end{tabular}
  \caption{\label{eq:expclassification}
   Classification of  datasets into  process types.
  }
\end{table}

Next, we formulate a variety of prescriptions of how to construct
Eq.~(\ref{eq:covth}) by picking a set of scale variations and
correlation patterns.
A simple possibility is the 3-point prescription, in which we vary coherently
both scales (thus setting $k_F=k_R$) by a fixed amount about the central
value, independently for each process.
More sophisticated  prescriptions are constructed by varying
the two scales independently, but by the same amount, and assuming that while $\mu_R$ is
only correlated within a given process, $\mu_F$ is fully
correlated among processes.
This assumption is based on the observation that
$\mu_F$ variations estimate the MHOU in the evolution equations,
which are universal (process-independent).

We then proceed to the validation of the resulting covariance matrices at
NLO.
We use the same experimental data and theory calculations
as in the \texttt{NNPDF3.1} $\alpha_s$
study~\cite{Ball:2018iqk} with two minor differences:
the value of the lower kinematic
cut has been increased from
$Q_{\text{min}}^2=2.69$~GeV$^2$ to
$13.96$~GeV$^2$ in order to ensure the validity of the
perturbative QCD expansion when scales
are varied downwards, and the HERA $F_2^b$ and
fixed-target Drell-Yan cross sections have been removed, for technical
reasons related to difficulties in implementing scale variation.
In total we then have $N_{\text{dat}}=2819$ data points.
The theory covariance matrix $S_{ij}$ has been constructed
by means of the \texttt{ReportEngine} software~\cite{zahari_kassabov_2019_2571601}
taking as input the
scale-varied NLO theory cross sections $T_i(k_F,k_R)$, provided
by \texttt{APFEL}~\cite{Bertone:2013vaa}
for the DIS structure functions
and by \texttt{APFELgrid}~\cite{Bertone:2016lga} combined with
\texttt{APPLgrid}~\cite{Carli:2010rw} for the hadronic
cross sections.

\begin{figure}[!h] 
    \begin{center}
    \makebox{\includegraphics[width=0.99\columnwidth]{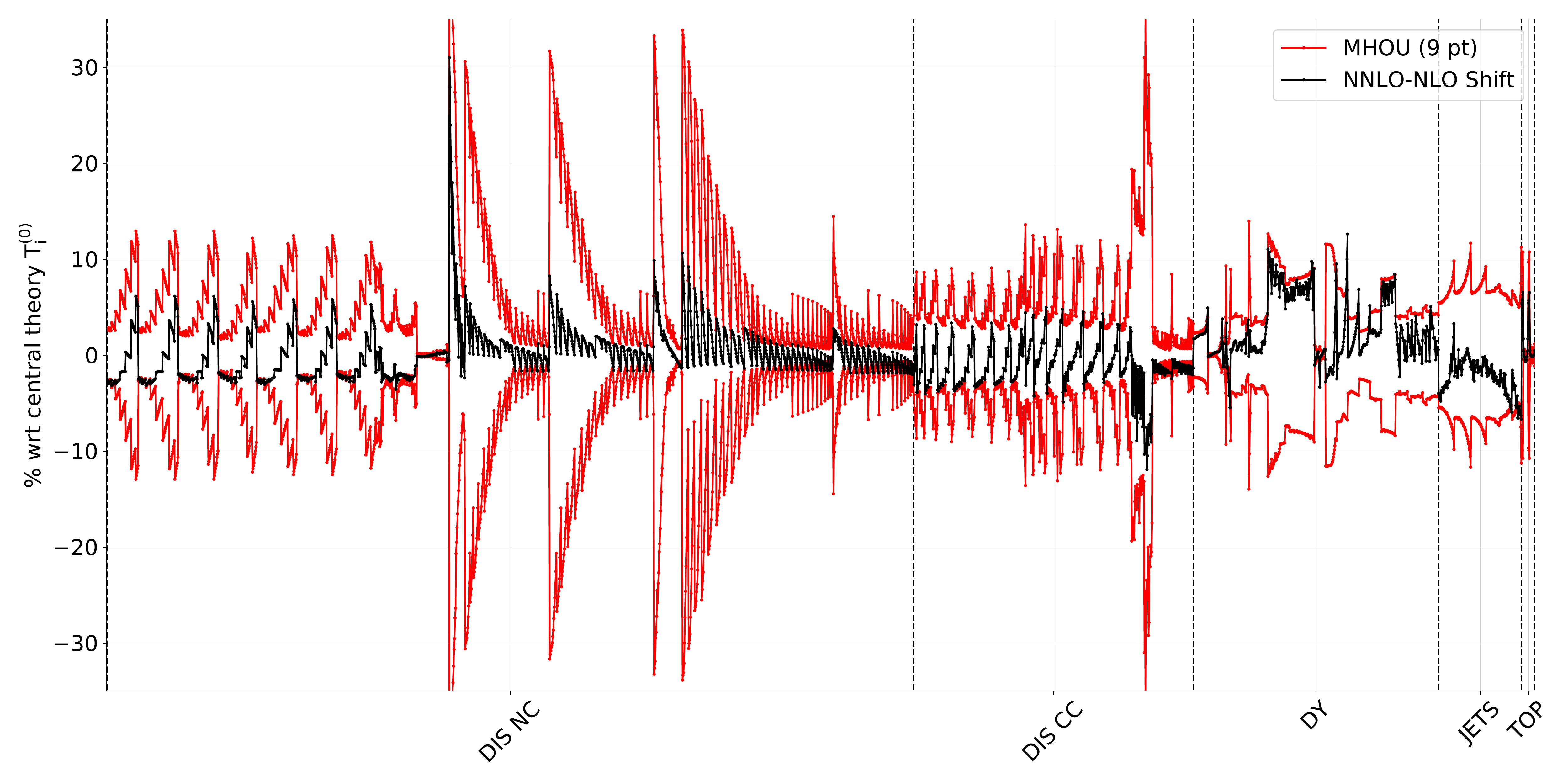}}
    \end{center}
  \caption{The relative uncertainties $\sigma_i$
    (9-point prescription) on
    the 2819 data points used in the PDF fit,
     compared to the known NLO-NNLO relative shifts $\delta_i$ in
     theory prediction.
  }
  \label{fig:shift_diag_cov_comparison}
\end{figure}

Since for the processes in Table~\ref{eq:expclassification} the
NNLO predictions are
known, we can then validate the NLO covariance matrix against the known
NNLO result.
For this exercise, a common input NLO PDF is used in both cases.
In order to validate the diagonal elements of $S_{ij}$, which correspond to the
overall size of the MHOU, we first normalise it to the central theory prediction,
$\widehat{S}_{ij}=S_{ij}/T^{(0)}_iT^{(0)}_j$.
Then we compare
in Fig.~\ref{fig:shift_diag_cov_comparison} the relative uncertainties,
$\sigma_i=\sqrt{\widehat{S}_{ii}}$ to the relative shifts between predictions at
NLO and NNLO,
$\delta_{i}=(T^{(0),{\text{nnlo}}}_{i}-T^{(0),{\text{nlo}}}_{i})/T^{(0),{\text{nlo}}}_i$,
for each of the $N_{\text{dat}}=2819$ cross sections.
In all cases, $\delta_{i}$  turns out to be smaller or comparable
to $\sigma_i$, showing that this prescription provides a good
(if somewhat conservative) estimate of the diagonal theory uncertainties.

\begin{figure}[!h] 
    \begin{center}
    \makebox{\includegraphics[width=0.85\columnwidth]{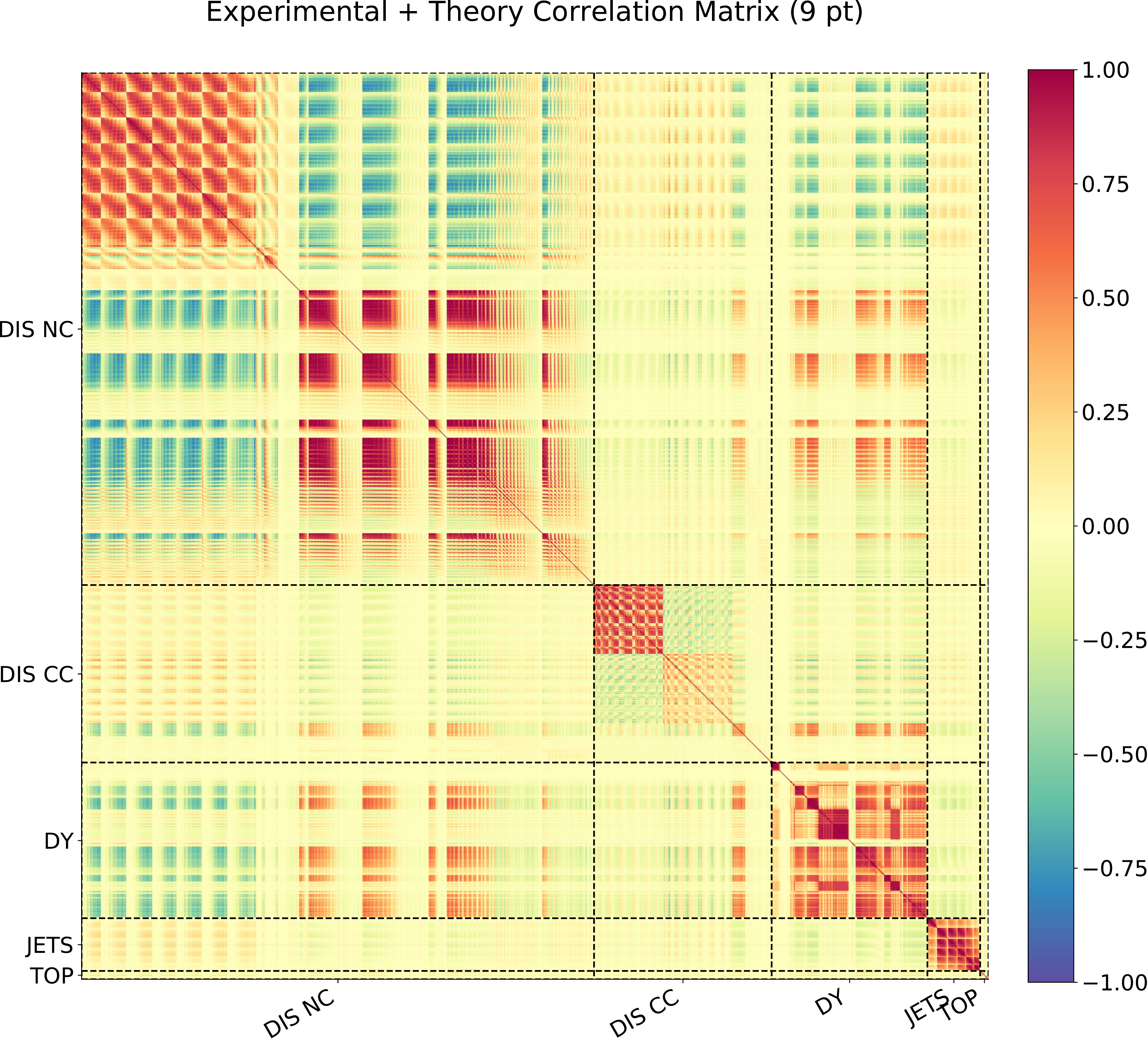}}
   \end{center}
  \caption{The combined experimental
    and theoretical (9-point) correlation matrix for the $N_{\text{dat}}$ cross sections
    in the fit.
  }
  \label{fig:default_theory0_plot_expcorrmat_heatmap}
\end{figure}

Adding the theory covariance matrix $S_{ij}$ to the experimental covariance matrix $C_{ij}$, while increasing the diagonal uncertainty on each individual prediction, also (and perhaps more importantly) introduces a set of theory-induced correlations between different experiments and processes, even when the experimental data points are uncorrelated. This is illustrated in Fig.~\ref{fig:default_theory0_plot_expcorrmat_heatmap}, showing the combined experimental and theoretical (9-point) correlation matrix: it is clear that substantial correlations appear even between experimentally unrelated measurements.

%
\begin{table}[!h] 
  \centering
  \renewcommand*{\arraystretch}{1.3}
  \begin{tabular}{|c|C{1.6cm}|C{2.2cm}|C{2.2cm}|}
    \hline
    & $C$    &  $C+ S^{\text{(3pt)}}$ &  $C+ S^{\text{(9pt)}}$\\
    \hline
    $\chi^2$ in Eq.~(\ref{eq:th_chi2})  & 1.139 &  1.139 & 1.109  \\
    $\phi$ in Eq.~(\ref{eq:phi_estimator})  & 0.314 &  0.394 & 0.415  \\
    \hline
  \end{tabular}
  \caption{\label{eq:chi2table} The
    central $\chi^2$ per data point and the average uncertainty
    reduction $\phi$ for the 3-point and 9-point fits.
  }
\end{table}
%
We can now proceed to a NLO global PDF determination with a theory covariance
matrix $S_{ij}$ computed using the 9-point prescription.
From the point of view of the \texttt{NNPDF} fitting methodology, the addition of
the theory contribution to the covariance matrix does not
entail any changes:
we follow the  procedure of Ref.~\cite{Ball:2014uwa},
but with the covariance matrix $C_{ij}$ now replaced by $C_{ij}+S_{ij}$,
both in the Monte Carlo replica generation and in the fitting.
In Table~\ref{eq:chi2table} we show some fit quality estimators
for the resulting PDF sets obtained using only the experimental
covariance matrix, and then also the theory covariance matrix
with two different prescriptions.
In particular, we show the
$\chi^2$ per data point
and the $\phi$ estimator~\cite{Ball:2014uwa} defined as:
\begin{equation} \label{eq:phi_estimator}
  \phi = \sqrt{\braket{\chi^2_{\text{exp}}[T_i]}-\chi^2_{\text{exp}}[\braket{T_i}]}
\end{equation}
where by $\chi^2_{\text{exp}}[T_i]$ we denote the value of the $\chi^2$ computed using the $i$-th PDF replica, and only including the experimental covariance matrix. We refer to the average over PDF replicas by $\braket{}$. The $\phi$ estimator therefore gives the ratio of the uncertainty in the predictions using
the output PDFs to that of the original data, averaged in quadrature 
over all data.
The quality of the fit is improved by the inclusion
of the MHOU, with the 9-point prescription performing
rather better than 3-point.
Interestingly, $\phi$ only increases by around 30\% 
when one includes the theory covariance
matrix, much less than the 70\% 
one would expect taking into account the
relative size of the NLO MHOU and experimental uncertainties . This 
means that in the region of
the data, taking the MHOU into account increases the PDF
uncertainties only rather moderately. This suggests that the addition of
the MHOU is resolving some of the
tension between data and theory, so that the larger overall uncertainty is
partly compensated by the improved fit quality.

\begin{figure}[!h] 
  \begin{center}
    \makebox{\includegraphics[width=\textwidth]{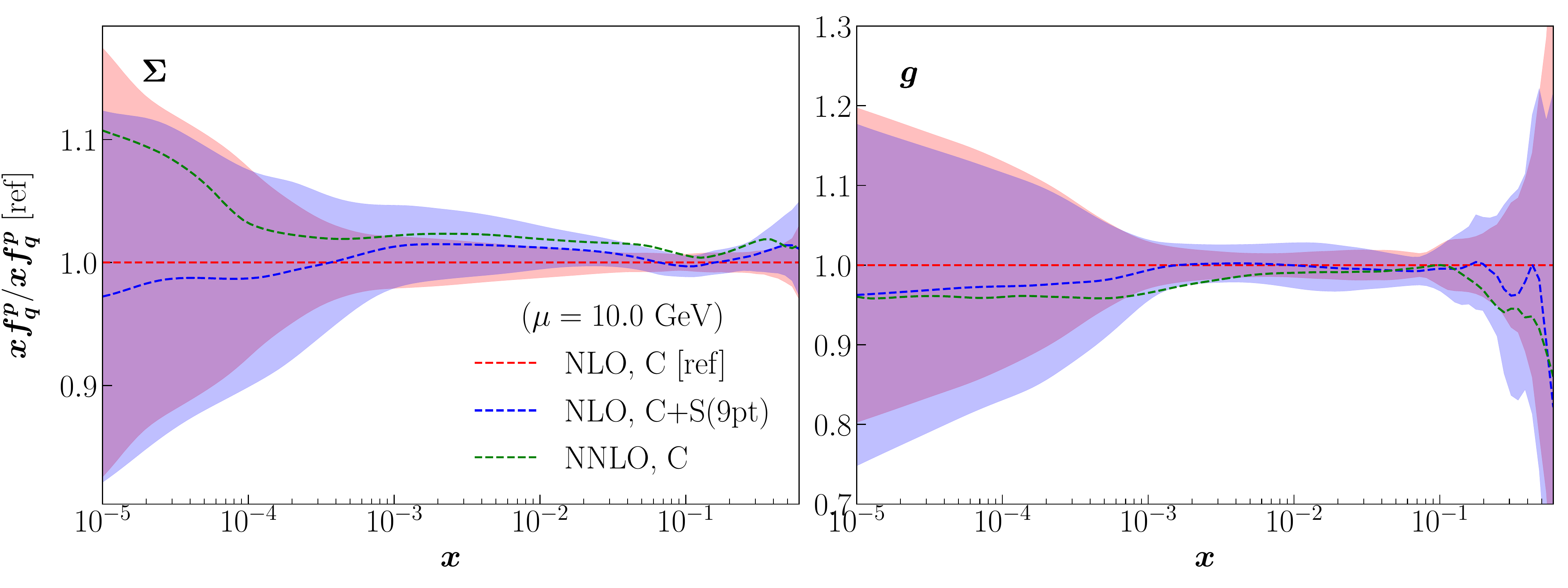}}
   \end{center}
  \caption{The quark singlet and gluon PDFs with their 1-$\sigma$ uncertainty band from the \texttt{NNPDF3.1} NLO fits
    without (C) and with (C+S) the MHOU (9-points) in the
    covariance matrix at $Q=10$ GeV, normalised
    to the former. The central NNLO result is also shown.
  }
  \label{fig:Global-NLO-CovMatTH-EXP-vsTH}
\end{figure}

In Fig.~\ref{fig:Global-NLO-CovMatTH-EXP-vsTH}
we compare at $\mu=10$ GeV the gluon and quark singlet PDFs obtained at NLO with (C+S) and without (C)
theory covariance matrix, normalised
to the latter. We also show the
central NNLO result when the theory covariance matrix is not included.
Three features of this comparison are apparent. First, when including
the MHOU, the increase in PDF uncertainty in the data region is quite
moderate, in agreement with the $\phi$ values of Tab.~\ref{eq:chi2table}.
Second, the NLO-NNLO shift is fully
compatible with the overall uncertainty.
Finally, also the central value is
modified by the inclusion of $S_{ij}$ in the fit, as the balance between different data sets adjusts according to their relative theoretical precision.
Interestingly, the central prediction shifts towards the known NNLO
result, showing that,
thanks to the inclusion of the MHOU, the overall fit quality has improved.



In summary, the analysis presented is the first global PDF analysis
that accounts for the MHOU
associated to the fixed order QCD perturbative calculations used in the fit.
The inclusion of the MHOU shifts central values by an amount that is
not negligible on the scale of the PDF uncertainty,
moving the NLO result towards the result of the NNLO fit.
PDF uncertainties increase moderately,
because of the improvement of fit quality due to the rebalancing of datasets
according to their theoretical precision. Note that
for this to be effective, the correlations in $S_{ij}$ play a crucial
role.


\subsection{Single jet and dijet production } \label{s1:PDF_NNLO_jet}
The inclusive jet cross section is the simplest hadron collider
observable with a purely strongly interacting final state. It is one of the most
promising and appropriate observables for precision QCD studies, such as the
determination of the PDFs and of the strong coupling constant $\alpha_s$.
The computation of its next-to-next-to-leading order (NNLO) QCD corrections
was completed~\cite{Ridder:2013mf,Currie:2013dwa,Currie:2016bfm,Czakon:2019tmo}, and opens
up the possibility of doing precision phenomenology with jet
observables. Whereas single-inclusive
jets have been used for the determination of 
proton PDFs~\cite{Gao:2017yyd}
for over thirty years~\cite{Martin:1987vw}, there is a number of
unsettled theoretical issues related to the definition of their associated observables.

The simplest inclusive observable, the single-inclusive jet cross section~\cite{Aversa:1988fv,Ellis:1988hv} displayed in Sect.~\ref{s2:Flavour_separation},
has the undesirable feature
of being non-unitary: each event is counted more than once, so 
the integral of the differential cross section is not equal to the
total cross section. The dijet cross section is free of this issue and
it appears to be especially well-suited for PDF 
determination~\cite{Giele:1994xd}. However, for 
this observable several 
scale choices are possible, because the more complex nature of the final state 
offers a wide choice of dimensionful kinematic variables;
consequently, the significant scale dependence of NLO results has so far 
effectively prevented the use of this observable for PDF determination.

The availability of NNLO calculations has opened up the possibility of
settling these issues, though their full understanding has posed a
theoretical challenge, with the single-inclusive jet and dijet
observables presenting different features. On the one hand, the issue
of scale choice for the dijet observable has been essentially settled by
the NNLO computation, with the scale dependence being under control at
NNLO and the dijet invariant mass $m_{jj}$ emerging as the preferred choice.
On the other hand, the single-inclusive jet cross section has shown a 
dependence on the choice of scale which is not significantly reduced from NLO 
to NNLO~\cite{Currie:2017ctp}, so that the understanding of the
perturbative behaviour, the scale dependence~\cite{Currie:2018xkj},
and even the appropriate definition~\cite{Cacciari:2019qjx} of this 
observable are non-trivial. 
A careful analysis reveals that the apparent lack 
of improvement of scale stability from NLO to NNLO is due to an accidental 
NLO scale cancellation which occurs for particular values of the jet 
radius~\cite{Dasgupta:2016bnd,Cacciari:2019qjx}. The
persistence of a dependence on the central scale choice at
NNLO can in turn be understood as a consequence of infrared sensitivity, which
is aggravated by particular scale choices~\cite{Currie:2018xkj}.
It then appears that the non-unitary definition of the observable is in
fact necessary for perturbative stability, with dijets offering
essentially the only viable unitary stable alternative~\cite{Cacciari:2019qjx}.
From these studies the partonic transverse energy \(\displaystyle \widehat{H}_T = \sum_{i \in \text{partons}} p_{T,i} \) emerges as the 
optimal scale choice~\cite{Currie:2018xkj} for the calculation
of single-inclusive jet cross sections.

To address these issues we consider the complete inclusive jet~\cite{Aad:2014vwa,Aaboud:2017dvo,
Chatrchyan:2012bja,Khachatryan:2016mlc}
and dijet\cite{Aad:2013tea,Chatrchyan:2012bja,Sirunyan:2017skj} dataset
from ATLAS and CMS at $\sqrt{s}=7$ and 8 TeV, while keeping the rest of the global
\texttt{NNPDF3.1} dataset and adopting the same general PDF fitting methodology. More details on the datasets are given in Sect.~2 of Ref.~\mycite{AbdulKhalek:2020jut}.
In Fig.~\ref{fig:jetsdijets}, we study the impact of adding separately the single-inclusive jet and dijet cross sections to the \texttt{NNPDF3.1} global dataset. In each case, we assess the fit quality and the impact of the data on the PDFs, at various perturbative orders. For that we have performed six independent PDF determinations, where we compare our baseline \texttt{NNPDF3.1} (1) without jets against (2) with all single-inclusive jets and (3) with all dijets both at NLO and NNLO QCD at a scale of $\mu=100\,\text{GeV}$. Only the impact of QCD corrections is displayed in this thesis but a thorough study of the inclusion of
electroweak (EW) corrections to jet predictions and the sensitivity of results to the 
treatment of experimental correlated systematic uncertainties has been performed in Ref.~\mycite{AbdulKhalek:2020jut}. It is however fair to mention that the general conclusions, which I will elaborate next, are not altered by the additional consideration of EW corrections as these are subdominant w.r.t the QCD ones.

The NNLO QCD corrections are computed with \textsc{NNLOjet}~\cite{Gehrmann-DeRidder:2019ibf}
and are included by means of K-factors discussed in Sect.~\ref{s2:NNLO_corrections}. More details and plots on the latter can be found in Sect.~3 of Ref.~\mycite{AbdulKhalek:2020jut}. The fit quality of the different determination is given in
Table~\ref{tab:NNPDF31_jet_chi2s}. In this and all subsequent tables and plots ``jets'' is
short for single-inclusive jets. Based on the $\chi^2$ values from Tables~\ref{tab:NNPDF31_jet_chi2s} and
the PDF comparisons in 
Figs.~\ref{fig:jetsdijets}-\ref{fig:jetsdijets}, our conclusions are the following:

\begin{figure}[!t] 
  \begin{center}
    \makebox[\textwidth]{\includegraphics[width=\textwidth]{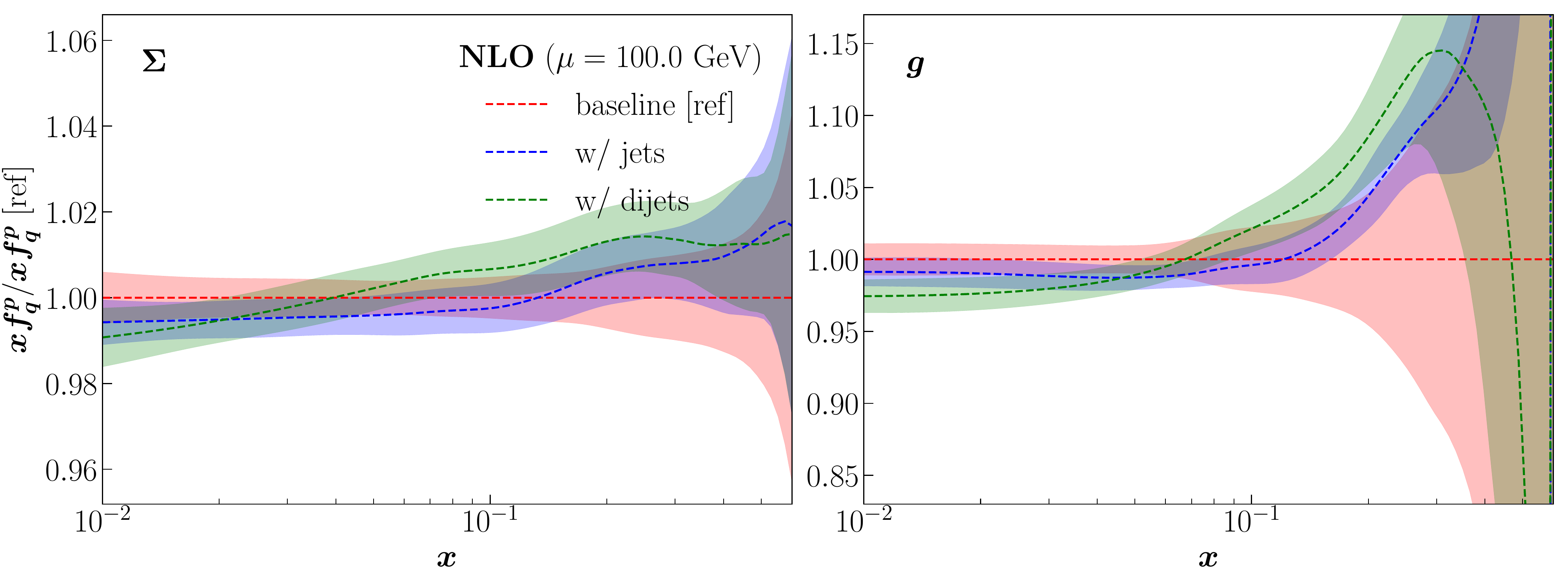}}\\
    \makebox[\textwidth]{\includegraphics[width=\textwidth]{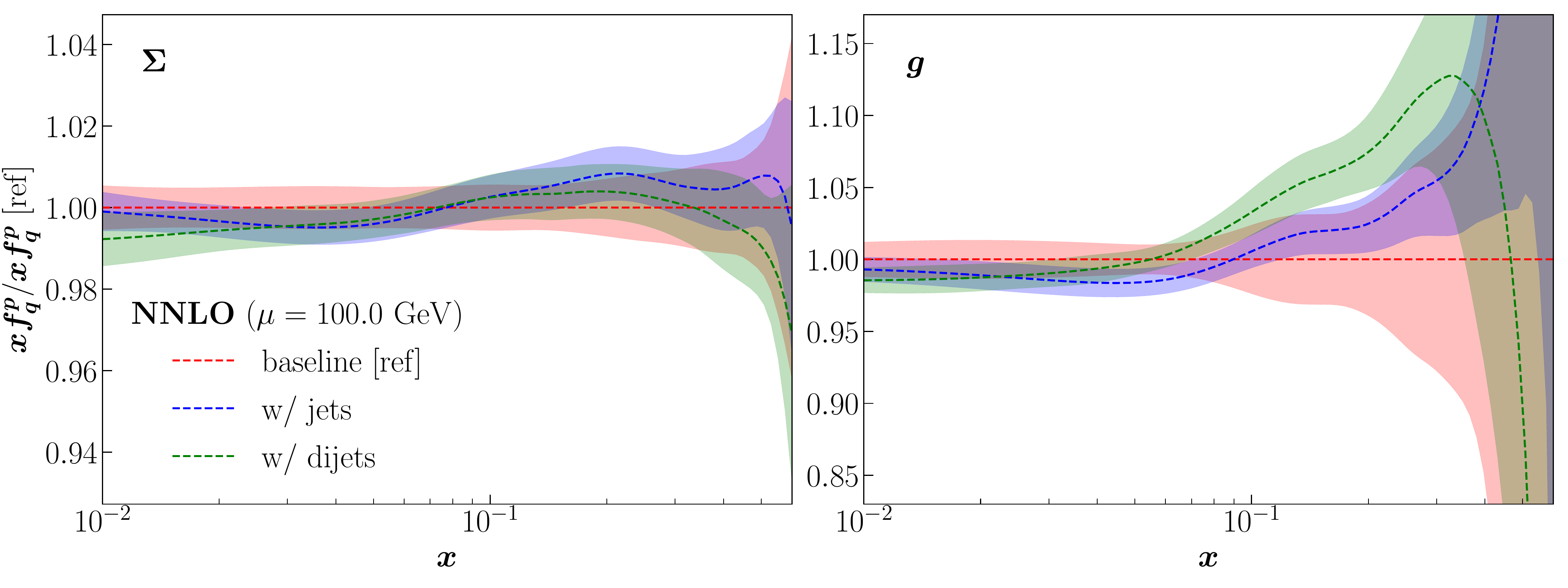}}\\
    \end{center}
  \caption{Comparison between the baseline fit with no jet data and the fit with all single-inclusive jet and dijet data at NLO (first row) and NNLO QCD (second row). The bands correspond to the 1-$\sigma$ uncertainty.}
  \label{fig:jetsdijets}
\end{figure}
\begin{enumerate}

\item Concerning the relative impact on PDFs of single-inclusive jets and
dijets:
 \begin{enumerate}
\item The effect on PDFs of the inclusion of jet and dijet data in the
  \texttt{NNPDF3.1} global dataset is qualitatively the
  same. Namely, they mainly affect the gluon, by leading to an enhancement
of its central value in the region $0.1\lesssim x \lesssim 0.4$, accompanied by a 
suppression in the region $0.01\lesssim x \lesssim 0.1$. The
suppression is by about  1\%, while the enhancement at the peak,
localized at  $x\simeq 0.3$  is by about 2.5\%
for single-inclusive jets, but stronger, by about 7.5\% for dijets. An
enhanced gluon is also present in the CT18 PDF determination, which includes the 8~TeV CMS single-inclusive jet data, and whose gluon PDF
is consistent with our result within its rather larger uncertainty.

\item The inclusion of either single-inclusive or dijets leads to a
  reduction in the gluon uncertainty, with a somewhat stronger
  reduction observed for single-inclusive jets. It should be noted
  in this respect that for the most accurate 8~TeV dijet dataset, 
  only CMS data
  are currently available. The constraining power of the dijet dataset
  is consequently at present more limited than that of the
  single-inclusive jet dataset.

\item The inclusion of single-inclusive jet or dijet data does not lead to a 
deterioration in the description of the rest of the data in comparison to 
the baseline fit: almost all $\chi^2$ values for other datasets are
unchanged. This shows that the single-inclusive and dijet data are not
only consistent with each other, but also with the rest of the global
dataset, and their impact on the gluon central value, accompanied by a
reduction in uncertainty, corresponds to a genuine addition of new
information in the fit. Indeed, a comparative
assessment of the impact of jet, Z $p_T$ and top production data on
the gluon distribution in Ref.~\cite{Nocera:2017zge} showed good
consistency, specifically wih the top data also leading to an
enhancement of the gluon in the $x\gsim 0.1$ region.
\end{enumerate}

\begin{table}
\centering
\begin{tabular}{lcccc|cccc}  \toprule
Accuracy           &   \multicolumn{4}{c}{NLO} &   \multicolumn{4}{c}{NNLO} \\\midrule  
Dataset                       & $n_{\text{dat}}$ &  baseline     & jets     & dijets  & $n_{\text{dat}}^*$  & baseline  & jets & dijets\\
DIS NC                        &      2113        &  1.19  &  1.20  &  1.23    &  2103           &  1.18  &  1.19  & 1.19  \\
DIS CC                        &      989         &  1.04  &  1.07  &  1.10    &  989            &  1.08  &  1.09  & 1.08  \\
Drell-Yan                     &      567         &  1.35  &  1.35  &  1.35    &  580            &  1.37  &  1.33  & 1.31  \\
$Z$ $p_T$                     &      120         &  1.97  &  2.04  &  2.16    &  120            &  1.03  &  1.06  & 1.11  \\
Top pair                      &      25          &  1.12  &  2.27  &  1.53    &  25             &  1.07  &  1.55  & 1.39  \\
Jets                     &      629         & [1.52] &  1.21  & [1.56]   &  629            & [2.63] &  1.78  & [2.04] \\
Dijets                   &      266         & [4.17] & [3.99] &  2.67    &  266            & [3.52] & [2.40] & 1.81  \\
    Total                     &                  &  1.20  & 1.21   &  1.33    &                 & 1.18   &  1.27   &  1.22   \\
\bottomrule
\end{tabular}
\caption{The $\chi^2$ per data point for all fits considered.
Results are shown
for all datasets, aggregated by process type. For jets data, results are
shown both for the sets included in each fit, and also for those not
included, enclosed in square brackets.
The number of data points in each
dataset is also shown. ($^*$): At NNLO and due to slightly different pertrubative order dependent kinematic cuts, we end up with different number of data points for some datasets.}
\label{tab:NNPDF31_jet_chi2s}
\end{table}

\vspace{30pt}
\item Concerning relative fit quality:

\begin{enumerate}

\item The quality of the fit at NNLO to single-inclusive jet data ($\chi^2/N=1.78$) and dijet
  data ($\chi^2/N=1.81$) when each of them is fitted is comparable. The quality of
  the fit to dijets when single inclusive jets are fitted ($\chi^2/N=2.40$) is somewhat worse than for single-inclusive
  jets when fitting dijets ($\chi^2/N=2.04$). This confirms
  the full consistency of the two datasets, with a marginal preference for dijets.

\item The fit including dijet data is also somewhat more
  internally consistent than the fit including single-inclusive jet
  data. Indeed, the $\chi^2$ per data point of the global fit is closer to one
  ($1.22$ vs $1.27$), and also, the $\chi^2$ for individual datasets is generally  
  better. In particular, this happens for top production data,
  also sensitive to the large-$x$ gluon. It is unclear whether
  this is due to a greater theoretical accuracy of the NNLO dijet
  observable, or to better quality of the dijet data (specifically a
  better control of correlated systematics). However, the issue is
  phenomenologically immaterial, given that the shape 
  and size of the data to theory ratio are qualitatively comparable for all of 
  the jet and dijet data, regardless whereof dataset is actually fitted.
\end{enumerate}

\item Concerning relative perturbative stability:

\begin{enumerate}

 \item When fitting the dijet data, we find that the overall fit quality to the fitted data
   improves from NLO to NNLO ($\chi^2/N=1.33$ at NLO vs.
   $1.22$ at NNLO), but the fit quality to the single-inclusive jet
   data actually deteriorates from NLO to NNLO (from $\chi^2/N=1.21$ to
   $1.27$). But, perhaps surprisingly, the fit quality to the
   dijet data, not fitted, does improve  (from $\chi^2=3.99$ at NLO
   to $\chi^2=2.40$ at NNLO). Whereas this shows a good theoretical
   consistency of the dijet data, it is unclear whether the lack of
   improvement of the single-inclusive jet data is due to a less
   stable perturbative behaviour of the jet observable, or to issues
   with data.

  \item The fit quality to
    all other data included in the global datasets deteriorates at NLO
    when including jet data. At NNLO, when dijets are fitted the global
    fit quality significantly improves and becomes almost the same as that of the baseline
    ($\chi^2=1.22$, in comparison to $\chi^2=1.18$ of the baseline)
    while for the fit to single-inclusive jets it does not
    improve. The greater deterioration of fit quality at NLO for dijets
    can be understood as a consequence of the fact, observed in point
    1.a above, that dijets have a greater pull on the gluon: hence
    missing NNLO corrections lead to a stronger loss of accuracy. The
    lack of improvement in the description of single-inclusive jets
    shows again that this observable seems to be somewhat less
    well-behaved, either for theoretical or experimental reasons.
\end{enumerate}
\end{enumerate}

We generally conclude that single-inclusive jets and dijets are
mutually consistent and at NNLO consistent with the global dataset and
have a similar impact on the gluon. The dijet observable has a better
behaved perturbative behaviour and a stronger pull on the gluon PDF
and it appears to be marginally preferable, though 
it leads to a less pronounced decrease of the gluon uncertainty,
possibly because ATLAS dijet measurements are not yet
available at 8~TeV, while single-inclusive jet measurements are available both
from ATLAS and CMS.
  \chapter{Nuclear PDFs from lepton-nucleus scattering}
\label{chap:nNNPDF10}
\vspace{-1cm}
\begin{center}
  \begin{minipage}{1.\textwidth}
      \begin{center}
          \textit{This chapter is based on my results that are presented in Refs.~\mycite{Khalek:2018bbv,AbdulKhalek:2019mzd}}.
      \end{center}
  \end{minipage}
  \end{center}
  
\myparagraph{Introduction} The longitudinal distribution of unpolarised partons within nuclei, described by nuclear PDFs (nPDFs)~\cite{Zurita:2018vrs,Paukkunen:2018kmm,Rojo:2019uip,Ethier:2020way}, 
is significantly modified w.r.t. the free-nucleon PDFs~\cite{Gao:2017yyd} as a result of different non-perturbative QCD dynamics (see Sect.~\ref{s2:Nuclear_modification}).
Understanding the theoretical mechanisms that generate such dynamics remains an
open challenge, thus my focus in the next chapter will be on the phenomenological determination of nPDFs that incorporate them.
%
%
%

\myparagraph{Motivation} Precise extractions of nPDFs are not only crucial to study the strong interaction and the validity of factorisation theorems
in confined nuclei, but are also necessary to model
the initial state of heavy ion collisions which aim to characterise the Quark-Gluon
Plasma (QGP)~\cite{Abreu:2007kv,Adams:2005dq} using hard probes.

Furthermore, nPDFs contribute to global QCD analyses of the proton structure~\cite{Ball:2014uwa,Harland-Lang:2014zoa,Dulat:2015mca,Alekhin:2017kpj,Hou:2019efy} through the inclusion of neutrino structure function data collected in collisions involving heavy nuclear targets. These measurements on nuclear targets provide important information on the quark flavour separation and strangeness in the proton~\cite{Ball:2018twp}.
Given the current precision of proton PDF fits, neglecting the nuclear uncertainties associated with neutrino-nucleus scattering may not be well justified any longer, as opposed to the situation some years ago~\cite{Ball:2009mk}.

Unfortunately, the determination
of the nuclear PDFs is hampered by the rather limited experimental data sets
available.
While nPDF analyses are based on a significantly reduced number of data sets
compared to the free-nucleon case, the situation has improved
in recent years with the availability of hard-scattering 
cross section data from proton-lead collisions at the LHC for processes such as jet,
W and Z, and heavy quark production~\cite{Adam:2015hoa,Adam:2016dau,Adam:2015xea,Aad:2016zif,Chatrchyan:2014hqa,Zhu:2015kpa,TheATLAScollaboration:2015lnm,Aad:2015gta,Khachatryan:2015pzs,Khachatryan:2015hha,CMS:2015lca,Adam:2016mkz,Adam:2015qda,Abelev:2014hha,Khachatryan:2015sva,Khachatryan:2015uja,Aaij:2017gcy,Aaij:2019lkm}.
%
%

%
These collider measurements can clarify several
poorly understood aspects of nuclear PDFs, such as the quark 
flavour dependence of nuclear effects
and the nuclear modifications of the gluon distribution (see discussion in Sect.~\ref{s2:Nuclear_modification}).
Several studies have indeed demonstrated the valuable constraints 
that can be provided for the nuclear PDFs from proton-lead 
collisions at the LHC, see \textit{e.g.} Refs.~\cite{Eskola:2016oht,Kusina:2016fxy,
Kusina:2017gkz,Armesto:2015lrg,Eskola:2019dui}.
Since measurements of neutral-current (NC) deep-inelastic scattering (DIS) nuclear structure
functions on isoscalar targets
are only sensitive to a single quark PDF combination (as we will see in this chapter),
one needs to rely on the information provided by independent processes
to disentangle quark and antiquarks of different flavours.
The main options that are available to accomplish this are
neutrino-induced charged current (CC) DIS cross sections on heavy nuclear targets,
sensitive to different quark combinations w.r.t the NC case, and
electroweak gauge boson production at the LHC, which I explore in Chapter~\ref{chap:nNNPDF20} and \ref{chap:nNNPDF30}. We note that the compatibility of the former in global nPDF analyses has been subject to many independent investigations by different groups~\cite{Schienbein:2007fs,Kovarik:2010uv,Alvarez-Ruso:2017oui, Paukkunen:2013grz, Paukkunen:2010hb}.

In fact, several groups have recently presented determinations
of nPDFs using
different input data sets, theoretical assumptions,
and methodological settings~\cite{Hirai:2007sx, Eskola:2008ca, Eskola:2009uj,deFlorian:2003qf,deFlorian:2011fp,Eskola:2016oht,Kovarik:2015cma,Khanpour:2016pph,AbdulKhalek:2019mzd,Walt:2019slu}.
From the methodological point of view, there exist two
primary limitations that affect the separation between  
quark and antiquark flavours in nPDF extractions.
The first, is the reliance on \textit{ad-hoc} theoretical
assumptions required to model the dependence of the nuclear
modifications on both the parton momentum fraction $x$ per nucleon (defined in Sect.~\ref{s2:Nuclear_modification}) and atomic mass
number $A$, where in some cases the expected behaviour is hard-coded in
the nPDF parameterisation.
The second, is the lack of consistency between the nuclear PDF
determination and that of the corresponding proton baseline, to which the former should
reduce to in the $A\to 1$ limit in terms of central values and uncertainties. 
This consistency is particularly important given that the precision
LHC data impose stringent
constraints on the quark flavour separation for the proton PDFs,
for example via measurements 
of inclusive W and Z production characterised by per-mille level uncertainties.
Ensuring that the LHC constrains on the proton PDF baseline
are appropriately propagated to the nPDF determination for $A>2$ is 
therefore critical.
Finally, PDF uncertainties are often estimated using the Hessian method,
which is restricted to a Gaussian approximation
with \textit{ad-hoc} tolerances, introducing 
a level of arbitrariness in their statistical
interpretation.

Motivated by these limitations and the need for a reliable and consistent
determination of nuclear PDFs and their uncertainties,
my main work during this PhD was to develop a framework to perform a first nPDF analysis
based on the \texttt{NNPDF} methodology~\cite{Forte:2002fg,DelDebbio:2004xtd,DelDebbio:2007ee,Ball:2008by,Rojo:2008ke,Ball:2009qv,Ball:2010de,Ball:2011mu,Ball:2011uy,Ball:2012cx} dubbed \texttt{nNNPDF}.

\myparagraph{Overview on the next chapters} In this chapter, I present \texttt{nNNPDF1.0}, the first nPDF set release that is inferred from NC DIS data at NLO and NNLO QCD accuracy. In that respect, I introduce the inference framework that will be built upon for the subsequent two chapters~\ref{chap:nNNPDF20} and \ref{chap:nNNPDF30}. Being inferred only from NC DIS, the \texttt{nNNPDF1.0} sets suffer from a lack of flavour separation since effectively only 2 flavours (quark singlet and the gluon) can be constrained. This is remedied in Chapter~\ref{chap:nNNPDF20}, with the NLO \texttt{nNNPDF2.0} sets that are inferred from the NC DIS data augmented by CC DIS and electroweak bosons production from the LHC. This new release holds many methodological updates w.r.t. the first one, including an updated proton baseline, a new cross section positivity constraint as well as an extended set of 6 parameterised flavours among others. Finally in Chapter~\ref{chap:nNNPDF30}, I present the results of the ongoing efforts towards \texttt{nNNPDF3.0}, in which on one hand we aim to include NNLO QCD corrections in association with the processes considered in \texttt{nNNPDF2.0} and on the other hand, we also aim to include new processes that will particularly constrain the gluon nPDF such as single inclusive jet and dijet production among others. 

\myparagraph{Notations and conventions} Let us begin by establishing the nPDF notation and conventions
that will be used throughout this chapter and the next two~\ref{chap:nNNPDF20} and \ref{chap:nNNPDF30}.
Parton distributions can be parametrised in 
a number of different bases, all of which are
related by linear transformations.
Two popular ones are the flavour basis, corresponding
to the individual quark and anti-quark PDFs, and the evolution basis,
given by the eigenvectors of the DGLAP
evolution equations~\cite{Ball:2014uwa} (see Sect.~\ref{s2:DGLAP}).
Expressed in terms of the elements of the flavour basis, the evolution basis
distributions are given by Eq.~(\ref{eq:Evolution_basis}).
Although the results of a nPDF analysis should be
independent of the basis choice for the parameterisation, 
some bases offer practical advantages, for example in 
the implementation of the sum rules which are discussed in Sect.~\ref{s2:Physical_constraints}.

We define $f^{(N/A)}$ to be the PDF for the 
flavour $f$ associated to the average nucleon $N$ bound in
a nucleus with atomic number $Z$ and mass number $A$.
This object can be written as:
\begin{equation}
\label{eq:qNAdefinition}
f^{(N/A)}(x,\mu_0) = \frac{Z}{A} f^{(p/A)}(x,\mu_0)  + \lp 1-\frac{Z}{A} \rp f^{(n/A)}(x,\mu_0) \, ,
\end{equation}
where $f^{(p/A)}$ and $f^{(n/A)}$ represent the PDFs of a proton and a neutron,
respectively, bound in the same nucleus of mass number $A$.
Assuming isospin symmetry, the PDFs of the neutron in Eq.~(\ref{eq:qNAdefinition})
can be expressed in terms of the proton PDFs as follows:
\begin{eqnarray}
u^{(n/A)}(x,\mu_0) = d^{(p/A)}(x,\mu_0) \, , \quad && \nonumber
\bar{u}^{(n/A)}(x,\mu_0) = \bar{d}^{(p/A)}(x,\mu_0) \\
d^{(n/A)}(x,\mu_0) = u^{(p/A)}(x,\mu_0)\, , \quad \label{eq:isospin}  &&
\bar{d}^{(n/A)}(x,\mu_0) = \bar{u}^{(p/A)}(x,\mu_0) \\
s^{(n/A)}(x,\mu_0) = s^{(p/A)}(x,\mu_0)\, , \quad  &&  \nonumber
g^{(n/A)}(x,\mu_0) = g^{(p/A)}(x,\mu_0) \, .
\end{eqnarray}
Using the relations above, the strange and gluon distributions
of the average bound nucleon $f^{(N/A)}$
and bound proton $f^{(p/A)}$ become equivalent, while 
the up and down flavoured distributions of the average 
bound nucleon are instead linear combinations 
of the bound proton PDFs with coefficients depending on the values of $A$ and $Z$.

\myparagraph{Outline} In Sect.~\ref{s1:nNNPDF10_framework}, I discuss the framework of the first analysis \texttt{nNNPDF1.0}~\cite{AbdulKhalek:2019mzd}, where only NC DIS function measurements are considered, and computed up to NNLO in QCD.
In Sect.~\ref{s1:nNNPDF10_results}, I present the main results of the \texttt{nNNPDF1.0} analysis including an assessment of the inference quality, a detailed discussion on the main features of the nPDF sets and the results of different methodological validation tests.

\section{\texttt{\texttt{nNNPDF1.0}} framework} 
\label{s1:nNNPDF10_framework}

In this section, I present the framework of the first release of nPDFs
based on the \texttt{NNPDF} methodology: \texttt{nNNPDF1.0}.
This analysis is based on NC DIS
structure function data and is performed up to NNLO
in QCD calculations with heavy quark mass effects.
For the first time in the \texttt{NNPDF} fits,
the $\chi^2$ minimization was achieved using
stochastic gradient descent
with reverse-mode automatic 
differentiation (backward propagation).
%
%
The \texttt{nNNPDF1.0} distributions satisfy the boundary condition whereby the
NNPDF3.1 proton PDF central values and uncertainties are reproduced at $A=1$,
which introduces important constraints particularly for low-$A$ nuclei.
This framework constitutes the first nPDF determination obtained
using a Monte Carlo methodology consistent with that of state-of-the-art
proton PDF fits, and provided the foundation for a subsequent
global nPDF analyses including also proton-nucleus
data (see Chapter~\ref{chap:nNNPDF20} and \ref{chap:nNNPDF30}).

In Sect.~\ref{s2:nNNPDF10_data}, I present
the experimental data used in this analysis, namely
ratios of NC DIS structure functions.
In Sect.~\ref{s2:nNNPDF10_parameterisation}, I introduce the parameterisation adopted in \texttt{nNNPDF1.0} in terms of NNs and preprocessing exponents (to control the small- and large-$x$ nPDF behaviours) in light of the nPDF flavour sensitivity manifested by the NC DIS data. 
In Sect.~\ref{s2:nNNPDF10_protonBC}, I discuss the constraint on nPDFs to reproduce the free nucleon PDF in the limit of the atomic mass $A=1$ and how it is implemented by means of the Lagrange multiplier method. I then outline in Sect.~\ref{s2:nNNPDF10_minimisation} the minimisation procedure adopted including the SGD algorithm and the cross validation method.

\subsection{Experimental data} \label{s2:nNNPDF10_data}

\begin{figure}[ht]
\begin{center}
  \includegraphics[width=0.90\textwidth]{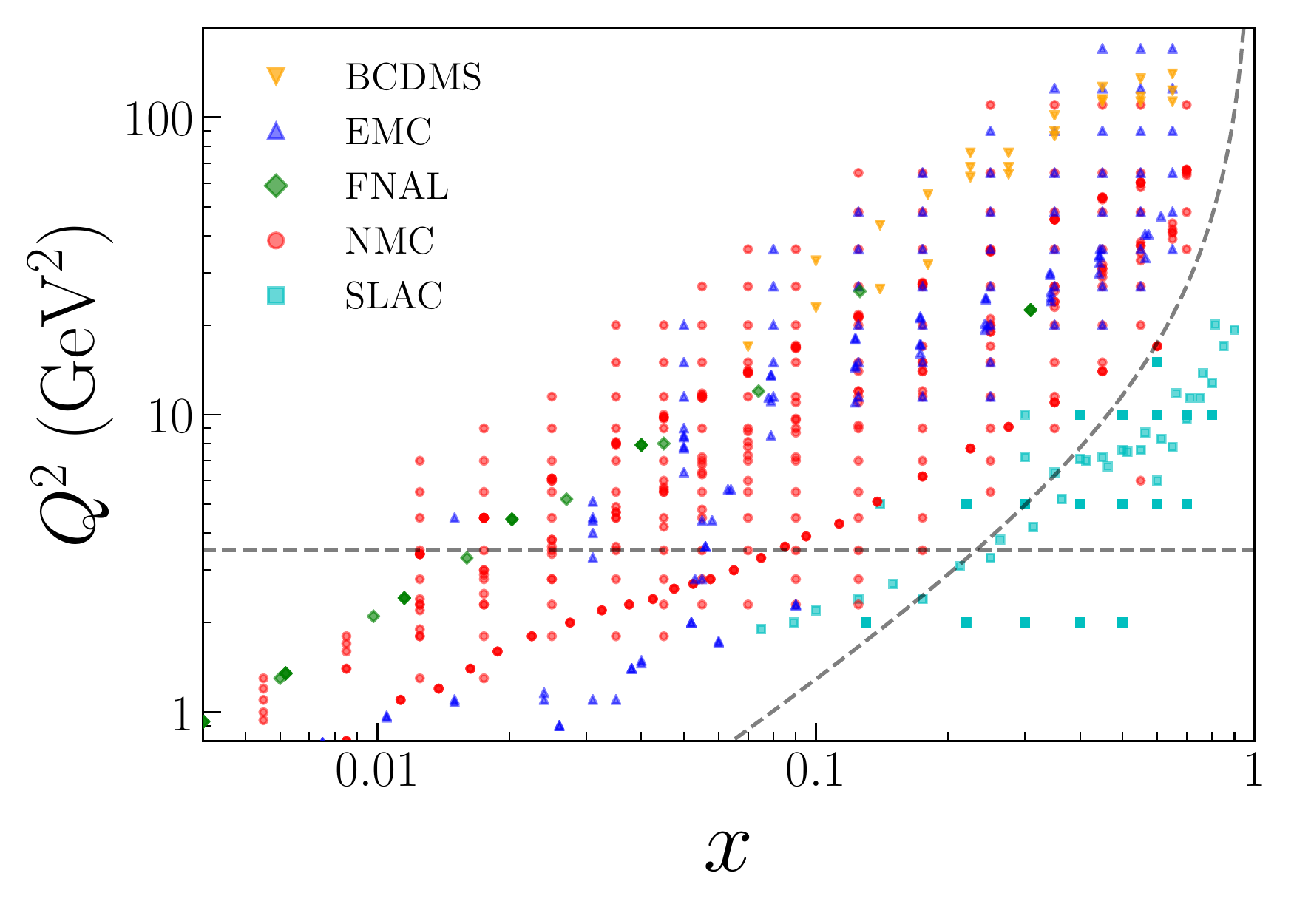}
 \end{center}
\vspace{-0.3cm}
\caption{\small Kinematical coverage in the $(x,Q^2)$ plane
  of the DIS NC nuclear structure function data included in \texttt{nNNPDF1.0},
  as summarised in Table~\ref{tab:nNNPDF10_data}.
  The horizontal dashed and curved dashed lines correspond
  to $Q^2 = 3.5$ GeV$^2$ and $W^2 = 12.5$ GeV$^2$, respectively,
  which are the kinematic cuts imposed in this analysis.
  \label{figkinplot}
}
\end{figure}

In this analysis,
we include all available inclusive DIS measurements
of NC structure functions on nuclear targets.
In particular, we use data from the
EMC~\cite{Aubert:1987da,Ashman:1988bf,Arneodo:1989sy,Ashman:1992kv}, 
NMC~\cite{Amaudruz:1995tq,Arneodo:1995cs,Arneodo:1996rv,Arneodo:1996ru},
and BCDMS~\cite{Alde:1990im,Benvenuti:1987az} experiments at CERN, E139 measurements from 
SLAC~\cite{PhysRevD.49.4348}, and E665 data
from Fermilab~\cite{Adams:1992vm}.
The measurements of nuclear structure functions
are typically presented as ratios of the form:
\begin{equation}
\label{eq:Sect2Rf2}
R_{F_2}\lp x, Q^2, A_1, A_2\rp \equiv
\frac{F_2(x,Q^2,A_2)}{F_2(x,Q^2,A_1)} \, ,
\end{equation}
where $A_1$ and $A_2$
are the atomic mass numbers of the two different nuclei.
Some of the experimental measurements included in this analysis
are presented instead as ratios of DIS cross-sections.

the maximum value of the momentum transfer $Q^2$
in the \texttt{nNNPDF1.0} input dataset is $Q^2_{\rm max} \simeq 200$ GeV$^2$ 
(see Fig.~\ref{figkinplot}).
Given that $Q^2_{\rm max}\ll M_Z^2$,
the contribution from the parity-violating $xF_3$ structure functions
and the contributions to $F_2$ and $F_L$ arising
from $Z$ boson exchange can be safely neglected.
Therefore, for the kinematic range relevant to the description
of available nuclear DIS data, Eq.~(\ref{eq:Master_DIS_xsec})
simplifies to:
\begin{equation}
\label{eq:sec2fullcrosssection2}
\frac{d^2 \sigma^{{\rm NC},l^\pm}}{dx dQ^2}(x,Q^2,A) =
\frac{2\pi \alpha^2}{xQ^4} Y_+F_2^{\rm NC}(x,Q^2,A) \lc 1
 - \frac{y^2}{1+(1-y)^2}\,\frac{F_L^{\rm NC}(x,Q^2,A)}{F_2^{\rm NC}(x,Q^2,A)}\ \rc \, ,
 \end{equation}
 where only the photon-exchange contributions
 are retained for the $F_2$ and $F_L$ structure functions.
 In Eq.~(\ref{eq:sec2fullcrosssection2})
 we have isolated the dominant $F_2$ dependence, since the second
 term is typically rather small.
 Note that since the center of mass energy of the lepton-nucleon
 collision $\sqrt{s}$ is determined by:
 \begin{equation}
 s = \lp k\,+\,P\rp^2 \simeq 2 k\cdot P = \frac{Q^2}{xy} \, ,
 \end{equation}
 where hadron and lepton masses have been neglected, measurements
 with the same values for $x$ and $Q^2$ but different center of mass energies 
 $\sqrt{s}$ will lead to a different value of the prefactor
 in front of the $F_L/F_2$ ratio in Eq.~(\ref{eq:sec2fullcrosssection2}),
 allowing in principle the separation of the two structure functions
 as in the free proton case.

Therefore, one should in principle account for the contributions from the
longitudinal structure function $F_L$ to cross-section ratios measured
by experiment.
However, it is well known that the ratio $F_L/F_2$ exhibits a very weak dependence
with $A$~\cite{Amaudruz:1992wn,Dasu:1988ru}, and
therefore the second term in Eq.~(\ref{eq:sec2fullcrosssection2})
cancels out to a good approximation when taking
ratios between different nuclei.
In other words, we can exploit the fact that:
\begin{equation}
\frac{d^2 \sigma^{\rm NC}(x,Q^2,A_2)/dxdQ^2}{
d^2 \sigma^{\rm NC}(x,Q^2,A_1)/dxdQ^2} \simeq
\frac{F_2(x,Q^2,A_2)}{F_2(x,Q^2,A_1)} = R_{F_2}\lp x, Q^2, A_1, A_2\rp \, ,
\end{equation}
in which then the ratios
of DIS cross-sections for $Q \ll M_Z$ in the form
of Eq.~(\ref{eq:sec2fullcrosssection2}) are equivalent
to ratios of the $F_2$ structure functions.
Lastly, it is important to note that whenever 
the nuclei involved in the 
measurements are not isoscalar,
the data is corrected to give isoscalar ratios and an 
additional source of systematic error is added as a
result of this conversion.

summarised in Table~\ref{tab:nNNPDF10_data} are the different types of nuclei 
measured by the experiments included in the \texttt{nNNPDF1.0} analysis.
  For each dataset, we indicate the nuclei $A_1$ and $A_2$
  that are used to construct the
  structure function ratios in Eq.~\ref{eq:Sect2Rf2}, 
  quoting
  explicitly the corresponding atomic mass numbers.
  We also display the number of data points that
  survive the baseline kinematical cuts, and give
  the corresponding publication references.

\begin{table}
  \centering
  \small
   \renewcommand{\arraystretch}{1.09}
\begin{tabular}{c c c c}
Experiment & ${\rm A}_1/{\rm A}_2$ & ${\rm N}_{\rm dat}$ & Reference\\
\toprule
  SLAC E-139 & $^4$He/$^2$D & 3 & \cite{Gomez:1993ri} \\
  NMC 95, re. & $^4$He/$^2$D & 13 & \cite{Amaudruz:1995tq}\\
\midrule
  NMC 95 & $^6$Li/$^2$D & 12 & \cite{Arneodo:1995cs}\\
\midrule
  SLAC E-139 & $^9$Be/$^2$D & 3 & \cite{Gomez:1993ri}\\
  NMC 96 & $^9$Be/$^{12}$C & 14 & \cite{Arneodo:1996rv}\\
\midrule
  EMC 88, EMC 90 & $^{12}$C/$^2$D & 12 & \cite{Ashman:1988bf,Arneodo:1989sy}\\
  SLAC E-139 & $^{12}$C/$^2$D & 2 & \cite{Gomez:1993ri}\\
  NMC 95, NMC 95, re.  & $^{12}$C/$^2$D & 26 & \cite{Arneodo:1995cs,Amaudruz:1995tq}\\
  FNAL E665 & $^{12}$C/$^2$D & 3 & \cite{Adams:1995is}\\
  NMC 95, re. & $^{12}$C/$^6$Li & 9 & \cite{Amaudruz:1995tq}\\
\midrule
  BCDMS 85 & $^{14}$N/$^2$D & 9 & \cite{Alde:1990im}\\
\midrule
  SLAC E-139 & $^{27}$Al/$^2$D & 3 & \cite{Gomez:1993ri}\\
  NMC 96 & $^{27}$Al/$^{12}$C & 14 & \cite{Arneodo:1996rv}\\
\midrule
  SLAC E-139 & $^{40}$Ca/$^2$D & 2 & \cite{Gomez:1993ri}\\
  NMC 95, re. & $^{40}$Ca/$^2$D & 12 & \cite{Amaudruz:1995tq}\\
  EMC 90 & $^{40}$Ca/$^2$D & 3 & \cite{Arneodo:1989sy}\\
  FNAL E665 & $^{40}$Ca/$^2$D & 3 & \cite{Adams:1995is}\\
  NMC 95, re. & $^{40}$Ca/$^6$Li & 9 & \cite{Amaudruz:1995tq}\\
  NMC 96 & $^{40}$Ca/$^{12}$C & 23 & \cite{Arneodo:1996rv}\\
\midrule
  EMC 87 & $^{56}$Fe/$^2$D & 58 & \cite{Aubert:1987da}\\
  SLAC E-139 & $^{56}$Fe/$^2$D & 8 & \cite{Gomez:1993ri}\\
  NMC 96 & $^{56}$Fe/$^{12}$C & 14 & \cite{Arneodo:1996rv}\\
  BCDMS 85, BCDMS 87 & $^{56}$Fe/$^2$D & 16 & \cite{Alde:1990im,Benvenuti:1987az}\\
\midrule
  EMC 88, EMC 93 & $^{64}$Cu/$^2$D & 27 & \cite{Ashman:1988bf,Ashman:1992kv}\\
\midrule
  SLAC E-139 & $^{108}$Ag/$^2$D & 2 & \cite{Gomez:1993ri}\\
\midrule
 EMC 88 & $^{119}$Sn/$^2$D & 8 & \cite{Ashman:1988bf}\\
 NMC 96, $Q^2$ dependence  & $^{119}$Sn/$^{12}$C & 119 & \cite{Arneodo:1996ru}\\
\midrule
 FNAL E665 & $^{131}$Xe/$^2$D & 4 & \cite{Adams:1992vm}\\
\midrule
  SLAC E-139 & $^{197}$Au/$^2$D & 3 & \cite{Gomez:1993ri}\\
\midrule
 FNAL E665 & $^{208}$Pb/$^2$D & 3 & \cite{Adams:1995is}\\
 NMC 96 & $^{208}$Pb/$^{12}$C & 14 & \cite{Arneodo:1996rv}\\
 \midrule
 \midrule
 {\bf Total NC DIS} & & {\bf 451} & \\
\bottomrule
\end{tabular}
\vspace{4mm}
\caption{\small The
neutral-current nuclear deep-inelastic
  input datasets included in \texttt{nNNPDF1.0}.
  For each dataset, we indicate the nuclei $A_1$ and $A_2$
  involved, the number of data points that
  satisfy the baseline kinematical cuts,
  and the publication reference.
}
\label{tab:nNNPDF10_data}
\end{table}

In Fig.~\ref{figkinplot} we show the
kinematical coverage in the $(x,Q^2)$ plane
of the DIS nuclear data included in \texttt{nNNPDF1.0}.
To minimise the contamination from low-scale non-perturbative corrections and 
higher-twist effects, and also to remain
consistent with the baseline proton PDF analysis (to be discussed
in Sect.~\ref{s2:nNNPDF10_minimisation}), we impose the same kinematical cuts on $Q^2$ and
the invariant final state mass squared $W^2=(P+q)^2$ as in the 
NNPDF3.1 global fit~\cite{Ball:2017nwa},
namely:
 \begin{equation}
 \label{eq:kincuts}
 Q^2 \ge Q^2_{\rm min}=3.5~{\rm GeV}^2 \, , \qquad 
 W^2 \ge W^2_{\rm min}=12.5~{\rm GeV}^2 \, ,
\end{equation}
which are represented by the 
dashed lines in Fig.~\ref{figkinplot}.
In Table~\ref{tab:kincuts}, we compare our kinematics cuts in $W^2$ and
 $Q^2$ to those implemented in the \texttt{nCTEQ15} and \texttt{EPPS16} fits.
We find that our cuts are very similar to those of
the \texttt{nCTEQ15} analysis~\cite{Kovarik:2015cma}, and as a result
our NC DIS nuclear structure function dataset is similar
to that used in their analysis.
On the other hand, our choice of both the $Q^2_{\rm min}$ and $W^2_{\rm min}$ 
cut is more stringent than that made in the \texttt{EPPS16} analysis \cite{Eskola:2016oht},
where they set $Q^2_{\rm min} = 1.69$~GeV$^2$ and do not impose any cut
in $W^2$.

\begin{table}
  \centering
     \renewcommand{\arraystretch}{1.90}
  \begin{tabular}{cccc}
    &  \texttt{nNNPDF1.0}  &   \texttt{nCTEQ15}  &  \texttt{EPPS16} \\
    \toprule
    $W^2_{\rm min}$  &  12.5 GeV$^2$  &  12.25 GeV$^2$  &  n/a \\
    \midrule
    $Q^2_{\rm min}$  & 3.5 GeV$^2$  &  4 GeV$^2$  &  1.69 GeV$^2$ \\
\bottomrule
  \end{tabular}
  \vspace{0.3cm}
  \caption{\label{tab:kincuts} The kinematics cuts in $W^2$ and
    $Q^2$ imposed in the \texttt{nNNPDF1.0} analysis compared to those
    used in the \texttt{nCTEQ15} and \texttt{EPPS16} fits.
  }
  \end{table}

After imposing the kinematical cuts in Eq.~(\ref{eq:kincuts}),
we end up with $N_{\rm dat}=451$ data points.
As indicated in Table~\ref{tab:nNNPDF10_data}, around half of these points
correspond to ratios of heavy nuclei with respect to
to deuterium, namely $R_{F_2}(A_1,A_2=2)$ in the notation
of Eq.~(\ref{eq:Sect2Rf2}).
For the rest of the data points, the values of $A_1$ and $A_2$
both correspond to heavier nuclei, with $A_2 \ge 6$.
It is worth noting that the measurements
from the NMC collaboration contain a significant amount of points
for which the carbon structure function is in the denominator,
$R_{F_2}(A_1,A_2=12)$.
In particular, we have $N_{\rm dat}=119$ data points
for the $Q^2$ dependence of the tin to carbon
ratio, $R_{F_2}(119,12)$.
These measurements provide valuable constraints on the $A$ dependence of the 
nuclear PDFs, since nuclear effects enter both the numerator
and denominator of Eq.~(\ref{eq:Sect2Rf2}).

Concerning the treatment of the experimental uncertainties,
we account for all correlations among data points whenever 
this information is provided
by the corresponding experiments. The covariance matrix is constructed based on Eq.~(\ref{eq:t0_covmat}).
For all of the measurements listed in 
Table~\ref{tab:nNNPDF10_data}, the detailed
break-up of the experimental systematic errors is not available
(in most cases these partially or totally cancel out when taking
ratios of observables), and the only systematic error that
enters the $t_0$ covariance matrix Eq.~(\ref{eq:t0_covmat}) is
the multiplicative normalization error.

Note that all the datasets listed in Table~\ref{tab:nNNPDF10_data} 
largely took place before it became clear that a detailed break-up of the 
systematic errors and their correlations is essential to fully exploit the information
contained in these kinds of measurements.
Recently, the state of affairs has improved
in this respect with the availability of LHC measurements 
of hard probes in p+Pb collisions, where the full covariance matrix 
is often made available.

\subsection{Parameterisation} \label{s2:nNNPDF10_parameterisation}

By means of the {\tt APFELgrid} formalism and the FK tables detailed in Sect.~\ref{s2:fast_evolution}, we can express any DIS structure function
in terms of the nPDFs at the initial evolution scale $Q_0^2$ using
Eq.~(\ref{eq:ev_interp}).
In this analysis as well as in Chapters~\ref{chap:nNNPDF20} and \ref{chap:nNNPDF30}, the FK tables are computed
up to NNLO in the QCD coupling expansion, with heavy quark effects evaluated
by the FONLL general-mass variable flavor number scheme~\cite{Forte:2010ta}.
Specifically, we use the FONLL-B scheme for the NLO fits and the FONLL-C for
the NNLO fits.
The value of the strong coupling constant is set to be $\alpha_s(m_Z)=0.118$, consistent
with the PDG average~\cite{Tanabashi:2018oca} and with recent high-precision
determinations~\cite{Ball:2018iqk,Verbytskyi:2019zhh,Bruno:2017gxd,Zafeiropoulos:2019flq}
(see~\cite{Pich:2018lmu} for an overview).
Our variable flavor number scheme has a maximum of $n_f=5$ active quarks,
where the heavy quark pole masses are taken to be $m_c=1.51$ GeV and $m_b=4.92$ GeV
following the Higgs Cross-Sections Working Group recommendations~\cite{deFlorian:2016spz}.
The charm and bottom PDFs are generated dynamically from the gluon and the light
quark PDFs starting from the thresholds $\mu_c=m_c$ and $\mu_b=m_b$.

In principle, one would need to parameterise 7 independent PDFs: the
up, down, and strange quark and antiquark PDFs and the gluon.
Another two input PDFs would be required if in addition the 
charm and anti-charm PDFs
are also parameterised, as discussed in~\cite{Ball:2016neh}.
However, given that our input dataset  in this analysis is restricted
to DIS NC structure functions, a full quark flavour separation
of the fitted nPDFs is not possible.

\myparagraph{Quark flavour decomposition}
Let us start by discussing the specific quark flavour decomposition that is adopted
in the \texttt{nNNPDF1.0} fit by considering the NC DIS structure function $F_2(x,Q^2,A)$
at leading order in terms of the nPDFs.
This decomposition is carried out for $Q^2 < m_c^2$ and therefore the charm PDF is absent.
In this case, one finds for the $F_2$ structure function:
\begin{equation}
  F^{(\rm LO)}_2 (x,Q^2,A) =  x \sum_{i=1}^{n_f}e_i^2f_i^+(x,Q^2,A) = x
  \lc \frac{4}{9} u^+(x,Q^2,A) + \frac{1}{9}\lp d^+ + s^+\rp (x,Q^2,A) \rc \, ,
\end{equation}
where for consistency the DGLAP evolution has been performed at LO, and the 
quark and antiquark PDF combinations are given by:
\begin{equation}
f_i^\pm (x,Q^2,A) \equiv \lc f_i (x,Q^2,A) \, \pm \, \bar{f}_i (x,Q^2,A)\rc \,.
\qquad  i = u,d,s \, .
\end{equation}
In this analysis, we will work in the PDF evolution basis, which is defined
as the basis composed by the eigenstates of the DGLAP evolution equations.
If we restrict ourselves to the  $Q < m_c$ ($n_f=3$) region,
the quark combinations are defined in this basis as:
\begin{eqnarray}
         \Sigma(x,Q^2,A) & \equiv& \sum_{i=1}^{n_f=3}f^+_i(x,Q^2,A)\qquad \text{(quark singlet)} \, ,\\
   \label{eq:def}      T_{3}(x,Q^2,A) & \equiv& \lp u^+ - d^+\rp (x,Q^2,A)\qquad \text{(quark triplet) }  \, , \\
        T_{8}(x,Q^2,A) & \equiv& \lp u^+ + d^+ - 2s^+\rp (x,Q^2,A) \qquad \text{(quark octet)}  \, .
 \end{eqnarray}
It can be shown that the NC DIS structure functions depend only on these
three quark combinations: $\Sigma$, $T_3$, and $T_8$.
Other quark combinations in the evolution basis, such as the valence distributions
$V=u^-+d^-+s^-$ and $V_3=u^--d^-$, appear only at the level
of charged-current structure functions, as well as in hadronic observables
such as $W$ and $Z$ boson production.

In the evolution basis, the $F_2$ structure function for a proton and a neutron target
at LO in the QCD expansion can be written as:
\begin{eqnarray}
  \label{eq:F2_p_lo}
    F^{({\rm LO}),p}_2 (x,Q^2) &=& x \lc \frac{2}{9} \Sigma + \frac{1}{6} T_3 + \frac{1}{18}T_8 \rc \, , \\
    F^{({\rm LO}),n}_2 (x,Q^2) &=& x \lc  \frac{2}{9} \Sigma - \frac{1}{6} T_3 + \frac{1}{18}T_8 \rc \, . \nonumber
\end{eqnarray}
Therefore, since the nuclear effects are encoded in the nPDFs,
the structure function for a nucleus with atomic number $Z$ and mass number $A$
will be given by a simple sum of the proton and neutron structure functions:
\begin{equation}
  \label{eq:F2_A}
  F^{({\rm LO})}_2 (x,Q^2,A) = \frac{1}{A}\left( Z F_2^{({\rm LO}),p}(x,Q^2) + (A-Z) F_2^{({\rm LO}),n}(x,Q^2) \right)\\ .
\end{equation}
Inserting the decomposition of Eq.~(\ref{eq:F2_p_lo}) into Eq.~(\ref{eq:F2_A}), we find:
\begin{equation}
\label{eq:F2_p_lo_v2}
F^{({\rm LO})}_2 (x,Q^2,A)= x \left[ \frac{2}{9} \Sigma - \left(\frac{Z}{3A}-\frac{1}{6}\right) T_3 + \frac{1}{18}T_8\right](x,Q^2,A) \, .
\end{equation}
Note that nuclear effects, driven by QCD, are electric-charge blind and therefore
depend only on the total number of nucleons $A$ within a given nuclei, in addition to
$x$ and $Q^2$.
The explicit dependence on $Z$ in Eq.~(\ref{eq:F2_p_lo_v2}) arises from QED effects, since
the virtual photon $\gamma^*$ in the deep-inelastic scattering couples more strongly
to up-type quarks ($|e_q|=2/3$) than to down-type quarks ($|e_q|=1/3$).

From Eq.~(\ref{eq:F2_p_lo_v2}) we see that at LO the  $F_2^p$ structure function
in the nuclear case depends on three independent quark combinations: the
total quark singlet $\Sigma$, the quark triplet $T_3$, and the quark octet $T_8$.
However, the dependence on the non-singlet triplet combination is very weak, since 
its coefficient is given by:
\begin{equation}
\left(\frac{Z}{3A}-\frac{1}{6}\right) = \left(\frac{Z}{3(2Z+\Delta A)}-\frac{1}{6}\right)
\simeq -\frac{\Delta A}{12Z} \, ,
\end{equation}
where $\Delta A \equiv A - 2Z$ quantifies the deviations from nuclear isoscalarity ($A=2Z$).
This coefficient is quite small for nuclei in which data is available,
and in most cases nuclear structure functions are corrected
for non-isoscalarity effects.
In this work, we will assume $\Delta A=0$ such that we have only
isoscalar nuclear targets.
The dependence on $T_3$ then drops out and the nuclear structure function
$F_2$ at LO is given by:
\begin{equation}
\label{eq:f2lo_v3}
F^{({\rm LO})}_2 (x,Q^2,A)= x \lc  \frac{2}{9} \Sigma + \frac{1}{18}T_8\rc(x,Q^2,A) \,,
\end{equation}
where now the only relevant quark combinations are the quark singlet $\Sigma$
and the quark octet $T_8$.
Therefore, at LO, NC structure function measurements
on isoscalar targets below
the $Z$ pole can only constrain a single quark combination, namely:
\begin{equation}
\label{eq:singlequarkcombination}
F^{({\rm LO})}_2 (x,Q^2,A) \propto \lp \Sigma +\frac{1}{4} T_8\rp(x,Q^2,A) \,.
\end{equation}

At NLO and beyond, the dependence on the gluon PDF enters
and the structure function Eq.~(\ref{eq:f2lo_v3}) becomes:
\begin{equation}
\label{eq:f2lo_v4}
F^{({\rm NLO})}_2 (x,Q^2,A)= C_{\Sigma}\otimes \Sigma(x,Q^2,A) + C_{T_8}\otimes T_8(x,Q^2,A)
+ C_{g}\otimes g(x,Q^2,A) \, ,
\end{equation}
where $C_{\Sigma}$, $C_{T_8}$, and $C_{g}$ are the coefficient functions associated
with the singlet, octet, and gluon respectively.
In principle one could aim to disentangle $\Sigma$ from $T_8$ due to 
their different $Q^2$ behaviour, but in practice this is not possible
given the limited kinematical coverage of the available experimental data.
Therefore, only the $\Sigma+ T_8/4$ quark combination is effectively constrained
by the experimental data used in this analysis, as indicated by
Eq.~(\ref{eq:singlequarkcombination}).

\begin{figure}[ht]
\begin{center}
  \includegraphics[width=0.9\textwidth]{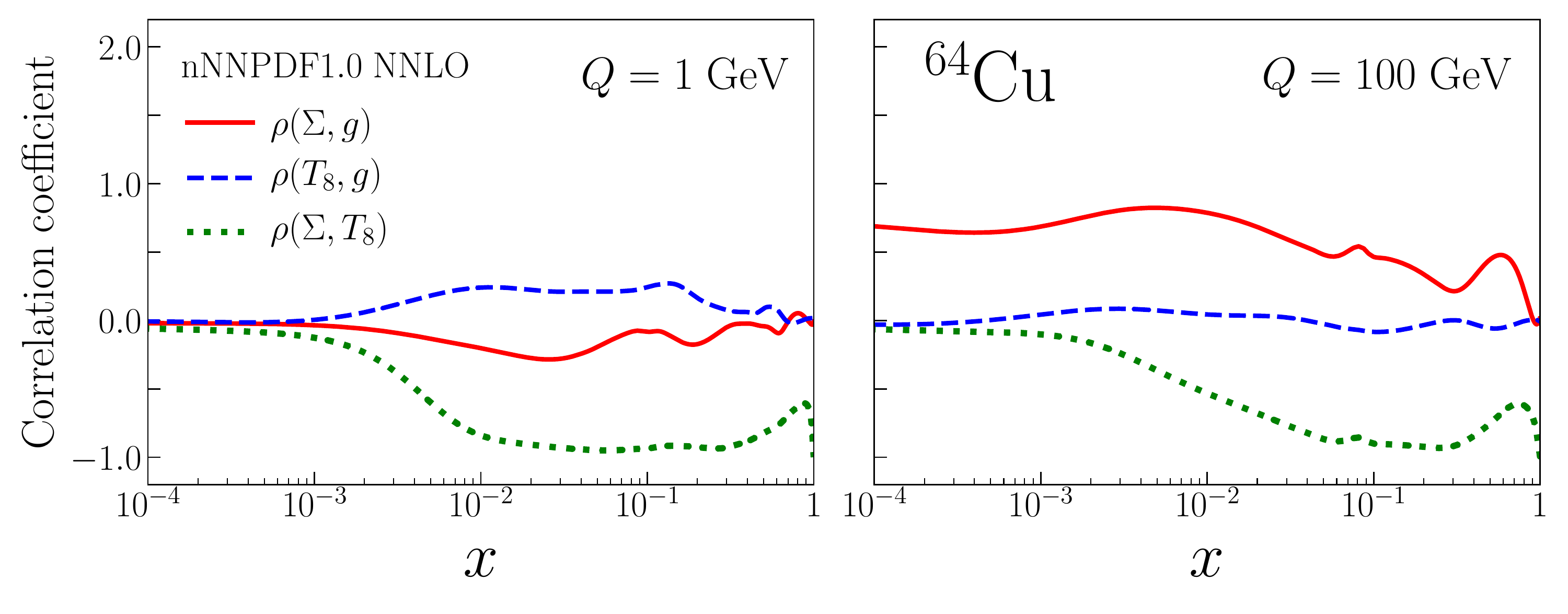}
 \end{center}
\vspace{-0.3cm}
\caption{\small The correlation coefficient $\rho=\langle \left(f_i - \langle f_i \rangle\right)\left(f_j - \langle f_j \rangle\right)\rangle /  \left(\sigma_i \sigma_j\right)$ between the
  the quark singlet $\Sigma$ and gluon $g$ (solid red line), 
  the quark octet $T_8$ and $g$ (dashed blue line),
  and between $\Sigma$ and $T_8$ (dotted green line).
  The coefficients are computed with $N_{\rm rep}=200$ replicas
  of the copper ($A=64$)
  \texttt{nNNPDF1.0} NNLO set
  at $Q=1$ GeV (left) and $Q=100$ GeV (right).
}
\label{fig:correlations}
\end{figure}

Putting together all of this information, we will consider the following
three independent PDFs at the initial parameterisation scale $Q_0$:
\begin{itemize}

\item the total quark singlet   $\Sigma(x,Q^2_0,A) = \sum_{i=1}^{3}f^+_i(x,Q^2_0,A)$,

\item the quark octet   $T_{8}(x,Q^2,A) = \lp u^+ + d^+ - 2s^+\rp (x,Q^2,A)$,

\item and the gluon nPDF $g(x,Q_0,A)$.

\end{itemize}
In Sect.~\ref{s2:nNNPDF10_parameterisation} we discuss the parameterisation of these three
nPDFs using neural networks.
In Fig.~\ref{fig:correlations} we show the results for
the correlation coefficient between the
nPDFs that are parameterised in the \texttt{nNNPDF1.0} fit (presented
in Sect.~\ref{s2:nNNPDF10_determination}),
specifically the NNLO set for copper ($A=64$) nuclei.
  The nPDF correlations are computed at both $Q=1$ GeV and $Q=100$ GeV,
  the former of which contains experimental data in the region $ 0.01 \lsim x \lsim 0.4$ (illustrated in Fig.~\ref{figkinplot}).
 In the data region, there is a strong anti-correlation
  between $\Sigma$ and $T_8$, consistent with
  Eq.~(\ref{eq:singlequarkcombination}) which implies that only their weighted
  sum can be constrained.
  As a result, we will show in the following sections only results of the 
combination $\Sigma+T_8/4$ which can be meaningfully
determined given our input experimental data.
 From Fig.~\ref{fig:correlations}, one can also observe the strong correlation between $\Sigma$ and $g$
  for $x\lsim 0.01$ and $Q=100$ GeV, arising from the fact that
  these two PDFs are coupled via the DGLAP evolution equations 
  as opposed to $T_8$ and $g$ where the correlation is very weak.
    
As mentioned in Sect.~\ref{s2:nNNPDF10_data}, the non-perturbative 
distributions that enter the collinear factorisation framework
in lepton-nucleus scattering
are the PDFs of a nucleon within an isoscalar nucleus with atomic mass 
number $A$, $f_i(x,Q^2,A)$.
While the dependence of the nPDFs on the scale $Q^2$ is determined
by the perturbative DGLAP evolution equations,
the dependence on both Bjorken-$x$ and the atomic
mass number $A$ is non-perturbative and needs
to be extracted from experimental data through
a global analysis.\footnote{See~\cite{Lin:2017snn}
  for an overview of recent efforts in the first-principle
  calculations of PDFs by means of lattice QCD.}
Taking into account the flavour decomposition
presented in Eq.~\ref{eq:def},
we are required to parameterise the $x$ and $A$ dependence
of the quark singlet $\Sigma$, the quark octet $T_8$, and the gluon $g$,
as indicated
by Eq.~(\ref{eq:f2lo_v3}) at LO and by Eq.~(\ref{eq:f2lo_v4})
for NLO and beyond.

\begin{figure} 
  \floatbox[{\capbeside\thisfloatsetup{capbesideposition={right,top},capbesidewidth=0.6\textwidth}}]{figure}[\FBwidth]
  {\caption{The NN architecture $\{3,\,25,\,3\}$ used in \texttt{\texttt{nNNPDF1.0}} having 3 input nodes $x$, $\ln{(1/x)}$ and $A$, one hidden layer of 25 neurons with a sigmoid activation function and 3 output nodes $N_i(x)$ of the considered nPDF flavours: $g$, $\Sigma$ and $T_8$ with a linear activation function. An overall of 178 free parameters (weights and biases).}\label{fig:nNNPDF10_architecture}}
  {\vspace{-1cm}\includegraphics[width=0.35\textwidth]{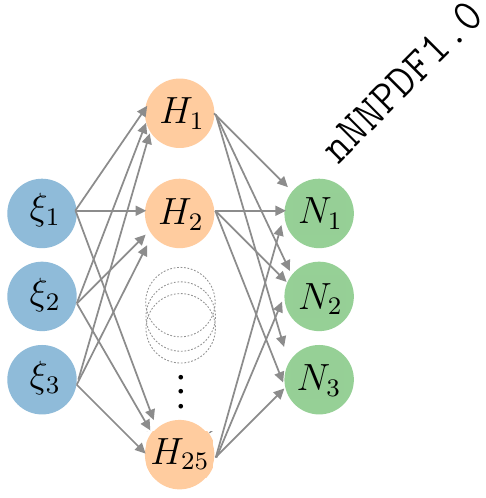}} 
  \vspace{-0.3cm}
  \end{figure}

The three distributions listed above are parameterised
at the input scale $\mu_0$ by the output of a
neural network $N_f$ multiplied by an $x$-dependent polynomial functional form.
In the \texttt{\texttt{NNPDF3.1}} analysis (see Chapter~\ref{chap:PDF}), a different multi-layer feed-forward
neural network was used for each of
the parameterised PDFs so that three independent
neural networks would be required (see Fig.~\ref{fig:NNPDF31_architecture}).
However, in this work we use instead a single artificial neural network
consisting of an input layer, one hidden layer, and an output layer. 
In Fig.~\ref{fig:nNNPDF10_architecture}
we display a schematic representation of the architecture
of the feed-forward neural network used in the present analysis.
The input layer 
contains three neurons which take as input the values
of the momentum fraction $x$, $\ln(1/x)$, and 
atomic mass number $A$, respectively.
The subsequent hidden layer contains 
25 neurons, which feed into the final output layer of three neurons, 
corresponding to the three fitted distributions $\Sigma, T_8$ and $g$. 
A sigmoid activation function is used for the neuron activation in the first
two layers, while a linear activation is used for the output layer. This latter 
choice ensures that the network output will not be bounded and
can take any value required to reproduce experimental data. 
The output from the final layer of neurons is then used to construct the full
parameterisation:
\begin{eqnarray}
x\Sigma(x,\mu_0,A) &=&x^{-\alpha_\Sigma} (1-x)^{\beta_\sigma} \,N_1(x,A) \, , \nonumber \\
xT_8(x,\mu_0,A) &=&x^{-\alpha_{T_8}} (1-x)^{\beta_{T_8}} \,N_2(x,A) \, , \label{eq:param} \\
xg(x,\mu_0,A) &=&B_gx^{-\alpha_g} (1-x)^{\beta_g} \,N_3(x,A) \, , \nonumber
\end{eqnarray}
where $N_i$ represent the values of the $i$-th neuron's activation state
in the third and final layer of the neural network.

%
Overall, there are a total of $N_{\rm par}=178$ free parameters
(weights and thresholds) in the neural network represented in Fig.~\ref{fig:nNNPDF10_architecture}.
These are supplemented
by the normalisation coefficient $B_g$
for the gluon nPDF and by the six preprocessing exponents $\alpha_f$ and $\beta_f$.
The latter are fitted simultaneously with the network parameters, while
the former is fixed by the momentum sum rule, described in more detail below.
Lastly, the input scale $\mu_0$ is set to 1 GeV to
maintain consistency 
with the settings of the
baseline proton fit, chosen to be the \texttt{NNPDF3.1} set with perturbative charm.

\myparagraph{Sum rules}
Since the nucleon energy must be distributed among 
its constituents in a way that ensures energy conservation,
the PDFs are required to obey the momentum sum rule outlined in Sect.~\ref{s2:Physical_constraints}.
In this analysis, similar to \texttt{\texttt{NNPDF3.1}}, Eq.~(\ref{eq:NNPDF_sumrules}) is applied by setting the overall 
normalisation of the gluon nPDF to:
\begin{equation}
\label{eq:NormG}
B_g(A) = \frac{1 - \int_0^1 dx\, x\Sigma(x,\mu_0,A)}{\int_0^1 dx\, xg(x,\mu_0,A)}. 
\end{equation}
where the denominator of Eq.~(\ref{eq:NormG}) is computed using 
Eq.~(\ref{eq:param}) and setting $B_g=1$.
Note that this expression needs only to be implemented at the input scale 
$\mu_0$, since the properties of DGLAP evolution guarantees 
that it will also be satisfied for any $Q > \mu_0$.
Since the momentum sum rule requirement must
be satisfied for any value of $A$,
the normalisation factor for the gluon distribution $B_g$ needs to be
computed separately for each
value of $A$ given by the experimental data (see Table~\ref{tab:nNNPDF10_data}).

In addition to the momentum sum rule, 
nPDFs must satisfy other sum rules such as those for the valence distributions (see Sect.~\ref{s2:Physical_constraints}).
These valence sum rules involve quark combinations
which are not relevant for the description of NC DIS structure
functions, and therefore do not need to be used in the present analysis.

\myparagraph{Preprocessing}
The polynomial preprocessing functions $x^{-\alpha_f}(1-x)^{\beta_f}$ in Eq.~(\ref{eq:param}) have
long been 
known to approximate well the general asymptotic behaviour of the PDFs
at small and large $x$~\cite{Ball:2016spl}.
Therefore, they help to increase 
the efficiency of parameter optimisation since the neural
networks have to learn smoother functions.
Note that the preprocessing exponents $\alpha_f$ and $\beta_f$
are independent of $A$, implying that the entire $A$ dependence of the input
nPDFs will arise from the output of the neural networks.

In previous \texttt{\texttt{NNPDF}} analyses, the preprocessing exponents $\alpha_f$ and $\beta_f$ were 
fixed to a randomly chosen value from a range that was determined iteratively.
Here instead we will fit their values for each Monte Carlo replica, so that they
are treated simultaneously with the weights and thresholds of the
neural network.
The main advantage of this approach is that one does not need to iterate the fit to find
the optimal range for the exponents, since now their best-fit values are automatically
determined for each replica.

Based on basic physical requirements, as well as
on empirical studies, we impose some additional constraints
on the range of allowed values that the exponents $\alpha_f$ and $\beta_f$
can take.
More specifically, we restrict the parameter values to:
\begin{equation}
\label{eq:preprocessing}
\alpha_f \in [-5,1] \, , \qquad \beta_f \in [0,10] \, ,\qquad f=\Sigma,T_8,g \, .
\end{equation}
Concerning the large-$x$ exponent $\beta_f$, the lower limit
in Eq.~(\ref{eq:preprocessing}) guarantees that the nPDFs vanish 
in the elastic limit $x\to 1$; the upper limit follows from the observation that it
is unlikely for the nPDFs to be more strongly suppressed at large $x$.~\cite{Ball:2016spl}.
With respect to the small-$x$ exponent $\alpha_f$, the upper limit follows
from the nPDF integrability condition, given that for $\alpha_f > 1$
the momentum integral Eq.~(\ref{eq:MSR}) becomes divergent.

In addition to the conditions encoded in Eq.~(\ref{eq:preprocessing}), we 
also set $\beta_\Sigma = \beta_{T_8}$,
namely we assume that the two quark distributions $\Sigma$ and $T_8$ share
a similar large-$x$ asymptotic behaviour.
The reason for this choice is two-fold.
First, we know that these two distributions are
highly (anti-) correlated for
NC nuclear DIS observables (see Eq.~(\ref{eq:singlequarkcombination})).
Secondly, the large-$x$ behaviour of these distributions is expected to be approximately
the same, given that the strange distribution $s^+$ is known to be suppressed
at large $x$ compared to the corresponding $u^+$ and $d^+$ distributions.
In any case, it is important to emphasize that
the neural network has the ability to compensate 
for any deviations in the shape of the preprocessing function, 
and therefore can in principle
distinguish any differences
between $\Sigma$ and $T_8$ in the large-$x$ region.

\begin{figure}[ht]
\begin{center}
  \includegraphics[width=0.9\textwidth]{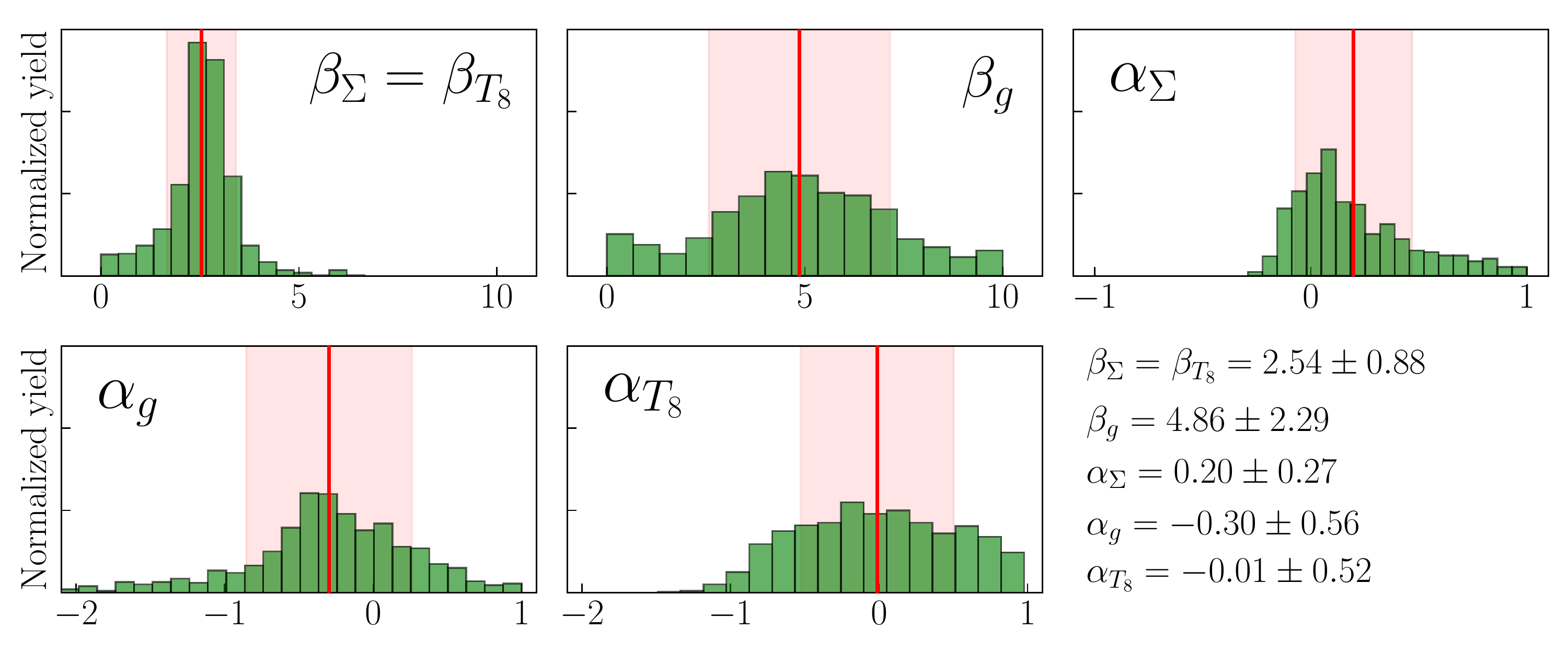}
 \end{center}
\vspace{-0.5cm}
\caption{\small The probability distribution of the fitted preprocessing
  exponents computed with the $N_{\rm rep}=1000$ replicas of the \texttt{nNNPDF1.0}
  NLO set.
  The vertical red line indicates the mean value and the transparent red band
  the 1-$\sigma$ range corresponding to each exponent.
}
\label{fig:preprocessing}
\vspace{-0.3cm}
\end{figure}

To illustrate the results of fitting the small and
large-$x$ preprocessing exponents, we display in Fig.~\ref{fig:preprocessing}
the probability distributions associated with the $\alpha_f$ and $\beta_f$
exponents computed using the $N_{\rm rep}=1000$ replicas of the \texttt{nNNPDF1.0}
NLO set, to be discussed in Sect.~\ref{s2:nNNPDF10_determination}.
Here the mean value of each exponent is marked by the solid red line, 
and the transparent 
red band describes the 1-$\sigma$ deviation.
Note that these exponents are restricted to vary only in the interval given by
Eq.~(\ref{eq:preprocessing}).
Interestingly, the resulting distributions for each of the $\alpha_f$ and $\beta_f$
exponents turn out to be quite different, for instance $\beta_{\Sigma}$ is Gaussian-like
while $a_{\Sigma}$ is asymmetric.

\subsection{Free proton baseline}
\label{s2:nNNPDF10_protonBC}
An important physical requirement that must be satisfied by the nPDFs
is that they should reproduce the $x$ dependence
of the PDFs corresponding to isoscalar free nucleons when evaluated at $A=1$.
Therefore, the following boundary conditions needs to be satisfied
for all values of $x$ and $Q^2$:
\begin{equation}
\label{eq:constraintprotonPDFs}
f(x,Q,A=1) = \frac{1}{2}\lc f^{p}(x,Q^2) + f^{n}(x,Q^2) \rc \, , \quad f=\Sigma,T_8,g \, ,
\end{equation}
where $f^p$ and $f^n$ indicate the parton distributions of the free proton
and neutron, respectively, and are related by isospin symmetry (which is assumed
to be exact).
As opposed to other approaches adopted
in the literature, we do not implement Eq.~(\ref{eq:constraintprotonPDFs})
at the nPDF parameterisation level,
but rather we impose it as a restriction in the allowed parameter space at the
level of $\chi^2$ minimisation, as will be discussed below.
Our strategy has the crucial advantage that it assures that both central values
and uncertainties of the free-nucleon PDFs will be reproduced by the \texttt{nNNPDF1.0}
nuclear set in the $A\to 1$ limit.


Having described the strategy
for the nPDF parameterisation in terms
of neural networks, we 
turn now to discuss how the best-fit values of these parameters, namely the weights and thresholds of the neural network and the preprocessing exponents $\alpha_f$ and $\beta_f$, are determined.
%
In this analysis, the best-fit parameters are determined from the minimisation
of a $\chi^2$ function defined as:
\begin{eqnarray}
\label{eq:chi2_nNNPDF10}
\chi^2 &\equiv & \sum_{i,j=1}^{N_{\rm dat}} \lp R_i^{\rm (exp)}-R_i^{\rm (th)}(\{ f_m\})\rp
 \lp {\rm cov_{t_0}}\rp_{ij}^{-1} \lp R_j^{\rm (exp)}-R_j^{\rm (th)}(\{ f_m\})\rp \\
&+&
 \lambda \sum_{m=g,\Sigma,T_8}\sum_{l=1}^{N_x} \lp f_m(x_l,\mu_0,A=1) - f_m^{(p+n)/2}(x_l,\mu_0) \rp^2 \, .
 \nonumber
 \end{eqnarray}
Here,
  $R_i^{\rm (exp)}$ and $R_i^{\rm (th)}(\{ f_m\})$ stand for the experimental
 data and the corresponding theoretical predictions for the
 nuclear ratios, respectively, the latter of which depend on the
 nPDF fit parameters. 
 The $t_0$ covariance matrix ${\rm cov_{t_0}}$ has been defined in
 Eq.~(\ref{eq:t0_covmat}), and $N_{\rm dat}$ stands for the total
 number of data points included in the fit.
 Therefore, the first term above is the same
 as in previous \texttt{\texttt{NNPDF}} fits.
 %

%
The second term in Eq.~(\ref{eq:chi2_nNNPDF10}) is a new feature in \texttt{nNNPDF1.0}.
It corresponds to the Lagrange multiplier method (see Sect.~\ref{s2:likelihood}) that forces
the fit to satisfy the boundary condition in Eq.~(\ref{eq:constraintprotonPDFs}),
namely that the fitted nPDFs for $A=1$ reproduce the PDFs of an isoscalar
free nucleon constructed as the average of the proton and neutron PDFs.
 In order to impose this constraint in a fully consistent way, it is necessary
 for the proton PDF baseline to have been determined using theoretical settings
 and a fitting methodology that best match those of the current
 nPDF analysis.
 This requirement is satisfied by the \texttt{NNPDF3.1} global analysis~\cite{Ball:2017nwa}  based on a wide range
 of hard-scattering processes together with higher-order QCD calculations.
 Crucially, \texttt{NNPDF3.1} shares most of the methodological choices of \texttt{nNNPDF1.0}
 such as the use of neural networks for the PDF
 parameterisation and of the Monte Carlo replica method
 for error propagation and estimation.

 As can be seen from Eq.~(\ref{eq:chi2_nNNPDF10}), this constraint is only 
 imposed at the initial scale $\mu_0$. 
 This is all that is required, since the properties of DGLAP 
 evolution will result in distributions at $Q > \mu_0$ that 
 automatically satisfy the constraint.
 The $A=1$ boundary condition is then constructed with a grid of $N_x=60$ values of $x$, where
 $10$ points are distributed logarithmically from $x_{\rm min}=10^{-3}$ to $x_{\rm mid}=0.1$ and
 $50$ points are 
 linearly distributed from $x_{\rm mid}=0.1$ to $x_{\rm max}=0.7$.
 
 Note that in the low-$x$ region the coverage of this constraint
 is wider than that of the available nuclear data (see Fig.~\ref{figkinplot}).
 Since proton
 PDF uncertainties, as a result of including HERA structure function data,
 are more reduced at small $x$ than in the corresponding
 nuclear case, the constraint in Eq.~(\ref{eq:chi2_nNNPDF10}) introduces
 highly non-trivial information regarding the shape of the nPDFs within
 and beyond the experimental data region.
 Moreover, we have also verified that the constraint can also be applied down
 to much smaller values of $x$, such as $x_{\rm min}=10^{-5}$, by
 taking as a proton baseline one of the NNPDF3.0 sets which include LHCb charm production
 data~\cite{Gauld:2015yia,Gauld:2016kpd,Bertone:2018dse}, as
 will be demonstrated in App.~\ref{s2:nNNPDF10_methodologyresults}.

 It is important to emphasize that the boundary condition, Eq.~(\ref{eq:constraintprotonPDFs}),
 must be satisfied both for the PDF central values and for the corresponding uncertainties.
 Since proton PDFs are known to much higher precision than
 nPDFs, imposing this condition introduces a significant amount of new information
 that is ignored in most other nPDF analyses.
 In order to ensure that PDF uncertainties are also reproduced in
 Eq.~(\ref{eq:constraintprotonPDFs}), for each \texttt{nNNPDF1.0} fit
  we randomly choose a replica from the 
  \texttt{NNPDF3.1} proton global fit in Eq.~(\ref{eq:chi2_nNNPDF10}).
  Since we are performing a large $N_{\rm rep}$ number of fits to estimate the uncertainties
  in \texttt{nNNPDF1.0}, the act of randomly choosing a different proton PDF baseline each time 
  guarantees that the necessary information contained in \texttt{NNPDF3.1} will propagate into the nPDFs.
Finally, we fix the hyper-parameter to $\lambda = 100$, which is 
found to be the optimal setting together with the choice of architecture to yield $A=1$ 
distributions that best describe the central values and uncertainties of \texttt{NNPDF3.1}.

\subsection{Minimisation strategy} \label{s2:nNNPDF10_minimisation}
Having defined our $\chi^2$ function in Eq.~(\ref{eq:chi2_nNNPDF10}), we now move
to present our procedure to determine the best-fit values
of the parameters associated with each Monte Carlo replica.
This procedure begins by sampling the initial values
of the fit parameters.
Concerning the preprocessing
exponents $\alpha_f$ and $\beta_f$, they are sampled from a uniform prior
in the range $\alpha_f \in [-1,1]$ and $\beta_f \in [1,10]$ for all fitted distributions.
Note that these initial ranges are contained within the ranges from Eq.~(\ref{eq:preprocessing}) 
in which the exponents are allowed to vary.
Since the neural network can accommodate changes in the PDF shapes from
the preprocessing exponents, we find the choice of the prior range from which
$\alpha_f$ and $\beta_f$ are initially sampled does not affect the resulting distributions. 
In the end, the distributions of $\alpha_f$ and $\beta_f$ do not exhibit flat behaviour, as 
is shown in Fig.~(\ref{fig:preprocessing}).

Concerning the initial sampling of the neural network parameters, we use Xavier initialization~\cite{GlorotAISTATS2010},
which samples from a normal distribution with a mean of zero and standard 
deviation that is dependent on the specific architecture of the network.
Furthermore, the initial 
values of the neuron activation are dropped and re-chosen if they are outside two standard 
deviations.
Since a sigmoid activation function is used for the hidden layer, this truncation of the sampling
distribution ensures the neuron input to be around the origin where the derivative is largest, 
allowing for more efficient network training.

The most significant difference between the fitting methodology used
in \texttt{nNNPDF1.0} as compared to previous \texttt{NNPDF} studies is the choice
of the optimisation algorithm for the $\chi^2$ minimisation.
In the most recent unpolarized~\cite{Ball:2017nwa} and polarized~\cite{Nocera:2014gqa}
proton PDF analysis
based on the \texttt{NNPDF} methodology, an in-house Genetic Algorithm (GA)
was employed for the $\chi^2$ minimisation, while for the \texttt{NNFF} fits
of hadron fragmentation functions~\cite{Bertone:2018ecm} the related
Covariance Matrix Adaptation -
Evolutionary Strategy (CMA-ES) algorithm was used (see also~\cite{Rojo:2018qdd}).
In both cases, the optimisers require as input only the local values
of the $\chi^2$ function for different points in the parameter space, but never use
the valuable information contained in its gradients.

In the \texttt{nNNPDF1.0} analysis,
we utilise for the first time gradient descent with backward propagation (see Sect.~\ref{s2:minimisation}), 
the most widely used training technique for neural networks (see also~\cite{Forte:2002fg}).
The main requirement
to perform gradient descent is to be able to efficiently compute the gradients of the cost
function Eq.~(\ref{eq:chi2_nNNPDF10}) with respect to the fit parameters.
Such gradients can in principle be computed analytically, by exploiting
the fact that the relation between the structure functions and the input
nPDFs at $\mu_0$ can be compactly expressed in terms of a matrix multiplication
within the {\tt APFELgrid} formalism as indicated by Eq.~(\ref{eq:ev_interp}).
One drawback of such approach is that the calculation of the gradients
needs to be repeated whenever the structure of the $\chi^2$ is modified.
For instance, different analytical expressions for the gradients are required if uncertainties
are treated as uncorrelated and added in quadrature as opposed to 
the case in which systematic correlations are taken into account.

Rather than following this path, in \texttt{nNNPDF1.0} we have implemented 
backward propagation neural network training using reverse-mode automatic differentiation
in {\tt TensorFlow}, a highly efficient and accurate
method to automatically compute the 
gradients of any user-defined cost function.
As a result, the use of automatic differentiation makes it significantly easier to explore optimal
settings in the model and extend the analysis to include other types of observables
in a global analysis. 

One of the drawbacks of the gradient descent approach, which is partially
avoided by using GA-types of optimisers, is the
risk of ending up trapped in local minima.
To ensure that such situations are avoided as much as possible, in \texttt{nNNPDF1.0}
we use the Adaptive Moment Estimation (ADAM) algorithm~\cite{DBLP:journals/corr/KingmaB14} 
to perform stochastic gradient descent (SGD).
The basic idea here is to perform the training
on randomly chosen subsets of the input experimental data,
which leads to more frequent parameter updates (see Sect.~\ref{s2:minimisation}).

In this analysis, most of the ADAM hyper-parameters 
are set to be the default values given by the algorithm, which have been 
tested on various machine learning problems. 
This includes
the initial learning rate of the parameters, $\eta = 0.001$, the 
exponential decay rate of the averaged squared gradients from past iterations, 
$\beta_2 = 0.999$, and a smoothing parameter $\epsilon = 10^{-8}$. 
However,
we increase the exponential decay rate of the mean of previous
gradients, $\beta_1 = 0.9 \to 0.99$, 
which can be interpreted more simply as the descent momentum. 
This choice was observed to improve the performance of the minimisation
overall, as it exhibited quicker exits from local minima and increased
the rate of descent. 

Given that our neural-network-based parameterisation
of the nPDFs, Eq.~(\ref{eq:param}), can be shown to be highly redundant
for the current input dataset (see also App.~\ref{s2:nNNPDF10_methodologyresults}),
we run the risk of fitting
the statistical fluctuations in the data rather than the underlying physical law.
To prevent such overfitting,
we have implemented the look-back cross-validation stopping criterion
presented for the first time in \texttt{NNPDF} fits (see Sect.~\ref{s2:neural_networks}).
The basic idea of this algorithm is to separate the input dataset into disjoint training
and validation datasets (randomly chosen replica by replica),
minimise the training set $\chi^2$ function, $\chi^2_{\rm tr}$, and stop the training
when the validation $\chi^2$, $\chi^2_{\rm val}$, reveals a global minimum.
In this analysis, the data is partitioned 50\%/50\% to construct each
of the two sets, except for experiments with 5 points
or less which are always included in the training set. 

The final fits are chosen to satisfy simultaneously the following conditions:
\begin{eqnarray}
\label{eq:chi2cuts}
&\chi^2_{\rm tr}/{N_{\rm tr}} < 5,\\ \nonumber
&\chi^2_{\rm val}/{N_{\rm val}} < 5,\\
&\chi^2_{\rm penalty}/(3N_x) < 5, \nonumber
\end{eqnarray}
where $N_{\rm tr}$ and $N_{\rm val}$ are the number of data points in the training
and validation sets, respectively, and $\chi^2_{\rm penalty}$ corresponds to the 
second term in Eq.~\ref{eq:chi2_nNNPDF10}. 
Upon reaching the above conditions during $\chi^2$ minimisation, checkpoints 
are saved for every 100 iterations. A fit is then terminated 
when a smaller value for the validation 
$\chi^2$ is not obtained after 
$5\times 10^4$ iterations, or when the fit has proceeded $5\times 10^5$
iterations (early stopping).
The former is set to allow sufficient time to 
escape local minima, and the latter is set due to the SGD algorithm, which 
can fluctuate the parameters around the minimum indefinitely. 
In either case the fit is considered successful, and the parameters 
that minimise $\chi^2_{\rm val}$ are selected as the best-fit parameters
(look-back).

A qualitative estimate for improvement in performance that has been achieved as a result of using the SGD w.r.t. GAs is presented in Sect.~3.2 of Ref.~\mycite{AbdulKhalek:2019mzd}.
\section{Results}
\label{s1:nNNPDF10_results}

In this section, I present the main results this analysis, namely
the \texttt{nNNPDF1.0} sets.
In Sect.~\ref{s2:nNNPDF10_datavstheory}, I discuss the quality of the inference by comparing 
the resulting structure function ratios with
experimental data.
In Sect.~\ref{s2:nNNPDF10_determination}, I present the main features
of the \texttt{nNNPDF1.0} sets, as well as a comparison with
the recent \texttt{EPPS16} and \texttt{nCTEQ15} nPDF analyses.
I also assess the stability of the results w.r.t. the 
perturbative order, which are generated using NLO
and NNLO QCD theories.
I provide in the App.~\ref{s2:nNNPDF10_methodologyresults}, a few methodological 
validation tests. These include the stability of the results w.r.t. variations
of the network architecture and the quantification of the the $A=1$ boundary condition impact.

\subsection{Inference quality} \label{s2:nNNPDF10_datavstheory}

To begin, we list in Table~\ref{tab:nNNPDF10_chi2}
the absolute and normalised values of the $\chi^2$ for
each of the input datasets (see Table~\ref{tab:nNNPDF10_data}) and for the total dataset.
The values are given for both the NLO and NNLO fits.
In total, there are $N_{\rm dat}=451$ data points that survive the kinematic
cuts and result in the overall value $\chi^2/N_{\rm dat}=0.68$, indicating
an excellent agreement between the experimental data and the 
theory predictions.
Moreover, we find that the fit quality is quite similar between 
the NLO and NNLO results.
The fact that we obtain an overall $\chi^2/N_{\rm dat}$ less than one can
be attributed to the absence of information on the correlations between experimental
systematics,
leading to an overestimation of the total error.  
%

\begin{table}
  \centering
  \small
   \renewcommand{\arraystretch}{1.09}
   \begin{tabular}{c c c | c c | c c}
     \multirow{2}{*}{Experiment} &
     \multirow{2}{*}{$~~~{\rm A}_1/{\rm A}_2~~~$} &
     \multirow{2}{*}{${N}_{\rm dat}$} & \multicolumn{2}{c|}{NLO} & \multicolumn{2}{c}{NNLO}\\
          &  &  & $\chi^2$  &   $\chi^2/N_{\rm dat}$ & $\chi^2$  &   $\chi^2/N_{\rm dat}$ \\
 \toprule
  SLAC E-139 & $^4$He/$^2$D & 3 & 1.49  &  0.50   & 1.50   & 0.50   \\ 
  NMC 95, re. & $^4$He/$^2$D & 13 & 12.81    & 1.0   & 12.79   & 0.98    \\ 
\midrule
  NMC 95 & $^6$Li/$^2$D & 12 & 10.96    &  0.91  & 10.50   &  0.88  \\  
\midrule
  SLAC E-139 & $^9$Be/$^2$D & 3 & 2.91    &  0.97  &  2.91  & 0.97   \\  
  NMC 96 & $^9$Be/$^{12}$C & 14 & 4.03    & 0.29   & 4.06   &  0.29  \\  
\midrule
  EMC 88, EMC 90 & $^{12}$C/$^2$D & 12 & 12.98    & 1.08   &  13.04  & 1.09   \\  
  SLAC E-139 & $^{12}$C/$^2$D & 2 & 0.65    &  0.33  & 0.74   & 0.37   \\  
  NMC 95, NMC 95, re.   & $^{12}$C/$^2$D & 26 & 25.12    &  0.97  & 24.81   &  0.95   \\  
  FNAL E665 & $^{12}$C/$^2$D & 3 & 3.13    &  1.04  &  3.13  &  1.04  \\ 
  NMC 95, re. & $^{12}$C/$^6$Li & 9 & 6.62    &  0.74   & 6.25   &  0.69  \\  
\midrule
  BCDMS 85 & $^{14}$N/$^2$D & 9 & 11.10    &  1.23  &  11.16  & 1.24   \\  
\midrule
  SLAC E-139  & $^{27}$Al/$^2$D & 3 & 0.52    &  0.17  & 0.65   & 0.22   \\  
  NMC 96 & $^{27}$Al/$^{12}$C & 14 & 4.34    &  0.31  &  4.31  & 0.31   \\  
\midrule
  SLAC E-139 & $^{40}$Ca/$^2$D & 2 & 2.79    & 1.40   &  2.95  &  1.48  \\ 
  NMC 95, re. & $^{40}$Ca/$^2$D & 12 & 11.75    & 0.98   &  11.86  &  0.99  \\  
  EMC 90 & $^{40}$Ca/$^2$D & 3 & 4.11    & 1.37   & 4.09   & 1.36   \\  
  FNAL E665 & $^{40}$Ca/$^2$D & 3 & 5.07    & 1.69   & 4.77   &  1.59  \\  
  NMC 95, re. & $^{40}$Ca/$^6$Li & 9 & 2.18    &  0.24  &  2.05  &  0.23  \\ 
  NMC 96 & $^{40}$Ca/$^{12}$C & 23 & 13.20    & 0.57   &  13.26  & 0.58   \\  
\midrule
  EMC 87 & $^{56}$Fe/$^2$D & 58 & 36.89    &  0.63  &  37.12  & 0.64   \\  
  SLAC E-139 & $^{56}$Fe/$^2$D & 8 & 11.01    &  1.38  & 11.20   &  1.4  \\  
  NMC 96 & $^{56}$Fe/$^{12}$C & 14 & 9.21    &   0.66 & 9.00   & 0.64   \\  
  BCDMS 85, BCDMS 87 & $^{56}$Fe/$^2$D & 16 & 9.48    & 0.6   & 9.53   & 0.6   \\  
\midrule
  EMC 88, EMC 93 & $^{64}$Cu/$^2$D & 27 & 12.56    & 0.47   & 12.63   &  0.47  \\  
\midrule
  SLAC E-139 & $^{108}$Ag/$^2$D & 2 & 1.04    &  0.52  &  1.04  & 0.52   \\  
\midrule
 EMC 88 & $^{119}$Sn/$^2$D & 8 & 17.77    &  2.22  & 17.71   & 2.21   \\  
 NMC 96, $Q^2$ dependence & $^{119}$Sn/$^{12}$C & 119 & 59.24    & 0.50   & 58.28   & 0.49   \\  
\midrule
 FNAL E665 & $^{131}$Xe/$^2$D & 4 & 1.47    & 0.37   & 1.45   & 0.36    \\  
\midrule
  SLAC E-139 & $^{197}$Au/$^2$D & 3 & 2.46    &  0.82  & 2.33   & 0.78   \\ 
\midrule
 FNAL E665 & $^{208}$Pb/$^2$D & 3 & 4.97    & 1.66   & 5.10   & 1.7   \\  
 NMC 96 & $^{208}$Pb/$^{12}$C & 14 & 5.23    &  0.37   & 5.61   & 0.4   \\  
 \midrule
 \midrule
 {\bf Total} & & {\bf 451} & {\bf 307.1}   & {\bf 0.68}   & {\bf 305.82}   & {\bf 0.68}    \\ 
\bottomrule
\end{tabular}
\vspace{4mm}
\caption{\small Same as Table~\ref{tab:nNNPDF10_data}, now indicating
  the absolute and normalised values of the $\chi^2$ for
  each of the input datasets as well as for the total dataset.
  Listed are the results for both the NLO and NNLO \texttt{nNNPDF1.0} sets.
}
\label{tab:nNNPDF10_chi2}
\end{table}

\begin{figure}[ht]
\begin{center}
  \includegraphics[width=0.99\textwidth]{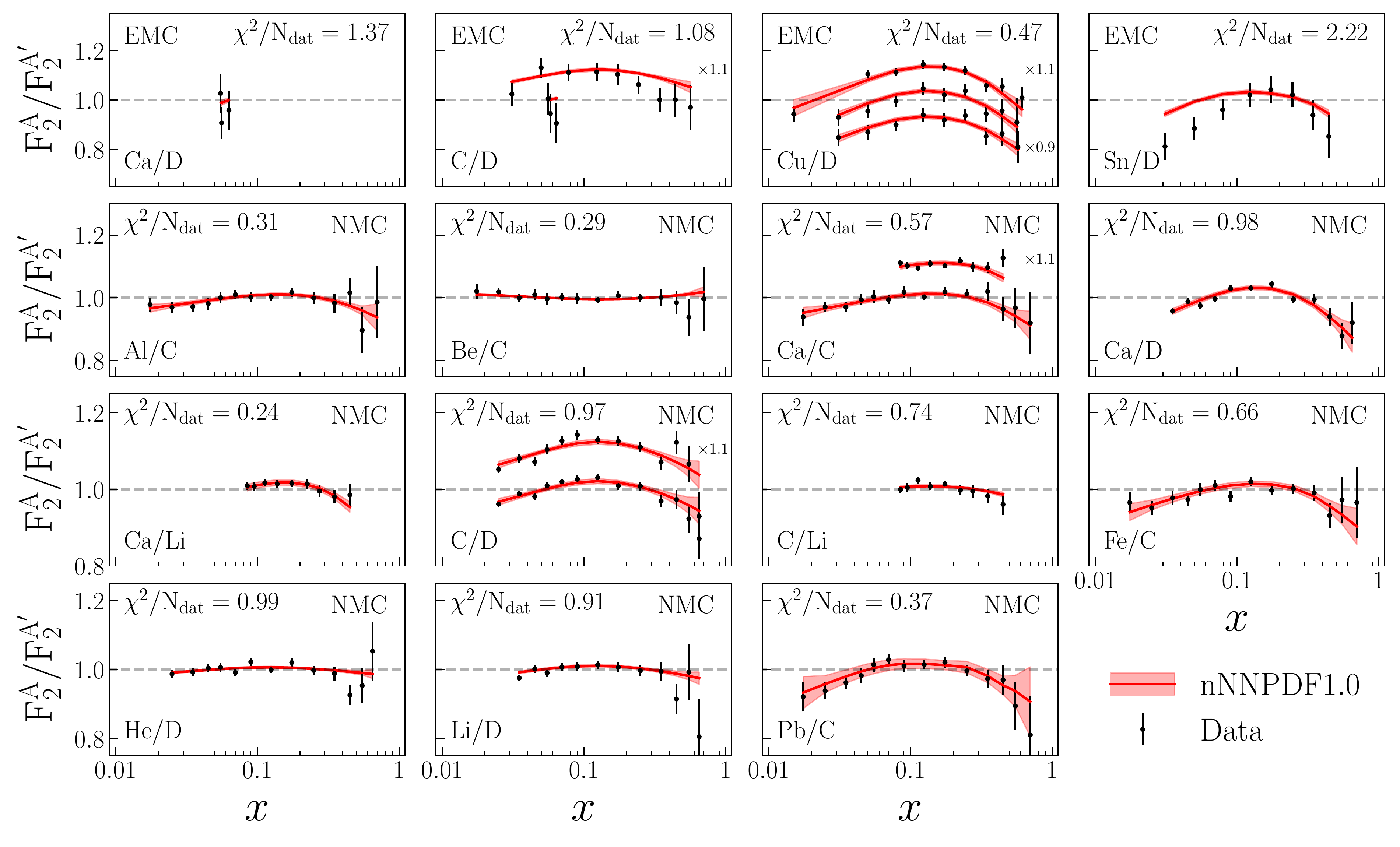}
 \end{center}
\vspace{-0.3cm}
\caption{\small Comparison between the experimental data on the structure
  function ratios $F_2^A/F_2^{A'}$ and the corresponding theoretical predictions
  from the \texttt{nNNPDF1.0} NLO fit (solid red line and shaded band) for the measurements 
  provided by the EMC
  and NMC experiments.
  The central values of the experimental data points have been shifted
  by an amount determined by the 
  multiplicative systematic uncertainties and their nuisance parameters,
  and the data errors are defined by adding in quadrature the 
  uncorrelated uncertainties.
  Also indicated are the $\chi^2/N_{\rm dat}$ values for each of the datasets.
}
\label{figDvT1}
\end{figure}

\begin{figure}[ht]
\begin{center}
  \includegraphics[width=0.99\textwidth]{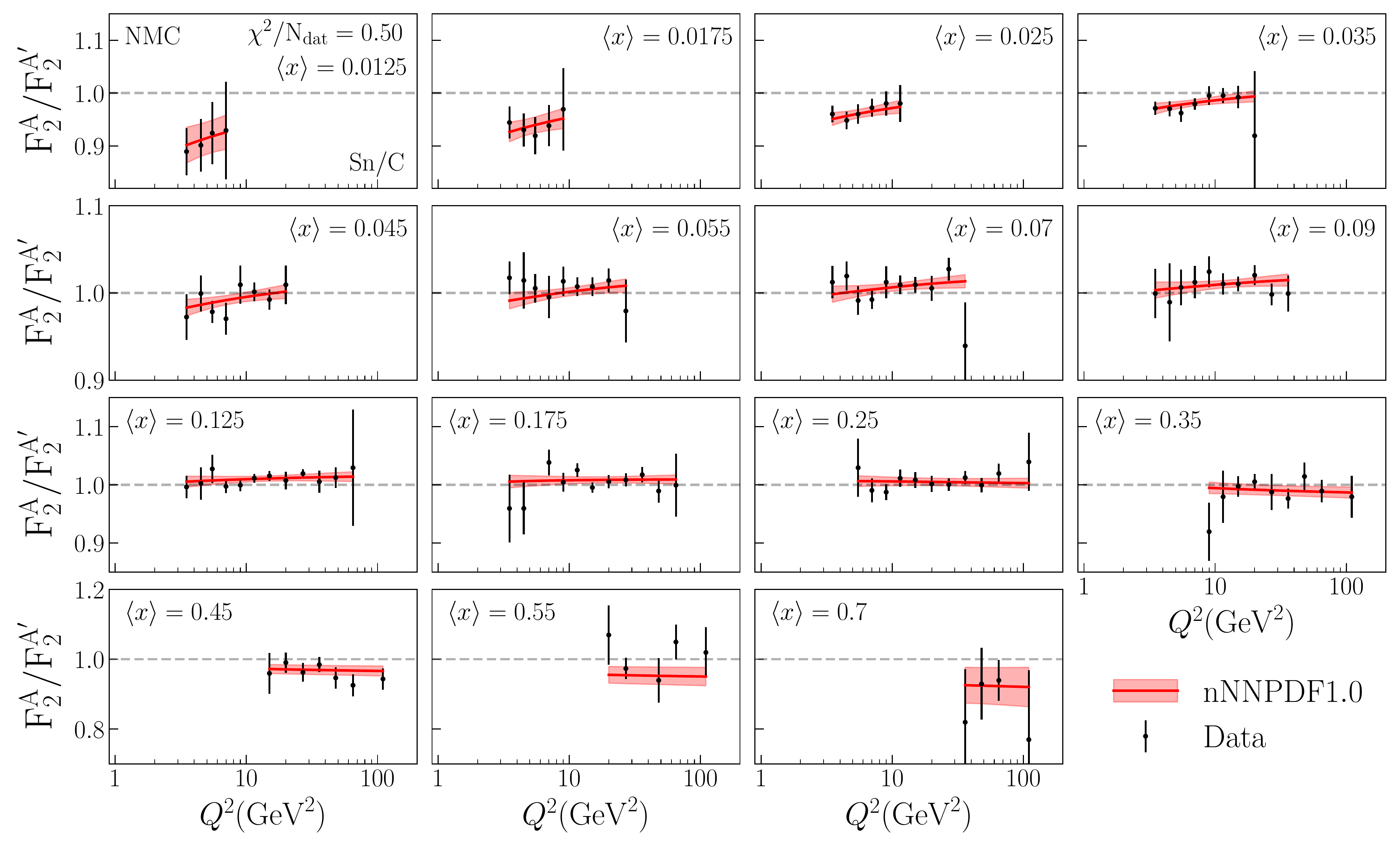}
 \end{center}
\vspace{-0.3cm}
\caption{\small Same as Fig~\ref{figDvT1} but for the 
$Q^2$-dependent structure
function ratio $F_2^{\rm Sn}/F_2^{\rm C}$ provided by the NMC experiment.}
\label{figDvT2}
\end{figure}

\begin{figure}[ht]
\begin{center}
  \includegraphics[width=0.99\textwidth]{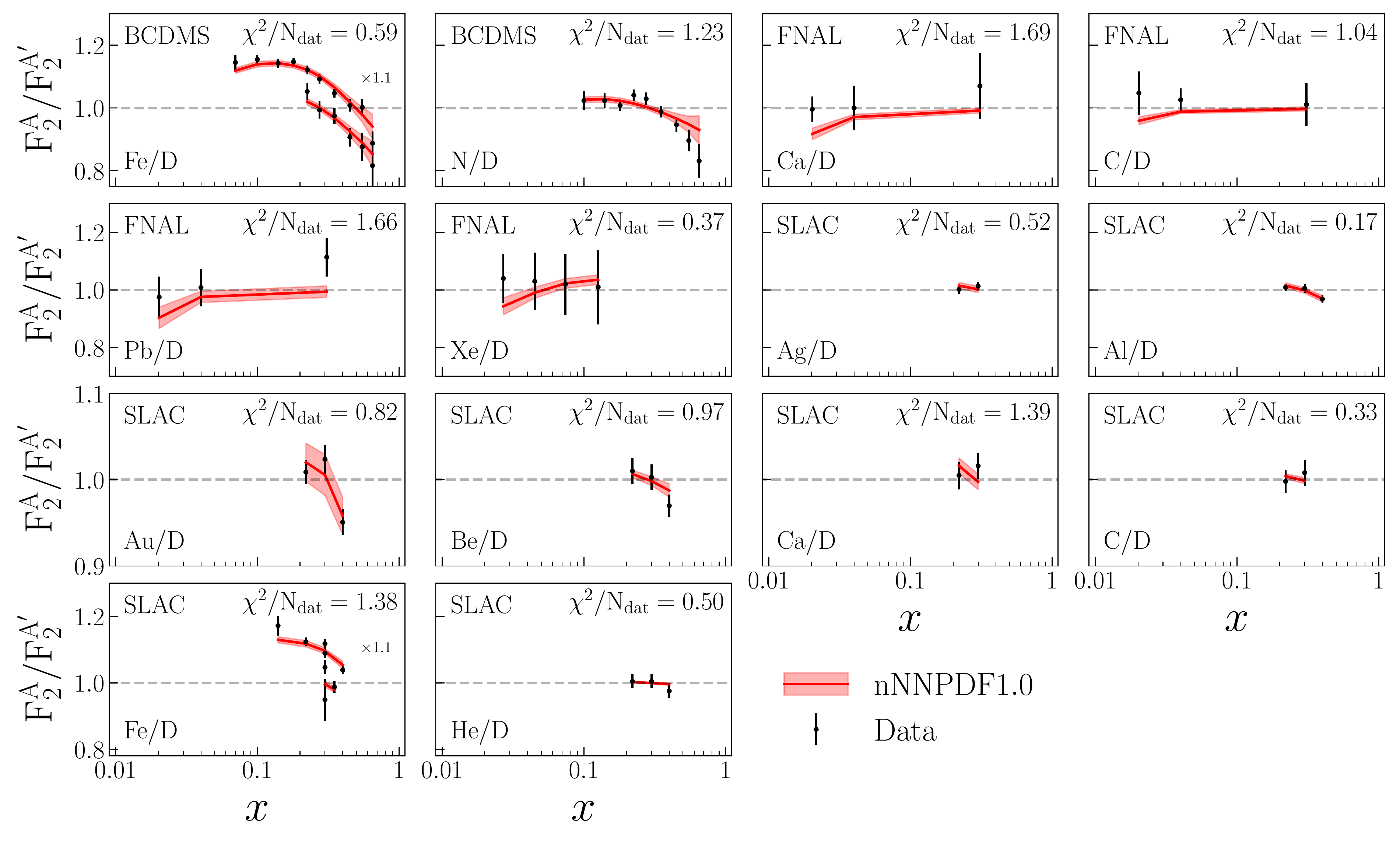}
 \end{center}
\vspace{-0.3cm}
\caption{\small Same as Fig~\ref{figDvT1} but for the data provided by the BCDMS, 
FNAL E665, and SLAC-E139 experiments.}
\label{figDvT3}
\end{figure}

At the level of individual datasets, we find in most cases a good 
agreement between the experimental measurements and
the corresponding theory calculations, with many
$\chi^2/N_{\rm dat} \lsim 1$ both at NLO and at NNLO.
The agreement is slightly worse for the ratios Ca/D and 
Pb/D from FNAL E665, as well as the Sn/D ratio from EMC, all of
which have $\chi^2/N_{\rm dat} \ge 1.5$.
The apparent disagreement of these datasets
can be more clearly understood with the visual
comparison between data and theory.
In Fig.~\ref{figDvT1} we display
the structure
function ratios $F_2^A/F_2^{A'}$ measured by the EMC and 
NMC experiments 
and the corresponding theoretical predictions
from the \texttt{nNNPDF1.0} NLO fit.
Furthermore, in Figs.~\ref{figDvT2} and~\ref{figDvT3} we show
the corresponding comparisons for the
$Q^2$-dependent structure
function ratio $F_2^{\rm Sn}/F_2^{\rm C}$ provided by the NMC experiment,
and the data provided by the BCDMS, 
FNAL E665, and SLAC-E139 experiments, respectively.

In the comparisons shown in Figs.~\ref{figDvT1}--\ref{figDvT3},
the central values of the experimental data points have been shifted
by an amount determined by the 
multiplicative systematic uncertainties and their nuisance parameters, while
uncorrelated uncertainties are added in quadrature to define the total error bar.
We also indicate in each panel the value of $\chi^2/N_{\rm dat}$,
which include the quadratic penalty as a result 
of shifting the data to its corresponding value displayed in the figures.
The quoted $\chi^2$ values
therefore coincide with those of Eq.~(\ref{eq:chi2_nNNPDF10}) without
the $A=1$ penalty term.
Lastly, the theory predictions are computed at each $x$ and $Q^2$ bin
given by the data,
and its width corresponds to the 1-$\sigma$ deviation of the observable
using the \texttt{nNNPDF1.0} NLO set with $N_{\rm rep} = 200$ replicas.
Note that in some panels, the theory curves (and the corresponding data points)
are shifted by an arbitrary factor to improve visibility.
  
As expected by the $\chi^2$ values listed in Table~\ref{tab:nNNPDF10_chi2}, 
the experimental measurements agree well with
the structure function ratios computed using the \texttt{nNNPDF1.0} sets, apart
from the three observables mentioned previously.
For the FNAL data, the disagreement comes from
datasets that contain a total of 3 data points with larger
uncertainties than other experimental measurements,
and therefore do not significantly impact
the fit results. 

A similar argument can be made for the Sn/D ratio
from the EMC experiment, which has 
$\chi^2/N_{\rm dat} = 2.22$.
Here the lack of agreement between theory and data 
can be traced 
to the
low-$x$ region of the structure function ratio.
Such a deviation can also be seen in the recent nCTEQ 
and EPPS analyses, and can be attributed to a possible tension with the 
$Q^2$ dependent ratio Sn/C presented in Fig.~\ref{figDvT2}. 
While the 
comparison here is with carbon and not deuterium, the nuclei are relatively 
close in mass number and therefore the effects in the ratio are expected 
to be similar.
On the other hand, the data show a roughly $\sim15-20\%$ 
difference between EMC's Sn/D and NMC's Sn/C at $x\sim0.03$. Since 
the NMC data have smaller uncertainties than EMC, its influence on the 
fit is much stronger, driving the disagreement with EMC Sn/D at low $x$.
Overall, the agreement with NMC data is excellent, including the $Q^2$ 
dependent Sn/C data presented in Fig.~\ref{figDvT2}.

From the data versus theory comparisons, the various
nuclear effects encoded in the structure
function ratios can clearly be observed.
At small $x$ the structure functions exhibit shadowing, namely the 
depletion of $F_2(x,Q,A)$
compared to its free-nucleon counterpart (or compared
to lighter nuclei).
At larger $x$ the well known EMC effect is visible, resulting
in ratios below unity.
Between these two regimes, one finds an enhancement of the
nuclear structure functions.
However, we do not observe the Fermi motion effect, 
which gives $R_{F_2} > 1$ for large $x$ 
and increases rapidly in the $x\to1$ limit.
This is due simply to the kinematic $W^2$ cut illustrated
in Fig.~\ref{figkinplot}, which
removes much of the large-$x$ data.
Note that although the three nuclear regimes are visible at the
structure function level, such effects may not be reflected at 
the level of PDF ratios, as we will highlight in the following section.

\subsection{The \texttt{nNNPDF1.0} determination} \label{s2:nNNPDF10_determination}
With the agreement between data and theory established, we present
now the results for the NLO nPDF sets.
Later,
we will assess the perturbative stability of the results by comparing to
the corresponding NNLO fit.
Unless otherwise indicated, the results presented in this section
are generated with $N_{\rm rep} = 1000$ Monte Carlo
replicas.

\begin{figure}[ht]
\begin{center}
  \includegraphics[width=0.90\textwidth]{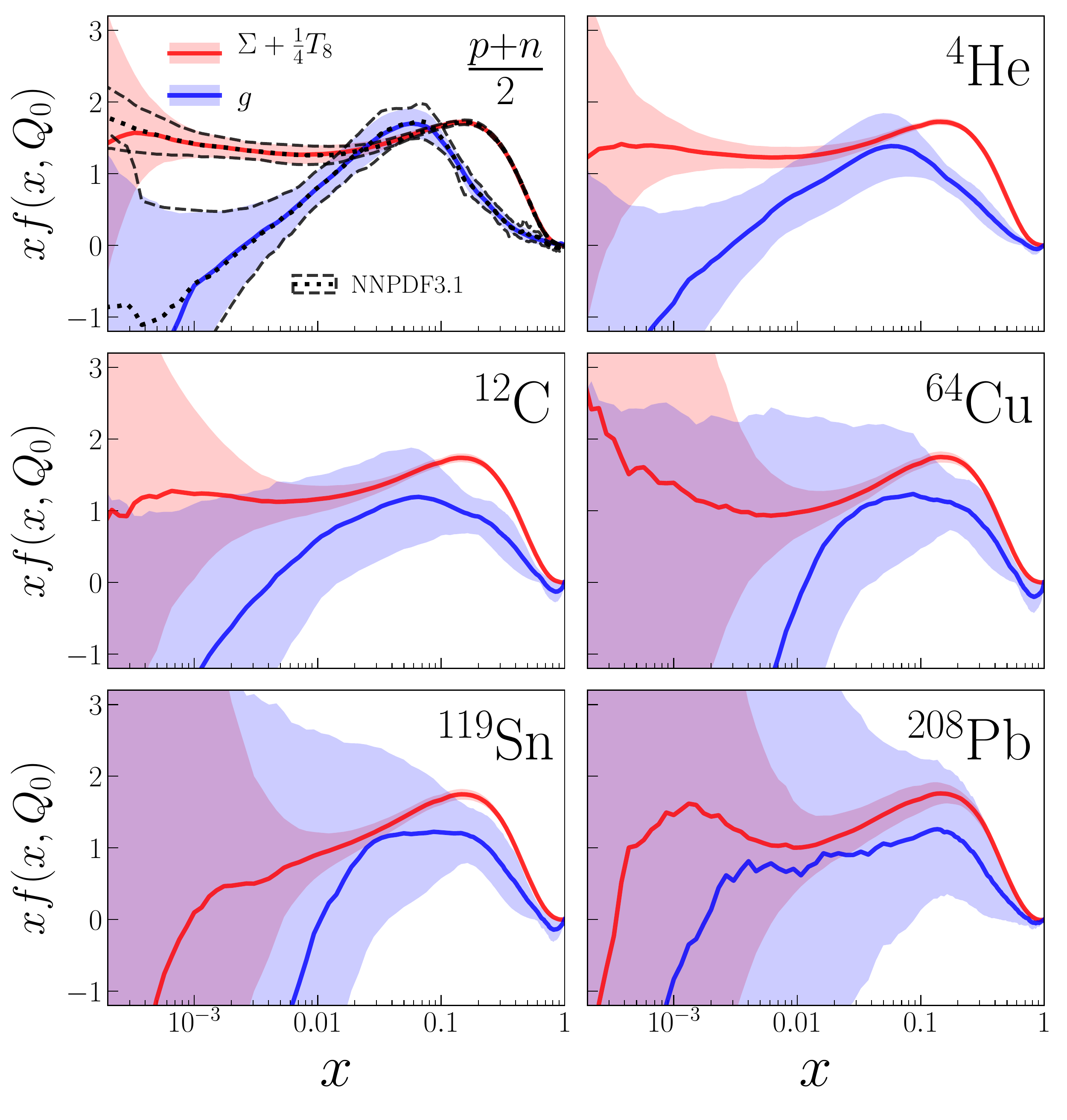}
 \end{center}
\vspace{-0.5cm}
\caption{\small The \texttt{nNNPDF1.0} NLO set as a function of $x$
  at the input scale $Q_0=1$ GeV for different values of $A$.
  We show the central value for the gluon $g$ (solid blue line) 
  and the quark combination $\Sigma+T_8/4$ (solid red line)
  for $A=1$ (isoscalar nucleon), $A=4$ (He),
  $A=12$ (C), $A=64$ (C), $A=119$ (Sn), and $A=208$ Pb.
  The corresponding uncertainties (shaded bands) correspond 
  to the 90\% confidence level intervals.
  In the case of $A=1$ we also show the central value of the baseline free-nucleon
  PDF set, NNPDF3.1 (black dotted line), and its uncertainties (black dashed lines).
}
\label{figPDFsQ0}
\end{figure}

To begin, we show in Fig.~\ref{figPDFsQ0} the \texttt{nNNPDF1.0} NLO set as a function of $x$
at the input scale $Q_0=1$ GeV for different values of $A$.
In this figure,
the nPDF uncertainty bands are computed as the 90\% confidence level
intervals, with the central value being taken as the midpoint of the corresponding
range.
The confidence levels presented here follow that of previous NNPDF studies~\cite{Ball:2010de}
and are computed in the following way.
For a given $x$, $Q$, and $A$, we have $N_{\rm rep}$ values of
a particular nPDF flavour $f_{k}(x,Q,A)$.
The replicas are then ordered such that:
\begin{equation}
    f_1 \le f_2 \le \ldots \le
    f_{N_{\rm rep}-1} \le f_{N_{\rm rep}} \,. 
\end{equation}
Finally, we remove symmetrically $(100-X)\%$ of the replicas
with the highest and lowest values.
 The resulting interval defines the $X\%$ confidence level
 for the nPDF $f(x,Q,A)$ for a given value of $x$, $Q$, and $A$.
 In other words, a 90\% CL interval 
 is obtained by
 keeping the central 90\% replicas, leading to:
 \begin{equation}
 \label{eq:res90CL}
 \lc f_{0.05\,N_{\rm rep}}, f_{0.95\, N_{\rm rep}}  \rc \, .
 \end{equation}

 The rationale for estimating the nPDF
 uncertainties as 90\% CL intervals, as opposed to the standard deviation,
 is that it turns out that the \texttt{nNNPDF1.0} probability
 distribution is not well described by a Gaussian, in particular when ratios
 between different nuclei $A$ are taken.
 Therefore, the variance $\sigma^2$ may not be the best estimator for the
 level of fluctuations in the distribution.
 While deviations from the Gaussian approximation in the proton case are
 moderate, there are several reasons why the nPDFs may behave differently.
 First of all, there is a limited amount of experimental information, especially
 for the gluon.
 Secondly, imposing the $A=1$ boundary condition skews the $A$ dependence
 of the distribution.
 Lastly, even if the resulting nPDFs do follow a Gaussian distribution, in general 
 their ratio between different values of $A$ will not.
 Therefore, in Fig.~\ref{figPDFsQ0}, and in the remaining figures of this analysis,
 the uncertainties will be presented as the 90\% CL defined above.

We also show in Fig.~\ref{figPDFsQ0} the results
of the baseline free-nucleon PDF set, NNPDF3.1, compared
to the nuclear parton distributions evaluated at $A=1$.
As can be observed, there is an excellent match between both the central
values and the PDF uncertainties of \texttt{nNNPDF1.0} and those
of NNPDF3.1 in the region of $x$ where the boundary condition is imposed,
$10^{-3} \le x \le 0.7$.
This agreement demonstrates that the quadratic penalty
in Eq.~(\ref{eq:chi2_nNNPDF10}) is sufficient to achieve
its intended goals.
In App.~\ref{s2:nNNPDF10_methodologyresults} we will discuss the importance
of implementing such a constraint, particularly
for light nuclei.

From Fig.~\ref{figPDFsQ0}, we can also see that the PDF uncertainties
increase as we move towards larger values of $A$, in particular for the
gluon nPDF.
Recall that the latter is only constrained indirectly from inclusive DIS data via
DGLAP evolution effects.
On the other hand, the quark combination $\Sigma + T_8/4$ turns
out to be reasonably well
constrained for $x\gsim 10^{-2}$, since this is the combination directly related
to the nuclear structure function $F_2(x,Q^2,A)$.
For both the gluon and the quark nuclear distributions, the PDF uncertainties
diverge in the small-$x$ extrapolation region, the beginning of which
varies with $A$.
For example, the extrapolation region for the quarks in Sn ($A=119$)
is found to be $x\lsim 5 \times 10^{-3}$, while for the gluon
PDF uncertainties become very large already for $x\lsim 5 \times 10^{-2}$.

\begin{figure}[ht]
  \begin{center}
    \includegraphics[width=0.90\textwidth]{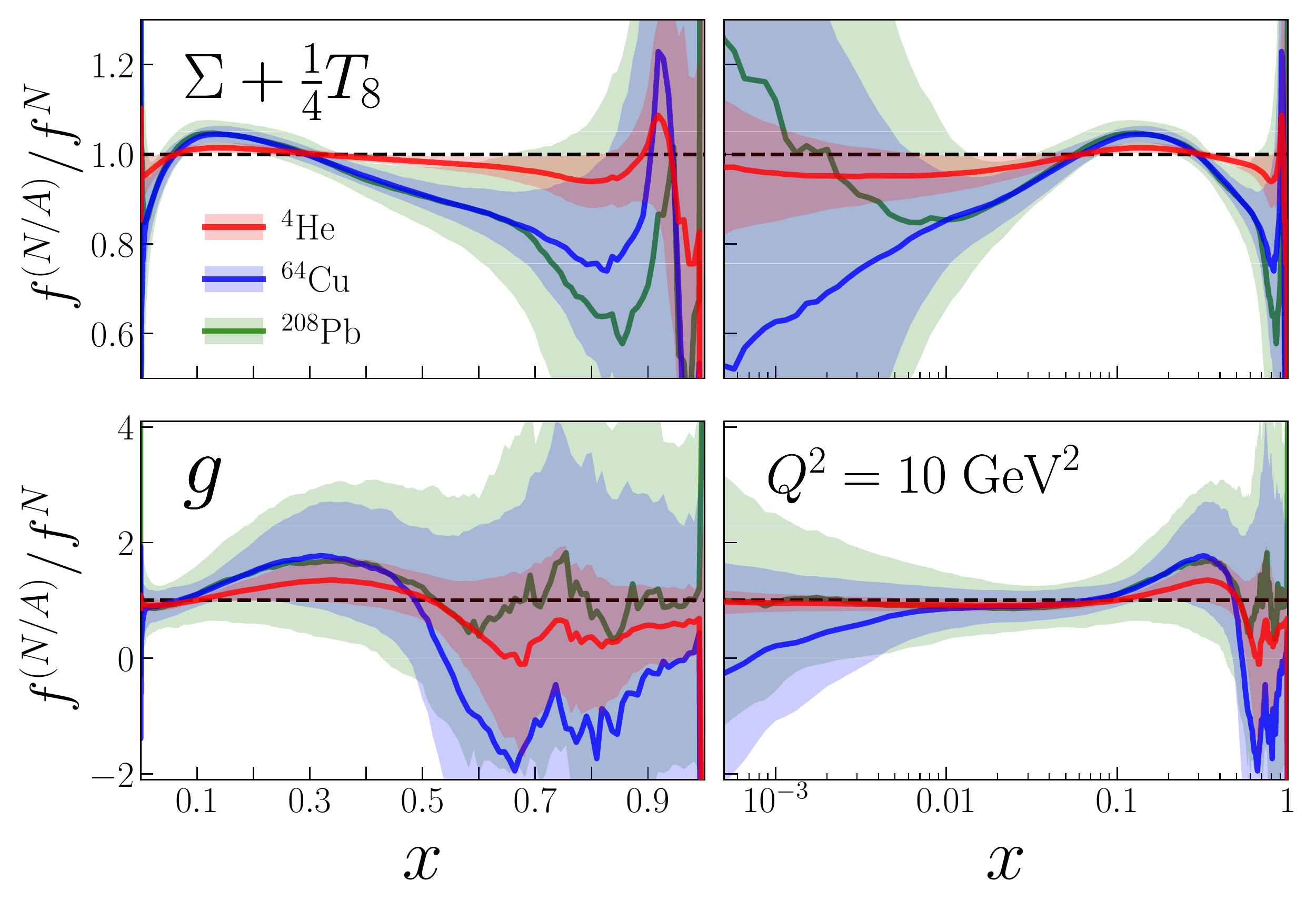}
   \end{center}
  \vspace{-0.3cm}
  \caption{\small Ratios of the \texttt{\texttt{nNNPDF1.0}} NLO
    distributions normalised to the $A=1$ result.
    The central values (solid lines) and uncertainties (shaded bands) for
    the quark combination $\Sigma + \frac{1}{4}T_8$ (top panels) and
    gluon (bottom panels) are shown at $Q^2 = 10$ GeV$^2$ 
    for $^4$He (red), $^{64}$Cu (blue),
    and $^{208}$Pb (green) nuclei.
  }
  \label{figPDFsRatio}
  \end{figure}

Next, we illustrate in Fig.~\ref{figPDFsRatio} the \texttt{nNNPDF1.0}
PDFs normalised by the $A=1$ distributions.
Here the results for He ($A=4$), Cu ($A=64$),
and Pb ($A=208$) nuclei are shown for $Q^2 = 10$ GeV$^2$.
With this comparison, we can assess whether the different nuclear
effects introduced previously are visible at the nPDF level,
since Eq.~(\ref{eq:param}) is analogous to the structure function ratios
displayed in Figs.~\ref{figDvT1}--\ref{figDvT3}.

When evaluating ratios of nPDFs between different values of $A$,
it is important to account for the correlations between the numerator
and denominator.
These correlations stem from the fact that nPDFs at two values of $A$ are
related by the common underlying parameterisation, Eq.~(\ref{eq:param}),
and therefore are not independent.
This can be achieved by computing the ratio $R_f$ for each of the $N_{\rm rep}$
Monte Carlo replicas of the fit as follows:
\begin{equation}
\label{eq:nPDFratiosErr}
R_f^{(k)} =
\frac{f^{(N/A)(k)}(x,Q^2,A)}{f^{(N)(k)}(x,Q^2)} \,
\end{equation}
and then evaluating the 90\% CL interval following the procedure that leads
to Eq.~(\ref{eq:res90CL}).

Note that a rather different result from that
of Eq.~(\ref{eq:nPDFratiosErr}) would be obtained if either the correlations
between different $A$ values were ignored (and thus the PDF uncertainties in numerator
and denominator of Eq.~(\ref{eq:nPDFratiosErr}) are added in quadrature) or if the uncertainties
associated to the $A=1$ denominator were not considered.
%

From Fig.~\ref{figPDFsRatio}, we can see that for the relevant quark combination
$\Sigma+T_8/4$ in $A=64$ and $A=208$ nuclei, it is possible to identify the same three
types of nuclear effects that were present at the structure function level.
In particular, the anti-shadowing and EMC effects are most evident, where 
the deviation from unity is outside the 90\% CL range. 
Moreover, shadowing behaviour appears briefly in the region $x\simeq0.01$, 
particularly for copper nuclei,
before the uncertainties grow quickly in the extrapolation region.
On the other hand, the nuclear effects appear to be negligible
for all $x$ in helium nuclei within the present uncertainties. 

The situation is much worse for the nuclear gluons,
where the ratio $R_f=f^{(N/A)}/f^N$ is consistent with one
within the uncertainties for all values of $x$.
This indicates that
using only neutral-current DIS nuclear structure functions, there is limited
information that one can extract
about the nuclear modifications of the gluon PDF.
Here we find no evidence for gluon shadowing,
and the ratio $R_f$ is consistent with one for $x\lsim 0.1$.
The only glimpse of a non-trivial nuclear modification
of the gluon nPDF is found for Cu ($A=64$), where between
$x\simeq 0.1$ and $x\simeq 0.3$ there appears to be an enhancement
reminiscent of the anti-shadowing effect.

The comparisons of Fig.~\ref{figPDFsRatio}
demonstrate that, without additional experimental input, we are rather 
far from being able to probe in detail the nuclear modifications
of the quark and gluon PDFs, particularly for the latter case.
We will highlight in Sect.~\ref{s1:EIC} how the present situation would
be dramatically improved with an Electron Ion Collider,
allowing us to pin down nuclear PDFs
in a wider kinematic range and with much better precision.

\begin{figure}[ht]
  \begin{center}
    \includegraphics[width=0.90\textwidth]{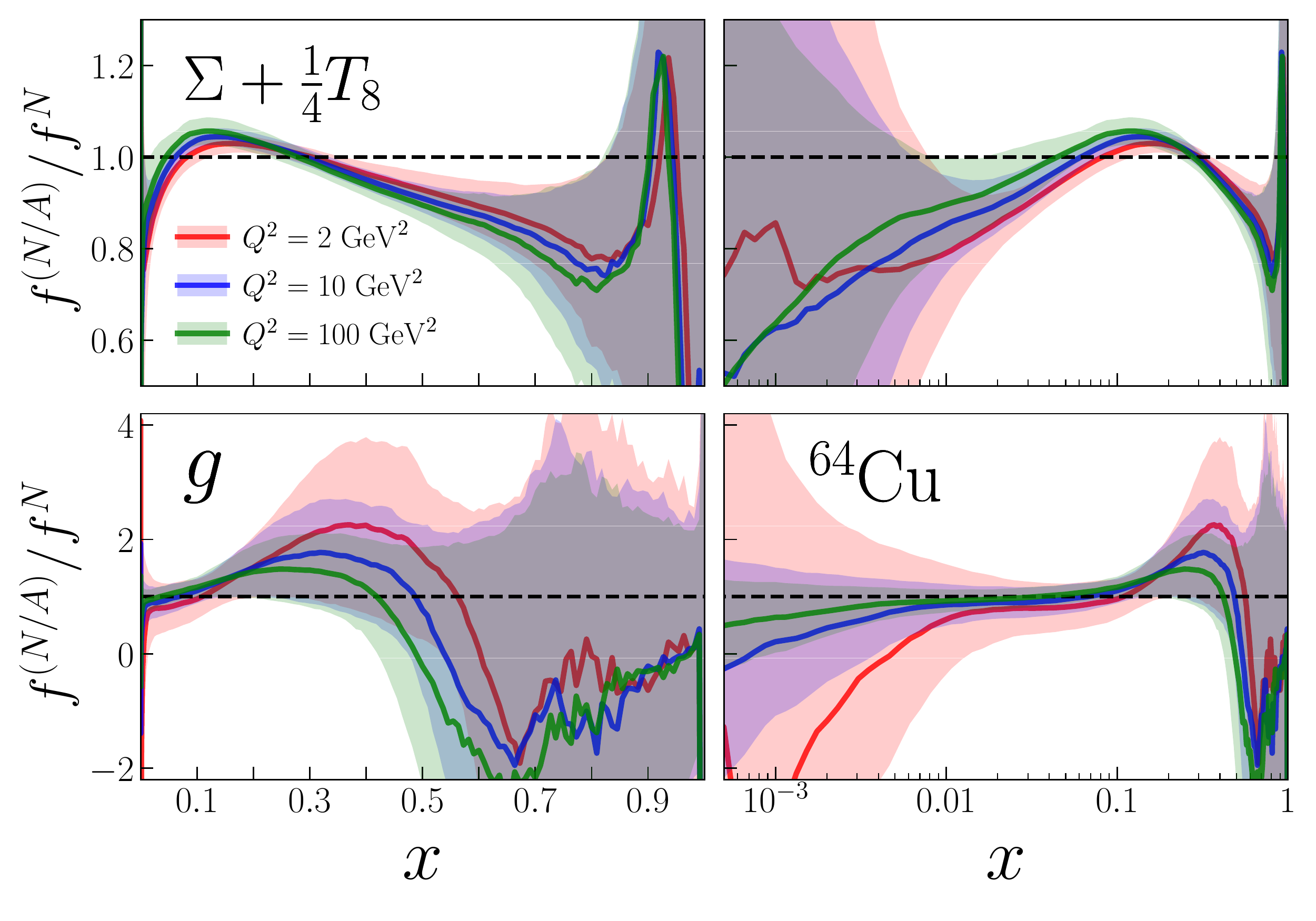}
   \end{center}
  \vspace{-0.3cm}
  \caption{\small Same as Fig.~\ref{figPDFsRatio}, but now
    for the dependence of the nuclear modifications of $^{64}$Cu
    on the momentum transfer $Q^2$.
    The ratios are given for $Q^2=1$ GeV$^2$ (red), 
    10 GeV$^2$ (blue), and 100 GeV$^2$ (green).
  }
  \label{figPDFsQ2}
  \end{figure}

\myparagraph{The scale dependence of the nuclear modifications}
In Fig.~\ref{figPDFsQ2}, we show a similar
comparison as that of Fig.~\ref{figPDFsRatio}, but now
for the $Q^2$ dependence of the nuclear modifications in $^{64}$Cu.
More specifically,
we compare the results of \texttt{nNNPDF1.0},
normalised as before
to the $A=1$ distribution, for $Q^2=2$ GeV$^2$, 10 GeV$^2$, and 100 GeV$^2$.
We can observe in this case how nPDF uncertainties are reduced
when the value of $Q^2$ is increased.
This effect is particularly dramatic for the gluon
in the small-$x$ region, but is also visible for the quark
distributions.
This feature is a direct consequence of the structure of DGLAP
evolution, where at small $x$ and large $Q^2$ the results
tend to become insensitive of the specific boundary condition
at low scales as a result of double asymptotic scaling~\cite{das}.

It is important to point out that, by the same token, the sensitivity to nuclear
modifications is also reduced when going from low to high $Q^2$ in
the small-$x$ region.
Indeed, we can see from Fig.~\ref{figPDFsQ2} that the ratios $R_f$ move closer to one at
small $x$ as $Q$ is increased.
However, this is not the case for medium and large $x$, where DGLAP
evolution effects are milder.
Therefore, nuclear effects in this region can be accessible
using probes both at low and high momentum transfers.
The comparisons in  Fig.~\ref{figPDFsQ2} highlight that
the best sensitivity for nuclear modifications present
in the small-$x$ region arises from low-scale observables,
while for medium and large-$x$ modifications there is also
good sensitivity at high scales.

\begin{figure}[ht]
\begin{center}
  \includegraphics[width=0.90\textwidth]{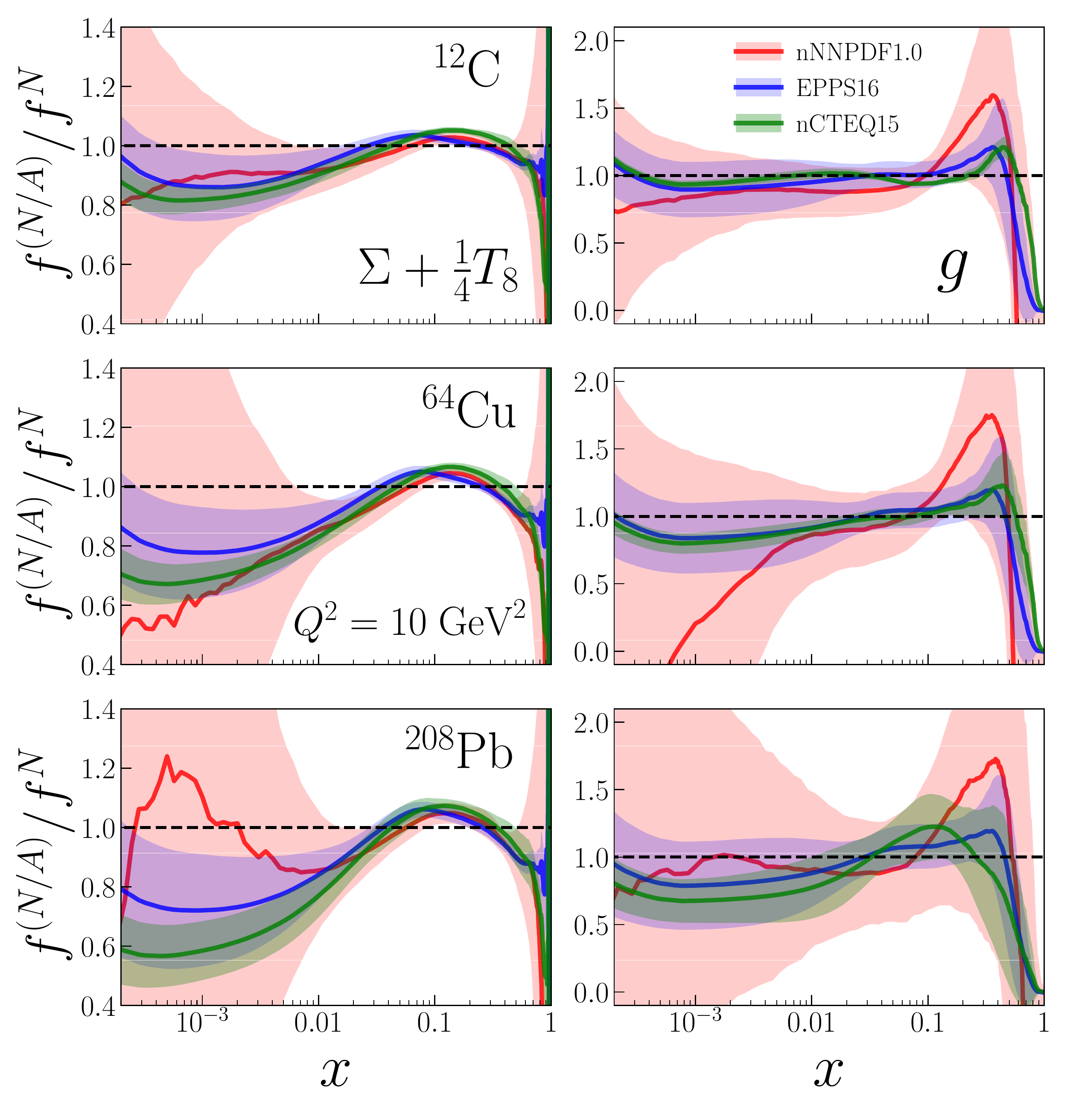}
 \end{center}
\vspace{-0.3cm}
\caption{\small Comparison between the \texttt{nNNPDF1.0}, \texttt{EPPS16} and \texttt{nCTEQ15} fits at NLO
  for $Q^2 = 10$ GeV$^2$.
  The quark combination $\Sigma + \frac{1}{4}T_8$ (left panels) and gluon (right panels)
  are normalised to the central value of each group's proton PDF baseline, and
  are shown for $^{12}$C (top panels), $^{64}$Cu (middle panels), and 
  $^{208}$Pb (bottom panels) nuclei.
  The uncertainties (shaded bands) correspond to the 90\% CL ranges computed with
  the corresponding prescription for each fit.
}
\label{figPDFsComp}
\end{figure}

\myparagraph{Comparison with \texttt{EPPS16} and \texttt{nCTEQ15}}
We now turn to compare the \texttt{nNNPDF1.0} nuclear PDFs with other recent
analyses.
Here we restrict our comparison to the \texttt{EPPS16} and \texttt{nCTEQ15} fits, given that they
are the only recent nPDF sets available in {\tt LHAPDF}.
In Fig.~\ref{figPDFsComp}, we display the \texttt{nNNPDF1.0} NLO distributions
together with \texttt{EPPS16} and \texttt{nCTEQ15} at $Q^2 = 10$ GeV$^2$ for three different nuclei: 
$^{12}$C, $^{64}$Cu, and $^{208}$Pb.
The three nPDF sets have all been normalised to the central value
of their respective proton PDF baseline to facilitate the comparison.
For the \texttt{nNNPDF1.0} results, the uncertainties are computed 
as before but without including the correlations with the $A=1$ distribution.
Lastly, the PDF uncertainties for \texttt{EPPS16} and \texttt{nCTEQ15} correspond to the
90\% CL ranges computed using the standard Hessian prescription.
%

From this comparison, there are a number of interesting similarities and
differences between the three nPDF fits.
First of all, the
three nuclear regimes discussed in Sect.~\ref{s2:Nuclear_modification}, namely shadowing, anti-shadowing,
and the EMC effect, are visible between the three sets for the quark combination $\Sigma+T_8/4$.
Interestingly, in the data region the PDF uncertainties for this quark
combination are similar between the different analyses.
Much larger differences are found in the small-$x$ and large-$x$ extrapolation
regions, particularly for \texttt{nCTEQ15}, where the uncertainties are smaller.
Note that the different approaches for uncertainty estimation 
have noticeable physical consequences.
For instance,
it would appear that there is rather strong
evidence for quark shadowing down to $x\simeq 10^{-4}$ for the \texttt{nCTEQ15} result, 
while for \texttt{nNNPDF1.0},
the nuclear modifications are consistent with zero within uncertainties for
$x\lsim 10^{-2}$.

Concerning the nuclear modifications of the gluon PDF,
here we can perceive large differences at the level of PDF errors,
with \texttt{nCTEQ15} exhibiting the smallest uncertainties and \texttt{nNNPDF1.0} the largest.
While \texttt{nCTEQ15} indicates some evidence of small-$x$ gluon shadowing, this evidence
is absent from both \texttt{nNNPDF1.0} and \texttt{EPPS16}.
Moreover, the three sets find some preference for a mild enhancement of the gluon at
large $x$, but the PDF uncertainties prevent making any definite
statement.
Overall, the various analyses agree well within the large
uncertainties for $x\gsim 0.3$.

While it is beyond the scope to pin down the origin
of the differences between the three nPDF analyses, one known 
reason is the choice of nPDF parameterisation
together with the method of imposing the $A\to 1$ boundary condition.
Recall that in \texttt{nNNPDF1.0} we adopt a model-independent parameterisation
based on neural networks, Eq.~(\ref{eq:param}), with the boundary
condition imposed at the optimisation level in Eq.~(\ref{eq:chi2_nNNPDF10}).
In the \texttt{EPPS16} analysis, the bound nucleus PDFs are instead defined 
relative to a free nucleon baseline
(CT14) as:
\begin{equation}
f_i^{(N/A)}(x,Q^2,A) = R_i^A(x,Q^2)\,f_i^{(N)}(x,Q^2) \, ,
\end{equation}
where the nuclear modification factors are parameterized at the input evolution scale
$R_i^A(x,Q_0^2)$ with piece-wise polynomials that hard-wire some of the
theoretical expectations discussed in Sect.~\ref{s2:Nuclear_modification}.
In this approach, the information contained in PDF uncertainties of the free nucleon baseline is not
exploited to constrain the nPDFs.

In the \texttt{nCTEQ15} analysis, the nuclear PDFs are parameterized
by a polynomial functional form given by:
\begin{equation}
f_i^{p/A}(x,Q^2,A) = c_0\,x^{c_1}\,(1-x)^{c_2}\,e^{c_3\,x}(1+e^{c_4}x)^{c_5} \, ,
\end{equation}
where the coefficients $c_k(A)$ encode all the $A$ dependence.
During the fit, these coefficients are constrained in a way that
for $A=1$ they reproduce the central value of the
the CTEQ6.1-like fit of Ref.~\cite{Owens:2007kp}.
Note here that in the \texttt{nCTEQ15} fit the baseline proton set
does not include the experimental measurements that have become
available in the last decade, in particular
 the information provided by the high-precision
LHC data and the HERA combined structure functions.
Moreover, as in the case of \texttt{EPPS16}, the information about the PDF uncertainties
in the free-nucleon case is not exploited to constrain the nPDF errors.

While these methodological choices are likely to explain the bulk of the differences
between the three analyses, a more detailed assessment could only be obtained
following a careful benchmarking exercise along the lines of those
performed for the proton PDFs~\cite{Butterworth:2015oua,Ball:2012wy,Botje:2011sn,Alekhin:2011sk}.

\begin{figure}[ht]
  \begin{center}
    \includegraphics[width=0.90\textwidth]{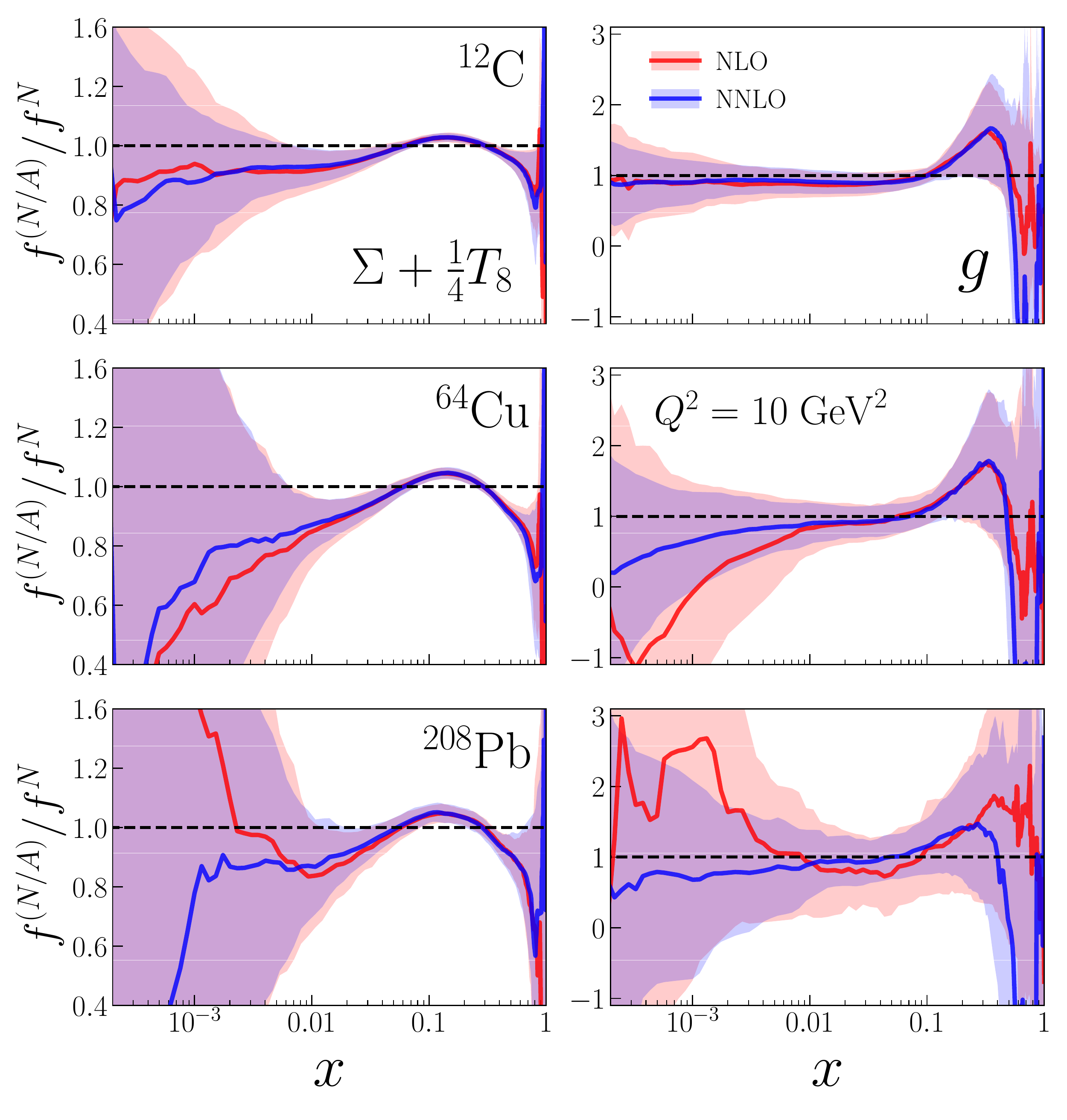}
   \end{center}
  \vspace{-0.6cm}
  \caption{\small
    Same as Fig.~\ref{figPDFsRatio}, but now comparing the results
    of the \texttt{nNNPDF1.0} fits between NLO and NNLO.
  }
  \label{figPDFsCompPTO}
  \end{figure}

\myparagraph{Perturbative stability}
To conclude the discussion of the main properties of the \texttt{nNNPDF1.0}
fits, in Fig.~\ref{figPDFsCompPTO} we compare 
the NLO and NNLO nuclear ratios $R_f$
for the same three nuclei as in Fig.~\ref{figPDFsComp}.
The ratios are constructed using the $A=1$ distributions
from their respective perturbative order PDF set using $N_{\rm rep}=200$
replicas.
In terms of central values, we can see that the NLO and
NNLO fit results are consistent within the 90\% CL uncertainty band.
The regions where the differences between the two perturbative orders are the largest
turn out to be the small- and large-$x$ extrapolation regions, in particular as $A$ is increased.

Another difference between the NLO and NNLO \texttt{nNNPDF1.0} fits concerns the size of the
PDF uncertainty band.
We find that for the gluon nPDF, the NNLO fit leads to a slight decrease
in uncertainties, perhaps due to the improved overall fit consistency when higher-order
theoretical calculations are used.
This effect is more distinct for the gluon distribution of $A=64$ and $A=208$ nuclei,
while it is mostly absent for $A=12$.
The apparent reduction of uncertainties, together with marginally better $\chi^2$
values (see Table~\ref{tab:nNNPDF10_chi2}), suggests that the NNLO fit is only slightly 
preferred over the NLO one.
That said, the difference is unlikely
to have significant phenomenological implications given the current level of 
uncertainties.
%

  \chapter{Nuclear PDFs from proton-nucleus scattering}
\label{chap:nNNPDF20}
\vspace{-1cm}
\begin{center}
  \begin{minipage}{1.\textwidth}
      \begin{center}
          \textit{This chapter is based on my results that are presented in Refs.~\mycite{AbdulKhalek:2020yuc,Khalek:2021ulf}}.
      \end{center}
  \end{minipage}
  \end{center}

\myparagraph{Motivation} In this chapter, I present the \texttt{nNNPDF2.0} analysis~\mycite{AbdulKhalek:2020yuc}, successor of \texttt{nNNPDF1.0} that was presented in Chapter~\ref{chap:nNNPDF10}.
This study focuses on the determination of the quarks and anti-quark nuclear PDFs,
with emphasis on their flavour separation.
Since measurements of neutral-current (NC) deep-inelastic scattering (DIS) nuclear structure
functions on isoscalar targets
are only sensitive to a single quark PDF combination,
one needs to rely on the information provided by independent processes
to disentangle quark and antiquarks of different flavours.
The main options that are available to accomplish this goal are the measurements of 
neutrino-induced charged current (CC) DIS on heavy nuclear targets and the
electroweak gauge boson production at the LHC.

We therefore complement our previous \texttt{nNNPDF1.0} analysis of NC DIS nuclear structure
functions with CC inclusive and charm-tagged measurements from fixed-target
neutrino experiments as well as with inclusive W
and Z production cross-sections in proton-lead collisions
from ATLAS and CMS at $\sqrt{s}=5.02$ TeV (Run 1) and 8.16 TeV (Run 2).
The $A=1$ proton PDF baseline used in the present analysis
is defined to be a variant of the NNPDF3.1 fit that excludes 
heavy nuclear target data.
%

\myparagraph{Outline} In Sect.~\ref{s1:nNNPDF20_framework}, I discuss the extended framework of \texttt{nNNPDF2.0}~\mycite{AbdulKhalek:2020yuc}, covering new features like an updated proton baseline and a cross section positivity constraint on the nPDFs.
In Sect.~\ref{s1:nNNPDF20_results}, I present the main results of the \texttt{nNNPDF2.0} analysis including an assessment of the inference quality, a detailed discussion on the main features of the nPDF sets, an examination of the momentum and valence integrals as well as a validation of the positivity constraint.

\section{\texttt{nNNPDF2.0} framework} 
\label{s1:nNNPDF20_framework}
The \texttt{nNNPDF2.0} framework is based on the \texttt{nNNPDF1.0} one but extended to cover additional features which I mainly focus on discussing in this section. 
In Sect.~\ref{s2:nNNPDF20_data}, I present
the experimental observables used in this analysis, highlighting in particular the new electroweak gauge boson production data from the LHC.
In Sect.~\ref{s2:nNNPDF20_updates}, I describe new aspects of the \texttt{nNNPDF2.0} inference methodology, focusing in particular on the differences and improvements w.r.t. the \texttt{nNNPDF1.0} analysis introduced in Chapter~\ref{chap:nNNPDF10}.

\subsection{Experimental data}
\label{s2:nNNPDF20_data}

In this section, I provide the details on the experimental measurements 
used as input for the \texttt{nNNPDF2.0} determination.
An emphasis is made in particular 
on the new data sets that are added with 
respect to those that were present in \texttt{nNNPDF1.0}.
I then discuss the theoretical calculations corresponding to these data sets 
and their numerical implementation in our fitting framework.

Common to the previous \texttt{nNNPDF1.0} analysis are the nuclear 
NC DIS measurements listed in Table~\ref{tab:nNNPDF10_data}.
%
%
%
The DIS kinematic cuts are the same as in our previous study, 
i.e. $Q^2 = 3.5$ GeV$^2$ and $W^2 = 12.5$ GeV$^2$,
consistent with the \texttt{NNPDF3.1} proton PDF baseline 
used to satisfy our boundary condition.
Note that all NC DIS measurements listed in Table~\ref{tab:nNNPDF10_data} 
are provided in terms of ratios of structure functions between two 
different nuclei.
In most cases the denominator is given by the deuterium structure 
function, but ratios to carbon and lithium are also provided.
As we will discuss in Sect.~\ref{s2:nNNPDF20_updates}, our fitting approach 
parameterises the PDFs entering the absolute structure functions 
for each value of $A$, after which their ratios are constructed.

The remaining input data which is newly added to our \texttt{nNNPDF2.0} 
analysis is presented in terms of absolute cross-sections, without 
normalising to any baseline nucleus.
We list these data in Table~\ref{tab:nNNPDF20_data}, divided into two 
categories: CC neutrino DIS reduced cross-sections on 
nuclear targets and leptonic rapidity distributions in electroweak gauge boson production 
from proton-lead collisions at the LHC.
The neutrino and anti-neutrino reduced cross-sections are further 
separated into inclusive cross-sections from 
CHORUS~\cite{Onengut:2005kv} and charm-tagged cross-sections 
from NuTeV~\cite{Goncharov:2001qe}.
The LHC measurements are divided into data from ATLAS and from 
CMS from the Run 1 and Run 2 data-taking periods.
In this table we also indicate the total number of data points included 
in the fit, combining the NC and CC cross-sections measurements 
with the LHC data. In total, the \texttt{nNNPDF2.0} global fit contains 
$N_{\rm dat}=1467$ data points.

\begin{table}[t]
  \centering
  \small
   \renewcommand{\arraystretch}{1.45}
\begin{tabular}{c c c c}
Experiment & ${\rm A}$ & ${\rm N}_{\rm dat}$ & Reference\\
\toprule
CHORUS $\nu$  & 208  &  423  & \cite{Onengut:2005kv}  \\
CHORUS $\bar{\nu}$  & 208  &  423  &  \cite{Onengut:2005kv} \\
NuTeV $\nu$  & 56  &  39  &   \cite{Goncharov:2001qe}\\
NuTeV $\bar{\nu}$  & 56  &  37  & \cite{Goncharov:2001qe}   \\
{\bf Total CC DIS} & & {\bf 922} & \\
\midrule
CMS $W^{\pm}$ $\sqrt{s}=8.16$ TeV  &  208  & 48     & \cite{Sirunyan:2019dox}   \\
CMS $W^{\pm}$ $\sqrt{s}=5.02$ TeV  &  208  & 20     & \cite{Khachatryan:2015hha}   \\
CMS $Z$ $\sqrt{s}=5.02$ TeV  &  208  &  12    & \cite{Khachatryan:2015pzs}   \\
ATLAS $Z$ $\sqrt{s}=5.02$ TeV  &  208  & 14     & \cite{Aad:2015gta}   \\
{\bf Total LHC} & & {\bf 94} & \\
\midrule
{\bf Total } & & {\bf 1467} & \\
\bottomrule
\end{tabular}
\vspace{4mm}
\caption{\small \label{tab:nNNPDF20_data} Same as Table~\ref{tab:nNNPDF10_data}
  for the new datasets that have been added to \texttt{nNNPDF2.0}.
  As opposed to the NC structure function measurements,
  these datasets are presented as absolute distributions
  rather than as as cross-sections ratios.
  We also indicate the total number of data points in the fit,
  combining the NC and CC structure functions with the LHC data. 
}
\end{table}


Starting with the CC measurements from CHORUS,
we fit the inclusive neutrino and anti-neutrino 
double-differential cross-sections, $d^2\sigma^{\nu N}/dxdQ^2$.
After imposing kinematic cuts, the dataset consists of 
$N_{\rm dat}=846$ data points equally distributed between neutrino 
and anti-neutrino beams.
The fitted cross-sections are not corrected for non-isoscalarity 
of the lead target, and therefore the corresponding theory 
calculations take into account effects related to the 
difference between $Z=82$ and $Z=A/2=104$.
The situation is therefore different from the treatment 
of NC nuclear structure functions, where the experimental results
are presented with non-isoscalar
effects already subtracted, as discussed in~\cite{AbdulKhalek:2019mzd}.

In addition to the CHORUS reduced cross-sections, nNNPDF2.0 also 
includes the NuTeV di-muon cross-sections from 
neutrino-iron scattering.
Dimuon events in neutrino DIS are associated with the 
W$^\pm + s~(d) \to c$ scattering process, where the charm 
quark hadronises into a charmed meson and then decays 
into a final state containing a muon.
This process is dominated by the strange-initiated contributions 
since other initial states are CKM-suppressed, thus providing 
direct sensitivity to the strange quark nuclear PDF.
In fact, the NuTeV di-muon data are known to play an important 
role in studies of proton strangeness in global QCD analyses.
After kinematic cuts, we end up with $N_{\rm dat}=39$ and 37 
data points for the neutrino and the anti-neutrino 
cross-sections, respectively.
Together with the CHORUS cross section data, the 
CC measurements comprise a majority of the
input dataset with a total of $N_{\rm dat}=922$ data points.

Moving now to the LHC electroweak gauge boson cross-sections,
we consider in this work the four data sets that are listed in 
Table~\ref{tab:nNNPDF20_data}.
Three of the data sets come from the Run 1 data-taking period,
corresponding to a per-nucleon center-of-mass energy 
of $\sqrt{s}=5.02$ TeV.
These are the ATLAS Z rapidity distributions~\cite{Aad:2015gta} 
and the CMS W$^\pm$~\cite{Khachatryan:2015hha} and 
Z~\cite{Khachatryan:2015pzs} rapidity distributions,
which contain $N_{\rm dat}=14$, 20, and 12 data points respectively.
Note that ATLAS does not have a published measurement of the 
W$^\pm$ rapidity distributions from Run 1 and that only preliminary 
results have been presented~\cite{ATLAS-CONF-2015-056}.

In the same way as the CC reduced cross-sections, the LHC measurements 
of electroweak gauge boson production are provided as absolute 
distributions.
In this case, however,
it is possible to construct new observables with the LHC W$^\pm$ and 
Z production data that might be beneficial for nPDF determinations.
For example, the \texttt{EPPS16} analysis composed and analysed the 
forward-to-backward ratio, where cross sections at 
positive lepton rapidities are divided by the ones at negative rapidities.
Nevertheless, in this work we choose only to work with the absolute 
rapidity distributions presented in the experimental publications.

In addition to the three Run 1 results, we add also for the first 
time in an nPDF analysis measurements from Run 2 corresponding 
to a per-nucleon center-of-mass energy of $\sqrt{s}=8.16$ TeV.
More specifically, the measurements correspond to W$^+$ and 
W$^-$ leptonic rapidity distributions~\cite{Sirunyan:2019dox} from CMS, which 
provide an additional $N_{\rm dat}=48$ data points.
The fact that the amount of data is more than doubled compared 
to the corresponding Run 1 measurements is a consequence of the 
increase in the CoM energy as well as the higher integrated luminosity. 
In particular, the Run 2 measurements are based on $\mathcal{L}=173$ 
nb$^{-1}$ compared to $\mathcal{L}=34.6$ nb$^{-1}$ available from Run 1.
The CMS Run 2 data are therefore expected to provide 
important constraints on the quark flavour separation of the nuclear PDFs.

As opposed to the situation in proton-proton collisions, the LHC gauge boson production measurements from proton-lead collisions do not provide information on the correlation 
between experimental systematic uncertainties.
For this reason, we construct the total experimental error by adding the 
various sources of uncertainty in quadrature.
The only source of systematic error which is kept as fully correlated 
among all the data bins of a given dataset is the overall normalisation uncertainty.
Note that this normalisation uncertainty is correlated within a single 
experiment and LHC data-taking period, but elsewhere is uncorrelated 
between different experiments.

\begin{figure}[ht]
\begin{center}
  \includegraphics[width=0.8\textwidth]{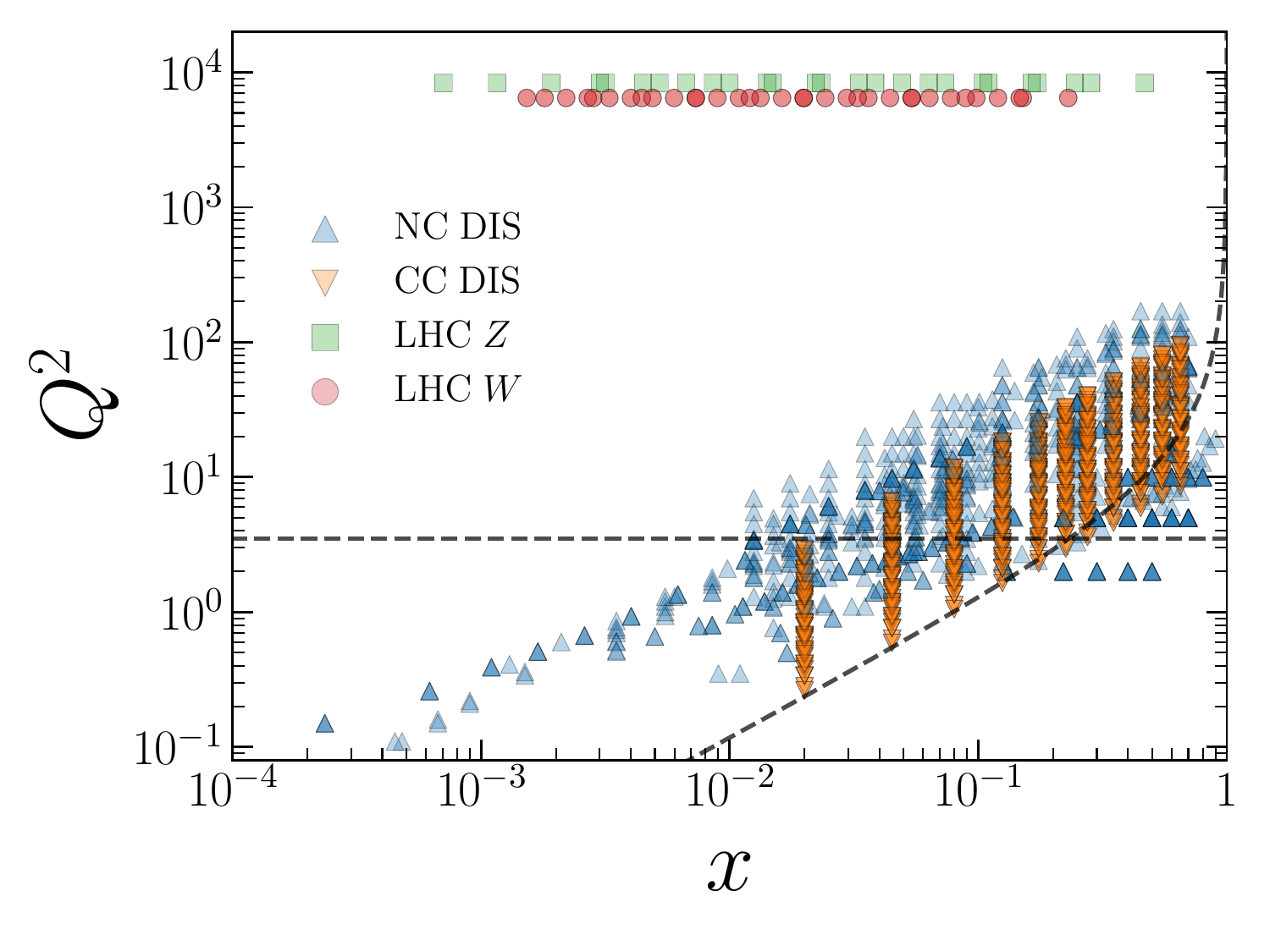}
 \end{center}
\vspace{-0.9cm}
\caption{\small Kinematical coverage in $x$ and $Q^2$ of the data points 
included in the \texttt{nNNPDF2.0} determination.
  The horizontal dashed and curved dashed lines correspond to 
  $Q^2 = 3.5$ GeV$^2$ and $W^2 = 12.5$
  GeV$^2$, respectively, which are the kinematic cuts imposed in this 
  analysis.
  For each LHC measurement, there are two values of $x$ associated 
  with leading order kinematics of proton-lead scattering being displayed.
  \label{fig:nNNPDF20_kinplot}
}
\end{figure}

In order to illustrate the coverage of the experimental data that is 
included in \texttt{nNNPDF2.0} and summarised in Tables~\ref{tab:nNNPDF10_data} 
and~\ref{tab:nNNPDF20_data}, we display in Fig.~\ref{fig:nNNPDF20_kinplot} their kinematical 
range in the $(x,Q^2)$ plane.
Here the horizontal dashed and curved dashed lines correspond to 
the kinematic cuts of $Q^2 = 3.5$ GeV$^2$ and 
$W^2 = 12.5$ GeV$^2$, respectively.
In addition, we show for each hadronic data point the 
two values of $x$ corresponding to the 
parton momentum fractions of the incoming 
proton and lead beams, computed to leading order.

There are several interesting observations that one can make regarding 
Fig.~\ref{fig:nNNPDF20_kinplot}.
First of all, the LHC proton-lead measurements significantly extend the kinematic 
coverage of the fixed-target
DIS reduced cross-sections, both in terms of $x$ and $Q^2$. 
In particular, the LHC data reside at $Q^2$ values that are orders of magnitude 
larger while the coverage in partonic
momentum fraction is extended down to $x \simeq 10^{-3}$.  
Secondly, the CC reduced cross-sections have a similar coverage compared 
to the NC ones, providing sensitivity to different quark and antiquark 
combinations across the shared medium- to large-$x$ region.
Finally, the kinematics of the LHC W and Z measurements largely 
overlap. The ability to describe them simultaneously can therefore 
demonstrate the compatibility between the experimental data and theoretical 
calculations.

\myparagraph{DIS structure functions.}
For the NC DIS structure functions we use the same 
theoretical settings as in \texttt{nNNPDF1.0}, i.e. the structure functions 
are evaluated at NLO using {\tt APFEL}~\cite{Bertone:2013vaa} 
in the FONLL-B general-mass variable flavour number 
scheme~\cite{Forte:2010ta}.
The value of the strong coupling constant 
is taken to be the same as in the \texttt{NNPDF3.1} proton PDF fit, 
$\alpha_S(m_Z)=0.118$, as well as the charm and bottom 
mass thresholds $m_c=1.51$ GeV and $m_b=4.92$ GeV, respectively.
The charm PDF is generated perturbatively by the DGLAP evolution equations
and is thus absent from the $n_f=3$ scheme.
Lastly, the structure functions are processed by the 
{\tt APFELgrid}~\cite{Bertone:2016lga} fast interpolation tables
which allow for efficient evaluations during the PDF fit.

Concerning the CC neutrino reduced cross-sections, 
most of the theory settings are shared with their NC
counterparts.
The main difference is that the heavy quark contributions in the
CC predictions at NLO are accounted for in the FONLL-A scheme 
instead to maintain consistency with the proton baseline. 
Massive $\mathcal{O}\lp \alpha_s^2\rp$ corrections to charm 
production in CC DIS have been presented in 
Ref.~\cite{Berger:2016inr}, and subsequently used to study their 
impact in the determination of the strange content of the nucleon 
in Ref.~\cite{Gao:2017kkx}.
Further details about the implementation of heavy quark mass 
corrections in the NNPDF framework for charged-current scattering can be found in 
Ref.~\cite{Ball:2011uy}.

\begin{figure}[ht]
\begin{center}
  \includegraphics[width=0.75\textwidth,angle=0]{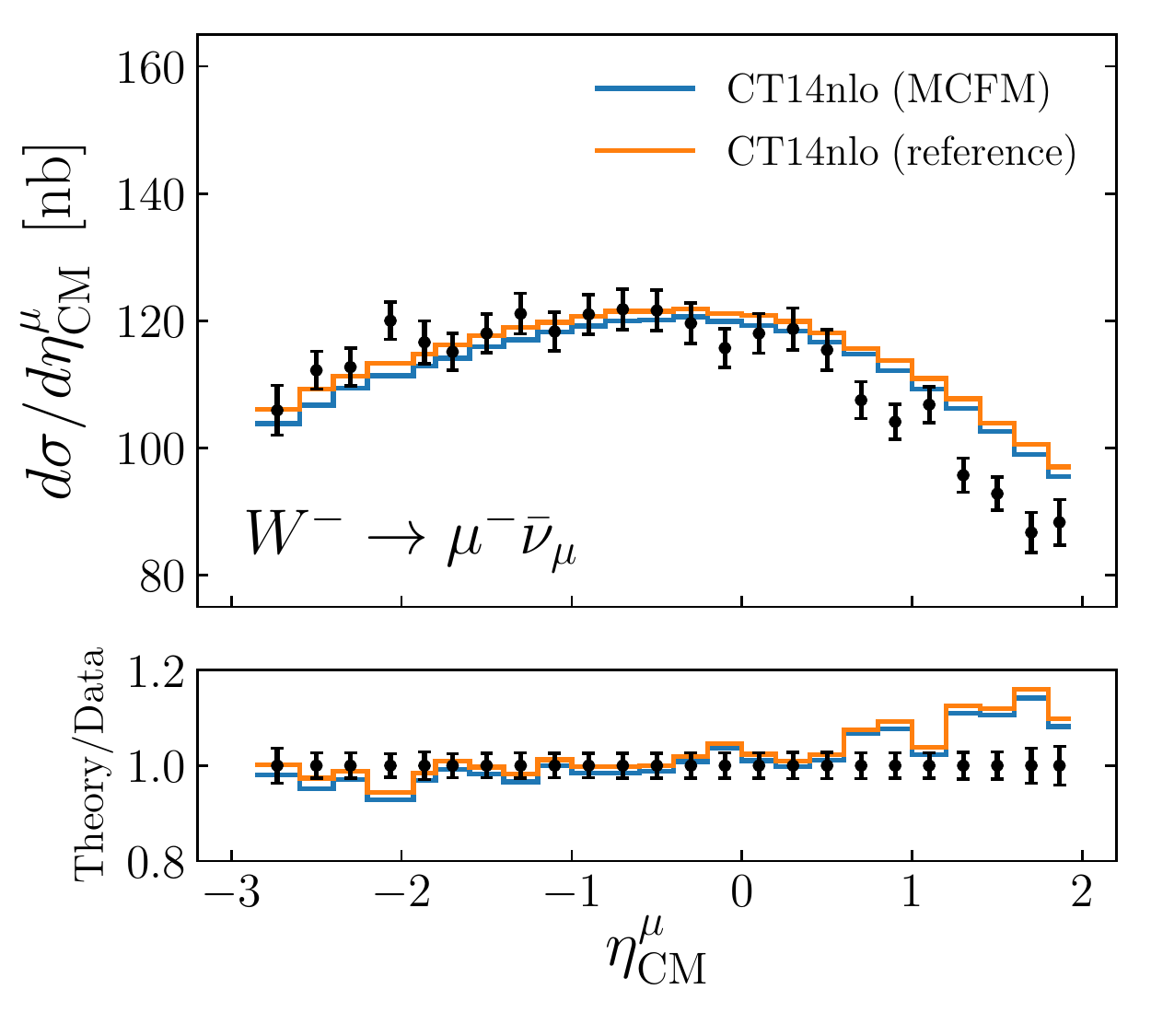}
 \end{center}
\vspace{-0.75cm}
\caption{\small The leptonic rapidity distributions for W$^-$ boson production
  at $\sqrt{s}=8.16$ TeV in the center-of-mass frame.
  Our {\tt MCFM}-based  calculation is compared
  with the theory predictions
  provided in~\cite{Sirunyan:2019dox}
  both as absolute cross-sections (upper) and as
  ratios to the experimental data (lower panel).
  In both cases the CT14 NLO proton PDF set is adopted and
  nuclear corrections are neglected.
  \label{fig:benchmarking}
}
\end{figure}

\myparagraph{Hadronic cross-sections.}
The rapidity distributions from W and Z boson production 
in proton-lead collisions are evaluated at NLO using 
{\tt MCFM}~\cite{Boughezal:2016wmq} v6.8 interfaced with
{\tt APPLgrid}~\cite{Carli:2010rw}.
We have ensured that the numerical integration uncertainties 
in the {\tt MCFM} calculations are always much smaller than 
the corresponding experimental errors.
Furthermore, our calculations are benchmarked with the 
reference theoretical values whenever provided by the 
corresponding experimental publications.
To illustrate this benchmarking, we display in 
Fig.~\ref{fig:benchmarking} the muon rapidity distributions for 
W$^-$ boson production at $\sqrt{s}=8.16$ TeV in the 
center-of-mass frame.
Here we compare our {\tt MCFM}-based calculation with the theory 
predictions provided in Ref.~\cite{Sirunyan:2019dox} at the level 
of absolute cross-sections (upper panel) and also as ratios to the 
central experimental values (lower panel).
In both cases, the CT14 NLO proton PDF set is adopted as input and nuclear corrections 
are neglected.
As can been seen by the figure, there is good agreement at the 
$\sim$1\% level between our calculations and the reference results 
provided in the CMS paper, with residual differences likely
to be traced back to a different choice of electroweak scheme.
Similar agreement is obtained for the rest of hadronic data sets 
included in the present analysis.

%
Since the fast interpolation grids are computed in the 
center-of-mass frame of the proton-lead collision,
rapidity bins that are given in the laboratory frame $\eta_{\rm lab}$
are shifted to the center-of-mass frame $\eta_{\rm CM}$ when
required. 
This shift is given by $\eta_{\rm lab} = \eta_{\rm CM}+0.465$
both at $\sqrt{s}=5.02$ and 8.16 center-of-mass energies.
Lastly, we note that the same theoretical settings were used for the 
evaluation of W and Z production in proton-proton collisions
for the baseline \texttt{NNPDF3.1} fit.

\subsection{Updates relative to \texttt{nNNPDF1.0}}
\label{s2:nNNPDF20_updates}

The cross-sections for hard scattering processes
involving heavy nuclei can be expressed 
either in terms of $f^{(N/A)}$ or $f^{(p/A)}$.
%
%
One can therefore
choose to parameterise either the PDFs of
the average bound nucleon or those 
of the bound proton in a global nPDF 
analysis.
In this study we choose the latter option for two main reasons.
First, the connection with the free-proton boundary condition 
is more straightforward.
In addition, the valence sum rules 
for non-isoscalar nuclei are independent of $A$ and $Z$.
If instead $f^{(N/A)}$ is parameterised, one of the valence sum rules
would depend on the value of the $Z/A$ ratio and thus be different for each nuclei, 
making it inconvenient from the parameterisation point of view.

The relation between $f^{(N/A)}$ and $f^{(p/A)}$ is trivial also for
PDF combinations that comprise the evolution basis.
Consider for example the total quark singlet, where the flavour combination 
is the same in the proton and in the neutron, i.e. $\Sigma^{(p/A)}=\Sigma^{(N/A)}$.
From Eq.~(\ref{eq:qNAdefinition}), it simply follows that $\Sigma^{(N/A)}=\Sigma^{(p/A)}$.
The same equivalence holds also for $V$ and $T_8$.
However, the distinction is important for $T_3$ and $V_3$, for which we have:
\begin{equation}
V_3^{(N/A)} = \frac{Z}{A} V_3^{(p/A)}   + \lp 1-\frac{Z}{A}\rp   V_3^{(n/A)}
= \lp  \frac{2Z}{A} -1\rp  V_3^{(p/A)}  \, ,
\end{equation}
\begin{equation}
T_3^{(N/A)} = \frac{Z}{A} T_3^{(p/A)}   + \lp 1-\frac{Z}{A}\rp  T_3^{(n/A)}
= \lp \frac{2Z}{A} -1\rp  T_3^{(p/A)}  \, ,
\end{equation}
so there is an overall rescaling factor of $(2Z/A-1)$ between $f^{(N/A)}$ and $f^{(p/A)}$.
The main consequence of this relation is highlighted by 
assuming an isoscalar nucleus, with $Z=A/2$.
In this case, $V_3^{(N/A)}=T_3^{(N/A)}=0$ 
while their bound proton counterparts are different from zero.
Unless otherwise indicated, the nPDFs discussed in this section will
always correspond to those of the bound proton.

\myparagraph{Fitting basis and functional form}
In our previous \texttt{nNNPDF1.0} analysis, we parameterised only 
three independent evolution basis distributions at the initial scale $Q_0$,
namely the total quark singlet $\Sigma(x,Q_0)$,
the quark octet $T_{8}(x,Q_0)$, and the gluon nPDF $g(x,Q_0)$.
From the LO expression of  Eq.~(\ref{eq:F2_p_lo}), it is clear that
NC structure functions are  sensitive only to a 
specific combination of $\Sigma$ and $T_8$ for isoscalar nuclei, 
in particular $\Sigma+T_8/4$, while the contribution
proportional to $T_3$ vanishes. 
In other words,  $\Sigma$ and
$T_8$ are strongly anti-correlated and only the combination 
$\Sigma + T_8/4$ can be meaningfully determined from the data.

The picture is quite different in the present study,
where the availability of charged current DIS data and electroweak gauge
boson production cross-sections in proton-lead collisions allow additional 
elements of the evolution PDF basis to be parameterised (see Sect.~\ref{s2:Flavour_separation}).
If non-isoscalar effects are neglected,
there is only a single distribution to be added
to our evolution basis choice, namely the total valence quark
combination $V =  u^- + d^- $.
However, non-isoscalar corrections are necessary for 
the targets considered in this analysis, particularly for lead.
In this case, the quark triplet
$T_3 = u^+ - d^+  $ and the valence triplet $V_3= u^- - d^- $ 
must also be parameterised.
Note that since $T_3$ and $V_3$ correspond to bound protons, they will be different
from zero even for isoscalar nuclei. However, in such cases their contribution
to the scattering cross-section vanishes and therefore the data provides no constraint
on these combinations.

\begin{figure}[ht]
  \floatbox[{\capbeside\thisfloatsetup{capbesideposition={right,top},capbesidewidth=0.5\textwidth}}]{figure}[\FBwidth]
  {\caption{Similarly to Fig.~\ref{fig:nNNPDF10_architecture}, the NN architecture $\{3,\,25,\,6\}$ used in \texttt{nNNPDF2.0} having 3 input nodes $x$, $\ln{(1/x)}$ and $A$, one hidden layer of 25 neurons with a sigmoid activation function and 6 output nodes with a linear activation function designating $N_i(x)$ of the considered flavours: $\Sigma$, $T_3$, $T_8$, $V$, $V_3$ and $g$. An overall of 256 free parameters (weights and biases).}\label{fig:nNNPDF20_architecture}}
  {\vspace{-1.0cm} \includegraphics[width=0.45\textwidth]{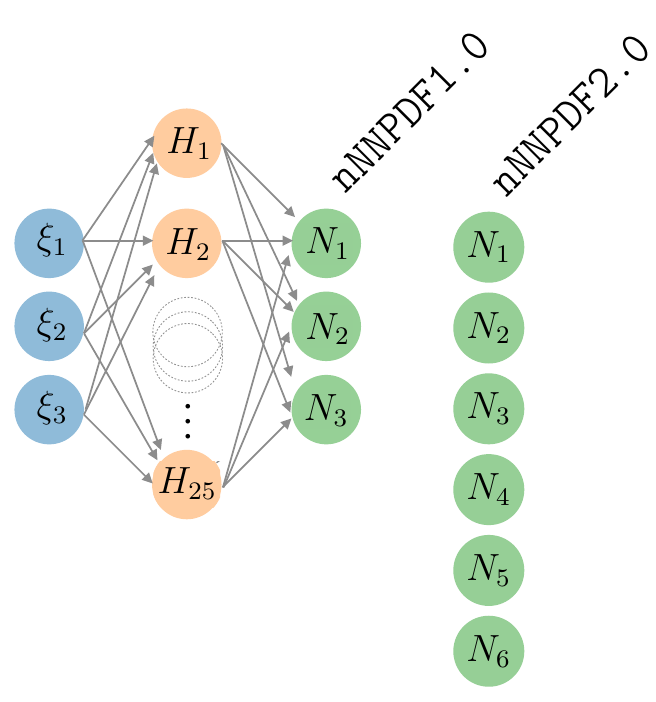}} 
  \end{figure}

Putting together these considerations, in this work we parameterise six independent
PDF combinations in the evolution basis as follows
\begin{eqnarray}
x\Sigma^{(p/A)}(x,Q_0) &=&x^{\alpha_\Sigma} (1-x)^{\beta_\Sigma} N_1(x,A) \, , \nonumber \\
xT_3^{(p/A)}(x,Q_0) &=&x^{\alpha_{T_3}} (1-x)^{\beta_{T_3}} N_2(x,A) \, , \nonumber \\
xT_8^{(p/A)}(x,Q_0) &=&x^{\alpha_{T_8}} (1-x)^{\beta_{T_8}} N_3(x,A) \, , \label{eq:param2} \\
xV^{(p/A)}(x,Q_0) &=&B_{V}x^{\alpha_V} (1-x)^{\beta_V} N_4(x,A) \, , \nonumber \\
xV_3^{(p/A)}(x,Q_0) &=&B_{V_3}x^{\alpha_{V_3}} (1-x)^{\beta_{V_3}} N_{5}(x,A) \nonumber\, ,  \\
xg^{(p/A)}(x,Q_0) &=&B_gx^{\alpha_g} (1-x)^{\beta_g} N_6(x,A) \, . \nonumber
\end{eqnarray}
In these expressions, $N_i(x,A)$ stands for the value of the
neuron in the output layer of the neural network associated to each specific
distribution.
In Fig.~\ref{fig:nNNPDF20_architecture}
we display a schematic representation of the architecture
of the feed-forward neural network used in the present and previous analysis.
As was done in \texttt{nNNPDF1.0}, we use a single artificial neural network
consisting of an input layer, one hidden layer with sigmoid activation
function, and an output layer with linear activation function.
The input layer contains three neurons that take as input the values of the
momentum fraction $x$, $\ln(1/x)$, and atomic mass number $A$,
respectively.
Since the hidden layer contains 25 neurons,
there are a total of $N_{\rm par}=256$ free parameters
(weights and thresholds) in the neural network used to model our nPDFs.
\vspace{1.0cm} 

The neural-net parameterisation in Eq.~(\ref{eq:param2}) is then
complemented by three normalisation coefficients 
$B_g$, $B_V$, and $B_{V_3}$ which are fixed by the sum rules,
and by twelve preprocessing exponents $\alpha_f$ and $\beta_f$
which are fitted simultaneously with the network parameters.
Since our proton baseline is a variant of the NNPDF3.1 global
NLO fit~\cite{Ball:2017nwa}
with perturbative charm, we adopt for consistency the
same parameterisation scale of $Q_0=1$ GeV.

It is important to emphasize here that the parameterisation
in Eq.~(\ref{eq:param2}) is valid from $A=1$ (free-proton)
up to $A=208$ (lead).
As a result, the \texttt{nNNPDF2.0} analysis incorporates an independent
determination of the free-proton PDFs, where agreement
with the proton PDF baseline is enforced by means of a boundary 
condition as explained below.
This is a relevant distinction, implying that the $A=1$
PDF can by construction differ slightly from 
our proton baseline, for example as a result of positivity 
constraints that are more general in the former case, or
by new information contained in the LHC proton-lead cross-sections.

\myparagraph{Sum rules}
For every nucleus, we assume that the fitted 
nuclear PDFs satisfy the same valence and momentum sum 
rules as in the proton case.
Similarly to \texttt{nNNPDF1.0}, the sum rules are implemented via an overall normalisation
factor in the nPDF parameterisation, 
which are adjusted each time the neural network parameters are 
modified in order to ensure that the sum rules remain satisfied.
%
%
First, energy conservation leads to the momentum sum rule constraint:
\begin{equation}
\label{eq:MSR}
\int_0^1 dx \,x \left(\Sigma^{(p/A)}(x,Q_0) + g^{(p/A)}(x,Q_0)\right) = 1 \, , \quad \forall \, A \, ,
\end{equation} 
which is implemented according to Eq.~(\ref{eq:NormG}).
The additional rules in \texttt{nNNPDF2.0} are the three valence sum 
rules that follow from the valence quark quantum numbers of the proton:
\begin{equation}
\int_0^1 dx~ \left(u^{(p/A)}(x,Q_0) - \bar{u}^{(p/A)}(x,Q_0)\right) = 2 \, ,\quad \forall \, A \label{eq:valencesr1}\\
\int_0^1 dx~ \left(d^{(p/A)}(x,Q_0) - \bar{d}^{(p/A)}(x,Q_0)\right) = 1 \, ,\quad \forall \, A \label{eq:valencesr2}\\
\int_0^1 dx~ \left(s^{(p/A)}(x,Q_0) - \bar{s}^{(p/A)}(x,Q_0)\right) = 0 \, ,\quad \forall \, A \label{eq:valencesr3}
\end{equation}
where the final relation is trivially satisfied due to our inherent flavour assumption
of $s=\bar{s}$.
To implement the former two valence sum rules in our analysis,
we first must derive the corresponding constraints in the evolution basis. 
Adding Eqs.~(\ref{eq:valencesr1}) and~(\ref{eq:valencesr2}) results in
\begin{align}
\label{eq:valencesr4}
\int_0^1 dx~ &\left(u^{(p/A)}(x,Q_0) - \bar{u}^{(p/A)}(x,Q_0)+ d^{(p/A)}(x,Q_0) - \bar{d}^{(p/A)}(x,Q_0)\right) = \nonumber \\
&\int_0^1 dx~ V^{(p/A)}(x,Q_0) = 3\, , \quad \forall \, A \, .
\end{align}
This condition can then be implemented in the same way as the 
momentum sum rule, namely by setting the overall normalisation 
factor of $V$ as:
\begin{equation}
\label{eq:NormV}
B_V(A) = \frac{3}{\int_0^1 dx\, V^{(p/A)}(x,Q_0,A)}\, ,
\end{equation}
where the denominator of Eq.~(\ref{eq:NormV}) is evaluated using 
Eq.~(\ref{eq:param2}) and setting $B_V=1$.
The second valence sum rule in the evolution basis 
is the one related to the quark valence triplet $V_3$.
Subtracting Eq.~(\ref{eq:valencesr2}) from~(\ref{eq:valencesr1})
gives
\begin{align}
\label{eq:valencesr5}
\int_0^1 dx~ &\left(u^{(p/A)}(x,Q_0,A) - \bar{u}^{(p/A)}(x,Q_0,A) - d^{(p/A)}(x,Q_0,A) + \bar{d}^{(p/A)}(x,Q_0,A)\right) = \nonumber\\ 
&\int_0^1 dx~ V_3^{(p/A)}(x,Q_0,A) = 1 \, , \quad \forall \, A \, .
\end{align}
which again is imposed by setting:
\begin{equation}
\label{eq:NormV3}
B_{V_3}(A) = \frac{1}{\int_0^1 dx\, V_3^{(p/A)}(x,Q_0,A)}\, ,
\end{equation}
where the denominator of Eq.~(\ref{eq:NormV3}) is evaluated using 
Eq.~(\ref{eq:param2}) with $B_{V_3}=1$.

In this analysis, 
the normalisation pre-factors $B_g(A)$, $B_V(A)$, and $B_{V_3}(A)$ 
are computed using the trapezoidal rule for
numerical integration between
$x_{\rm min}=10^{-9}$ and $x_{\rm max}=1$
each time the fit parameters are 
updated by the minimization procedure.
With a suitable choice of the ranges for the preprocessing
exponents (see discussion below), we guarantee that each quark
combination satisfies the corresponding physical integrability conditions.
Lastly, we have confirmed that individual replicas
satisfy the sum rules with a precision
of a few per-mille or better.

An interesting question in the context of nuclear
global QCD analyses is the extent to which
theoretical constraints such as the sum rules are satisfied by the
experimental data when not explicitly imposed.
In fact, it was shown in Ref.~\cite{Ball:2011uy} that the momentum
sum for the free proton agrees with the QCD expectation 
within $\sim 1\%$ in this scenario.
Here we will revisit this analysis for the nuclear case,
and will present in Sect.~\ref{s2:nNNPDF20_sumrules} variants of
the \texttt{nNNPDF2.0} fit where either the momentum
sum rule or the valence sum rule, is not enforced.
Interestingly, we will find that the experimental data
is in agreement with sum rule expectations, 
albeit within larger uncertainties, 
demonstrating the remarkable consistency of the 
nuclear global QCD analysis.

\myparagraph{Update on the free proton baseline}
As in Sect.~\ref{s2:nNNPDF10_protonBC}, we again implement
the condition that the proton PDF baseline, obtained with consistent theoretical
and methodological choices, is reproduced when $A\rightarrow1$.
This condition should be constructed to 
match the free-proton distributions
not only in terms of central values
but also at the level of PDF uncertainties.
In other words, it should allow a full propagation of the information 
contained in the proton baseline,
which is particularly important to constrain the 
nPDFs of relatively light nuclei.
%
In this analysis the proton baseline, $f^{(p)}(x,Q_0)$, is taken to be a variant of the
NNPDF3.1 NLO fit~\cite{Ball:2017nwa} with perturbative charm, 
where the neutrino DIS cross-sections
from NuTeV and CHORUS are removed along with the di-muon production
measurements in proton-copper collisions from the E605 experiment~\cite{Moreno:1990sf}.
As such, the proton baseline not only avoids double counting of the CC 
DIS data but also excludes constraints from heavy nuclear target
data where nuclear effects are neglected.
This choice is different to that used for \texttt{nNNPDF1.0}, where 
the global NNPDF3.1 fit was used and double-counting of 
experimental data was not an issue.
%

\begin{figure}[ht]
\begin{center}
  \includegraphics[width=0.99\textwidth]{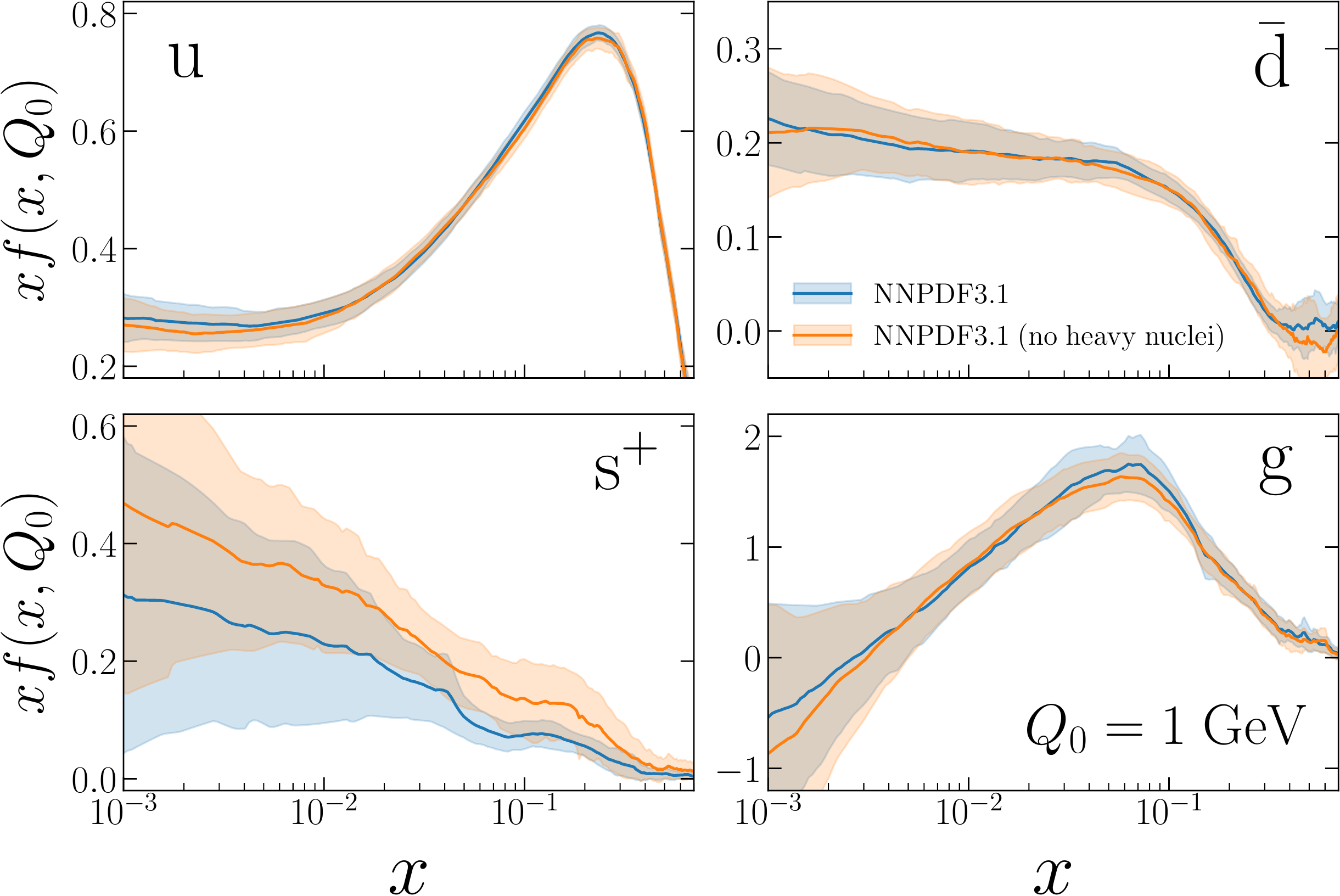}
 \end{center}
\vspace{-0.5cm}
\caption{\small A comparison between the global NNPDF3.1 free-proton analysis
  with its variant with heavy nuclear data excluded.
  We show results for the up quark, down antiquark, total strangeness, and the gluon at $Q_0=1$ GeV.
  The comparison is presented for
  the range of $x$ for which the proton boundary condition is implemented
  in \texttt{nNNPDF2.0}.
  The PDF uncertainty bands correspond to 90\% CL intervals.
}
\label{fig:NNPDF31_BC_comp}
\end{figure}

In order to illustrate the differences between the free-proton
boundary condition used in \texttt{nNNPDF1.0} and that employed in the present
analysis, we compare in Fig.~\ref{fig:NNPDF31_BC_comp} the NNPDF3.1 NLO global
and no heavy nuclear fits at the initial parameterisation scale of $Q_0=1$ GeV.
Displayed are the gluon, up quark, down sea quark,
and total strange PDFs in the
range of $x$ with which the proton boundary condition 
is constrained by Eq.~(\ref{eq:positivity})
Here one can see that removing the heavy nuclear data 
from NNPDF3.1 results in a moderate
increase of the PDF uncertainties associated to the quarks
as well as an upward shift of the central value of
the total strange distribution.
The former effect is primarily a consequence of 
information loss on quark flavour separation with
the removal of neutrino scattering measurements.
The strangeness feature, on the other hand, 
arises due to the absence of sensitivity 
from the NuTeV neutrino di-muon cross-sections, 
resulting in an upward pull by the ATLAS $W,Z$ 2011 rapidity distributions 
which are known to produce an enhanced strange
with respect to the up and down quark sea. 
The results of Fig.~\ref{fig:NNPDF31_BC_comp} highlight the importance of a 
consistent choice of the free-proton baseline in order to draw solid conclusions 
on the nuclear modifications, for example those associated to the nucleon's 
strange content.

In order to ensure that all central values
and PDF uncertainties are reproduced,
we select a different replica from the NNPDF3.1 
proton baseline when constructing the $\chi^2$ given by Eq.~(\ref{eq:positivity})
for each replica of \texttt{nNNPDF2.0}. 
Since we perform a large $N_{\rm rep}$ number of fits to 
estimate the uncertainties in \texttt{nNNPDF2.0}, 
we are able to propagate the necessary information 
contained in NNPDF3.1 to the resulting nPDFs in a robust manner.
Lastly, we note that the boundary condition term in Eq.~(\ref{eq:positivity}) is the only place in the analysis
where the free-proton NNPDF3.1 baseline is inserted.
In other parts of the fit where a free-nucleon PDF is required, for example
in the theoretical predictions of the proton-lead scattering cross-sections,
the \texttt{nNNPDF2.0} set with $A=1$ is used instead.

\myparagraph{Cross-section positivity constraint}
While PDFs are scheme-dependent and thus not necessarily positive-definite
beyond leading order in perturbative QCD, physical cross-sections constructed from 
them are scheme independent and should be positive-definite in the region of validity
of the perturbative expansion.\footnote{A recent study~\cite{Candido:2020yat} suggests, however,
that from a practical point of view PDFs in the $\overline{\rm MS}$-scheme
should also satisfy positivity beyond the LO approximation in the perturbative region.}
In the NNPDF family of proton PDF fits, the requirement that cross-sections 
remain positive is implemented by adding to the $\chi^2$ a penalty term 
in the presence of negative cross-section values~\cite{Ball:2014uwa}.
The cross-sections that enter this penalty term
correspond to theoretical predictions based on 
pseudodata generated for representative 
processes that are directly sensitive
to a sufficient number of PDF combinations.

In the \texttt{nNNPDF1.0} analysis, cross-section positivity was not imposed
and led to some observables, such as the longitudinal structure
function $F_L(x,Q^2)$, becoming negative at small-$x$ values
outside the data region.
To improve the methodological consistency
with the free-proton baseline, in \texttt{nNNPDF2.0} we impose the positivity of physical
cross-sections for all nuclei used in the fit 
by adding a suitable penalty to the figure of merit on top of the proton boundary penalty term as follows:
\begin{align}
\label{eq:positivity}
\chi^2 = \chi^2_\text{exp} &+
\lambda_{\rm BC} \sum_{f}\sum_{i=1}^{N_x} \lp q_f^{(p/A)}(x_i,Q_0,A=1)
- q_f^{(p)}(x_i,Q_0) \rp^2 \\
&+ \lambda_{\rm pos} \sum_{l=1}^{N_{\rm pos}}\sum_{j=1}^{N_A} \sum_{i_l=1}^{N_{\rm dat}^{(l)}} {\rm max}\lp 0, -\mathcal{F}_{i_l}^{(l)}(A_j) \rp \, ,
\end{align}
for $N_{\rm pos}$ positivity observables $\mathcal{F}^{(l)}$. 
Each of the observables
contain $N_{\rm dat}^{(l)}$ kinematic points that are chosen to
cover an adequately large region of phase space
relevant to various PDF combinations, 
as discussed in length in the App.~\ref{s1:positivity}.
The computed observables are summed over
all $N_A$ nuclei for which we have experimental data, as
listed in Tables~\ref{tab:nNNPDF10_data} and~\ref{tab:nNNPDF20_data}, 
as well as for the free-proton at $A=1$.
Finally, the value of the hyper-parameter $\lambda_{\rm pos} = 1000$ is 
determined by manual inspection of the optimisation process and is 
chosen so that positivity is satisfied without distorting the training 
on the real experimental data.\footnote{In future work
it might be advantageous to determine dynamically the fit hyper-parameters 
such as $\lambda_{\rm BC}$ and $\lambda_{\rm pos}$ using 
the hyper-optimisation method presented in Ref.~\cite{Carrazza:2019mzf}.}

\section{Results}
\label{s1:nNNPDF20_results}

In this section, I present the main results of this analysis, the \texttt{nNNPDF2.0} determination of nPDFs. In Sect.~\ref{s2:nNNPDF20_datavstheory}, I first start by discussing the quality of the latter w.r.t. experimental data, focusing mainly on the LHC electroweak gauge boson production cross sections. In Sect.~\ref{s2:nNNPDF20_determination}, I study the behaviour of nuclear modification ratios across different nuclei, how \texttt{nNNPDF2.0} compares to \texttt{EPPS16} and \texttt{nCTEQ15} sets and finally its implications on the nuclear strangeness.  Afterwards, in Sect.~\ref{s2:nNNPDF20_sumrules}, I discuss the role that the valence and momentum sum rules play in the global nPDF determination by presenting two variants of the \texttt{nNNPDF2.0} determinations in which one of the two sum rules is not imposed. 

I discuss the assessment of positivity in App.~\ref{s1:positivity}, where I demonstrate that the \texttt{nNNPDF2.0} inference satisfies the positivity of physical cross-sections in the kinematic range where experimental
data is available. In addition, I contrast in App.~\ref{s2:nNNPDF20_comparisonwith10} this analysis with its predecessor, \texttt{nNNPDF1.0} and trace the origin of observed differences via a series of fits with systematic changes.

\subsection{Inference quality} \label{s2:nNNPDF20_datavstheory}

We begin by discussing the fit quality which is assessed across the
various data sets and quantified by the $\chi^2$ figure of merit.
A comparison is then made using the \texttt{nNNPDF2.0} predictions with the LHC
weak gauge boson production measurements.
Following this, we take a closer look on the \texttt{nNNPDF2.0} parton
distributions and the corresponding ratios to the free-nucleon case.
Lastly, we investigate the sensitivity of the nuclear modification
factors with respect to the atomic mass $A$, in particular on the sea
quarks and strangeness, and compare our results with those from the
\texttt{EPPS16} analysis.

%
In Tables~\ref{tab:nNNPDF20_chi2} and~\ref{tab:nNNPDF20_chi2_2} we collect the values of
the $\chi^2$ per data point for all the data sets included in the
\texttt{nNNPDF2.0} analysis, i.e. the neutral and charged current DIS structure
functions as well as gauge boson production measurements at the LHC.
We compare the \texttt{nNNPDF2.0} results with a variant fit where we exclude all
LHC data sets (DIS only) and with \texttt{EPPS16}.\footnote{For the \texttt{EPPS16}
calculation we use CT14nlo as the free-proton PDF set for consistency.}
Values in italics indicate predictions for data sets
not included in the corresponding fits.
The numbers presented in Tables~\ref{tab:nNNPDF20_chi2} and~\ref{tab:nNNPDF20_chi2_2}
contain only the contribution to the $\chi^2$ associated with the
experimental data, and do not include penalty from the proton boundary
condition or positivity constraint (the latter of which vanishes for all
final \texttt{nNNPDF2.0} replicas anyway).
Moreover, while we use the $t_0$ prescription~\cite{Ball:2009qv} during
the optimisation to avoid the D'Agostini bias, the quoted numbers
correspond to the experimental definition of the $\chi^2$
instead~\cite{Ball:2012wy}, in which the central experimental value is
used to compute the correlated multiplicative uncertainties.

\begin{table}
  \centering
  \footnotesize
    \renewcommand{\arraystretch}{1.25}
\begin{tabular}{|l|c|C{2.7cm}|C{2.5cm}|C{2.5cm}|}
\toprule
 \multicolumn{2}{|c|}{} & \texttt{nNNPDF2.0} (DIS) & \texttt{nNNPDF2.0} & \texttt{EPPS16nlo} \\ \midrule
Dataset & $N_{\rm dat}$ &$\chi^ 2/N_{\rm dat}$ &$\chi^ 2/N_{\rm dat}$ &$\chi^ 2/N_{\rm dat}$ \\ \hline
${\rm NMC \,\, (He/D)}$ & 13 & 1.11 & 1.129 & 0.829 \\ \hline
${\rm SLAC \,\, (He/D)}$ & 3 & 0.623 & 0.638 & 0.152 \\ \hline
${\rm NMC \,\, (Li/D)}$ & 12 & 1.083 & 1.166 & 0.74 \\ \hline
${\rm SLAC \,\, (Be/D)}$ & 3 & 1.579 & 1.719 & 0.098 \\ \hline
${\rm EMC \,\, (C/D)}$ & 12 & 1.292 & 1.321 & 1.174 \\ \hline
${\rm FNAL \,\, (C/D)}$ & 3 & 0.932 & 0.838 & 0.985 \\ \hline
${\rm NMC \,\, (C/D)}$ & 26 & 2.002 & 2.171 & 0.872 \\ \hline
${\rm SLAC \,\, (C/D)}$ & 2 & 0.286 & 0.251 & 1.075 \\ \hline
${\rm BCDMS \,\, (N/D)}$ & 9 & 2.439 & 2.635 & n/a \\ \hline
${\rm SLAC \,\, (Al/D)}$ & 3 & 0.606 & 0.864 & 0.326 \\ \hline
${\rm EMC \,\, (Ca/D)}$ & 3 & 1.72 & 1.722 & 1.82 \\ \hline
${\rm FNAL \,\, (Ca/D)}$ & 3 & 1.253 & 1.194 & 1.354 \\ \hline
${\rm NMC \,\, (Ca/D)}$ & 12 & 1.503 & 1.747 & 1.772 \\ \hline
${\rm SLAC \,\, (Ca/D)}$ & 2 & 0.82 & 0.771 & 1.642 \\ \hline
${\rm BCDMS \,\, (Fe/D)}$ & 16 & 2.244 & 2.743 & \textit{0.765} \\ \hline
${\rm EMC \,\, (Fe/D)}$ & 58 & 0.827 & 0.875 & 0.445 \\ \hline
${\rm SLAC \,\, (Fe/D)}$ & 8 & 2.171 & 2.455 & 1.06 \\ \hline
${\rm EMC \,\, (Cu/D)}$ & 27 & 0.523 & 0.572 & 0.714 \\ \hline
${\rm SLAC \,\, (Ag/D)}$ & 2 & 0.667 & 0.691 & 1.595 \\ \hline
${\rm EMC \,\, (Sn/D)}$ & 8 & 2.197 & 2.248 & 2.265 \\ \hline
${\rm FNAL \,\, (Xe/D)}$ & 4 & 0.414 & 0.384 & n/a \\ \hline
${\rm SLAC \,\, (Au/D)}$ & 3 & 1.216 & 1.353 & 1.916 \\ \hline
${\rm FNAL \,\, (Pb/D)}$ & 3 & 2.243 & 2.168 & 2.044 \\ \hline 
${\rm NMC \,\, (Be/C)}$ & 14 & 0.268 & 0.269 & 0.27 \\ \hline
${\rm NMC \,\, (C/Li)}$ & 9 & 1.063 & 1.117 & 0.9 \\ \hline
${\rm NMC \,\, (Al/C)}$ & 14 & 0.345 & 0.354 & 0.396 \\ \hline
${\rm NMC \,\, (Ca/C)}$ & 23 & 0.468 & 0.44 & 0.564 \\ \hline
${\rm NMC \,\, (Fe/C)}$ & 14 & 0.663 & 0.667 & 0.751 \\ \hline
${\rm NMC \,\, (Sn/C)}$ & 119 & 0.607 & 0.638 & 0.626 \\ \hline
${\rm NMC \,\, (Ca/Li)}$ & 9 & 0.259 & 0.276 & 0.15 \\ 
\bottomrule
\end{tabular}
\vspace{0.3cm}
\caption{The values of the $\chi^2$ per data point for the
DIS neutral current structure function
  datasets included in \texttt{nNNPDF2.0}.
  We compare the \texttt{nNNPDF2.0} global and DIS-only
  results with those obtained using \texttt{EPPS16} as input for the theory predictions.
  \label{tab:nNNPDF20_chi2}
}
\end{table}

\begin{table}[t]
  \centering
  \small
  \renewcommand{\arraystretch}{1.30}
  \begin{tabular}{|l|c|C{2.9cm}|C{2.5cm}|C{2.5cm}|}
    \toprule
  \multicolumn{2}{|c|}{} & \texttt{nNNPDF2.0} (DIS) & \texttt{nNNPDF2.0} & \texttt{EPPS16nlo} \\ \midrule
Dataset & $N_{\rm dat}$ &$\chi^ 2/N_{\rm dat}$ &$\chi^ 2/N_{\rm dat}$ &$\chi^ 2/N_{\rm dat}$ \\ \hline
${\rm NuTeV \,\, (\bar{\nu} Fe)}$ & 37 & 0.946 & 1.094 & \textit{0.639} \\ \hline
${\rm NuTeV \,\, (\nu Fe)}$ & 39 & 0.287 & 0.264 & \textit{0.381} \\ \hline
${\rm CHORUS \,\, (\bar{\nu} Pb)}$ & 423 & 0.938 & 0.97 & 1.107 \\ \hline
${\rm CHORUS \,\, (\nu Pb)}$ & 423 & 1.007 & 1.015 & 1.024 \\ \hline \hline
${\rm ATLAS^{5TEV} \,\, Z}$ & 14 & \textit{1.469} & 1.134 & 1.12 \\ \hline
${\rm CMS^{5TeV} \,\, W^{-}}$ & 10 & \textit{1.688} & 1.078 & 0.857 \\ \hline
${\rm CMS^{8TeV} \,\, W^{-}}$ & 24 & \textit{1.453} & 0.72 & \textit{0.825} \\ \hline
${\rm CMS^{5TeV} \,\, W^{+}}$ & 10 & \textit{2.32} & 1.125 & 1.211 \\ \hline
${\rm CMS^{8TeV} \,\, W^{+}}$ & 24 & \textit{3.622} & 0.772 & \textit{0.951} \\ \hline
${\rm CMS^{5TeV} \,\, Z}$ & 12 & \textit{0.58} & 0.52 & 0.639 \\ \hline \hline
{\bf Total} & {\bf 1467} & {\bf 1.013} & {\bf 0.976} &  \\ \bottomrule 
\end{tabular}
\vspace{0.3cm}
\caption{Same as Table~\ref{tab:nNNPDF20_chi2} now
  for the new datasets
  included in \texttt{nNNPDF2.0}:
  charged current DIS structure functions
  and gauge boson production at the LHC.
  We also provide the values of $\chi^2/N_{\rm dat}$
  for the total dataset.
  Values in italics indicate predictions for datasets
  not included in the corresponding fit.
  \label{tab:nNNPDF20_chi2_2}
}
\end{table}


From the results of Table~\ref{tab:nNNPDF20_chi2} and~\ref{tab:nNNPDF20_chi2_2}, one can
see that the \texttt{nNNPDF2.0} determination achieves a satisfactory description
of all data sets included in this analysis.
A good $\chi^2$ is obtained in particular for the charged-current DIS
cross-sections and LHC gauge boson production distributions.
For instance, for the precise W boson rapidity distributions at
$\sqrt{s}=8.16$ TeV from CMS one finds $\chi^2/N_{\rm dat} = 0.74$ for
$N_{\rm dat}=48$ data points.
The corresponding predictions using \texttt{EPPS16}  (which do not include this dataset)
also lead to a good agreement with
$\chi^2/N_{\rm dat} = 0.88$.
The description of most neutral current DIS data sets is comparable
to that of \texttt{nNNPDF1.0}.
Some data sets, such as the SLAC iron structure functions,
are somewhat deteriorated with respect to \texttt{nNNPDF1.0},
possibly due to some mild tension with the CC cross-sections.
Further, we find that our resulting fit quality to the CC
deep-inelastic structure functions is similar to that obtained in the
corresponding proton PDF analysis~\cite{Ball:2017nwa}.

Overall, the resulting $\chi_{\rm tot}^2/N_{\rm dat}= 0.976$ for the $N_{\rm
dat}=1467$ data points included in the fit highlights the remarkable
consistency of the experimental data on nuclear targets and the
corresponding theory predictions based on the QCD factorisation
framework.
A similar total $\chi^2$ is obtained for the theoretical predictions
computed using \texttt{EPPS16} as input.
Note that while our parameterisation is more flexible 
than that used in the EPPS analysis, we also account for a number 
of constraints such as the positivity constraint and boundary condition, 
which restricts the range of functional forms available.
In any case, the fact that both global fits lead to comparable
$\chi^2$ values can be explained by the corresponding similarities at the nPDF level,
as will be shown below.

%
To facilitate our discussion regarding the comparison between
data and theory calculations, we first introduce here the conventions that we use to
define the nuclear modification factors.
Following the notations discussed in the introduction of this chapter, the Drell-Yan rapidity
  distributions in proton-nucleus collision can be expressed as:
  \begin{equation}
  \label{eq:DYdecomposition}
  \frac{d\sigma_{\rm DY}(y)}{dy} \equiv A\frac{d\sigma^{(N/A)}_{\rm DY}(y)}{dy} = Z\frac{d\sigma^{(p/A)}_{\rm DY}(y)}{dy}
  + \lp A-Z\rp\frac{d\sigma^{(n/A)}_{\rm DY}(y)}{dy} \, ,
  \end{equation}
  where the superindices $N/A$, $p/A$, and $n/A$ indicate respectively the collision
  between a proton with an average nucleon, a proton, or a neutron bound
  within a nucleus of atomic number $Z$ and mass number $A$.
  As in the case of the PDFs, the bound proton and nucleon
  cross-sections $\sigma_{\rm DY}^{(p/A)}$ and $\sigma_{\rm DY}^{(n/A)}$
  are related to each other via isospin symmetry.
  
  The expression in Eq.~(\ref{eq:DYdecomposition}) helps in emphasizing
  the two reasons why the Drell-Yan cross-sections will be different
  between pp and pA collisions.
  The first is due to the modifications of the bound proton PDFs in
  nuclei, namely the fact that $\sigma_{\rm DY}^{(p/A)} \ne \sigma_{\rm
  DY}^{(p)}$.
  Secondly, the predictions of pA collisions using non-isoscalar nuclei
  would still differ from those of pp reactions in the absence of
  nuclear modifications, i.e. $\sigma_{\rm DY}^{(p/A)}= \sigma_{\rm
  DY}^{(p)}$, as a consequence of the unequal amount of protons and
  neutrons in the target, resulting in $\sigma_{\rm DY}^{(N/A)} \ne \sigma_{\rm DY}^{(p)}$.
  
  Taking these considerations into account, one should define the
  nuclear modification factor in Drell-Yan proton-nuclear collisions as
   \begin{equation}
   \label{eq:RADY}
   R_A^{\rm DY}(y) \equiv \lp \frac{d\sigma^{(p/A)}_{\rm DY}(y)}{dy} +
   \lp \frac{A}{Z}-1\rp  \frac{d\sigma^{(n/A)}_{\rm DY}(y)}{dy} \rp \Bigg/ \lp 
   \frac{d\sigma^{(p)}_{\rm DY}(y)}{dy} + \lp \frac{A}{Z}-1\rp \frac{d\sigma^{(n)}_{\rm
   DY}(y)}{dy} \rp \, .
   \end{equation}
   With the above definition, one should find that $R_A^{\rm DY}(y) \ne
   1$ only in the presence of genuine nuclear corrections, that is, when
   $\sigma_{\rm DY}^{(p/A)} \ne \sigma_{\rm DY}^{(p)}$.
   The definition of Eq.~(\ref{eq:RADY}) differs from that of an
   observable frequently measured in proton-lead collisions, where the
   proton-nucleus cross-section is normalised to a proton-proton
   baseline,
    \begin{eqnarray}
   \label{eq:RADYexp}
   R_{A, \rm exp}^{\rm DY}(y) &\equiv&
   \frac{d\sigma^{(N/A)}_{\rm DY}(y)}{dy}
    \Bigg/ \frac{d\sigma^{(p)}_{\rm DY}(y)}{dy} \\ &=&  \nonumber \lp  \frac{Z}{A}
      \frac{d\sigma^{(p/A)}_{\rm DY}(y)}{dy} + \lp 1- \frac{Z}{A}\rp
      \frac{d\sigma^{(n/A)}_{\rm DY}(y)}{dy} \rp \Bigg/ \lp 
      \frac{d\sigma^{(p)}_{\rm DY}(y)}{dy} \rp \, .
   \end{eqnarray}
   As  explained above, in proton-nuclear collisions one will find
   $R_{A, \rm exp}^{\rm DY}(y) \ne 1$ for non-isoscalar targets
   even when $\sigma_{\rm
     DY}^{(p/A)}= \sigma_{\rm DY}^{(p)}$ due to the imbalance between
   the number of protons and neutrons.
   In this section, we will exclusively use the definition of
   Eq.~(\ref{eq:RADY}) when evaluating the theoretical predictions of
   nuclear modification ratios in Drell-Yan distributions.
   
\begin{figure}[ht]
\begin{center}
  \includegraphics[width=0.99\textwidth]{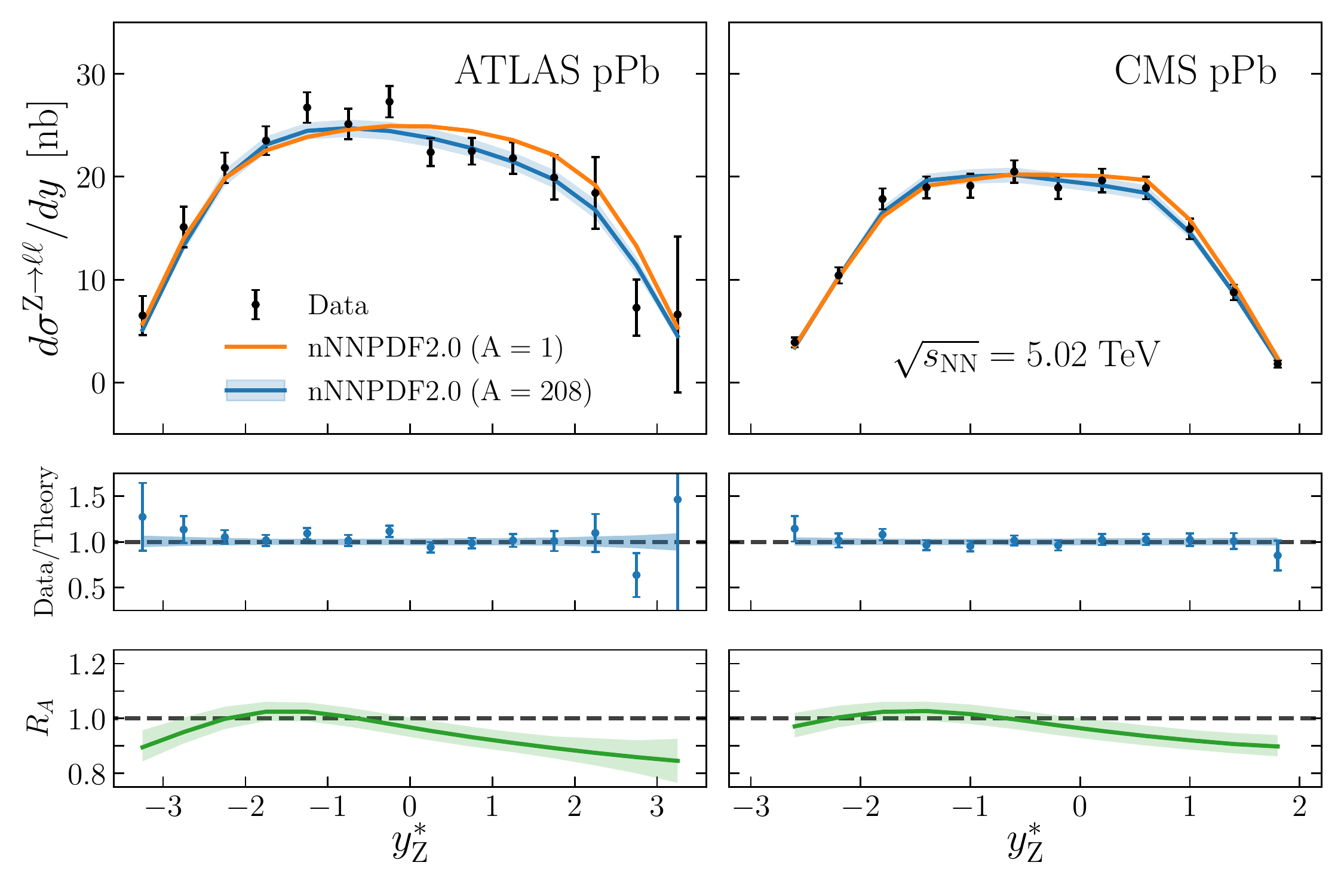}
 \end{center}
\vspace{-0.8cm}
\caption{ Comparison between the ATLAS (left) and CMS (right panel)
  measurements of Z boson production in proton-lead collisions at
  $\sqrt{s}=5.02$ TeV with the theoretical predictions using \texttt{nNNPDF2.0}
  as input.
  We show the predictions obtained both for $A=1$ and $A=208$, in the
  later case including the 90\% confidence level band.
  From top to bottom, the three panels display the absolute
  cross-sections, the ratio between data and theory, and the nuclear
  modification $R_A(y)$.
  \label{nNNPDF20comp_data_1}
}
\end{figure}

In Fig.~\ref{nNNPDF20comp_data_1} we display the comparison between the
ATLAS and CMS measurements of Z boson production in proton-lead collisions at
$\sqrt{s}=5.02$ TeV with the theoretical predictions using \texttt{nNNPDF2.0}.
  The theory calculations were computed using
  Eq.~(\ref{eq:DYdecomposition}) with the \texttt{nNNPDF2.0} $A=1$ and $A=208$
  distributions, in the latter case also including the corresponding
  90\% CL uncertainty band.
  Note that for the theory cross-section obtained with the $A=1$ PDFs,
  nuclear effects are absent since it corresponds to the free proton
  distributions.
  From top to bottom, the three panels display the absolute
  cross-sections as a function of the Z rapidity $y$, the ratio
  between data and theory, and the nuclear modification factor ratio
  $R_A(y)$ defined by Eq.~(\ref{eq:RADY}).
  
  Here the ATLAS and CMS measurements of the $Z$ rapidity
  distributions are both provided in the center-of-mass
  reference frame of the proton-lead collision.
  The CMS absolute cross-sections are lower than ATLAS due to the
  different kinematical selection cuts.
  We also note that for these data sets, as well as for the rest of LHC
  measurements, forward rapidities correspond to the direction of the
  incoming proton.
  From the comparisons in Fig.~\ref{nNNPDF20comp_data_1} we can see that
  the theory predictions are in good agreement with the experimental
  data.
Interestingly, the $R_A(y)$ ratios exhibit a strong preference for
nuclear modifications, especially at forward rapidities which correspond
to small values in $x$ for the bound nucleons.
As will be discussed below, this behaviour can be explained at the level
of the nuclear PDFs by a notable shadowing effect at small-$x$ for up
and down quarks and antiquarks.

\begin{figure}[ht]
  \begin{center}
    \includegraphics[width=0.99\textwidth]{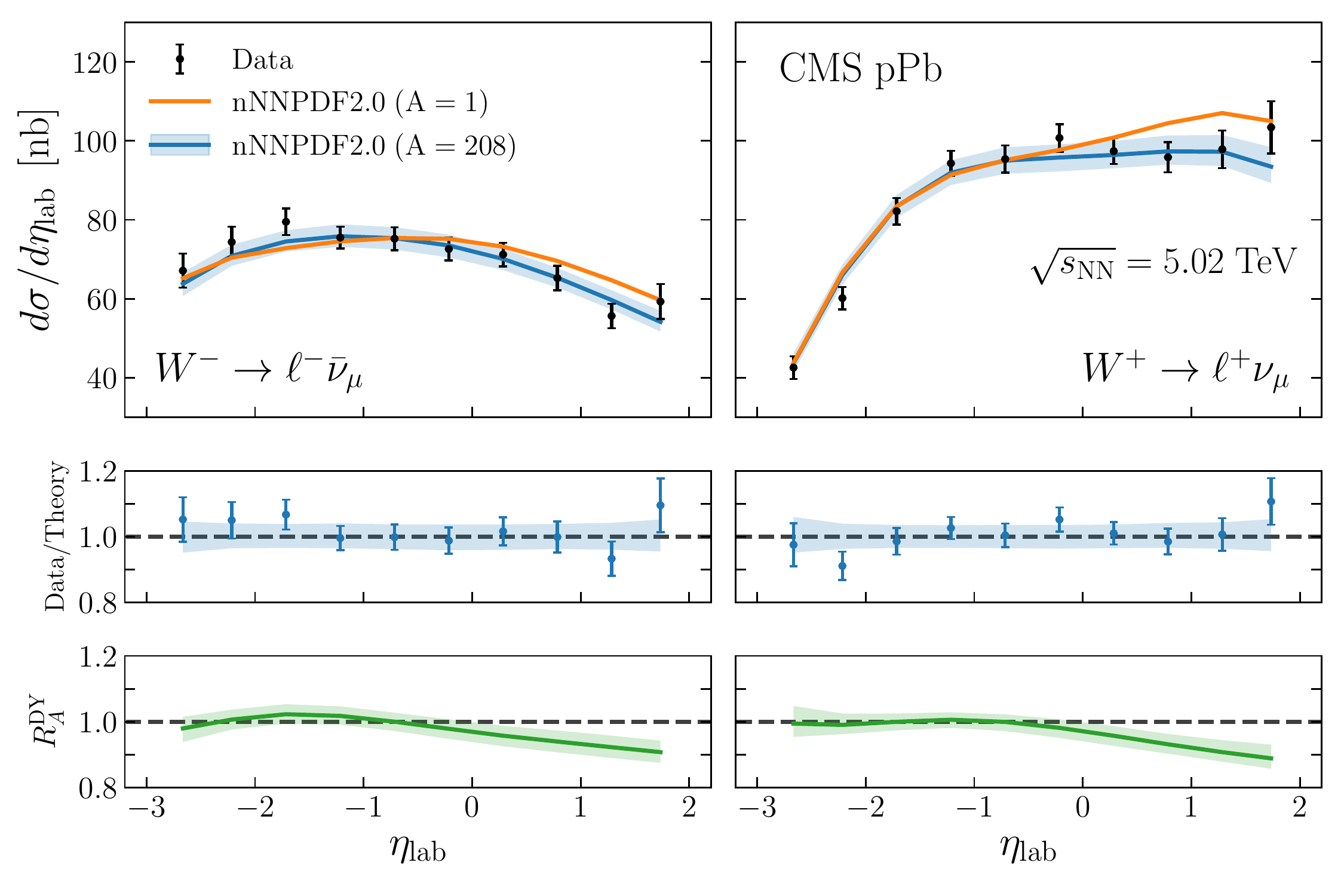}
   \end{center}
  \vspace{-0.8cm}
  \caption{Same as Fig.~\ref{nNNPDF20comp_data_1} now for the CMS muon
    rapidity distributions in W$^-$ and W$^+$ production at
    $\sqrt{s}=5.02$ TeV.
    \label{nNNPDF20comp_data_2}
  }
  \end{figure}

\begin{figure}[ht]
\begin{center}
  \includegraphics[width=0.99\textwidth]{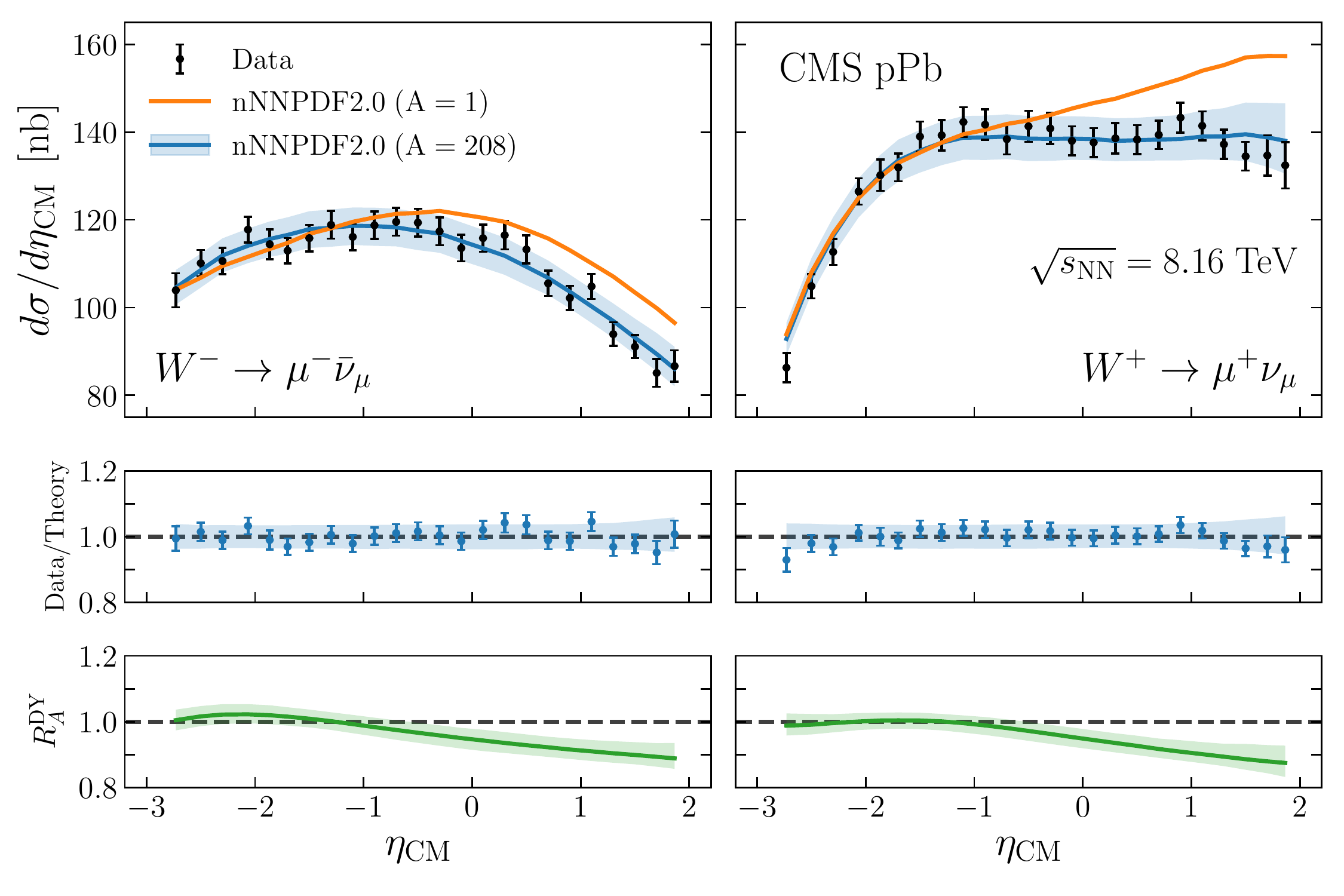}
 \end{center}
\vspace{-0.8cm}
\caption{Same as Fig.~\ref{nNNPDF20comp_data_1} now for the CMS muon
  rapidity distributions in W$^-$ and W$^+$ production at
  $\sqrt{s}=8.16$ TeV.
  \label{nNNPDF20comp_data_3}
}
\end{figure}

The corresponding comparisons for the CMS muon rapidity distributions in
W$^-$ and W$^+$ production at $\sqrt{s}=5.02$ TeV and 8.16 TeV are
displayed in Figs.~\ref{nNNPDF20comp_data_2}
and~\ref{nNNPDF20comp_data_3} respectively.
The results are presented as functions of the rapidity of the charged
lepton from the W boson decay in the laboratory frame.
In all cases the theoretical predictions based on \texttt{nNNPDF2.0} provide a
satisfactory description of the experimental data.
It is interesting to note that for the high-statistics CMS measurement at
8 TeV, the $A=208$ prediction is remarkably better than the free-proton
one, particularly at forward rapidities where one is sensitive to the
small-$x$ region of the bound nucleons.
This feature highlights how the CMS 8 TeV W production data provides
direct evidence for the nuclear modifications of valence and sea quark
distributions.


\subsection{The \texttt{nNNPDF2.0} determination} \label{s2:nNNPDF20_determination}
\begin{figure}[ht]
\begin{center}
  \includegraphics[width=0.99\textwidth]{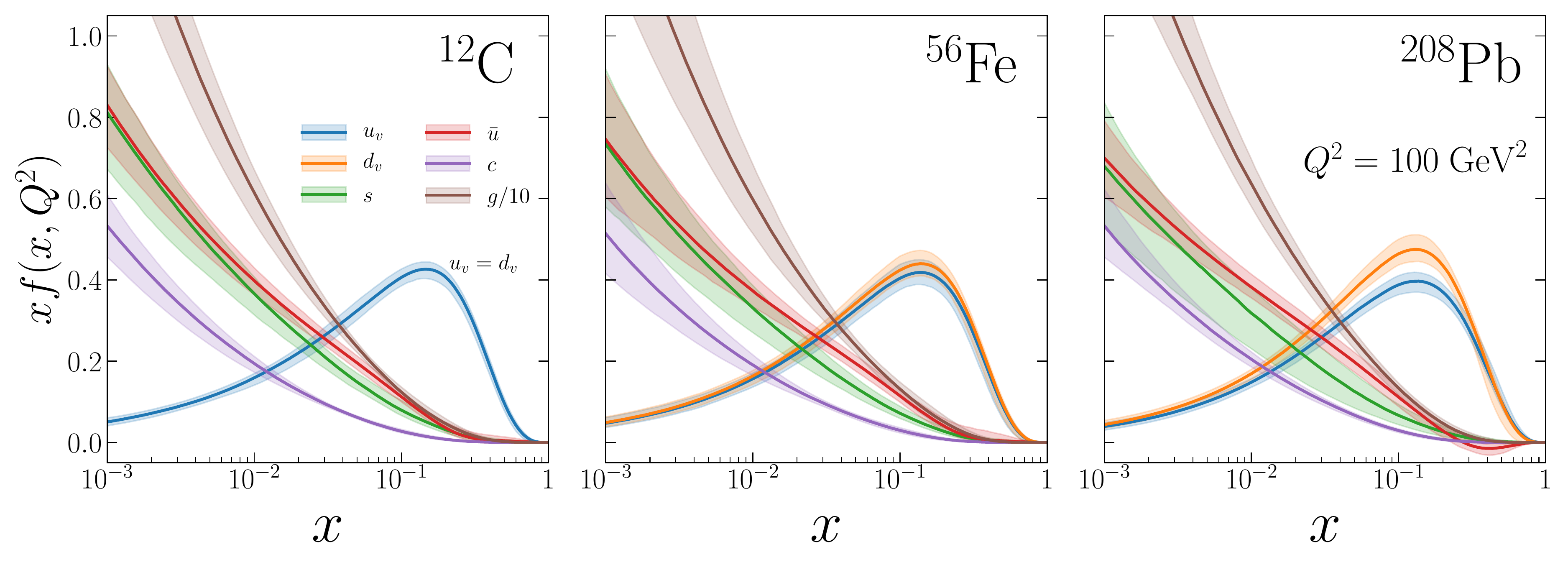}
 \end{center}
\vspace{-0.6cm}
\caption{The \texttt{nNNPDF2.0}  set of nuclear PDFs for $^{12}$C, $^{56}$Fe, and
  $^{208}$Pb at the scale $Q=10$ GeV.
  We display the gluon, the up and down valence quarks (which coincide
  for isoscalar nuclei), as well as the down, strange, and charm sea
  quark distributions.
  The bands indicate the 90\% CL uncertainty range.
  \label{fig:nNNPDF20_PDFs}
}
\end{figure}

In Fig.~\ref{fig:nNNPDF20_PDFs} we display the \texttt{nNNPDF2.0} set of nuclear
PDFs for three different nuclei, $^{12}$C, $^{56}$Fe, and $^{208}$Pb,
constructed using Eq.~(\ref{eq:qNAdefinition}) at a scale of $Q=10$ GeV.
Specifically, we display the gluon, the up and down valence quarks, and
the down, strange, and charm sea quark distributions.
For isoscalar nuclei such as $^{12}$C, the up and down valence
distributions are equivalent, $u_v^{(N/A)}=d_v^{(N/A)}$, as well as the
up and down sea PDFs, $\bar{u}^{(N/A)}=\bar{d}^{(N/A)}$, as a result of
Eq.~(\ref{eq:qNAdefinition}).

From the comparisons in Fig.~\ref{fig:nNNPDF20_PDFs}, we can see that
the nuclear PDFs exhibit a moderate dependence on the atomic number $A$.
The resulting pattern of PDF uncertainties can partly be
explained by the input data.
For example, nPDF uncertainties on strangeness are smaller in $^{12}$C
and $^{56}$Fe compared to $^{208}$Pb, due to the impact of the proton
boundary condition and the NuTeV di-muon data, respectively.
We also observe that the PDF uncertainties on the gluon (and
correspondingly on the dynamically generated charm PDF) at medium and
small-$x$ are larger in iron than in carbon and lead.
While the gluon uncertainties for carbon are largely determined by the impact of
the free-proton boundary condition, those on lead nuclei can
likely be attributed to the LHC measurements of W and Z production
and the large amount of charged-current DIS data, which indirectly
provide constraints via DGLAP evolution.

Fig.~\ref{fig:nNNPDF20_PDFs} also shows that in the case of $^{208}$Pb,
there is a clear difference between the $d_v$ and $u_v$ distributions
due to the non-isoscalar nature of the nucleus, where $d_v$ is larger due to
the significant neutron excess in lead.
The fact that $u_v$ and $d_v$ do not overlap within the 90\% CL bands in
a wide range of $x$ highlights how a careful treatment of the quark and
antiquark flavour separation is essential in order to describe the
precise data available on lead targets, especially the weak boson
production measurements in proton-lead collisions at the LHC.

To further illustrate the features of the \texttt{nNNPDF2.0} determination, it is useful to study
them in terms of ratios with respect to the corresponding free-nucleon
baseline.
In the following we will define the nuclear modification ratios of PDFs
for a nucleus with mass number $A$ as,
\begin{equation}
\label{eq:nuclearratioRv3}
R_f^{A}(x,Q^2) \equiv \frac{Z  q^{(p/A)}_f(x,Q^2) + (A-Z)
  q^{(n/A)}_f(x,Q^2) }{Z q^{(p)}_f(x,Q^2) + (A-Z) q^{(n)}_f(x,Q^2)} \, .
\end{equation}
When evaluating Eq.~(\ref{eq:nuclearratioRv3}), it is important to
account both for the uncertainties associated to the nuclear and free-proton
PDFs.
In the case of a Monte Carlo set such as \texttt{nNNPDF2.0}, this entails
evaluating $R_f^{A}$ for each of the $N_{\rm rep}$ replicas and then
determining the resulting median and 90\% CL interval.

\begin{figure}[ht]
\begin{center}
  \includegraphics[width=0.93\textwidth]{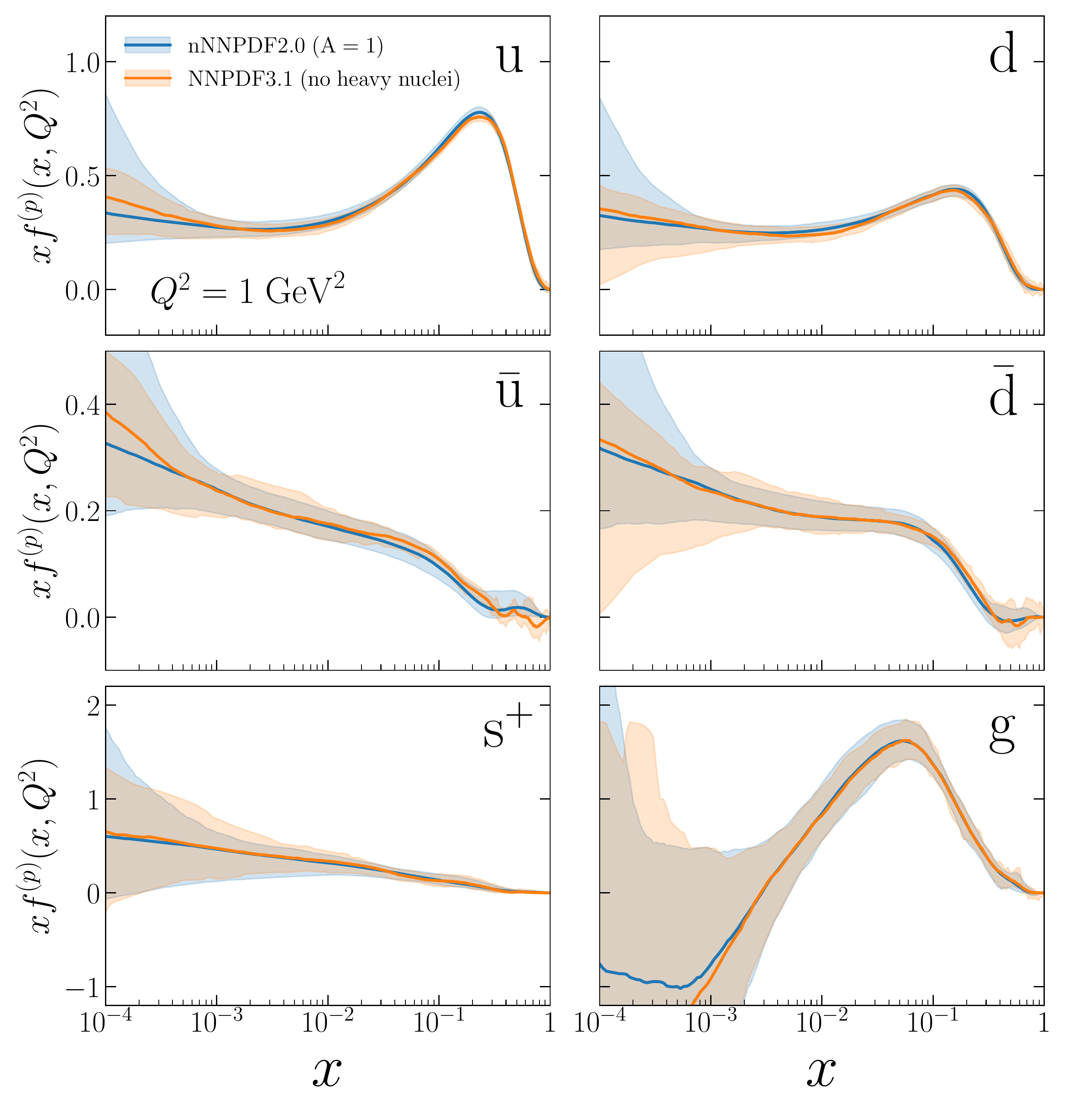}
 \end{center}
\vspace{-0.8cm}
\caption{Comparison of the \texttt{nNNPDF2.0}  parton distributions for $A=1$
  with the NNPDF3.1
  baseline used for the boundary condition
  at the parametrisation scale $Q_0=1$ GeV.
  \label{fig:nNNPDF20_vs_BC}
}
\end{figure}

In Fig.~\ref{fig:nNNPDF20_vs_BC} we show the \texttt{nNNPDF2.0} 
distributions  for $A=1$ that enter the denominator of Eq.~(\ref{eq:nuclearratioRv3}).
They are compared with the NNPDF3.1
  proton baseline used for implementation the boundary condition 
  at the input parametrisation scale  $Q_0=1$ GeV.
Overall, there is very good agreement between our $A=1$ result
and the proton boundary condition, particularly in the region
of $x$ where the constraint is being imposed, $10^{-3} < x < 0.7$. 
It is important to emphasize that due to the \texttt{nNNPDF2.0} $A=1$ set being 
determined not only by the boundary condition but also by the positivity constraints and  the LHC
cross-sections, one expects some moderate differences with the NNPDF3.1 proton baseline.
The most notable differences indeed are found in the $\bar{u}$
and $\bar{d}$ PDFs at medium to large $x$, where
the newly added DY positivity observables for 
$\bar{u}d$ and $u\bar{d}$ quark combinations, 
as well as the LHC proton-lead data, play a significant role. 
Nevertheless, the level of agreement
reported in Fig.~\ref{fig:nNNPDF20_vs_BC} is quite remarkable
and highlights how the \texttt{nNNPDF2.0} determination manages to take into account the extensive
information provided by the global analysis of free-proton
structure.

\begin{figure}[ht]
\begin{center}
  \includegraphics[width=0.93\textwidth]{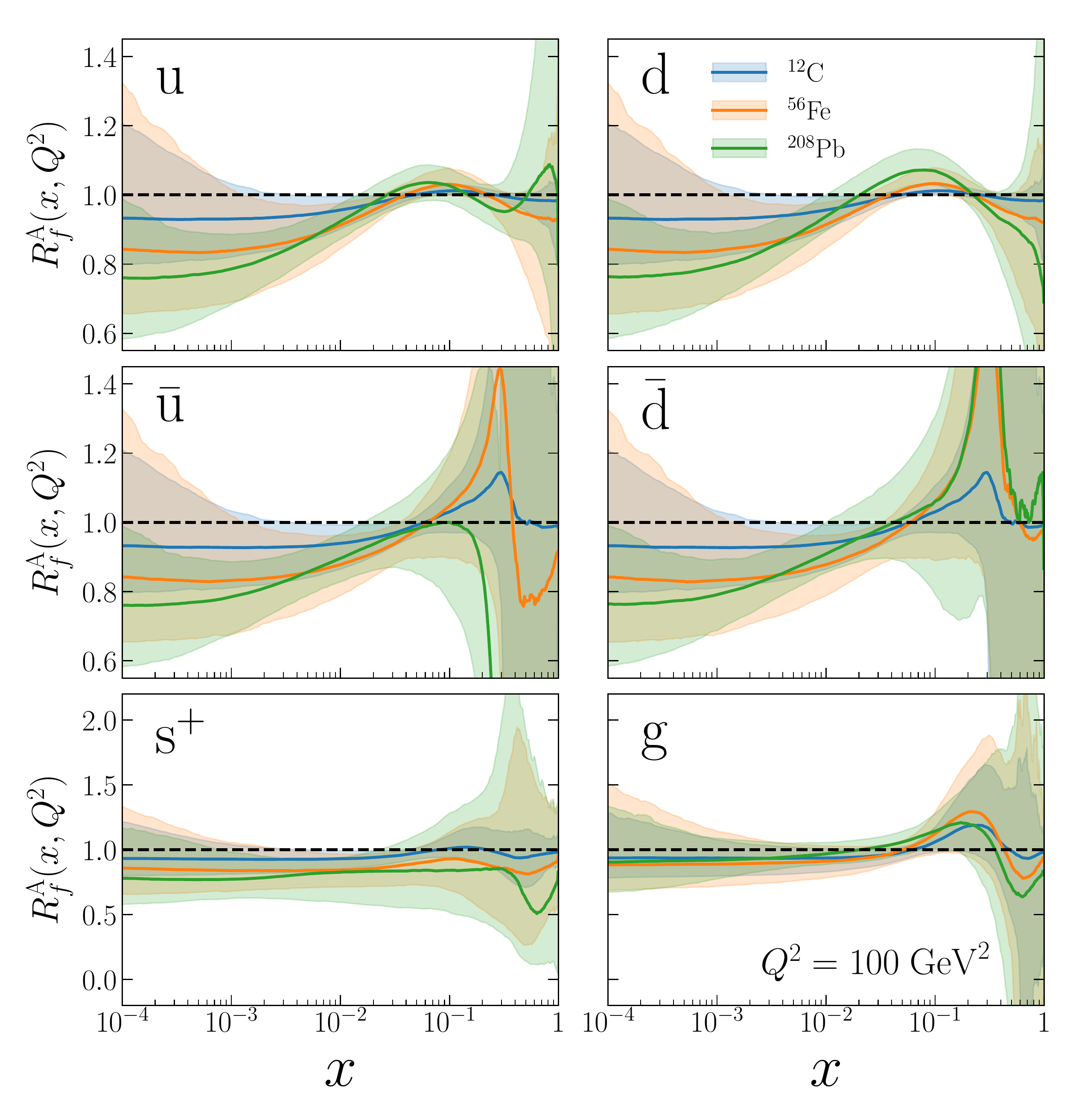}
 \end{center}
\vspace{-0.8cm}
\caption{Comparison of the nuclear PDF ratios,
  Eq.~(\ref{eq:nuclearratioRv3}), for three different nuclei, $^{12}$C,
  $^{56}$Fe, and $^{208}$Pb, at $Q=10$ GeV.
  From top to bottom we show the up and down quarks, the corresponding
  antiquarks, the total strangeness, and the gluon.
  The bands indicate the 90\% confidence level intervals and take into
  account the correlations with the proton baseline used for the
  normalisation.
  \label{fig:R_A}
}
\end{figure}

In Fig.~\ref{fig:R_A} we display the nuclear PDF ratios, defined 
by Eq.~(\ref{eq:nuclearratioRv3}), for the same parton flavours
as in Fig.~\ref{fig:nNNPDF20_vs_BC}.
Here the ratios for $^{12}$C, $^{56}$Fe, and $^{208}$Pb nuclei 
are compared at $Q=10$ GeV. 
The shaded bands indicating the 90\% confidence level intervals 
take into account also the correlations with the proton baseline.
  Overall, the comparison highlights the dependence on the central value
  and uncertainties of the nuclear ratios $R_f^{A}$ as the value of $A$
  is varied from lighter to heavier nuclei.

 For the up and down quark nPDFs in Fig.~\ref{fig:R_A}, we can see that the
 shadowing effects become more prominent at small-$x$ as $A$ increases,
 with the central value reaching $R_f^{A}\simeq 0.75$ at $x=10^{-4}$ for
 the lead ratios.
Interestingly, the nPDF uncertainties on the quarks for $x\lsim 10^{-2}$
are reduced in lead as compared to the lighter nuclei.
This is a consequence of the constraints provided by the LHC data, as
will be shown in App.~\ref{s2:nNNPDF20_comparisonwith10}.
In the large-$x$ region, deviations from the $R_f^{A}=1$ scenario
(no nuclear corrections)
 appear more prominent for the quarks and antiquarks of heavier
nuclei.

Turning now to the valence quarks, one can distinguish the shadowing and
anti-shadowing regions for all values of $A$, though nuclear effects in
carbon are quite small.
While one generally finds a suppression $R_f^{A} < 1$ at large $x$ that
is consistent with the EMC-effect expectation, the position of the
so-called "EMC minimum" is not universal or even guaranteed at the nPDF
level.
For the anti-quarks, the only region where a well-defined qualitative
behaviour is observed is the small-$x$ shadowing region, while at
large-$x$ the nPDF uncertainties are too large to draw any solid
conclusion.
Finally, we observe that the nuclear modifications on the gluon nPDF are
rather stable as $A$ is varied.

\begin{figure}[ht]
\begin{center}
  \includegraphics[width=0.93\textwidth]{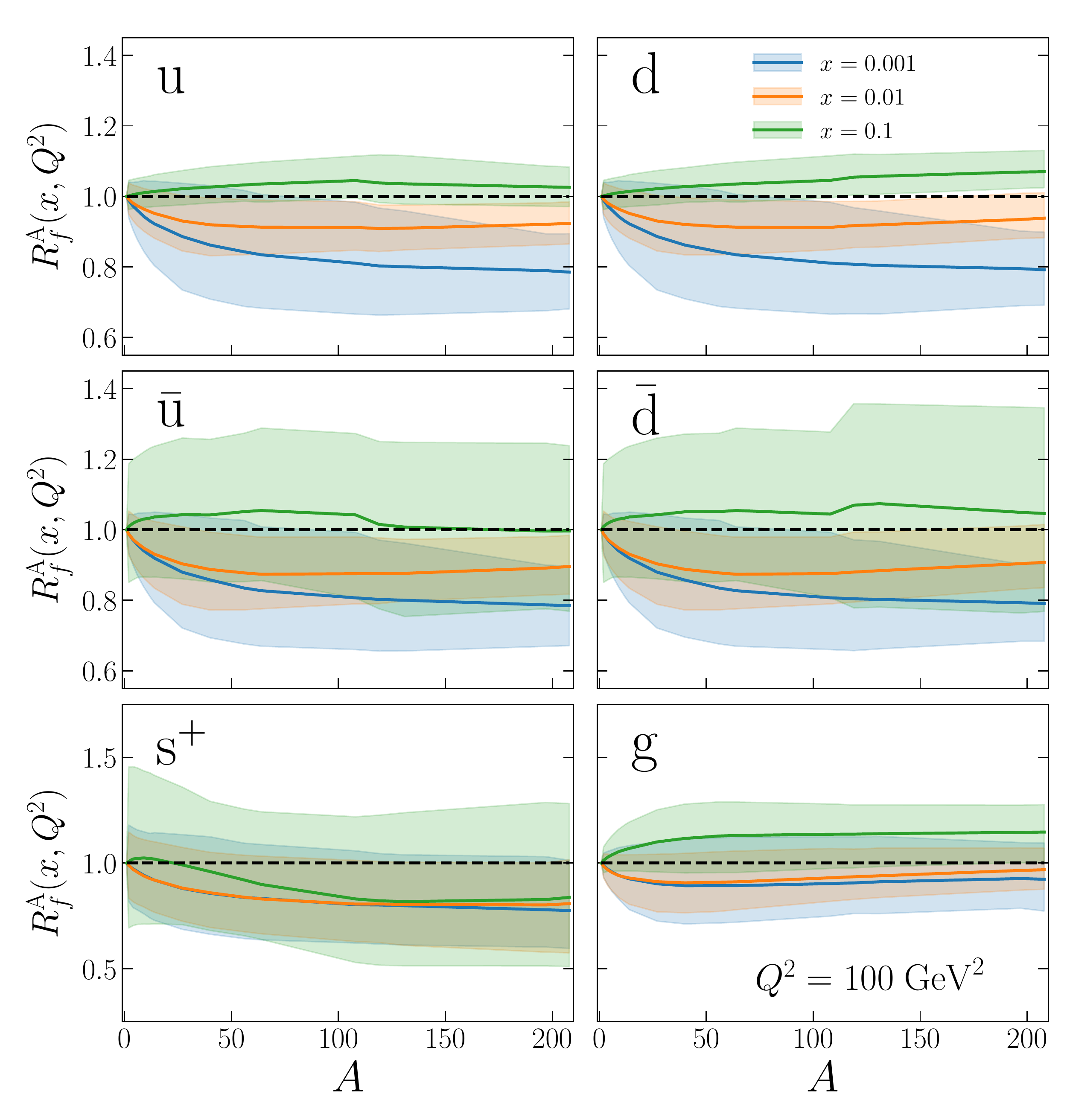}
 \end{center}
\vspace{-0.8cm}
\caption{Comparison of the nuclear PDF ratios,
  Eq.~(\ref{eq:nuclearratioRv3}) as a function of atomic mass $A$ 
  for three different $x$ values, $x=0.001$,
  $x=0.01$, and $x=0.1$, at $Q=10$ GeV.
  From top to bottom we show the up and down quarks, the corresponding
  antiquarks, the total strangeness, and the gluon.
  The bands indicate the 90\% confidence level intervals and take into
  account the correlations with the proton baseline used for the
  normalisation.
  \label{R_A_x}
}
\end{figure}

To illustrate further the dependence of nuclear modifications on the atomic
mass $A$, we show in Fig.~\ref{R_A_x} the ratios $R_f$ as a function
of $A$ for three values of $x$ at $Q=10$ GeV.
Overall, the ratios are stable and exhibit smooth dependence for all parton
flavours across the $A$ values that are included in the fit.
For the up and down quark ratios, the transition from anti-shadowing
to shadowing is clearly visible as $x$ is decreased from 0.1 to 0.001.
We can also see a notable improvement in
PDF uncertainties as $A$ approaches 208 (lead) at low $x$ as 
result of the constraints from LHC measurements.
Similar behaviour is seen also for the anti-quarks, although with
larger uncertainties.
On the other hand, the strange quark modifications are stable 
across both $x$ and $A$, with significantly larger errors
at $x=0.1$. 
Lastly, the gluon distribution also displays smooth dependence
on $A$ for the modification ratios.
Here the notable shift from anti-shadowing to shadowing is
also seen, although the ratios all agree with unity within the
90\% CL range.

\paragraph{Nuclear strangeness}
The strangeness content of the proton in unpolarised PDF fits has
attracted a lot of attention recently.
Traditionally, the determination of $s(x,Q^2)$ in global proton PDF fits
has been dominated by the constraints provided by charm production in
neutrino
DIS~\cite{Bazarko:1994tt,chorus-dimuon,MasonPhD,Mason:2007zz,Samoylov:2013xoa}.
These measurements suggest that the strange sea is suppressed compared
to its up and down quark counterparts, favouring values of around $r_s
\simeq 0.5$ when expressed in terms of the strangeness ratio defined as
\begin{equation}
\label{eq:strangenessratio}
r_s(x,Q^2) \equiv
\frac{s(x,Q^2)+\bar{s}(x,Q^2)}{\bar{u}(x,Q^2)+\bar{d}(x,Q^2)} \, .
\end{equation}
Other strange-sensitive processes agree qualitatively with the
constraints on $r_s$ provided by the neutrino DIS data, such as W
production in association with charm quarks~\cite{Stirling:2012vh} from
CMS~\cite{Chatrchyan:2013uja,CMS-PAS-SMP-18-013,Sirunyan:2018hde} and
ATLAS 7 TeV~\cite{Aad:2014xca}, and semi-inclusive deep-inelastic
scattering (SIDIS)~\cite{Airapetian:2013zaw,Borsa:2017vwy,Sato:2019yez}.
However, the  ATLAS measurements of the leptonic rapidity distributions
in inclusive W and Z production at 7
TeV~\cite{Aad:2012sb,Aaboud:2016btc} exhibit instead a strong preference
for a symmetric strange sea with $r_s\simeq 1$.
One should point out that general considerations based on perturbative
DGLAP evolution imply that $r_s \to 1$ at large $Q$ and small-$x$, but
at low $Q$ and medium/large-$x$ the value of $r_s$ is dictated by
non-perturbative dynamics.

\begin{figure}[ht]
\begin{center}
  \includegraphics[width=0.99\textwidth]{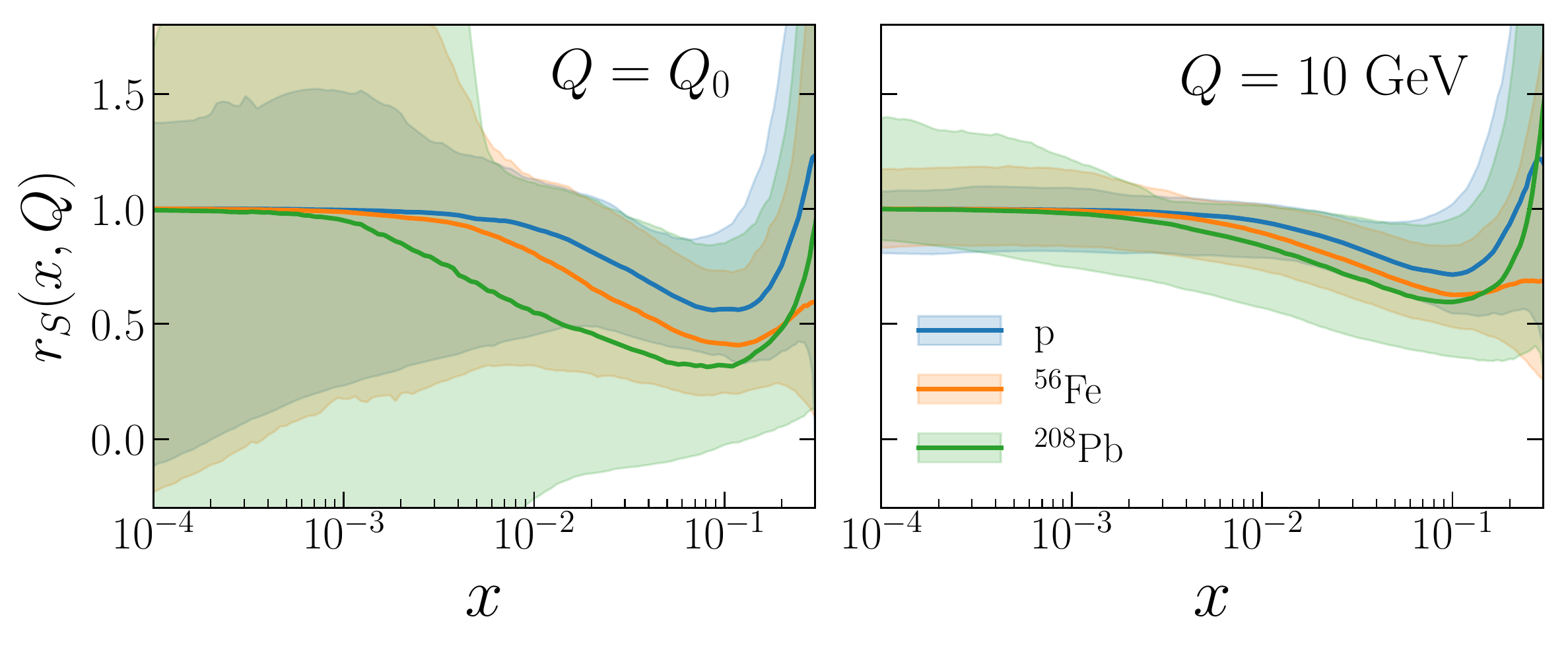}
 \end{center}
\vspace{-0.75cm}
\caption{The strangeness ratio $r_s(x,Q^2)$, defined in
  Eq.~(\ref{eq:strangenessratio}), in \texttt{nNNPDF2.0} comparing the results
  for the free-proton baseline, $^{56}$Fe, and $^{208}$Pb at both the
  input parametrisation scale $Q_0=1$ GeV (left) and at a higher scale
  $Q=10$ GeV (right plot).
  \label{fig:R_S}
}
\end{figure}

As was motivated in Ref.~\cite{Ball:2018twp}, it is important to
carefully assess the nuclear uncertainties associated to the nuclear
strangeness, given that these will potentially affect the determination
of the proton strangeness from global fits based on neutrino data.
 We display in Fig.~\ref{fig:R_S} the strangeness ratio $r_s(x,Q^2)$
  defined by Eq.~(\ref{eq:strangenessratio}) obtained with our \texttt{nNNPDF2.0}
  result for $^1$p, $^{56}$Fe, and $^{208}$Pb at both the input
  parameterisation scale $Q_0=1$ GeV and at a higher scale of $Q=10$
  GeV.

From the comparison in Fig.~\ref{fig:R_S} we find that at the input
parameterisation scale $r_s$ is particularly suppressed in the case of
lead, where the central value of \texttt{nNNPDF2.0} satisfies $r_s < 0.5$ for $5
\cdot 10^{-3} \lsim x \lsim 0.2$.
A similar preference for a suppressed strange sea, albeit less
pronounced, can be seen in iron nuclei.
In any case, the nPDF uncertainties affecting this ratio are rather large,
in particular for the heavier nuclei.
The fact that for $x \lsim 10^{-3}$ one obtains $r_s \simeq 1$ for all
three nuclei is a consequence of the parameterisation preprocessing,
whose ranges are chosen to ensure that in the small-$x$ extrapolation
region all quark and antiquark flavours behave in the same way (see
Figs.~\ref{fig:beta_exp} and~\ref{fig:alpha_exp}).
Once DGLAP evolution takes place, $r_s$ tends to become closer to unity
across a wider range in $x$, but even at the higher scale a suppressed
strangeness for $x\gsim 0.01$ is preferred for both iron and lead.

The results in Fig.~\ref{fig:R_S} suggest that including neutrino CC
structure functions such as CHORUS and NuTeV in proton PDF fits without
accounting for nuclear uncertainties might not be a justified
approximation, given the current precision that modern fits achieve.
It will be interesting nonetheless to determine the impact on the global NNPDF
proton PDF fits when \texttt{nNNPDF2.0}  is used to
account for nuclear uncertainties using the procedure outlined in
Ref.~\cite{Ball:2018twp}. 

\subsection{Comparison with \texttt{EPPS16} and \texttt{nCTEQ15}}
\begin{figure}[ht]
\begin{center}
  \includegraphics[width=0.95\textwidth]{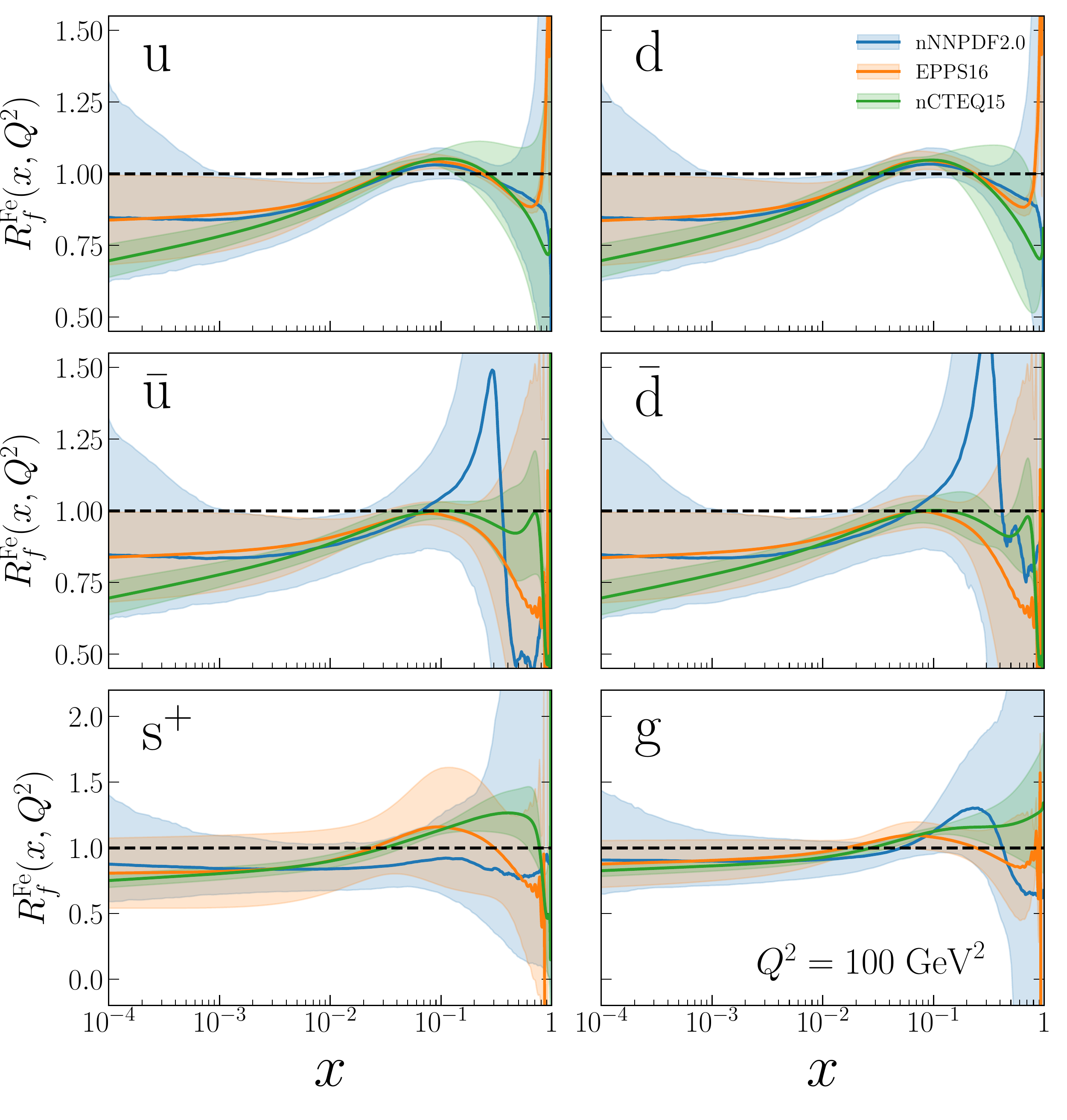}
 \end{center}
\vspace{-0.9cm}
\caption{The nuclear PDF modification ratios for iron, $R_{\rm
  Fe}(x,Q^2)$, as a function of $x$ at $Q=10$ GeV for \texttt{nNNPDF2.0},
  \texttt{EPPS16}, and \texttt{nCTEQ15}.
  From top to bottom we show the up and down quarks, the corresponding
  antiquarks, the total strangeness, and the gluon.
  The bands correspond to the 90\% CL uncertainties, and each nPDF set
  is normalised to its corresponding free-proton baseline as indicated
  by Eq.~(\ref{eq:nuclearratioRv3}).
  \label{fig:R_Fe}
}
\end{figure}

\begin{figure}[ht]
\begin{center}
  \includegraphics[width=0.95\textwidth]{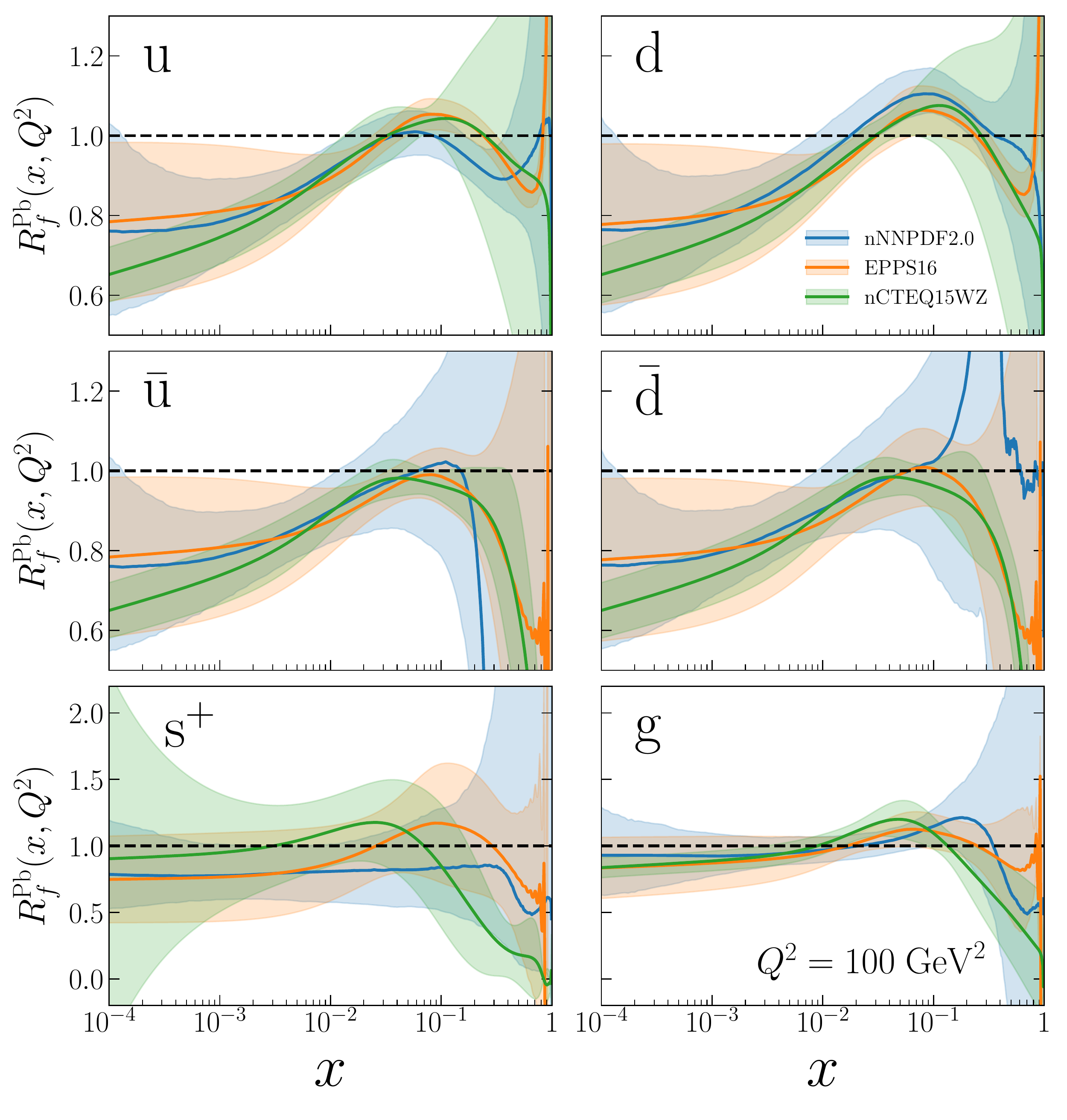}
 \end{center}
\vspace{-0.9cm}
\caption{Same as Fig.~\ref{fig:R_Fe} in the case of lead nuclei, $R_{\rm
Pb}(x,Q^2)$.
  \label{fig:R_Pb}
}
\end{figure}

In Figs.~\ref{fig:R_Fe} and~\ref{fig:R_Pb}, we display the nuclear PDF
modification ratios  at $Q=10$
GeV for iron, $R_f^{\rm Fe}(x,Q^2)$, and lead, $R_f^{\rm
Pb}(x,Q^2)$, for \texttt{nNNPDF2.0}, \texttt{EPPS16}, and \texttt{nCTEQ15}, 
each normalised to the corresponding free-proton baseline.
  As in Fig.~\ref{fig:R_A}, we show the up and down quarks, the
  corresponding antiquarks, the total strangeness, and the gluon.
  The bands again correspond to the 90\% CL uncertainties constructed
  using the appropriate prescription for each set.
  This means that for the other analyses, the error is computed by adding in
  quadrature the differences in value between the $N_{\rm eg}$
  eigenvectors of the Hessian set and the best fit result.\footnote{When
  computing PDF ratios with \texttt{EPPS16} and \texttt{nCTEQ15} we neglect 
  proton PDF uncertainties, since adding these errors in quadrature is likely to
  represent an overestimate given the missing mutual correlations.}

Beginning with the nPDF comparison for iron nuclei, we find that there
is good agreement between the results of \texttt{nNNPDF2.0} and \texttt{EPPS16} both in
terms of central values and of the nPDF uncertainties in the range of
$x$ for which experimental data is available.
Our result also agrees with the \texttt{nCTEQ} parameterisation 
within uncertainties, although the errors for \texttt{nCTEQ} are much smaller in the
low-$x$ region, displaying more significant shadowing effects.
Such differences in uncertainties between \texttt{nCTEQ} and the other analyses 
may be attributed to the choice of tolerance criteria in Hessian error 
propagation and/or the absence of proton PDF uncertainties.
Note also that the \texttt{nCTEQ15} analysis does not include any 
charged-current DIS or LHC data. The quark flavour separation 
is instead obtained from Drell-Yan di-lepton data at the Tevatron and 
inclusive pion production data at RHIC.
Our result displays more visible differences with the other analyses
in the small- and large-$x$ extrapolation regions, where our uncertainties
are larger as a result of a more flexible parameterisation.
The Fermi-motion-like growth of $R_u^{\rm Fe}$ and
$R_d^{\rm Fe}$ at very large $x$, which is built into the \texttt{EPPS16}
parameterisation, is absent in the \texttt{nNNPDF2.0} and \texttt{nCTEQ15} results.
There instead one finds a suppression compared to the free-proton case,
especially in the case of $R_d^{\rm Fe}$.
However, our result remains fully consistent with Fermi-motion
behaviour within the uncertainties.
As expected, the observed pattern of nuclear modifications is very
similar between up and down quarks and between the corresponding
antiquarks due to iron being nearly isoscalar.

Considering the behaviour of the sea quarks, \texttt{nNNPDF2.0} and \texttt{EPPS16} agree
well in terms of central values and uncertainties in the shadowing
region, $x\lsim 0.05$.
Here the \texttt{nCTEQ} result again shows more prominent shadowing
effects with smaller uncertainties.
There are more significant differences at large-$x$, where the
qualitative behaviour between the nPDF sets is the opposite:
\texttt{nNNPDF2.0} favours an enhancement compared to the free-proton baseline,
while \texttt{EPPS16} and \texttt{nCTEQ15} prefer instead a suppression.
In any case, the differences are well within the large uncertainty
bands, and additional data is needed to be able to ascertain the correct
behaviour in this region.
Note that at large-$x$ the free-proton baseline antiquarks are also
affected by large errors, complicating the interpretation of
$R_{\bar{u}}^{\rm Fe}$ and $R_{\bar{d}}^{\rm Fe}$.

Turning to the nuclear modification of the total strangeness, in
\texttt{nNNPDF2.0} we find a suppression of $\sim 20$\% compared to the proton
baseline in the relevant range of $x$.
This is consistent with studies of the interplay between the NuTeV
di-muon and the ATLAS W,Z 2011 data in proton global analyses, where
the latter data set largely suppress strangeness in contrast to the
former.
Such behaviour is not reported by \texttt{EPPS16}, which exhibits much larger nPDF
uncertainties that are likely due to the absence of the strange-sensitive
NuTeV cross-sections in their analysis. Furthermore, the ATLAS W,Z
2011 distributions are missing from the CT14 proton baseline used by
\texttt{EPPS16} (although these data have been accounted for in the recent CT18
release~\cite{Hou:2019efy}).

Interestingly, the \texttt{nCTEQ15} result also shows strange quark
suppression at low-$x$ 
with significantly smaller uncertainties, despite not including 
any strange-sensitive observables.
Furthermore, their analysis exhibits a clear anti-shadowing behaviour
in the intermediate $x$ region, in better agreement with \texttt{EPPS16}
central values.
In any case, all distributions remain in agreement within uncertainties and
additional strange-sensitive observables are needed to clarify the nuclear
modifications of the strange quark distribution.

Finally, concerning the gluon PDF we find from this comparison that in
the \texttt{nNNPDF2.0} fit there is little evidence for nuclear shadowing, with
$R_g^{\rm Fe}\simeq 1$ in the region $x\le 0.05$.
We also find that the nPDF uncertainties on the gluon are larger
compared to \texttt{EPPS16} by roughly a factor of two.
At larger values of $x$, the uncertainties increase significantly and
\texttt{nNNPDF2.0} prefers a suppressed central value, unlike 
the \texttt{nCTEQ15} result.
We note that none of the analyses include direct constraints on the
large-$x$ nuclear gluons, hence the sizeable nPDF uncertainties,
although available data on dijet and photon production could improve
this situation.

In the corresponding comparison for lead nuclei, displayed in
Fig.~\ref{fig:R_Pb}, one observes a number of similarities and
differences with respect to the nPDFs of iron.
Concerning the up and down quarks, we find our \texttt{nNNPDF2.0} result provides
significant evidence for shadowing at small-$x$.
For instance, at $x\simeq 5\times 10^{-3}$ we obtain $R_u^{\rm Pb} \ne 1$ at the
four-sigma level or higher.
Interestingly, nPDF uncertainties in the shadowing region are up to a
factor of two smaller in \texttt{nNNPDF2.0} than in \texttt{EPPS16}.
Here we illustrate the ratio comparison with the updated 
 nCTEQ15WZ analysis~\cite{Kusina:2020lyz}, 
 which included constraints from W and Z boson 
 production from proton-lead collisions at the LHC. In this case, the uncertainties
 are more comparable to the \texttt{EPPS16} result, but still smaller than \texttt{nNNPDF2.0}
 at low $x$.

While in all cases anti-shadowing at $x\simeq 0.1$ is observed, the
larger $x$ qualitative behaviour is different between the analyses,
with \texttt{EPPS16} finding the (built-in) EMC suppression followed by
Fermi-motion rise while in \texttt{nNNPDF2.0} and nCTEQ15WZ
the pattern of nuclear modifications is different.
In any case, the agreement between the central values of \texttt{nNNPDF2.0} and
the other analyses for $R_u^{\rm Pb}$ and $R_d^{\rm Pb}$ in the region of $x\lsim
0.3$ is quite remarkable given the very different methodologies employed in
each study.

Concerning the nuclear modifications of the sea quarks in lead nuclei,
one finds a similar qualitative behaviour as in the case of iron.
For $x\lsim 0.1$ there is good agreement between the central values of
$R_{\bar{u}}^{\rm Pb}$ and $R_{\bar{d}}^{\rm Pb}$ between \texttt{EPPS16} and
\texttt{nNNPDF2.0}, with the latter exhibiting smaller uncertainties.
The nCTEQ15WZ result again displays more significant sea 
quark modifications at low-$x$, but resides in agreement with our result
within the 90\% error.
The sets are also notably different for $x\gsim 0.1$, where
\texttt{EPPS16} and nCTEQ15WZ
predict a EMC-like suppression common to $\bar{u}$ and $\bar{d}$
while \texttt{nNNPDF2.0} favours a large suppression for $\bar{u}$ but an
enhancement for $\bar{d}$.
However, the large nPDF uncertainties in this region prevent any
definitive conclusions, as the two fits are fully consistent
within the 90\% CL bands.
One possible source for the differences could be in the respective
free-proton counterparts, where large-$x$ antiquarks are poorly known.
For the total strangeness, the behaviour of $R_{s^+}^{\rm Pb}$ is similar
to that of iron, where \texttt{nNNPDF2.0} predicts a suppression more or
less independent of $x$, with rather larger uncertainties for \texttt{EPPS16}
compared to our \texttt{nNNPDF2.0} result due to the missing constraints from the
NuTeV di-muon cross-sections.
On the other hand, the nCTEQ15WZ distribution displays much 
different behaviour than the ratio for iron nuclei constructed with the preceding
\texttt{nCTEQ} results, since the former analysis contains some 
strange sensitivity from the W/Z LHC data. 
In this case, the strange uncertainties are comparable in 
size to our result but shows much less suppression. Moreover, the nCTEQ15WZ
shows a strong EMC-like effect in the region of $x\sim0.2-0.3$.

Finally, regarding the nuclear modifications of the lead gluon PDF
illustrated in the bottom right panel of Fig.~\ref{fig:R_Pb}, we again
find that $R_g^{\rm Pb}$ agrees with unity across all relevant $x$.
Here the initial-scale differences are washed out partially by DGLAP
evolution, but clearly the shadowing in the nuclear gluons is less
apparent than for the quarks, aside from the \texttt{nCTEQ15}
result which clearly shows shadowing behaviour.
Although the nPDF uncertainties increase at large $x$ due to the lack of
direct constraints, the qualitative behaviour of $R_g^{\rm Pb}$ differs
between the various PDF determinations.

\subsection{The momentum and valence integrals in nuclei}
\label{s2:nNNPDF20_sumrules}

As was discussed in Sect.~\ref{s2:nNNPDF20_updates}, we impose three sum
rules in the \texttt{nNNPDF2.0} determination, namely the momentum sum rule,
Eq.~(\ref{eq:MSR}), and the two valence sum rules,
Eqs..~(\ref{eq:valencesr4}) and~(\ref{eq:valencesr5}).
These constraints are satisfied by adjusting the overall prefactors
$B_f$ in  Eq.~(\ref{eq:param2}) for the gluon $g$, the total valence
$V$, and the valence triplet $V_3$ distributions, respectively.
Furthermore, they are independently imposed for each value of $A$ for
which there is available experimental data.

Here we investigate the role played by these sum rules in the global
nPDF determination.
In particular, we address whether or not the physical requirements of
energy and valence quark number conservation are satisfied by the
phenomenological fit to experimental data (within uncertainties) when
the sum rules are not explicitly imposed.
Recently, theoretical arguments have been put forward that the momentum
sum rule for nucleons in nuclei might not hold~\cite{Brodsky:2019jla}.
Motivated in this respect, we have carried out a similar study to the
one presented in Ref.~\cite{Ball:2011uy}, where global proton PDF fits without
imposing the momentum sum rule were performed.
In the proton case, while the LO prediction for the momentum integral
was expected to be far from the QCD expectation, both the NLO
and NNLO fits exhibited remarkable agreement at the $\simeq 1\%$ level~\cite{Ball:2011uy}.

We have therefore produced two variants of the \texttt{nNNPDF2.0} analysis, each
based on $N_{\rm rep}=250$ replicas, where either the momentum sum rule
or the total valence sum rule is not imposed.
Afterwards, we evaluate in each case the corresponding momentum and
total valence integrals, defined as
\begin{equation}
I_{\rm M}(A) \equiv \int_0^1 dx \,x \left(\Sigma^{(p/A)}(x,Q_0) +
g^{(p/A)}(x,Q_0)\right) \label{eq:MSRintegral} \\
I_{\rm V}(A) \equiv \int_0^1 dx~ V^{(p/A)}(x,Q_0) \label{eq:valencesr4integral}
\end{equation}
and assess whether or not they are in agreement with the QCD
expectations, namely $I_{\rm M}(A)=1$ and $I_{V}(A)=3$
respectively.
One should note that the momentum and valence sum rules are already
satisfied at the level of the proton boundary condition, and thus some
constraints are expected to be propagated to the lighter nuclei.
However, the analysis of Ref.~\cite{Ball:2011uy} demonstrates that
results would be largely unchanged if the momentum and valence sum rules
would have been excluded also from the free-proton baseline.

\begin{figure}[ht]
\begin{center}
  \includegraphics[width=0.9\textwidth]{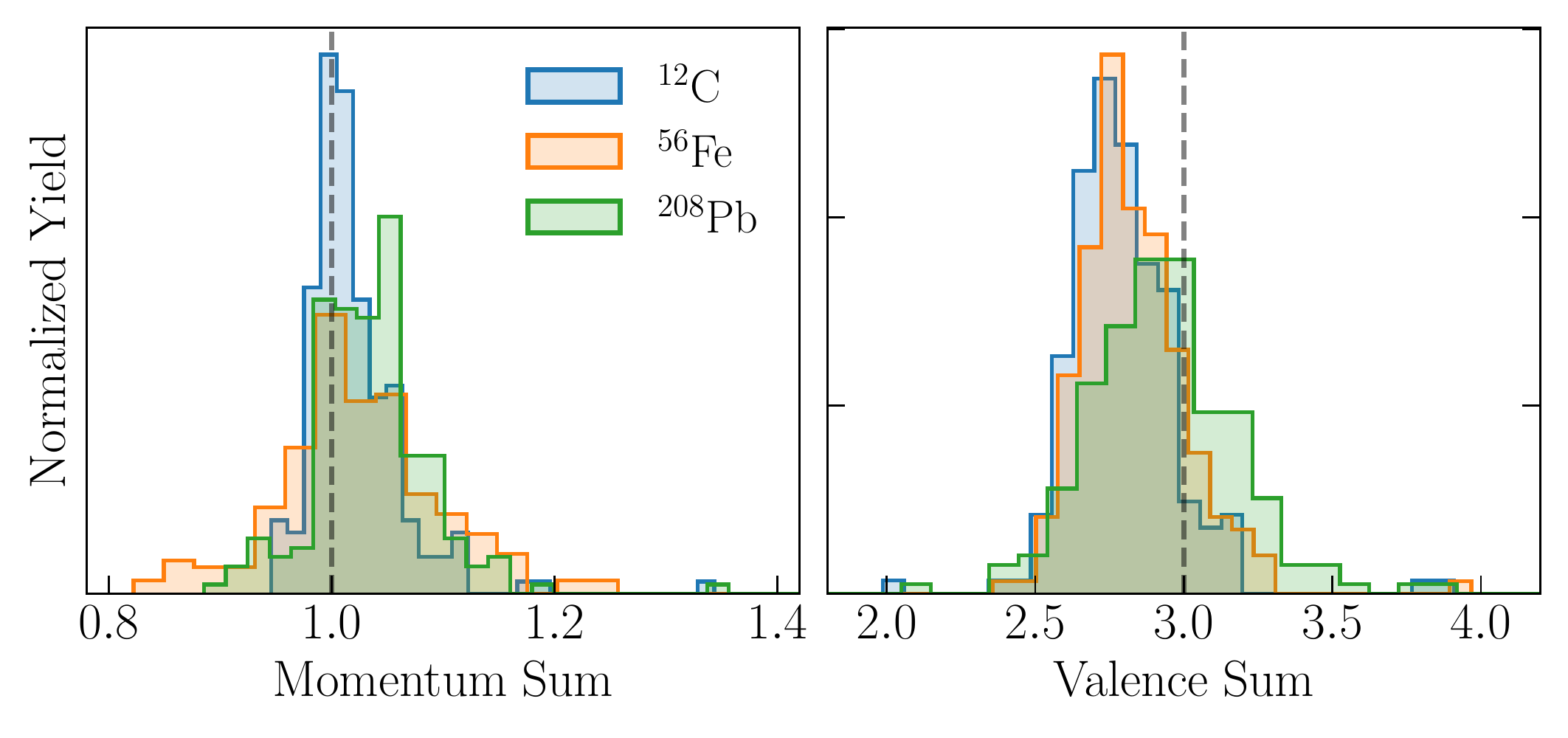}
 \end{center}
\vspace{-0.67cm}
\caption{\small The distribution of the momentum (left) and total
  valence (right panel) integrals, Eqs..~(\ref{eq:MSRintegral})
  and~(\ref{eq:valencesr4integral}) respectively, in the variants of the
  \texttt{nNNPDF2.0} determination where the corresponding sum rules have not
  been explicitly imposed.
  We show the relative frequency of the momentum and valence integral
 for three representative nuclei: $^{12}$C, $^{56}$Fe, and
 $^{208}$Pb.
 The dashed vertical line indicates the corresponding the QCD expectations,
 $I_{\rm M}(A)=1$ and $I_{\rm V}(A)=3$.
 The associated 90\% CL ranges are reported in
 Table~\ref{tab:momintegral}.}
\label{fig:MomIntegral_noMSR}
\end{figure}

In Fig.~\ref{fig:MomIntegral_noMSR} we display the distribution of the
momentum and valence integrals, Eqs..~(\ref{eq:MSRintegral})
and~(\ref{eq:valencesr4integral}), respectively, in the variants of the
\texttt{nNNPDF2.0} fit where the corresponding sum rules are not being explicitly
imposed.
  We show the relative frequency of the momentum and valence integral
 for three representative nuclei: $^{12}$C, $^{56}$Fe, and $^{208}$Pb.
 The dashed vertical line in
 Fig.~\ref{fig:MomIntegral_noMSR} indicates the QCD expectations for
 $I_{\rm M}(A)$ and $I_{\rm V}(A)$.
 The corresponding values for the 90\% confidence level intervals for
 each of these two integrals for the relevant values of $A$, as well as
 for the free-proton baseline $A=1$, can be found in
 Table~\ref{tab:momintegral}.
 
\begin{table}
  \renewcommand{\arraystretch}{1.35}
  \centering
  \begin{tabular}{|c|c|c|}
    \toprule
    $\qquad\qquad A\qquad \qquad$  &  $ \qquad\qquad I_{\rm M}(A)
    \qquad\qquad$  & $ \qquad\qquad I_{\rm V}(A) \qquad\qquad$  \\
    \midrule
    1  & $\lc 0.99, 1.06 \rc$ & $\lc 2.53, 3.12 \rc$   \\
    12  & $\lc 0.97, 1.10 \rc$ &  $\lc 2.56, 3.11 \rc$  \\
    56  & $\lc 0.90, 1.16 \rc$ &  $\lc 2.58, 3.16 \rc$  \\
    208  & $\lc 0.94, 1.12 \rc$ & $\lc 2.54, 3.34 \rc$   \\
    \bottomrule
  \end{tabular}
  \vspace{0.3cm}
  \caption{\label{tab:momintegral}
 The 90\% CL ranges for the momentum and valence integrals,
  Eqs..~(\ref{eq:MSRintegral}) and~(\ref{eq:valencesr4integral}),
  in the variants of the \texttt{nNNPDF2.0} fits whether either one
  or the other sum rule is not imposed. }
\end{table}

From the results presented in Fig.~\ref{fig:MomIntegral_noMSR} and
Table~\ref{tab:momintegral} one finds that the momentum integral is in
agreement with the QCD expectation, $I_{\rm M}(A)=1$, within
uncertainties for all nuclei.
In the case of $^{12}$C for example, one finds that $0.97 \lsim I_{\rm M}
\lsim 1.10$ at the 90\% confidence level, with somewhat larger
uncertainties for the heavier nuclei.
Even for lead, where the proton boundary condition has little effect,
the median of the distribution is reasonably close to the QCD
expectation.
The uncertainties on $I_{\rm M}$ are larger in the nuclear
PDF analysis than the $\simeq 1\%$ error
found in the proton case~\cite{Ball:2011uy}, as expected since the experimental data for
nuclear collisions is far less abundant and further distributed between
different nuclei.
Nevertheless, the overall consistency
with the QCD expectations
is quite compelling.
Note also that here the proton boundary condition is imposed
only for $x\ge 10^{-3}$, and therefore our prediction for $I_{\rm M}(A=1)$
is expected to be less accurate compared to the 
proton global analysis case.

\begin{figure}[ht]
\begin{center}
  \includegraphics[width=0.95\textwidth]{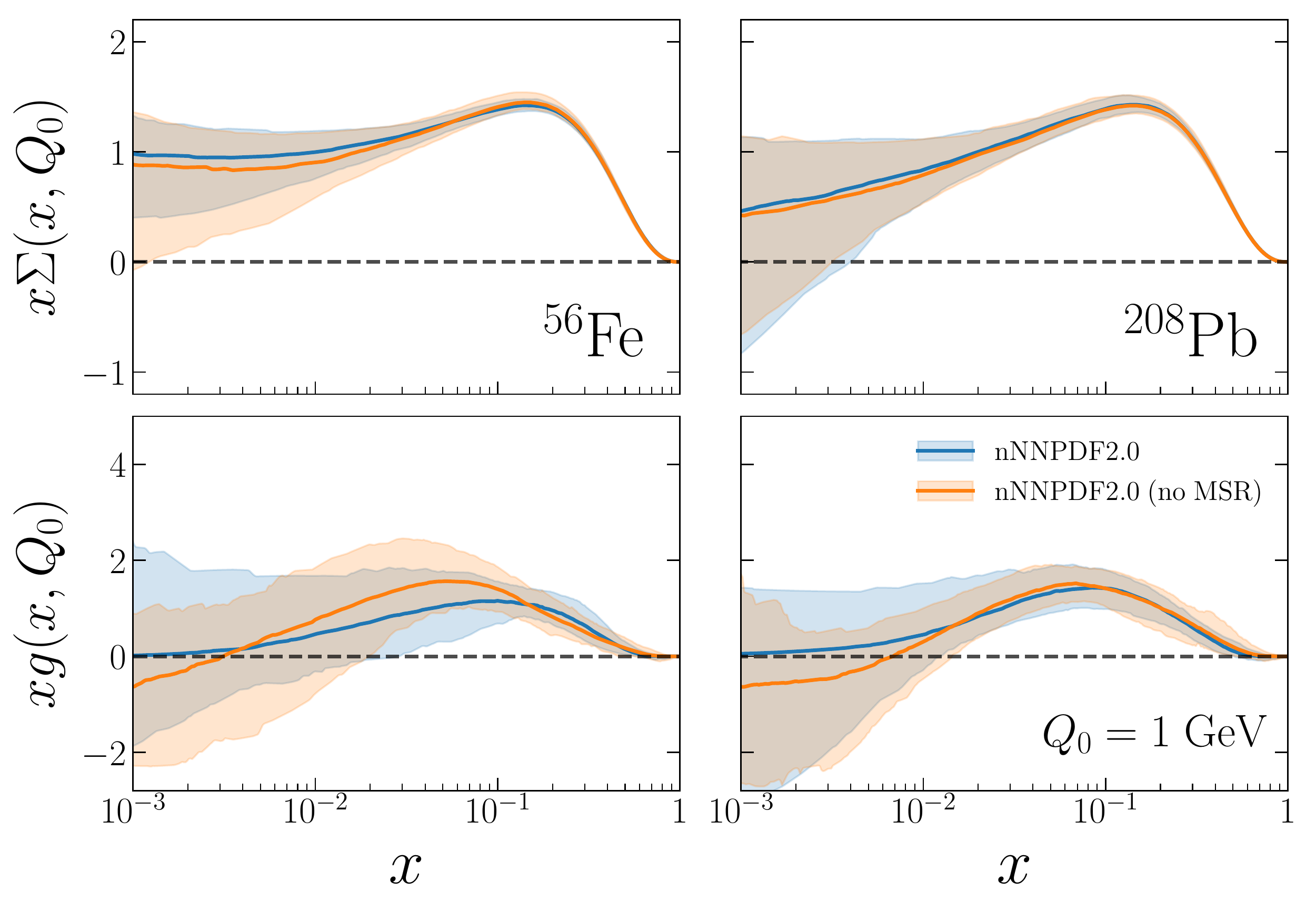}
 \end{center}
\vspace{-0.75cm}
\caption{\small Comparison of the baseline \texttt{nNNPDF2.0} fit at $Q_0=1$ GeV
  with the variant in which the momentum sum rule is not being imposed.
  We show the total quark singlet (left) and the gluon (right) for
  $^{56}$Fe (upper) and $^{208}$Pb (lower panels).}
\label{fig:PDFcomp_noMSR}
\end{figure}

The result that the momentum integral agrees with the theoretical
predictions for all nuclei is a non-trivial validation of the global
nuclear PDF analysis framework based on the QCD factorisation hypothesis.
It further demonstrates the robustness of our fitting methodology, in
that the resulting nPDFs are reasonably stable regardless of whether or
not the momentum sum rule is imposed during the fit.
To illustrate better this latter point, in Fig.~\ref{fig:PDFcomp_noMSR} we
provide a comparison between the baseline \texttt{nNNPDF2.0} fit at $Q_0=1$ GeV
with the variant in which the momentum sum rule is not being imposed.
  We show the total quark singlet and the gluon for both $^{56}$Fe  and
  $^{208}$Pb.
  Recall that the momentum sum rule is used to fix the overall gluon
  normalisation in Eq.~(\ref{eq:param2}).
  In the case of lead, where the experimental constraints are relatively
  abundant, we find that both the singlet and the gluon are reasonably
  similar irrespective of whether or not the momentum sum rule is
  imposed.
  The momentum sum rule plays a larger role in iron, especially in
  reducing the gluon nPDF uncertainties, but interestingly the central
  value of the all distributions is quite stable when comparing the two
  fits.
 This stability is consistent with the results reported in
 Fig.~\ref{fig:MomIntegral_noMSR}.

The main conclusions are qualitatively similar for a fit in which the
total valence sum rule has not been imposed.
Results of this fit are displayed in the right panel of
Fig.~\ref{fig:MomIntegral_noMSR}, where the normalised frequency of the
total valence integral are shown for the same three nuclei discussed
previously.
 The corresponding 90\% CL intervals are also reported in
 Table~\ref{tab:momintegral}.
Similar to the momentum sum results, we find that for the valence
integral the fit results agree with the QCD expectations within
uncertainties.
The preferred value of the valence integral (median) turns out to be
$I_{\rm V} \simeq 2.8$ irrespective of $A$.
This implies that even when Eq.~(\ref{eq:valencesr4integral}) is not
imposed explicitly, the experimental measurements favour the QCD
prediction within 5\% for all values of $A$ relevant for the present
study.
The value for the $\chi^2$ per degree of freedom
is somewhat smaller (0.953) than in the baseline (0.976).
We have also verified that, in a similar way as in
Fig.~\ref{fig:PDFcomp_noMSR}, the resulting nPDFs are reasonably stable
regardless of whether or not the valence sum rule is imposed.

Putting together the results of these two exercises, one can conclude
that the fit results are relatively stable in the nNNPDF framework even
in the absence of the sum rules, consistent with the fact that
experimental data and the QCD expectations based on the factorisation
theorem are in agreement with each other for hard-scattering collisions
involving heavy nuclei.

  \chapter{Towards nuclear PDFs with NNLO QCD corrections}
\label{chap:nNNPDF30}
\vspace{-1cm}
\begin{center}
  \begin{minipage}{1.\textwidth}
      \begin{center}
          \textit{This chapter is based on my work in progress towards the} \texttt{nNNPDF3.0} \textit{release~\mycite{AbdulKhalek:2021xxx1}}.
      \end{center}
  \end{minipage}
  \end{center}

\myparagraph{Motivation} In this chapter, I discuss the results achieved to date as part of the upcoming release \texttt{nNNPDF3.0}. On top of deep-inelastic scattering and electroweak gauge bosons production considered in \texttt{nNNPDF2.0}, the new release will include a number of new processes that constrain the nuclear gluons: single jet and dijet cross-sections from ATLAS and CMS, direct photon production from ATLAS, and charm production by the LHC experiments. Furthermore, NNLO perturbative QCD calculations will be exploited for all processes included in the fit. The proton baseline, which nNNPDF3.0 reduces to in the limit A=1, is a variant of the \texttt{NNPDF3.1} proton PDFs, which includes the dijet production data discussed in Sect.~\ref{s1:PDF_NNLO_jet} constraining the gluon PDF.

\myparagraph{Outline} In Sect.~\ref{s1:nNNPDF30_framework}, I discuss the \texttt{nNNPDF3.0} framework that shares all the aspects of the \texttt{nNNPDF2.0} one but extend it to include new experimental data, K-factors to include the NNLO QCD corrections and an updated proton baseline.
In Sect.~\ref{s1:nNNPDF30_results}, I present the results of including two of the CMS data sets at NNLO in QCD, the dijet production in pPb collisions as well as the Z-boson production, both at 5 TeV. 

\section{\texttt{nNNPDF3.0} Framework} \label{s1:nNNPDF30_framework}
The nPDFs parameterisation and constraints in \texttt{nNNPDF3.0} are identical to those of \texttt{nNNPDF2.0} (see Sect.~\ref{s1:nNNPDF20_framework}).
However, in this new framework, we focus on including the QCD NNLO corrections to all the processes already included in \texttt{nNNPDF2.0} and the new experimental data that will be considered.

In Sect.~\ref{s2:nNNPDF30_data}, I outline the different new data sets focusing on the ones that already had been already analysed in Sect.~\ref{s1:nNNPDF30_results}.
In Sect.~\ref{s2:nNNPDF30_updates}, I discuss the new proton baseline that includes all the dijet data considered in Sect.~\ref{s1:PDF_NNLO_jet}, namely the 7 and 8 TeV from ATLAS and CMS at both NLO and NNLO. I also investigate the inclusion of the 5 TeV absolute pp dijet spectra in light of the K-factors provided for this data set.


\subsection{Experimental data}
\label{s2:nNNPDF30_data}

In this section, I provide details on the experimental measurements 
used as input for the \texttt{nNNPDF3.0} determination.
An emphasis is made in particular 
on the data sets that are already analysed.
Common to the previous \texttt{nNNPDF2.0} analysis are the nuclear 
NC DIS measurements listed in Table~\ref{tab:nNNPDF10_data}, the CC neutrino DIS reduced cross-sections on nuclear targets as well as the leptonic rapidity distributions in electroweak gauge boson production from proton-lead collisions at the LHC listed in Table~\ref{tab:nNNPDF20_data}.

The new foreseen experimental data includes measurement of prompt photon production in 8 TeV pPb collisions~\cite{Aaboud:2019tab} and inclusive jet production in 5 TeV pPb collisions~\cite{Aad_2015} with ATLAS, dijet production in 5 TeV pPb collision with CMS~\cite{Sirunyan_2018}, Z-boson production in 8 TeV pPb collisions~\cite{Acharya:2020puh} with ALICE. 
In addition, the plan is to include either by means of an inference or reweighting (see Sect.~\ref{s2:monte_carlo}), measurements of heavy-flavour mesons production at the LHC. These include prompt $D^0$ meson production in 5 TeV pPb collisions~\cite{Aaij:2017gcy} and prompt/non-prompt $J/\Psi$ production in 8 TeV pPb collisions~\cite{Aaij:2017cqq} with LHCb, quarkonium production in 5 TeV pPb collisions~\cite{Aaboud:2017cif} with ATLAS and $D$-meson production in 5 TeV pPb collisions~\cite{Adam:2016ich} with ALICE.
Finally, we also intend to explore the large-$x$ high-precision data of inclusive nuclear DIS cross sections~\cite{Schmookler:2019nvf} measured by the CLAS/JLab collaboration in light of potential higher-twist contributions.
The majority of the data mentioned is expected to provide unprecedented constraints on the lead gluon down to low values of $x\sim 10^{-5}$ and potentially less for some data processes~\cite{Kusina:2017gkz,Eskola:2019dui}. 

The focus in this chapter will be on both dijet production in 5 TeV pPb collision~\cite{Sirunyan_2018} and Z-boson production in 5 TeV pPb collision~\cite{Khachatryan:2015pzs} with CMS. The Z-boson data is discussed in Sect.~\ref{s2:nNNPDF20_data}. The dijet data is provided in terms of pseudorapidity distributions of dijets ($\eta_{\text{dijet}} = (\eta_1+\eta_2)/2$) as functions of their average transverse momentum ($p_{T,\text{dijet}}^{\text{avg}}=(p_{T,1}+p_{T,2})/2 \sim Q$) for both pp ($\mathcal{L}_{\text{int}}=27.4\text{ pb}^{-1}$) and pPb ($\mathcal{L}_{\text{int}}=35\text{ nb}^{-1}$) collisions. Three observables are provided for this data set: the absolute pp and pPb spectra as well as their ratio $\frac{\text{pPb}}{\text{pp}}$. The spectra are defined as the number of dijet per bin of $\eta_{\text{dijet}}$ and $p_{T,\text{dijet}}^{\text{avg}}$, normalised by the pseudorapidity-integrated number of dijets in the associated $p_{T,\text{dijet}}^{\text{avg}}$ bin as follows:
\begin{equation}
    {\displaystyle d\left(\frac{1}{N^{\text{col}}_{\text{dijet}}}\frac{dN^{\text{col}}_{\text{dijet}}}{d\eta_{\text{dijet}}}\right)\bigg/dp_{T,\text{dijet}}^{\text{avg}}} \qquad \text{col}=\text{pp, pPb}
\end{equation}
The fiducial cuts~\footnote{We note that the minimal $p_T$ for the leading jet in Ref.~\cite{Sirunyan_2018} is incorrect and the correct cuts are the ones mentioned above.} associated with this measurement are a minimal $p_T$ of $30$ and $20$ GeV for the leading and subleading jets respectively, a distance parameter $R=0.3$ for the anti-$k_T$ recombination algorithm as well as a $2\pi/3$ absolute difference of azimuthal angles between the leading and subleading jets. The only uncertainties associated with this data set are a statistical and systematic uncorrelated uncertainties that will be added in quadrature during the inference.

Based on LO kinematics (see the jet production part in Sect.~\ref{s1:QCD_summary}) we can estimate the coverage of this data in terms of the scaling variables $x_{1,2}$ to be:
\begin{equation} \label{eq:dijet_coverage}
    x_{1,2} = \frac{p_{T,\text{dijet}}^{\text{avg}}}{\sqrt{s_{\text{NN}}}}e^{\pm y} \simeq [5\times 10^{-4}, 1]\quad \text{for CMS dijet} \begin{cases} \sqrt{s_{\text{NN}}}&=5020 \text{ GeV}\\ p_{T,\text{dijet}}^{\text{avg}} &\simeq [55,400] \text{ GeV}\\ y &\simeq [-3,3] \end{cases}
\end{equation}

In Sect.~\ref{s1:nNNPDF30_results}, I discuss my completed analysis of dijet production in 5 TeV pPb collisions with CMS~\cite{Sirunyan_2018} both at NLO and NNLO in QCD. This data set is new w.r.t. \texttt{nNNPDF2.0}. I also present the work done to date on the NNLO QCD calculations of the Z-boson production in 5 TeV pPb collisions with CMS~\cite{Khachatryan:2015pzs} that has already been included in \texttt{nNNPDF2.0}. The latter will eventually be followed by the rest of the \texttt{nNNPDF2.0} data sets listed in Table~\ref{tab:nNNPDF20_data}.

\subsection{Updates relative to \texttt{nNNPDF2.0}}
\label{s2:nNNPDF30_updates}
Throughout the previous chapters, I emphasized on the importance the proton PDF baseline has in constraining nPDFs. Moreover, I highlighted on how crucial is for them to be consistent, \text{i.e.} same theoretical settings, no double-counting in the data they're inferred from, etc. In Chapter~\ref{chap:nNNPDF10} and since only nuclear NC DIS were considered in \texttt{nNNPDF1.0}, the proton baseline was chosen to be \texttt{NNPDF3.1}~\cite{Ball:2017nwa} (see Sect.~\ref{s2:nNNPDF10_minimisation}). In Chapter~\ref{chap:nNNPDF20}, nuclear CC DIS and electroweak gauge boson production were added and therefore a variant of \texttt{NNPDF3.1} without CHORUS and NuTeV was determined and used as a baseline for \texttt{nNNPDF2.0} (see Sect.~\ref{s2:nNNPDF20_updates}).

For \texttt{nNNPDF3.0}, I include on top of the \texttt{nNNPDF2.0} proton baseline, the dijet\cite{Aad:2013tea,Chatrchyan:2012bja,Sirunyan:2017skj} data sets
from ATLAS and CMS at $\sqrt{s}=7$ and 8 TeV reviewed in Sect.~\ref{s1:PDF_NNLO_jet} based on Ref.~\mycite{AbdulKhalek:2020jut}. 
In this respect, two new proton baselines are inferred, one with NLO QCD corrections and the other with NNLO ones implemented by means of K-factors (see Sect.~\ref{s2:NNLO_corrections}). I will refer to the new baselines with an asterisk. For instance, \texttt{nNNPDF2.0}$^*$($^1p$) is the new proton baseline and \texttt{nNNPDF2.0}($^1p$) the old one (as defined in Sect.~\ref{s2:nNNPDF20_updates}).
The reason behind the exclusion of the 5 TeV pp dijet spectra~\cite{Sirunyan_2018} in the proton baseline is due to our inability to describe it by means of NNLO K-factors, which is discussed in the following. 

\begin{figure}[ht]
    \begin{center}
    \includegraphics[width=0.99\textwidth]{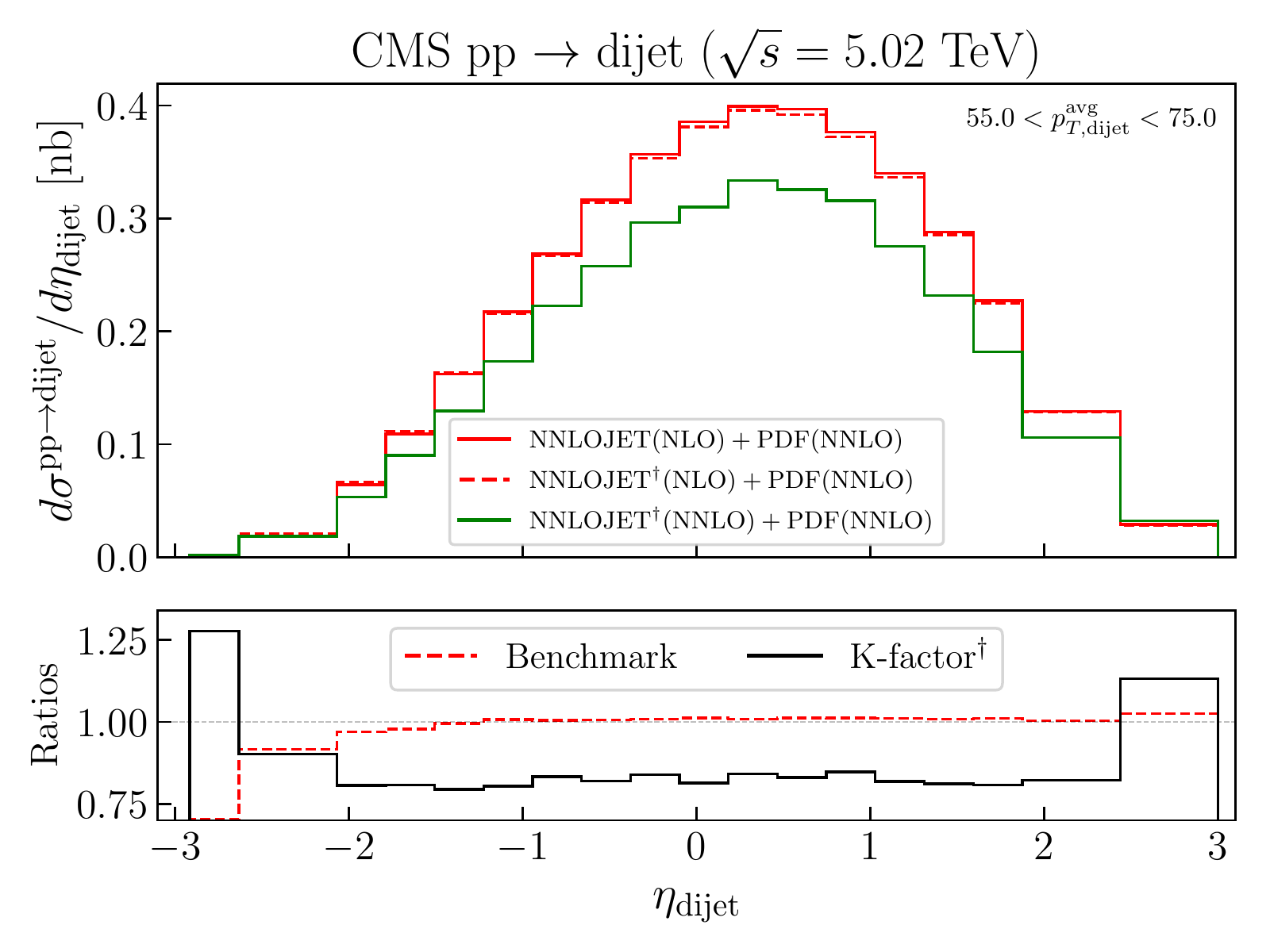}
    \end{center}
    \vspace{-0.8cm}
    \caption{The QCD calculations of the lowest $p_{T,\text{dijet}}^{\text{avg}}$ bin of the 5 TeV pp dijet spectra from the CMS data set.
        The NLO calculations (solid red histogram) and the reference NLO (dashed red histogram) and NNLO (solid green histogram) calculations using \textsc{NNLOJET} in the upper-panel. The ratio of NLO calculation to the reference ones (dashed red histogram) and NNLO QCD K-factors (solid black histogram) Eq.~\ref{eq:kfactor}. The dagger ($\dagger$) refers to any calculations performed by my co-authors in Ref.~\mycite{AbdulKhalek:2020jut}.
    \label{fig:nNNPDF20_CMS_DIJET_kfac}
    }
    \end{figure}

As mentioned in Sect.~\ref{s1:PDF_NNLO_jet}, the inclusion of the 5 TeV pp dijet spectra from CMS in a global PDF fit relies on the NLO and NNLO QCD corrections as computed with \textsc{NNLOJET}~\cite{Gehrmann-DeRidder:2019ibf}. Although fast interpolation grids are possible to produce for NLO matrix elements, it is not the case for the NNLO ones which are included by means of K-factors discussed in Sect.~\ref{s2:NNLO_corrections}.
In Fig.~\ref{fig:nNNPDF20_CMS_DIJET_kfac}, I consider a representative bin in $p_{T,\text{dijet}}^{\text{avg}}$ (the lowest among the five available) of the 5 TeV pp dijet spectra numerator (cross section).
In the upper-panel I compare my NLO calculations (solid red histogram) using \textsc{NNLOJET} to the independent calculations at NLO (dashed red histogram) and NNLO (solid green histogram) performed by my co-authors of Ref.~\mycite{AbdulKhalek:2020jut} and referred to with a dagger ($\dagger$) in the figure. In the lower-panel I validate the benchmarking (dashed red histogram) where I plot the ratio of my calculations at NLO to the independent calculations. The K-factor (solid black histogram) are also presented. All of these calculations are performed using the NNLO PDF set \texttt{NNPDF31\_nnlo\_as\_0118}~\cite{Ball:2017nwa}.

\begin{figure}[ht]
    \begin{center}
    \includegraphics[width=0.99\textwidth]{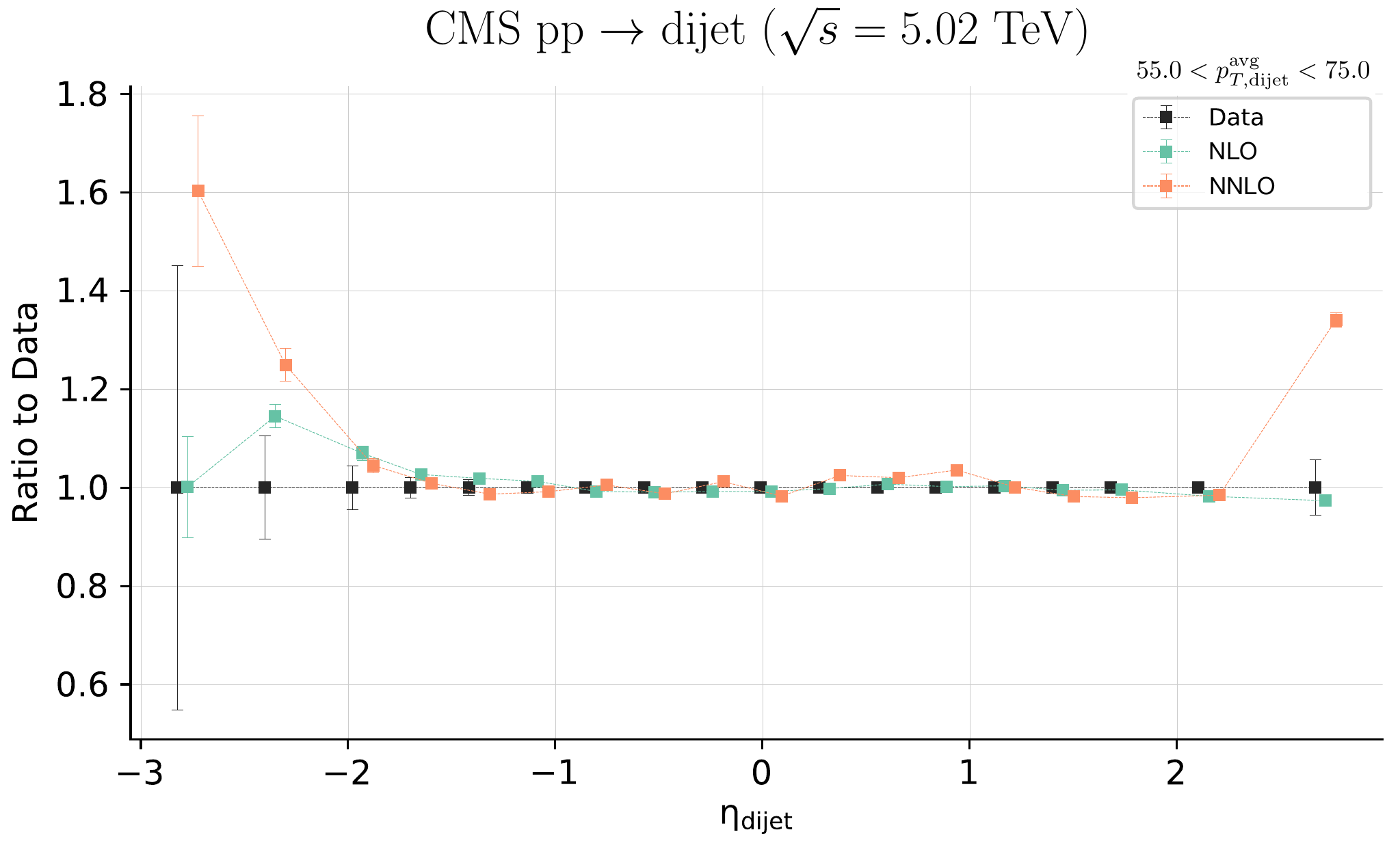}
    \end{center}
    \vspace{-0.8cm}
    \caption{The comparison of \texttt{nNNPDF2.0}$^*$($^1p$) at NLO (green) and at NNLO (orange) with fitted CMS 5 TeV pp dijet spectra to the data (black), all normalised to the data points.
    \label{fig:CMS2JET_pp_datavstheory}
    }
    \end{figure}

The successful benchmark of the NLO calculations enables the computations of the interpolation grids, which in turn allows for a global inference using the 5 TeV pp dijet spectra from CMS both at NLO and NNLO (by means of the K-factors).
Table~\ref{tab:nNNPDF20_newbaseline_chi2_1} describes the fit-quality of the new proton baseline \texttt{nNNPDF2.0}$^*$($^1p$) both at NLO and NNLO. I denote by \textbf{w/}(\textbf{w/o}) when the CMS 5 TeV pp dijet is considered(not considered) on top of the ATLAS and CMS at $\sqrt{s}=7$ and 8 TeV data sets.
The proton baseline fits without the CMS 5 TeV data set follow the same set of conclusions in Sect.~\ref{s1:PDF_NNLO_jet} that remain intact with the exclusion of the CHORUS and NuTeV data sets, as well as a lower initial scale ($\mu_0=1$ GeV) a perturbative charm PDF. However, when included at NLO, the description of this data set seems to improve from a $\chi^2$ per data point of $5.87$ to $2.51$ which is between the $\chi^2$ of the CMS at $\sqrt{s}=7$ and 8 TeV data sets. Nevertheless, the inclusion of this data set deteriorates the global $\chi^2$, which per data point goes from $1.37$ to $1.42$. At NNLO, the fit quality of this data set also improves upon its inclusion ($12.04$ to $6.91$), but its description is not satisfactory due to the significantly large $\chi^2$. The main contribution to this large $\chi^2$ comes from the extreme pseudorapidity that are harder to fit as shown in the representative Fig.~\ref{fig:CMS2JET_pp_datavstheory}. For this reason, the proton baseline of \texttt{nNNPDF3.0} is restricted only to the dijet data sets considered in Sect.~\ref{s1:PDF_NNLO_jet}, namely the ATLAS and CMS at $\sqrt{s}=7$ and 8 TeV ones.
\begin{table}
    \centering
    \begin{tabular}{lccccc}  \toprule
        & & \multicolumn{2}{c}{NLO} & \multicolumn{2}{c}{NNLO}\\
        data set              & $N_{\text{dat}}$ &  \textbf{w/o}      & \textbf{w/}  &  \textbf{w/o}     & \textbf{w/}\\
        \midrule
        ATLAS 7 TeV             &    90    &   1.03  & 1.01 &   1.98   &    1.91  \\
        CMS   7 TeV             &    54    &   1.58  & 2.03 &   1.75   &    1.92  \\
        CMS   8 TeV             &    122   &   3.87  & 3.61 &   1.48   &    1.55  \\
        \textbf{CMS   5 TeV}    &    \textbf{85}    &  \textbf{[5.87]}  & \textbf{2.51} &  \textbf{[12.04]}    &    \textbf{6.91}  \\
        \midrule
         Total                    &          &  1.37   & 1.42     &   1.24   &   1.41      \\
\bottomrule
\end{tabular}
\caption{Similar to Table~\ref{tab:NNPDF31_jet_chi2s}, the $\chi^2$ per data point for the two new \texttt{nNNPDF3.0} proton baselines: \texttt{nNNPDF2.0}$^*$($^1p$) at NLO and NNLO.
Results are shown
for the dijet data sets together with the number of data points in each
data set.}
\label{tab:nNNPDF20_newbaseline_chi2_1}
\end{table}

Now that we have defined our new \texttt{nNNPDF3.0} proton baselines, we compare it to the \texttt{nNNPDF2.0} one. In Table~\ref{tab:nNNPDF20_newbaseline_chi2_2}, I start by comparing the fit quality of the \texttt{nNNPDF2.0} nPDF sets on the CMS $\frac{\text{pPb}}{\text{pp}}$ ratio data using the old and new proton baselines (without fitting the data set). The fact that the $\chi^2$ per data point improved (from $3.342$ to $3.145$) merely due to the new proton baseline, highlights on one hand the importance of the proton PDF contribution to the heavy-ion observables and on the other hand, the fact that both ATLAS and CMS data sets at $\sqrt{s}=7$ and 8 TeV dijet provide information that helps describing the $\frac{\text{pPb}}{\text{pp}}$ CMS 5 TeV spectra.
\begin{table}
\centering
\renewcommand{\arraystretch}{1.20}
\begin{tabular}{c|ccc}
Data set & $N_{\text{dat}}$ &  \texttt{nNNPDF2.0}  & \texttt{nNNPDF2.0}$^*$  \\
& & \multicolumn{2}{c}{NLO} \\
\toprule
CMS dijet $\frac{\text{pPb}}{\text{pp}}$ 5 TeV & 84 & [3.342]  & [3.145] \\
\bottomrule
\end{tabular}
\vspace{0.3cm}
\caption{\small The $\chi^2$ per data point calculated for the $\frac{\text{pPb}}{\text{pp}}$ CMS 5 TeV spectra using the new proton baseline \texttt{nNNPDF2.0}$^*$($^1p$) and the \texttt{nNNPDF2.0} lead nuclear PDF at NLO. Both values are enclosed in square brackets as the data set is not included in the fit.\label{tab:nNNPDF20_newbaseline_chi2_2}
}
\end{table}

\begin{figure}
\begin{center}
\includegraphics[width=0.99\textwidth]{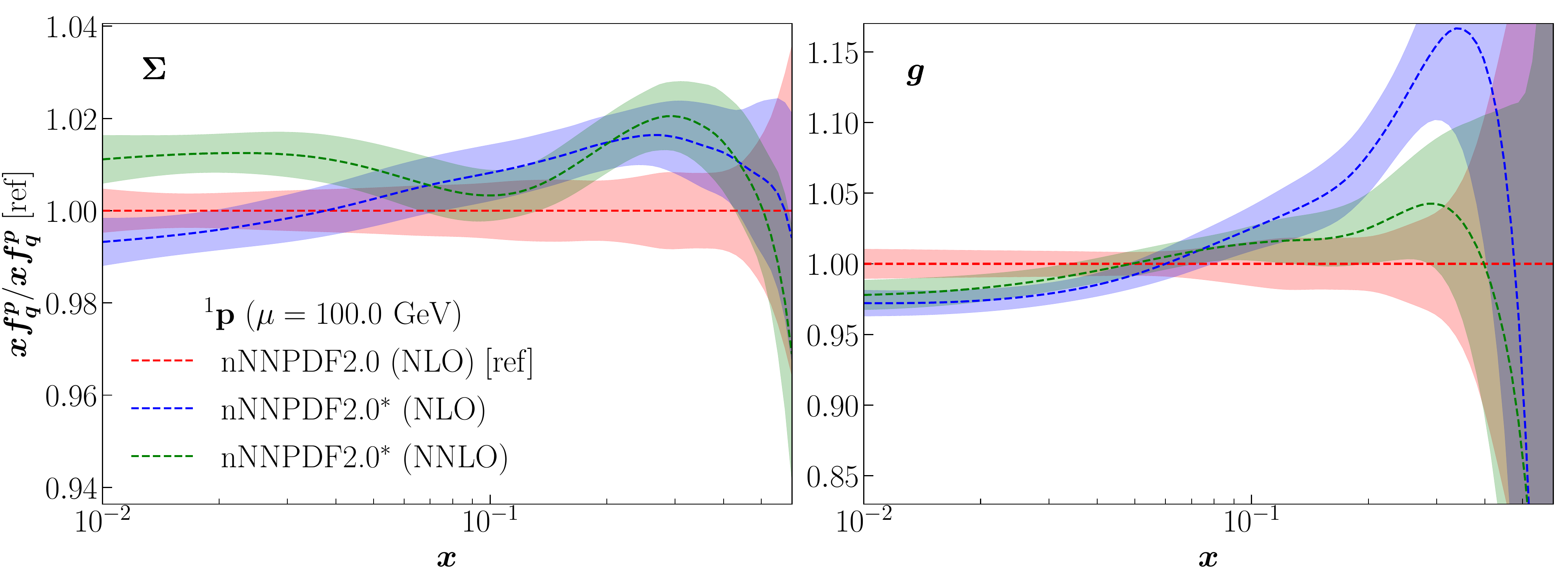}
\end{center}
\vspace{-0.8cm}
\caption{Similar to Fig.~\ref{fig:jetsdijets}, a comparison between the new proton baseline \texttt{nNNPDF2.0}$^*$($^1p$) fitted to the ATLAS and CMS dijet at $\sqrt{s}=7$ and 8 TeV and the old \texttt{nNNPDF2.0} proton baseline. The bands correspond to the 1-$\sigma$ uncertainty.
\label{fig:nNNPDF20_newbaseline}
}
\end{figure}

Finally In Fig.~\ref{fig:nNNPDF20_newbaseline}, I compare the proton PDF baselines themselves. In analogy to Fig.~\ref{fig:jetsdijets}, I restrict the flavours to the singlet $\Sigma$ and the gluon and the $x$-range to $[10^{-2},0.6]$ which the \texttt{NNPDF3.1} set is most sensitive to. As expected, the PDFs show more or less similar trend to those observed in Fig.~\ref{fig:jetsdijets}, which can be summarised mainly in terms of a reduction of gluon uncertainties at large-$x$. Note however that the comparison with Fig.~\ref{fig:jetsdijets} should only be qualitative, in particular for $\Sigma$, as the reference proton PDF in Fig.~\ref{fig:nNNPDF20_newbaseline} is different (not including nuclear CC DIS).

\section{Results}
\label{s1:nNNPDF30_results}
In this section, I discuss the results achieved so far towards the foreseen global analysis \texttt{nNNPDF3.0}. As mentioned in Sect.~\ref{s2:nNNPDF30_data}, the focus will be on the dijet production in 5 TeV pPb collision~\cite{Sirunyan_2018} and Z-boson production in 5 TeV pPb collision~\cite{Khachatryan:2015pzs} with CMS at NNLO.

In Sect.~\ref{s2:nNNPDF30_dijetresults}, I start by presenting the impact of including the CMS dijet $\frac{\text{pPb}}{\text{pp}}$ at 5 TeV on the \texttt{nNNPDF2.0}$^*$ fit\footnote{The \texttt{nNNPDF2.0}$^*$ fit is based on the \texttt{nNNPDF2.0} data sets listed in Sect.~\ref{s2:nNNPDF20_data} and an updated proton baseline fitted to the ATLAS and CMS dijet measurements at $\sqrt{s}=7$ and 8 TeV.} at NLO. I perform a comparison on the level of the fit-quality, theory predictions relative to data and the lead nPDFs. Subsequently, I consider the NNLO case where at first, I gauge the impact of the NNLO QCD corrections on nPDFs by comparing an NLO and NNLO set of nPDFs fitted to DIS (NC and CC) and dijet data. The reason behind this restriction is due to the fact that the K-factors corresponding to the electroweak gauge bosons production data sets are yet to be computed and validated, thus beyond the scope of this thesis. Secondly, I draw some observations on the impact of the dijet data set w.r.t to DIS on nPDFs at NNLO.
In Sect.~\ref{s2:nNNPDF30_Zbosonresults}, I augment the previous DIS and dijet data sets with the CMS Z-boson production at 5 TeV at NNLO (already included in \texttt{nNNPDF2.0} at NLO). This inclusion is based on a low-statistics K-factors that I have computed by means of the \textsc{MATRIX} Monte Carlo based framework~\cite{Grazzini:2017mhc}. 

\subsection{CMS dijet production in 5 TeV pPb}
\label{s2:nNNPDF30_dijetresults}

In Sect.~\ref{s2:nNNPDF30_updates}, I showed that the QCD NNLO K-factors did not lead to a satisfactory description of the CMS 5 TeV absolute pp dijet spectra (see Table~\ref{tab:nNNPDF20_newbaseline_chi2_1}). This turns out to be also the case, even at NLO, for the absolute pPb dijet spectra when added to \texttt{nNNPDF2.0}. For that reason, we focus only on the ratio $\frac{\text{pPb}}{\text{pp}}$ data that we find to be describable at both NLO and NNLO with a satisfactory $\chi^2$ value. 

\myparagraph{Comparison with \texttt{nNNPDF2.0}$^*$} 
In order to gauge the impact of the CMS dijet $\frac{\text{pPb}}{\text{pp}}$ data set w.r.t. \texttt{nNNPDF2.0}$^*$, I perform a fit at NLO and assess the fit quality in Table~\ref{tab:nNNPDF20_CMS2JET_chi2}. Note that the fit quality of the rest of the data sets (DIS and electroweak gauge bosons production) are very comparable to those quoted in Table~\ref{tab:nNNPDF20_chi2} and \ref{tab:nNNPDF20_chi2_2}, thus are omitted. This can only mean that this new data set is not in tension with any of the prior data sets considered in \texttt{nNNPDF2.0}. Additionally, Table~\ref{tab:nNNPDF20_CMS2JET_chi2} shows that this new data set is well described when included as a ratio $\frac{\text{pPb}}{\text{pp}}$ as opposed to the absolute pPb spectra. Upon fitting this data set the $\chi^2$ per data point value goes from a value of $3.145$ to $1.644$ with a global $\chi^2$ per data point of $1.0$. This is mainly due to the cancellation of low-statistics effect and uncertainties from the extreme dijet pseudorapidity bins that we observed in Fig.~\ref{fig:CMS2JET_pp_datavstheory}.
\begin{table}
    \centering
    \renewcommand{\arraystretch}{1.20}
    \begin{tabular}{c|ccc}
        Data set & $N_{\text{dat}}$ &  \texttt{nNNPDF2.0}$^*$   & \texttt{nNNPDF2.0}$^*$ + CMS dijet \\
        & & \multicolumn{2}{c}{NLO} \\
        \toprule
        CMS dijet $\frac{\text{pPb}}{\text{pp}}$ 5 TeV & 84 & [3.145]  & 1.644 \\
        \midrule
        Total & 1551 & [1.192]  & 1.0 \\
    \bottomrule
    \end{tabular}
    \vspace{0.3cm}
    \caption{\small Comparison between the $\chi^2$ per data point of \texttt{nNNPDF2.0}$^*$ and a new NLO inference including the $\frac{\text{pPb}}{\text{pp}}$ CMS 5 TeV spectra. Values enclosed in square brackets are of the data set that is not included in the fit.
    \label{tab:nNNPDF20_CMS2JET_chi2}
    }
    \end{table}

\begin{figure} 
    \vspace{1cm}\hspace{-2.5cm}
    \floatbox[{\capbeside\thisfloatsetup{capbesideposition={right,bottom},capbesidewidth=0.42\textwidth}}]{figure}[\FBwidth]
    {\hspace{-8.5cm}\caption{Comparison between the NLO theory predictions of the CMS dijet $\frac{\text{pPb}}{\text{pp}}$ data from both (\texttt{nNNPDF2.0}$^*$) and (\texttt{nNNPDF2.0}$^*$ + CMS dijet) fits and the data for all bins of $\eta_{\text{dijet}}$ and $p_{T,\text{dijet}}^{\text{avg}}$.\vspace{2cm}}\label{fig:nNNPDF20_CMS2JET_datavstheory}}
    {\includegraphics[width=1.15\textwidth]{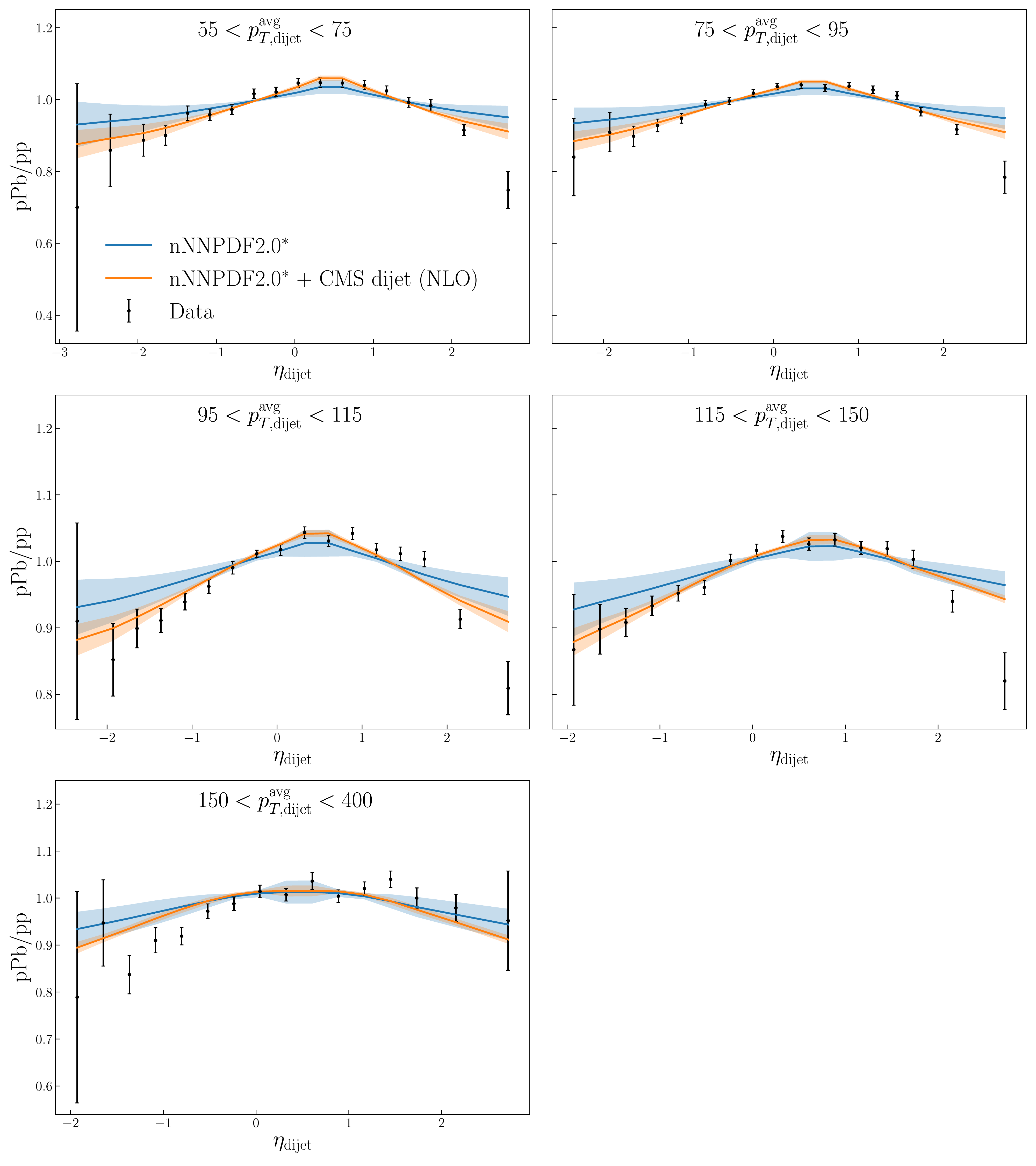}} 
    \end{figure}

In Fig.~\ref{fig:nNNPDF20_CMS2JET_datavstheory}, I compare the theory predictions of the CMS dijet $\frac{\text{pPb}}{\text{pp}}$ data in all bins of $\eta_{\text{dijet}}$ and $p_{T,\text{dijet}}^{\text{avg}}$. One can directly notice that the last $2$ extreme positive $\eta_{\text{dijet}}$ bins in the first $4$ bins of $p_{T,\text{dijet}}^{\text{avg}}$ are the most difficult to fit. Therefore, they must hold the major contribution to the $\chi^2$ per data point of $1.644$. In fact, as we can observe in Table.~\ref{tab:nNNPDF20_CMS2JET_chi2_cut}, upon removing the bins with $\eta_{\text{dijet}}>2.7$ in all bins of $p_{T,\text{dijet}}^{\text{avg}}$, the $\chi^2$ for the dijet data set reduces from $1.644$ to $1.334$ and the global one from $1.0$ to $0.982$. 
%
\begin{table}[h]
    \centering
    \renewcommand{\arraystretch}{1.20}
    \begin{tabular}{c|cc|cc}
        Data set & $N_{\text{dat}}$ &  CMS dijet & $N_{\text{dat}}$ & CMS dijet ($\eta_{\text{dijet}}<2.7$) \\
        & & NLO & & NLO\\
        \toprule
        CMS dijet $\frac{\text{pPb}}{\text{pp}}$ 5 TeV & 84  & 1.644 & 79 & 1.334\\
        \midrule
        Total & 1551 & 1.0 & 1546 & 0.982\\
    \bottomrule
    \end{tabular}
    \vspace{0.3cm}
    \caption{\small Same as Table~\ref{tab:nNNPDF20_CMS2JET_chi2} but $\chi^2$ calculated with $\eta_{\text{dijet}}<2.7$ in all bins of $p_{T,\text{dijet}}^{\text{avg}}$.
    \label{tab:nNNPDF20_CMS2JET_chi2_cut}
    }
    \end{table}

\begin{figure}[!ht]
\begin{center}
    \includegraphics[width=0.99\textwidth]{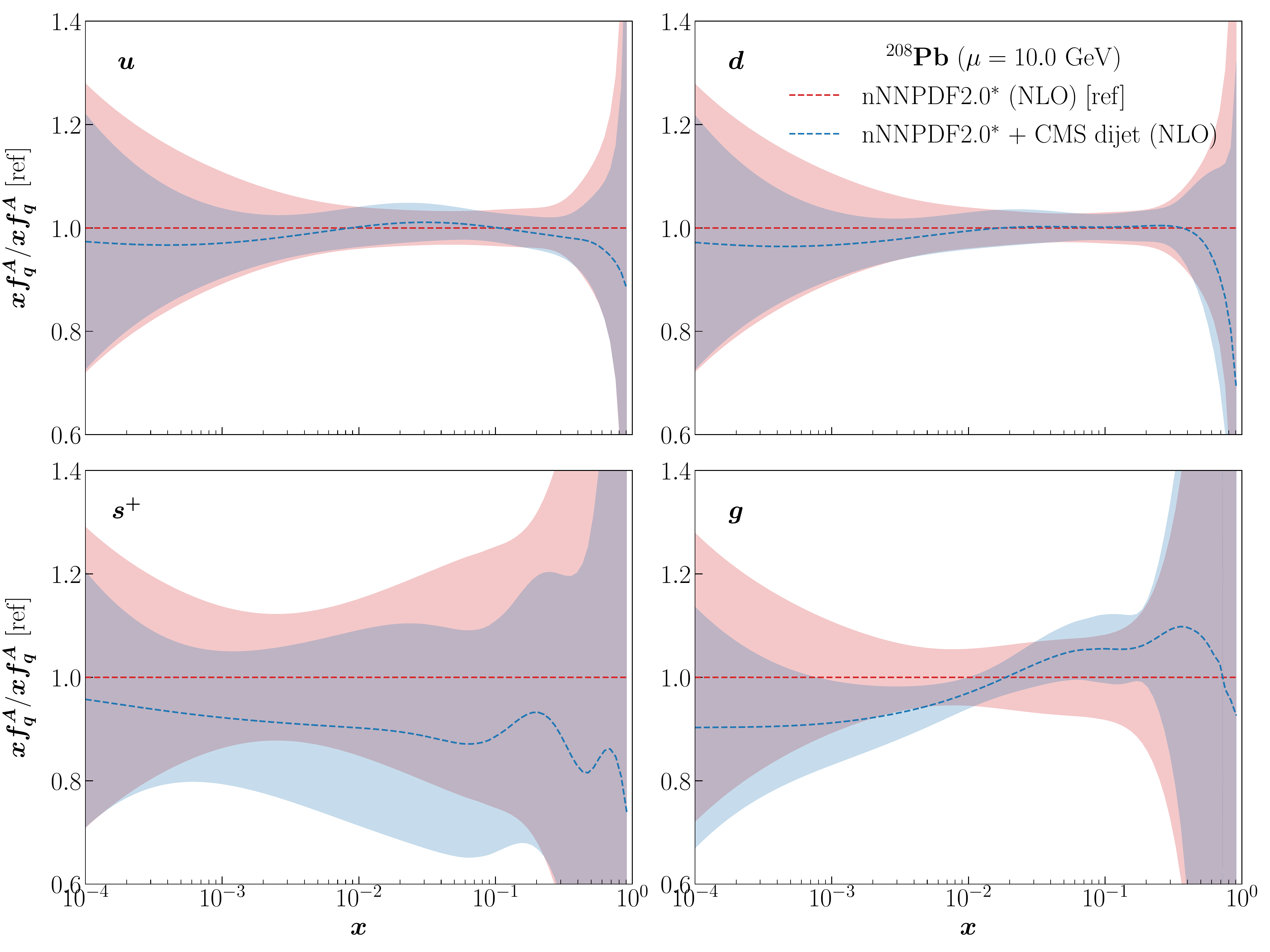}
    \end{center}
\vspace{-0.8cm}
\caption{\small Comparison of the up, down, strange and antistrange combination and gluon NLO nPDFs as a ratio of a fit augmented by the CMS dijet $\frac{\text{pPb}}{\text{pp}}$ data to the \texttt{nNNPDF2.0}-like fit with the new baseline (called \texttt{nNNPDF2.0}$^*$). The bands correspond to the 1-$\sigma$ uncertainty.
    \label{fig:nNNPDF20_CMS2JET}
}
\end{figure}

Finally, in Fig.~\ref{fig:nNNPDF20_CMS2JET}, I compare the nPDFs obtained from the \texttt{nNNPDF2.0}$^*$ fit and a similar fit augmented by the CMS dijet $\frac{\text{pPb}}{\text{pp}}$ data. Although \texttt{nNNPDF2.0}$^*$ contains a handful of hadronic data (listed in Table~\ref{tab:nNNPDF20_data}) the new CMS dijet data set provides distinct information, particularly in the $x$-range defined in Eq.~\ref{eq:dijet_coverage}. Perhaps the most prominent impact is on the gluon, where it's suppressed for $x \leq 10^{-2}$ and enhanced for $ 10^{-2} \leq x \leq 0.3$. The rest of the plotted flavours, in particular the combination $s^{+}=s+\bar{s}$, also manifest a suppression to accommodate the new data, however their central values remains within uncertainties of \texttt{nNNPDF2.0}$^*$.

\myparagraph{Impact of dijet production on nPDFs at NNLO}
As mentioned in the end of Sect.~\ref{s2:NNLO_corrections}, currently and in order to include any hadronic process including a nuclear target (pA) at NNLO, the approximation ${\displaystyle K_{\text{NNLO}}^{\text{QCD, pA}} \approx K_{\text{NNLO}}^{\text{QCD, pp}}}$ is used (purely a technical limitation that we hope to overcome in the near future). The only currently available NNLO calculations available, as mentioned in the introduction of this chapter are the analytical NNLO calculations for DIS (NC and CC) as well as the approximate K-factors for dijet and Z-boson production from 5 TeV pPb with CMS. The Z-boson production will be investigated next in Sect.~\ref{s2:nNNPDF30_Zbosonresults}. However, for the CMS dijet $\frac{\text{pPb}}{\text{pp}}$ data, this approximation leads to a complete cancellation of K-factors due to the ratio in the observable as follows:
\begin{align}
    \left(\frac{\text{pPb}}{\text{pp}}\right)_{\text{NNLO}} &\equiv \left(\frac{1}{N^{\text{pPb}}}\frac{dN_{\text{dijet}}^{\text{pPb}}}{d\eta_{\text{dijet}}}\right)_{\text{NNLO}}\bigg/\left(\frac{1}{N^{\text{pp}}}\frac{dN_{\text{dijet}}^{\text{pp}}}{d\eta_{\text{dijet}}}\right)_{\text{NNLO}} \\
    & \approx \frac{K^{\text{QCD, pPb}}_{\text{NNLO}}}{K^{\text{QCD, pp}}_{\text{NNLO}}} \left(\frac{1}{N^{\text{pPb}}}\frac{dN_{\text{dijet}}^{\text{pPb}}}{d\eta_{\text{dijet}}}\right)_{\text{NLO}}\bigg/\left(\frac{1}{N^{\text{pp}}}\frac{dN_{\text{dijet}}^{\text{pp}}}{d\eta_{\text{dijet}}}\right)_{\text{NLO}}\\
    & \sim \left(\frac{1}{N^{\text{pPb}}}\frac{dN_{\text{dijet}}^{\text{pPb}}}{d\eta_{\text{dijet}}}\right)_{\text{NLO}}\bigg/\left(\frac{1}{N^{\text{pp}}}\frac{dN_{\text{dijet}}^{\text{pp}}}{d\eta_{\text{dijet}}}\right)_{\text{NLO}}
\end{align}
In Table~\ref{tab:NNLO_CMS2JETchi2}, I compare the quality of a NNLO fit with DIS augmented by the CMS dijet ratio, a similar NLO fit (in order to assess the impact of NNLO QCD corrections) and another NNLO fit with DIS only (in order to assess the impact of the dijet production ratio on nPDFs at NNLO). I first note that a satisfactory $\chi^2$ is achieved for both NLO and NNLO fits, showing that the NNLO QCD corrections included in the DIS processes did not lead to a tension with the CMS dijet data. Second, the values of $\chi^2$ associated with the DIS data sets are very comparable at both NLO and NNLO (slight improvement w.r.t. NLO) to the values obtained in Table~\ref{tab:nNNPDF10_chi2} thus are omitted. The only mild deterioration in the $\chi^2$ occurs when transitioning from NLO to NNLO for the CMS dijet data set ($1.556$ to $1.737$ per data point). Compared to Table~\ref{tab:nNNPDF20_CMS2JET_chi2}, the $\chi^2$ per data point of the dijet data set seems to mildly improve at NLO, reducing from $1.644$ to $1.556$ when the LHC electroweak gauge bosons production data is not considered. This reduction however is small enough, supporting the earlier conclusion that the dijet data set is not in tension with the rest of the LHC data sets. Compared to the DIS only fit, the $\chi^2$ values are significantly different showing that the fit to the currently available NC and CC DIS data could not predict well the CMS dijet data set. This point emphasize on the importance of being inclusive in all possible type of processes when analysing PDFs. 
\begin{table}
    \centering
    \renewcommand{\arraystretch}{1.20}
    \begin{tabular}{c|cccc}
        Data set & $N_{\text{dat}}$ &  DIS  & \multicolumn{2}{c}{DIS + CMS dijet} \\
        & & NNLO & NLO & NNLO  \\
        \toprule
        CMS dijet $\frac{\text{pPb}}{\text{pp}}$ 5 TeV & 84 & [12.155] & 1.556 & 1.737 \\
        \midrule
        Total & 1457 & [1.561] & 0.974 & 1.014\\
    \bottomrule
    \end{tabular}
    \vspace{0.3cm}
    \caption{\small  Comparison between the $\chi^2$ per data point of a NNLO fit to DIS augmented by the CMS dijet ratio, a similar fit at NLO and another NNLO fit with DIS only. The fit quality of the rest of the DIS data sets are very comparable to those quoted in Table~\ref{tab:nNNPDF20_chi2}, thus are omitted. Values enclosed in square brackets are of the data set that is not included in the fit. \label{tab:NNLO_CMS2JETchi2}
    }
    \end{table}
In Table~\ref{tab:NNLO_CMS2JETchi2_cuts} and in analogy with Table~\ref{tab:nNNPDF20_CMS2JET_chi2_cut}, we observe that upon removing the bins with $\eta_{\text{dijet}}>2.7$ in all bins of $p_{T,\text{dijet}}^{\text{avg}}$, the $\chi^2$ for the dijet data set reduces by a value of $\simeq 0.3$ for both NLO and NNLO.
\begin{table}
\centering
\renewcommand{\arraystretch}{1.20}
\begin{tabular}{c|ccc|ccc}
    Data set & $N_{\text{dat}}$ & \multicolumn{2}{c}{DIS + CMS dijet} & $N_{\text{dat}}$ & \multicolumn{2}{c}{DIS + CMS dijet($\eta_{\text{dijet}}<2.7$)} \\
    &  & NLO & NNLO & & NLO & NNLO \\
    \toprule
    CMS dijet $\frac{\text{pPb}}{\text{pp}}$ 5 TeV & 84 & 1.556 & 1.737 & 79 & 1.283 & 1.417\\
    \midrule
    Total & 1457 & 0.974 & 1.014 & 1452 & 0.957 & 0.994\\
\bottomrule
\end{tabular}
\vspace{0.3cm}
\caption{\small Same as Table~\ref{tab:NNLO_CMS2JETchi2} but $\chi^2$ calculated with $\eta_{\text{dijet}}<2.7$ in all bins of $p_{T,\text{dijet}}^{\text{avg}}$. \label{tab:NNLO_CMS2JETchi2_cuts}
}
\end{table}

\begin{figure} 
    \vspace{1cm}\hspace{-2.5cm}
    \floatbox[{\capbeside\thisfloatsetup{capbesideposition={right,bottom},capbesidewidth=0.46\textwidth}}]{figure}[\FBwidth]
    {\hspace{-8.6cm}\caption{Comparison between the theory predictions of the CMS dijet $\frac{\text{pPb}}{\text{pp}}$ data from a NNLO fit to DIS augmented by the CMS dijet ratio (DIS+CMS dijet (NNLO)), a similar fit at NLO (DIS+CMS dijet (NLO)), another NNLO fit with DIS only (DIS (NNLO)) and the data for all bins of $\eta_{\text{dijet}}$ and $p_{T,\text{dijet}}^{\text{avg}}$.\vspace{1cm}}\label{fig:NNLO_CMS2JET_datavstheory}}
    {\includegraphics[width=1.15\textwidth]{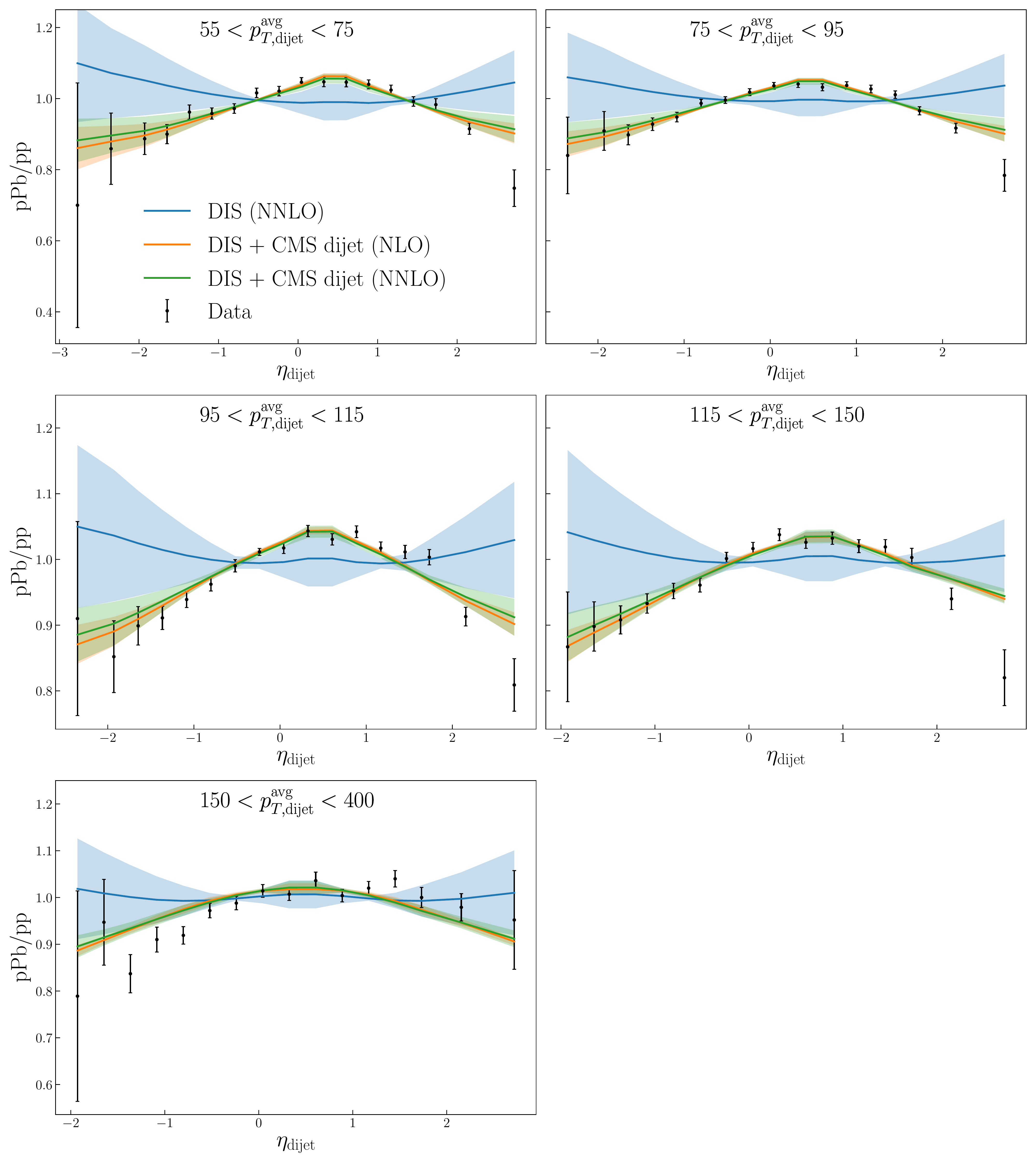}} 
    \end{figure}

In Fig.~\ref{fig:NNLO_CMS2JET_datavstheory}, I compare the mentioned three fits on the level of theory predictions relatively to the data. The difference between the NLO and NNLO predictions of the DIS+CMS dijet fit are noticeable, even though mild. However, the predictions of the DIS only fit misses completely the pattern of the CMS dijet data as opposed to the \texttt{nNNPDF2.0}$^*$ fit shown in Fig.~\ref{fig:nNNPDF20_CMS2JET_datavstheory}. This shows the complementarity of the electroweak gauge bosons production data sets and dijet one and perhaps what is remarkable is that the latter brings the predictions to an uncertainty very comparable to the one from the \texttt{nNNPDF2.0}$^*$ fit (\textit{i.e.} all the electroweak gauge bosons production data sets combined). We therefore deduce that the CMS dijet $\frac{\text{pPb}}{\text{pp}}$ data provide vital constraint on nPDFs, competing with the LHC electroweak gauge bosons production data.

\begin{figure}[ht]
\begin{center}
    \includegraphics[width=0.99\textwidth]{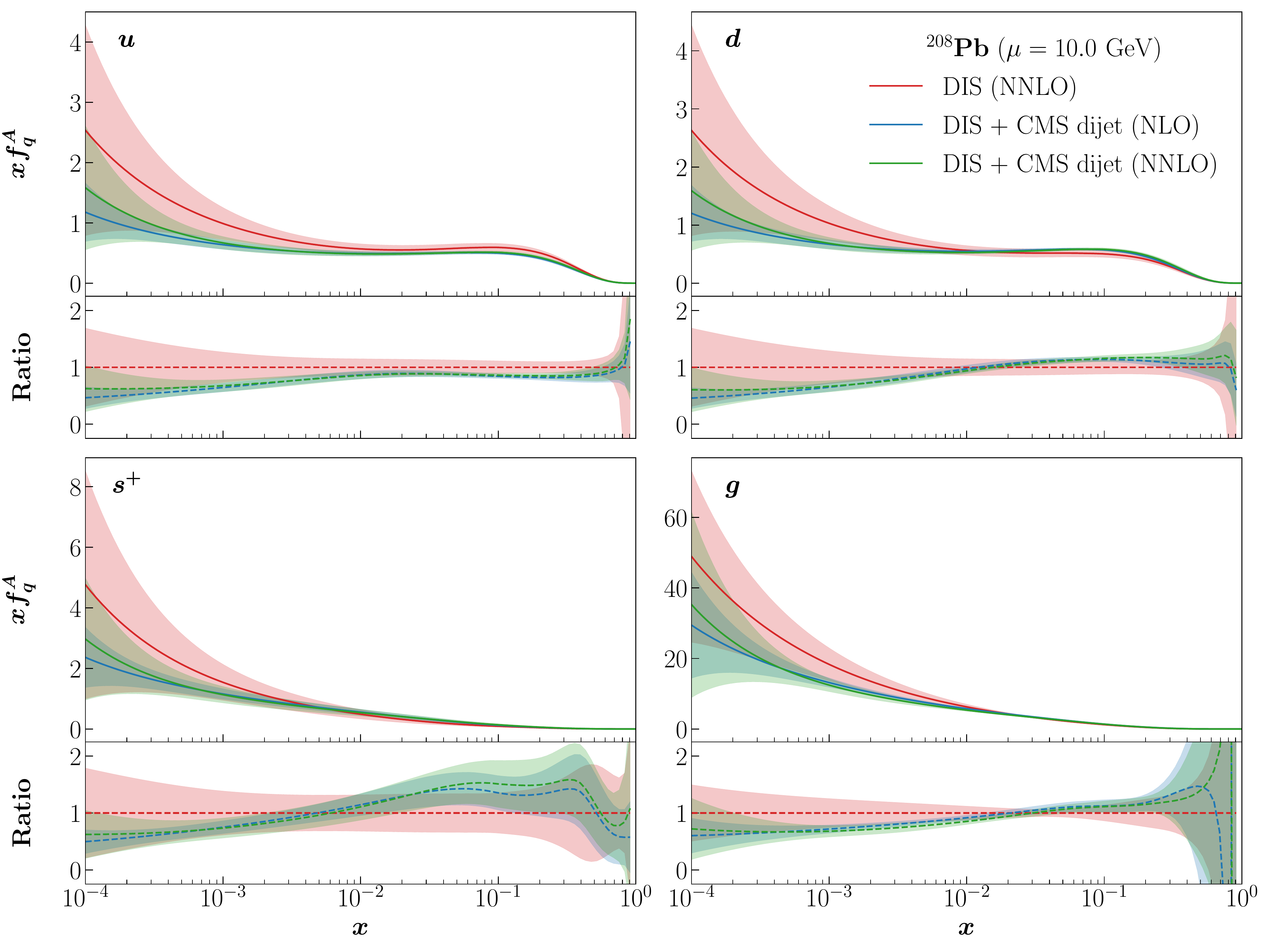}
    \end{center}
\vspace{-0.8cm}
\caption{ Comparison of the up, down, strange and antistrange combination and gluon NLO absolute nPDFs and ratio of nPDFs from a NNLO fit to DIS augmented by the CMS dijet ratio (DIS+CMS dijet (NNLO)), a similar fit at NLO (DIS+CMS dijet (NLO)) to a NNLO fit with DIS only (DIS (NNLO)). The bands correspond to the 1-$\sigma$ uncertainty.
\label{fig:NNLO_CMS2JET}
}
\end{figure}

The sizeable impact of the dijet data set is further highlighted in Fig.~\ref{fig:NNLO_CMS2JET} on the level of nPDFs. One of the main features of this data seems to be manifested by a general suppression across all displayed flavours for $x \leq 10^{-2}$ and an enhancement above this value (except for the $u$ distribution) w.r.t to a fit to DIS only. The impact of this data on the uncertainties is quite significant when compared to a DIS only fit across all the $x$-range denoted by Eq.~\ref{eq:dijet_coverage}. The NNLO QCD corrections included only for DIS seems to play a mild impact in the (DIS + CMS dijet (NNLO)) fit, pulling slightly upward the distributions. 

\subsection{CMS Z-boson production in 5 TeV pPb}
\label{s2:nNNPDF30_Zbosonresults}
In this section, I investigate my first attempt to compute the NNLO QCD K-factor corresponding to the CMS Z-boson production in 5 TeV pPb collision~\cite{Khachatryan:2015pzs} data set using the \textsc{MATRIX}~\cite{Grazzini:2017mhc} framework. \textsc{MATRIX} is a computational framework allowing the evaluation of fully differential cross sections for a wide class of processes at hadron colliders in NNLO QCD. The processes are $2\rightarrow 1$ and $2\rightarrow 2$ hadronic reactions involving Higgs and vector bosons in the final state. \textsc{MATRIX} is based on the Monte Carlo (MC) integrator \textsc{MUNICH}~\cite{Grazzini:2017mhc}, which provides a fully automated computation of arbitrary SM NLO processes.

\begin{figure}[ht]
    \begin{center}
    \includegraphics[width=0.99\textwidth]{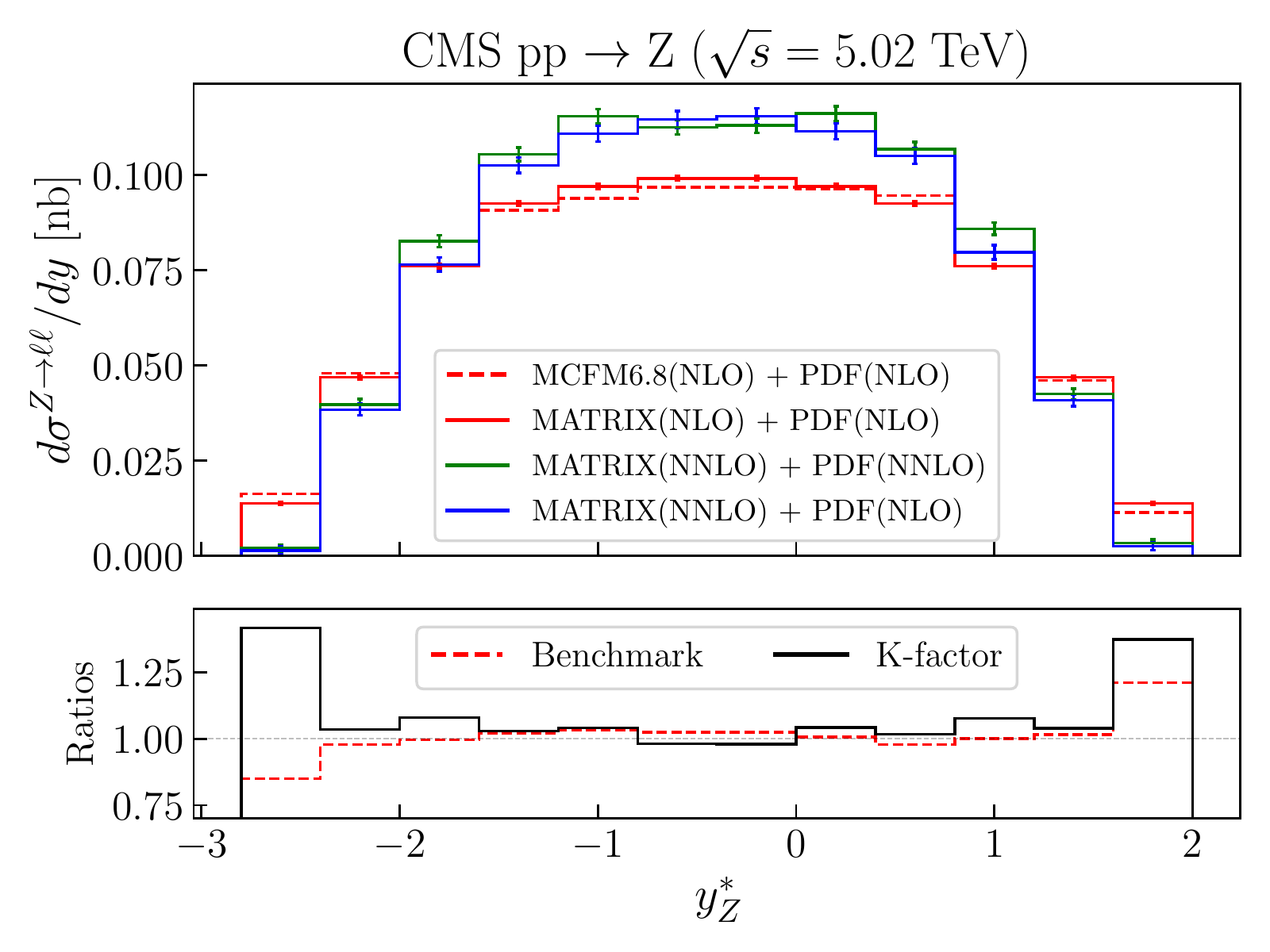}
    \end{center}
    \vspace{-0.8cm}
    \caption{ The QCD calculations corresponding to the CMS Z-boson production in 5 TeV pp collision data set.
    In the upper panel, the histograms displayed are the NLO calculations (solid red histogram), the reference NLO (dashed red histogram), the NNLO calculations with NLO PDFs  (solid blue histogram) and NNLO PDFs (solid green histogram). The PDF sets used are the new proton baseline \texttt{nNNPDF2.0}$^*$($^1$p). In the lower panel, the ratio of NLO calculation to the reference ones (dashed red histogram) and NNLO QCD K-factors (solid black histogram) Eq.~\ref{eq:kfactor} are displayed.
    \label{fig:nNNPDF20_CMS_Z_kfac}
    }
    \end{figure}

As mentioned in Sect.~\ref{s2:NNLO_corrections}, and assumed in the previous section, it is only possible to compute $K^{\text{QCD, pp}}_{\text{NNLO}}$ with \textsc{MATRIX} currently, therefore neglecting the nuclear effects in the ratio of NNLO/NLO cross section w.r.t. the NNLO effects. For that purpose, I present in Fig.~\ref{fig:nNNPDF20_CMS_Z_kfac} the $K^{\text{QCD, pp}}_{\text{NNLO}}$ corresponding to the CMS Z-boson production data set. I start by validating my theoretical settings by performing a benchmark at NLO between \textsc{MATRIX} and the independent \textsc{MCFM6.8} MC generator~\cite{Campbell:2015qma,Boughezal:2016wmq} used in $\text{nNNPDF2.0}$ to compute all the NLO interpolation tables. An agreement of up to $1\%$ is achieved at NLO between the to MC generators for all Z-boson rapidity in the pPb center-of-mass frame $y^*_Z$ bins except for the two extreme negative and positive bins that suffers from low number of events (low-statistics). Once validated, the NNLO predictions are produced using the same settings. In order to compute the K-factors defined in Eq.~\ref{eq:kfactor}, one has to compute twice the predictions, once with NLO PDF (that will serve as the denominator of the K-factor) and once with a NNLO PDF (that will serve as the numerator) both with NNLO hard matrix-element. The PDF used to compute this K-factor is the baseline \texttt{nNNPDF2.0}$^*$($^1p$) defined in Sect.~\ref{s2:nNNPDF30_updates}. The NNLO QCD corrections (see the K-factor in the lower panel of Fig.~\ref{fig:nNNPDF20_CMS_Z_kfac}) seems to vary mildly couple of percents in the central rapidity region, however blows up in for the same two extreme negative and positive rapidity bins that, as mentioned before, suffer from low statistics of simulated events by the MC. Admitting the fact that this low-statistics K-factor needs to be reproduced for a higher number of simulated events and therefore higher accuracy in the future, we still carry on the analysis to at least study the impact of the QCD corrections on the predictions in the central rapidity bins where they are more reliable. 

To that end, I perform two new fits. One at NNLO including all the DIS data in Table~\ref{tab:nNNPDF10_data} associated with predictions at NNLO, the CMS dijet $\frac{\text{pPb}}{\text{pp}}$ data with predictions at NLO augmented by the CMS Z-boson production supplemented by the K-factor displayed in Fig.~\ref{fig:nNNPDF20_CMS_Z_kfac} and a second similar fit but at NLO.
In Table~\ref{tab:nNNPDF20_CMSZ_DIJET_chi2}, I compare the $\chi^2$ per data point for both of these fits. The first thing to notice is the significant deterioration of the CMS Z-boson data set $\chi^2$ at NNLO ($1.431$) w.r.t. the NLO one ($0.76$) which is solely due to the inaccurate estimation of QCD correction in two low-statistics extreme rapidity bins as argued before. The description of the CMS dijet data set seems mildly affected by this inaccuracy.
\begin{table}
\centering
\renewcommand{\arraystretch}{1.20}
\begin{tabular}{c|ccc}
& $N_{\text{dat}}$ &  \multicolumn{2}{c}{DIS+ CMS dijet + CMS Z }    \\
& & NLO & NNLO \\
\toprule
CMS dijet $\frac{\text{pPb}}{\text{pp}}$ 5 TeV & 84 & 1.591 & 1.638 \\
\midrule
CMS Z pPb 5 TeV & 12 & 0.76 & 1.431 \\
\midrule
Total & 1469 & 1.0 & 0.995 \\
\bottomrule
\end{tabular}
\vspace{0.3cm}
\caption{\small Comparison between the $\chi^2$ per data point of a NLO and NNLO fit to DIS, CMS dijet $\frac{\text{pPb}}{\text{pp}}$ ratio and Z-boson production. \label{tab:nNNPDF20_CMSZ_DIJET_chi2}
}
\end{table}

\begin{figure}[t]
\begin{center}
\includegraphics[width=0.99\textwidth]{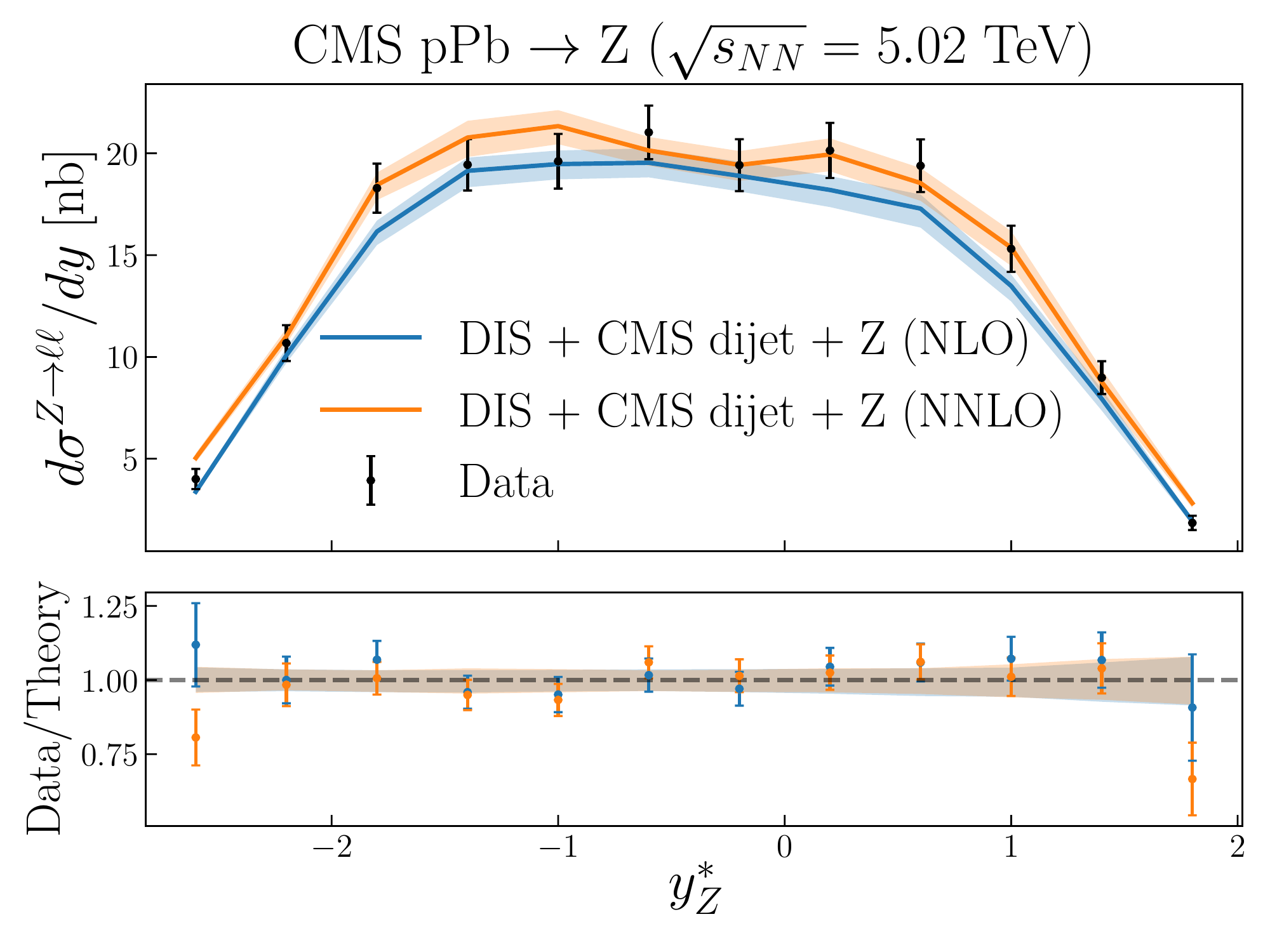}
\end{center}
\vspace{-0.8cm}
\caption{ Comparison between the theory predictions of the CMS Z-boson production data from a NLO and NNLO fit to DIS, CMS dijet $\frac{\text{pPb}}{\text{pp}}$ ratio and Z-boson production and the data in bins of Z-boson rapidity in the pPb center-of-mass frame $y^*_Z$.
\label{fig:nNNPDF20_CMSZ_DIJET}
}
\end{figure}

In Fig.~\ref{fig:nNNPDF20_CMSZ_DIJET}, we compare the predictions of the two fits w.r.t. to the data. Aside the two problematic bins that the NNLO fit overshoots, it is remarkable to see that the NNLO predictions shift upwards, closer to the data points, the rest of the Z-boson rapidity in the pPb center-of-mass frame $y^*_Z$. We therefore deduce that, although with the K-factor approximation and the inaccuracy of the low-statistics rapidity bins, QCD NNLO corrections will have a significant contribution to the predictions shifting the latter in the direction of the data.
I finally note that the effort towards refining this chapter's results are currently undergoing. In particular, those related to the the K-factor approximation, the Z-boson production K-factors as well as the rest of the electroweak gauge bosons production data sets in Table~\ref{tab:nNNPDF20_data} and the inclusion of new data mentioned in Sect.~\ref{s2:nNNPDF30_data}. Once validated, the upcoming results of a global \texttt{nNNPDF3.0} analysis at NNLO will take part in the upcoming Ref.~\mycite{AbdulKhalek:2021xxx1}.
  \chapter{Pion fragmentation functions}
\label{chap:FF}
\vspace{-1cm}
\begin{center}
  \begin{minipage}{1.\textwidth}
      \begin{center}
          \textit{This chapter is based on my results that are presented in Refs.~\mycite{Khalek:2021gxf}}.
      \end{center}
  \end{minipage}
  \end{center}
  
\myparagraph{Introduction} Unpolarised collinear fragmentation functions
(FFs)~\cite{Metz:2016swz} encode the non-perturbative mechanism,
called hadronisation, that leads a fast on-shell parton (a
quark or a gluon) to inclusively turn into a fast hadron moving along
the same direction. In the framework of quantum chromodynamics (QCD),
FFs are a fundamental ingredient to compute the cross section for any
process that involves the measurement of a hadron in the final
state. Among these processes are single-hadron production in
electron-positron annihilation (SIA), semi-inclusive deep-inelastic
scattering (SIDIS), and proton-proton collisions (see also Sect.~\ref{s2:QCD_fragmentation}).

\myparagraph{Motivation} An analysis of the measurements for one or more of these processes allows for a
phenomenological determination of FFs. Measurements are compared to the
predictions obtained with a suitable parametrisation of the FFs, which is then
optimised to achieve the best global description possible. The determination
of the FFs has witnessed a remarkable progress in the last years, in particular
of the FFs of the pion, which is the most copiously produced hadron.
Three aspects have been investigated separately:
the variety of measurements analysed, the accuracy of the theoretical settings
used to compute the predictions, and the sophistication of the methodology
used to optimise FFs. In the first respect, global determinations of
FFs including recent measurements for all of the three processes
mentioned above have become available~\cite{deFlorian:2014xna};
in the second respect, determinations of FFs accurate to
next-to-next-to-leading order
(NNLO)~\cite{Anderle:2015lqa,Bertone:2017tyb} or including all-order
resummation~\cite{Anderle:2016czy} have been presented, albeit based
on SIA data only; in the third respect, determinations of FFs using
modern optimisation techniques that minimise parametrisation
bias~\cite{Bertone:2017tyb}, or attempting a simultaneous
determination of the parton distribution functions
(PDFs)~\cite{Moffat:2021dji}, have been performed. These three aspects
have also been investigated for the FFs of the
kaon~\cite{Sato:2016wqj, deFlorian:2017lwf,Bertone:2017tyb,Ethier:2017zbq,
  Sato:2019yez,Moffat:2021dji}.

\myparagraph{Outline} In this chapter, I present a determination of the FFs of charged pions,
called {\tt MAPFF1.0}~\mycite{Khalek:2021gxf}, in which the most updated SIA and SIDIS
measurements are analysed up to next-to-leading order (NLO) accuracy in
perturbative QCD. 
In Sect.~\ref{s1:MAPFF10_framework} 
In Sect.~\ref{s1:MAPFF10_results}, I present the results of
our analysis, where we assess the interplay between SIA and SIDIS data sets
and the stability of FFs upon variation of the kinematic
cuts.

\section{\texttt{MAPFF1.0} framework}
\label{s1:MAPFF10_framework}
The focus in \texttt{MAPFF1.0} framework is on a proper statistical treatment of
experimental uncertainties and of their correlations in the
representation of FF uncertainties, and on the efficient minimisation
of model bias in the optimisation of the FF parametrisation. These
goals are achieved by means of a fitting methodology that is inspired
by the framework developed by the NNPDF Collaboration for the
determination of the proton
PDFs~\cite{Ball:2008by,Ball:2012cx,Ball:2014uwa,Ball:2017nwa,Nocera:2014gqa,
  Ball:2013lla}, nuclear
PDFs~\mycite{AbdulKhalek:2019mzd,AbdulKhalek:2020yuc}, and
FFs~\cite{Bertone:2017tyb,Bertone:2018ecm}.  The framework combines
the Monte Carlo sampling method to map the probability density
distribution from the space of data to the space of FFs and neural
networks (NNs) to parametrise the FFs with minimal bias.  In comparison to
previous work~\cite{Bertone:2017tyb,Bertone:2018ecm}, the input data set in \texttt{MAPFF1.0} is extended to SIDIS and the NN is
optimised by means of a gradient descent algorithm that makes use of
the knowledge of the analytic derivatives of the NN
itself~\mycite{AbdulKhalek:2020uza}.

In Sect.~\ref{s2:MAPFF10_data}, I
introduce the data sets used in this analysis, their features, and the
criteria applied to select the data points. In Sect.~\ref{s2:MAPFF10_theory},
I discuss the setup used to compute the theoretical
predictions, focusing on the description of SIDIS multiplicities. In
Sect.~\ref{s2:MAPFF10_parameterisation}, I illustrate the methodological framework
adopted in our analysis, specifically the treatment of the
experimental uncertainties and the details of the NN
parametrisation. 
\subsection{Experimental data}
\label{s2:MAPFF10_data}

This analysis is based on a comprehensive set of measurements of
pion-production cross sections in electron-positron SIA and in
lepton-nucleon SIDIS.  In the first case, the data corresponds to the
sum of the cross sections for the production of positively and
negatively charged pions, differential with respect to either the
longitudinal momentum fraction $z$ carried by the fragmenting parton
or the momentum of the measured pion $p_\pi$; the differential cross section is
usually normalised to the total cross section (see Sect.~2.2 in
Ref.~\cite{Bertone:2017tyb} for details). In the second case, the data
corresponds to the hadron multiplicity, that is the SIDIS cross
section normalised to the corresponding inclusive DIS cross section
(see Sect.~\ref{s2:MAPFF10_theory} for details). Multiplicities are measured separately
for the production of positively and negatively charged pions.

In the case of SIA, we consider measurements performed at CERN
(ALEPH~\cite{Buskulic:1994ft}, DELPHI~\cite{Abreu:1998vq} and
OPAL~\cite{Akers:1994ez}), DESY
(TASSO~\cite{Brandelik:1980iy,Althoff:1982dh, Braunschweig:1988hv}),
KEK (BELLE~\cite{Leitgab:2013qh} and TOPAZ~\cite{Itoh:1994kb}), and
SLAC (BABAR~\cite{Lees:2013rqd}, TPC~\cite{Aihara:1988su} and
SLD~\cite{Abe:2003iy}). These experiments cover a range of
centre-of-mass energies between $\sqrt{s}\sim 10$~GeV and
$\sqrt{s}= M_Z$, where $M_Z$ is the mass of the $Z$ boson. In the case
of SIDIS, we consider measurements performed at CERN by
COMPASS~\cite{Adolph:2016bga} and at DESY by
HERMES~\cite{Airapetian:2012ki}. The COMPASS experiment utilises a
muon beam with an energy $E_\mu=160$~GeV and a $^6$LiD target. The
HERMES experiment utilises electron and positron beams with an energy
$E_e=27.6$~GeV and hydrogen or deuterium target. Both experiments
measure events within a specific fiducial region. The features of SIA
and SIDIS data are summarised in Tab.~2.1 of
Ref.~\cite{Bertone:2017tyb} and in Tab.~\ref{tab:SIDIS},
respectively. Specific choices that concern some of the available data
sets are discussed below.

\begin{table}[!t]
  \centering
  \small
  \renewcommand{\arraystretch}{1.0}
  \begin{tabularx}{\textwidth}{cccccc}
    \toprule Data set & Ref.  & $N_{\rm dat}$ & Targets &
    $E_{\rm beam}$ [GeV] & Fiducial cuts\\
    \midrule COMPASS & \cite{Adolph:2016bga} & 314 (622) & $^6$LiD &
    160 & $W\geq 5$~GeV, $ 0.1 \leq y\leq 0.7$ \\
    HERMES & \cite{Airapetian:2012ki} & \ \ \ 8 (72) & H, $^2$H &
    27.6 & $W\geq \sqrt{10}$~GeV, $0.1 \leq
    y\leq 0.85$\\
    \bottomrule
  \end{tabularx}
  \caption{\small A summary of the features of the SIDIS data included
    in this analysis. For each of the two data sets we indicate the
    reference, the number of data points after (before) kinematic
    cuts, the target, the beam energy, and the experimental cuts on
    the invariant mass of the final state $W$ and of the inelasticity
    $y$ that define the fiducial region.}
  \label{tab:SIDIS}
\end{table}

Concerning the BELLE experiment, we use the measurement corresponding
to an integrated luminosity
$\mathcal{L}_{\text{int}}=68$~fb$^{-1}$~\cite{Leitgab:2013qh} in spite of the
availability of a more recent measurement based on a larger
luminosity $\mathcal{L}_{\text{int}}=558$~fb$^{-1}$~\cite{Seidl:2020mqc}. Because
of the reduced statistical uncertainties (due to the higher luminosity
of the data sample), the second measurement is significantly more
precise than the first. Therefore, the ability to describe this data
set in a global analysis of FFs crucially depends on the control of
the systematic uncertainties. At present, such a control is
unfortunately lacking. Examples are the unrealistically large
asymmetry of the uncertainties (mainly due to the Pythia tune to
correct for initial-state radiation effects) and the unknown degree of
uncertainty correlation across data points. For these reasons, we were
not able to achieve an acceptable description of the data set of
Ref.~\cite{Seidl:2020mqc}, which we exclude in favour of that of
Ref.~\cite{Leitgab:2013qh}. We multiply all data points by a factor
$1/c$, with $c=1.65$. This is required to correct the data for the
fact that a kinematic cut on radiative photon events was applied to
the data sample instead of unfolding the radiative QED effects, see
Ref.~\cite{Leitgab:2013qh} for details.

Concerning the BABAR experiment, two sets of data are available, based
on {\it prompt} and {\it conventional} yields. The difference between
the two consists in the fact that the latter includes all decay
products with lifetime $\tau$ up to $3\times 10^{-1}$~s, while the
former includes only primary hadrons or decay products from particles
with $\tau\lesssim 10^{-11}$~s.  The conventional cross sections are
about 5-15\% larger than the prompt ones.  Although the conventional
sample was derived by means of an analysis which is closer to that
adopted in other experiments, it turns out to be accommodated in the
global fit worse than its prompt counterpart. We therefore include the
prompt cross section in our baseline fit. The same choice was made in
similar analyses~\cite{deFlorian:2014xna,Sato:2016wqj,Bertone:2017tyb}.

For DELPHI and SLD, in addition to the inclusive measurements, we also
include flavour-tagged measurements, whereby the production of the
observed pion has been reconstructed from the hadronisation of all
light quarks ($u$, $d$, $s$) or of an individual $b$ quark. These
measurements are unfolded from flavour-enriched samples by means of
Monte Carlo simulations and are therefore affected by additional
model uncertainties.  Similar samples for the $c$ quark have been
measured by SLD~\cite{Abe:2003iy}. However,
these are not included because we found it difficult to obtain an
optimal description of them in the fit (see
Sect.~\ref{s2:MAPFF10_fitquality}). The OPAL experiment has also measured
completely separated flavour-tagged probabilities for a quark to
hadronise in a jet containing a pion~\cite{Abbiendi:1999ry}.  The interpretation of these measurements is ambiguous in
perturbative QCD beyond leading order, therefore they are not included
in this analysis.

The HERMES multiplicities are presented for various projections of the
fully differential measurement in $P_{h\perp}$, $x$, $z$, and $Q^2$:
these are respectively the transverse component of the hadron momentum
$p_\pi$, the momentum fractions carried by the struck and the
fragmenting parton, and the virtuality of the incoming photon. We use
the projected measurement provided as a function of $Q^2$ and $z$ in
single bins in $x$. We discard the bins with $z<0.2$, which are used
to control the model dependence of the smearing-unfolding procedure,
and with $z>0.8$, which lie in the region where the fractional
contribution from exclusive processes becomes sizeable.

\begin{figure}[!t]
 \begin{center}
   \includegraphics[width=0.75\textwidth]{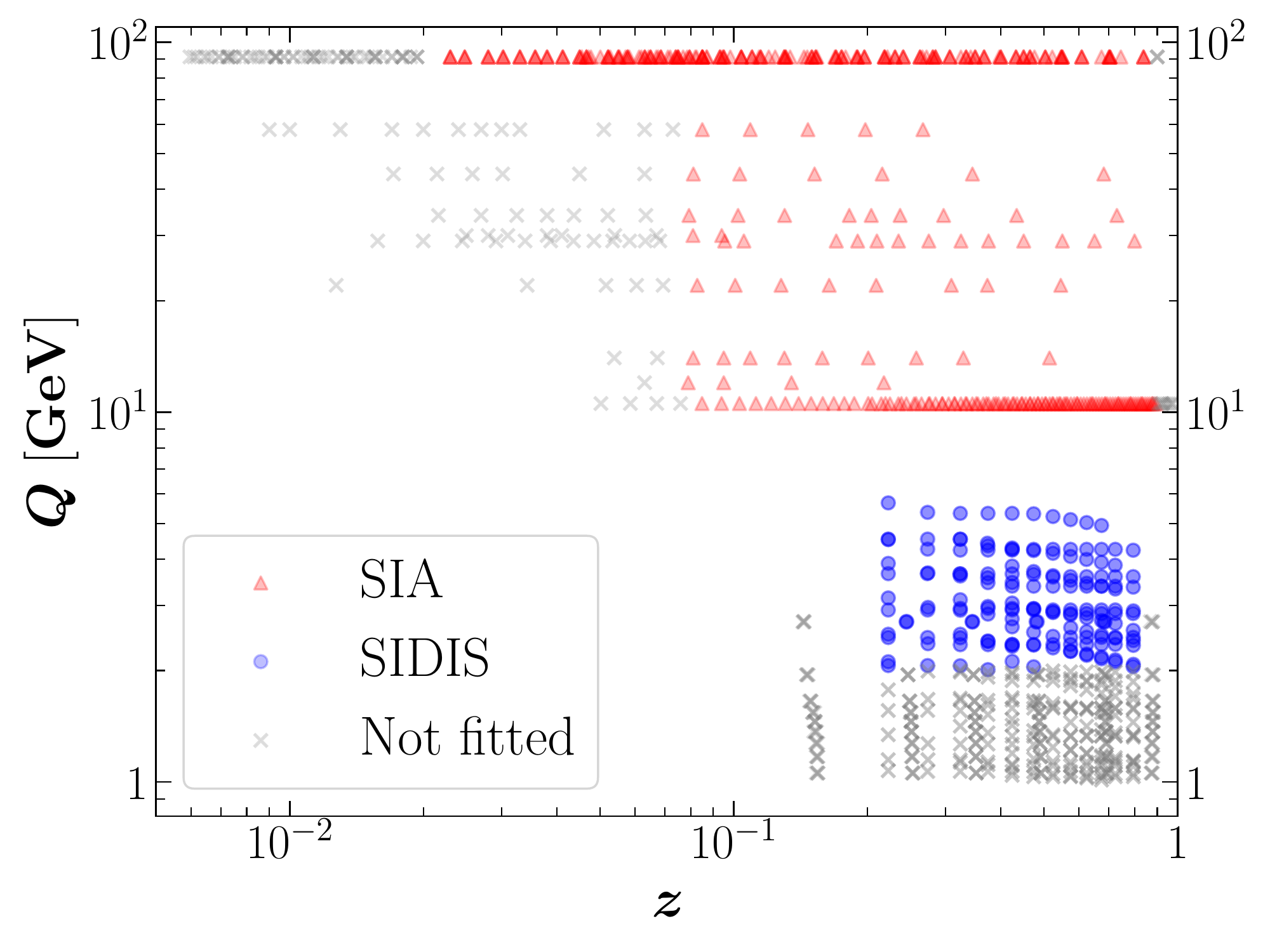}\\
 \end{center}
 \caption{\small Kinematic coverage in the $(z, Q)$ plane of the data
   set included in this analysis. Data points are from SIA (red) and
   SIDIS (blue); gray points are excluded by kinematic cuts.}
 \label{fig:kinplot}
\end{figure}

The kinematic coverage of the data sets included in this analysis is
displayed in Fig.~\ref{fig:kinplot}. As is apparent, SIA and SIDIS data
sets cover two different regions in $Q$: the former range from the
centre-of-mass energy of the $B$-factory measurements, $Q\sim 10$~GeV,
to that of LEP measurements, $Q=M_Z$; the latter, instead, lie at
lower energy scales, $Q\sim 1 - 6$~GeV.
The two data sets are nevertheless complementary. On the one hand,
SIDIS data widens the $Q$ lever arm needed to determine
the gluon FF from perturbative evolution effects. On the other hand,
SIDIS data provides a direct constraint on individual quark and
antiquark FFs, that are otherwise always summed in SIA data.
As expected from kinematic considerations, experiments at higher
centre-of-mass energies provide data at smaller values of $z$.

Kinematic cuts are applied to select only the data points for
which perturbative fixed-order predictions are reliable. For SIA, we
retain the data points with $z$ in the range
$[z_{\rm min}, z_{\rm max}]$; the values of $z_{\rm min}$ and
$z_{\rm max}$ are chosen as in Ref.~\cite{Bertone:2017tyb}:
$z_{\rm min}=0.02$ for experiments at a centre-of-mass energy of $M_Z$
and $z_{\rm min}=0.075$ for all other experiments; $z_{\rm max}=0.9$
for all experiments. For SIDIS, we retain the data points satisfying
$Q>Q_{\rm cut}$, with $Q_{\rm cut}=2$~GeV. This choice maximises the
number of data points included in the fit without spoiling its quality
and follows from a study of the stability of the fit upon choosing
different values of $Q_{\rm cut}$, see Sect.~\ref{s2:MAPFF10_determination}. 
Overall, after kinematic cuts, we consider $N_{\rm dat}=699$ data
points in our baseline fit almost equally split between SIA
($N_{\rm dat}=377$) and SIDIS ($N_{\rm dat}=322$). Each of the two
processes is dominated by measurements coming respectively from LEP
and $B$-factories, and from COMPASS. 

Information on correlations of experimental uncertainties is taken
into account whenever available. Specifically, for the BABAR
measurement, which is provided with a breakdown of bin-by-bin
correlated systematic uncertainties, for the HERMES measurement,
which is provided with a set of covariance matrices accounting for
correlations of statistical uncertainties obtained from the unfolding
procedure, and for the COMPASS measurement, which is provided with a correlated
systematic uncertainty. Normalisation uncertainties available for the BELLE,
BABAR, TASSO, ALEPH and SLD experiments are assumed to be fully correlated
across all data points in each experiment. If the degree of correlation of
systematic uncertainties is not known, we sum them in quadrature with
the statistical uncertainties. Finally, we symmetrise the systematic uncertainties
reported by the BELLE experiment as described in Ref.~\cite{DAgostini:2004kis}.

\subsection{Theoretical setup}
\label{s2:MAPFF10_theory}

In this section we discuss the theoretical setup used to compute the theoretical
predictions for the SIDIS multiplicities corresponding to the measurements
performed by COMPASS and HERMES. The computation of SIA cross sections and 
the evolution of FFs closely follow Refs.~\cite{Bertone:2017tyb,
  Bertone:2015cwa} and are therefore not discussed here.

We consider the inclusive production of a charged pion, $\pi^\pm$,
in lepton-nucleon scattering as summarised in Sect.~\ref{s2:Flavour_separation}.
We use the \texttt{NNPDF31\_nlo\_pch\_as\_0118}~\cite{Ball:2017nwa} as a
reference PDF set. In Sect.~\ref{s2:MAPFF10_parameterisation} we will explain how the
PDF uncertainty is propagated into SIDIS observables and in
Sect.~\ref{s2:MAPFF10_determination} we will discuss the impact of alternative PDF sets
on the determination of FFs.

The coefficient functions $C$ in Eq.~\eqref{eq:f1sidis} admit the
usual perturbative expansion
\begin{equation}\label{eq:pertexpC}
C(x,z,Q) = \sum_{n=0}\left(\frac{\alpha_s(Q)}{4\pi}\right)^nC^{(n)}(x,z)\,,
\end{equation}
where $\alpha_s$ is the running strong coupling for which we choose
$\alpha_s(M_Z)=0.118$ as a reference value. Presently,
the full set of coefficient functions for both $F_2$ and $F_L$ is only
known up to $\mathcal{O}(\alpha_s)$, \textit{i.e.}
NLO~\cite{Altarelli:1979kv,Graudenz:1994dq}. Explicit $x$- and $z$-space
expressions up to this order can be found for instance in
Ref.~\cite{deFlorian:1997zj} and are implemented in the public code
{\tt APFEL++}~\cite{Bertone:2013vaa, Bertone:2017gds}. A subset of the
$\mathcal{O}(\alpha_s^2)$, \textit{i.e.} NNLO, corrections has been recently
presented in Refs.~\cite{Anderle:2016kwa,Anderle:2018qrw}. However, as long as
the full set of $\mathcal{O}(\alpha_s^2)$ corrections are not known, NNLO
accuracy cannot be attained. For this reason in this analysis
we limit ourselves to NLO accuracy that amounts to considering the
first two terms in the sum in Eq.~\eqref{eq:pertexpC}. For
consistency, also the $\beta$-function and the splitting functions
responsible for the evolution of the strong coupling $\alpha_s$ and of
the FFs, respectively, are computed to NLO accuracy.

So far no heavy-quark mass corrections have been computed for
SIDIS. Therefore, our determination of FFs relies on the so-called
zero-mass variable-flavour-number scheme (ZM-VFNS). In this scheme all
active partons are treated as massless but a partial heavy-quark mass
dependence is introduced by requiring that sub-schemes with different
numbers of active flavours match at the heavy-quark thresholds. Here
we choose $m_c = 1.51$~GeV and $m_b=4.92$~GeV for the charm and bottom
thresholds, respectively, as in the
\texttt{NNPDF31\_nlo\_pch\_as\_0118} PDF set.  In view of the fact
that intrinsic heavy-quark FFs play an important role, it is important
to stress that in our approach inactive-flavour FFs, such as the charm
and bottom FFs below the respective thresholds, are \textit{not} set
to zero. On the contrary, they are allowed to be different from zero
but are kept constant in scale below threshold, \textit{i.e.}  they do
not evolve. This has the consequence that heavy-quark FFs do
contribute to the computation of cross sections also below their
threshold. However, in the specific case of SIDIS this contribution is
suppressed by the PDFs and only appears at NLO.\footnote{This is
  strictly true when using a PDF set that does not include any
  intrinsic heavy-quark contributions as we do here.}

A property of the expressions for the perturbative
coefficient functions $C$ is that the functions $C^{(n)}(x,z)$ with
$n=0,1$ (see Eq.~\eqref{eq:pertexpC}) are bilinear combinations of
single-variable functions:
\begin{equation}
C^{(n)}(x,z) = \sum_t c_tO_t^{(1)}(x)O_t^{(2)}(z)\,,
\end{equation}
where $c_t$ are numerical coefficients. This property enables one to
decouple the double convolution integral in
Eq.~\eqref{eq:doubleconvolution} into a linear combination of single
integrals
\begin{equation}
C^{(n)}(x,z)\otimes f(x)\otimes D(z) = \sum_t
c_t\left[O_t^{(1)}(x)\otimes f(x)\right]\left[O_t^{(2)}(z)\otimes D(z)\right]\,.
\end{equation}
This observation allowed us to considerably speed up the numerical
computation of the SIDIS cross sections.

In order to benchmark the implementation in {\tt APFEL++} used for the
fits and based on the $x$- and $z$-space expressions of
Ref.~\cite{deFlorian:1997zj}, we have carried out a totally
independent implementation of the SIDIS cross section based on the
Mellin-moment expressions~\cite{Stratmann:2001pb,
  Anderle:2012rq}. We made the
Mellin-space version of the cross section publicly available through
the code {\tt MELA}~\cite{Bertone:2015cwa}. The outcome of the
benchmark was totally satisfactory in that in the kinematic region
covered by HERMES and COMPASS the agreement between {\tt APFEL++} and
{\tt MELA} was well below the per-mil level.

The quantity actually measured by both the HERMES and COMPASS
experiments is not an absolute cross section, Eq.~\eqref{eq:sidis2}, but
rather an integrated multiplicity defined as
\begin{equation}\label{eq:multiplicities}
  \frac{dM}{dz} = \left. \left[\int_{Q_{\rm min}}^{Q_{\rm max}}dQ \int_{x_{\rm
          min}}^{x_{\rm max}}dx \int_{z_{\rm min}}^{z_{\rm
          max}}dz\frac{d^3\sigma}{dx dQ dz}\right]\right/
  \left[\Delta z \int_{Q_{\rm min}}^{Q_{\rm max}}dQ \int_{x_{\rm min}}^{x_{\rm max}}dx\frac{d^2\sigma}{dx dQ}\right]\,,
\end{equation}
where the integration bounds define the specific kinematic bin and
$\Delta z = z_{\rm max}-z_{\rm min}$. The denominator is given by the
DIS cross section inclusive w.r.t. the final state that is thus
independent from the FFs. Despite NNLO and heavy-quark mass
corrections are known for the inclusive DIS cross sections, we use the
ZM-VFNS at NLO also in the denominator of
Eq.~\eqref{eq:multiplicities} to match the accuracy of the
numerator. However, we have checked that including NNLO and/or
heavy-quark mass corrections into the inclusive DIS cross section
makes little difference on the determination of FFs.

While the multiplicities measured by the HERMES experiment are binned
in the variables $\{x, Q^2, z\}$, exactly matching the quantity in
Eq.~\eqref{eq:multiplicities}, those measured by the COMPASS
ex\-pe\-ri\-ment are binned in the variables $\{x, y, z\}$ with $y$
defined in Sect.~\eqref{s2:Physical_constraints}. In this case, theoretical
predictions are obtained after adjusting the integration bounds in $Q$
and $x$ in Eq.~\eqref{eq:multiplicities} that become
\begin{equation}
Q_{\rm min} = \sqrt{x_{\rm min}y_{\rm min} s}\,,\qquad Q_{\rm max} = \sqrt{x_{\rm max}y_{\rm max} s}\,,
\end{equation}
and
\begin{equation}\label{eq:compassrepl}
x_{\rm min} \rightarrow {\rm max}\left[x_{\rm min},\frac{Q^2}{sy_{\rm
    max}}\right]\qquad x_{\rm max} \rightarrow {\rm min}\left[x_{\rm max},\frac{Q^2}{sy_{\rm
    min}}\right]\,,
\end{equation}
with $y_{\rm min}$ and $y_{\rm max}$ the bin bounds in $y$. Moreover,
both HERMES and COMPASS measure cross sections within a specific
fiducial region given by
\begin{equation}
W=\sqrt{\frac{(1-x)Q^2}{x}} \geq W_{\rm low}\,,\quad y_{\rm low} \leq
y\leq y_{\rm up}\,,
\end{equation}
with the values of $W_{\rm low}$, $y_{\rm low}$, and $y_{\rm up}$
reported in Tab.~\ref{tab:SIDIS}. These constraints reduce the phase
space of some bins placed at the edge of the fiducial region. The net
effect is that of replacing the $x$ integration bounds in
Eq.~\eqref{eq:multiplicities} with
\begin{equation}
  x_{\rm min}\rightarrow\overline{x}_{\rm min} = {\rm max}\left[x_{\rm min},\frac{Q^2}{sy_{\rm
        up}}\right]\quad\mbox{and}\quad x_{\rm max}\rightarrow\overline{x}_{\rm max} = {\rm min}\left[x_{\rm max},\frac{Q^2}{sy_{\rm
        low}},\frac{Q^2}{Q^2+W_{\rm low}^2}\right]\,,
\end{equation}
with $x_{\rm min}$ and $x_{\rm max}$ to be interpreted as in
Eq.~(\ref{eq:compassrepl}) in the case of COMPASS. We stress that in
our determination of FFs all the integrals in
Eq.~\eqref{eq:multiplicities} are duly computed during the fit. The
effect of computing these integrals, in comparison to evaluating the
cross sections at the central point of the bins, is modest for COMPASS
but sizeable for HERMES.  However, in both cases the integration
contributes to achieving a better description of the data.

Both the HERMES and COMPASS experiments measure multiplicities for
$\pi^+$ and $\pi^-$ separately. However, $\pi^+$ and $\pi^-$ are
related by charge conjugation. In practice, this means that it is
possible to obtain one from the other by exchanging quark and
antiquark distributions and leaving the gluon unchanged:
\begin{equation}
D_{q(\overline{q})}^{\pi^-}(x,Q) =
D_{\overline{q}(q)}^{\pi^+}(x,Q)\,,\quad D_{g}^{\pi^-}(x,Q) =
D_{g}^{\pi^+}(x,Q)\,.
\end{equation}
In this analysis, we use this symmetry to express the $\pi^-$ FFs in
terms of the $\pi^+$ ones and effectively only extract the latter.

We finally note that part of the HERMES measurements and all of the
COMPASS ones are performed on isoscalar targets (deuterium for HERMES
and lithium for COMPASS, see Table~\ref{tab:SIDIS}).
To account for this we have adjusted the PDFs
of the target by using SU(2) isospin symmetry to deduce the neutron
PDFs from the proton ones, which simply amounts to exchanging (anti) up
and (anti) down PDFs, and taking the average between proton and neutron
PDFs. No nuclear corrections are taken into account; no target mass corrections
are considered either, given the complexity to consistently account for them
together with final-state hadron mass corrections~\cite{Guerrero:2015wha}.

\subsection{Parameterisation}
\label{s2:MAPFF10_parameterisation}

The statistical framework that we adopted for the inference of the
\texttt{MAPFF1.0} FFs from experimental data relies on the Monte Carlo
sampling method explained in Sect.~\ref{s2:monte_carlo}.
The Monte Carlo method aims at propagating the experimental
uncertainties into the space of parameters defined in our case by a
NN. In order to do so, we generate $N_{\text{rep}}$
replicas of the data (see Eq.~(\ref{eq:MCgen})) that we use independently for the inference. In the case of
SIDIS, a different proton PDF replica taken at random from the
\texttt{NNPDF31\_nlo\_pch\_as\_0118} set is associated to each replica
data replica. This ensures that the PDF uncertainty is propagated
into the FF uncertainty.

In order to choose the best set of independent FF combinations that define our
parametrisation basis, we study three different cases.
\begin{enumerate}
\item \textbf{11 independent flavours}. This is the most general case
  implied by Eq.~(\ref{eq:f1sidis}) where one aims at disentangling
  all FF flavours and the gluon FF. This parametrisation is overly redundant in
  that the data set used is not able to constrain all 11 combinations.

\item \underline{\textbf{7 independent flavours}}. The sea
  distributions are assumed to be partially symmetric. Specifically,
  $D^{\pi^+}_q=D^{\pi^+}_{\bar{q}}$, for $q=s,c,b$, and
  $D^{\pi^+}_d=D^{\pi^+}_{\bar{u}}$. By doing so, we reduce the number
  of independent distributions down to 7. We observe that under these
  assumptions the quality of the fit does not significantly
  deteriorate with respect to the most general case discussed
  above. In particular, we find it to be the best solution in terms of
  generality and accuracy and therefore we adopt it as our baseline
  parametrisation.

\item \textbf{6 independent flavours}. The approximate SU(2) isospin
  symmetry would suggest that the additional constraint
  $D^{\pi^+}_{u} = D^{\pi^+}_{\bar{d}}$ may hold, further lowering the
  number of independent combinations down to 6. However, it turns out
  that this additional assumption leads to a deterioration
  of the quality of the fit therefore we dropped it.
\end{enumerate}
Finally, the set of 7 independent FF combinations parametrised in our
fit are:
\begin{equation} \label{eq:param_flavours}
\{D^{\pi^+}_u, \,\,\, D^{\pi^+}_{\bar{d}}, \,\,\, D^{\pi^+}_{d}=D^{\pi^+}_{\bar{u}}, \,\,\, D^{\pi^+}_s=D^{\pi^+}_{\bar{s}}, \,\,\, D^{\pi^+}_c=D^{\pi^+}_{\bar{c}}, \,\,\, D^{\pi^+}_b=D^{\pi^+}_{\bar{b}}, \,\,\, D^{\pi^+}_g \}\,.
\end{equation}

\begin{figure}[ht]
  \floatbox[{\capbeside\thisfloatsetup{capbesideposition={right,top},capbesidewidth=0.6\textwidth}}]{figure}[\FBwidth]
  {\caption{The NN architecture $\{1,\,20,\,7\}$ used in \texttt{MAPFF1.0} having 1 input nodes designating the kinematic variable $z$, one hidden layer with 20 neurons having a sigmoid activation function and the output nodes with a linear activation function designating $N_i(x)$ of some FF flavour $i$. An overall of 187 free parameters (weights and biases).}\label{fig:MAPFF_architecture}}
  {\vspace{-0.5cm}\includegraphics[width=0.35\textwidth]{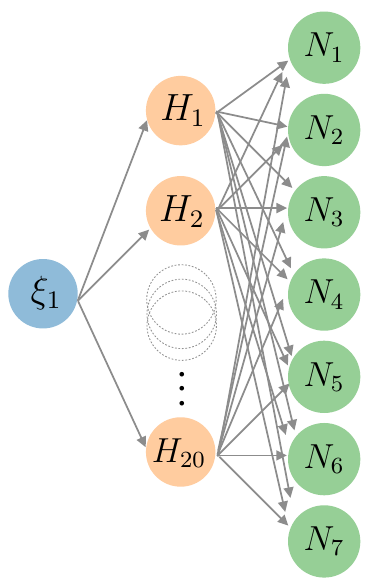}} 
  \end{figure}

The parametrisation is introduced at the initial scale
$\mu_0 = 5\,\text{GeV}$ and consists of a single one-layered
feed-forward NN $N_i(z;\bm{\theta})$, where
$\bm{\theta}$ denotes the set of parameters. This network, as displayed in Fig.~\ref{fig:MAPFF_architecture} has one
input node corresponding to the momentum fraction $z$, 20 intermediate
nodes with a sigmoid activation function, and 7 output nodes with a
linear activation function corresponding to the flavour combinations
in Eq.~(\ref{eq:param_flavours}). This architecture $[1,\,20,\,7]$
amounts to a total of 187 free parameters. We do not include any
power-like function to control the low- and high-$z$ behaviours,
however we do impose the kinematic constraint $D_i^{\pi^+}(z=1)=0$ by
simply subtracting the NN itself at $z=1$ as done in
Ref.~\cite{Bertone:2017tyb}. Moreover, we constrain the FFs to be
positive-definite by squaring the outputs. This choice is motivated by
the fact that allowing for negative distributions leads to FFs that
may become unphysically negative. Our parametrisation finally reads:
\begin{equation}
    zD^{\pi^+}_i(z, \mu_0 = 5\,\text{GeV}) = \Big(N_i(z;\bm{\theta}) - N_i(1;\bm{\theta})\Big)^2\,,
\end{equation}
where the index $i$ runs over the combinations in
Eq.~(\ref{eq:param_flavours}).

The fit is performed by maximising the log-likelihood
$\mathcal{L}\left(\bm{\theta}|\bm{x}^{(k)}\right)$, which is the
probability of observing a given data replica $\bm{x}^{(k)}$ given the
set of parameters $\bm{\theta}^{(k)}$. Together with the multivariate
Gaussian assumption, this is equivalent to minimising the
$\chi^2$ defined in Eq.~(\ref{eq:chi2_general}) as:
\begin{equation}
\chi^{2(k)} \equiv  \left(\bm{T}(\bm{\theta}^{(k)})- \bm{x}^{(k)}\right)^T \cdot \bm{C}^{-1}\cdot \left(\bm{T}(\bm{\theta}^{(k)})- \bm{x}^{(k)}\right)\,,
\end{equation}
where $\bm{T}(\bm{\theta}^{(k)})$ is the set of theoretical
predictions with the NN parametrisation as input. We set
$T_i(\bm{\theta}^{(k)})=\mu_i$ if point $i$ does not satisfy the
kinematic cuts defined in Sect.~\ref{s2:MAPFF10_data} or if it belongs to
the validation set. This last procedure promotes the likelihood into a conditional one (see Sect.~\ref{s2:likelihood}), that upon maximisation, leads to a set of parameters inferred from a subset of data points that survive the cuts while accounting for correlations from those who do not.

We adopt the cross-validation procedure in order to avoid
\textit{overfitting} our FFs (see Sect.~\ref{s2:neural_networks}). For each data replica, data sets
amounting to more than 10 data points are randomly split into
\textit{training} and \textit{validation} subsets, each containing
half of the points, and only those in the training set are used in the
fit. Data sets with 10 or less data points are instead fully included
in the training set. The $\chi^2$ of the validation set is monitored
during the minimisation of the training $\chi^2$ and the fit is
stopped when the validation $\chi^2$ reaches its absolute minimum.
Replicas whose total $\chi^2$ per point, \textit{i.e.} the $\chi^2$
computed over all data points in the fit, is larger than 3 are
discarded.

The minimisation algorithm adopted for our fit is a
\textit{trust-region} algorithm, specifically the Levenberg-Marquardt
algorithm as implemented in the \texttt{Ceres Solver}
code~\cite{ceres-solver}. This is an open source \texttt{C++} library
for modelling and solving large optimisation problems. The
neural-network parametrisation and its analytical derivatives with
respect to the free parameters are provided by
\texttt{NNAD}~\mycite{AbdulKhalek:2020uza} (see also Sect.~\ref{s2:neural_networks}), an open source \texttt{C++}
library that provides a fast implementation of an arbitrarily large
feed-forward NN and its analytical derivatives.

\section{Results}
\label{s1:MAPFF10_results}

In this section I present the main results of this analysis. In
Sect.~\ref{s2:MAPFF10_fitquality}, I discuss the quality of our baseline fit
that we dub {\tt MAPFF1.0}. In Sect.~\ref{s2:MAPFF10_determination}, I illustrate its
features: comparing our FFs to other recent determinations and  
studying the stability of the fit upon the choice of input PDFs and of
the parametrisation scale $\mu_0$. Finally, in Sect.~\ref{s2:MAPFF10_dataimpact}, I discuss the impact of some specific data sets: the origin of
the difficulty in fitting SIA charm-tagged data, the
impact of the SIDIS data on FFs, and finally the dependence
of the fit quality on the low-$Q$ cut on the SIDIS data.

\subsection{Inference quality}
\label{s2:MAPFF10_fitquality}

Tab.~\ref{tab:chi2s} reports the value of the $\chi^2$ per data point for
the individual data sets included in the {\tt MAPFF1.0} fit along with the
number of data points $N_{\rm dat}$ that pass the kinematic cuts discussed
in Sect.~\ref{s2:MAPFF10_data}. The table also reports the partial $\chi^2$ values
of the SIDIS and SIA data sets separately as well as the global one.

\begin{table}
\renewcommand{\arraystretch}{1.1}
\centering
\small
\begin{tabular}{lccccccccc}
  \toprule
  Experiment & $\chi^2$ per point & $N_{\rm dat}$ after cuts \\
  \midrule
  HERMES $\pi^-$ deuteron & 0.60 &  2 \\
  HERMES $\pi^-$ proton & 0.02 &  2 \\
  HERMES $\pi^+$ deuteron & 0.30 &  2 \\
  HERMES $\pi^+$ proton & 0.53 &  2 \\
  COMPASS $\pi^-$ & 0.80 &  157 \\
  COMPASS $\pi^+$ & 1.07 &  157 \\
  \midrule
  Total SIDIS & 0.78 & 322 \\
  \midrule
  BELLE $\pi^\pm$ & 0.09 &  70 \\
  BABAR prompt $\pi^\pm$ & 0.90 &  39 \\
  TASSO 12 GeV $\pi^\pm$ & 0.97 &  4 \\
  TASSO 14 GeV $\pi^\pm$ & 1.39 &  9 \\
  TASSO 22 GeV $\pi^\pm$ & 1.85 &  8 \\
  TPC $\pi^\pm$ & 0.22 &  13 \\
  TASSO 30 GeV $\pi^\pm$ & 0.34 &  2 \\
  TASSO 34 GeV $\pi^\pm$ & 1.20 &  9 \\
  TASSO 44 GeV $\pi^\pm$ & 1.20 &  6 \\
  TOPAZ $\pi^\pm$ & 0.28 &  5 \\
  ALEPH $\pi^\pm$ & 1.29 &  23 \\
  DELPHI total $\pi^\pm$ & 1.29 &  21 \\
  DELPHI $uds$ $\pi^\pm$ & 2.84 &  21 \\
  DELPHI bottom $\pi^\pm$ & 1.67 &  21 \\
  OPAL $\pi^\pm$ & 1.72 &  24 \\
  SLD total $\pi^\pm$ & 1.14 &  34 \\
  SLD $uds$ $\pi^\pm$ & 2.05 &  34 \\
  SLD bottom $\pi^\pm$ & 0.55 &  34 \\
  \midrule
  Total SIA & 1.10 & 377 \\
  \midrule
\textbf{Global data set} & \textbf{0.90} & \textbf{699} \\
  \bottomrule
\end{tabular}
\caption{\small Values of the $\chi^2$ per data point for the
  individual data sets included in the {\tt MAPFF1.0} analysis. The
  number of data points $N_{\rm dat}$ that pass kinematic cuts and the
  SIDIS, SIA, and global $\chi^2$ values are also displayed.}
\label{tab:chi2s}
\end{table}

The global $\chi^2$ per data point, equal to 0.90, indicates a general
very good description of the entire data set. A comparable fit quality
is observed for both the SIDIS and SIA sets separately, with
collective $\chi^2$ values equal to 0.78 and 1.10, respectively. A
closer inspection of Tab.~\ref{tab:chi2s} reveals that an acceptable
description is achieved for all of the individual data sets. Some
particularly small $\chi^2$ values are also obtained. This is the case
of the HERMES $\pi^-$ and BELLE data. In the case of HERMES, the
smallness of the $\chi^2$ is not statistically significant given that
only two data points survive the cuts. The smallness of the $\chi^2$
of BELLE is instead well-known and follows from an overestimate of the
systematic uncertainties~\cite{deFlorian:2014xna, Hirai:2016loo,
  Bertone:2017tyb, Gamberg:2021lgx}.

\begin{figure}
  \begin{center}
  \includegraphics[width=0.49\textwidth]{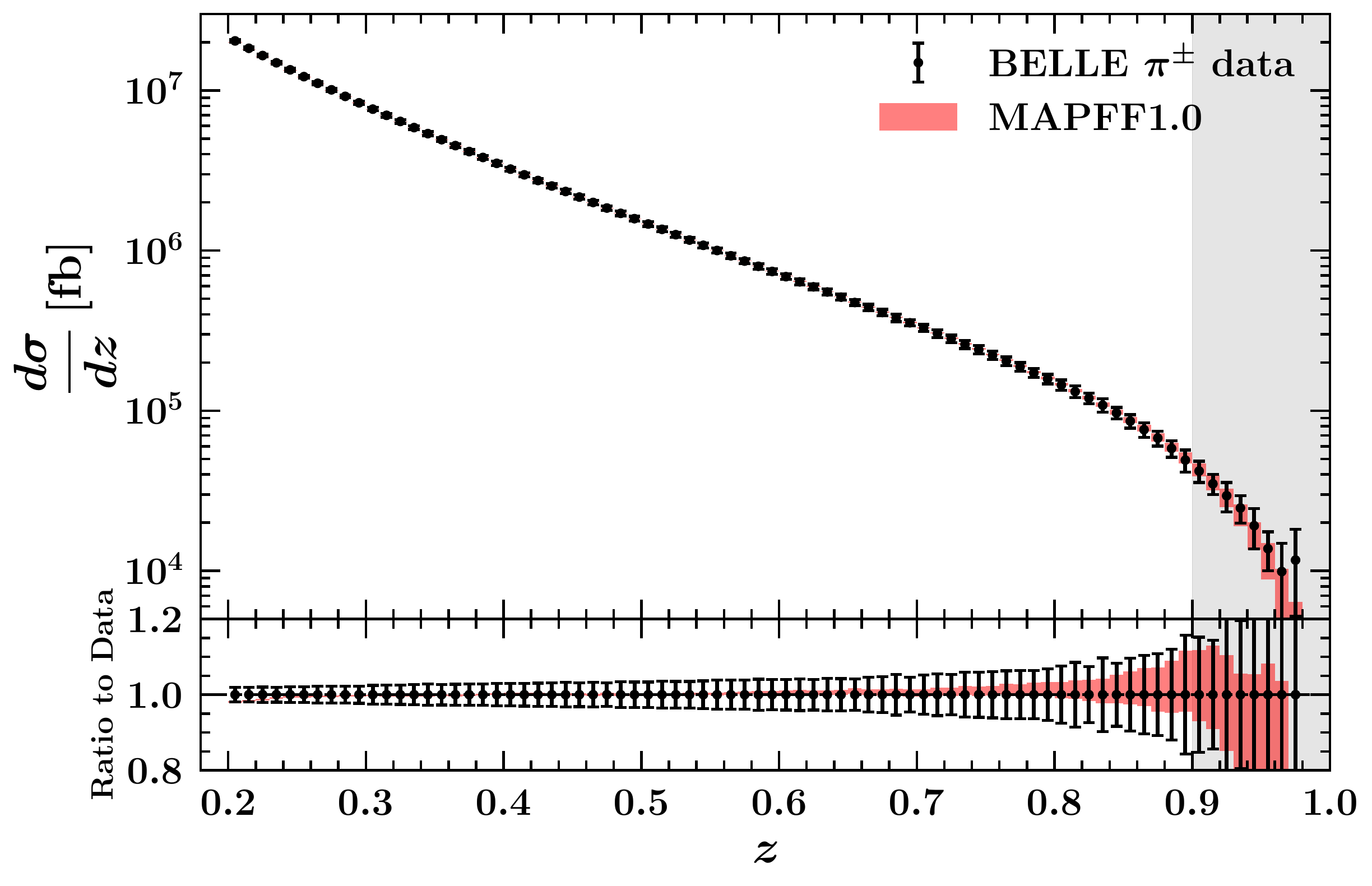}
  \includegraphics[width=0.49\textwidth]{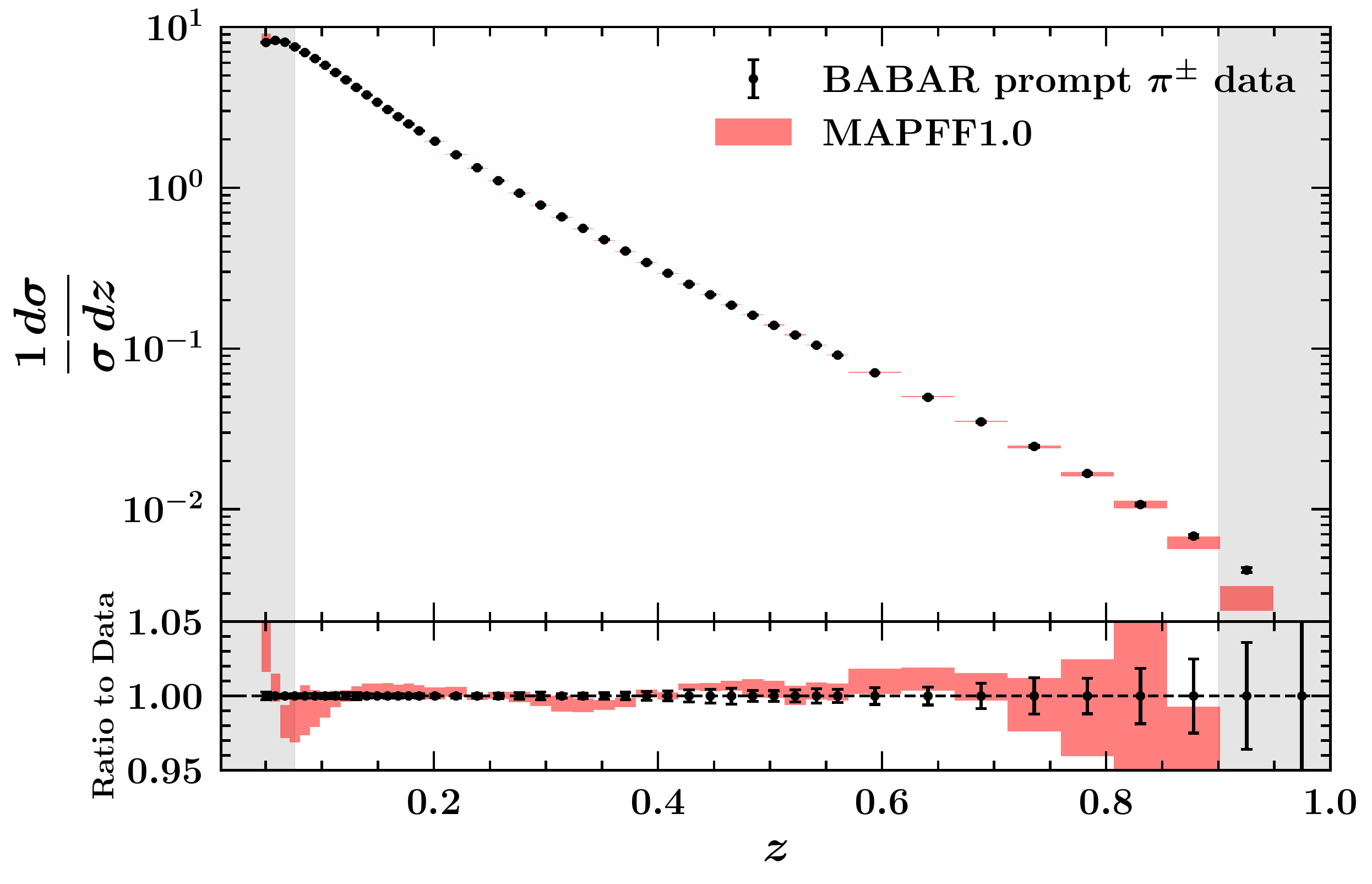}\\
  \includegraphics[width=0.49\textwidth]{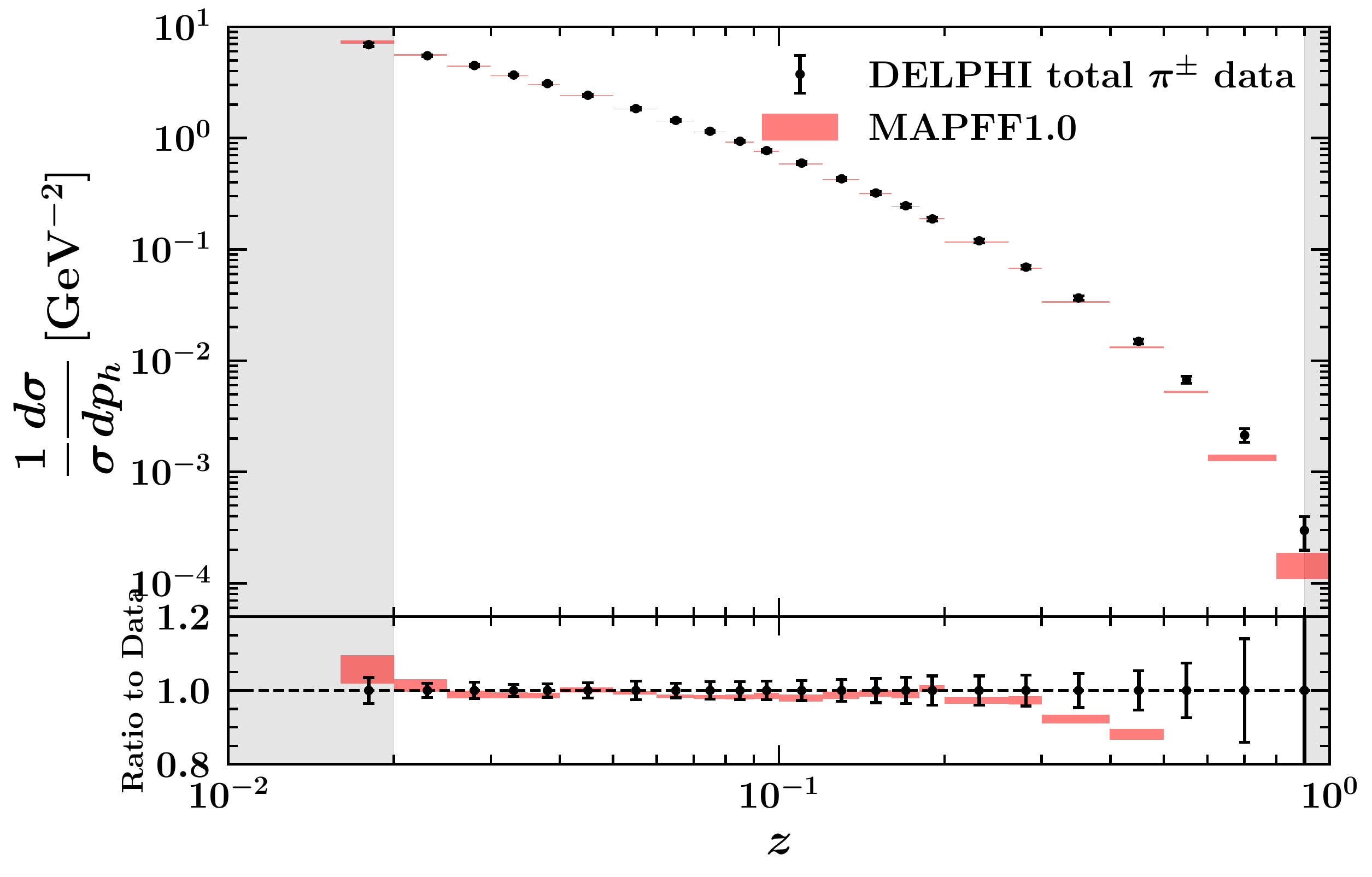}
  \includegraphics[width=0.49\textwidth]{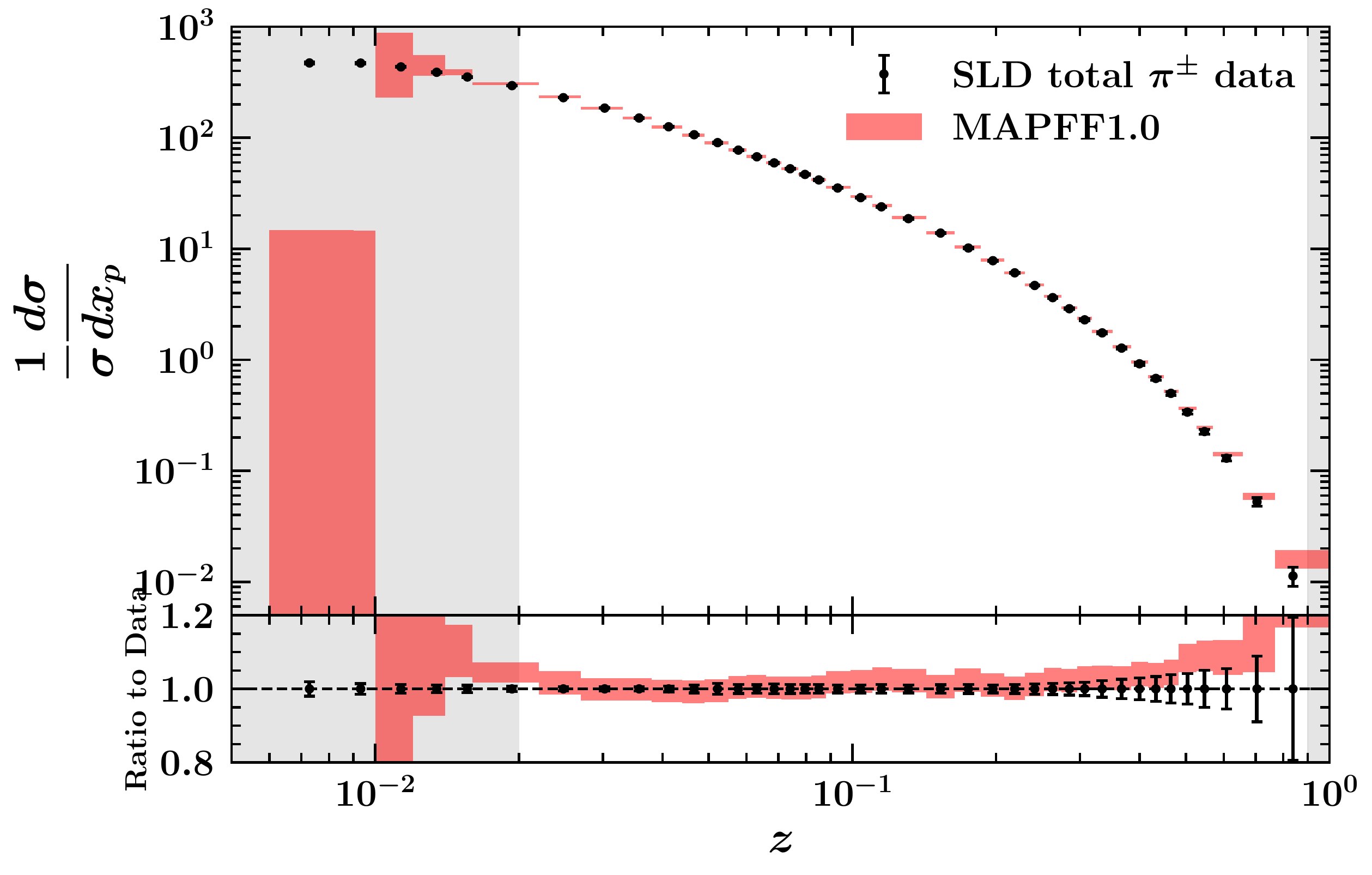}\\
  \end{center}
  \vspace{-0.8cm}
  \caption{\small Top row: data-theory comparison for the $B$-factory
    experiments BELLE (left) and BABAR (right) at
    $\sqrt{s}\simeq10.5$~GeV.  Bottom row: representative data-theory
    comparison for SIA data sets at $\sqrt{s} = M_Z$ from DELPHI
    (left) and SLD (right). The shaded regions are excluded from the
    fit.}
  \label{fig:BfactoriesMZexps}
\end{figure}

It is instructive to look at the comparison between data and
predictions obtained with the {\tt MAPFF1.0} FFs for some selected
data sets. The top row of Fig.~\ref{fig:BfactoriesMZexps} shows the
comparison for the $B$-factory experiments BELLE and BABAR at
$\sqrt{s}\simeq 10.5$ GeV, while the bottom row shows the comparison
for two representative data sets at $\sqrt{s}=M_Z$, \textit{i.e.} the
total cross sections from DELPHI and SLD. The upper panels display the
absolute distributions while the lower ones the ratio to the
experimental central values. The shaded areas indicate the regions
excluded from the fit by the kinematic cuts. In order to facilitate
the visual comparison, predictions are shifted to account for
correlated systematic uncertainties~\cite{Pumplin:2002vw}, when
present.  Consistently with the $\chi^2$ values reported in
Tab.~\ref{tab:chi2s}, the description of these data sets is very good
within cuts.

\begin{figure}
  \begin{center}
    \includegraphics[width=\textwidth]{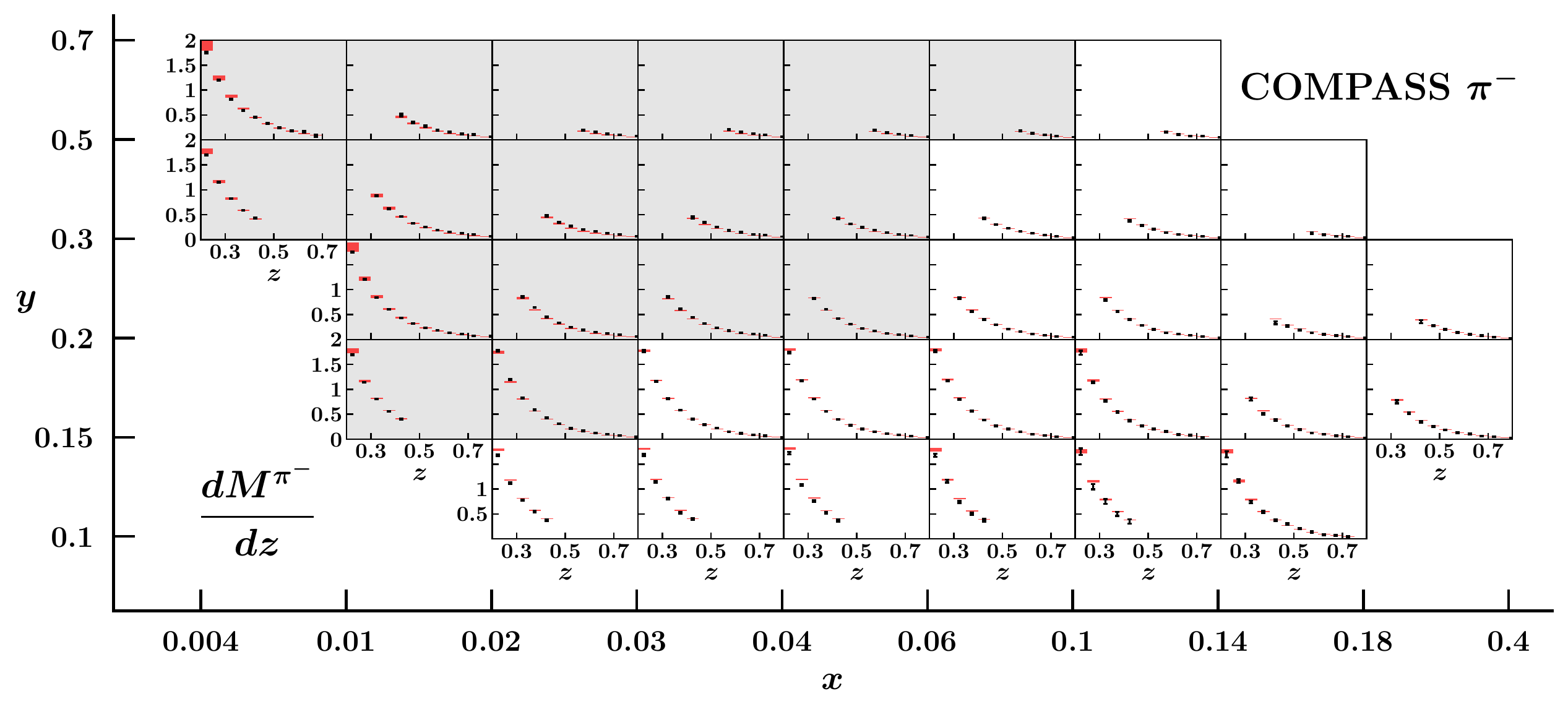}\\
  \end{center}
  \vspace{-0.8cm}
  \caption{\small Data-theory comparison for the $\pi^-$
    multiplicities from COMPASS. The shaded bins are excluded from the
    fit due to the cut in $Q$.}
  \label{fig:COMPASSall}
\end{figure}

Fig.~\ref{fig:COMPASSall} shows the data-theory comparison for the
COMPASS $\pi^-$ multiplicities. Each panel
displays a distribution in $z$ corresponding to a bin in $x$ and
$y$. As above, theoretical predictions have been shifted to ease the
visual comparison. The grey-shaded panels are not fitted because they
do not fulfil the cut in $Q$ discussed in Sect.~\ref{s2:MAPFF10_data}. Once
again, the goodness of the $\chi^2$ values in Tab.~\ref{tab:chi2s} is
reflected in a general very good description of the data. The analogous
plot for $\pi^+$ multiplicities looks qualitatively similar to
Fig.~\ref{fig:COMPASSall}, therefore it is not shown.

\subsection{The \texttt{MAPFF1.0} determination}
\label{s2:MAPFF10_determination}
I now present our FFs. I first compare them with other FF sets, then discuss the impact of some relevant theoretical choices.

\begin{figure}
  \begin{center}
    \includegraphics[width=0.9\textwidth]{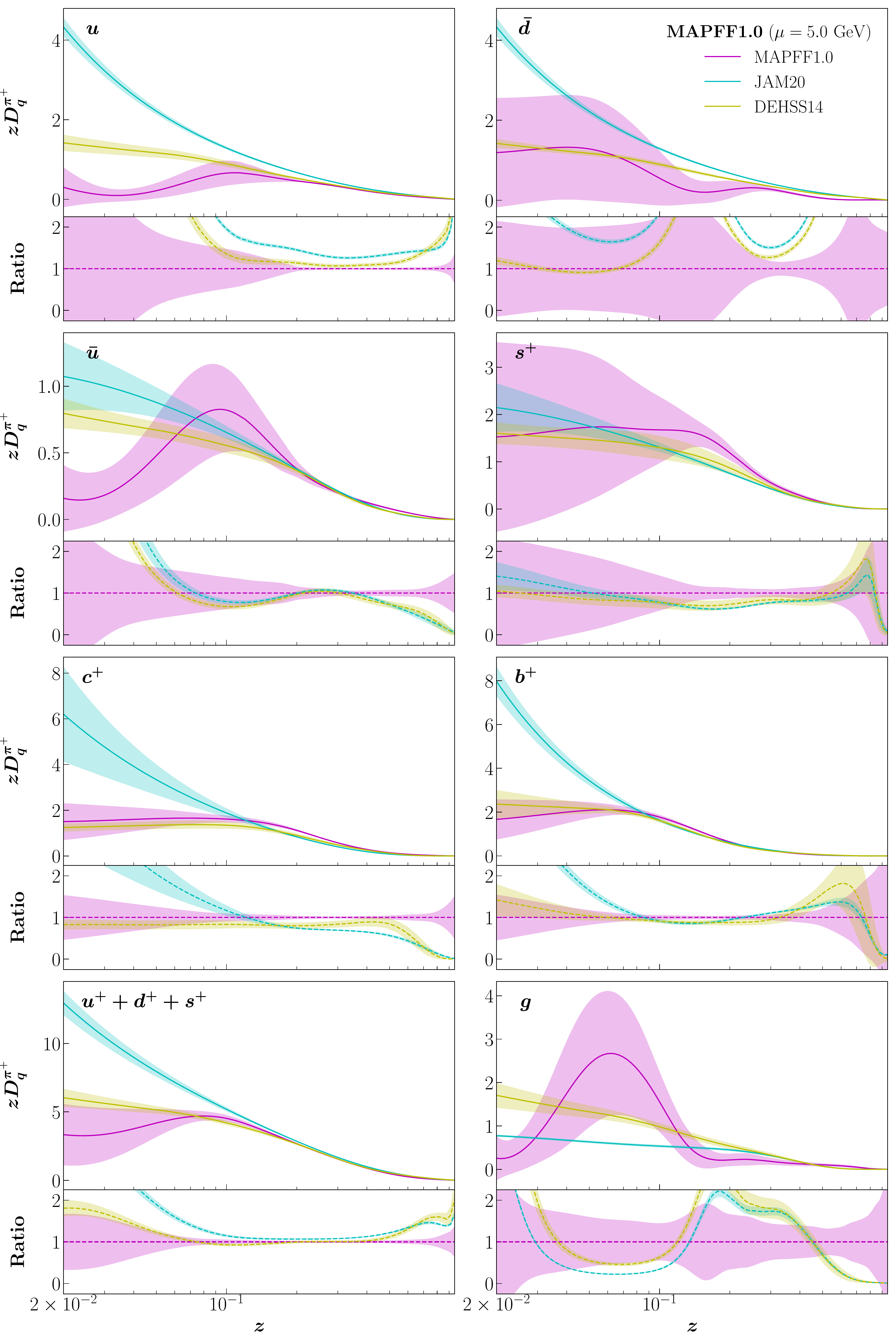}\\
    \end{center}
    \vspace{-0.8cm}
    \caption{\small Comparison of the \texttt{MAPFF1.0},
      \texttt{DEHSS14}~\cite{deFlorian:2014xna}, and
      \texttt{JAM20}~\cite{Moffat:2021dji} FFs. We display the
      $D_u^{\pi^+}$, $D_{\bar d}^{\pi^+}$, $D_{\bar u}^{\pi^+}$,
      $D_{s^+}^{\pi^+}$, $D_{c^+}^{\pi^+}$, $D_{b^+}^{\pi^+}$,
      $D_{u^++d^++s^+}^{\pi^+}$ and $D_g^{\pi^+}$ FFs at
      $\mu=5$~GeV. For each FF we plot the absolute distributions in
      the upper panel and their ratio to the central value of the
      \texttt{MAPFF1.0} set in the lower one.}
  \label{fig:Baseline}
\end{figure}

\myparagraph{Comparison with other FF sets}
In Fig.~\ref{fig:Baseline}, I compare the $\pi^+$ FFs obtained from
our baseline fit, \texttt{MAPFF1.0}, to those from
\texttt{JAM20}~\cite{Moffat:2021dji} and
\texttt{DEHSS14}~\cite{deFlorian:2014xna}. All three sets include a
similar SIA and SIDIS data set. However, the \texttt{JAM20} set also
includes inclusive deep-inelastic scattering and fixed-target
Drell-Yan measurements that are used to simultaneously determine PDFs,
while the \texttt{DEHSS14} set also includes pion production
measurements in proton-proton collisions.

In the case of $D_u^{\pi^+}$ and $D_{\bar d}^{\pi^+}$, as is clear
from the upper row of Fig.~\ref{fig:Baseline}, \texttt{JAM20} assumes
SU(2) isospin symmetry which results in
$D_u^{\pi^+} = D_{\bar{d}}^{\pi^+}$. \texttt{DEHSS14} instead assumes
$D_u^{\pi^+} + D_{\bar{u}}^{\pi^+}\propto D_{{d}}^{\pi^+} +
D_{\bar{d}}^{\pi^+}$ where the proportionality factor is a
$z$-independent constant that parameterises any possible isospin
symmetry violation. As explained in Sect.~\ref{s2:MAPFF10_parameterisation}, in
\texttt{MAPFF1.0} we parametrise $D_u^{\pi^+}$ and
$D_{\bar{d}}^{\pi^+}$ independently, thus allowing for a $z$-dependent
isospin symmetry violation which however turned out not to be
significant. For $z\lesssim 0.1$, where experimental data is sparse
(see Fig.~\ref{fig:kinplot}), the relative uncertainty on the
$D_u^{\pi^+}$ and $D_{\bar{d}}^{\pi^+}$ distributions from
\texttt{MAPFF1.0} is larger than that of the corresponding
distributions from \texttt{DEHSS14} and \texttt{JAM20}.  At large $z$,
we observe a suppression of the \texttt{MAPFF1.0} FFs
w.r.t. \texttt{DEHSS14} and \texttt{JAM20} for both $D_u^{\pi^+}$ and
$D_{\bar{d}}^{\pi^+}$. This suppression is compensated by an
enhancement of the sea quarks as I will further discuss below.

In the case of the sea FFs, we find a
good agreement at very large $z$ with both \texttt{DEHSS14} and
\texttt{JAM20} for $D_{s^+}^{\pi^+}$ and $D_{b^+}^{\pi^+}$. At low
$z$, on the one hand, we observe that the sea FFs from
\texttt{DEHSS14}, except $D_{\bar{u}}^{\pi^+}$, and the
$D_{s^+}^{\pi^+}$ FF from \texttt{JAM20} are within the
\texttt{MAPFF1.0} uncertainties. On the other hand, \texttt{JAM20} and
\texttt{DEHSS14} show an enhancement for the rest of the FFs
w.r.t.~\texttt{MAPFF1.0}.

We note that a very good agreement at intermediate $z$ is found for
the light singlet $D_{u^+}^{\pi^+}+D_{d^+}^{\pi^+}+D_{s^+}^{\pi^+}$
combination. This particular combination is most sensitive to
inclusive SIA data and the agreement reflects the fact that all three
collaborations are able to describe this data comparably well. The
gluon FF of \texttt{MAPFF1.0} is affected by large uncertainties. This
is a consequence of using observables that are not directly sensitive
to this distribution. The \texttt{MAPFF1.0}, \texttt{JAM20}, and
\texttt{DEHSS14} gluon FFs remain fairly compatible within
uncertainties.

Finally, we observe that most of the distributions of the
\texttt{MAPFF1.0} set present a turn-over in the region $ 0.1 \lesssim
z \lesssim 0.2$ that is absent in the other two sets. This feature can
be ascribed to the fact that \texttt{MAPFF1.0} implements cuts on the
minimum value of $z$ that are generally lower than those used by the
other two collaborations (see Ref.~\cite{Bertone:2017tyb} for a
detailed study).

\myparagraph{Impact of theoretical choices}
We have studied the stability of our FFs upon the input PDF set used
to compute SIDIS multiplicities. 
In order to assess the impact
of the PDF uncertainty, we performed an additional fit using the
central PDF member of the \texttt{NNPDF31\_nlo\_pch\_as\_0118} set
for all Monte Carlo replicas. We found that neglecting the PDF
uncertainty has a very small impact on the FFs. In addition, we
studied the dependence of the FFs on the specific PDF set by
performing two additional fits using the central member of the
\texttt{CT18NLO}~\cite{Hou:2019efy} and
\texttt{MSHT20nlo\_as118}~\cite{Bailey:2020ooq} sets. Also in this
case, we found that the difference at the level of FFs was very mild.
In fact, a
reduced sensitivity to the treatment of PDFs was to be expected
because our FFs depend on them only through the SIDIS measurements
that are delivered as multiplicities for both by COMPASS and
HERMES. For this particular observable, see
Eq.~\eqref{eq:multiplicities}, PDFs enter both the numerator and the
denominator, hence the sensitivity of the observable to the PDFs
largely cancels out.

We have also studied the stability of our FFs upon the choice of the
parametrisation scale $\mu_0$. To this purpose, we have repeated our
baseline fit by lowering the value of $\mu_0$ from 5~GeV to 1~GeV.
This led to almost identical FFs. We stress that the possibility to
freely choose the parameterisation scale, no matter whether above or
below the heavy quark thresholds, is due to the fact that we do not
set inactive-flavour FFs to zero below their respective threshold (see
Sect.~\ref{s2:MAPFF10_theory}).

\subsection{Impact of the data}
\label{s2:MAPFF10_dataimpact}

I now justify our exclusion of the SIA charm-tagged data from the fit
as well as our choice of the cut on the virtuality $Q^2$ for the SIDIS
data. I also discuss the separate impact of COMPASS and HERMES on the
FFs.

\begin{figure}
  \begin{center}
    \includegraphics[width=\textwidth]{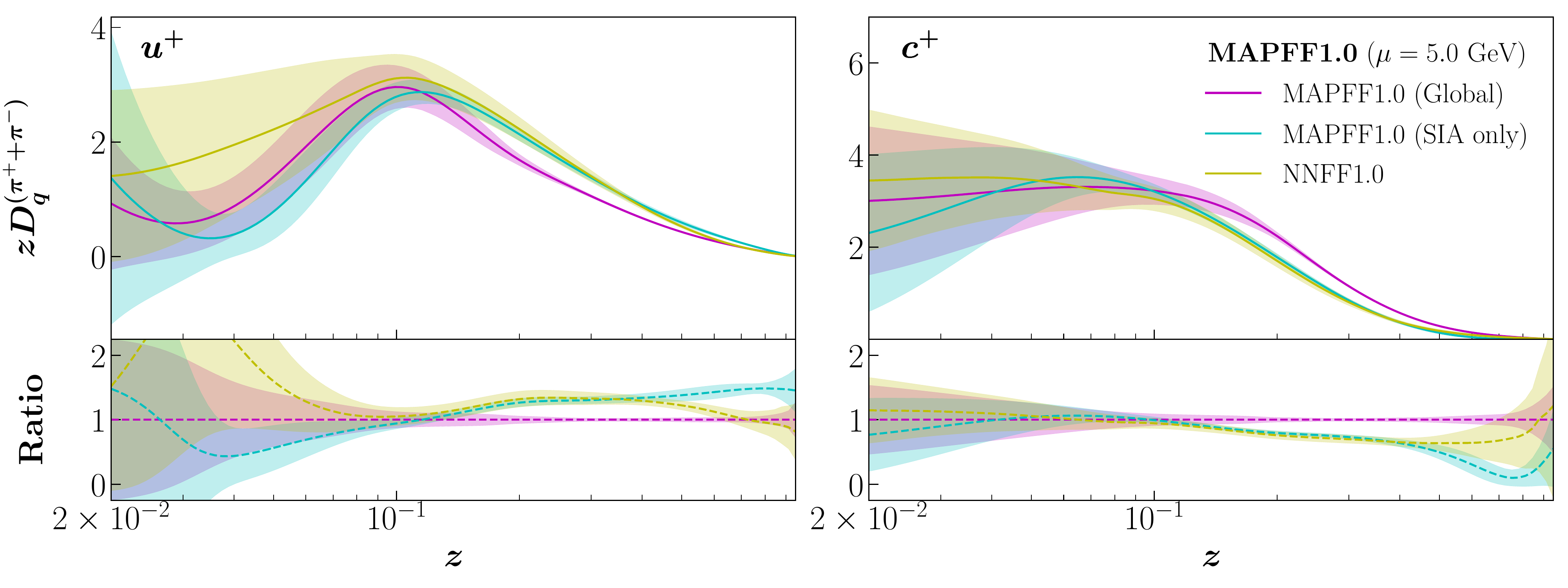}
  \end{center}
  \vspace{-0.8cm}
  \caption{\small Comparison between the baseline \texttt{MAPFF1.0}
    fit (which includes SIA and SIDIS data), a variant of the {\tt
      MAPFF1.0} fit to SIA data only, and
    \texttt{NNFF1.0}~\cite{Bertone:2017tyb} (which includes only SIA
    data). The left (right) plot shows the $D_{u^{+}}^{\pi^++\pi^-}$
    ($D_{c^{+}}^{\pi^++\pi^-}$) distribution at $\mu=5$~GeV. The upper
    panels display the absolute distributions while the lower ones
    their ratio to the central value of the baseline \texttt{MAPFF1.0}
    fit.}
  \label{fig:SIAonly}
\end{figure}

\myparagraph{Data compatibility}
As mentioned in Sect.~\ref{s2:MAPFF10_data}, we did not include the SLD
charm-tagged measurements because we have not been able to achieve an
acceptable description for this particular data set. Specifically, we
found that its inclusion causes a general deterioration of the fit
quality with a $\chi^2$ per data point of SLD-charm itself exceeding
6. We have identified the origin of this behaviour in an apparent
tension between the SLD charm-tagged and the COMPASS measurements. As
a matter of fact, if the COMPASS data is excluded from the fit, the
SLD charm-tagged data can be satisfactorily fitted. More precisely we
observe that the inclusion of COMPASS on top of SIA data leads to a
suppression of the
$D_{u^+}^{\pi^++\pi^-}=D_{u}^{\pi^+}+D_{\bar
  u}^{\pi^+}+D_{u}^{\pi^-}+D_{\bar u}^{\pi^-}$ distribution for
$z\gtrsim 0.1$ as compared to a fit to SIA data only. This behaviour
is visible in the left panel of Fig.~\ref{fig:SIAonly} where the
$D_{u^{+}}^{\pi^++\pi^-}$ distribution for the sum of positively and
negatively charged pions is displayed at $\mu=5$~GeV for the following
FF sets: the baseline \texttt{MAPFF1.0} fit (which includes SIA and
SIDIS data), a \texttt{MAPFF1.0}-like fit to SIA data only, and the
NNFF1.0 fit~\cite{Bertone:2017tyb} (which includes SIA data only). We
see that at intermediate values of $z$ the SIA-only {\tt MAPFF1.0} fit
and NNFF1.0 fit are in good agreement, while the baseline
\texttt{MAPFF1.0} fit is suppressed. As a consequence of this
suppression, the
$D_{c^{+}}^{\pi^++\pi^-}=D_{c}^{\pi^+}+D_{\bar
  c}^{\pi^+}+D_{c}^{\pi^-}+D_{\bar c}^{\pi^-}$ distribution of the
global \texttt{MAPFF1.0} fit gets enhanced to accommodate the
inclusive SIA data. This effect is visible in the right panel of
Fig.~\ref{fig:SIAonly} that shows for the $D_{c^{+}}^{\pi^++\pi^-}$
distribution a good agreement between the SIA-only fit and NNFF1.0
with the global \texttt{MAPFF1.0} fit being generally harder for
$z\gtrsim 0.1$.  This enhancement of the $D_{c^{+}}^{\pi^++\pi^-}$
distribution deteriorates the description of the SLD charm-tagged
data.  This is not surprising in that charm-tagged observables are
naturally sensitive to the charm FFs. We interpret the suppression of
$D_{u^{+}}^{\pi^++\pi^-}$ and the consequent enhancement of
$D_{c^{+}}^{\pi^++\pi^-}$ as an effect of the COMPASS data. We
conclude that the COMPASS and SLD charm-tagged data are in tension and
we decided to keep the former and drop the latter from our global fit.

We finally note that the suppression of the
$D_{u^{+}}^{\pi^++\pi^-}$ distribution also leads to a deterioration
of the description of the $uds$-tagged measurements from both DELPHI
and SLD that feature a $\chi^2$ per data point of respectively 2.84
and 2.05 (see Tab.~\ref{tab:chi2s}). However, this deterioration is
milder than that of the SLD charm-tagged data, thus we opted for
keeping the $uds$-tagged measurements in the fit.

\begin{figure}
  \begin{center}
    \includegraphics[width=0.9\textwidth]{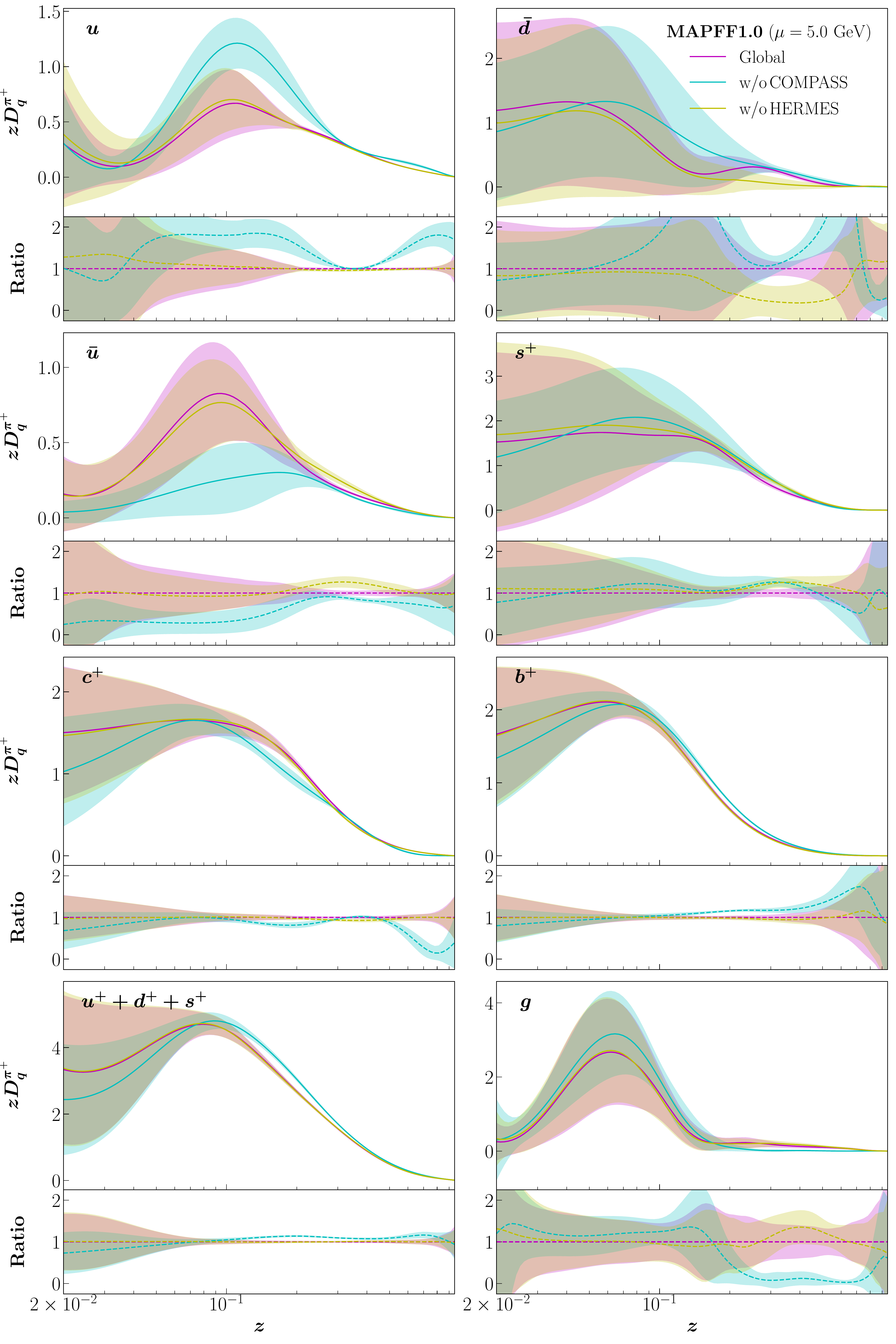}
  \end{center}
  \vspace{-0.8cm}
  \caption{\small Comparison between three variant of the
    \texttt{MAPFF1.0} fit: a fit to the global data set, a fit without
    the COMPASS data, and a fit without the HERMES data. The format of
    the plot is as in Fig.~\ref{fig:Baseline}.}
  \label{fig:DatasetsVariation}
\end{figure}

\myparagraph{Impact of SIDIS data}
In this section, I discuss the impact of the individual SIDIS data sets
included in our analysis. To this purpose, we have repeated our
baseline fit by removing either the COMPASS or the HERMES
measurements. The three fits are compared in
Fig.~\ref{fig:DatasetsVariation}.

We note that the number of data points that survive the kinematic cuts
defined in Sect.~\ref{s2:MAPFF10_data} is 8 for HERMES and 314 for COMPASS.
Despite the limited amount of data points, HERMES still provides a sizeable
constraint in the region of its coverage. As shown in
Fig.~\ref{fig:DatasetsVariation} the impact of the HERMES data in the region
$0.2 \lesssim z \lesssim 0.6$ can be summarised as follows:
\begin{itemize}
\item Overruling the COMPASS data for $D_{\bar{d}}^{\pi^+}$ and partly
  for $D_{\bar{u}}^{\pi^+}$. When excluding HERMES, these FFs are
  respectively suppressed by approximately 2-$\sigma$ and enhanced by
  1-$\sigma$. 
\item Competing with the COMPASS data for $D_{s^+}^{\pi^+}$ because
  both datasets have a comparable impact on this FF combination.
\item Overruled by the COMPASS data for the remaining FFs, namely
  $D_{u}^{\pi^+}$, $D_{c^+}^{\pi^+}$, $D_{b^+}^{\pi^+}$,
  $D_{g}^{\pi^+}$ and for the combination
  $D_{u^+ + d^+ + s^+}^{\pi^+}$, as their trend in the global fit
  follows that of the fit without HERMES.
\end{itemize}
We finally note that, as expected, the three fits display a similar
behaviour in the extrapolation regions for both the central value and
the uncertainty.

\myparagraph{Impact of SIDIS energy scale kinematic cut}
As discussed in Sect.~\ref{s2:MAPFF10_data}, we included in our fit only
SIDIS data whose value of $Q$ is larger than $Q_{\rm cut}=2$~GeV. The
reason for excluding low-energy data stems from the fact that, as $Q$
decreases, higher-order perturbative corrections become increasingly
sizeable until eventually predictions based on NLO calculations become
unreliable. Therefore, $Q_{\rm cut}$ has to be such that NLO accuracy
provides an acceptable description of the data included in the fit.
\begin{figure}[htb]
  \begin{center}
    \includegraphics[width=0.65\textwidth]{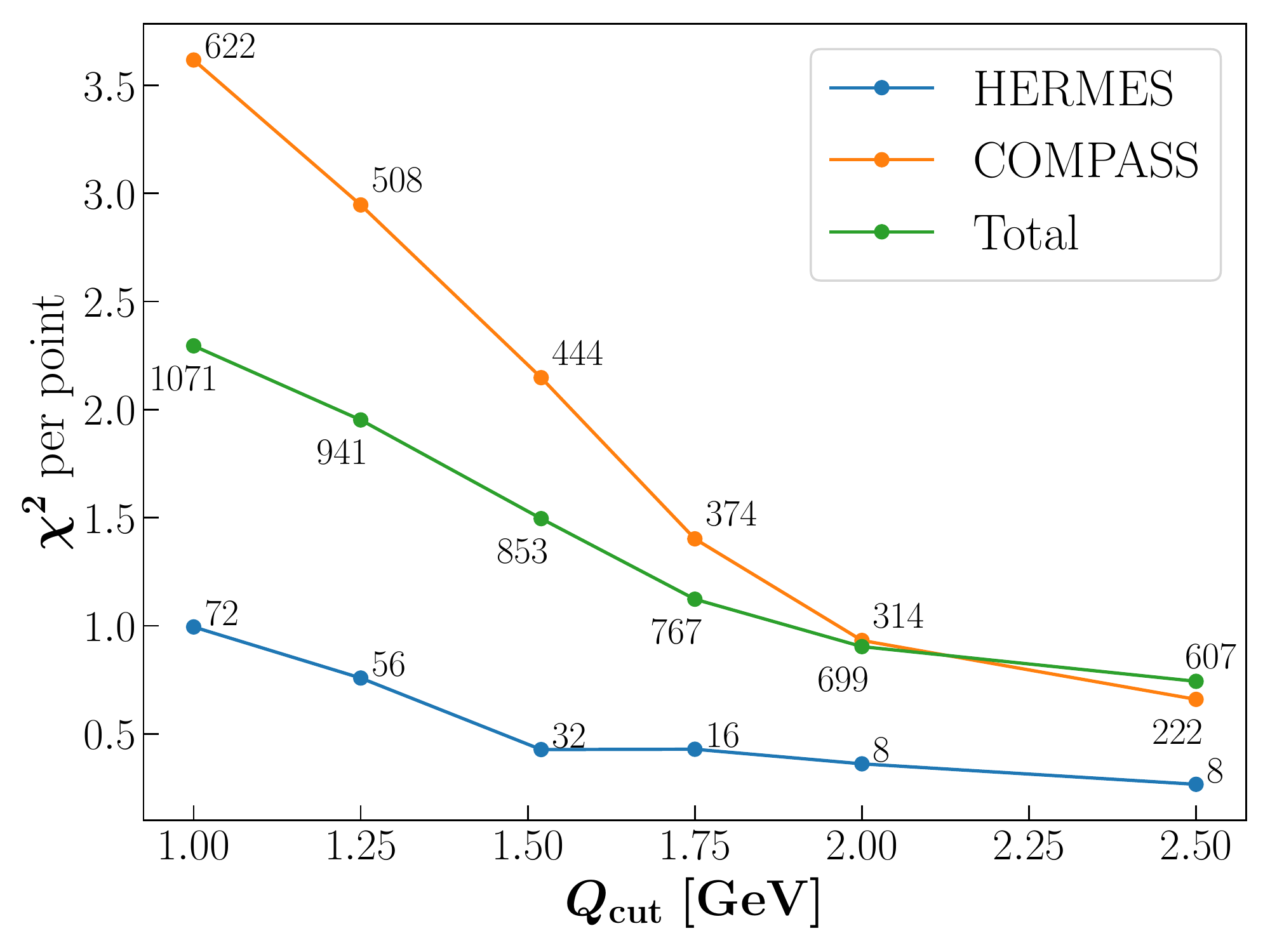}
    \end{center}
    \vspace{-0.8cm}
    \caption{\small Behaviour of the $\chi^2$ per data point as a
      function of the cut on $Q$, $Q_{\rm cut}$, applied to the SIDIS
      data. The $\chi^2$ is computed for the total \texttt{MAPFF1.0}
      data set (green curve), for the COMPASS data set (orange curve),
      and for the HERMES data set (blue curve). For each value of
      $Q_{\rm cut}$ considered, the plot also displays the number of
      data points $N_{\rm dat}$ that pass the cut.}
  \label{fig:SIDISQcut}
\end{figure}

Our particular choice is informed by studying the dependence of the
fit quality on the value of $Q_{\rm cut}$. To this purpose, we have
repeated our baseline analysis by varying the value of $Q_{\rm cut}$
in the range $[1.00,2.50]$~GeV. Fig.~\ref{fig:SIDISQcut} shows the
behaviour of the $\chi^2$ per data point for HERMES, COMPASS, and for
the total data set as functions of $Q_{\rm cut}$. For each point in
$Q_{\rm cut}$ the number of data points surviving the cut is also
displayed. As expected, the $\chi^2$ is a decreasing function of
$Q_{\rm cut}$ confirming the fact that perturbation theory works
better for larger values of the hard scale $Q$. However, while HERMES
can be satisfactorily described down to $Q_{\rm cut}=1$~GeV with a
$\chi^2$ that never exceeds one, the COMPASS $\chi^2$ quickly
deteriorates reaching a value as large as 3.5 at $Q_{\rm
  cut}=1$~GeV. Given the large size of the COMPASS data set, this
deterioration drives the total $\chi^2$ that also becomes
significantly worse as $Q_{\rm cut}$ decreases. Based on
Fig.~\ref{fig:SIDISQcut}, we have chosen $Q_{\rm cut}=2$~GeV for our
baseline fit because it guarantees an appropriate description not only
of the COMPASS data, but also of the entire data set.

  \chapter{Impact of future colliders}
\label{chap:Impact}
\vspace{-1cm}
\begin{center}
  \begin{minipage}{1.\textwidth}
      \begin{center}
          \textit{This chapter is based on my results in Refs.~\mycite{AbdulKhalek:2019mps,Khalek:2018mdn,Azzi:2019yne,AbdulKhalek:2020yuc,Cepeda:2019klc,Khalek:2021ulf}}.
      \end{center}
  \end{minipage}
  \end{center}

\myparagraph{Outline} In this chapter, I review the constraints on PDFs and nPDFs induced by pseudodata projected for several future-colliders.
In Sect.~\ref{s1:PDF_Impact_studies}, I discuss two different future high energy programs at the LHC, the High Luminosity Large Hadron Collider in Sect.~\ref{s2:PDF_HLLHC} and the Large Hadron Electron Collider in Sect.~\ref{s2:PDF_LHeC}. The pseudodata for these two colliders, covers different type of processes and is generated according to various considerations and scenarios of projected uncertainties. Based on these projections, I discuss the impact on the proton PDFs gauged by means of the Hessian profiling method (see Sect.~\ref{s2:hessian_method}). The proton PDF sets used for these two analysis are the \texttt{PDF4LHC15} hessian sets.
In Sect.~\ref{s1:nNNPDF20_ImpactStudies}, I discuss the various phenomenological
applications of the \texttt{nNNPDF2.0} determination, 
then, I revisit the potential of the FoCal upgrade to the ALICE
detector and finally, I provide the predictions based on \texttt{nNNPDF2.0} for inclusive hadron
production in proton-nuclear collisions.
In Sect.~\ref{s1:EIC}, I present an update to the initial studies of the Electron-Ion Collider (EIC) neutral-current DIS pseudodata based on \texttt{\texttt{nNNPDF1.0}} and presented in Ref.~\mycite{AbdulKhalek:2019mzd}. In the updated analysis, we quantify the impact of unpolarised lepton-proton and
lepton-nucleus inclusive DIS cross section
measurements from the the official release of the EIC projections~\mycite{AbdulKhalek:2021gbh}. To this purpose, we base our analysis~\mycite{Khalek:2021ulf} on a self-consistent set of proton (\texttt{NNPDF3.1}) and nuclear (\texttt{nNNPDF2.0}) global PDF determinations.

\section{Proton PDFs for the future Large Hadron Collider} \label{s1:PDF_Impact_studies}

Measurements of cross sections provide fundamental tests of theoretical predictions. Higher precision of both the experimental measurements and the theoretical predictions is required in order to determine fundamental parameters of the Standard Model and to discover beyond the Standard Model phenomena.
At the LHC, the precision of cross section measurements is limited by the uncertainty of the integrated luminosity, currently about 2\%. A target uncertainty of 1\% has been set for the High Luminosity phase, which is expected to be achieved by a combination of improved detector instrumentation and refined analysis techniques~\mycite{Khalek:2018mdn, Azzi:2019yne, Cepeda:2019klc} starting its data acquisition in 2027.

In parallel, the Large Hadron Electron Collider (LHeC)~\cite{AbelleiraFernandez:2012cc,AbelleiraFernandez:2012ty,Klein:2018rhq}, is a facility that would run concurrently with the HL--LHC and be based on a new purpose--built detector at the designed interaction point.
The kinematic range covered by the LHeC would extend that of HERA~\cite{Abramowicz:2015mha,Abramowicz:1900rp,Aaron:2009af,Abramowicz:2014zub}, the first electron-proton collider,  by a factor of twenty in $Q^2$ and $x$ and projected to exceed the integrated HERA luminosity by two orders of magnitude. The LHeC physics programme is devoted to an exploration of the energy frontier, complementing the HL-LHC and possibly resolving the observation of new phenomena based on the specifics of deep inelastic electron-proton scattering at energies extending to beyond a TeV~\mycite{AbdulKhalek:2019mps, AbelleiraFernandez:2012cc}.

In Sect.~\ref{s2:PDF_HLLHC}, we focus on the the High-Luminosity (HL) phase of the LHC, manifested by a major upgrade of the accelerator and detector systems. More details can be found in Refs.~\mycite{Khalek:2018mdn, Azzi:2019yne, Cepeda:2019klc}.
In Sect.~\ref{s2:PDF_LHeC}, we extended the previous study of the HL--LHC impact to include, both individually and in combination, the impact from the Large Hadron electron Collider (LHeC), a high-energy lepton-proton and lepton-nucleus collider based at CERN. More details can be found in Ref.~\mycite{AbdulKhalek:2019mps, AbelleiraFernandez:2012cc}.

\subsection{High Luminosity Large Hadron Collider } \label{s2:PDF_HLLHC}
The High Luminosity phase (HL-LHC) is expected to deliver
a $14$ TeV center-of-mass energy with an integrated luminosity of around $\mathcal{L}=3$ ab$^{-1}$ to ATLAS and CMS and around $\mathcal{L}=0.3$ ab$^{-1}$ to LHCb.
This phase will improve the precision measurements of the Higgs boson properties, the heavy vector bosons masses and mixing angles, as well as extend the sensitivity to physics beyond the Standard model. More opportunities are summarised and detailed in
the CERN Yellow Report~\mycite{Azzi:2019yne, Cepeda:2019klc}.

These measurements are often limited by the uncertainties on PDFs. Therefore, to gauge their reduction, we considered a number of PDF-sensitive
processes presented in Table~\ref{tab:data_HLLHC}. For which, we have
generated HL--LHC pseudodata based on the projected center-of-mass energy and luminosity mentioned above.
Statistical uncertainties are evaluated
from the expected number of events per bin,
taking into account branching ratios and
acceptance corrections determined
from the corresponding reference analysis.
Systematic uncertainties are taken to be those of the
13~(8) TeV baseline analyses and then rescaled appropriately as shown in Eq.~(\ref{eq:totalExpError})
We consider various scenarios for the reduction of systematic errors,
from a more conservative one to a more optimistic one.

%
%
Theoretical predictions are computed at NLO in the QCD and the central value of the pseudodata initially coincides with
the corresponding prediction obtained with the \texttt{PDF4LHC15}  NNLO set as input.
Subsequently, this central value is fluctuated
according to the prescription in Eq.~(\ref{eq:pseudodataGen}) by the corresponding experimental uncertainties.
%

For the HL--LHC, we consider three sources of systematic uncertainties:
\begin{itemize}
  \item $\delta^\text{exp}_{\text{tot},i}$ is the total
(relative) experimental uncertainty corresponding to a specific bin $i$
(excluding the luminosity and normalization uncertainties).
\item $\delta^\text{exp}_{\mathcal L}$
is the luminosity uncertainty, which is fully
correlated among all the pseudodata bins of the same experiment
(but uncorrelated among different experiments).
We take this luminosity uncertainty to be $\delta^\text{exp}_{\mathcal L}=1.5\%$
for the three LHC experiments.
\item $\delta^\text{exp}_{\mathcal N}$
are possible additional normalization uncertainties as in the case of
$W$ boson production in association with charm quarks.
\end{itemize}

If $\sigma_i^{\text{th}}$ is the theoretical cross-section
for  bin $i$ of a given process,
then the central value of
the HL--LHC pseudodata $\sigma_i^\text{exp}$ is constructed by means of
\begin{equation}
\label{eq:pseudodataGen}
\sigma_i^\text{exp} = \sigma_i^\text{th} \times \left( 1 + r_i\cdot \delta^\text{exp}_{\text{tot},i}
 + \lambda\cdot \delta^\text{exp}_{\mathcal L}+ s\cdot \delta^\text{exp}_{\mathcal N}\right)
\end{equation}
where $r_i$, $\lambda$, and $s$ are univariate Gaussian random numbers.
The total experimental uncertainty $\delta^\text{exp}_{\text{tot},i}$ is defined as
\begin{equation}
\label{eq:totalExpError}
\delta^\text{exp}_{\text{tot},i} \equiv \left( \left( \delta^\text{exp}_{\text{stat},i}\right) ^2 +
\left( f_\text{corr}\times f_\text{red}\times
\delta^\text{exp}_{\text{sys},i}\right)^2 \right)^{1/2}
\end{equation}
In this expression, the relative statistical error $\delta^\text{exp}_{\text{stat},i}$ is
computed as
\begin{equation}
\label{eq:acceptance}
\delta^\text{exp}_{\text{stat},i} = \left( f_\text{acc} \times N_{\text{ev},i}\right)^{-1/2} \, ,
  \end{equation}
  where $N_{\text{ev},i}=\sigma_i^\text{th} \times \mathcal{L}$ is the expected number
  of events in bin $i$ at the HL--LHC with $\mathcal{L}=3~(0.3)$ ab$^{-1}$ for ATLAS and CMS (LHCb).
  In Eq.~(\ref{eq:acceptance}),
  $f_\text{acc}\le 1$
  is an acceptance correction which accounts for the fact that, for some of the processes
  considered, such as top quark pair production, there is a finite experimental acceptance
  for the final state products
  and/or one needs to include the effects of branching fractions.
  The value of
  $f_\text{acc}$ is determined by extrapolation using the reference dataset,
  except for forward $W$+charm production (where there is no baseline measurement)
  where the acceptance is set to $f_\text{acc}=0.3$, due dominantly to the $c$--jet tagging efficiency.  
  
  In Eq.~(\ref{eq:totalExpError}),   $\delta^\text{exp}_{\text{sys},i}$ indicates the total
  systematic error of bin $i$ taken from the reference LHC measurement at either 8 TeV
  or 13 TeV; and $f_\text{red}\le 1$ is a correction factor that accounts for the fact that on average systematic
  uncertainties will decrease at the HL--LHC as compared to Run-2 due to both detector
  improvements and the enlarged dataset for calibration.
  Finally, $f_\text{corr}$ represents
  an effective correction factor that accounts
  for the fact that data with correlated systematics may be more constraining than the same
  data where each source of error is simply added in quadrature,
  as we do in this analysis.
  We show in Table~\ref{tab:data_HLLHC} the value of $f_\text{corr}$ determined by means of available LHC measurements
  for which the full information on correlated systematics
  is available.

\begin{table}[htb]
    \centering
    \small
    \begin{tabular}{cccccc}  \toprule
        Process    &   Kinematics  &   $N_\text{ dat}$  &  
        $f_\text{ corr}$  &  $f_\text{ red}$ &  Baseline  \\
  \toprule
  \multirow{3}{*}{$Z$ $p_T$}  &    $20\,\text{GeV}\leq p_T^{ll} \leq 3.5$ TeV           &
  \multirow{3}{*}{338} &   \multirow{3}{*}{0.5}  & \multirow{3}{*}{$( 0.4, 1)$}
  & \multirow{3}{*}{\cite{Aad:2015auj} (8 TeV)} \\
    &    $12\,\text{GeV}\leq m_{ll} \leq 150$GeV           &              & & \\
     &    $|y_{ll}|\leq 2.4$            &                    &   &     \\
  \midrule
  \multirow{2}{*}{high-mass Drell-Yan}  &   $p_T^{l1(2)}\ge 40(30)\,\text{GeV}$           &  \multirow{2}{*}{32}         &        \multirow{2}{*}{0.5}      
  &       \multirow{2}{*}{$( 0.4, 1)$}        &    \multirow{2}{*}{\cite{Aad:2016zzw} (8 TeV)}     \\
  &  $|\eta^l|\leq 2.5$, $m_{ll}\ge 116\,\text{GeV}$   & & & &  \\
  \midrule
  top quark pair  &     $m_{t\bar{t}}\simeq 5$ TeV, $|y_t|\leq 2.5$          &       52                   & 0.5
  &         $( 0.4, 1)$      &   \cite{Aad:2015mbv} (8 TeV)    \\
  \midrule
  \multirow{2}{*}{$W$+charm (central)}  &        $p_T^\mu \ge26\,\text{GeV}$, $p_T^c \ge5\,\text{GeV}$     &  \multirow{2}{*}{12}         &        \multirow{2}{*}{0.5}       
  &       \multirow{2}{*}{$( 0.2, 0.5)$}       &    \multirow{2}{*}{\cite{CMS-PAS-SMP-17-014} (13 TeV)}     \\
  &  $|\eta^\mu|\leq 2.4$    & & & &\\
  \midrule
  \multirow{3}{*}{$W$+charm (forward)}  &      $p_T^\mu \ge20\,\text{GeV}$, $p_T^c \ge20\,\text{GeV}$           &          \multirow{3}{*}{12}     &        \multirow{3}{*}{0.5}   
  &      \multirow{3}{*}{$( 0.4, 1)$}            &   \multirow{3}{*}{LHCb projection}         \\
  &   $p_T^{\mu+c} \ge20\,\text{GeV}$    & & & & \\
  &  $2\leq \eta^\mu \leq 5$, $2.2\leq \eta^c \leq 4.2$     & & & & \\
  \midrule
  Direct photon  &     $E_T^\gamma \lsim 3$ TeV, $|\eta_{\gamma}|\leq 2.5$          & 118              &      0.5        
  &    \multirow{1}{*}{$( 0.2, 0.5)$}           &   \cite{Aaboud:2017cbm} (13 TeV)      \\
  \midrule
  \multirow{3}{*}{Forward $W,Z$}  &  $p_T^{l}\ge 20\,\text{GeV}$, $2.0\leq \eta^l\leq 4.5$           &  \multirow{3}{*}{90}         &        \multirow{3}{*}{0.5}     
  &       \multirow{3}{*}{$( 0.4, 1)$}        &    \multirow{3}{*}{\cite{Aaij:2015zlq} (8 TeV)}     \\
  &    $2.0\leq y_{ll}\leq 4.5$  & & & & \\
  &    $60\,\text{GeV}\leq m_{ll}\leq 120\,\text{GeV}$  & & & & \\
  \midrule
  Inclusive jets  &       $|y| \leq 3$, $R = 0.4$      &       58        &      0.5        &
  \multirow{1}{*}{$( 0.2, 0.5)$} 
  &   \cite{Aaboud:2017wsi}    (13 TeV)               \\
  \bottomrule
  Total   &    &   712 &   &   &   \\
  \bottomrule
\end{tabular}
\caption{
Summary of the features of the HL--LHC pseudo-data generated for the present
study.
For each process we indicate the kinematic coverage, the number of pseudo-data
points used across all detectors $N_\text{ dat}$, the values of the correction factors
$f_\text{ corr}$ and $f_\text{ red}$; and finally the reference from the 8 TeV or
13 TeV measurement used as baseline to define the binning and the systematic
uncertainties of the HL--LHC pseudo-data, as discussed in the text.
}
\label{tab:data_HLLHC}
\end{table}

\begin{figure}[!h]
  \begin{center}
    \makebox[\textwidth]{\includegraphics[width=0.9\textwidth]{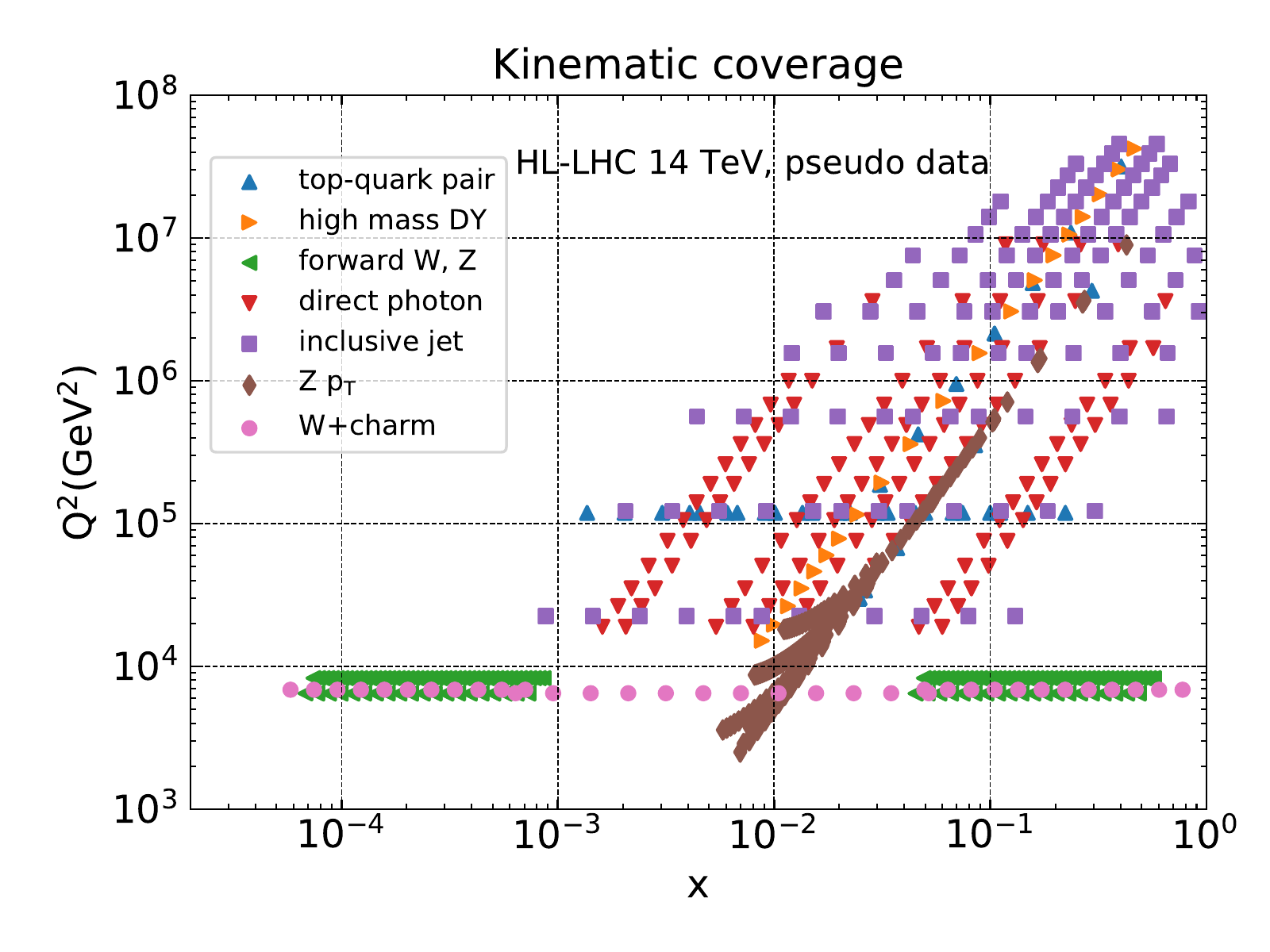}}
  \end{center}
  \vspace{-0.8cm}
\caption{
  The kinematical coverage in the $(x,Q^2)$ plane of the
  HL--LHC pseudodata included in this analysis.}
 \label{fig:kinHLLHC}
\end{figure}

In Fig.~\ref{fig:kinHLLHC} we show  the kinematical coverage in
the $(x,Q^2)$ plane of the
  HL--LHC pseudodata included in this analysis.
  %
  %
  We assume $x_1=x_2$ if rapidities are not specified for the
  final states. 
  We see that the HL--LHC pseudodata covers a wide kinematic region,
  including the large momentum transfers up to $Q^2\simeq 6^2$ TeV$^2$, as well
  as the large-$x$ region, with several different processes.
  Specifically, the input pseudodata spans
  the range $6\times 10^{-5} \lsim x \lsim 0.7$ and
  $40^2~\text{GeV}^2 \lsim Q^2 \lsim 7^2~\text{TeV}^2$ in
  the $(x,Q^2)$ kinematic plane.
  Note that the LHCb measurements are instrumental to constrain
  the small-$x$ region, $6\times 10^{-5} \lsim x \lsim 10^{-3}$, beyond
  the acceptance of ATLAS and CMS.
  
Subsequently, we quantify the constraints of the HL--LHC pseudodata on the symmetric Hessian\footnote{See Eq.~(\ref{eq:hessian_errors}) and Ref.~\cite{Buckley:2014ana} for the definition.} \texttt{PDF4LHC15}  PDF
set. The analysis is performed by means of the Hessian Profiling
method~\cite{Paukkunen:2014zia, Schmidt:2018hvu} (See
Sect.~\ref{s2:hessian_method}).
In our analysis, we use $T=3$ which roughly corresponds to the average tolerance
determined dynamically in the \texttt{CT14}  and \texttt{MMHT14}  analyses. We then combine the complete set
of HL--LHC pseudodata listed in Table~\ref{tab:data_HLLHC} to produce the final profiled
PDF sets, which quantify the impact of future HL--LHC measurements on our knowledge of the
quark and gluon structure of the proton. We have also performed a systematic study of the
constraints on the PDFs from the individual pseudodatas in Table~\ref{tab:data_HLLHC}
which can be found in Sect.~3 of Ref.\mycite{Khalek:2018mdn}.
As the list of processes for which pseudodata has been generated is by no means complete,
the results presented here should be understood as a conservative estimate (that is, an
upper bound) for the potential impact of HL--LHC measurements on PDFs.

In Table~\ref{tab:Scenarios} we list the three scenarios for the systematic uncertainties
of the HL--LHC pseudodata that we assume in the present analysis.
%
\begin{table}[htb]
    \centering
    \renewcommand{\arraystretch}{1.20}
    \begin{tabular}{c|c|c|c|c}
      Scenario    &   $f_\text{red}$ (8 TeV)  & $f_\text{red}$ (13 TeV) &   \texttt{ LHAPDF} set  &
      Comments \\
      \toprule
       A          &   1   &  0.5   & \texttt{ PDF4LHC\_nnlo\_hllhc\_scen1}  & Conservative \\
      \midrule
      B         &   0.7   &  0.36  & \texttt{ PDF4LHC\_nnlo\_hllhc\_scen2}  & Intermediate \\
       \midrule
        C          &   0.4   &  0.2  &  \texttt{ PDF4LHC\_nnlo\_hllhc\_scen3}  & Optimistic \\
   \bottomrule
    \end{tabular}
    \vspace{0.3cm}
    \caption{\small \label{tab:Scenarios}
      The three scenarios for the systematic uncertainties of the HL--LHC pseudo-data
      that we assume in the present study.
      These scenarios, ranging from conservative to optimistic, differ among them in
      the reduction factor $f_\text{red}$, Eq.~(\ref{eq:totalExpError}),
      applied to the systematic errors of the reference
      8 TeV or 13 TeV measurements.
      We also indicate in each case the name of the corresponding \texttt{ LHAPDF} grid.
    }
  \end{table}
These scenarios, ranging from more conservative to more optimistic, differ among them in
the reduction factor $f_\text{red}$, Eq.~(\ref{eq:totalExpError}), applied to the
systematic errors of the reference 8 TeV or 13 TeV measurements.
In particular, in the optimistic scenario we assume a reduction of the systematic errors
by a factor 2.5 (5) as compared to the reference 8 TeV (13 TeV) measurements, while for
the conservative scenario we assume no reduction in systematic errors with respect to 8
TeV reference.
We also indicate in each case the name of the corresponding \texttt{LHAPDF} grid.
Reassuringly, as we show below, the qualitative results of our study depend only mildly in
the specific assumption for the values of $f_\text{red}$.

\begin{figure}[!h]
  \begin{center}
    \makebox[\textwidth]{\includegraphics[width=\textwidth]{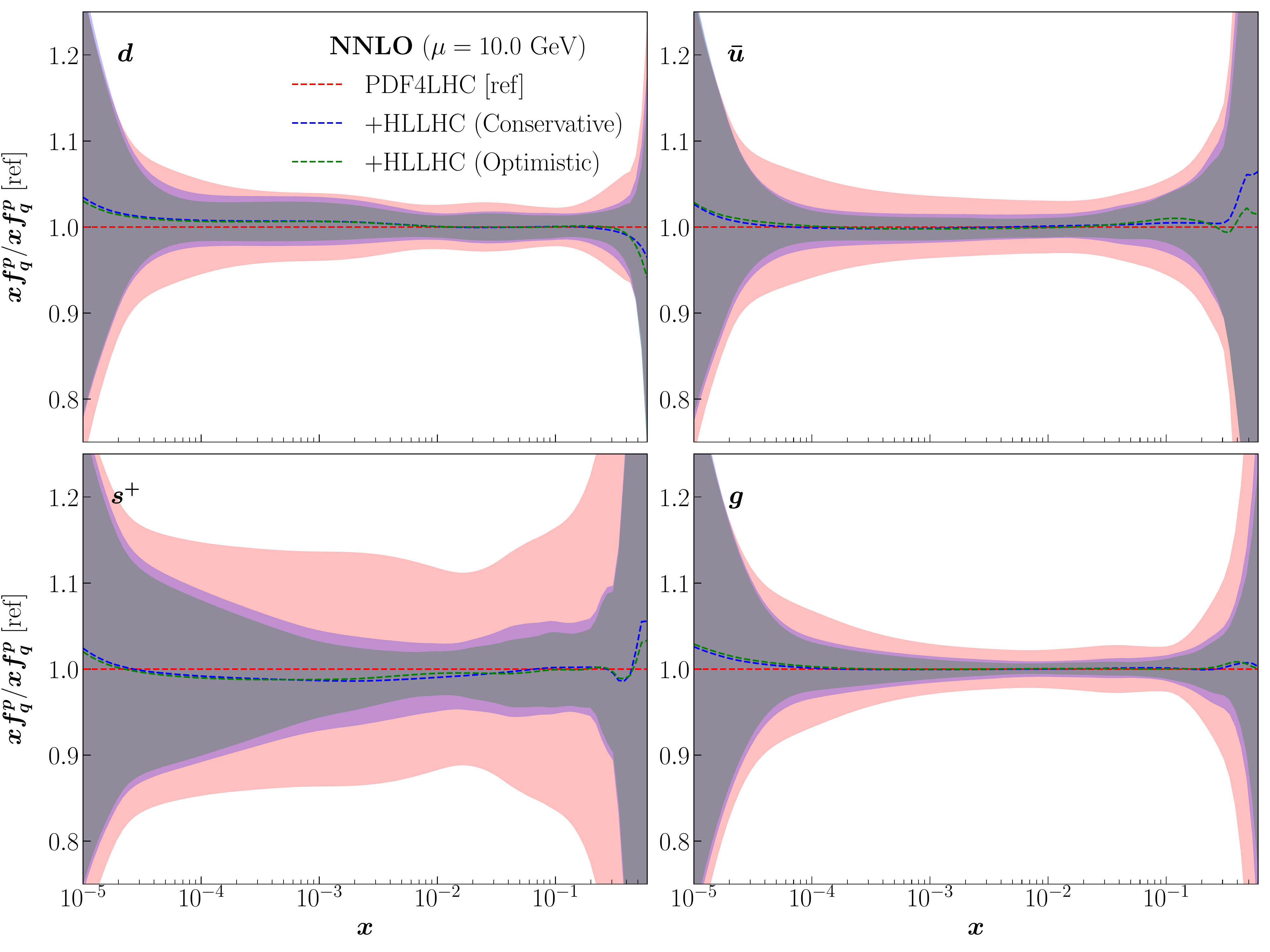}}
    \end{center}
    \vspace{-0.8cm}
    \caption{Comparison of the \texttt{PDF4LHC15}  at NNLO set with the HL-LHC profiled set in the conservative and optimistic scenarios, defined in Table~\ref{tab:Scenarios}. We show the down quark, up anti-quark, total strangeness and gluon at $Q$=10 GeV, normalised to the central value of the \texttt{PDF4LHC15}  baseline. The bands correspond to the one-sigma PDF uncertaintes. On the right, same as left but relative uncertainties are shown.}
  \label{fig:HLLHC_PDFs_Q10}
\end{figure}

In Fig.~\ref{fig:HLLHC_PDFs_Q10} we present a comparison of the baseline \texttt{PDF4LHC15}  set with the
profiled sets based on HL--LHC pseudodata from scenarios A (conservative) and C
(optimistic) as defined in Table~\ref{tab:Scenarios}.
Specifically, we show the down quark, up anti-quark, total strangeness and gluon at $Q=10$ GeV, normalised
to the central value of the \texttt{PDF4LHC15}  baseline.
In this comparison, the bands correspond to the one-sigma PDF uncertainties.

First of all, we observe that the impact of the HL--LHC pseudodata is reasonably similar
in the conservative and optimistic scenarios. Therefore we don't show the intermediate
one.
This is not so surprising, as we have explicitly chosen those datasets which will benefit
from a significant improvement in statistics, and these tend to lie in kinematic regions
where the PDFs themselves are generally less well determined.
Therefore, the dominant reason for the observed reduction of PDF uncertainties is the
increased statistics and the corresponding extended kinematic reach that becomes
available at the HL--LHC, rather than the specific assumptions about the systematic
uncertainties. This demonstrates that our results are robust against the details of the
projections of how the experimental systematic uncertainties will be reduced in the
HL--LHC era.

From Fig.~\ref{fig:HLLHC_PDFs_Q10} we observe a marked reduction of the PDF uncertainties in
all cases.
This reduction is specially significant for the gluon and the sea quarks, for the reason
that these are currently affected by larger uncertainties than in the case of the
valence quarks.
In the case of the gluon PDF, there is an improvement of uncertainties in the complete
relevant range of momentum fraction $x$.
This is a direct consequence of the fact that we have included here several HL--LHC
processes that have direct sensitivity to the gluon content of the proton, including
jet, direct photon, and top quark pair production, as well as the transverse momentum of
$Z$ bosons.

Another striking feature of Fig.~\ref{fig:HLLHC_PDFs_Q10} concerns the strange PDF.
In this case, the PDF uncertainties are reduced by almost a factor 4, from around 15\%
to a few percent, in a wide region of $x$.
This result highlights the importance of the $W$+charm measurements at the HL--LHC,
specially those in the forward region by LHCb, which
represent a unique handle on the poorly known strange content of the proton.
In turn, such improved understanding of the strange PDF will feed into a reduction of
theory uncertainties in crucial HL--LHC measurements such as those of $M_W$ or
$\sin^2\theta_W$.

We conclude that, even in the most conservative scenario, we find that
HL--LHC measurements can reduce PDF uncertainties by at least a factor between 2 and 3 as
compared to the current \texttt{PDF4LHC15}  baseline.
The PDF constraining information from the HL--LHC is expected to be specially significant
for gluon- and for strange-initiated processes.
This improved knowledge of the quark and gluon structure of the proton which will become
possible at the HL--LHC will directly benefit a number of phenomenologically important
processes which we present in Ref.~\cite{}, thanks to the reduction of the associated theoretical errors.
For instance, the PDF uncertainties in Higgs production in gluon fusion can be reduced
down to $\lsim 2\%$ for the entire range of Higgs transverse momenta accessible at the
HL--LHC.
Likewise, PDF uncertainties in high-mass supersymmetric particle production can be
decreased by up to a factor 3 as compared with the current situation.
This improvement should strengthen the bounds derived in the case of null searches, or
even facilitate their characterisation in the case of an eventual discovery.
Similar improvements are found for Standard Model process, for example dijet production,
which provides a unique opportunity to measurement the running of the strong coupling
constant at the TeV scale.
The results of this study are made publicly available in the \texttt{LHAPDF6}
format~\cite{Buckley:2014ana}, with the grid names listed in Table~\ref{tab:Scenarios}
for the three scenarios that have been considered.
%

\subsection{Large Hadron Electron Collider } \label{s2:PDF_LHeC}
In the previous section, we discussed the HL-LHC potential to constrain the PDFs~\mycite{Khalek:2018mdn}
by using projected measurements for a range of SM processes,
from Drell-Yan to top quark pair and jet production.
We found that PDF uncertainties on LHC processes can be reduced by a factor between two and five,
depending on the specific flavour combination and on the assumptions about the experimental systematic uncertainties.
Our PDF projections have already been used in a number of related HL--LHC studies,
as reported in~\mycite{Azzi:2019yne,Cepeda:2019klc}.

A quite distinct possibility to improve our understanding of proton structure
is the proposal to collide high energy electron and positron beams with the hadron beams from the HL--LHC.
This facility, known as the Large Hadron Electron Collider (LHeC)~\cite{AbelleiraFernandez:2012cc,AbelleiraFernandez:2012ty,Klein:2018rhq}, would run concurrently with the HL--LHC and be based on a new purpose--built detector at the designed interaction point.
A key outcome of the LHeC operations would be a significantly larger and higher--energy dataset of lepton--proton collisions in comparison to the existing HERA structure function measurements~\cite{Abramowicz:2015mha}.
Indeed, the latter to this day form the backbone of all PDF determinations~\cite{Harland-Lang:2014zoa,Dulat:2015mca,Abramowicz:2015mha,Alekhin:2017kpj,Ball:2017nwa}, and thus the LHeC would
provide the opportunity to greatly extend the precision and reach of HERA data in both $x$ and $Q^2$,
highlighting its potential for PDF constraints.
Moreover, these measurements
would be taken in the relatively clean environment of lepton--proton collisions, where the corresponding
theoretical predictions are known to a very high level of precision.
It should also be emphasized here that the LHeC would reach studies beyond the proton structure, such as the characterisation
of the Higgs sector or the study of cold nuclear matter in the small-$x$ region, where new QCD dynamical
regimes such as saturation are expected to appear.

Quantitative PDF projection studies based on LHeC pseudodata have been presented
previously~\cite{AbelleiraFernandez:2012cc,AbelleiraFernandez:2012ty,Paukkunen:2017phq,Cooper-Sarkar:2016udp},
where a sizeable reduction in the PDF uncertainties is reported.
These LHeC PDF projections are based upon the HERAPDF-like input PDF parameterisation and flavour assumptions~\cite{Abramowicz:2015mha},
with some additional freedom in the input parametrisation
added in the most recent studies~\cite{LHeCtalk1,LHeCtalk2}.
However, different
results may be obtained if a more flexible parameterisation or alternative flavour assumptions are used,
as shown for example in the study of~\cite{Ball:2017otu} carried out in the \texttt{NNPDF} framework.
In addition, the interplay of these constraints from the LHeC with the expected sensitivity from the HL--LHC~\mycite{Khalek:2018mdn} has not yet been studied.
Thus in this section we discuss how the projected sensitivity of a state-of-the-art
global PDF determination
will improve with data from the LHeC, and how this will complement the information
contained in the measurements in $pp$ collisions provided by the HL--LHC.

We follow the strategy presented in the previous section (Ref.~\mycite{Khalek:2018mdn} for more details), starting from the
\texttt{PDF4LHC15} ~\cite{Gao:2013bia,Carrazza:2015aoa,Butterworth:2015oua} baseline set, and quantify the expected impact of the LHeC measurements both individually and combined
with the information provided by the HL--LHC. 
As we will demonstrate, the expected constraints from the LHeC are significant and fully complementary
with those from the HL--LHC.
When included simultaneously, a significant reduction in PDF uncertainties in the entire relevant
kinematical range for the momentum fraction $x$ is achieved, with beneficial implications for LHC phenomenology. Additional methodological investigation has been performed in Ref.~\mycite{AbdulKhalek:2019mps, AbelleiraFernandez:2012cc} where we assess the impact of adopting a more restrictive input parameterisation as our baseline PDF set.
%

For the LHeC pseudodata, we use the most recent publicly available official
LHeC projections~\cite{mklein} (see also~\cite{Klein:1564929} for
further details) for electron
and positron neutral-current (NC) and charged-current (CC) scattering.
The main features of the pseudodata sets we consider are summarised in Table~\ref{tab:lhecdat}, along with the corresponding integrated luminosities and kinematic reach. 
While the nominal high energy data ($E_p=7$ TeV) provides the dominant PDF constraints, the
lower energy ($E_p=1$ TeV) data extends the acceptance to higher $x$ and provides a handle on the longitudinal structure function, $F_L$, and hence the gluon PDF (we note that further variations in the electron and/or proton energy will provide additional constraints on this, as well as on other novel low--$x$ QCD phenomena). 
The charm and bottom heavy quark NC structure function pseudodata provide additional constraints on the gluon.
In addition, charm production in $e^- p$ CC scattering provides important information on the anti-strange quark distributions
via the $\overline{s}+ W \to \overline{c}$ process. We do not include charm production in $e^+ p$ CC scattering, as the corresponding projections are not currently publicly available, though this would provide an additional constraint on the strange quark PDF.
We apply a kinematic cut of $Q \ge 2$ GeV to ensure that the fitted
data lie in the range where perturbative QCD calculations can be reliably
applied.
%

\begin{table}[!h]
    \centering
    \renewcommand{\arraystretch}{1.70}
    \small
    \begin{tabular}{c|c|c|c|c}
      \toprule
      Observable    &  $E_p$ &  Kinematics  &    $N_\text{dat}$  & $\mathcal{L}_\text{int}$ [$\text{ab}^{-1}$] \\
  \toprule
  $\tilde{\sigma}^\text{NC}$ ($e^- p$)  & 7 TeV    &  $5\times10^{-6}\le x \le 0.8$,
  $5 \le Q^2 \le 10^6$ $\text{GeV}^2$          &
  150  &  1.0 \\
  \midrule
  $\tilde{\sigma}^\text{CC}$ ($e^- p$) & 7 TeV    &  $8.5 \times 10^{-5}\le x \le 0.8$,
  $10^2 \le Q^2 \le 10^6$ $\text{GeV}^2$            &
  114   &  1.0 \\
  \midrule
  $\tilde{\sigma}^\text{NC}$ ($e^+ p$)  & 7 TeV    &  $5\times10^{-6}\le x \le 0.8$,
  $5 \le Q^2 \le 5\times 10^5$ $\text{GeV}^2$          &
  148   &  0.1 \\
  \midrule
  $\tilde{\sigma}^\text{CC}$ ($e^+ p$)  & 7 TeV    &  $8.5 \times 10^{-5}\le x \le 0.7$,
  $10^2 \le Q^2 \le 5\times 10^5$ $\text{GeV}^2$            &
  109   &  0.1 \\
  \midrule
  $\tilde{\sigma}^\text{NC}$ ($e^- p$) & 1 TeV    &  $5\times10^{-5}\le x \le 0.8$,
  $2.2 \le Q^2 \le 10^5$ $\text{GeV}^2$          &
  128   &  0.1 \\
  \midrule
  $\tilde{\sigma}^\text{CC}$ ($e^- p$) & 1 TeV    &  $5\times 10^{-4}\le x \le 0.8$,
  $10^2 \le Q^2 \le 10^5$ $\text{GeV}^2$            &
  94   &  0.1 \\
  \midrule
  $F_{2}^{c,\text{NC}}$ ($e^- p$) & 7 TeV    &  $7\times 10^{-6}\le x \le 0.3$,
  $4 \le Q^2 \le 2\times 10^5$ $\text{GeV}^2$            &
  111  &  1.0 \\
  \midrule
  $F_2^{b,\text{NC}}$ ($e^- p$) & 7 TeV    &  $3\times 10^{-5}\le x \le 0.3$,
  $32 \le Q^2 \le 2\times 10^5$ $\text{GeV}^2$            &
  77  &  1.0 \\
  \midrule
  $F_2^{c,\text{CC}}$ ($e^- p$) & 7 TeV    &  $10^{-4}\le x \le 0.25$,
  $10^2 \le Q^2 \le 10^5$ $\text{GeV}^2$            &
  14  &  1.0 \\
  \bottomrule
  Total   &   &  &   945  &  \\
  \bottomrule
    \end{tabular}
    \vspace{0.3cm}
    \caption{\small \label{tab:pseudo-dataLHeC}
      Overview of the main features of the LHeC pseudo-data~\cite{mklein}  included in our PDF projections.
      For each process, we indicate the kinematic coverage, the integrated
      luminosity, the proton energy, and the number of pseudo-data
      points, $N_\text{dat}$, after the $Q\ge 2$ GeV kinematic cut.
      Note that in all cases the incoming lepton energy is fixed to be
      $E_l=60$ GeV.
      We ignore the effect of the incoming lepton beam polarization.
    }\label{tab:lhecdat}
  \end{table}

\begin{figure}[!htb]
  \begin{center}
    \makebox[\textwidth]{\includegraphics[width=0.9\textwidth]{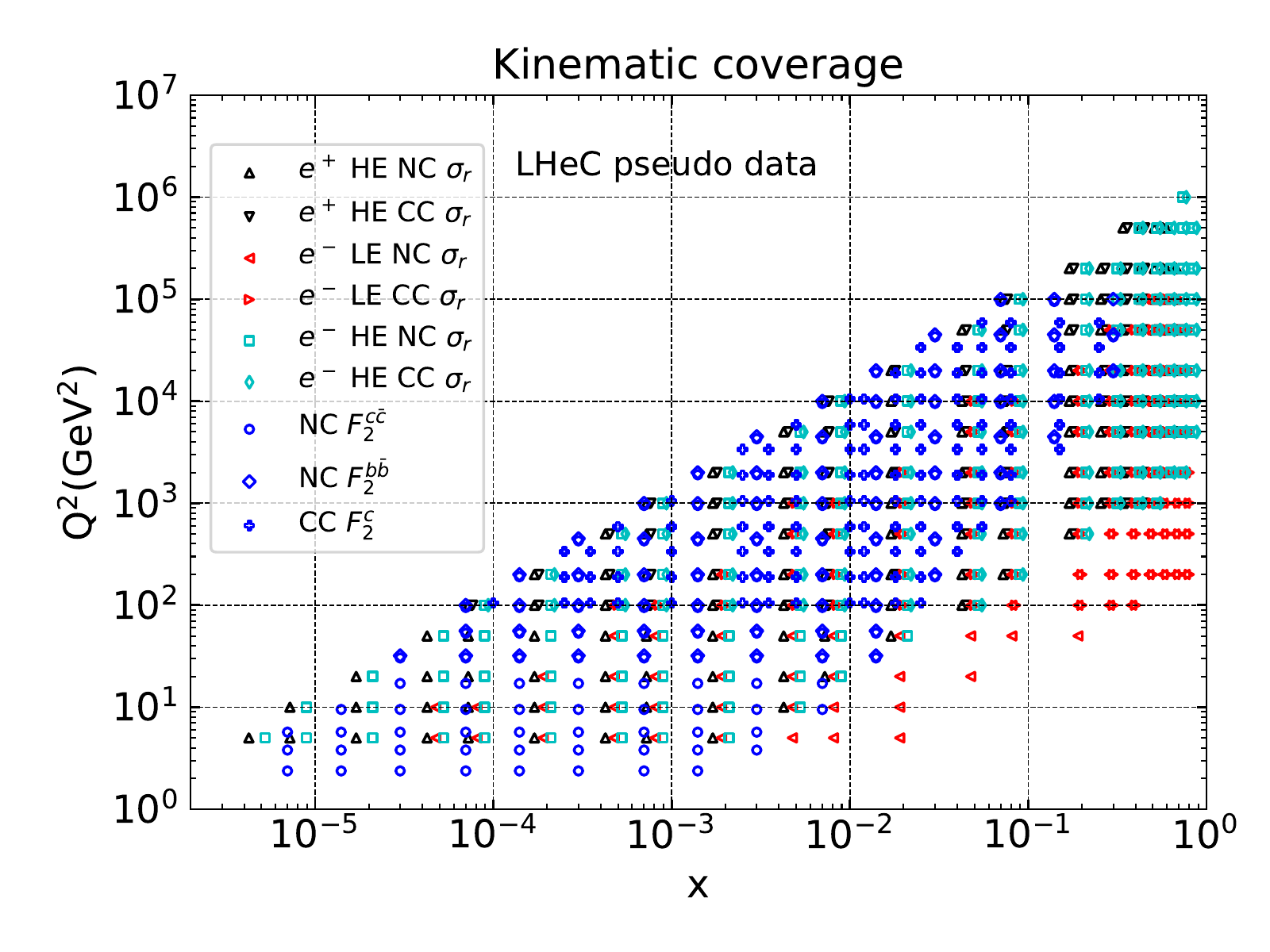}}
  \end{center}
  \vspace{-0.8cm}
\caption{
  The kinematic coverage in the $(x,Q^2)$ plane of the LHeC pseudodata~\cite{mklein} included in this analysis.}
 \label{fig:kinLHeC}
\end{figure}

The kinematic reach in  the $(x,Q^2)$ plane of the LHeC pseudodata is shown in Fig.~\ref{fig:kinLHeC}.
%
The reach in the perturbative region ($Q\ge 2$ GeV)
is well below $x \approx 10^{-5}$ and extends up to $Q^2 \approx 10^6$ $\text{GeV}^2$
(that is, $Q\simeq 1$ TeV), increasing the HERA coverage
by over an order of magnitude in both cases, via the factor $\sim$ 4
increase in
the collider centre-of-mass energy $\sqrt{s}$.
Due to the heavy quark tagging requirements, the reach for semi-inclusive structure functions
only extends up to $x\simeq 0.3$ in the large-$x$ region.

\begin{figure}[!h]
  \begin{center}
    \makebox[\textwidth]{\includegraphics[width=\textwidth]{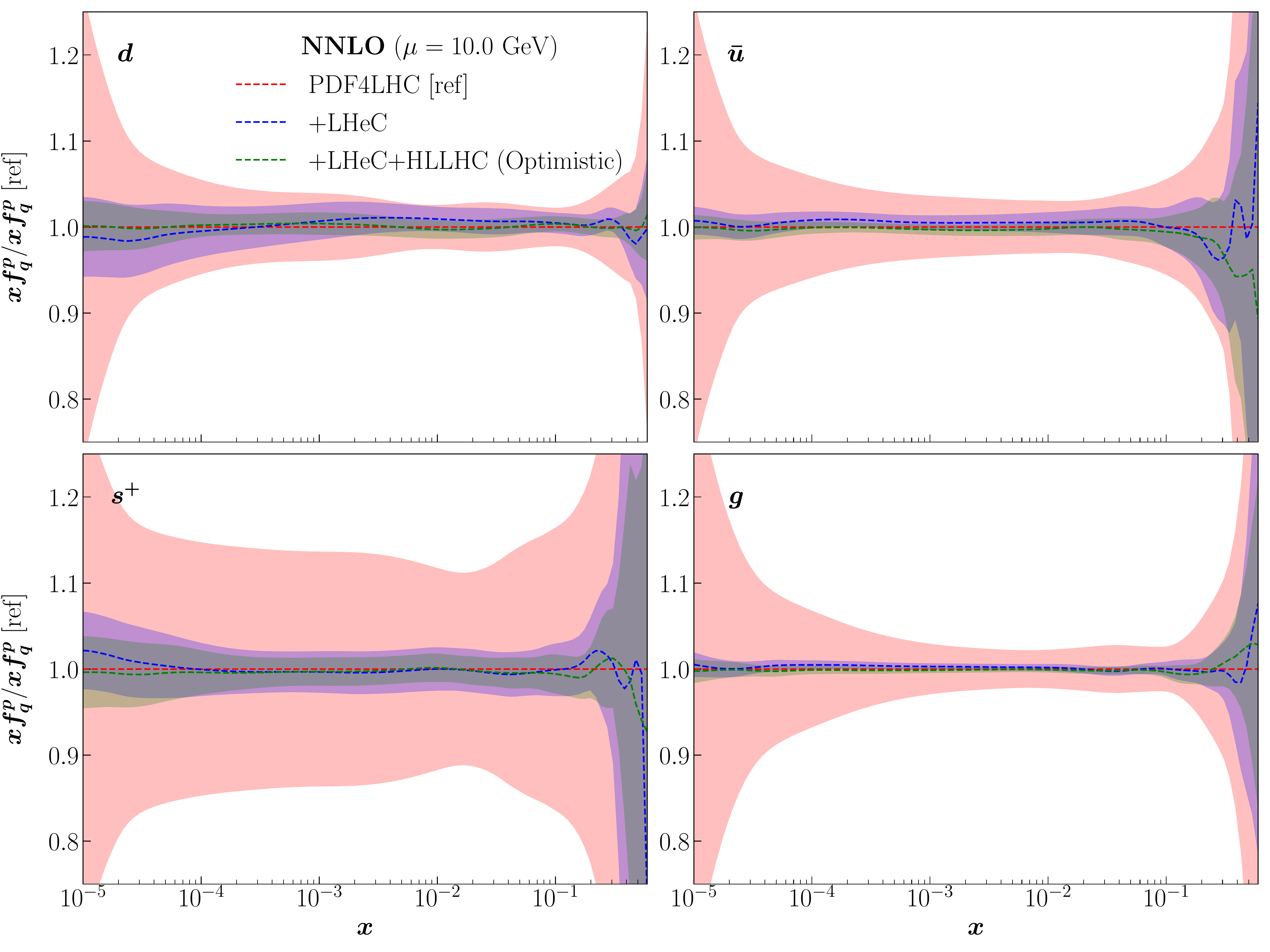}}
    \end{center}
    \vspace{-0.8cm}
    \caption{On the left, Comparison of the \texttt{PDF4LHC15}  at NNLO set with the LHeC profiled set, the HL-LHC profiled set in optimistic scenario (see Table~\ref{tab:Scenarios}) and their combinations. We show the down quark, up anti-quark, total strangeness and gluon at $Q$=10 GeV, normalised to the central value of the \texttt{PDF4LHC15}  baseline. The bands correspond to the one-sigma PDF uncertainties. On the right, same as left but relative uncertainties are shown.}
  \label{fig:LHeCandHLLHC_PDFs_Q10}
\end{figure}

In this analysis, we use $T=3$ same as in the HL--LHC study, except that in this case we don't need for uncertainty-tuners ($f_\text{corr}$,$f_\text{red}$ and $f_\text{acc}$) as we're using the official LHeC projections. 
The resulting profiled PDF set is shown in Fig.~\ref{fig:LHeCandHLLHC_PDFs_Q10}.
We find that
the LHeC and HL--LHC datasets both place significant constraints on the PDFs, with some differences
depending on the kinematic region or the specific flavour combination being considered.
Most importantly, we deduce that these are rather complementary: while the LHeC places the most significant constraint at low to intermediate $x$ in general (though in the latter case the HL--LHC impact is often comparable in size), at high $x$ the HL--LHC places the dominant constraint on the gluon and strangeness, while the LHeC dominates for the up and down quarks.
Moreover, when both the LHeC and HL--LHC pseudodata are simultaneously included
in the fit, all PDF flavours can be constrained across a wide range of $x$, providing a strong motivation
to exploit the  input for PDF fits from both experiments, and therefore for realising the LHeC itself.

Finally, a few important caveats concerning this exercise should be mentioned.
First, the processes included for both the LHeC and HL--LHC, while broad in scope, are by no means exhaustive.
Most importantly, no jet production were included for the LHeC, which
would certainly improve the constraint on the high-$x$ gluon. In addition, the inclusion of charm production in $e^+ p$ CC scattering would further constrain the strange quark.
In the case of the HL--LHC, only those processes which provide an impact at high-$x$ were included, and hence the lack of constraint at low-$x$ that is observed occurs essentially by construction.
In particular, there are a number of processes that will become available
with the legacy HL--LHC dataset, or indeed those in the current LHC dataset that are not currently included in global fits, but which can in principle constrain the low-$x$ PDFs, from low mass Drell--Yan to inclusive $D$ meson production~\cite{Zenaiev:2015rfa,Gauld:2016kpd} and exclusive vector meson photo-production~\cite{Jones:2016ldq}, though here the theory is not available at the same level of precision to the LHeC case.



\section{Nuclear PDFs for the future Large Hadron Collider} 
\label{s1:nNNPDF20_ImpactStudies}
In this section, I discuss various phenomenological
applications of the \texttt{nNNPDF2.0} determination.
%
%
%
%
I start by discussing the potential of the FoCal upgrade to the ALICE
detector in constraining the small-$x$ gluon nuclear PDF using
measurements of direct photon production in the forward region.
Afterwards, I provide the predictions based on \texttt{nNNPDF2.0} for inclusive hadron
production in proton-nuclear collisions, a process that can constrain both the
quark and gluon nuclear PDFs as well as the corresponding fragmentation
functions in vacuum and in medium.

\myparagraph{Isolated photon production in pA collisions with FoCal}
%
Current measurements of direct photon production at the LHC, such as
those discussed above from the ATLAS collaboration~\cite{Aaboud:2019tab}
as well as related measurements from CMS and
ALICE~\cite{Acharya:2020sxs}, are restricted to the central rapidity
region.
The reason is that this is the only region instrumented with
electromagnetic calorimeters and thus suitable to identify photons.
A measurement of isolated photon production in the forward region,
however, is also highly interesting for nPDF studies.
Not only would such measurements provide direct access to the
poorly-known gluon nuclear modifications at small-$x$, but 
it would also allow testing for the possible onset of QCD non-linear
dynamics~\cite{Benic:2016uku}.

With this motivation, a new forward calorimeter extension of the ALICE
detector, dubbed
FoCal~\cite{vanderKolk:2020fqo,ALICECollaboration:2719928}, has been
proposed.
Both the acceptance and instrumentation of this detector have been optimized
to provide access to the nuclear PDFs at low scales and small momentum
fractions via the measurement of isolated photon production at low
transverse momenta and forward rapidities in proton-ion collisions.
The FoCal is proposed for installation during the Long Shutdown 3
(2025-2026) phase of the LHC.

The impact of future FoCal measurements on the small-$x$ nuclear PDFs
was first studied in Ref.~\cite{vanLeeuwen:2019zpz}.
In that analysis, pseudodata based on the expected kinematical reach
and experimental uncertainties for FoCal was generated and used to
constrain the \texttt{nNNPDF1.0} determination by means of the Bayesian
reweighting method~\cite{Ball:2010gb,Ball:2011gg}.
It was found that the FoCal measurement would constrain the nuclear
gluon modifications down to $x\simeq 10^{-5}$, leading to an uncertainty
reduction by up to an order of magnitude as compared to the baseline
fit.
These results indicated a comparable or superior constraining power on
the small-$x$ nPDFs when compared to related projections from future
facilities, such as the Electron Ion Collider~\cite{Accardi:2012qut}.

Motivated by the new and improved projections for the FoCal 
pseudodata that have recently became available, 
we revisit their impact on nuclear PDFs
using the present nPDF determination. 
In this case, the \texttt{nNNPDF2.0} PDFs represent a more realistic 
baseline since they provide a robust quark flavour
separation with a better handle on the gluon.
Moreover, the positivity of physical cross-sections is guaranteed, 
a constraint that helps to reduce the small-$x$ nuclear PDF uncertainties.

\begin{figure}[!h]
\begin{center}
\includegraphics[width=0.49\textwidth]{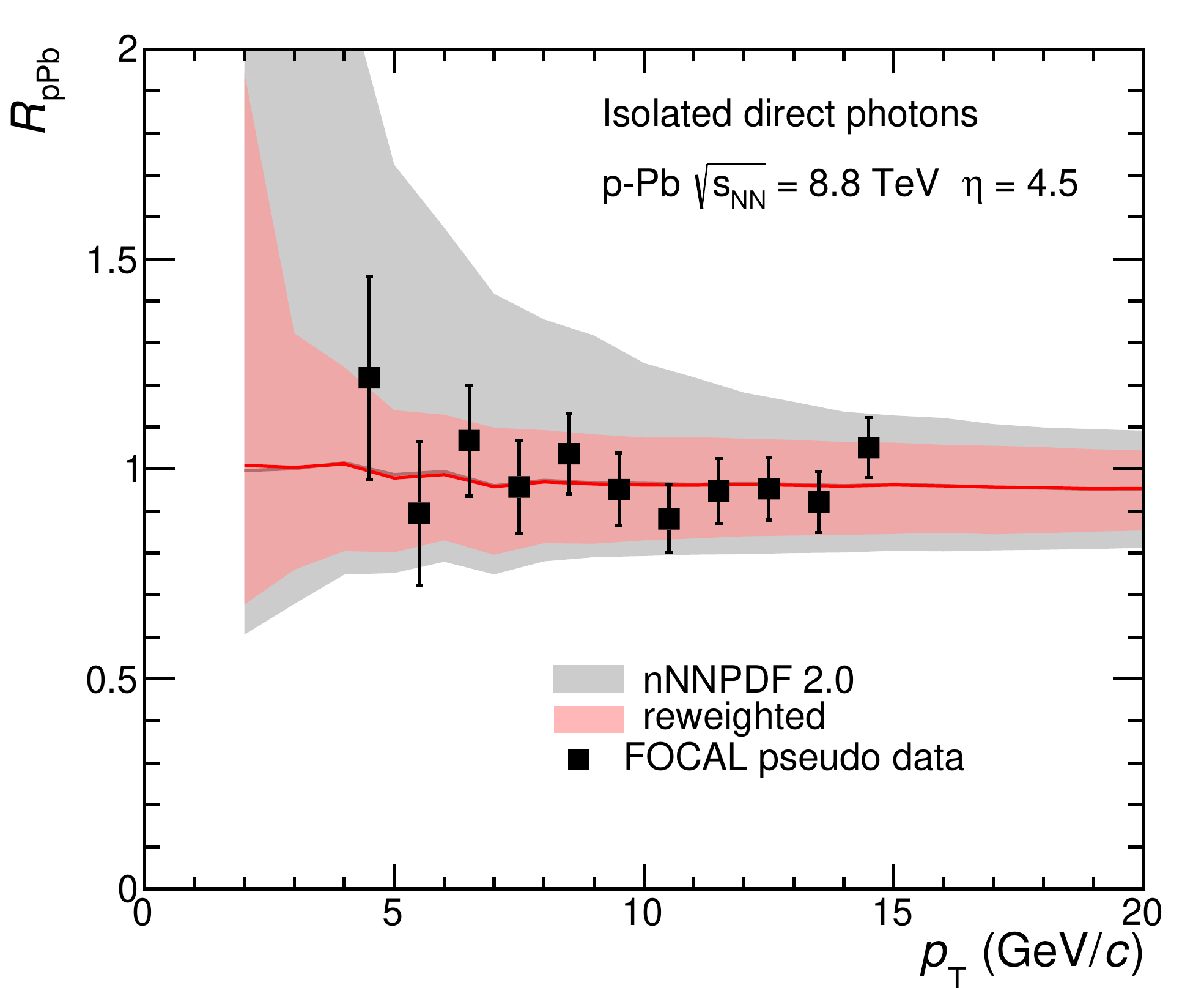}
\includegraphics[width=0.49\textwidth]{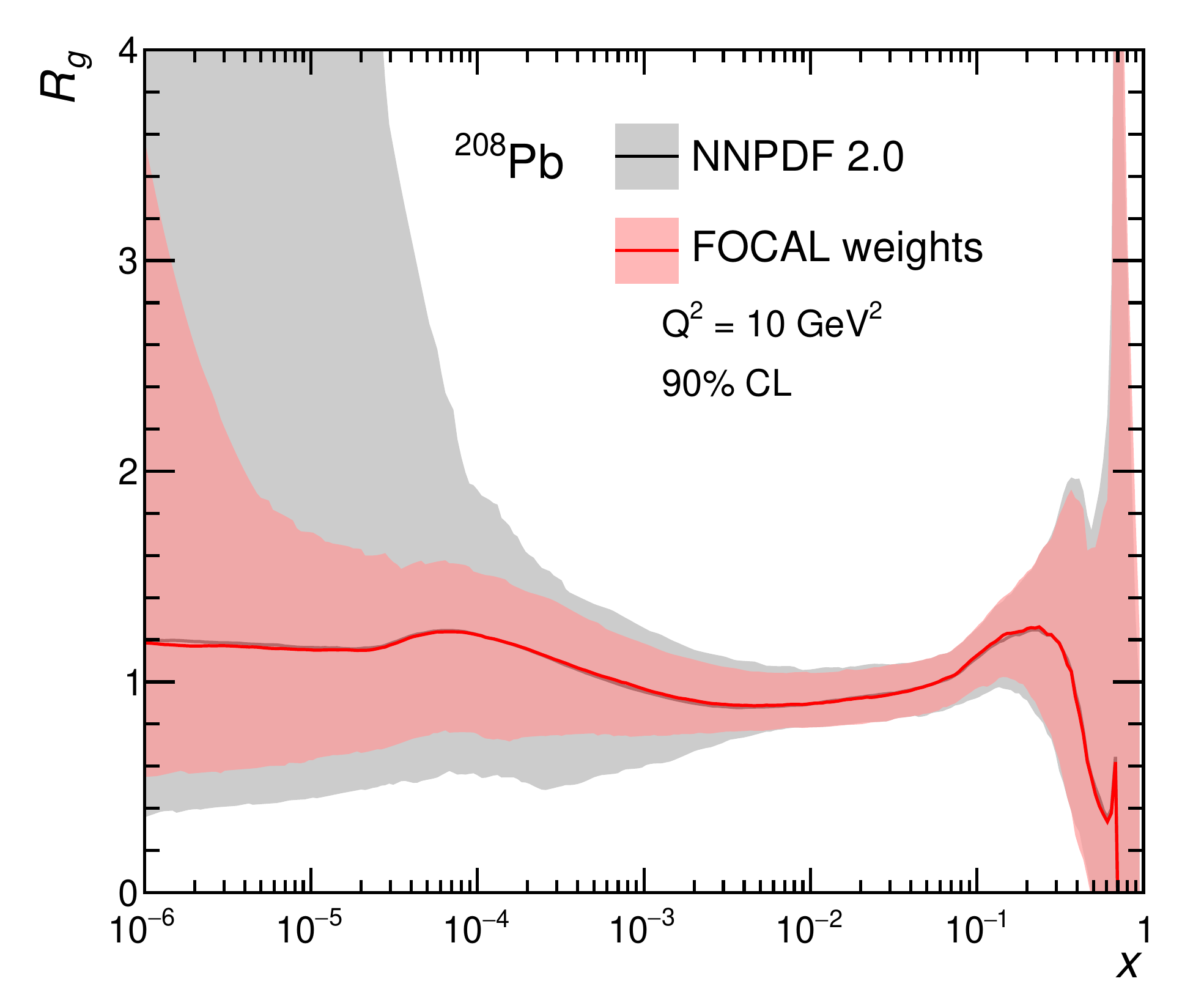}
  \end{center}
  \vspace{-0.8cm}
\caption{Left: the nuclear modification factor $R_{\rm pPb}(p_T^\gamma)$
  for direct photon production in pPb collisions at $\sqrt{s}=8.8$ TeV
  for a rapidity of $\eta_\gamma=4.5$ as a function of the photon
  transverse momentum $p_T^\gamma$.
  The theoretical predictions are compared with the FoCal pseudodata
  for two sets of input nPDFs: the original \texttt{nNNPDF2.0} set, and the
  variant that has been reweighted with with FoCal projections.
  Here the FoCal pseudodata assumes the central value of the \texttt{nNNPDF2.0}
  prediction.
  Right: the gluon nuclear modification factor $R_g(x,Q)$ for $Q^2=10$
  GeV$^2$ for both the original and the reweighted \texttt{nNNPDF2.0} fits.
  The nPDF uncertainties correspond in both cases to the 90\% confidence
  level intervals.
  \label{fig:focal}
}
\end{figure}

For this study we have adopted the same settings as
in Ref.~\cite{vanLeeuwen:2019zpz} and computed NLO QCD predictions with a
modified version of {\tt INCNLO} that benefits from improved numerical
stability at forward rapidities~\cite{Helenius:2014qla}.
Theoretical predictions for FoCal cross-sections have been computed with
$N_{\rm rep}=400$ replicas of \texttt{nNNPDF2.0}, which are subsequently used to
account for the impact of the FoCal pseudodata by means of Bayesian
reweighting.\footnote{We are grateful to Marco van Leeuwen for providing
us with the results presented here.}
Fig.~\ref{fig:focal} displays the nuclear modification factor $R_{\rm
  pPb}(p_T^\gamma)$ for direct photon production in pPb collisions at
  $\sqrt{s}=8.8$ TeV for a rapidity of $\eta_\gamma=4.5$ as a function
  of the photon's transverse momentum $p_T^\gamma$.
  The theoretical predictions based on NLO QCD theory are compared with
  the FoCal pseudodata for two sets of input nPDFs: the original
  \texttt{nNNPDF2.0} set, and the variant that has been reweighted with the FoCal
  projections.
  Here the central value of the FoCal pseudodata has been
  chosen to be the same as that of the \texttt{nNNPDF2.0} prediction.
  In the right panel of Fig.~\ref{fig:focal} we show the gluon nuclear
  modification factor $R_g(x,Q)$ for $Q^2=10$ GeV$^2$ for both the
  original and the reweighted \texttt{nNNPDF2.0} fits.
  In all cases, the nPDF uncertainty bands correspond to the 90\%
  confidence level intervals.

From the results of Fig.~\ref{fig:focal}, one finds that the FoCal
measurements would still impact the uncertainties of the nuclear gluon
modifications at small-$x$, especially in the upper limit of the
uncertainty band.
The effective number of replicas in this case is $N_{\rm
eff}=345$.
Note that \texttt{nNNPDF2.0} exhibits a preference for $R_{\rm pPb}\simeq 1$,
and thus shadowing is not favoured in the gluon sector, consistent
with the results reported in Fig.~\ref{R_A}.
On the other hand, \texttt{nNNPDF2.0} does not contain any dataset with
particular sensitivity to the nuclear gluon modifications, implying that
the projections for the impact of FoCal in the global nPDF analysis
could be somewhat over-optimistic.

\begin{figure}[!h]
  \begin{center}
    \includegraphics[width=0.49\textwidth]{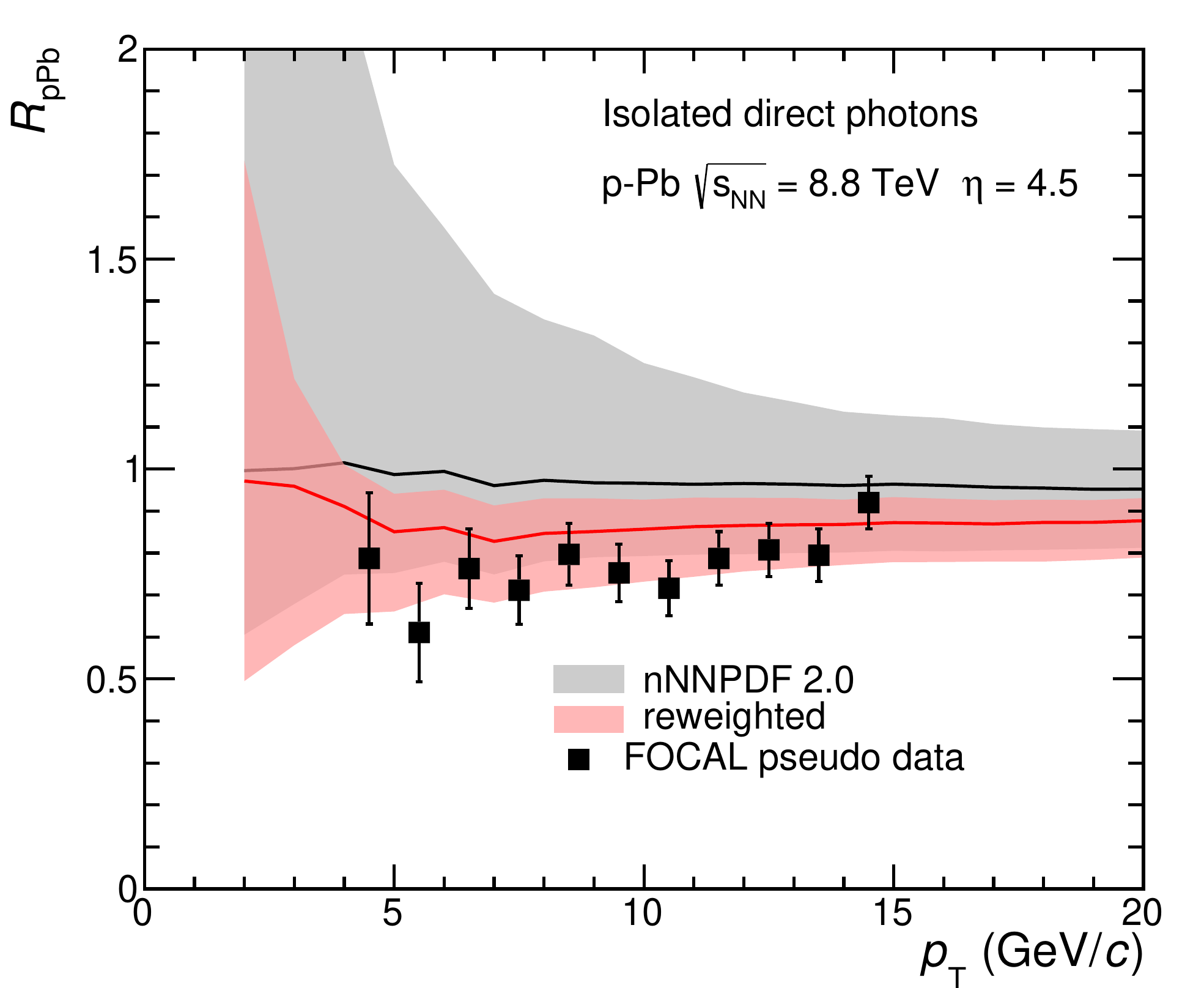}
    \includegraphics[width=0.49\textwidth]{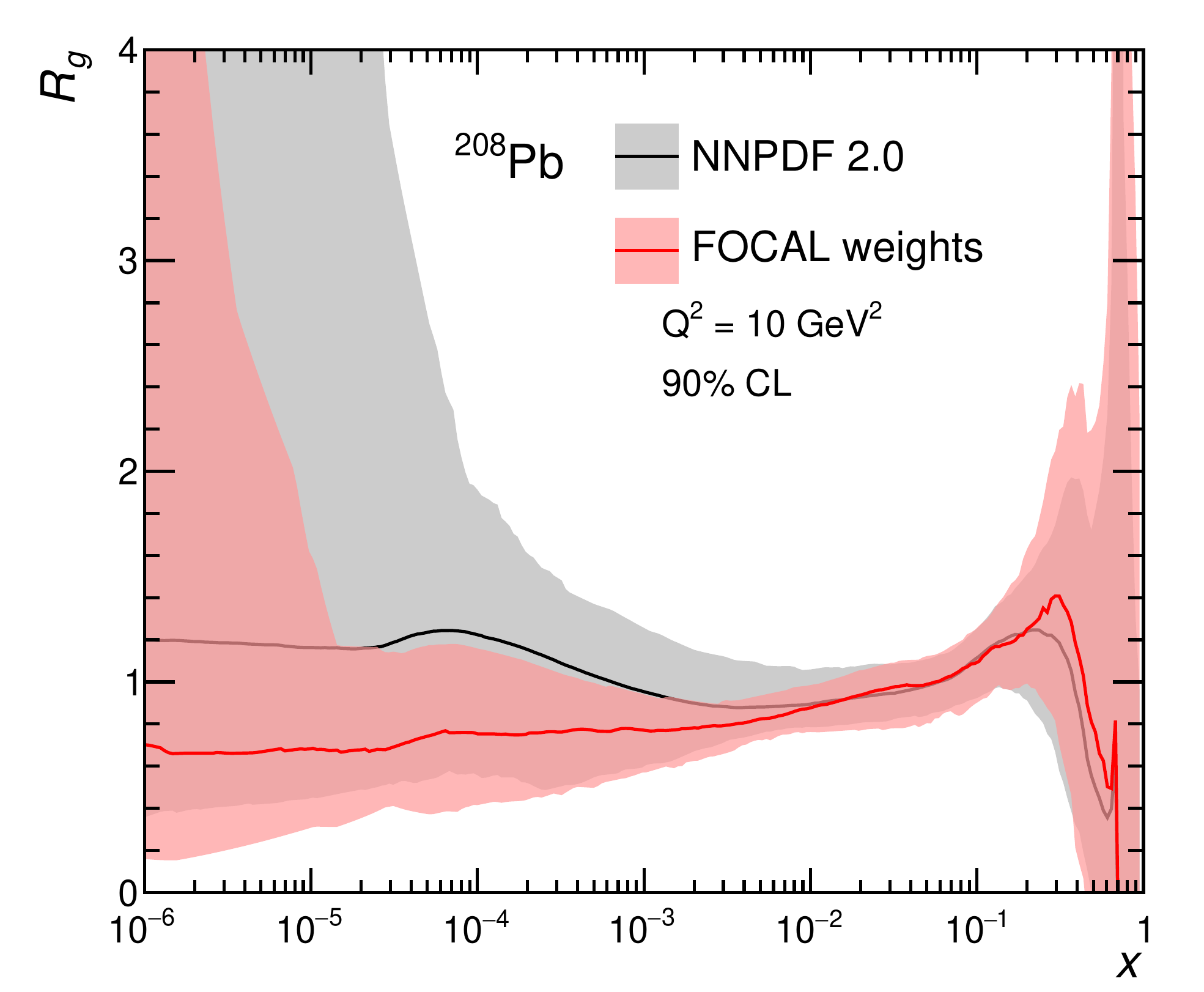}
    \end{center}
    \vspace{-0.8cm}
\caption{Same as Fig.~\ref{fig:focal} now for the case where the FoCal
  pseudodata has been generated under the assumption that $R_{\rm pPb}
  \simeq 0.6$ rather than based on the \texttt{nNNPDF2.0} central value.
  \label{fig:focal2}
}
\end{figure}

Crucially, however, we have assumed in this exercise that the central value of the
FoCal measurement would be unchanged compared to the initial baseline
prediction.
In Fig.~\ref{fig:focal2} we display instead the results of the
reweighting for a scenario in which the FoCal pseudodata 
have a value of $R_{\rm pPb} \simeq 0.6$.
  In this case, the effective number of replicas is much smaller, $N_{\rm
  eff}=117$, indicating that the FoCal data are adding a significant
  amount of new information to the global fit.
  Here the resulting value for the gluon nuclear modification
  ratio at small-$x$ would be $R_g\simeq 0.7$.\footnote{
  Note that the reweighting technique may lead to unreliable 
  uncertainty bands when using data values that fall 
  outside the predictions produced by the prior.}
  Therefore, this analysis indicates that FoCal measurements could be
  sensitive either to the gluon shadowing effects or to possible
  non-linear QCD dynamics.
  To disentangle one from the other, a dedicated analysis of the
  $\chi^2$ and nPDF behaviour in the small-$x$ region would
  be required, following the approach developed in Ref.~\cite{Ball:2017otu}. 

\myparagraph{Inclusive hadron production in pA collisions}
%
The inclusive production of pions and kaons in hadronic collisions
provides information not only on the initial state (parton
distribution functions) but also on the final-state hadronisation
mechanism of partons into hadrons.
The latter is described by the fragmentation functions (FFs), which are
extracted from experimental data by means of a global analysis akin to
that of the PDFs~\cite{deFlorian:2014xna,Sato:2019yez,d'Enterria:2013vba,Bertone:2018ecm,Bertone:2017tyb,Albino:2005me,Hirai:2007cx,Khalek:2021gxf}.
Likewise, in proton-nuclear collisions the production of identified
hadrons can provide information on the initial state nuclear PDFs
as well the parton-to-hadron hadronisation in the
presence of cold nuclear matter effects.

\begin{figure}[!h]
\begin{center}
  \includegraphics[width=0.99\textwidth]{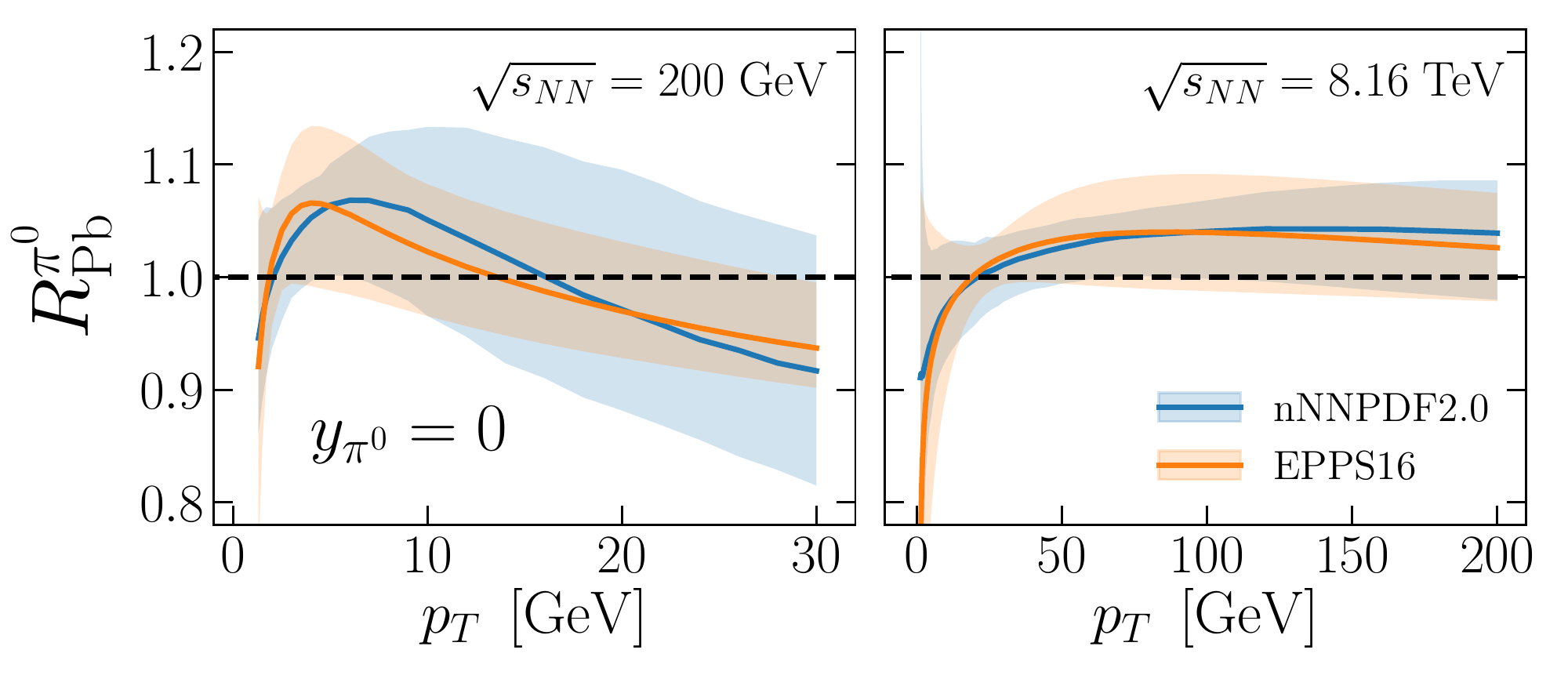}
 \end{center}
 \vspace{-0.8cm}
\caption{\small The nuclear modification ratio $R_{\rm Pb}^{\pi^0}$ for
  the production of neutral pions in proton-lead collisions as a
  function of the pion transverse momentum $p_T$.
  We provide theoretical predictions based on NLO QCD and the DSS14
  hadron fragmentation functions both for \texttt{nNNPDF2.0} and \texttt{EPPS16}, with the
  corresponding nPDF uncertainties in each case.
  Results are provided for the RHIC kinematics (left), corresponding to
  $\sqrt{s}=200$ GeV, and for the LHC kinematics (right), where
  $\sqrt{s}=8.16$ TeV, and in both cases pions are produced centrally,
  $|y_{\pi^0}|=0$.
  \label{fig:plot_R_all}
}
\end{figure}

In Fig.~\ref{fig:plot_R_all}, we display the nuclear modification ratio
  $R_{\rm Pb}^{\pi^0}$ for the production of neutral pions in
  proton-lead collisions as a function of the pion transverse momentum
  $p_T$.
  The theoretical calculations
  are based on NLO QCD and use the DSS14
  hadron fragmentation functions~\cite{deFlorian:2014xna}
  for both the \texttt{nNNPDF2.0} and \texttt{EPPS16} predictions.\footnote{We are grateful to Ilkka Helenius for providing us with the results of this calculation.}
  Moreover, the central values and 90\% CL uncertainties 
  are provided for RHIC kinematics, corresponding to
  $\sqrt{s}=200$ GeV, and for LHC kinematics, where $\sqrt{s}=8.16$
  TeV.
  In both cases, the pions are assumed to be measured at central
  rapidities, $y_{\pi^0}= 0$. 
  See Refs.~\cite{d'Enterria:2013vba,Albacete:2017qng} for additional details
  regarding the theoretical calculation of inclusive pion production in
  hadronic collisions.

From Fig.~\ref{fig:plot_R_all} we can see that the \texttt{nNNPDF2.0} prediction
for $R_{\rm Pb}^{\pi^0}$ is consistent with unity within uncertainties
for all values of $p_T$ both at RHIC and LHC kinematics.
At RHIC kinematics, we find that the ratio is less than one at the
smallest $p_T$ values, becomes $R > 1$ between $p_T=3$ and 17 GeV, and
then goes back to $R<1$.
Since inclusive hadron production is dominated by quark-gluon
scattering, in particular the scattering of valence quarks for
neutral pion production, this behaviour is consistent with the results shown in
Fig.~\ref{fig:R_Pb}. 
From low to high $p_T$, one moves from the
shadowing region to the anti-shadowing enhancement, and ends
in the region sensitive to EMC suppression.
A similar explanation can be made for the trends in $R_{\rm Pb}^{\pi^0}$
at the LHC kinematics. 
However, here the ratio
$p_T/\sqrt{s}$ does not become large enough to reach the EMC region, and
thus the ratio remains larger than one for most of the $p_T$ range as a result 
of anti-shadowing effects.
Lastly, the \texttt{EPPS16} predictions agree with the \texttt{nNNPDF2.0}
result well within uncertainties, reflecting the underlying consistency 
at the nPDF level.

Overall, the results of Fig.~\ref{fig:plot_R_all} confirm that inclusive hadron
production in proton-nucleus collisions can provide 
a handle on the nuclear PDF modifications at medium and large-$x$, 
although an optimal interpretation of the experimental data can only be 
achieved by the simultaneous determination of the nPDFs together 
with the hadron fragmentation functions.

\section{Proton and nuclear PDFs for the Electron-Ion Collider} 
\label{s1:EIC}
\myparagraph{Motivation} The construction of an Electron-Ion Collider
(EIC)~\cite{Accardi:2012qut,Aschenauer:2017jsk} has been recently approved by
the United States Department of Energy at Brookhaven National Laboratory, and
could record the first scattering events as early as 2030. By colliding
(polarised) electron or positron beams with proton or ion beams for a wide range of
center-of-mass energies, the EIC will perform key measurements to investigate
quantum chromodynamics (QCD) at the intensity frontier. These measurements will
be fundamental to understand how partons are distributed in position and
momentum spaces within a proton, how the proton spin originates from the spin
and the dynamics of partons, how the nuclear medium modifies partonic
interactions, and whether gluons saturate within heavy nuclei.

\myparagraph{Outline} In this section we focus on one important class of EIC measurements, namely
inclusive cross sections for unpolarised lepton-proton and lepton-nucleus
deep-inelastic scattering (DIS). In particular we study how such data could
improve the determination of the unpolarised proton and nuclear parton
distribution functions (PDFs)~\cite{Ethier:2020way} by incorporating suitable
pseudodata in a self-consistent set of PDF fits based on the NNPDF methodology
(see Ref.~\cite{Ball:2014uwa} and references therein for a comprehensive
description). The unique ability of an EIC to measure inclusive DIS cross
sections consistently for the proton and a wide range of nuclei will be
exploited also to update the proton PDFs used as a boundary condition in the
nuclear PDF fit. This feature distinguishes our analysis from previous
studies~\cite{Aschenauer:2017oxs,AbdulKhalek:2019mzd}, and may be extended to a
simultaneous determination of proton and nuclear PDFs in the future. The
results presented in this work integrate those contained in Sects.~7.1.1 and
7.3.3 of the upcoming EIC Yellow Report~\mycite{AbdulKhalek:2021gbh}. They systematically
account for the impact of projected inclusive DIS measurements at an EIC on the
unpolarised proton PDFs for the first time (for projected semi-inclusive DIS
measurements see Ref.~\cite{Aschenauer:2019kzf}), and supersede a previous
NNPDF analysis of the impact of EIC measurements on nuclear
PDFs~\mycite{AbdulKhalek:2019mzd}. Similar studies for polarised PDFs have been
performed elsewhere~\cite{Aschenauer:2012ve,Aschenauer:2013iia,
  Aschenauer:2015ata,Aschenauer:2020pdk}, including in the NNPDF
framework~\cite{Ball:2013tyh}.

We first describe how EIC pseudodata
are generated. We then study how they affect the proton and nuclear PDFs once
they are fitted. Lastly, we illustrate how an updated determination of nuclear
PDFs can affect QCD at the cosmic frontier, in particular predictions for
the interactions of highly-energetic neutrinos with matter as they propagate
through Earth towards large-volume detectors.

\myparagraph{Pseudodata generation}
In this analysis we use the same pseudodata as in the EIC Yellow
Report~\mycite{AbdulKhalek:2021gbh}, see in particular Sect.~8.1. In the case of lepton-proton
DIS, they consist of several sets of data points corresponding to either the
neutral-current (NC) or the charged-current (CC) DIS reduced cross sections,
$\sigma^\text{NC}$ and $\sigma^\text{CC}$, respectively. See, \textit{e.g.}, Eqs.~(7)
and (10) in Ref.~\cite{Ball:2008by} for their definition.
Both electron and positron beams are considered, for various forecast energies
of the lepton and proton beams. In the case of lepton-nucleus DIS, the
pseudodata correspond only to NC DIS cross sections, see, \textit{e.g.}, the
discussion in Sect.~2.1 of Ref.~\mycite{AbdulKhalek:2019mzd} for their definition.
Both electron and positron beams are considered in conjunction with a deuteron
beam; only an electron beam is instead considered for other ions, namely
$^4$He, $^{12}$C, $^{40}$Ca, $^{64}$Cu, and $^{197}$Au.
A momentum transfer $Q^2>1$~GeV$^2$, a squared invariant mass of the system
$W^2>10$~GeV$^2$ and a fractional energy of the virtual particle exchanged in
the process $0.01\leq y \leq 0.95$ are assumed in all of the above cases,
consistently with the detector requirements outlined in Sect.8.1
of~\mycite{AbdulKhalek:2021gbh}.

The pseudodata distribution is assumed to be multi-Gaussian, as in the case
of real data. It is therefore uniquely identified by a vector of mean values
$\bm{\mu}$ and a covariance matrix $\bm{C}$, for which the following
assumptions are made. The mean values correspond to the theoretical expectations
$\bm{t}$ of the DIS cross sections obtained with a \textit{true} underlying set of
PDFs, and smeared by normal random numbers $\bm{r}$ sampled from the Cholesky decomposition $\bm{L}$ of the covariance matrix such that $\bm{\mu = t + L\cdot r}$ (see Sect.~\ref{s2:monte_carlo}). Specifically we use a recent
variant~\cite{Faura:2020oom} of the NNPDF3.1 determination~\cite{Ball:2017nwa},
and the \texttt{nNNPDF2.0} determination~\mycite{AbdulKhalek:2020yuc}, for proton and
nuclear PDFs, respectively. The covariance matrix is made up of three
components, which correspond to a statistical uncertainty, an additive
uncorrelated systematic uncertainty, and a multiplicative correlated
systematic uncertainty. The statistical uncertainty is determined by assuming
an integrated luminosity $\mathcal{L}$ of 100 fb$^{-1}$ for electron-proton
NC and CC DIS, and of 10 fb$^{-1}$ in all other cases. The systematic
uncertainties are instead determined with the \textsc{djangoh} event
generator~\cite{Charchula:1994kf}, which contains the Monte Carlo program
\textsc{heracles}~\cite{Kwiatkowski:1990es} interfaced to
\textsc{lepto}~\cite{Ingelman:1996mq}. These pieces of software collectively allow
for an account of one-loop electroweak radiative corrections and radiative
scattering. The Lund string fragmentation model, as implemented in
\textsc{pythia/jetset} (see, \textit{e.g.}, Ref.~\cite{Sjostrand:2019zhc} and
references therein) is used to obtain the complete hadronic final state.
The non-perturbative proton and nuclear PDF input is made available to
\textsc{djangoh} by means of numeric tables corresponding to the relevant NC and
CC DIS structure functions, which were generated with
\textsc{apfel}~\cite{Bertone:2013vaa} in the format of
\textsc{lhapdf}~\cite{Buckley:2014ana} grids. The optimal binning of the pseudodata
is determined accordingly.

The complete set of pseudodata considered in this work is summarized in
Table~\ref{tab:EIC_pseudodata}. For each pseudodata set, we indicate the
corresponding DIS process, the number of data points $N_\text{dat}$ before (after)
applying kinematic cuts (see below), the energy of the lepton and of the proton
or ion beams $E_\ell$ and $E_p$, the center-of-mass energy $\sqrt{s}$, the
luminosity $\mathcal{L}$, and the relative uncorrelated and correlated
systematic uncertainties (in percentage) $\sigma_u$ and $\sigma_c$.
Two different scenarios, called \textit{optimistic} and \textit{pessimistic}
henceforth, are considered, which differ for the number of data points and
for the size of the projected systematic uncertainties. In the case of NC
cross sections, the uncorrelated uncertainty was estimated to be 1.5\% (2.3\%)
in the optimistic (pessimistic) scenario. These uncertainties originated from a 1\%
uncertainty on the radiative corrections, and a 1\% (2\%) uncertainty due to
detector effects. The normalization uncertainty was set to 2.5\% (4.3\%). This
included a 1\% uncertainty on the integrated luminosity and a 2\% (4\%)
uncertainty due to detector effects. In the case of CC cross sections,
an uncorrelated uncertainty of 2\% was used in both the optimistic and
pessimistic scenarios, while a normalization uncertainty of 2.3\% (5.8\%)
was used in the optimistic (pessimistic) scenario. This uncertainty includes
contributions from luminosity, radiative corrections and simulation errors.

Estimating systematic uncertainties for an accelerator and a detector which
have not yet been constructed is particularly challenging. The percentages
given in Table~\ref{tab:EIC_pseudodata} build upon the experience of previous
experiments (primarily those at HERA) as well as simulation studies performed
using the EIC Handbook detector and the current EIC detector
matrix~\cite{AbdulKhalek:2021gbh}. Relative systematic uncertainties are
estimated to be independent from the values of $x$ and $Q^2$, in contrast to
statistical uncertainties. For NC pseudodata (with $\mathcal{L}$=100~fb$^{-1}$),
systematic uncertainties are significantly larger than statistical
uncertainties in much of the probed kinematic phase space, see \textit{e.g.}
Figs.~7.1 and 7.67 in~\mycite{AbdulKhalek:2021gbh}. Conversely, for CC pseudodata,
systematic uncertainties are comparable to statistical uncertainties for most
of the measured kinematic space.

\begin{table}
  \footnotesize
  \centering
  \renewcommand{\arraystretch}{1.3}
  \begin{tabularx}{\linewidth}{XXrccrrcc}
    \toprule
    & \multicolumn{2}{l}{\ DIS process}
    & $N_\text{dat}$
    & $E_\ell\times E_p$~[GeV]
    & $\sqrt{s}$~[GeV]
    & $\mathcal{L}$~[fb$^{-1}$]
    & $\sigma_u$~[\%]
    & $\sigma_c$~[\%] \\
    \midrule
      1 
    & $e^-p$
    & CC
    & 89(89)/89(89)
    & 18$\times$275
    & 140.7
    & 100
    & 2.0/2.0
    & 2.3/5.8 \\
      2
    & $e^+p$
    & CC
    & 89(89)/89(89)
    & 18$\times$275
    & 140.7
    & 10
    & 2.0/2.0
    & 2.3/5.8 \\
    \midrule
      3
    & $e^-p$
    & NC
    & 181(140)/131(107)
    & 18$\times$275
    & 140.7
    & 100
    & 1.5/2.3
    & 2.5/4.3 \\
      4 
    & 
    & 
    & 126(81)/91(70)
    & 10$\times$100
    & 63.2
    & 100
    & 1.5/2.3
    & 2.5/4.3 \\
      5
    & 
    & 
    & 116(68)/92(66)
    & 5$\times$100
    & 44.7
    & 100
    & 1.5/2.3
    & 2.5/4.3 \\
      6
    & 
    & 
    & 87(45)/76(45)
    & 5$\times$41
    & 28.6
    & 100
    & 1.5/2.3
    & 2.5/4.3 \\
      7
    & $e^+p$
    & NC
    & 181(140)/131(107)
    & 18$\times$275
    & 140.7
    & 10
    & 1.5/2.3
    & 2.5/4.3 \\
      8 
    & 
    & 
    & 126(81)/91(70)
    & 10$\times$100
    & 63.2
    & 10
    & 1.5/2.3
    & 2.5/4.3 \\
      9
    & 
    & 
    & 116(68)/92(66)
    & 5$\times$100
    & 44.7
    & 10
    & 1.5/2.3
    & 2.5/4.3 \\
      10  
    & 
    & 
    & 87(45)/76(45)
    & 5$\times$41
    & 28.6
    & 10
    & 1.5/2.3
    & 2.5/4.3 \\
    \midrule
      11
    & $e^-d$
    & NC
    & 116(92)/116(92)
    & 18$\times$110
    & 89.0
    & 10
    & 1.5/2.3
    & 2.5/4.3 \\
      12 
    &
    &
    & 107(83)/107(83)
    & 10$\times$110
    & 66.3
    & 10
    & 1.5/2.3
    & 2.5/4.3 \\
      13  
    &
    &
    & 76(45)/76(45)
    & 5$\times$41
    & 28.6
    & 10
    & 1.5/2.3
    & 2.5/4.3 \\
      14
    & $e^+d$
    & NC
    & 116(92)/116(92)
    & 18$\times$110
    & 89.0
    & 10
    & 1.5/2.3
    & 2.5/4.3 \\
      15
    &
    &
    & 107(83)/107(83)
    & 10$\times$110
    & 66.3
    & 10
    & 1.5/2.3
    & 2.5/4.3 \\
      16 
    &
    &
    & 76(45)/76(45)
    & 5$\times$41
    & 28.6
    & 10
    & 1.5/2.3
    & 2.5/4.3 \\
    \midrule
      17  
    & $e^-{^4}$He
    & NC
    & 116(92)/116(92)
    & 18$\times$110
    & 89.0
    & 10
    & 1.5/2.3
    & 2.5/4.3 \\
      18  
    &
    &
    & 107(83)/107(83)
    & 10$\times$110
    & 66.3
    & 10
    & 1.5/2.3
    & 2.5/4.3 \\
      19  
    &
    &
    & 76(45)/76(45)
    & 5$\times$41
    & 28.6
    & 10
    & 1.5/2.3
    & 2.5/4.3 \\
      20  
    & $e^-{^{12}}$C
    & NC
    & 116(92)/116(92)
    & 18$\times$110
    & 89.0
    & 10
    & 1.5/2.3
    & 2.5/4.3 \\
      21  
    &
    &
    & 107(83)/107(83)
    & 10$\times$110
    & 66.3
    & 10
    & 1.5/2.3
    & 2.5/4.3 \\
      22  
    &
    &
    & 76(45)/76(45)
    & 5$\times$41
    & 28.6
    & 10
    & 1.5/2.3
    & 2.5/4.3 \\
      23  
    & $e^-{^{40}}$Ca
    & NC
    & 116(92)/116(92)
    & 18$\times$110
    & 89.0
    & 10
    & 1.5/2.3
    & 2.5/4.3 \\
      24  
    &
    &
    & 107(83)/107(83)
    & 10$\times$110
    & 66.3
    & 10
    & 1.5/2.3
    & 2.5/4.3 \\
      25  
    &
    &
    & 76(45)/76(45)
    & 5$\times$41
    & 28.6
    & 10
    & 1.5/2.3
    & 2.5/4.3 \\
      26  
    & $e^-{^{64}}$Cu
    & NC
    & 116(92)/116(92)
    & 18$\times$110
    & 89.0
    & 10
    & 1.5/2.3
    & 2.5/4.3 \\
      27  
    &
    &
    & 107(83)/107(83)
    & 10$\times$110
    & 66.3
    & 10
    & 1.5/2.3
    & 2.5/4.3 \\
      28  
    &
    &
    & 76(45)/76(45)
    & 5$\times$41
    & 28.6
    & 10
    & 1.5/2.3
    & 2.5/4.3 \\
      29  
    & $e^-{^{197}}$Au
    & NC
    & 116(92)/116(92)
    & 18$\times$110
    & 89.0
    & 10
    & 1.5/2.3
    & 2.5/4.3 \\
      30  
    &
    &
    & 107(83)/107(83)
    & 10$\times$110
    & 66.3
    & 10
    & 1.5/2.3
    & 2.5/4.3 \\
      31  
    &
    &
    & 76(45)/76(45)
    & 5$\times$41
    & 28.6
    & 10
    & 1.5/2.3
    & 2.5/4.3 \\    
    \bottomrule
  \end{tabularx}

  \vspace{0.3cm}
  \caption{The EIC pseudodata sets considered in this work. For each of them we
    indicate the corresponding DIS process, the number of data points
    $N_\text{dat}$ in the optimistic/pessimistic scenarios before (after)
    kinematic cuts, the energy of the lepton and of the proton or ion beams
    $E_\ell$ and $E_p$, the center-of-mass energy $\sqrt{s}$, the integrated
    luminosity $\mathcal{L}$, and the relative uncorrelated and correlated
    systematic uncertainties (in percentage) $\sigma_u$ and $\sigma_c$
    in the optimistic/pessimistic scenarios.}
  \label{tab:EIC_pseudodata}
\end{table}

\begin{figure}[!h]
 \centering
 \includegraphics[width=\textwidth,clip=true,trim=0 0cm 0 0cm]{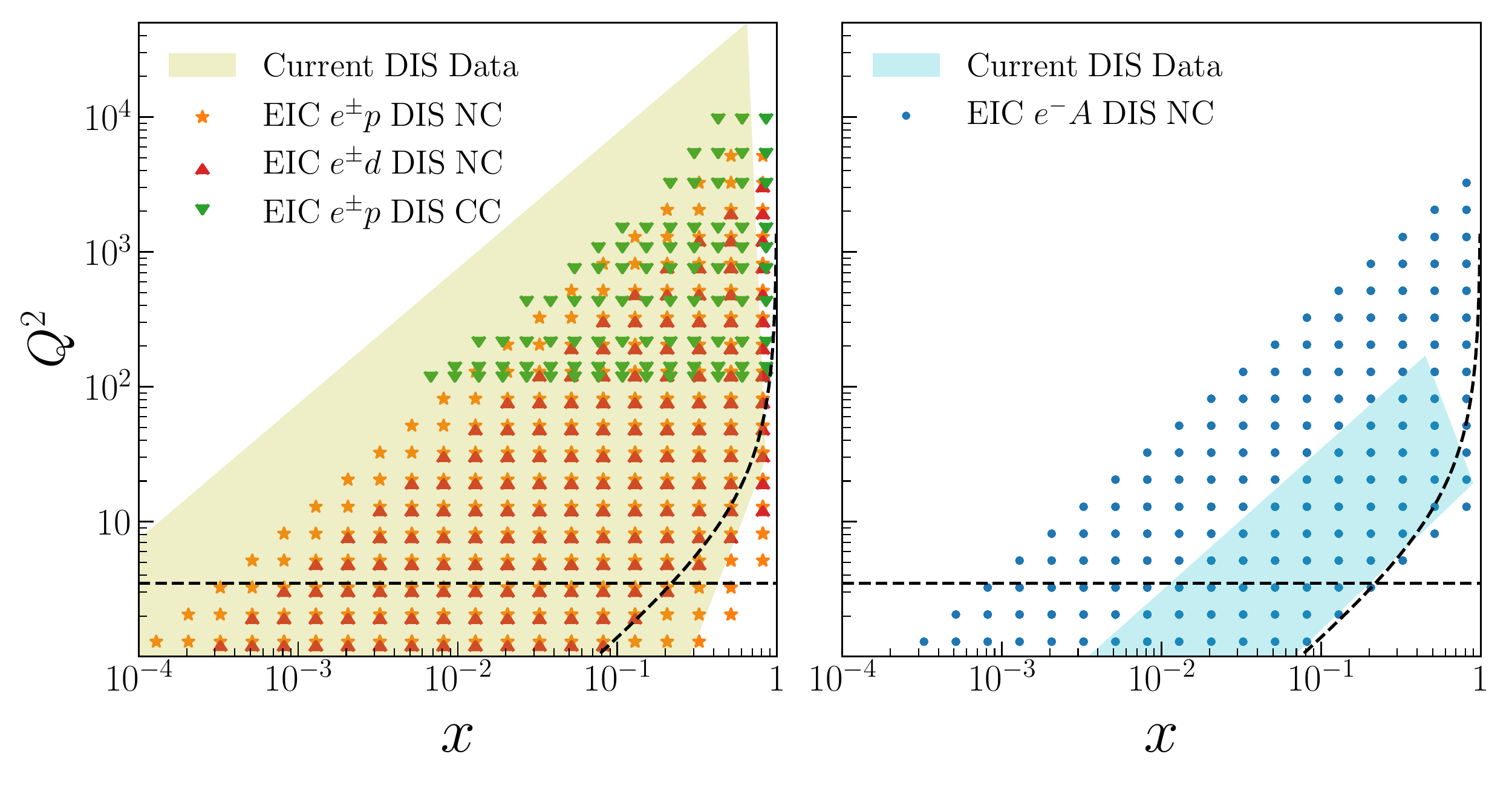}
 \vspace{-0.8cm}
 \caption{The expected kinematic coverage in the $(x,Q^2)$ plane of the EIC
   pseudodata for lepton-proton or lepton-deuteron (left) and lepton-nucleus
   (right panel) collisions, see Table~\ref{tab:EIC_pseudodata}. Shaded areas
   indicate the approximate kinematic coverage of the available inclusive DIS
   measurements. The dashed lines denote the kinematic cuts used in the PDF
   fits, $Q^2 \ge 3.5$ GeV$^2$ and $W^2 \ge 12.5$ GeV$^2$.}
 \label{fig:EIC_kinematics}
\end{figure}

The kinematic coverage of the EIC pseudodata in the ($x,Q^2$) plane is displayed
in Fig.~\ref{fig:EIC_kinematics} for the optimistic scenario. Pseudodata for
lepton-proton and lepton-deuteron are separated from pseudodata for electron-ion
collisions via different panels. The approximate coverage of currently available
inclusive DIS measurements is shown as a shaded area. Dashed lines correspond
to the kinematic cuts used in the PDF fits described below. From
Fig.~\ref{fig:EIC_kinematics}, we already can appreciate the relevance of the EIC
for the determination of nuclear PDFs. In this case, the EIC measurements extend
the kinematic reach of DIS by more than one order of magnitude in both $x$ and
$Q^2$. In the case of proton PDFs, instead, the EIC measurements mostly
overlap with those already available, in particular from HERA, except for a
slightly larger extension at very high $x$ and $Q^2$.

\myparagraph{Fitting procedure}
We include the pseudodata in the series of fits summarized in
Table~\ref{tab:fits}. All these fits use the NNPDF methodology. Because nuclear
PDFs are correlated with proton PDFs (the former should reduce to the latter in
the limit $A\to 1$, where $A$ is the nucleon number), and because the EIC
measurements of Table~\ref{tab:EIC_pseudodata} will affect both, we determine them
sequentially.
\begin{table}
  \footnotesize
  \centering
  \renewcommand{\arraystretch}{1.3}
\begin{tabularx}{\linewidth}{lX}
  \toprule
  Fit ID & Description \\
  \midrule
    NNPDF3.1+EIC (optimistic)
  & Same as the \texttt{base} fit of~\cite{Faura:2020oom} augmented with
    the $e^\pm p$ (CC and NC) and $e^\pm d$ (NC) EIC pseudodata sets
    for the optimistic scenario. \\
    NNPDF3.1+EIC (pessimistic)
  & Same as NNPDF3.1+EIC (optimistic), but with EIC pseudodata sets for the
    pessimistic scenario. \\
  \midrule
    NNPDF3.1\_pch+EIC (optimistic)
  & Same as the proton baseline fit of~\cite{AbdulKhalek:2020yuc} augmented with
    the $e^\pm p$ (CC and NC) pseudodata sets for the optimistic scenario. \\
    NNPDF3.1\_pch+EIC (pessimistic)
  & Same as NNPDF3.1\_pch+EIC (optimistic), but with EIC pseudodata sets for
    the pessimistic scenario. \\
    nNNPDF2.0+EIC (optimistic)
  & Same as the nuclear fit of~\cite{AbdulKhalek:2020yuc} augmented with the $e^-A$ (NC) pseudodata sets (with $A=^2$d,$^4$He,$^{12}$C,
    $^{40}$Ca, $^{64}$Cu and $^{197}$Au for the optimistic scenario. \\
    nNNPDF2.0+EIC (pessimistic)
  & Same as nNNPDF2.0+EIC (optimistic), but with EIC pseudodata sets for
    the pessimistic scenario. \\
  \bottomrule
\end{tabularx}

  \caption{A summary of the fits performed in this study, see text for details.}
  \label{tab:fits}
\end{table}

First, we focus on the proton PDFs, and perform the NNPDF3.1+EIC optimistic and
pessimistic fits. These are a rerun of the \texttt{base} fit of
Ref.~\cite{Faura:2020oom}, which is now augmented with the $e^\pm p$ (CC and NC)
and $e^\pm d$ (NC) EIC pseudodata sets for the optimistic and pessimistic
scenarios. As in Ref.~\cite{Ball:2017nwa,Faura:2020oom}, they are all
made of $N_\text{rep}=100$ Monte Carlo replicas. After kinematic cuts, the fits
include a total of 5264 (5172) data points in the optimistic (pessimistic)
scenario, out of which 1286 (1194) are EIC pseudodata and 3978 are real data
(see Ref.~\cite{Faura:2020oom} for details). Kinematic cuts are the same as
in Ref.~\cite{Ball:2017nwa,Faura:2020oom}, specifically
$Q^2>3.5$~GeV$^2$ and $W^2>12.5$~GeV$^2$. These cuts, which serve the purpose
of removing a kinematic region in which potentially large higher-twist and
nuclear effects may spoil the accuracy of the PDF analysis, are more
restrictive than those used to generate the pseudodata. This fact is however
not contradictory, and reproduces what customarily happens with real data, when
different kinematic cuts are used in the experimental analysis and in a fit.
These fits are accurate to next-to-next-to-leading order (NNLO) in perturbative
QCD, they utilize the FONLL scheme~\cite{Forte:2010ta,Ball:2015tna,Ball:2015dpa}
to treat heavy quarks, and they include a parameterisation of the charm PDF on
the same footing as the lighter quark PDFs. In comparison to the original
NNPDF3.1 fits~\cite{Ball:2017nwa}, a bug affecting the computation of
theoretical predictions for charged-current DIS cross sections has been
corrected, the positivity of the $F_2^c$ structure function has been enforced,
and NNLO massive corrections~\cite{Berger:2016inr,Gao:2017kkx} have been
included in the computation of neutrino-DIS structure functions.

We then focus on nuclear PDFs, and perform the NNPDF3.1\_pch+EIC and
\texttt{nNNPDF2.0}+EIC optimistic and pessimistic fits. These are a rerun of the
proton and nuclear baseline determinations of Ref.~\mycite{AbdulKhalek:2020yuc},
augmented respectively with the $e^\pm p$ (CC and NC) and the $e^-A$ (NC),
$A=d$, $^4$He, $^{12}$C, $^{40}$Ca, $^{64}$Cu, and $^{197}$Au,
pseudodata sets for the optimistic and pessimistic scenarios.
As in Ref.~\mycite{AbdulKhalek:2020yuc}, the proton (nuclear) fits are made of
$N_\text{rep}=100$ ($N_\text{rep}=250$) Monte Carlo replicas. After kinematic 
cuts, the NNPDF3.1\_pch+EIC fits include a total of 4147 (4055) data points
in the optimistic (pessimistic) scenario, out of which 846 (754) are EIC
pseudodata and 3301 are real data (see Ref.~\mycite{AbdulKhalek:2020yuc} for
details). The nuclear fits include a total of 3007 data points, out of which
1540 are EIC pseudodata and 1467 are real data. Kinematic cuts are the same as
above, and are in turn equivalent to these used in
Refs.~\cite{Ball:2017nwa,AbdulKhalek:2020yuc}. These fits are accurate to
next-to-leading order (NLO) in perturbative QCD, and assume that charm is
generated perturbatively, consistent with Ref.~\mycite{AbdulKhalek:2020yuc}.

Although the proton and nuclear PDF fits are performed independently,
they remain as consistent as possible. Most importantly, the unique
feature of an EIC to measure DIS cross sections with a comparable accuracy and
precision for a wide range of nuclei and for the proton is
key to inform the fit of nuclear PDFs as much as possible. Not only
do the measurements on nuclear targets enter the fit directly, but also the
measurements on a proton target are first used to update the necessary
baseline proton PDF determination. This feature distinguishes our work from
previous similar studies~\cite{Aschenauer:2017oxs,AbdulKhalek:2019mzd}, where
only the effect of measurements on nuclear targets were taken into account in
the determination of nuclear PDFs. A simultaneous determination of proton and
nuclear PDFs might eventually become advisable at an EIC, should the
measurements be sufficiently precise to make an independent determination
less reliable.

We also note that the pseudodata sets for a deuteron target are alternatively
included in the fit of proton PDFs or in the fit of nuclear PDFs. To avoid
double counting, they are not included in the fit of proton PDFs used
as baseline for the fit of nuclear PDFs. This choice follows the common
practice to include fixed-target DIS data on deuteron targets in fits of proton
PDFs, as done, \textit{e.g.}, in NNPDF3.1 and in the variant fit used here to
generate the pseudodata. The reason being that they are essential to achieve a
good quark flavour separation. The EIC pseudodata sets for a deuteron target
are then treated, in the proton PDF fits performed here, similarly to the
fixed-target DIS data already included in NNPDF3.1. Specifically we assume
that nuclear corrections are negligible, and therefore we do not include them.
This assumption could be overcome by means of a simultaneous fit of proton and
nuclear PDFs, or by means of the iterative procedure proposed
in Ref.~\cite{Ball:2020xqw}, whereby proton and deuteron PDFs are determined by
subsequently including the uncertainties of each in the other. Any of these
approaches goes beyond the scope of this work, as they will have little
applicability in the context of pseudodata.

\myparagraph{Results} We now turn to discuss the results of the fits collected
in Table~\ref{tab:EIC_pseudodata}. As expected, the goodness of each fit 
measured by the $\chi^2$ per number of data points is comparable to that of the
fits used to generate the pseudodata. The description of each data set remains
unaltered within statistical fluctuations, and the $\chi^2$ per
number of data points for each of the new EIC pseudodata sets is of order one,
as it should by construction. In the following we therefore exclusively discuss
how the EIC pseudodata affect PDF uncertainties.

\begin{figure}[!h]
  \centering
  \includegraphics[width=\textwidth,clip=true,trim=1.5cm 0cm 1.5cm 2cm]{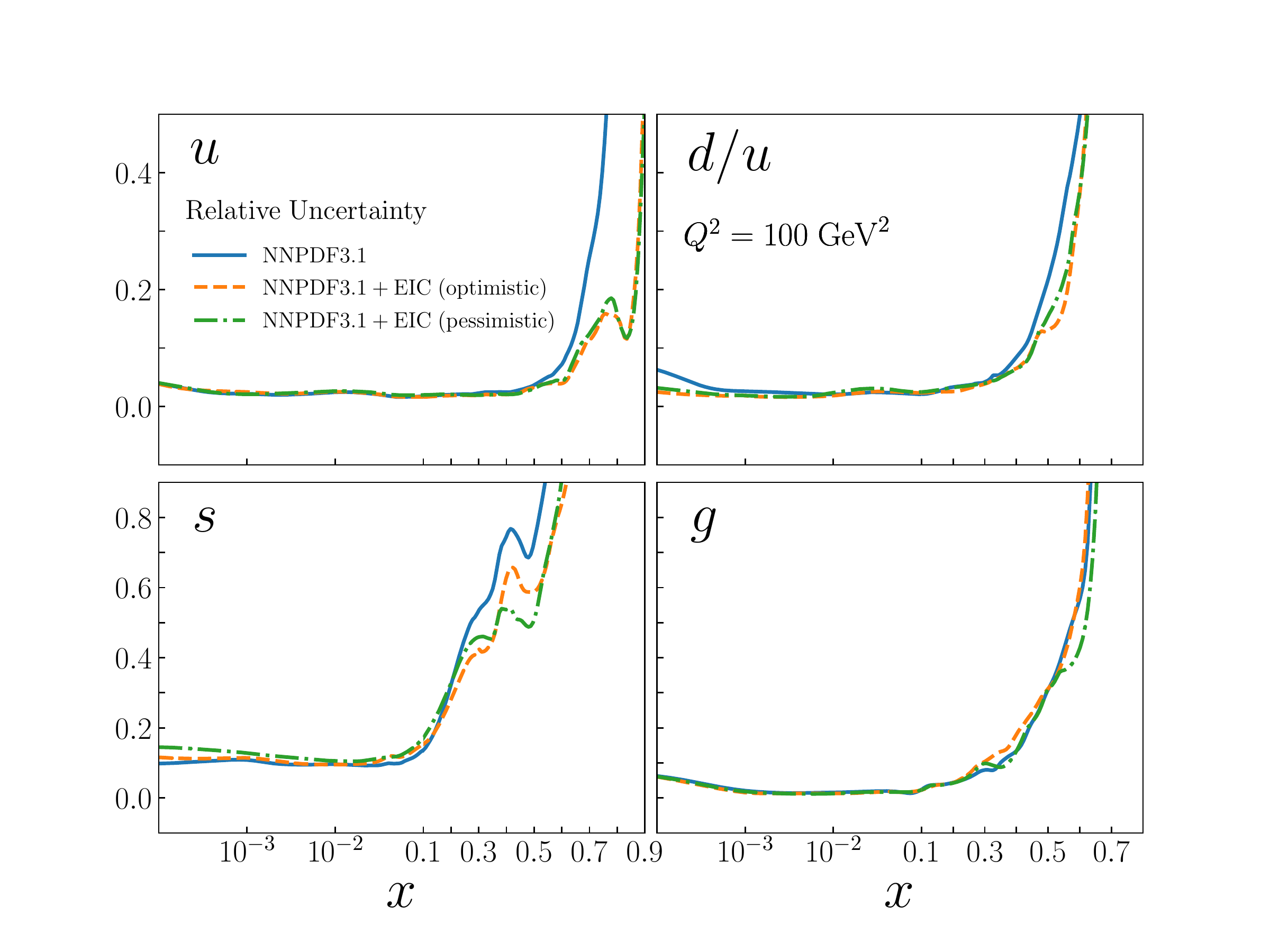}
  \vspace{-0.8cm}
  \caption{The relative uncertainty of the proton PDFs determined in the
  NNPDF3.1 fit variant used to generate the pseudodata, and in the
  NNPDF3.1+EIC fits, in the optimistic and pessimistic scenarios. Uncertainties
  correspond to one standard deviation and are computed as a function of $x$
  at $Q^2=100$~GeV$^2$. Only the subset of flavours (or flavour combinations) that
  are the most affected by the EIC pseudodata are shown, namely $u$,
  $d/u$, $s$ and $g$. Note the use of a log/linear scale on the $x$ axis.}
  \label{fig:ppdfs}
\end{figure}

In Fig.~\ref{fig:ppdfs} we show the relative uncertainty of the proton PDFs
in the NNPDF3.1 fit variant used to generate the pseudodata, and in the
NNPDF3.1+EIC fits, both for the optimistic and pessimistic scenarios. In each
case, uncertainties correspond to one standard deviation, and are computed
as a function of $x$ at $Q^2=100$~GeV$^2$. Only the subset of flavours (or
flavour combinations) that are the most affected by the EIC
pseudodata are shown: $u$, $d/u$, $s$ and $g$.

Fig.~\ref{fig:ppdfs} allows us to make two conclusions. First, the impact of the
EIC pseudodata is localised in the large-$x$ region, as expected from their
kinematic reach (see Fig.~\ref{fig:EIC_kinematics}). This impact is significant in
the case of the $u$ PDF, for which PDF uncertainties could be reduced by up to
a factor of two for $x\gtrsim 0.7$. The impact is otherwise moderate for the
$d/u$ PDF ratio (for which it amounts to an uncertainty reduction of about
one third for $0.5\lesssim x \lesssim 0.6$) and for the $s$ PDF (for which 
it amounts to an uncertainty reduction of about one fourth for
$0.3\lesssim x \lesssim 0.6$). The relative uncertainty of the gluon PDF, and
of other PDFs not shown in Fig.~\ref{fig:ppdfs}, remains unaffected.
These features rely on the unique ability of the EIC to perform precise DIS
measurements at large $x$ and large $Q^2$: their theoretical interpretation
remains particularly clean, as any non-perturbative large-$x$ contamination
due, \textit{e.g.}, to higher-twist effects, is suppressed. This possibility
distinguishes the EIC from HERA, which had a similar reach at high $Q^2$ but a
more limited access at large-$x$, and from fixed-target experiments
(including the recent JLab-12 upgrade~\cite{Dudek:2012vr}), which can access the
high-$x$ region only at small $Q^2$. Secondly, the impact of the EIC pseudodata
does not seem to depend on the scenario considered: the reduction of PDF
uncertainties remains comparable irrespective of whether optimistic or
pessimistic pseudodata projections are included in the fits. Because the two
scenarios only differ in systematic uncertainties, we conclude that it may be
sufficient to control these to the level of precision forecast
in the pessimistic scenario.

\begin{figure}[!h]
  \centering
  \includegraphics[width=\textwidth,clip=true,trim=1.5cm 2.5cm 1.5cm 3.5cm]{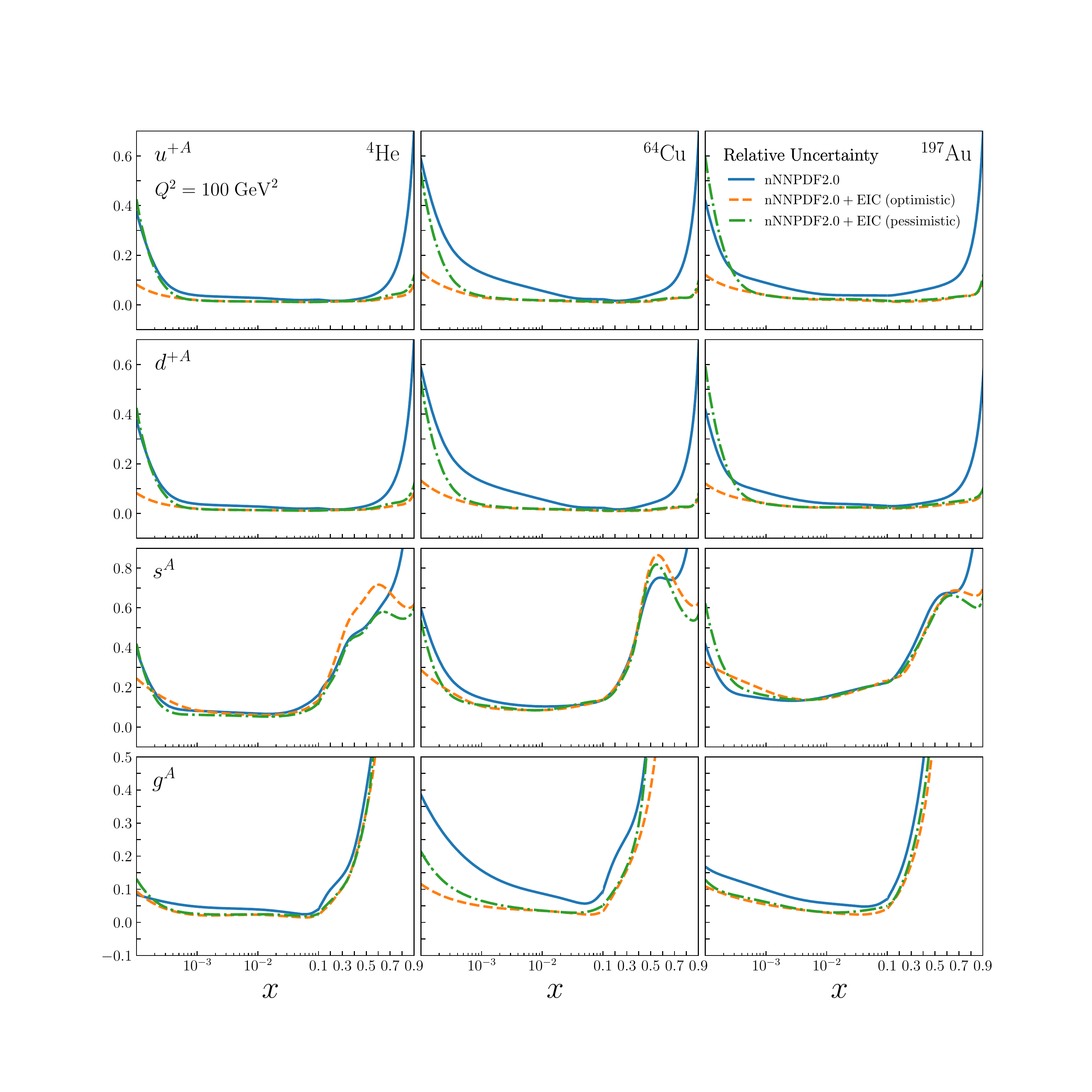}
  \vspace{-0.8cm}
  \caption{The relative uncertainty of the nuclear PDFs determined in the
  \texttt{nNNPDF2.0} fit used to generate the pseudodata, and in the \texttt{nNNPDF2.0}+EIC fits,
  in the optimistic and pessimistic scenarios. Uncertainties correspond to one
  standard deviation, and are computed as a function of $x$ at
  $Q^2$=100~GeV$^2$. Results are displayed for the ions with the lowest and
  highest atomic mass, $^4$He (left) and $^{197}$Au (right), and for an
  intermediate atomic mass ion, $^{64}$Cu (middle column), and only for the
  PDF flavours that are the most affected by the EIC pseudodata: $u$, $\bar{d}$,
  $s$ and $g$. Note the use of a log/linear scale on the $x$ axis.}
  \label{fig:npdfs}
\end{figure}

A similar behaviour is observed for the NNPDF3.1\_pch fits, which are therefore
not displayed. In Figs.~\ref{fig:npdfs} we show the relative uncertainty of the
nuclear PDFs in the \texttt{nNNPDF2.0} fit used to generate the pseudodata, and in the
\texttt{nNNPDF2.0}+EIC fits, both in the pessimistic and in the optimistic scenarios.
Uncertainties correspond to one standard deviation, and
are computed as a function of $x$ at $Q^2$=100~GeV$^2$. Results
are displayed for the ions with the lowest and highest atomic mass, $^4$He and
$^{197}$Au, and for an intermediate atomic mass ion, $^{64}$Cu,
and only for the PDF flavours that are the most affected by
the EIC pseudodata: $u$, $\bar{d}$, $s$ and $g$.

From Fig.~\ref{fig:npdfs} we observe a reduction of nuclear
PDF uncertainties, due to EIC pseudodata, that varies with the nucleus, the
$x$ region considered, and the PDF. Overall, the heavier the nucleus,
the largest the reduction of PDF uncertainties. This is a consequence of the
fact that nuclear PDFs are customarily parametrized as continuous functions
of the nucleon number $A$: nuclear PDFs for $^4$He, which differ from the proton
PDF boundary by a small correction, are better constrained than nuclear PDFs
for $^{197}$Au because proton data are more abundant than data for nuclei.
In this respect, the EIC will allow one to perform a comparatively accurate
scan of the kinematic space for each nucleus individually, and, as shown in
Fig.~\ref{fig:npdfs}, to determine the PDFs of all ions with a similar
precision. The reduction of PDF uncertainties is localised in the small-$x$
region, where little or no data are currently available
(see Fig.~\ref{fig:EIC_kinematics}), and
in the large-$x$ region, where nuclear PDF benefit from the increased precision
of the baseline proton PDFs. In the case of the gluon PDF, the reduction of
uncertainties is seen for the whole range in $x$. This is a consequence of the
extended data coverage in $Q^2$, which allows one to constrain the gluon PDF
even further via perturbative evolution. As observed in the case of
proton PDFs, the fits obtained upon inclusion of the EIC pseudodata do not
significantly differ whether the optimistic or the pessimistic scenarios are
considered, except for very small values of $x$. In this case the optimistic
scenario leads to a more marked reduction of PDF uncertainties, especially for
the total PDF combinations $u^+$ and $d^+$. This feature is mainly driven by the
smaller systematic uncertainties that affect the NC pseudodata in the optimistic
scenario (about 3\%) with respect to the pessimistic one (about 5\%), see
Table~\ref{tab:EIC_pseudodata}. That is aligned with the fact that the statistical
uncertainties are comparable between the two scenarios. 

\myparagraph{Implications for neutrino astrophysics} The reduction of PDF
uncertainties due to EIC pseudodata, in particular for
nuclear PDFs, may have important phenomenological implications. Not only at the
intensity frontier, \textit{e.g.} to characterise gluon saturation at small $x$,
but also at the energy frontier, \textit{e.g.}, for searches of new physics that
require a precise knowledge of PDFs at high $x$, and at the cosmic frontier,
\textit{e.g.}, in the detection of highly energetic neutrinos from astrophysical
sources. We conclude this section by focusing on this last aspect. Specifically
it was shown in Ref.~\cite{Garcia:2020jwr} that the dominant source of
uncertainty in the theoretical predictions for the cross section of
neutrino-matter interactions is represented by nuclear effects.
The corresponding NC and CC inclusive DIS cross sections may differ
significantly depending on whether they are computed for neutrino-nucleon or
neutrino-nucleus interactions. The uncertainty is larger in the latter case,
because nuclear PDFs are not as precise as proton PDFs, and is such that it
encompasses the difference in central values. We revisit this statement in
light of the precise nNNPDF3.0+EIC fits.
\begin{figure}[!h]
  \centering
  \includegraphics[width=\textwidth]{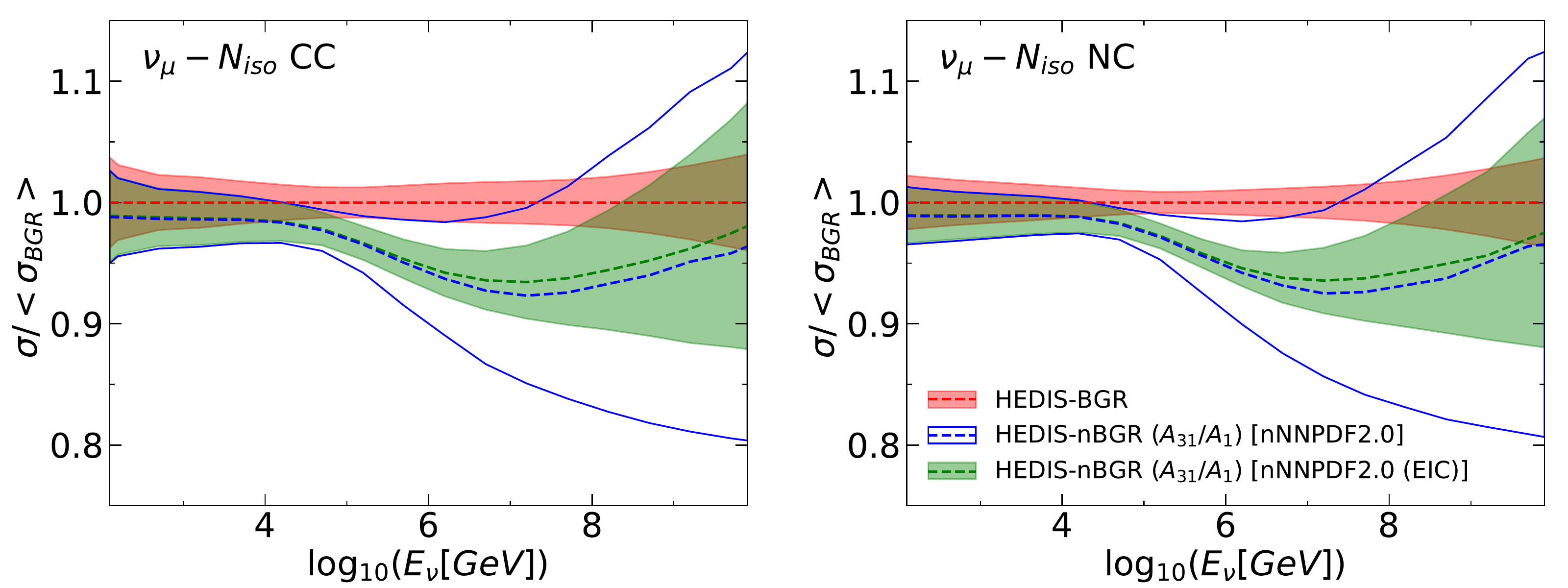}
  \vspace{-0.8cm}
  \caption{The CC (left) and NC (right) neutrino-nucleus DIS cross
  sections, with their one-sigma uncertainties, as a function of the neutrino
  energy $E_{\nu}$. Predictions correspond to the HEDIS-BGR
  computation~\cite{Garcia:2020jwr} with the proton PDF of~\cite{Gauld:2016kpd},
  and with the \texttt{nNNPDF2.0} and \texttt{nNNPDF2.0}+EIC nuclear PDFs. They are all
  normalised to the central value of the proton results. See text for details.}
  \label{fig:UHExsec}
\end{figure}
\begin{figure}[!h]
  \centering
  \includegraphics[width=\textwidth]{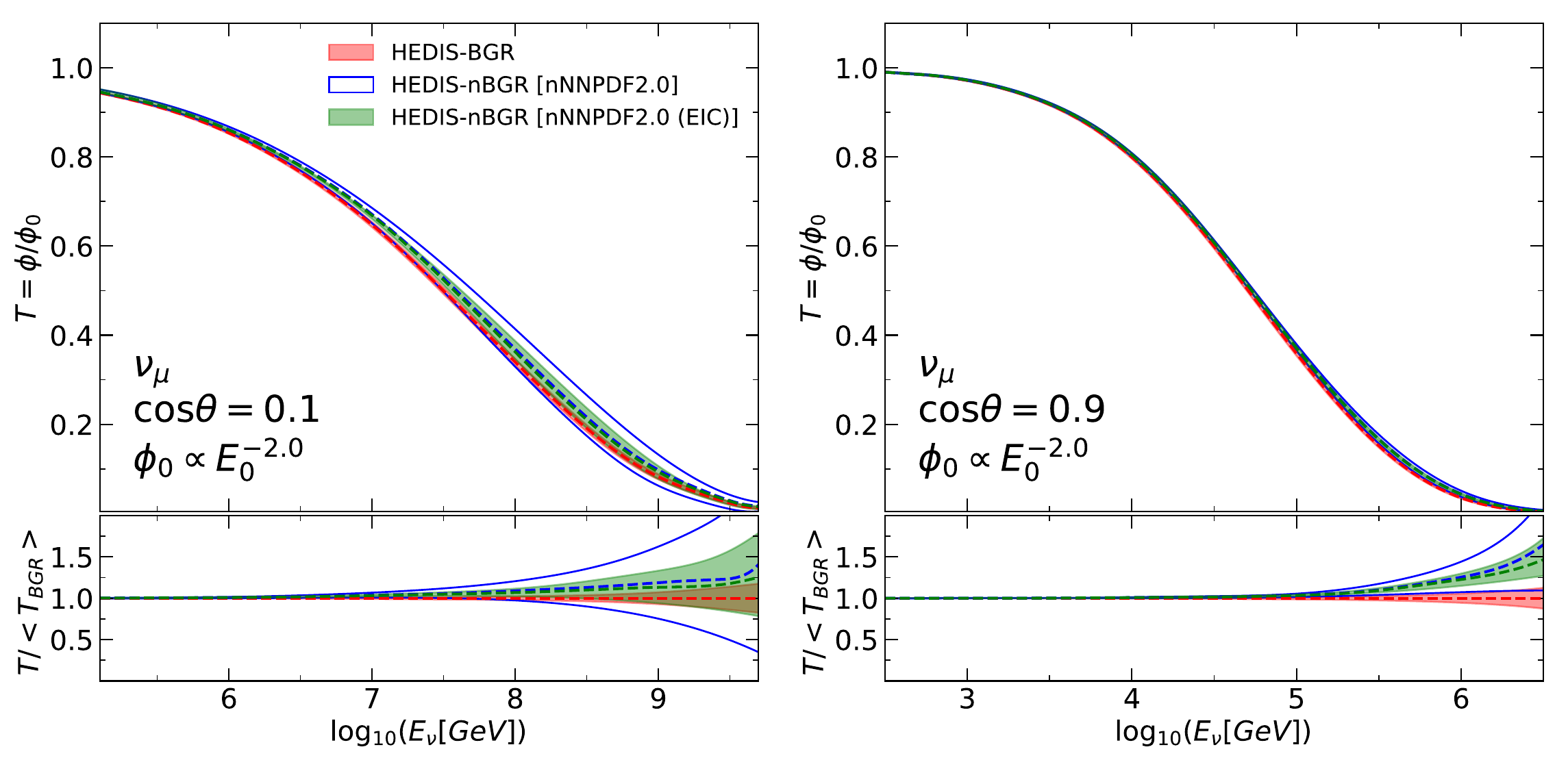}
  \vspace{-0.8cm}
  \caption{The transmission coefficient $T$ for muonic neutrinos as a function
  of the neutrino energy $E_\nu$ and for two values of the nadir angle $\theta$.
  Predictions correspond to the computation of~\cite{Garcia:2020jwr} with the
  proton PDF of~\cite{Gauld:2016kpd}, and with the \texttt{nNNPDF2.0} and \texttt{nNNPDF2.0}+EIC
  nuclear PDFs. They are all normalised to the central value of the proton
  results. See text for details.}
  \label{fig:attenuation}
\end{figure}

In Fig.~\ref{fig:UHExsec} we show the CC (left) and NC (right)
neutrino-nucleus inclusive DIS cross sections, with their one-sigma PDF
uncertainties, as a function of the neutrino energy $E_\nu$. Moreover, in
Fig.~\ref{fig:attenuation} we show the transmission coefficient $T$ for muonic
neutrinos, defined as the ratio between the incoming neutrino flux $\Phi_0$ and
the flux arriving at the detector volume $\Phi$ (see Eq.~(3.1) and the ensuing
discussion in Ref.~\cite{Garcia:2020jwr} for details); $T$ is displayed for two
values of the nadir angle $\theta$ as a function of the neutrino energy $E_\nu$.
In both cases, we compare predictions obtained with the calculation presented
in Refs.~\cite{Bertone:2018dse,Garcia:2020jwr} and implemented in
\textsc{hedis}~\cite{Brown:1971qr}. For a proton target the prediction is made
with the proton PDF set determined in Ref.~\cite{Gauld:2016kpd},
a variant of the NNPDF3.1 PDF set in which small-$x$ resummation
effects~\cite{Ball:2017otu} and additional constraints from $D$-meson
production measurements in proton-proton collisions at 5,7 and
13~TeV~\cite{Aaij:2013mga,Aaij:2015bpa,Aaij:2016jht} have been included.
This prediction is labeled HEDIS-BGR in
Figs.~\ref{fig:UHExsec}-\ref{fig:attenuation}. For a nuclear
target ($A=31$ is adopted as in Ref.~\cite{Garcia:2020jwr}), the prediction is
made alternatively with the \texttt{nNNPDF2.0} and the \texttt{nNNPDF2.0}+EIC (optimistic) PDFs.
The corresponding predictions are labeled HEDIS-nBGR [\texttt{nNNPDF2.0}] and HEDIS-nBGR
[\texttt{nNNPDF2.0} (EIC)] in Figs.~\ref{fig:UHExsec}-\ref{fig:attenuation}. Predictions
are all normalised to the central value of the proton result. In comparison to
\texttt{nNNPDF2.0}, the effect of the EIC pseudodata is seen to reduce the uncertainty
of the prediction for a nuclear target by roughly a factor of two for
$E_\nu\simeq 10^6$~GeV. The reduced uncertainty no longer encompasses the
difference between predictions obtained on a proton or on a nuclear target,
except in the case of an attenuation rate computed with a large nadir angle.
Furthermore, this reduction extends to much larger neutrino energy
($E_\nu \gtrsim 10^7$), beyond the EIC-sensitive $x$-region
of the PDFs. We believe this to be partly due to DGLAP evolution and sum rules
that smoothen the low-$x$ PDF behaviour, but also potentially a consequence of
the factorisation approximation used to account for nuclear corrections in the
ultra high-energy cross-sections highlighted by Eqs.~(5.2, 5.3) in
Ref.~\cite{Garcia:2020jwr}.

  \markboth{}{}
\chapter*{Summary}
\addcontentsline{toc}{chapter}{Summary}
\label{chap:Summary}

For this thesis, I lead the development of a new statistical framework called \texttt{nNNPDF}, inspired by the \texttt{NNPDF} methodology, and aimed at inferring nuclear parton distribution functions (nPDFs) from experimental data. Furthermore, I contributed to two different statistical frameworks aimed at inferring proton PDFs, called \texttt{NNPDF3.1}, and pion fragmentation functions (FFs), called \texttt{MAPFF1.0}. These three frameworks have the Monte Carlo method in common, which is adopted to propagate experimental uncertainties into the space of parameters. Moreover, they share artificial neural networks (NNs) as the basis for parameterisation. The latter is motivated by the fact that NNs can approximate any continuous function which complements the fact that the functional form of (n)PDFs and FFs is not determined by first-principles in QCD. In the following I outline the main findings of my research.

\subsection*{Proton PDFs} 
\myparagraph{\texttt{NNPDF3.1}}  In Chapter~\ref{chap:PDF}, I started by introducing the \texttt{NNPDF3.1} statistical framework, which was conceived by the \texttt{NNPDF} collaboration. 
I then presented my contribution to the first global proton PDF analysis that accounts for the missing higher order uncertainty (MHOU) associated to the fixed-order QCD calculations. We demonstrated that the MHOU can be included as a contribution to the covariance matrix which leads to a shift in the PDFs central values by an amount that is not negligible on the scale of the PDF uncertainty, moving the NLO result towards the NNLO one. Moreover, we found that PDF uncertainties increased moderately, because of the improvement of fit quality caused by the rebalancing of data sets according to their theoretical precision. 

Afterwards, I presented my contribution to the phenomenological investigation of inclusive jet production at the LHC exploiting NNLO QCD corrections. We first showed that NNLO corrections were crucial in order to ensure compatibility of the jet observables with the rest of the global data set. Second, we found a full consistency between the constraints imposed on PDFs, specifically the gluon, by single-inclusive jets and dijets. Finally, we concluded that the dijet observable has a more marked impact on the gluon central value and displays a better-behaved perturbative behaviour.

\subsection*{Nuclear PDFs} 
\myparagraph{\texttt{nNNPDF1.0}} In Chapter~\ref{chap:nNNPDF10}, I started by introducing the first release of the nuclear PDF framework, dubbed \texttt{nNNPDF1.0} that focuses on inferring nuclear PDFs based on lepton-nucleus neutral-current DIS scattering at both NLO and NNLO accuracy in QCD.
We found an excellent overall agreement with the fitted experimental data, with stable results with respect to the order of the perturbative QCD calculations.
While the quark distributions were reasonably well constrained for $x\gsim 10^{-2}$,
the nuclear gluon PDFs were affected by large uncertainties, in particular
for heavy nuclei.

From the methodological point of view, the main
improvement with respect to previous NNPDF fits has been
the implementation of {\tt TensorFlow} to perform 
stochastic gradient descent.
As opposed to other nPDF analyses, the
nNNPDF1.0 set was determined with the 
boundary condition imposed at the minimisation level so that
the baseline proton PDFs (NNPDF3.1) were reproduced 
both in terms of their central values and, more importantly, their uncertainties.
We concluded that this $A=1$ constraint results in
a significant reduction of the nPDF
uncertainties, especially for low-$A$ nuclei,
and therefore represents a vital ingredient
for any nPDF analysis. 

\myparagraph{\texttt{nNNPDF2.0}} In Chapter~\ref{chap:nNNPDF20}, I introduced the second release \texttt{nNNPDF2.0} that focused on complementing the lepton-nucleus neutral-current DIS data by charged-current ones and gauge boson production in
proton-lead collisions.
We demonstrated that a satisfactory description of all the fitted
data sets can be achieved, highlighting the reliability of the QCD
factorisation paradigm in the heavy nuclear sector.
%
%
Our results demonstrated significant nuclear effects
among the quark flavors in
nuclei, in particular a shadowing of the up and down
quark distributions in heavy nuclei such as lead.

Nuclear modifications were 
found also in the strangeness of 
heavier nuclei, which displayed a suppression with respect to the free proton
across a large region of $x$.
In addition, we showed that upon releasing the momentum
and valence sum rule constraints, 
the data prefer integral values that agree with QCD expectations
for all values of $A$.
The nNNPDF2.0 release was recently used in a proton global PDF fits to
estimate the theory uncertainties associated with neutrino scattering
data~\cite{Ball:2018twp,Ball:2020xqw}, and also in high-energy
astroparticle physics processes that involve hard scattering on
nuclei~\cite{Garcia:2020jwr,Bertone:2018dse}.

\myparagraph{\texttt{nNNPDF3.0}} In Chapter~\ref{chap:nNNPDF30}, I presented the results achieved to date as part of the upcoming release \texttt{nNNPDF3.0} where on top of the DIS and gauge boson production considered in \texttt{nNNPDF2.0} at NLO, I investigated the inclusion of dijet and Z-boson production up to NNLO accuracy in QCD. 
We demonstrated first that the inclusion of the absolute pp dijet spectra from CMS at $5$ TeV is not compatible with the \texttt{NNPDF3.1} global data set at NLO and get worsened at NNLO. We therefore considered an updated proton baseline w.r.t \texttt{nNNPDF2.0} that includes the dijet data set investigated only in Chapter~\ref{chap:PDF}.
We found that this new baseline provides a slightly improved prediction of the pPb/pp spectra ratio from CMS at $5$ TeV using \texttt{nNNPDF2.0}.

We then performed a fit with this data set on top of the \texttt{nNNPDF2.0} global data set and found a satisfactory global description at NLO. Moreover, we showed the constraint this data set has on the different nuclear PDF flavours, in particular the gluon of lead that was mostly affected at intermediate and low-$x$.
Finally, we gauged the impact of NNLO QCD corrections on both nuclear PDFs and predictions based on the available calculations, namely for DIS and the CMS dijet and Z-boson productions at $5$ TeV. Upon the inclusion of NNLO QCD corrections, the overall lepton-nucleus DIS description improved as it was shown in Chapter~\ref{chap:nNNPDF10} and the pPb/pp CMS dijet data showed a slight enhancement in all nuclear PDF flavours across the entire $x$-range. As for the CMS Z-boson data, the associated NNLO corrections suffered from a reduced accuracy in the extreme rapidity bins leading to a deterioration in the description of the data set. However, in the central rapidity bins, the theory predictions moved closer to the experimental central values.

\subsection*{Fragmentation functions}
\myparagraph{\texttt{MAPFF1.0}} In Chapter~\ref{chap:FF}, I started by introducing the first release of the FF framework, dubbed \texttt{MAPFF1.0} that focused on inferring charged pion FFs based on a broad data set
that includes single-inclusive annihilation and semi-inclusive deep-inelastic scattering data at NLO accuracy in QCD. 
The description of the global and single data sets included in the fits are
fully satisfactory.
We explored
different sets of independent FF combinations and selected the one
that implied a minimal set of restrictions in order to avoid any bias
deriving from over-restrictive flavour assumptions. This resulted in a
set of 7 positive-definite independent combinations fitted to
data.

Given the dependence of the SIDIS predictions on PDFs, we discussed the effect on FFs of
including the PDF uncertainty as well as that of using different PDF
sets. As expected on the basis of the particular structure of the
SIDIS observable being fitted (multiplicities), we found that the
resulting FFs were almost insensitive to the treatment of PDFs both in
terms of uncertainties and central values. 
Finally, we justified our particular
choice for the cut on the minimum value of $Q$, $Q_{\rm cut}=2$~GeV,
for the SIDIS data included in the fit. We argued that this particular
value guarantees a reliable applicability of NLO accurate predictions
to the SIDIS data set. As a matter of fact, $Q_{\rm cut}=2$~GeV allows
us to obtain a satisfactory description of the global and the COMPASS
data set.

\subsection*{Impact of future colliders} 
In Chapter~\ref{chap:Impact}, I presented all my contributions to the impact studies carried out on proton and nuclear PDFs from future colliders.
We demonstrated that even in the most conservative scenarios, the High Luminosity LHC (HL-LHC) can reduce the proton PDF uncertainties by at least a factor between 2 and 3 as compared to the \texttt{PDF4LHC15} baseline. We then extended the study to the Large Hadron electron Collider (LHeC) where we found that the latter complements the HL-LHC and constrains further the low and intermediate $x$ regions of the proton PDFs. Afterwards, we studied the impact on the small-$x$ nuclear gluon PDF from the forward isolated photons production at the FoCal upgrade of the
ALICE detector. We found that the latter projections would significantly constrain the gluon, particularly in the small-$x$ region.
We then analysed our theory predictions for
neutral pion production in proton-lead collisions at RHIC and LHC
center-of-mass energies, which showed a potential constraint on nuclear PDFs conditional to reliable fragmentation functions.

Finally, we quantified the impact that unpolarised lepton-proton and lepton-nucleus inclusive DIS cross section measurements at the future Electron-Ion Collider (EIC) will have on the unpolarised proton (\texttt{NNPDF3.1}) and nuclear PDFs (\texttt{nNNPDF2.0}). We found that the EIC could reduce the uncertainty of the light quark PDFs of the proton at large x, and, more significantly, the quark and gluon PDF uncertainties for nuclei in a wide range of atomic mass A values both at small and large x. We also illustrated how theoretical predictions obtained with nuclear PDFs constrained by EIC data would improve the modelling of the interactions of ultra-high energy cosmic neutrinos with matter. In particular we demonstrated that nuclear PDF uncertainties may no longer encompass the difference between predictions obtained on a proton and on a nuclear target. 

\subsection*{Future directions} 
In Chapter~\ref{chap:nNNPDF30}, I emphasized on the potential of the jet, heavy-flavour mesons and photon production data to provide unprecedented constraints on the gluon lead PDF down to values of $x\sim 10^{-5}$. I also presented the first results that takes into account NNLO QCD corrections associated to proton-nucleus processes and their current technical challenges. Our current short-term aim is to sort out the latter and achieve the first NNLO global nPDF determination complemented by processes sensitive to the gluon and based on the Monte Carlo methodology within \texttt{nNNPDF3.0}. 

As far as both PDFs and nPDFs are concerned, one might aim to achieve their simultaneous
determination from a universal analysis, thus bypassing the need to 
include proton information by means of the proton boundary condition penalty. 
In the same spirit of the QCD analyses of proton PDFs and fragmentation functions
presented in Refs.~\cite{Sato:2019yez,Ethier:2017zbq}, such an integrated fit
of proton and nuclear PDFs would ensure the ultimate theoretical and
methodological consistency of the determination of the nuclear
modifications of the free-nucleon quark and gluon structure.

As for FFs, a possible continuation to \texttt{MAPFF1.0} is a determination of the
charged kaon, proton/antiproton and charged unidentified hadron
FFs. In all these cases, SIA and SIDIS data is available, which would
allow the extraction of these sets of distributions in a very
similar manner as done here for pions. In this respect, a particularly
interesting measurement is that of the $K^-/K^+$ and $p/\overline{p}$
ratios recently presented by COMPASS~\cite{Akhunzyanov:2018ysf,Alexeev:2020jia}. These observables are affected by very small
systematic uncertainties and are thus promising to constrain kaon and
proton FFs.

Moreover, an accurate determination of the
collinear FFs is instrumental for a reliable
determination of transverse-momentum-dependent (TMD)
distributions. 
Specifically, since the description of
$p_{\text{T}}$-dependent SIDIS multiplicities at low values of
$p_{\text{T}}$, where $p_{\text{T}}$ is the transverse momentum of the
outgoing hadron, can be expressed in terms of TMD PDFs and TMD FFs
that in turn depend on their collinear counterpart. 
The HERMES~\cite{Airapetian:2012ki} and COMPASS~\cite{Adolph:2013stb}
experiments have measured this observable, which can be
used to extract TMD distributions extending, for example, the
analysis of TMD PDFs carried out in Ref.~\cite{Bacchetta:2019sam} to
the TMD FFs relying on the collinear FFs determined in \texttt{MAPFF1.0}.

\textcolor{white}{R}

\end{mainmatter}

\renewcommand{\thechapter}{\Alph{chapter}}
\renewcommand{\thesection}{\thechapter.\arabic{section}}
\renewcommand{\theequation}{\thesection.\arabic{equation}}
\begin{appendices}
  \chapter{Methodology elements and cross-checks}
\section{Methodological studies in \texttt{nNNPDF1.0}} \label{s2:nNNPDF10_methodologyresults}

In this section, I present some further studies using the \texttt{nNNPDF1.0} framework that demonstrate
the robustness of our fitting methodology.
In particular, in the following we discuss the stability of our results with respect to variations
of the neural network architecture and the role
of the $A=1$ boundary condition in constraining the nPDF
uncertainties.
For all results presented in this section, we use $N_{\rm rep} = 200$ Monte
Carlo replicas.

\myparagraph{Stability with respect to the network architecture}
As explained in Sect.~\ref{s2:nNNPDF10_parameterisation},
the \texttt{nNNPDF1.0} fits are based on a single neural network
with the $3-25-3$ architecture represented in Fig.~\ref{fig:nNNPDF10_architecture}.
This architecture is characterized by $N_{\rm par}=178$ free parameters, without
counting the preprocessing exponents.
We have verified that this choice of network architecture is redundant given our input
dataset, namely that the \texttt{nNNPDF1.0} results are stable if neurons are either
added or removed from the hidden layer of the network.
To illustrate this redundancy, here we compare fit results
using the standard $3-25-3$ architecture with that
using twice as many neurons
in the hidden layer, $3-50-3$.
The latter configuration is characterized by $N_{\rm par}=353$ free parameters,
which is enlarged by a factor two compared to the baseline fits.

\begin{figure}[ht]
  \begin{center}
    \includegraphics[width=0.90\textwidth]{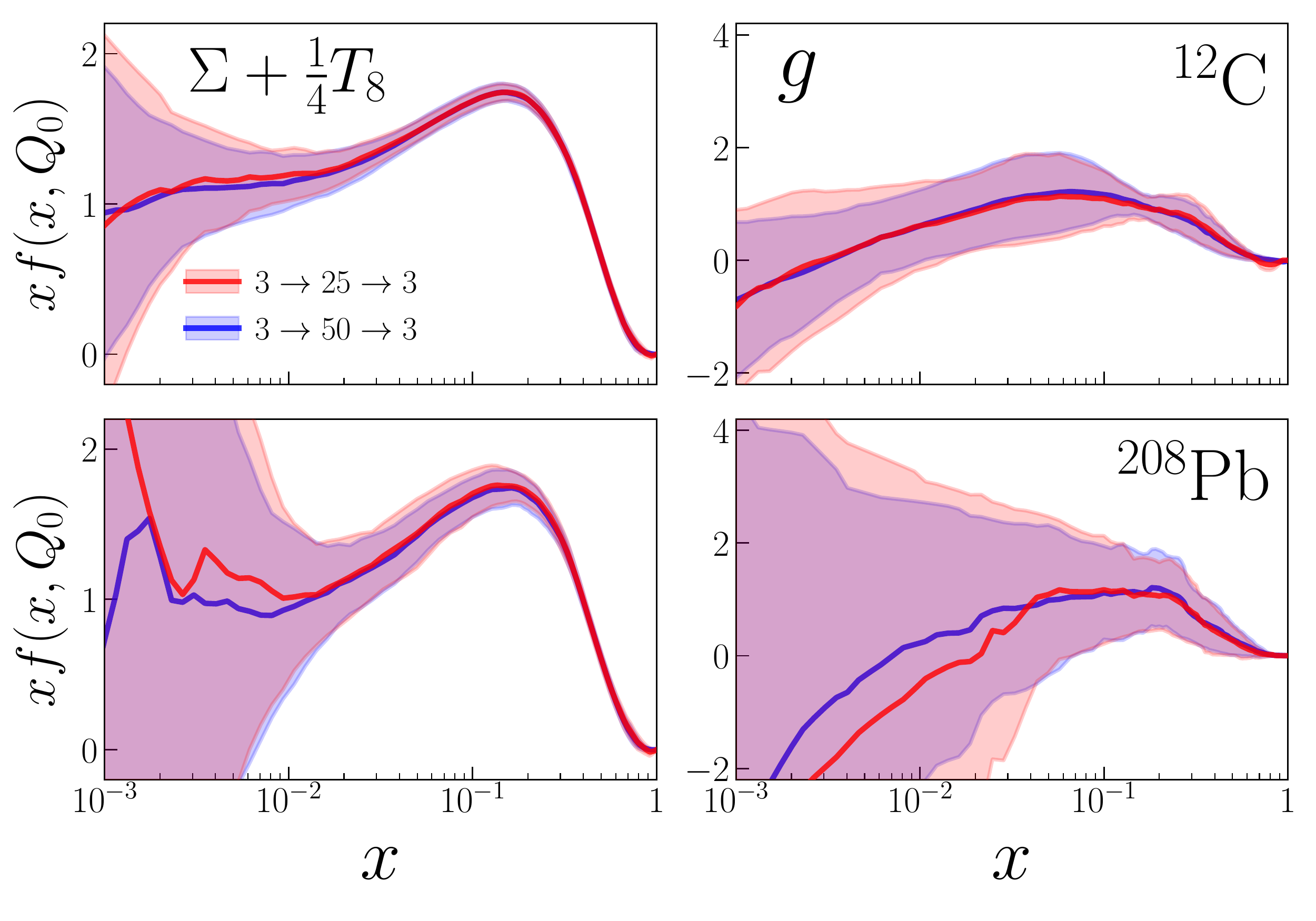}
   \end{center}
  \vspace{-0.6cm}
  \caption{\small Dependence of the \texttt{nNNPDF1.0} results
    at the input scale $Q_0=1$ GeV with respect
    to the choice of neural network architecture.
    We compare the baseline results 
    obtained with a $3-25-3$ architecture (solid red line and shaded band),
    with the corresponding ones using a $3-50-3$ architecture (solid blue line
    and shaded band) for $^{12}$C (top panels) and $^{208}$Pb (bottom panels)
    nuclei.
  }
  \label{fig:architectureVariation}
  \end{figure}

In Fig.~\ref{fig:architectureVariation}
the \texttt{nNNPDF1.0} results at the input scale $Q_0=1$ GeV for $^{12}$C
and $^{208}$Pb nuclei are shown with the two different architectures,
$3-25-3$ (baseline) and $3-50-3$.
We find that differences are very small and consistent with statistical fluctuations.
Given that now there are twice as many free parameters as in the baseline settings,
this stability demonstrates that our results are driven by the input experimental
data and not by methodological choices such as the specific network
architecture.
Furthermore, 
we have also verified that the outcome of the fits is similarly unchanged if a network
architecture with a comparable number of parameters but two hidden layers is used. 
  
\myparagraph{The role of the $A=1$ boundary condition}
Imposing the $A=1$ boundary condition Eq.~(\ref{eq:constraintprotonPDFs}) leads to
important constraints on both the central values and the uncertainties
of \texttt{nNNPDF1.0} fit, particularly for low values of the atomic mass number $A$.
Here we want to quantify this impact by comparing the baseline \texttt{nNNPDF1.0} results
with those of the corresponding fit where this boundary condition
has not been imposed.
This can be achieved by performing the fits with the
hyper-parameter $\lambda=0$ in Eq.~(\ref{eq:chi2_nNNPDF10}).
Note that in this case the behaviour of the fitted at nPDFs for $A=1$ is unconstrained,
since only experimental data with $A \ge 2$ is included in the fit.

\begin{figure}[ht]
  \begin{center}
    \includegraphics[width=0.93\textwidth]{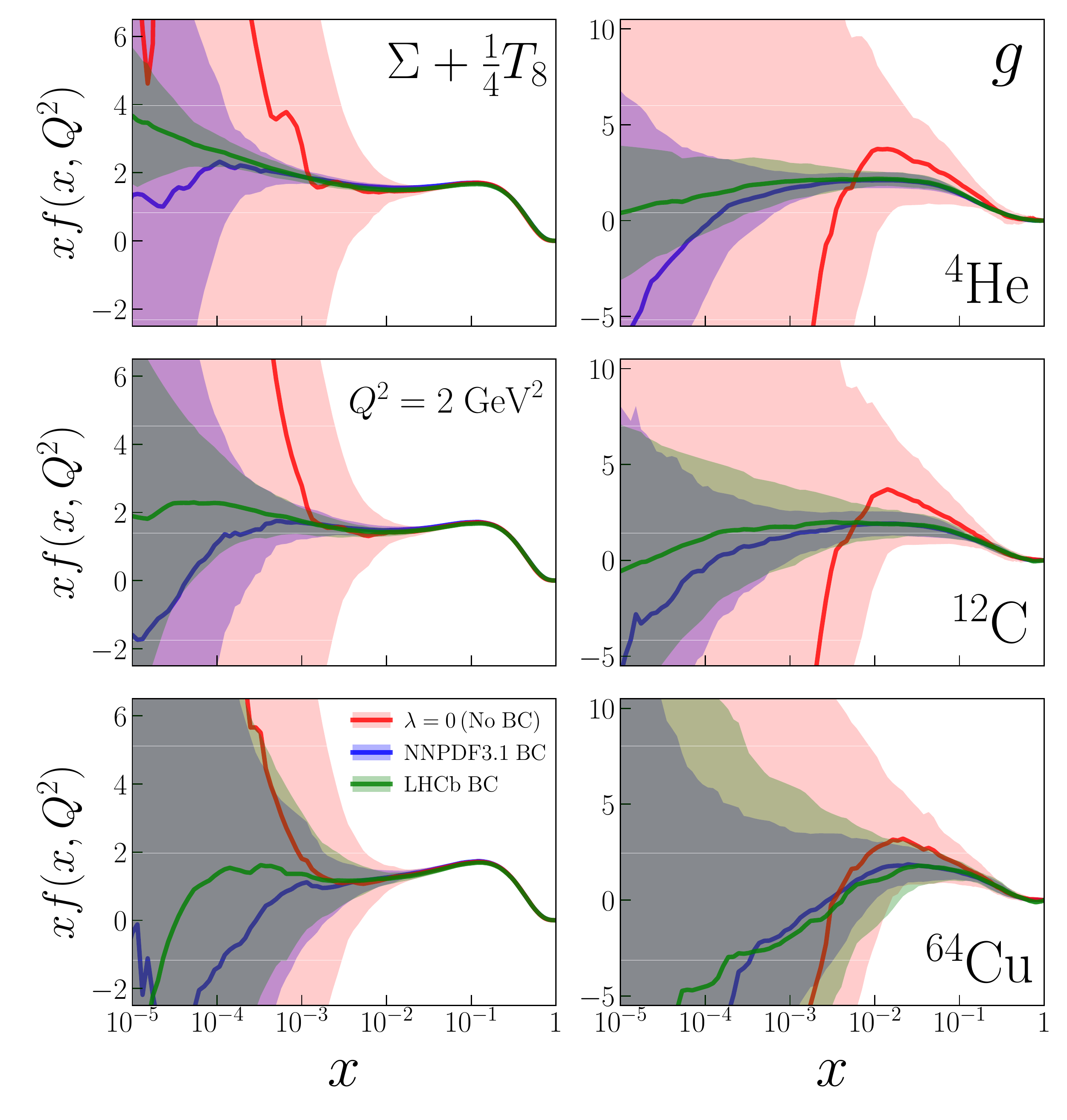}
   \end{center}
  \vspace{-0.7cm}
  \caption{\small Comparison of the \texttt{nNNPDF1.0} fits for different choices of
    the $A=1$ boundary condition (BC).
    The baseline result, which imposes NNPDF3.1 as boundary condition
    between $x=10^{-3}$ and $x=0.7$ (blue), 
    are shown together with two fit variants, one 
     produced using the NNPDF3.0+LHCb set
     as boundary condition down to $x=10^{-5}$ (green), 
     and another without the boundary condition by setting $\lambda=0$ in Eq.~(\ref{eq:chi2_nNNPDF10})
     (red).
    The central values (solid lines) and uncertainties (shaded bands) are given for the
    quark combination $\Sigma + \frac{1}{4}T_8$ (left panels) and gluon (right panels) at
     $Q^2=2$ GeV$^2$ for $A=4$ (top panels), $A=12$ (middle panels) and $A=64$ (bottom panels).
  }
  \label{fig:nPDFcomp_BCA_LHCb}
  \end{figure}

In Fig.~\ref{fig:nPDFcomp_BCA_LHCb},
we show a comparison between the \texttt{nNNPDF1.0} baseline, which imposes NNPDF3.1 as
the $A=1$ boundary condition
between $x=10^{-3}$ and $x=0.7$, in addition to a resulting
fit where this boundary condition is not implemented.
Moreover, we display the gluon and the $\Sigma+T_8/4$ quark
combination at $Q^2=2$ GeV$^2$ for $A=4, 12,$ and 64.
This comparison demonstrates a significant impact on \texttt{nNNPDF1.0}
resulting from the $A=1$ constraint, especially for helium and carbon 
nuclei where the PDF 
uncertainties are increased dramatically if no boundary condition is used.
The impact is more distinct for the gluon, where even for relatively
heavy nuclei such as $^{64}$Cu the boundary condition leads
to a reduction of the nPDF uncertainties by up to a factor two.
We can thus conclude that imposing consistently the $A=1$ limit 
using a state-of-the-art proton fit is an essential ingredient
of any nPDF analysis.

While the baseline \texttt{nNNPDF1.0} fits only constrain the $A=1$ distribution between 
$x=10^{-3}$ and $x=0.7$, one in principle could extend
the coverage of the boundary condition
down to smaller values of $x$ provided a reliable proton PDF baseline is used.
Indeed, it is possible to demonstrate that we can impose
the constraint down to much smaller values of $x$,
e.g. $x=10^{-5}$.
For this purpose, we perform a fit using instead for the boundary condition
the NNPDF3.0+LHCb NLO sets constructed in Ref.~\cite{Gauld:2015yia,Gauld:2016kpd}.
More specifically we use the set based on the $N_5$, $N_7$, and $N_{13}$ normalised
distributions of $D$ meson production in the forward region
at 5, 7, and 13 TeV.
The reason is that these sets exhibit reduced quark and gluon PDF uncertainties
down to $x\simeq 10^{-6}$, and therefore are suitable to constrain 
the small-$x$ nPDF uncertainties.

The comparison between the baseline \texttt{nNNPDF1.0} fit and
its LHCb variant is shown in 
Fig.~\ref{fig:nPDFcomp_BCA_LHCb}.
We now find a further reduction of the nPDF uncertainties at small-$x$, again
more notably for light nuclei.
In this case, the reduction of uncertainties is more distinct for the quarks,
which benefit from the very accurate determination of the proton's quark
sea at small-$x$ in NNPDF3.0+LHCb.
Note that, in turn, the improved nPDF errors at small-$x$ might lead
to increased sensitivity to effects such as shadowing
and evidence for non-linear evolution corrections.

\section{Preprocessing exponents in \texttt{nNNPDF2.0}}
The $x^{\alpha_f}(1-x)^{\beta_f}$ polynomial pre-factors
appearing in Eq.~(\ref{eq:param2}) are included to increase
the efficiency of the parameter optimisation, since they
approximate the general PDF behaviour 
in the small- and large-$x$ asymptotic limits~\cite{Ball:2016spl}.
Note that the exponents $\alpha_f$ and $\beta_f$
are $A$-independent, implying that $A$ dependence of the
nPDFs will arise completely from the output of the neural network.
As in the case of the \texttt{nNNPDF1.0} analysis, the values of $\alpha_f$
and $\beta_f$ are fitted for each Monte Carlo replica
on the same footing as the weights and thresholds of the
neural network.

The ranges of the preprocessing parameters are determined
both by physical considerations and by empirical observations.
First of all, the lower limit of the small-$x$ parameter is
set so that each PDF combination satisfies various
integrability conditions.
In particular, the non-singlet combinations 
$xV$, $xV_3$, $xT_3$, and $xT_8$
must tend to zero at small-$x$, else the valence sum rules
and other relations such as the Gottfried sum 
rule~\cite{Gottfried:1967kk,Forte:1992df,Abbate:2005ct} 
would be ill-defined.
Moreover, the singlet combinations $x\Sigma$ and $xg$ must be exhibit 
finite integrable behaviour as $x\rightarrow0$, 
otherwise the momentum integral cannot be computed.
Concerning the large-$x$ exponents $\beta_f$, the lower limits 
of their ranges ensure that PDFs vanish in the elastic limit, 
while the upper limit is determined largely from general arguments 
related to sum rule expectations. 
In general, however, the upper values of both 
$\alpha_f$ and $\beta_f$ are chosen to be sufficiently 
large to allow flexibility 
in exploring the parameter space while keeping 
fit efficiency optimal.

Under these considerations,
we restrict the parameter values for the pre-processing factors
during the fit to the following intervals,
\begin{align}
\label{eq:nNNPDF20_preprocessing}
\alpha_\Sigma & \in [-1,5]~~([-1,1]) \, , & \beta_\Sigma &\in [1,10]~~([1,5]) \, ,\nonumber \\
\alpha_g & \in [-1,5]~~([-1,1]) \, ,      & \beta_g &\in [1,10]~~([1,5]) \, ,\nonumber\\
\alpha_V & \in [0,5]~~([1,2]) \, ,        & \beta_V &\in [1,10]~~([1,5]) \, ,\\
\alpha_{T_8} & \in [-1,5]~~([-1,1]) \, ,  & \beta_{T_8} &\in [1,10]~~([1,5]) \, ,\nonumber\\
\alpha_{V_3} & \in [0,5]~~([1,2]) \, ,    & \beta_{V_3} &\in [1,10]~~([1,5]) \, , \nonumber\\
\alpha_{T_3} & \in [-1,5]~~([-1,1]) \, ,  & \beta_{T_3} &\in [1,10]~~([1,5]) \, ,\nonumber
\end{align}
where the ranges in parentheses are those used to randomly select the initial values
of $\alpha_f$ and $\beta_f$ at the start of the minimisation.
We do not impose any specific relation between the 
small- or large-$x$ exponents of the different quark combinations, 
so that each are fitted independently.
It is also worth emphasizing here that
the neural network has the ability to compensate 
for any deviations in the shape of the preprocessing function,
so the dependence on $x$ and $A$ of the nPDFs in the data region 
will be dominated by the neural network output.
This implies that the preprocessing exponents will primarily
affect the results in the extrapolation regions.

\begin{figure}[ht]
\begin{center}
  \includegraphics[width=0.75\textwidth]{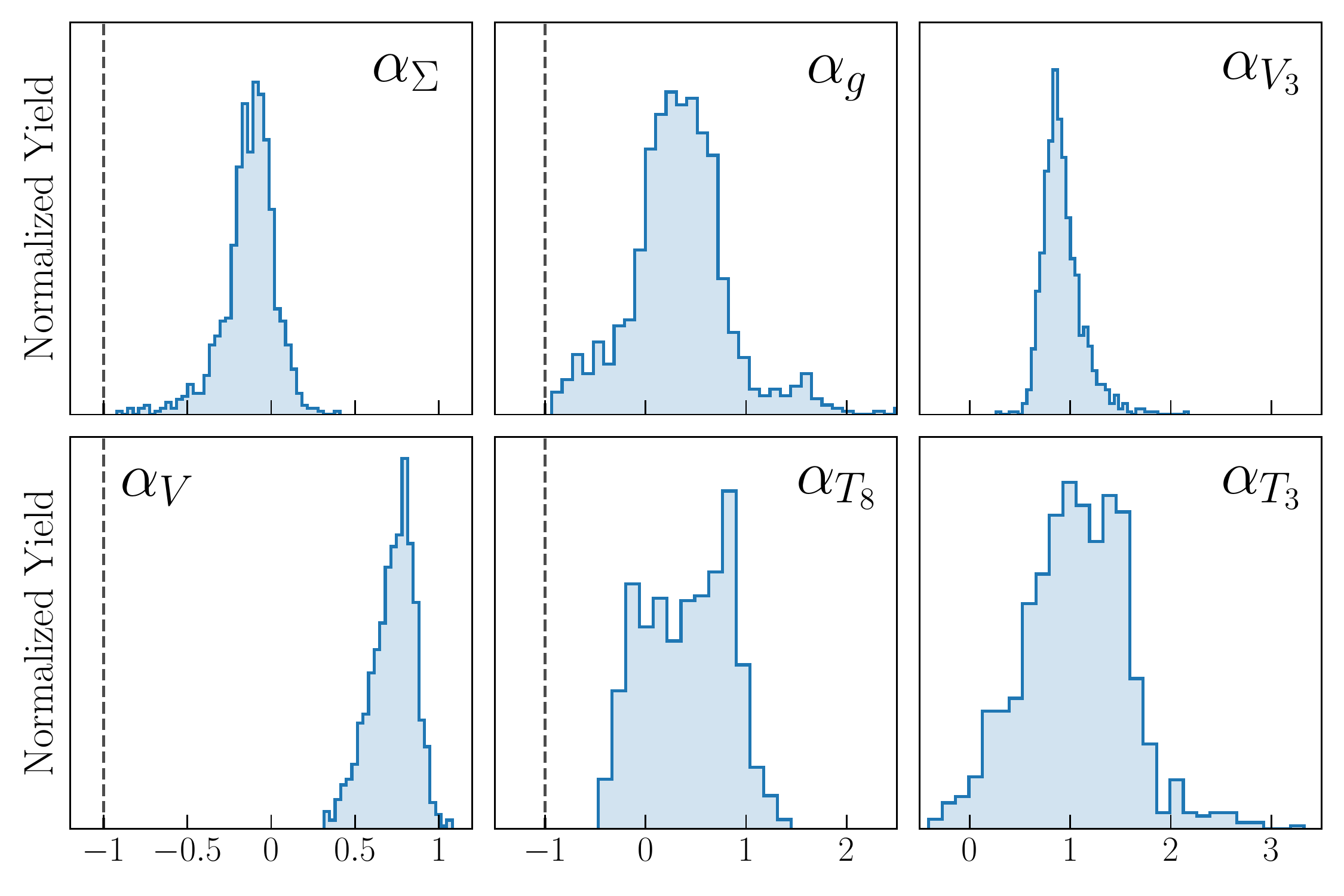}
 \end{center}
\vspace{-0.5cm}
\caption{\small The probability distribution associated to the fitted 
   small-$x$ preprocessing exponents $\alpha_f$ computed 
   with the $N_{\rm rep}=1000$ replicas of the \texttt{nNNPDF2.0} NLO set.
  The ranges for which these exponents
  are allowed to vary, Eq.~(\ref{eq:nNNPDF20_preprocessing}),
  are indicated by horizontal dashed lines.
  Note the difference in the $x$-axis range values for each column.
}
\label{fig:alpha_exp}
\end{figure}

\begin{figure}[ht]
\begin{center}
  \includegraphics[width=0.75\textwidth]{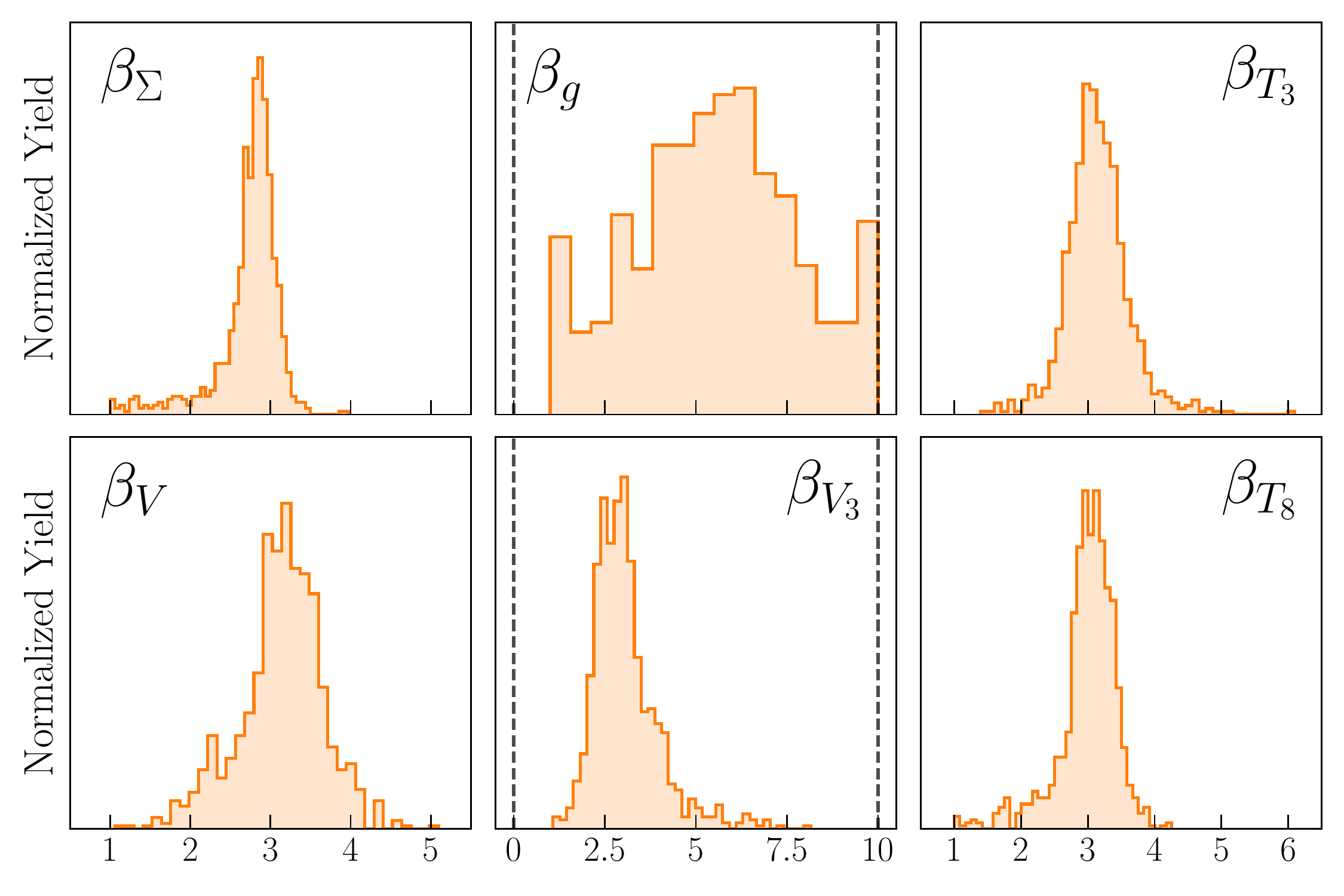}
 \end{center}
\vspace{-0.5cm}
\caption{\small Same as Fig.~\ref{fig:alpha_exp} for
   the fitted large-$x$ preprocessing 
   exponents $\beta_f$.
  Note the difference in the $x$-axis range values for each column.
}
\label{fig:beta_exp}
\end{figure}

To illustrate the values of the small and
large-$x$ preprocessing exponents preferred by the experimental
data, we display in Figs.~\ref{fig:alpha_exp} and~\ref{fig:beta_exp}
the probability distributions associated with the $\alpha_f$ and $\beta_f$
exponents, respectively, computed using the $N_{\rm rep}=1000$ replicas of the \texttt{nNNPDF2.0}
analysis.
Note how these exponents are restricted to lie in the interval given by
Eq.~(\ref{eq:nNNPDF20_preprocessing}).
For $T_3$ and $T_8$, we can see that despite
not imposing the strict integrability requirement that
$\alpha_f > 0$, it is still being satisfied for a large 
majority of the replicas, especially for $T_3$.
Interestingly, the gluon seems to prefer a valence-like
behaviour at small-$x$.
However, such behaviour is only observed 
at the parameterisation scale and as soon as 
$Q > Q_0$, DGLAP evolution drives it to its
expected singlet-like behaviour.

Concerning the behaviour of the large-$x$ exponents $\beta_f$, 
we find that they are reasonably well
constrained for the quark distributions,
where the best-fit values are located in a 
region close to $\beta_f \simeq 3$.
The fact that they share similar $\beta_f$ exponents
can be explained by the fact that in the large-$x$ region 
the quark combinations are dominated by valence 
components.
Interestingly, a best-fit value of $\beta_f \simeq 3$ for the 
valence quarks is consistent with the expectations from 
the QCD counting rules~\cite{Brodsky:1973kr}, as discussed in~\cite{Ball:2016spl}.
Furthermore, the best-fit value for $\beta_g$
is also consistent with the QCD counting rules 
prediction of $\beta_g \simeq 5$, although with
significant uncertainties.
The fact that $\beta_g$ is found to vary in a wide range
is due to the lack of direct constraints on the 
large-$x$ nuclear gluon PDF in the present analysis.

\section{Cross section positivity constraint}
\label{s1:positivity}

I dedicate this section to discuss the cross section positivity constraint imposed on both \texttt{NNPDF3.1} and \texttt{nNNPDF2.0} as a completion of their respective frameworks Sect.~\ref{s1:NNPDF31_framework} for \texttt{NNPDF3.1} and Sect.~\ref{s1:nNNPDF20_framework} for \texttt{nNNPDF2.0}.

\myparagraph{Positivity observables in \texttt{NNPDF3.1}}
All of the positivity constraints on the observables listed in Table~\ref{tab:NNPDF31_positivity} are imposed at $Q^2_{\text{pos}} = 5\,\text{GeV}^2$ and for $20$ points in $x \in [10^{-7},1]$. The structure of the DGLAP evolution ensures positivity at higher scales. More details are given in Sect.~3.2.3 of Ref.~\cite{Ball:2014uwa}.
\begin{table}[!h] 
  \centering
  \begin{tabular}{|c|c|}  
    \toprule
      Observable & LO expression \\ \cmidrule{1-2}
      $F_2^u(x,Q^2)$ & \(\displaystyle \propto u^+(x,Q^2) + \mathcal{O}(\alpha_s)\) \\
      $F_2^d(x,Q^2)$ & \(\displaystyle \propto d^+(x,Q^2) + \mathcal{O}(\alpha_s)\) \\
      $F_2^s(x,Q^2)$ & \(\displaystyle \propto s^+(x,Q^2) + \mathcal{O}(\alpha_s)\) \\
      $F_L(x,Q^2)$ & \(\displaystyle \propto C_g \otimes g(x,Q^2) + C_q \otimes q(x,Q^2) + \mathcal{O}(\alpha^2_s)\) \\\cmidrule{1-2}
      \(\displaystyle \frac{d^2 \sigma^{DY}_{u\bar{u}}}{dM^2dy}\) & \(\displaystyle \propto u(x_1,Q^2)\bar{u}(x_2,Q^2) + \mathcal{O}(\alpha_s)\) \\
      \(\displaystyle \frac{d^2 \sigma^{DY}_{d\bar{d}}}{dM^2dy}\) & \(\displaystyle \propto d(x_1,Q^2)\bar{d}(x_2,Q^2) + \mathcal{O}(\alpha_s)\) \\
      \(\displaystyle \frac{d^2 \sigma^{DY}_{s\bar{s}}}{dM^2dy}\) & \(\displaystyle \propto s(x_1,Q^2)\bar{s}(x_2,Q^2) + \mathcal{O}(\alpha_s)\) \\\cmidrule{1-2}
      \(\displaystyle  \frac{d\sigma_{gg}^H}{dy} \) & \(\displaystyle \propto g(x_1,M_H^2)g(x_2,M_H^2) + \mathcal{O}(\alpha^3_s) \quad M_H = Q_{pos}\)\\
  \bottomrule 
  \end{tabular}
  \caption{The LO expression of the observable used in \texttt{NNPDF3.1} to impose positivity.}
  \label{tab:NNPDF31_positivity}
  \end{table}

\myparagraph{Positivity observables in \texttt{nNNPDF2.0}}
In Table~\ref{tab:nNNPDF20_positivity} we list
the $\mathcal{F}^{(l)}$ processes 
used in the \texttt{nNNPDF2.0} analysis for which the positivity 
of physical cross sections is imposed using Eq.~(\ref{eq:positivity}).
For each observable, the LO expressions in terms of the average bound nucleon PDFs and
bound proton distributions are given together with the number of pseudodata points 
$N_{\rm dat}$ and the corresponding kinematic coverage.
Note that the LO expressions in Table~\ref{tab:nNNPDF20_positivity} are shown
for illustration purposes only, and in our analysis
these observables are computed using the full NLO formalism.

\begin{table}
  \footnotesize
  \begin{center}
    \renewcommand{\arraystretch}{1.55}
    \begin{tabular}{|c| c| c|c |}
      \toprule
      Observable  &  LO expression  &  $N_{\rm dat}$   &  Kinematic coverage \\
      \midrule
      \multirow{2}{*}{ $F_2^{u}(x,Q^2,A)$}  &  $\propto \lp u^{N/A}+\bar{u}^{N/A} \rp$  &   \multirow{2}{*}{20}  &
      $Q^2=5$ GeV\\
      \multirow{2}{*}{ }  &
      $\propto  \lc (Z/A)\lp u^{p/A}+\bar{u}^{p/A} \rp+ (1-Z/A)\lp d^{p/A}+\bar{d}^{p/A}\rp\rc $  & & $5\times 10^{-7}\le x \le 0.9$ \\
       \midrule
      \multirow{2}{*}{ $F_2^{d}(x,Q^2,A)$}  &  $\propto \lp d^{N/A}+\bar{d}^{N/A} \rp$  &   \multirow{2}{*}{20}  &
      $Q^2=5$ GeV\\
      \multirow{2}{*}{ }  &
      $\propto  \lc (Z/A)\lp d^{p/A}+\bar{d}^{p/A} \rp+ (1-Z/A)\lp u^{p/A}+\bar{u}^{p/A}\rp\rc $  & & $5\times10^{-7}\le x \le 0.9$ \\
       \midrule
      \multirow{2}{*}{ $F_2^{s}(x,Q^2,A)$}  &  $\propto \lp s^{N/A}+\bar{s}^{N/A} \rp$  &   \multirow{2}{*}{20}  &
      $Q^2=5$ GeV\\
      \multirow{2}{*}{ }  &
      $\propto  \lp s^{p/A}+\bar{s}^{p/A} \rp$   & & $5\times 10^{-7}\le x \le 0.7$ \\
       \midrule
      \multirow{2}{*}{ $F_L(x,Q^2,A)$}  & \multirow{2}{*}{sensitive to $xg(x,Q^2)$ (see text)}  &   \multirow{2}{*}{20}  &
      $Q^2=5$ GeV\\
      \multirow{2}{*}{ }  &
       & & $5\times 10^{-7}\le x \le 0.9$ \\
        \midrule
        \multirow{2}{*}{ $\sigma_{u\bar{u}}^{DY}(y,M^2,A)$}  &
        $\propto \lp  u^{p}(x_1)\times \bar{u}^{N/A}(x_2) \rp $  &   \multirow{2}{*}{20}  &
      $Q^2=5$ GeV\\
      \multirow{2}{*}{ }  &  $\propto \lp  u^{p}(x_1)\times \lp Z \bar{u}^{p/A}(x_2) + (A-Z)\bar{d}^{p/A}(x_2) \rp \rp $
      & & $10^{-2}\le x \le 0.9$ \\
        \midrule
        \multirow{2}{*}{ $\sigma_{d\bar{d}}^{DY}(y,M^2,A)$}  &
        $\propto \lp  d^{p}(x_1)\times \bar{d}^{N/A}(x_2) \rp $  &   \multirow{2}{*}{20}  &
      $Q^2=5$ GeV\\
      \multirow{2}{*}{ }  &  $\propto \lp  d^{p}(x_1)\times \lp Z \bar{d}^{p/A}(x_2) + (A-Z)\bar{u}^{p/A}(x_2) \rp \rp $
      & & $10^{-2}\le x \le 0.9$ \\
          \midrule
        \multirow{2}{*}{ $\sigma_{s\bar{s}}^{DY}(y,M^2,A)$}  &
        $\propto \lp  s^{p}(x_1)\times \bar{s}^{N/A}(x_2) \rp $  &   \multirow{2}{*}{20}  &
      $Q^2=5$ GeV\\
      \multirow{2}{*}{}  &$\propto \lp  s^{p}(x_1)\times \bar{s}^{p/A}(x_2) \rp $   
         & & $10^{-2}\le x \le 0.9$ \\
          \midrule
        \multirow{2}{*}{ $\sigma_{\bar{u}d}^{DY}(y,M^2,A)$}  &
        $\propto \lp  \bar{u}^{p}(x_1)\times d^{N/A}(x_2) \rp $  &   \multirow{2}{*}{20}  &
      $Q^2=5$ GeV\\
      \multirow{2}{*}{}  &   $\propto \lp  \bar{u}^{p}(x_1)\times \lp Z d^{p/A}(x_2) + (A-Z)u^{p/A}(x_2) \rp \rp $
      & & $10^{-2}\le x \le 0.9$ \\
       \midrule
        \multirow{2}{*}{ $\sigma_{\bar{d}u}^{DY}(y,M^2,A)$}  &
        $\propto \lp  \bar{d}^{p}(x_1)\times u^{N/A}(x_2) \rp $  &   \multirow{2}{*}{20}  &
      $Q^2=5$ GeV\\
      \multirow{2}{*}{}  &   $\propto \lp  \bar{d}^{p}(x_1)\times \lp Z u^{p/A}(x_2) + (A-Z)d^{p/A}(x_2) \rp \rp $
      & & $10^{-2}\le x \le 0.9$ \\
       \midrule
        \multirow{2}{*}{ $\sigma_{u\bar{s}}^{DY}(y,M^2,A)$}  &
        $\propto \lp  u^{p}(x_1)\times \bar{s}^{N/A}(x_2) \rp $  &   \multirow{2}{*}{20}  &
      $Q^2=5$ GeV\\
      \multirow{2}{*}{}  &   $\propto \lp  u^{p}(x_1)\times\bar{s}^{p/A}(x_2)    \rp $
      & & $10^{-2}\le x \le 0.9$ \\
       \midrule
        \multirow{2}{*}{ $\sigma_{\bar{u}s}^{DY}(y,M^2,A)$}  &
        $\propto \lp  \bar{u}^{p}(x_1)\times s^{N/A}(x_2) \rp $  &   \multirow{2}{*}{20}  &
      $Q^2=5$ GeV\\
      \multirow{2}{*}{}  &   $\propto \lp  \bar{u}^{p}(x_1)\times s^{p/A}(x_2)    \rp $ & & $10^{-2}\le x \le 0.9$ \\
      \bottomrule
    \end{tabular}
  \end{center}
  \vspace{-0.3cm}
  \caption{\label{tab:nNNPDF20_positivity}
    The processes used
    to impose the positivity of physical cross-sections by means
    of the constraint of Eq.~(\ref{eq:positivity}).
    For each process we indicate the corresponding LO expressions,
    the number of data points $N_{\rm dat}$, and the kinematic coverage
    spanned by the pseudodata.
  }
\end{table}

Here we consider two types of positivity observables.
The first type are the DIS structure functions $F_2^u$, $F_2^d$, 
$F_2^s$, and $F_L$.
The former three quantities, which contain only $u$, $d$, and $s$ 
contributions, respectively, are constructed to be positive-definite since there
exists consistent physical theories where the photon couples only to up-, down-,
or strange-type quarks.
To further clarify this point, 
consider such observables from an EFT perspective. 
For example, in the Standard Model Effective Field Theory (SMEFT)~\cite{Brivio:2017vri}
there are four-fermion operators at dimension 6 of the form:
\begin{equation}
\label{eq:4Fsmeft}
\mathcal{L}_{\rm SMEFT} \supset \frac{a_{q_il_j}}{\Lambda^2} \left( \bar{q}_i \Gamma_{\mu} q_i \right)
\left( \bar{l}_j \Gamma^{\mu} l_j \right),
\end{equation}
which describes the 4-point interaction of a quark of flavour $i$ with a lepton of type $j$. 
Here, $\Lambda$ corresponds to the energy scale in which the EFT expansion becomes invalid,
and $\Gamma^\mu$ represents the general Lorentz structure for the interaction.
Since any theory with a single non-zero coupling $a_{q_i l_j}$ is theoretically consistent~\cite{Carrazza:2019sec}, 
any cross sections computed with it must be positive definite, 
including the ones that lead to our positivity observables. 
Then one can take the $\Lambda \rightarrow \infty$ limit, in which
case the cross section will be vanishingly small, yet still remain positive, as we demand in 
our fit via the positivity constraint. 

While the positivity of the quark structure functions impact 
the various quark PDFs,
the longitudinal structure function $F_L$
largely impacts the nuclear gluon PDF since $F_L$ enters 
only at NLO and is dominated by the gluon contribution.
Together, we evaluate each of the structure functions on a grid of $N_{\rm}=20$ 
pseudodata points between $x=10^{-7}$ and $x=0.9$
at $Q=\sqrt{5}$ GeV, which is slightly above the input 
parameterisation scale $Q_0=1$ GeV to ensure
perturbative stability.

The second type of observable for which the
cross section positivity is imposed
is the double-differential Drell-Yan cross section.
In particular, we enforce the positivity of both neutral- and charged-current
Drell-Yan cross sections in pA scattering for specific combinations of 
quark-antiquark annihilation listed in Table~\ref{tab:nNNPDF20_positivity}.
At leading order, the Drell-Yan cross section
can be written schematically as:
\begin{equation}
\label{eq:DYpos}
\frac{d^2\sigma_{q_{f_1}\bar{q}_{f_2}}^{DY}}{dydQ^2} \propto \lp 
f_1^{(p)}(x_1,Q^2)  \bar{f_2}^{(p/A)}(x_2,Q^2) \rp \, ,
\end{equation}
where the momentum fractions $x_1$
and $x_2$ are related to the rapidity $y$ and invariant mass of the
final state $Q$ at at leading order by $x_{1,2} = \lp Q/\sqrt{s}\rp e^{\pm y}$.
Here we set $Q^2=5$ GeV$^2$ and
adjust the rapidity range and center-of-mass energy $\sqrt{s}$
so that the LO kinematic range for $x_1$ and $x_2$ correspond 
to $10^{-2} \le x_{1,2} \le 0.9$.

Note that positivity of Eq.~(\ref{eq:DYpos}) will affect also the 
fitted \texttt{nNNPDF2.0} $A=1$ distribution which enters as the free-proton PDF. 
While most of the positivity observables coincide with 
those included in the free-proton baseline from
which the $A=1$ distribution is derived, 
the $u\bar{d}$, $\bar{u}d$, $u\bar{s}$, and $\bar{u}s$
combinations of Table~\ref{tab:nNNPDF20_positivity} are
new in the \texttt{nNNPDF2.0} determination.
Consequently, we impose these new observables 
only for proton-iron and proton-lead collisions, 
the two nuclei for which experimental
data from charged-current DIS and Drell-Yan reactions
are analysed to study quark flavour separation.

In the following, we will demonstrate the
positivity of cross sections for all relevant processes
in the entire kinematical range.
We have verified that, in the absence of these constraints, 
the DIS structure functions
and the DY cross section will in general not satisfy positivity.

\myparagraph{Assessment of positivity in \texttt{nNNPDF2.0}}
%
As was discussed previously, we impose the
requirement that the cross sections of arbitrary physical processes are
positive-definite quantities.
This constraint is implemented by means of an additive penalty term in
the figure of merit, Eq.~(\ref{eq:positivity}).
Moreover, the penalty is constructed from the pseudodata summarized in
Table~\ref{tab:nNNPDF10_data}, which corresponds to lepton-nuclear
scattering structure functions and Drell-Yan cross sections in
proton-nucleus collisions.
Recall that the kinematics of the positivity pseudodata were chosen to
  cover those of the actual data used in the fit, see
  Fig.~\ref{figkinplot}.

\begin{figure}[ht]
\begin{center}
  \includegraphics[width=0.9\textwidth]{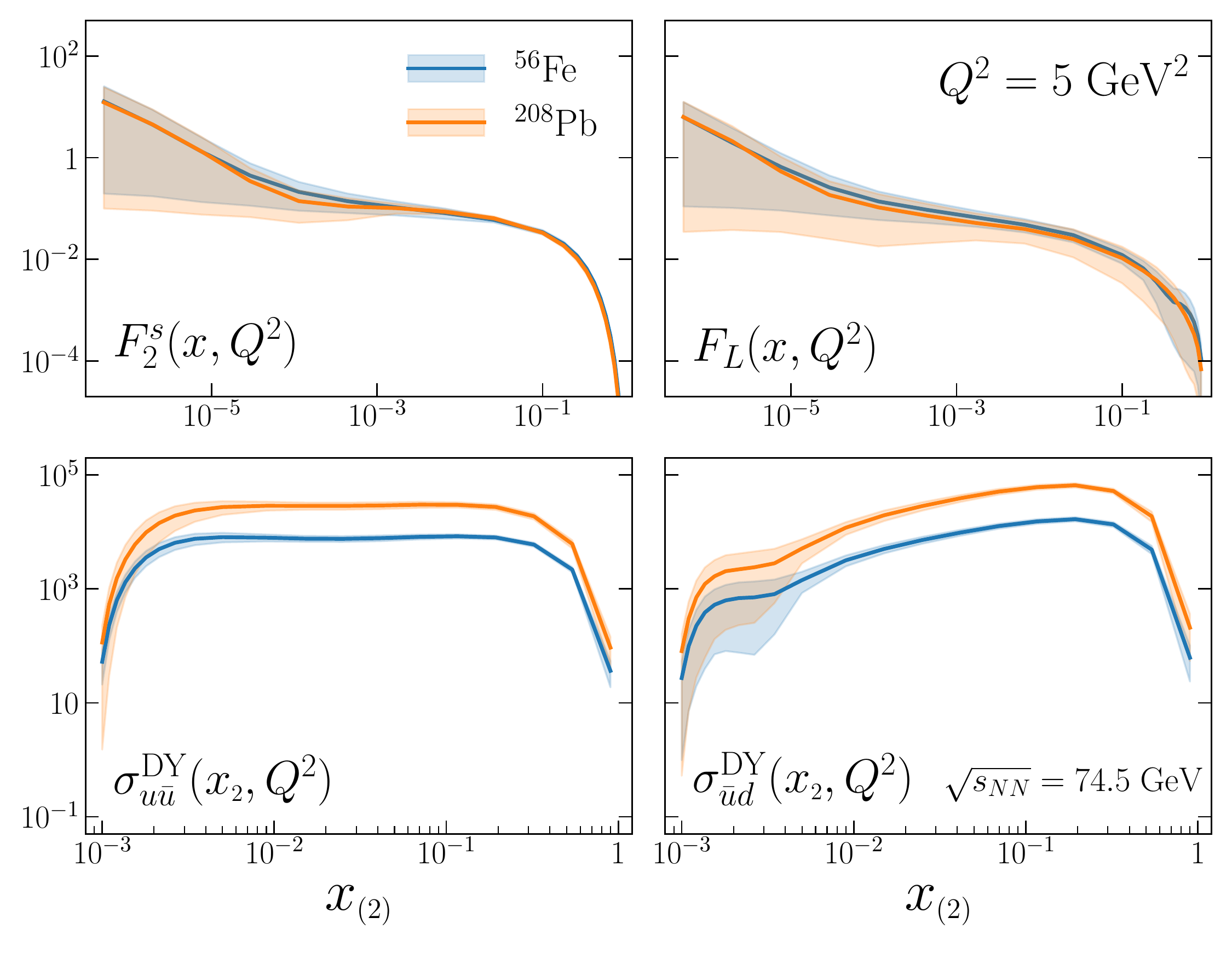}
 \end{center}
\vspace{-0.6cm}
\caption{A representative selection of the positivity observables used
in \texttt{nNNPDF2.0}.
  From top to bottom and from left to right, we show the DIS structure
  functions $F_2^s$ and $F_L$ and the Drell-Yan rapidity distributions
  $\sigma_{u\bar{u}}^{\rm DY}$ and $\sigma_{\bar{u}d}^{\rm DY}$.
  The bands indicate the 90\% confidence level interval.
  \label{fig:positivity}
}
\end{figure}

Here we want to demonstrate that the \texttt{nNNPDF2.0} determination indeed
satisfies these various positivity constraints.
In Fig.~\ref{fig:positivity} we display a representative selection of
the positivity observables imposed in \texttt{nNNPDF2.0}.
In particular, we show the DIS structure
functions $F_2^s(x,Q^2)$ and $F_L(x,Q^2)$, as well as the Drell-Yan
rapidity distributions $\sigma_{u\bar{u}}^{\rm DY}(y)$ and
$\sigma_{\bar{u}d}^{\rm DY}(y)$, where the bands indicates the 90\%
confidence level uncertainty interval.
  We use a scale of $Q^2=5$ GeV$^2$, which corresponds to the same scale
  in which Eq.~(\ref{eq:positivity}) is imposed.
  Furthermore, we provide the positivity predictions for both iron and
  lead nuclei.
  Note that since the Drell-Yan cross sections are not normalised by the
  value of $A$, the absolute magnitude of the two nuclei are different.
  Of course, the overall normalization is not relevant for the
  implementation of the positivity constraint.

The selection of positivity observables in Fig.~\ref{fig:positivity} is
representative since it contains one of the quark structure functions
($F_2^s$) constraining a $q^+$ combination, $F_L$ that is sensitive to
the gluon positivity, and a diagonal and off-diagonal DY cross section
which are relevant for different aspects of quark flavour separation
(confer also the expressions in Sect.~\ref{s2:Flavour_separation}).
Here the Drell-Yan cross sections are represented as a function of
$x_2$, which corresponds to the momentum fraction of the nuclear
projectile obtained using the LO kinematics defined in Sect.~\ref{s2:Flavour_separation}).
While the positivity constraint was only implemented for
$x_2\gsim10^{-2}$ with a per-nucleon center-of-mass energy of
$\sqrt{s}=23.5$ GeV, we illustrate instead the positivity for a choice
of kinematics that allow a reach to $x_2\sim10^{-3}$, with a per-nucleon
center-of-mass energy of $\sqrt{s}=74.5$ GeV.
As can be seen from Fig.~\ref{fig:positivity}, the \texttt{nNNPDF2.0}
determination satisfies the positivity of physical cross sections in the
entire kinematic range.
Here the nPDF uncertainty bands become larger near the kinematic
endpoints ($x=1$ for DIS and $x_2\simeq 10^{-3}$ for Drell-Yan), since
these correspond to regions of the phase space where experimental
constraints are scarce.
Recall that by virtue of DGLAP evolution properties, these results
ensure the cross sections involving higher momentum transfers, $Q^2 > 5$
GeV$^2$, will also be positive provided one maintains the initial coverage
in $x_2$.
Therefore, we conclude that while we have not explicitly imposed the positivity at
the level of the  nuclear PDFs, physical observables
constructed from \texttt{nNNPDF2.0}  are guaranteed to satisfy the positivity
requirement.

\section{Comparison between \texttt{nNNPDF2.0} and \texttt{nNNPDF1.0}}
\label{s2:nNNPDF20_comparisonwith10}

In this section we study the differences between the \texttt{nNNPDF1.0} and \texttt{nNNPDF2.0}
determinations by tracing back the impact of the various improvements in the latter
with a series of comparisons.
The goal of this exercise is to assess which of these differences can be
identified with specific methodological improvements, such as the
cross-section positivity constraint, and which ones are related to the
impact of the new experimental information, either the DIS charged
current structure functions or the LHC gauge boson production
measurements.

The starting point for this study will be a fit denoted \texttt{nNNPDF1.0r},
which has been obtained with the code used to produce \texttt{nNNPDF2.0} but
using the same theory, methodology settings, and input dataset as in the
\texttt{nNNPDF1.0} analysis.
The only differences at this level are related to optimisations and
improvements implemented in the code to speed up its performance.
We have verified that \texttt{nNNPDF1.0} and \texttt{nNNPDF1.0r} are statistically
indistinguishable, thus we can safely adopt the latter as baseline for
the comparisons in what follows.

\begin{figure}[ht]
\begin{center}
  \includegraphics[width=0.95\textwidth]{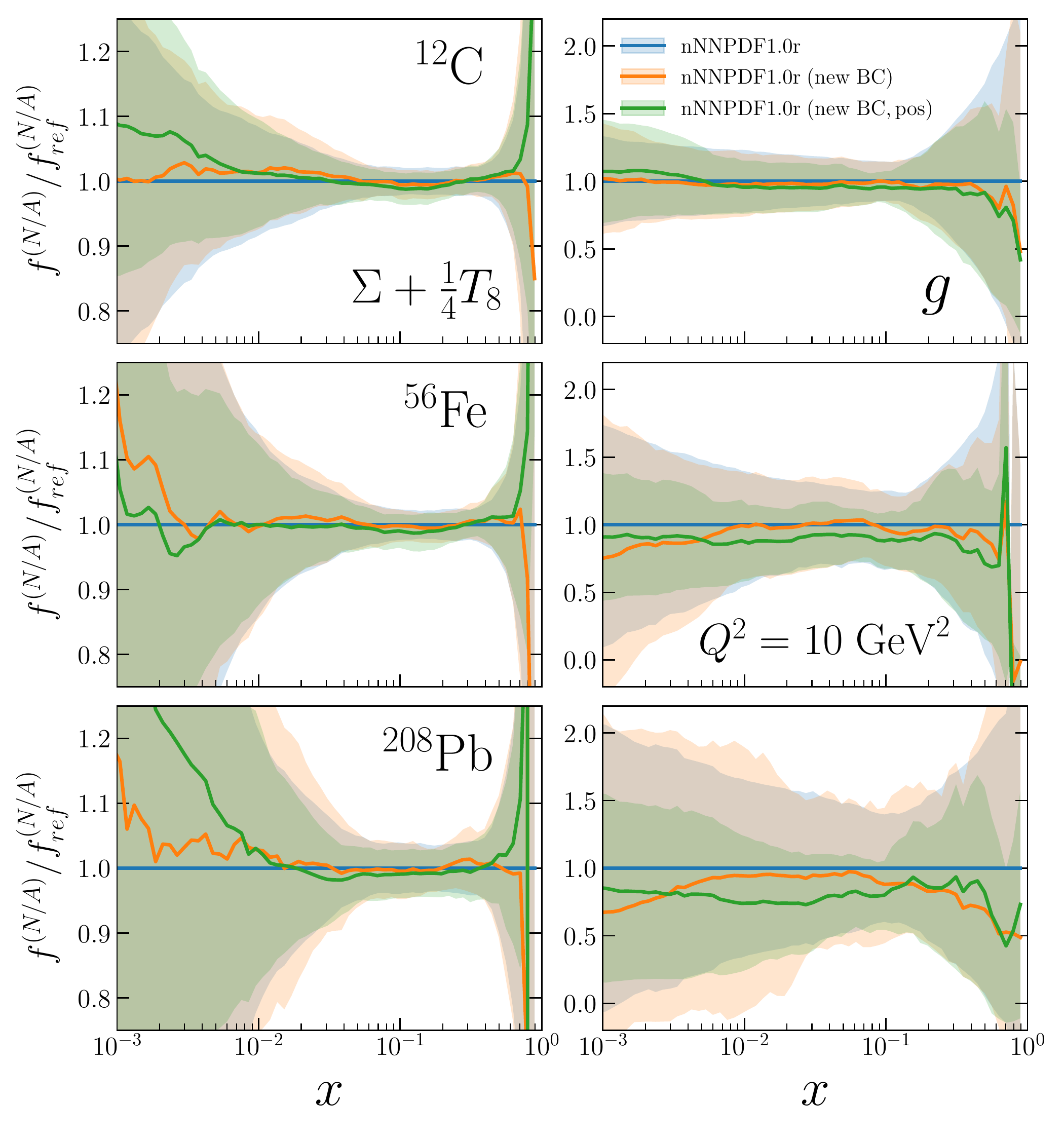}
 \end{center}
\vspace{-0.75cm}
\caption{Comparison between the \texttt{nNNPDF1.0r} baseline and two fit variants
  based on the same input dataset, one where the proton boundary
  condition has been updated and another where in addition the
  positivity of physical cross-sections has been imposed.
  We show the $\Sigma+T_8/4$ quark combination (left) and gluon (right)
  at $Q^2=10$ GeV$^2$ for three values of $A$.
  \label{nNNPDF10Ref}
}
\end{figure}

We have then produced several variants of this \texttt{nNNPDF1.0r} baseline, each
time adding one extra feature or dataset.
The first of these two variants is a fit where the proton boundary
condition has been updated to the no-nuclear NNPDF3.1 fit shown in
Fig.~\ref{fig:NNPDF31_BC_comp}.
The second is a fit where, in addition to the updated boundary
condition, the positivity of cross-sections has been imposed following
the procedure described in App.~\ref{s1:positivity}.
We display in Fig.~\ref{nNNPDF10Ref} the comparison between \texttt{nNNPDF1.0r}
and these two fit variants.
Since the isoscalar neutral-current DIS structure functions used
in \texttt{nNNPDF1.0} are primarily sensitive to the specific quark combination
$\Sigma+T_8/4$, we plot this together with the gluon distribution as a
function of $x$ at $Q^2=10$ GeV$^2$ for carbon, iron, and lead.

First, one can see from Fig.~\ref{nNNPDF10Ref} that the impact of the
new proton boundary condition in the nuclear fit is generally moderate
concerning the size of the uncertainty band.
There are some differences at the central value level for the small-$x$
quarks and for the nuclear gluon PDF of lead, but in both cases the
shifts are much smaller than the associated uncertainties.
This does not imply that using the updated proton boundary condition is
irrelevant for \texttt{nNNPDF2.0}, but rather that this choice is not
particularly impactful for the specific PDF
combinations that can be constrained by
the \texttt{nNNPDF1.0} dataset.
As shown in Fig.~\ref{fig:NNPDF31_BC_comp}, the differences between the
two variants of the proton boundary conditions are more distinguished
for the total strangeness compared to the other quark flavours.

On the other hand, imposing the positivity of the cross-sections leads
to more important differences.
This is not completely unexpected, since it is well known that in
general a model-independent (n)PDF analysis will lead to some
cross-sections being negative unless their positivity is explicitly
imposed.
In our case, one finds that there is not much difference in the quarks,
but there are clear changes for the nuclear gluons in iron and lead,
especially in the latter.
Here we see that imposing the positivity of cross-sections leads to a
significant reduction of the nPDF uncertainty band, which in the case of
lead can be up to a factor of two.

\begin{figure}[ht]
\begin{center}
  \includegraphics[width=0.95\textwidth]{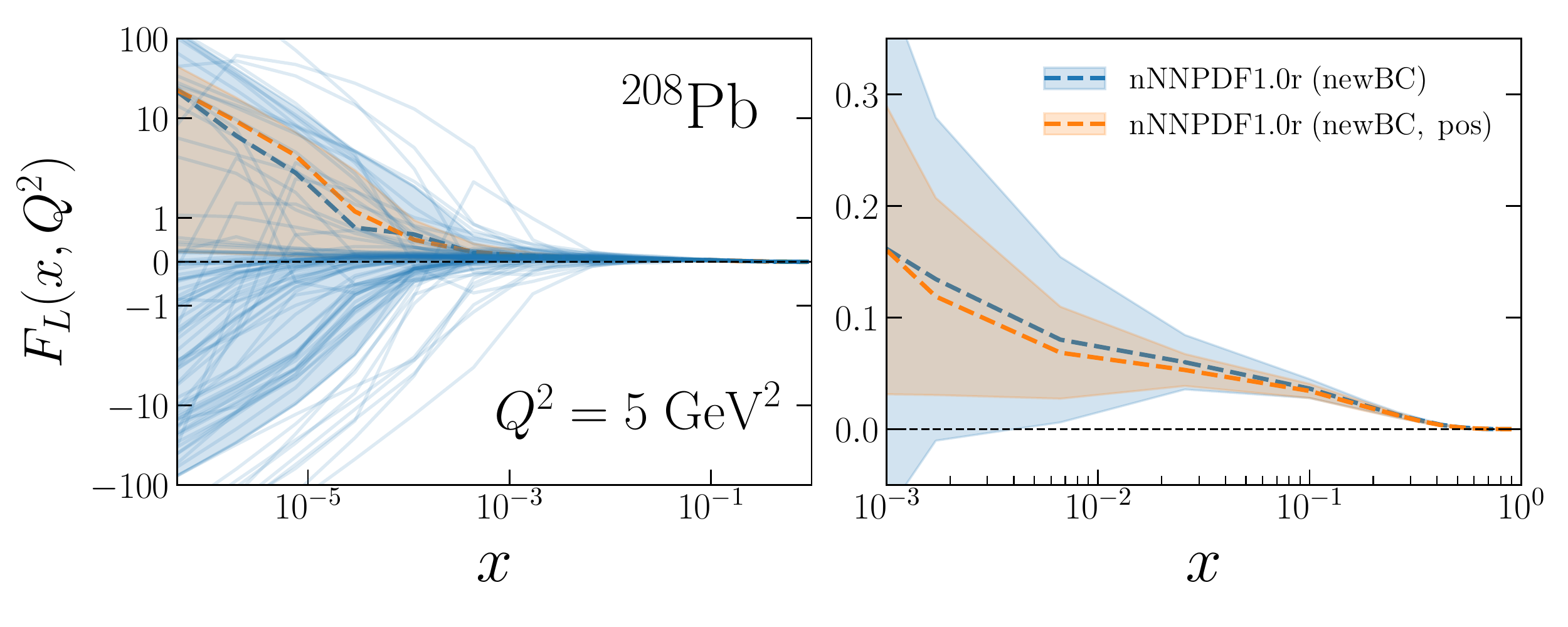}
 \end{center}
\vspace{-0.8cm}
\caption{The longitudinal structure function $F_L(x,Q^2)$ at the
  positivity scale $Q^2=5$ GeV.
  We compare the predictions of the \texttt{nNNPDF1.0r}(newBC) fits with and
  without the cross-section positivity constraint imposed.
  We show the extrapolation (left) and the data (right) regions, in the
  former case displaying also the predictions from the individual
  replicas in the  \texttt{nNNPDF1.0r}(newBC) fit that do not satisfy the
  positivity constraints.
   \label{fig:FLpos}
}
\end{figure}

To illustrate the impact of the cross-section positivity constraint, we
display in Fig.~\ref{fig:FLpos} the longitudinal structure function
$F_L(x,Q^2)$ at the positivity scale $Q^2=5$ GeV.
We compare the predictions of the \texttt{nNNPDF1.0r} fits including the updated
proton boundary condition with and without the cross-section positivity constraint imposed in
both the extrapolation and the data regions.
In the left panel, we display also the predictions from the individual
replicas of the \texttt{nNNPDF1.0r}(newBC) fit that do not
satisfy the cross-section positivity constraints.
Indeed, one can observe that many $F_L$ replicas become negative in some
region of $x$ unless this constraint is explicitly imposed, and that
removing them leads to a significant reduction of the nPDF
uncertainties, particularly in the small-$x$ region.
Interestingly, at medium-$x$ it is largely the upper (rather than the
lower) 90\% CL limit which is reduced by the positivity constraint: this
can be explained by the fact that the very negative $F_L$ replicas at
small-$x$ were actually higher than the median value at medium-$x$ in
order to satisfy the momentum sum rule.

\begin{figure}[ht]
\begin{center}
  \includegraphics[width=0.95\textwidth]{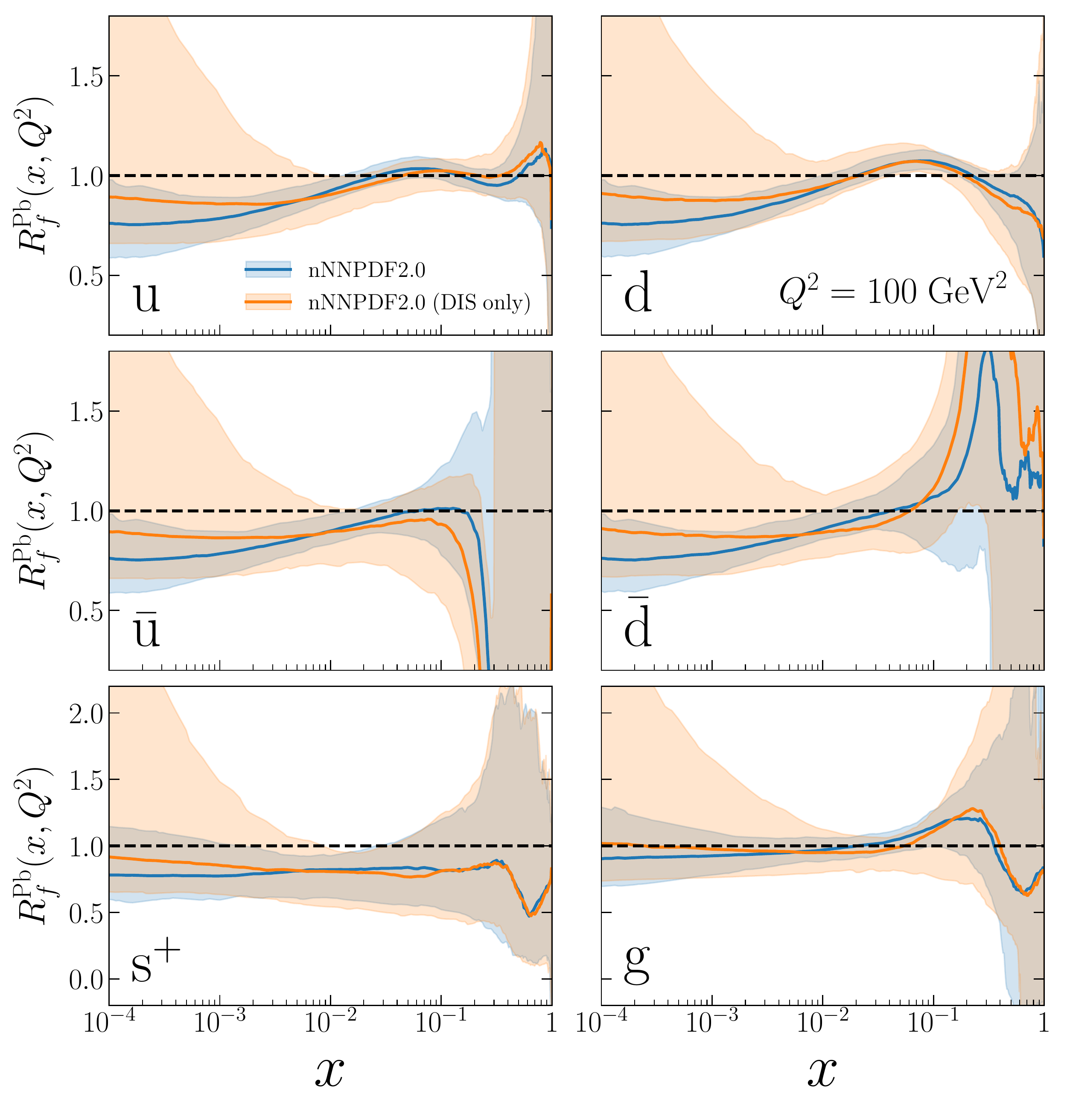}
 \end{center}
\vspace{-0.75cm}
\caption{Same as Fig.~\ref{fig:R_Pb} now comparing the \texttt{nNNPDF2.0}
  baseline results with those of a fit based on identical settings but
  restricted to a DIS-only dataset.
  \label{fig:R_Pb_DIS}
}
\end{figure}

Concerning the impact of the new datasets, a direct comparison of the
\texttt{nNNPDF1.0r}-like fits with those including CC DIS and LHC data is not
possible since as discussed in Sect.~\ref{s2:nNNPDF20_updates}
the input parameterisation basis and the flavour
assumptions are different.
However, we are still able to assess the relative contribution of the CC
structure functions and the LHC gauge boson cross-sections in
determining the \texttt{nNNPDF2.0} results.
In Fig.~\ref{fig:R_Pb_DIS} we display the nuclear modification ratios
for the nPDFs in lead, as was shown in Fig.~\ref{fig:R_Pb}, but now
comparing the \texttt{nNNPDF2.0} baseline results with those of a fit that is
restricted to DIS structure functions, including charged-current
scattering, and that uses identical theoretical and methodological
settings.

  One of the most remarkable features of this comparison is the sizeable
  impact that LHC measurements have in reducing the uncertainties of the
  nuclear PDFs.
  This effect is particularly significant for the gluon and for all
  quark flavours at $x\lsim 0.1$.
  On one hand, the LHC data clearly reveals the presence of nuclear
  shadowing at small-$x$ for both the valence and sea quarks, something
  which is not accessible in a DIS-only fit.
  This result is consistent with the nuclear modification ratios for the
  LHC Drell-Yan distributions reported {\it e.g.} in
  Fig.~\ref{nNNPDF20comp_data_3}.
  On the other hand, the impact of the LHC data on the central values
  and uncertainties of \texttt{nNNPDF2.0} at $x\gsim 0.1$ is less prominent,
  although in that region one also observes a reduction of the
  uncertainties.
  The fact that $R_{\bar{u}}\ll 1$ and $R_{\bar{d}}\gg 1$ for lead
  nuclei at large-$x$ is already present at the level of DIS-only fits
  implies that this trend is favoured by the CHORUS and NuTeV
  charged-current structure functions.
  \chapter{LHAPDF sets}

\subsection*{\texttt{nNNPDF1.0} nPDF sets}
The \texttt{nNNPDF1.0} NLO and NNLO
sets for different values of $A$, are available via 
the {\tt LHAPDF} library~\cite{Buckley:2014ana}, and have also been
linked to the NNPDF website:
\begin{center}
\url{http://nnpdf.mi.infn.it/for-users/nuclear-pdf-sets/}
\end{center}  
These {\tt LHAPDF} grid files contains $N_{\rm rep}=250$ replicas each,
which are fully correlated between different values of $A$ as discussed
in Sect.~\ref{s1:nNNPDF10_results}.

Moreover, due to the lack of
a complete quark flavour separation, additional
assumptions might be required when the \texttt{nNNPDF1.0} sets are used, in particular
for phenomenological applications in heavy-ion collisions.
To comply with the {\tt LHAPDF} format, we have assumed that
$u=d$ and that $\bar{u}=\bar{d}=\bar{s}=s$, namely a symmetric quark sea.
With this convention, the
only meaningfully constrained quark
combinations can be reconstructed using the flavour basis PDFs by means of
$\Sigma=2\,u+4\,\bar{u}$ and $T_8 = 2(u-\bar{u})$.

\subsection*{\texttt{nNNPDF2.0} nPDF sets}
The \texttt{nNNPDF2.0} determination is available in the {\tt LHAPDF6}
library~\cite{Buckley:2014ana} for all relevant nuclei from $A=1$ to
$A=208$.
The \texttt{nNNPDF2.0} sets are available both for the nPDFs of bound protons,
$f^{(p/A)}(x,Q^2)$, and those of bound nucleons, $f^{(N/A)}(x,Q^2)$,
following the conventions in Chapter.~\ref{chap:nNNPDF10}.
Each of these sets is composed by $N_{\rm rep}=250$ correlated replicas,
see Sect.~5 of~\cite{AbdulKhalek:2019mzd} for their usage prescriptions.
The naming convention used for the sets is the following:\\

\begin{center}
  \renewcommand{\arraystretch}{1.15}
\begin{tabular}{ll}
   $\qquad\qquad f^{(N/A)}(x,Q^2)$ &  $\qquad\qquad f^{(p/A)}(x,Q^2)$ \\
{\tt nNNPDF20\_nlo\_as\_0118\_N1}     &  {\tt nNNPDF20\_nlo\_as\_0118\_N1} \\
{\tt nNNPDF20\_nlo\_as\_0118\_D2}     &  {\tt nNNPDF20\_nlo\_as\_0118\_p\_A2\_Z1}\\
{\tt nNNPDF20\_nlo\_as\_0118\_He4}    &  {\tt nNNPDF20\_nlo\_as\_0118\_p\_A4\_Z2}\\
{\tt nNNPDF20\_nlo\_as\_0118\_Li6}    &  {\tt nNNPDF20\_nlo\_as\_0118\_p\_A6\_Z3}\\
{\tt nNNPDF20\_nlo\_as\_0118\_Be9}    &  {\tt nNNPDF20\_nlo\_as\_0118\_p\_A9\_Z4}\\
{\tt nNNPDF20\_nlo\_as\_0118\_C12}    &  {\tt nNNPDF20\_nlo\_as\_0118\_p\_A12\_Z6}\\
{\tt nNNPDF20\_nlo\_as\_0118\_N14}    &  {\tt nNNPDF20\_nlo\_as\_0118\_p\_A14\_Z7}\\
{\tt nNNPDF20\_nlo\_as\_0118\_Al27}   &  {\tt nNNPDF20\_nlo\_as\_0118\_p\_A27\_Z13}\\
{\tt nNNPDF20\_nlo\_as\_0118\_Ca40}   &  {\tt nNNPDF20\_nlo\_as\_0118\_p\_A40\_Z20}\\
{\tt nNNPDF20\_nlo\_as\_0118\_Fe56}   &  {\tt nNNPDF20\_nlo\_as\_0118\_p\_A56\_Z26}\\
{\tt nNNPDF20\_nlo\_as\_0118\_Cu64}   &  {\tt nNNPDF20\_nlo\_as\_0118\_p\_A64\_Z29}\\
{\tt nNNPDF20\_nlo\_as\_0118\_Ag108}  &  {\tt nNNPDF20\_nlo\_as\_0118\_p\_A108\_Z47}\\
{\tt nNNPDF20\_nlo\_as\_0118\_Sn119}  &  {\tt nNNPDF20\_nlo\_as\_0118\_p\_A119\_Z50}\\
{\tt nNNPDF20\_nlo\_as\_0118\_Xe131}  &  {\tt nNNPDF20\_nlo\_as\_0118\_p\_A131\_Z54}\\
{\tt nNNPDF20\_nlo\_as\_0118\_Au197}  &  {\tt nNNPDF20\_nlo\_as\_0118\_p\_A197\_Z79}\\
{\tt nNNPDF20\_nlo\_as\_0118\_Pb208}  &  {\tt nNNPDF20\_nlo\_as\_0118\_p\_A208\_Z82}\\
\end{tabular}
\end{center}

\noindent
Supplementary sets such as:
\begin{center}
  \renewcommand{\arraystretch}{1.15}
\begin{tabular}{ll}
{\tt nNNPDF20\_nlo\_as\_0118\_O16}    &  {\tt nNNPDF20\_nlo\_as\_0118\_p\_A16\_Z8}\\
{\tt nNNPDF20\_nlo\_as\_0118\_W184}  &  {\tt nNNPDF20\_nlo\_as\_0118\_p\_A184\_Z74}\\
\end{tabular}
\end{center}
and additional variants of the \texttt{nNNPDF2.0} NLO fit present in this work, such as the fits
without the momentum and valence sum rules and a $N_{\rm rep}=1000$ replica set
for lead, are available on the NNPDF collaboration website:
\begin{center}
\url{http://nnpdf.mi.infn.it/for-users/nnnpdf2-0/}
\end{center}

\subsection*{\texttt{MAPFF1.0} FF sets and code}
The entirety of the results presented in Chapter~\ref{chap:FF} have been obtained
using the public code available from:
\begin{center}
{\tt\href{https://github.com/MapCollaboration/MontBlanc}{https://github.com/MapCollaboration/MontBlanc}.}
\end{center}
On this website it is possible to find some documentation concerning
the code usage as well as the FF sets in the {\tt LHAPDF} format. We
provide three sets of FFs with $N_{\rm rep}=200$ replicas each and of
their average for the positively and negatively charged pions and for
their sum. These are correspondingly called {\tt MAPFF10NLOPIp}, {\tt
  MAPFF10NLOPIm}, and {\tt MAPFF10NLOPIsum} and are made available
via the {\tt LHAPDF} interface~\cite{Buckley:2014ana} at:
\begin{center}
{\tt \href{https://lhapdf.hepforge.org/}{https://lhapdf.hepforge.org/}.}
\end{center}

\subsection*{High-Luminosity LHC and Large Hadron electron collider}
The results of Sect.~\ref{s2:PDF_HLLHC} and \ref{s2:PDF_LHeC} are made
publicly available in the {\tt LHAPDF6} format~\cite{Buckley:2014ana} by means
of {\tt Zenodo} data repository:
\begin{center}
\url{https://zenodo.org/record/3250580}
  \end{center}
Specifically, the following PDF sets can be obtained from this repository:
\begin{center}
{\tt PDF4LHC15\_nnlo\_lhec}\\
{\tt PDF4LHC15\_nnlo\_hllhc\_scen1\_lhec}\\
{\tt PDF4LHC15\_nnlo\_hllhc\_scen2\_lhec}\\
{\tt PDF4LHC15\_nnlo\_hllhc\_scen3\_lhec}
\end{center}
where the first set corresponds to the profiling of PDF4LHC15 using the
entire LHeC dataset listed in Table~\ref{tab:lhecdat}, and the other three
correspond to the simultaneous profiling with both the LHeC and HL--LHC
pseudo--data, for the three different projections of the experimental systematic
errors.
In addition,
in the same repository one can also find
the corresponding projections based only on the HL--LHC pseudo--data:
 \begin{center}
{\tt PDF4LHC15\_nnlo\_hllhc\_scen1}\\
{\tt PDF4LHC15\_nnlo\_hllhc\_scen2}\\
{\tt PDF4LHC15\_nnlo\_hllhc\_scen3}
\end{center}

\subsection*{Electron-Ion collider}
The PDF sets discussed in Sect.~\ref{s1:EIC} are available in the
LHAPDF format~\cite{Buckley:2014ana} from the NNPDF website:
\begin{center}
\url{http://nnpdf.mi.infn.it/for-users/nnnpdf2-0eic/}
\end{center}

\newpage
\thispagestyle{empty}
\end{appendices}

\begin{backmatter}
\newpage
\markboth{}{}
\thispagestyle{plain}

\begin{acknowledgements}
\phantomsection
\addcontentsline{toc}{chapter}{Acknowledgements}
\vspace{-0.25cm}
I am very grateful to my supervisor, Juan Rojo, for giving me the opportunity to carry out four amazing years of research between Nikhef and Vrije Universiteit Amsterdam. Thank you for giving me all the space I needed to learn, specially at the start of my PhD; 
for your guidance on what projects you thought were best for me to be involved in;
for supporting my autonomy 
and encouraging me to attend and give talks in plenty of different conferences and schools across Europe; Finally for your patience with my stubbornness at times but also for your trust in my work and abilities.
I would also like to extend my gratitude to my co-supervisor, Piet Mulders, for always being available for questions and for his enthusiasm in approaching them. Many thanks to you and Juan for your critical reading and invaluable feedback on this thesis.

I would like to express my appreciation to the doctoral committee members for the time they spent reading this thesis. In particular, I would like to thank Dani\"{e}l Boer and Marco van Leeuwen for their vital remarks, which made the content more accurate. 

I would like to thank the NNPDF collaboration, in particular, Stefano Forte, Stefano Carrazza, Juan, Luigi, Richard, Maria and Jos\'{e} Ignacio for having me as a member. Thank you for involving me in a multitude of interesting projects, enlightening discussions and meetings during which I acquired quite a lot of coding skills and knowledge of QCD, statistics and machine learning. 
I am also grateful to Emanuele, Zahari, Emma, Luca, Rosalyn, Tommaso, Cameron and Michael for all the interesting discussions we had, technical help and the great times we spent during the NNPDF conferences.

I am deeply indebted to my friends and colleagues: Nathan Hartland, Valerio Bertone, Emanuele Nocera and Jacob Ethier, for playing such a decisive role in my learning process and completing all the projects we collaborated on. Thank you for hours and hours of whiteboard discussions, answering my naive questions, debugging codes, valuable advices and last but not least for all the off-topic discussions, lunches, jokes and beers we had. I was really fortunate to witness how each of you approached your work and how as a result I learned how to face mine. 

Nikhef was truly one of the most stimulating places I had the opportunity to work at. Avoiding the risk of naming and forgetting anyone, I sincerely thank each and every former and current member of the Nikhef theory group whom I had the pleasure of meeting, discussing and working with. Special thanks to Wouter, Bert, Marieke, Jos, Piet and Jan-Willem for their physics insights during our conversations at Nikhef, and to the former and current group leaders Eric Laenen and Robert Fleischer for their efforts on ensuring an exceptional level of cohesion within the group. Thanks to Bob van Eijk and my supervisors for being members of my C3 committee and expressing genuine care about the proceeding of my PhD. I am particularly grateful to Pedro and Lorenzo for their effortless friendship and sharing with me some of the best times in Amsterdam, Turin and San Teodoro. Thanks to Jorinde for the lovely long talks and walks in Amsterdam and Hamburg. Thanks to Darren for his sharp sense of humour and extreme kindness. Thanks to Gilberto, Michi, Rhorry, Franz, Rudy and Anastasiia for all the great physics/philosophical discussions/debates as well as their positive vibes. Thanks to Solange, Jort, Ruben, Eleftheria, Avanish and Tom for the pleasant coffee-break discussions and their great company in many academic events. 

These past four years would not have been as meaningful without the presence and support of my family and friends; I am ever grateful to Hanaa, my mother and source of inspiration (these are all her art-works on the cover layout!), determination and unconditional love; Rameh, my father and source of curiosity, untirability and sensitivity; Waed, my sister and my go-to person to relax, laugh, feel loved and breathe in critical times. Her family: Ihab, Chadi and Yara, that never fails to lift me instantly from any low point I am going through; Samah, my brother, friend and true definition of unlimited support. The completion of my dissertation would not have been possible without you. His family: Chema and (?), that is excitingly about to increase by one; Ihab and Chema, the lovely and indispensable extension of our small family; Mohamad, my old and true friend who is always there to give me the push I need, to go on dangerous hikes, enjoy cities, coffees, talks...; Jessica, with whom I shared a lot of memorable moments and who was able to alter my meaning of life; Finally, my cousins, friends and family back in Lebanon who despite their endurance, always make room for our overwhelming and unforgettable reunions.

\end{acknowledgements}

%
\bibliographystyle{JHEP}
\bibliography{RAK_PhDthesis}

\providecommand{\href}[2]{#2}\begingroup\raggedright\begin{thebibliography}{100}

\bibitem{Popper1966-POPOCA}
K.~R. Popper, {\em Of Clouds and Clocks}.
\newblock St. Louis, Washington University, 1966.

\bibitem{kragh2002paul}
H.~Kragh, \textit{Paul dirac: seeking beauty},  {\em Physics world} \textbf{15}
  (2002), no.~8 27.

\bibitem{beresford2010medical}
M.~J. Beresford, \textit{Medical reductionism: lessons from the great
  philosophers},  {\em QJM: An International Journal of Medicine} \textbf{103}
  (2010), no.~9 721--724.

\bibitem{thackray1966origin}
A.~W. Thackray, \textit{The origin of dalton's chemical atomic theory:
  Daltonian doubts resolved},  {\em Isis} \textbf{57} (1966), no.~1 35--55.

\bibitem{rocke2005search}
A.~J. Rocke, \textit{In search of el dorado: John dalton and the origins of the
  atomic theory},  {\em Social research} (2005) 125--158.

\bibitem{mendelejew1869beziehungen}
D.~Mendelejew, \textit{{\"U}ber die beziehungen der eigenschaften zu den
  atomgewichten der elemente},  {\em Zeitschrift f{\"u}r Chemie} \textbf{12}
  (1869), no.~5 405--406.

\bibitem{prout1815relation}
W.~Prout and T.~Thomson, {\em On the relation between the specific gravities of
  bodies in their gaseous state and the weights of their atoms}.
\newblock 1815.

\bibitem{prout1816correction}
W.~Prout, \textit{Correction of a mistake in the essay on the relation between
  the specific gravities of bodies in their gaseous state and the weights of
  their atoms},  {\em Ann. Philos} \textbf{7} (1816) 111--113.

\bibitem{rutherford2010collision}
E.~Rutherford, \textit{{Collision of \ensuremath{\alpha} particles with light
  atoms. IV. An anomalous effect in nitrogen}},  {\em Phil. Mag. Ser. 6}
  \textbf{37} (1919) 581--587.

\bibitem{chadwick1932existence}
J.~Chadwick, \textit{The existence of a neutron},  {\em Proceedings of the
  Royal Society of London. Series A, Containing Papers of a Mathematical and
  Physical Character} \textbf{136} (1932), no.~830 692--708.

\bibitem{yukawa1935interaction}
H.~Yukawa, \textit{On the interaction of elementary particles. i},  {\em
  Proceedings of the Physico-Mathematical Society of Japan. 3rd Series}
  \textbf{17} (1935) 48--57.

\bibitem{yukawa1937interaction}
H.~YUKAWA and S.~SAKATA, \textit{On the interaction of elementary particles
  ii},  {\em Proceedings of the Physico-Mathematical Society of Japan. 3rd
  Series} \textbf{19} (1937) 1084--1093.

\bibitem{yukawa1938interaction}
H.~YUKAWA, S.~SAKATA, and M.~TAKETANI, \textit{On the interaction of elementary
  particles. iii},  {\em Proceedings of the Physico-Mathematical Society of
  Japan. 3rd Series} \textbf{20} (1938) 319--340.

\bibitem{frisch1933magnetische}
R.~Frisch and O.~Stern, \textit{{\"U}ber die magnetische ablenkung von
  wasserstoffmolek{\"u}len und das magnetische moment des protons. i},  {\em
  Zeitschrift f{\"u}r Physik} \textbf{85} (1933), no.~1-2 4--16.

\bibitem{bacher1933note}
R.~Bacher, \textit{Note on the magnetic moment of the nitrogen nucleus},  {\em
  Physical Review} \textbf{43} (1933), no.~12 1001.

\bibitem{alvarez1940quantitative}
L.~W. Alvarez and F.~Bloch, \textit{A quantitative determination of the neutron
  moment in absolute nuclear magnetons},  {\em Physical review} \textbf{57}
  (1940), no.~2 111.

\bibitem{cowan1964electron}
C.~L. Cowan, \textit{Electron scattering and nuclear and nucleon structure. a
  collection of reprints with an introduction. robert hofstadter.},  1964.

\bibitem{gell1964schematic}
M.~Gell-Mann, \textit{A schematic model of baryons and mesons},  {\em Physics
  Letters} \textbf{8} (1964), no.~3 214--215.

\bibitem{zweig1964}
G.~Zweig, \textit{An su $ \_3 $ model for strong interaction symmetry and its
  breaking},  tech. rep., CM-P00042884, 1964.

\bibitem{bloom1969high}
E.~D. Bloom, D.~Coward, H.~DeStaebler, J.~Drees, G.~Miller, L.~W. Mo, R.~E.
  Taylor, M.~Breidenbach, J.~I. Friedman, G.~C. Hartmann, et~al.,
  \textit{High-energy inelastic e- p scattering at 6 and 10},  {\em Physical
  Review Letters} \textbf{23} (1969), no.~16 930.

\bibitem{Blumlein:2012bf}
J.~Blumlein, \textit{{The Theory of Deeply Inelastic Scattering}},  {\em Prog.
  Part. Nucl. Phys.} \textbf{69} (2013) 28--84,
  [\href{http://arxiv.org/abs/1208.6087}{{\texttt{arXiv:1208.6087}}}].

\bibitem{bjorken1969asymptotic}
J.~D. Bjorken, \textit{Asymptotic sum rules at infinite momentum},  {\em
  Physical Review} \textbf{179} (1969), no.~5 1547.

\bibitem{callan1969high}
C.~G. Callan~Jr and D.~J. Gross, \textit{High-energy electroproduction and the
  constitution of the electric current},  {\em Physical Review Letters}
  \textbf{22} (1969), no.~4 156.

\bibitem{feynman1969very}
R.~P. Feynman, \textit{Very high-energy collisions of hadrons},  {\em Physical
  Review Letters} \textbf{23} (1969), no.~24 1415.

\bibitem{feynman2018photon}
R.~P. Feynman, {\em Photon-hadron interactions}.
\newblock CRC Press, 2018.

\bibitem{nambu1966preludes}
Y.~Nambu, \textit{Preludes in theoretical physics},  {\em in honor of VF
  Weisskopf} (1966).

\bibitem{faddeev1967feynman}
L.~D. Faddeev and V.~N. Popov, \textit{Feynman diagrams for the yang-mills
  field},  {\em Physics Letters B} \textbf{25} (1967), no.~1 29--30.

\bibitem{hooft1971renormalization}
G.~Hooft, \textit{Renormalization of massless yang-mills fields},  {\em Nuclear
  physics: B} \textbf{33} (1971), no.~1 173--199.

\bibitem{fritzsch2002current}
H.~Fritzsch and M.~Gell-Mann, \textit{Current algebra: Quarks and what else?},
  {\em arXiv preprint hep-ph/0208010} (2002).

\bibitem{fritzsch1973advantages}
H.~Fritzsch, M.~Gell-Mann, and H.~Leutwyler, \textit{Advantages of the color
  octet gluon picture},  {\em Physics Letters B} \textbf{47} (1973), no.~4
  365--368.

\bibitem{gross1973ultraviolet}
D.~J. Gross and F.~Wilczek, \textit{Ultraviolet behavior of non-abelian gauge
  theories},  {\em Physical Review Letters} \textbf{30} (1973), no.~26 1343.

\bibitem{politzer1973reliable}
H.~D. Politzer, \textit{Reliable perturbative results for strong
  interactions?},  {\em Physical Review Letters} \textbf{30} (1973), no.~26
  1346.

\bibitem{csaji2001approximation}
B.~C. Cs{\'a}ji et~al., \textit{Approximation with artificial neural networks},
   {\em Faculty of Sciences, Etvs Lornd University, Hungary} \textbf{24}
  (2001), no.~48 7.

\bibitem{peskin2018introduction}
M.~Peskin, {\em An introduction to quantum field theory}.
\newblock CRC press, 2018.

\bibitem{schwartz2014quantum}
M.~D. Schwartz, {\em Quantum field theory and the standard model}.
\newblock Cambridge University Press, 2014.

\bibitem{AbdulKhalek:2020uza}
R.~Abdul~Khalek and V.~Bertone, \textit{{On the derivatives of feed-forward
  neural networks}},
  \href{http://arxiv.org/abs/2005.07039}{{\texttt{arXiv:2005.07039}}}.

\bibitem{AbdulKhalek:2019ihb}
\textbf{NNPDF} Collaboration, R.~Abdul~Khalek et~al., \textit{{Parton
  Distributions with Theory Uncertainties: General Formalism and First
  Phenomenological Studies}},  {\em Eur. Phys. J. C} \textbf{79} (2019), no.~11
  931, [\href{http://arxiv.org/abs/1906.10698}{{\texttt{arXiv:1906.10698}}}].

\bibitem{AbdulKhalek:2019bux}
\textbf{NNPDF} Collaboration, R.~Abdul~Khalek et~al., \textit{{A first
  determination of parton distributions with theoretical uncertainties}},  {\em
  Eur. Phys. J.} \textbf{C} (2019) 79:838,
  [\href{http://arxiv.org/abs/1905.04311}{{\texttt{arXiv:1905.04311}}}].

\bibitem{AbdulKhalek:2020jut}
R.~Abdul~Khalek et~al., \textit{{Phenomenology of NNLO jet production at the
  LHC and its impact on parton distributions}},  {\em Eur. Phys. J. C}
  \textbf{80} (2020), no.~8 797,
  [\href{http://arxiv.org/abs/2005.11327}{{\texttt{arXiv:2005.11327}}}].

\bibitem{Khalek:2018bbv}
\textbf{NNPDF} Collaboration, R.~Abdul~Khalek, J.~J. Ethier, and J.~Rojo,
  \textit{{Nuclear Parton Distributions from Neural Networks}},  {\em Acta
  Phys. Polon. Supp.} \textbf{12} (2019), no.~4 927,
  [\href{http://arxiv.org/abs/1811.05858}{{\texttt{arXiv:1811.05858}}}].

\bibitem{AbdulKhalek:2019mzd}
\textbf{NNPDF} Collaboration, R.~Abdul~Khalek, J.~J. Ethier, and J.~Rojo,
  \textit{{Nuclear parton distributions from lepton-nucleus scattering and the
  impact of an electron-ion collider}},  {\em Eur. Phys. J. C} \textbf{79}
  (2019), no.~6 471,
  [\href{http://arxiv.org/abs/1904.00018}{{\texttt{arXiv:1904.00018}}}].

\bibitem{AbdulKhalek:2020yuc}
R.~Abdul~Khalek, J.~J. Ethier, J.~Rojo, and G.~van Weelden, \textit{{nNNPDF2.0:
  quark flavor separation in nuclei from LHC data}},  {\em JHEP} \textbf{09}
  (2020) 183,
  [\href{http://arxiv.org/abs/2006.14629}{{\texttt{arXiv:2006.14629}}}].

\bibitem{Khalek:2021gxf}
R.~A. Khalek, V.~Bertone, and E.~R. Nocera, \textit{{Determination of
  unpolarized pion fragmentation functions using semi-inclusive
  deep-inelastic-scattering data}},  {\em Phys. Rev. D} \textbf{104} (2021),
  no.~3 034007,
  [\href{http://arxiv.org/abs/2105.08725}{{\texttt{arXiv:2105.08725}}}].

\bibitem{Khalek:2018mdn}
R.~Abdul~Khalek, S.~Bailey, J.~Gao, L.~Harland-Lang, and J.~Rojo,
  \textit{{Towards Ultimate Parton Distributions at the High-Luminosity LHC}},
  {\em Eur. Phys. J. C} \textbf{78} (2018), no.~11 962,
  [\href{http://arxiv.org/abs/1810.03639}{{\texttt{arXiv:1810.03639}}}].

\bibitem{Cepeda:2019klc}
M.~Cepeda et~al., \textit{{Report from Working Group 2}: {Higgs Physics at the
  HL-LHC and HE-LHC}},  {\em CERN Yellow Rep. Monogr.} \textbf{7} (2019)
  221--584,
  [\href{http://arxiv.org/abs/1902.00134}{{\texttt{arXiv:1902.00134}}}].

\bibitem{Azzi:2019yne}
P.~Azzi et~al., \textit{{Report from Working Group 1}: {Standard Model Physics
  at the HL-LHC and HE-LHC}},  {\em CERN Yellow Rep. Monogr.} \textbf{7} (2019)
  1--220,
  [\href{http://arxiv.org/abs/1902.04070}{{\texttt{arXiv:1902.04070}}}].

\bibitem{AbdulKhalek:2019mps}
R.~Abdul~Khalek, S.~Bailey, J.~Gao, L.~Harland-Lang, and J.~Rojo,
  \textit{{Probing Proton Structure at the Large Hadron electron Collider}},
  {\em SciPost Phys.} \textbf{7} (2019), no.~4 051,
  [\href{http://arxiv.org/abs/1906.10127}{{\texttt{arXiv:1906.10127}}}].

\bibitem{AbdulKhalek:2021gbh}
R.~Abdul~Khalek et~al., \textit{{Science Requirements and Detector Concepts for
  the Electron-Ion Collider: EIC Yellow Report}},
  \href{http://arxiv.org/abs/2103.05419}{{\texttt{arXiv:2103.05419}}}.

\bibitem{Khalek:2021ulf}
R.~A. Khalek, J.~J. Ethier, E.~R. Nocera, and J.~Rojo, \textit{{Self-consistent
  determination of proton and nuclear PDFs at the Electron Ion Collider}},
  {\em Phys. Rev. D} \textbf{103} (2021), no.~9 096005,
  [\href{http://arxiv.org/abs/2102.00018}{{\texttt{arXiv:2102.00018}}}].

\bibitem{AbdulKhalek:2021xxx2}
R.~Abdul~Khalek, V.~Bertone, D.~Pitonyak, A.~Prokudin, and N.~Sato,
  \textit{{New ideas for impact studies at the EIC}},
  \href{http://arxiv.org/abs/In preparation}{{\texttt{In preparation}}}.

\bibitem{sliney2016light}
D.~Sliney, \textit{What is light? the visible spectrum and beyond},  {\em Eye}
  \textbf{30} (2016), no.~2 222--229.

\bibitem{CHOPPIN2002192}
G.~R. CHOPPIN, J.-O. LILJENZIN, and J.~RYDBERG, \textit{Chapter 8 - detection
  and measurement techniques},  in {\em Radiochemistry and Nuclear Chemistry
  (Third Edition)} (G.~R. CHOPPIN, J.-O. LILJENZIN, and J.~RYDBERG, eds.),
  pp.~192--238.
\newblock Butterworth-Heinemann, Woburn, third edition~ed., 2002.

\bibitem{anderson1933positive}
C.~D. Anderson, \textit{The positive electron},  {\em Physical Review}
  \textbf{43} (1933), no.~6 491.

\bibitem{street1937new}
J.~C. Street and E.~Stevenson, \textit{New evidence for the existence of a
  particle of mass intermediate between the proton and electron},  {\em
  Physical Review} \textbf{52} (1937), no.~9 1003.

\bibitem{Hofstadter:1953zjy}
R.~Hofstadter, H.~R. Fechter, and J.~A. McIntyre, \textit{{High-Energy Electron
  Scattering and Nuclear Structure Determinations}},  {\em Phys. Rev.}
  \textbf{92} (1953), no.~4 978.

\bibitem{PhysRev.98.217}
R.~Hofstadter and R.~W. McAllister, \textit{Electron scattering from the
  proton},  {\em Phys. Rev.} \textbf{98} (Apr, 1955) 217--218.

\bibitem{PhysRev.102.851}
R.~W. McAllister and R.~Hofstadter, \textit{Elastic scattering of 188-mev
  electrons from the proton and the alpha particle},  {\em Phys. Rev.}
  \textbf{102} (May, 1956) 851--856.

\bibitem{Collins:1989gx}
J.~C. Collins, D.~E. Soper, and G.~F. Sterman, \textit{{Factorization of Hard
  Processes in QCD}},  {\em Adv. Ser. Direct. High Energy Phys.} \textbf{5}
  (1989) 1--91,
  [\href{http://arxiv.org/abs/hep-ph/0409313}{{\texttt{hep-ph/0409313}}}].

\bibitem{Altarelli:1977zs}
G.~Altarelli and G.~Parisi, \textit{{Asymptotic Freedom in Parton Language}},
  {\em Nucl. Phys. B} \textbf{126} (1977) 298--318.

\bibitem{Gross:1973ju}
D.~J. Gross and F.~Wilczek, \textit{{Asymptotically Free Gauge Theories - I}},
  {\em Phys. Rev. D} \textbf{8} (1973) 3633--3652.

\bibitem{Georgi:1951sr}
H.~Georgi and H.~D. Politzer, \textit{{Electroproduction scaling in an
  asymptotically free theory of strong interactions}},  {\em Phys. Rev. D}
  \textbf{9} (1974) 416--420.

\bibitem{Floratos:1977au}
E.~G. Floratos, D.~A. Ross, and C.~T. Sachrajda, \textit{{Higher Order Effects
  in Asymptotically Free Gauge Theories: The Anomalous Dimensions of Wilson
  Operators}},  {\em Nucl. Phys. B} \textbf{129} (1977) 66--88. [Erratum:
  Nucl.Phys.B 139, 545--546 (1978)].

\bibitem{GonzalezArroyo:1979df}
A.~Gonzalez-Arroyo, C.~Lopez, and F.~J. Yndurain, \textit{{Second Order
  Contributions to the Structure Functions in Deep Inelastic Scattering. 1.
  Theoretical Calculations}},  {\em Nucl. Phys. B} \textbf{153} (1979)
  161--186.

\bibitem{Floratos:1978ny}
E.~G. Floratos, D.~A. Ross, and C.~T. Sachrajda, \textit{{Higher Order Effects
  in Asymptotically Free Gauge Theories. 2. Flavor Singlet Wilson Operators and
  Coefficient Functions}},  {\em Nucl. Phys. B} \textbf{152} (1979) 493--520.

\bibitem{Furmanski:1980cm}
W.~Furmanski and R.~Petronzio, \textit{{Singlet Parton Densities Beyond Leading
  Order}},  {\em Phys. Lett. B} \textbf{97} (1980) 437--442.

\bibitem{Curci:1980uw}
G.~Curci, W.~Furmanski, and R.~Petronzio, \textit{{Evolution of Parton
  Densities Beyond Leading Order: The Nonsinglet Case}},  {\em Nucl. Phys. B}
  \textbf{175} (1980) 27--92.

\bibitem{GonzalezArroyo:1979he}
A.~Gonzalez-Arroyo and C.~Lopez, \textit{{Second Order Contributions to the
  Structure Functions in Deep Inelastic Scattering. 3. The Singlet Case}},
  {\em Nucl. Phys. B} \textbf{166} (1980) 429--459.

\bibitem{Floratos:1981hs}
E.~G. Floratos, C.~Kounnas, and R.~Lacaze, \textit{{Higher Order QCD Effects in
  Inclusive Annihilation and Deep Inelastic Scattering}},  {\em Nucl. Phys. B}
  \textbf{192} (1981) 417--462.

\bibitem{Hamberg:1991qt}
R.~Hamberg and W.~L. van Neerven, \textit{{The Correct renormalization of the
  gluon operator in a covariant gauge}},  {\em Nucl. Phys. B} \textbf{379}
  (1992) 143--171.

\bibitem{Moch:2004pa}
S.~Moch, J.~A.~M. Vermaseren, and A.~Vogt, \textit{{The Three loop splitting
  functions in QCD: The Nonsinglet case}},  {\em Nucl. Phys. B} \textbf{688}
  (2004) 101--134,
  [\href{http://arxiv.org/abs/hep-ph/0403192}{{\texttt{hep-ph/0403192}}}].

\bibitem{Vogt:2004mw}
A.~Vogt, S.~Moch, and J.~A.~M. Vermaseren, \textit{{The Three-loop splitting
  functions in QCD: The Singlet case}},  {\em Nucl. Phys. B} \textbf{691}
  (2004) 129--181,
  [\href{http://arxiv.org/abs/hep-ph/0404111}{{\texttt{hep-ph/0404111}}}].

\bibitem{Sheiman:1979ku}
J.~Sheiman, \textit{{Target Mass Corrections in the {QCD} Parton Model}},  {\em
  Nucl. Phys. B} \textbf{171} (1980) 445--470.

\bibitem{Ellis:1978ty}
R.~K. Ellis, H.~Georgi, M.~Machacek, H.~D. Politzer, and G.~G. Ross,
  \textit{{Perturbation Theory and the Parton Model in QCD}},  {\em Nucl. Phys.
  B} \textbf{152} (1979) 285--329.

\bibitem{Schienbein:2007gr}
I.~Schienbein et~al., \textit{{A Review of Target Mass Corrections}},  {\em J.
  Phys. G} \textbf{35} (2008) 053101,
  [\href{http://arxiv.org/abs/0709.1775}{{\texttt{arXiv:0709.1775}}}].

\bibitem{Moffat:2019qll}
E.~Moffat, T.~C. Rogers, W.~Melnitchouk, N.~Sato, and F.~Steffens,
  \textit{{What does kinematical target mass sensitivity in DIS reveal about
  hadron structure?}},  {\em Phys. Rev. D} \textbf{99} (2019), no.~9 096008,
  [\href{http://arxiv.org/abs/1901.09016}{{\texttt{arXiv:1901.09016}}}].

\bibitem{Bertone:2013vaa}
V.~Bertone, S.~Carrazza, and J.~Rojo, \textit{{APFEL: A PDF Evolution Library
  with QED corrections}},  {\em Comput.Phys.Commun.} \textbf{185} (2014) 1647,
  [\href{http://arxiv.org/abs/1310.1394}{{\texttt{arXiv:1310.1394}}}].

\bibitem{Hartland:2014nha}
N.~Hartland, \textit{{Proton structure at the LHC}},  other thesis, 11, 2014.

\bibitem{Buza:1995ie}
M.~Buza, Y.~Matiounine, J.~Smith, R.~Migneron, and W.~L. van Neerven,
  \textit{{Heavy quark coefficient functions at asymptotic values $Q^2 \gg
  m^2$}},  {\em Nucl. Phys. B} \textbf{472} (1996) 611--658,
  [\href{http://arxiv.org/abs/hep-ph/9601302}{{\texttt{hep-ph/9601302}}}].

\bibitem{Buza:1996wv}
M.~Buza, Y.~Matiounine, J.~Smith, and W.~L. van Neerven, \textit{{Charm
  electroproduction viewed in the variable flavor number scheme versus fixed
  order perturbation theory}},  {\em Eur. Phys. J. C} \textbf{1} (1998)
  301--320,
  [\href{http://arxiv.org/abs/hep-ph/9612398}{{\texttt{hep-ph/9612398}}}].

\bibitem{Cacciari:1998it}
M.~Cacciari, M.~Greco, and P.~Nason, \textit{{The P(T) spectrum in heavy flavor
  hadroproduction}},  {\em JHEP} \textbf{05} (1998) 007,
  [\href{http://arxiv.org/abs/hep-ph/9803400}{{\texttt{hep-ph/9803400}}}].

\bibitem{Forte:2010ta}
S.~Forte, E.~Laenen, P.~Nason, and J.~Rojo, \textit{{Heavy quarks in
  deep-inelastic scattering}},  {\em Nucl. Phys. B} \textbf{834} (2010)
  116--162,
  [\href{http://arxiv.org/abs/1001.2312}{{\texttt{arXiv:1001.2312}}}].

\bibitem{Thorne:1997uu}
R.~S. Thorne and R.~G. Roberts, \textit{{A Practical procedure for evolving
  heavy flavor structure functions}},  {\em Phys. Lett. B} \textbf{421} (1998)
  303--311,
  [\href{http://arxiv.org/abs/hep-ph/9711223}{{\texttt{hep-ph/9711223}}}].

\bibitem{Thorne:1997ga}
R.~S. Thorne and R.~G. Roberts, \textit{{An Ordered analysis of heavy flavor
  production in deep inelastic scattering}},  {\em Phys. Rev. D} \textbf{57}
  (1998) 6871--6898,
  [\href{http://arxiv.org/abs/hep-ph/9709442}{{\texttt{hep-ph/9709442}}}].

\bibitem{Thorne:2008xf}
R.~S. Thorne and W.~K. Tung, \textit{{PQCD Formulations with Heavy Quark Masses
  and Global Analysis}},  in {\em {HERA and the LHC: 4th Workshop on the
  Implications of HERA for LHC Physics}}, 9, 2008.
\newblock \href{http://arxiv.org/abs/0809.0714}{{\texttt{arXiv:0809.0714}}}.

\bibitem{Kramer:2000hn}
M.~Kr\"amer, F.~I. Olness, and D.~E. Soper, \textit{{Treatment of heavy quarks
  in deeply inelastic scattering}},  {\em Phys. Rev. D} \textbf{62} (2000)
  096007,
  [\href{http://arxiv.org/abs/hep-ph/0003035}{{\texttt{hep-ph/0003035}}}].

\bibitem{Collins:1978wz}
J.~C. Collins, F.~Wilczek, and A.~Zee, \textit{{Low-Energy Manifestations of
  Heavy Particles: Application to the Neutral Current}},  {\em Phys. Rev. D}
  \textbf{18} (1978) 242.

\bibitem{Collins:1998rz}
J.~C. Collins, \textit{{Hard scattering factorization with heavy quarks: A
  General treatment}},  {\em Phys. Rev. D} \textbf{58} (1998) 094002,
  [\href{http://arxiv.org/abs/hep-ph/9806259}{{\texttt{hep-ph/9806259}}}].

\bibitem{Webber:1999ui}
B.~R. Webber, \textit{{Fragmentation and hadronization}},  {\em Int. J. Mod.
  Phys. A} \textbf{15S1} (2000) 577--606,
  [\href{http://arxiv.org/abs/hep-ph/9912292}{{\texttt{hep-ph/9912292}}}].

\bibitem{Metz:2016swz}
A.~Metz and A.~Vossen, \textit{{Parton Fragmentation Functions}},  {\em Prog.
  Part. Nucl. Phys.} \textbf{91} (2016) 136--202,
  [\href{http://arxiv.org/abs/1607.02521}{{\texttt{arXiv:1607.02521}}}].

\bibitem{Bertone:2017gds}
V.~Bertone, \textit{{APFEL++: A new PDF evolution library in C++}},  {\em PoS}
  \textbf{DIS2017} (2018) 201,
  [\href{http://arxiv.org/abs/1708.00911}{{\texttt{arXiv:1708.00911}}}].

\bibitem{Aubert:1983xm}
\textbf{European Muon} Collaboration, J.~J. Aubert et~al., \textit{{The ratio
  of the nucleon structure functions $F2_n$ for iron and deuterium}},  {\em
  Phys. Lett. B} \textbf{123} (1983) 275--278.

\bibitem{Malace:2014uea}
S.~Malace, D.~Gaskell, D.~W. Higinbotham, and I.~Cloet, \textit{{The Challenge
  of the EMC Effect: existing data and future directions}},  {\em Int. J. Mod.
  Phys. E} \textbf{23} (2014), no.~08 1430013,
  [\href{http://arxiv.org/abs/1405.1270}{{\texttt{arXiv:1405.1270}}}].

\bibitem{Frankfurt:2012qs}
L.~Frankfurt and M.~Strikman, \textit{{QCD and QED dynamics in the EMC
  effect}},  {\em Int. J. Mod. Phys. E} \textbf{21} (2012) 1230002,
  [\href{http://arxiv.org/abs/1203.5278}{{\texttt{arXiv:1203.5278}}}].

\bibitem{PhysRevLett.103.202301}
J.~Seely et~al., \textit{New measurements of the european muon collaboration
  effect in very light nuclei},  {\em Phys. Rev. Lett.} \textbf{103} (Nov,
  2009) 202301.

\bibitem{Close:1989ca}
F.~E. Close, J.-w. Qiu, and R.~G. Roberts, \textit{{{QCD} Parton Recombination
  and Applications to Nuclear Structure Functions}},  {\em Phys. Rev. D}
  \textbf{40} (1989) 2820.

\bibitem{Catani:1996vz}
S.~Catani and M.~H. Seymour, \textit{{A General algorithm for calculating jet
  cross-sections in NLO QCD}},  {\em Nucl. Phys. B} \textbf{485} (1997)
  291--419,
  [\href{http://arxiv.org/abs/hep-ph/9605323}{{\texttt{hep-ph/9605323}}}].
  [Erratum: Nucl.Phys.B 510, 503--504 (1998)].

\bibitem{Bertone:2017tyb}
\textbf{NNPDF} Collaboration, V.~Bertone, S.~Carrazza, N.~P. Hartland, E.~R.
  Nocera, and J.~Rojo, \textit{{A determination of the fragmentation functions
  of pions, kaons, and protons with faithful uncertainties}},  {\em Eur. Phys.
  J. C} \textbf{77} (2017), no.~8 516,
  [\href{http://arxiv.org/abs/1706.07049}{{\texttt{arXiv:1706.07049}}}].

\bibitem{Campbell:2015qma}
J.~M. Campbell, R.~K. Ellis, and W.~T. Giele, \textit{{A Multi-Threaded Version
  of MCFM}},  {\em Eur. Phys. J. C} \textbf{75} (2015), no.~6 246,
  [\href{http://arxiv.org/abs/1503.06182}{{\texttt{arXiv:1503.06182}}}].

\bibitem{Boughezal:2016wmq}
R.~Boughezal, J.~M. Campbell, R.~K. Ellis, C.~Focke, W.~Giele, X.~Liu,
  F.~Petriello, and C.~Williams, \textit{{Color singlet production at NNLO in
  MCFM}},  {\em Eur. Phys. J. C} \textbf{77} (2017), no.~1 7,
  [\href{http://arxiv.org/abs/1605.08011}{{\texttt{arXiv:1605.08011}}}].

\bibitem{Gehrmann:2018szu}
T.~Gehrmann et~al., \textit{{Jet cross sections and transverse momentum
  distributions with NNLOJET}},  {\em PoS} \textbf{RADCOR2017} (2018) 074,
  [\href{http://arxiv.org/abs/1801.06415}{{\texttt{arXiv:1801.06415}}}].

\bibitem{Wobisch:2011ij}
\textbf{fastNLO} Collaboration, M.~Wobisch, D.~Britzger, T.~Kluge, K.~Rabbertz,
  and F.~Stober, \textit{{Theory-Data Comparisons for Jet Measurements in
  Hadron-Induced Processes}},
  \href{http://arxiv.org/abs/1109.1310}{{\texttt{arXiv:1109.1310}}}.

\bibitem{Carli:2010rw}
T.~Carli et~al., \textit{{A posteriori inclusion of parton density functions in
  NLO QCD final-state calculations at hadron colliders: The APPLGRID Project}},
   {\em Eur.Phys.J.} \textbf{C66} (2010) 503,
  [\href{http://arxiv.org/abs/0911.2985}{{\texttt{arXiv:0911.2985}}}].

\bibitem{Bertone:2016lga}
V.~Bertone, S.~Carrazza, and N.~P. Hartland, \textit{{APFELgrid: a high
  performance tool for parton density determinations}},  {\em Comput. Phys.
  Commun.} \textbf{212} (2017) 205--209,
  [\href{http://arxiv.org/abs/1605.02070}{{\texttt{arXiv:1605.02070}}}].

\bibitem{Campbell:2019dru}
J.~Campbell and T.~Neumann, \textit{{Precision Phenomenology with MCFM}},  {\em
  JHEP} \textbf{12} (2019) 034,
  [\href{http://arxiv.org/abs/1909.09117}{{\texttt{arXiv:1909.09117}}}].

\bibitem{Campbell:2006wx}
J.~M. Campbell, J.~W. Huston, and W.~J. Stirling, \textit{{Hard Interactions of
  Quarks and Gluons: A Primer for LHC Physics}},  {\em Rept. Prog. Phys.}
  \textbf{70} (2007) 89,
  [\href{http://arxiv.org/abs/hep-ph/0611148}{{\texttt{hep-ph/0611148}}}].

\bibitem{gelman2013bayesian}
A.~Gelman, J.~B. Carlin, H.~S. Stern, D.~B. Dunson, A.~Vehtari, and D.~B.
  Rubin, {\em Bayesian data analysis}.
\newblock CRC press, 2013.

\bibitem{mcelreath2020statistical}
R.~McElreath, {\em Statistical rethinking: A Bayesian course with examples in R
  and Stan}.
\newblock CRC press, 2020.

\bibitem{von1981probability}
R.~Von~Mises, {\em Probability, statistics, and truth}.
\newblock Courier Corporation, 1981.

\bibitem{Gao:2017yyd}
J.~Gao, L.~Harland-Lang, and J.~Rojo, \textit{{The Structure of the Proton in
  the LHC Precision Era}},  {\em Phys. Rept.} \textbf{742} (2018) 1--121,
  [\href{http://arxiv.org/abs/1709.04922}{{\texttt{arXiv:1709.04922}}}].

\bibitem{DelDebbio:2004xtd}
\textbf{NNPDF} Collaboration, L.~Del~Debbio, S.~Forte, J.~I. Latorre,
  A.~Piccione, and J.~Rojo, \textit{{Unbiased determination of the proton
  structure function F(2)**p with faithful uncertainty estimation}},  {\em
  JHEP} \textbf{03} (2005) 080,
  [\href{http://arxiv.org/abs/hep-ph/0501067}{{\texttt{hep-ph/0501067}}}].

\bibitem{Ball:2008by}
\textbf{NNPDF} Collaboration, R.~D. Ball, L.~Del~Debbio, S.~Forte, A.~Guffanti,
  J.~I. Latorre, A.~Piccione, J.~Rojo, and M.~Ubiali, \textit{{A Determination
  of parton distributions with faithful uncertainty estimation}},  {\em Nucl.
  Phys. B} \textbf{809} (2009) 1--63,
  [\href{http://arxiv.org/abs/0808.1231}{{\texttt{arXiv:0808.1231}}}].
  [Erratum: Nucl.Phys.B 816, 293 (2009)].

\bibitem{Ball:2009qv}
\textbf{NNPDF} Collaboration, R.~D. Ball, L.~Del~Debbio, S.~Forte, A.~Guffanti,
  J.~I. Latorre, J.~Rojo, and M.~Ubiali, \textit{{Fitting Parton Distribution
  Data with Multiplicative Normalization Uncertainties}},  {\em JHEP}
  \textbf{05} (2010) 075,
  [\href{http://arxiv.org/abs/0912.2276}{{\texttt{arXiv:0912.2276}}}].

\bibitem{Ball:2014uwa}
\textbf{NNPDF} Collaboration, R.~D. Ball et~al., \textit{{Parton distributions
  for the LHC Run II}},  {\em JHEP} \textbf{04} (2015) 040,
  [\href{http://arxiv.org/abs/1410.8849}{{\texttt{arXiv:1410.8849}}}].

\bibitem{Ball:2017nwa}
\textbf{NNPDF} Collaboration, R.~D. Ball et~al., \textit{{Parton distributions
  from high-precision collider data}},  {\em Eur. Phys. J.} \textbf{C77}
  (2017), no.~10 663,
  [\href{http://arxiv.org/abs/1706.00428}{{\texttt{arXiv:1706.00428}}}].

\bibitem{Sinervo:2003wm}
P.~Sinervo, \textit{{Definition and Treatment of Systematic Uncertainties in
  High Energy Physics and Astrophysics}},  {\em eConf} \textbf{C030908} (2003)
  TUAT004.

\bibitem{DAgostini:1993arp}
G.~D'Agostini, \textit{{On the use of the covariance matrix to fit correlated
  data}},  {\em Nucl. Instrum. Meth. A} \textbf{346} (1994) 306--311.

\bibitem{Ball:2013lla}
\textbf{NNPDF} Collaboration, R.~D. Ball, S.~Forte, A.~Guffanti, E.~R. Nocera,
  G.~Ridolfi, and J.~Rojo, \textit{{Unbiased determination of polarized parton
  distributions and their uncertainties}},  {\em Nucl. Phys. B} \textbf{874}
  (2013) 36--84,
  [\href{http://arxiv.org/abs/1303.7236}{{\texttt{arXiv:1303.7236}}}].

\bibitem{Nocera:2014gqa}
\textbf{NNPDF Collaboration} Collaboration, E.~R. Nocera, R.~D. Ball, S.~Forte,
  G.~Ridolfi, and J.~Rojo, \textit{{A first unbiased global determination of
  polarized PDFs and their uncertainties}},  {\em Nucl.Phys.} \textbf{B887}
  (2014) 276,
  [\href{http://arxiv.org/abs/1406.5539}{{\texttt{arXiv:1406.5539}}}].

\bibitem{Bertone:2018ecm}
\textbf{NNPDF} Collaboration, V.~Bertone, N.~P. Hartland, E.~R. Nocera,
  J.~Rojo, and L.~Rottoli, \textit{{Charged hadron fragmentation functions from
  collider data}},  {\em Eur. Phys. J.} \textbf{C78} (2018), no.~8 651,
  [\href{http://arxiv.org/abs/1807.03310}{{\texttt{arXiv:1807.03310}}}].

\bibitem{Sato:2016tuz}
\textbf{Jefferson Lab Angular Momentum} Collaboration, N.~Sato, W.~Melnitchouk,
  S.~E. Kuhn, J.~J. Ethier, and A.~Accardi, \textit{{Iterative Monte Carlo
  analysis of spin-dependent parton distributions}},  {\em Phys. Rev. D}
  \textbf{93} (2016), no.~7 074005,
  [\href{http://arxiv.org/abs/1601.07782}{{\texttt{arXiv:1601.07782}}}].

\bibitem{Sato:2016wqj}
N.~Sato, J.~J. Ethier, W.~Melnitchouk, M.~Hirai, S.~Kumano, and A.~Accardi,
  \textit{{First Monte Carlo analysis of fragmentation functions from
  single-inclusive $e^+ e^-$ annihilation}},  {\em Phys. Rev. D} \textbf{94}
  (2016), no.~11 114004,
  [\href{http://arxiv.org/abs/1609.00899}{{\texttt{arXiv:1609.00899}}}].

\bibitem{Ethier:2017zbq}
J.~J. Ethier, N.~Sato, and W.~Melnitchouk, \textit{{First simultaneous
  extraction of spin-dependent parton distributions and fragmentation functions
  from a global QCD analysis}},  {\em Phys. Rev. Lett.} \textbf{119} (2017),
  no.~13 132001,
  [\href{http://arxiv.org/abs/1705.05889}{{\texttt{arXiv:1705.05889}}}].

\bibitem{Barry:2018ort}
P.~C. Barry, N.~Sato, W.~Melnitchouk, and C.-R. Ji, \textit{{First Monte Carlo
  Global QCD Analysis of Pion Parton Distributions}},  {\em Phys. Rev. Lett.}
  \textbf{121} (2018), no.~15 152001,
  [\href{http://arxiv.org/abs/1804.01965}{{\texttt{arXiv:1804.01965}}}].

\bibitem{Moutarde:2019tqa}
H.~Moutarde, P.~Sznajder, and J.~Wagner, \textit{{Unbiased determination of
  DVCS Compton Form Factors}},  {\em Eur. Phys. J. C} \textbf{79} (2019), no.~7
  614, [\href{http://arxiv.org/abs/1905.02089}{{\texttt{arXiv:1905.02089}}}].

\bibitem{Cuic:2020iwt}
M.~\v{C}ui\'c, K.~Kumeri\v{c}ki, and A.~Sch\"afer, \textit{{Separation of Quark
  Flavors Using Deeply Virtual Compton Scattering Data}},  {\em Phys. Rev.
  Lett.} \textbf{125} (2020), no.~23 232005,
  [\href{http://arxiv.org/abs/2007.00029}{{\texttt{arXiv:2007.00029}}}].

\bibitem{press2007numerical}
W.~H. Press, H.~William, S.~A. Teukolsky, W.~T. Vetterling, A.~Saul, and B.~P.
  Flannery, {\em Numerical recipes 3rd edition: The art of scientific
  computing}.
\newblock Cambridge university press, 2007.

\bibitem{Ball:2011gg}
R.~D. Ball, V.~Bertone, F.~Cerutti, L.~Del~Debbio, S.~Forte, A.~Guffanti, N.~P.
  Hartland, J.~I. Latorre, J.~Rojo, and M.~Ubiali, \textit{{Reweighting and
  Unweighting of Parton Distributions and the LHC W lepton asymmetry data}},
  {\em Nucl. Phys. B} \textbf{855} (2012) 608--638,
  [\href{http://arxiv.org/abs/1108.1758}{{\texttt{arXiv:1108.1758}}}].

\bibitem{Pumplin:2001ct}
J.~Pumplin, D.~Stump, R.~Brock, D.~Casey, J.~Huston, J.~Kalk, H.~L. Lai, and
  W.~K. Tung, \textit{{Uncertainties of predictions from parton distribution
  functions. 2. The Hessian method}},  {\em Phys. Rev. D} \textbf{65} (2001)
  014013,
  [\href{http://arxiv.org/abs/hep-ph/0101032}{{\texttt{hep-ph/0101032}}}].

\bibitem{Paukkunen:2014zia}
H.~Paukkunen and P.~Zurita, \textit{{PDF reweighting in the Hessian matrix
  approach}},  {\em JHEP} \textbf{12} (2014) 100,
  [\href{http://arxiv.org/abs/1402.6623}{{\texttt{arXiv:1402.6623}}}].

\bibitem{Carrazza:2015aoa}
S.~Carrazza, S.~Forte, Z.~Kassabov, J.~I. Latorre, and J.~Rojo, \textit{An
  unbiased hessian representation for monte carlo pdfs},  {\em Eur. Phys. J.}
  \textbf{C75} (2015), no.~8 369,
  [\href{http://arxiv.org/abs/1505.06736}{{\texttt{arXiv:1505.06736}}}].

\bibitem{Schmidt:2018hvu}
C.~Schmidt, J.~Pumplin, C.~P. Yuan, and P.~Yuan, \textit{{Updating and
  Optimizing Error PDFs in the Hessian Approach}},
  \href{http://arxiv.org/abs/1806.07950}{{\texttt{arXiv:1806.07950}}}.

\bibitem{Albertsson:2018maf}
K.~Albertsson et~al., \textit{{Machine Learning in High Energy Physics
  Community White Paper}},  {\em J. Phys. Conf. Ser.} \textbf{1085} (2018),
  no.~2 022008,
  [\href{http://arxiv.org/abs/1807.02876}{{\texttt{arXiv:1807.02876}}}].

\bibitem{Schwartz:2021ftp}
M.~D. Schwartz, \textit{{Modern Machine Learning and Particle Physics}},
  \href{http://arxiv.org/abs/2103.12226}{{\texttt{arXiv:2103.12226}}}.

\bibitem{van2017neural}
F.~Van~Veen, \textit{Neural network zoo prequel: cells and layers},  {\em
  Retried from https://www. asimovinstitute. org/author/fjodorvanveen} (2017).

\bibitem{Butterworth:2015oua}
J.~Butterworth et~al., \textit{{PDF4LHC recommendations for LHC Run II}},  {\em
  J. Phys. G} \textbf{43} (2016) 023001,
  [\href{http://arxiv.org/abs/1510.03865}{{\texttt{arXiv:1510.03865}}}].

\bibitem{Carrazza:2019mzf}
S.~Carrazza and J.~Cruz-Martinez, \textit{{Towards a new generation of parton
  densities with deep learning models}},  {\em Eur. Phys. J. C} \textbf{79}
  (2019), no.~8 676,
  [\href{http://arxiv.org/abs/1907.05075}{{\texttt{arXiv:1907.05075}}}].

\bibitem{kingma2017adam}
D.~P. Kingma and J.~Ba, \textit{Adam: A method for stochastic optimization},
  2017.

\bibitem{tensorflow2015-whitepaper}
M.~Abadi, A.~Agarwal, et~al., \textit{{TensorFlow}: Large-scale machine
  learning on heterogeneous systems},  2015.
\newblock Software available from tensorflow.org.

\bibitem{levenberg1944method}
K.~Levenberg, \textit{A method for the solution of certain non-linear problems
  in least squares},  {\em Quarterly of applied mathematics} \textbf{2} (1944),
  no.~2 164--168.

\bibitem{marquardt1963algorithm}
D.~W. Marquardt, \textit{An algorithm for least-squares estimation of nonlinear
  parameters},  {\em Journal of the society for Industrial and Applied
  Mathematics} \textbf{11} (1963), no.~2 431--441.

\bibitem{ceres-solver}
S.~Agarwal, K.~Mierle, and Others, ``Ceres solver.''
  \url{http://ceres-solver.org}.

\bibitem{yang2020nature}
X.-S. Yang, {\em Nature-inspired optimization algorithms}.
\newblock Academic Press, 2020.

\bibitem{whitley1994genetic}
D.~Whitley, \textit{A genetic algorithm tutorial},  {\em Statistics and
  computing} \textbf{4} (1994), no.~2 65--85.

\bibitem{AbdulKhalek:2021xxx1}
R.~Abdul~Khalek, T.~Giani, E.~R. Nocera, and J.~Rojo, \textit{{nNNPDF3.0: A
  global analysis of nuclear parton distributions at NNLO}},
  \href{http://arxiv.org/abs/In preparation}{{\texttt{In preparation}}}.

\bibitem{VANDENBERGHEN2005157}
F.~{Vanden Berghen} and H.~Bersini, \textit{Condor, a new parallel, constrained
  extension of powell's uobyqa algorithm: Experimental results and comparison
  with the dfo algorithm},  {\em Journal of Computational and Applied
  Mathematics} \textbf{181} (2005), no.~1 157--175.

\bibitem{gavin2019levenberg}
H.~P. Gavin, \textit{The levenberg-marquardt algorithm for nonlinear least
  squares curve-fitting problems},  {\em Department of Civil and Environmental
  Engineering, Duke University http://people. duke. edu/\~{} hpgavin/ce281/lm.
  pdf} (2019) 1--19.

\bibitem{Ethier:2020way}
J.~J. Ethier and E.~R. Nocera, \textit{{Parton Distributions in Nucleons and
  Nuclei}},  {\em Ann. Rev. Nucl. Part. Sci.} \textbf{70} (2020) 43--76,
  [\href{http://arxiv.org/abs/2001.07722}{{\texttt{arXiv:2001.07722}}}].

\bibitem{Forte:2020yip}
S.~Forte and S.~Carrazza, \textit{{Parton distribution functions}},
  \href{http://arxiv.org/abs/2008.12305}{{\texttt{arXiv:2008.12305}}}.

\bibitem{deFlorian:2016spz}
\textbf{LHC Higgs Cross Section Working Group} Collaboration, D.~de~Florian
  et~al., \textit{{Handbook of LHC Higgs Cross Sections: 4. Deciphering the
  Nature of the Higgs Sector}},
  \href{http://arxiv.org/abs/1610.07922}{{\texttt{arXiv:1610.07922}}}.

\bibitem{Beenakker:2015rna}
W.~Beenakker, C.~Borschensky, M.~Kr\"amer, A.~Kulesza, E.~Laenen, S.~Marzani,
  and J.~Rojo, \textit{{NLO+NLL squark and gluino production cross-sections
  with threshold-improved parton distributions}},  {\em Eur. Phys. J. C}
  \textbf{76} (2016), no.~2 53,
  [\href{http://arxiv.org/abs/1510.00375}{{\texttt{arXiv:1510.00375}}}].

\bibitem{Ball:2018twp}
\textbf{NNPDF} Collaboration, R.~D. Ball, E.~R. Nocera, and R.~L. Pearson,
  \textit{{Nuclear Uncertainties in the Determination of Proton PDFs}},  {\em
  Eur. Phys. J.} \textbf{C79} (2019), no.~3 282,
  [\href{http://arxiv.org/abs/1812.09074}{{\texttt{arXiv:1812.09074}}}].

\bibitem{Ball:2020xqw}
R.~D. Ball, E.~R. Nocera, and R.~L. Pearson, \textit{{Deuteron Uncertainties in
  the Determination of Proton PDFs}},  {\em Eur. Phys. J. C} \textbf{81}
  (2021), no.~1 37,
  [\href{http://arxiv.org/abs/2011.00009}{{\texttt{arXiv:2011.00009}}}].

\bibitem{Bozzi:2015hha}
G.~Bozzi, L.~Citelli, and A.~Vicini, \textit{{Parton density function
  uncertainties on the W boson mass measurement from the lepton transverse
  momentum distribution}},  {\em Phys. Rev. D} \textbf{91} (2015), no.~11
  113005,
  [\href{http://arxiv.org/abs/1501.05587}{{\texttt{arXiv:1501.05587}}}].

\bibitem{Bozzi:2015zja}
G.~Bozzi, L.~Citelli, M.~Vesterinen, and A.~Vicini, \textit{{Prospects for
  improving the LHC W boson mass measurement with forward muons}},  {\em Eur.
  Phys. J. C} \textbf{75} (2015), no.~12 601,
  [\href{http://arxiv.org/abs/1508.06954}{{\texttt{arXiv:1508.06954}}}].

\bibitem{Bozzi:2011ww}
G.~Bozzi, J.~Rojo, and A.~Vicini, \textit{{The Impact of PDF uncertainties on
  the measurement of the W boson mass at the Tevatron and the LHC}},  {\em
  Phys. Rev. D} \textbf{83} (2011) 113008,
  [\href{http://arxiv.org/abs/1104.2056}{{\texttt{arXiv:1104.2056}}}].

\bibitem{Aaboud:2017svj}
\textbf{ATLAS} Collaboration, M.~Aaboud et~al., \textit{{Measurement of the
  $W$-boson mass in pp collisions at $\sqrt{s}=7$ TeV with the ATLAS
  detector}},  {\em Eur. Phys. J. C} \textbf{78} (2018), no.~2 110,
  [\href{http://arxiv.org/abs/1701.07240}{{\texttt{arXiv:1701.07240}}}].
  [Erratum: Eur.Phys.J.C 78, 898 (2018)].

\bibitem{Khachatryan:2014waa}
\textbf{CMS} Collaboration, V.~Khachatryan et~al., \textit{{Constraints on
  parton distribution functions and extraction of the strong coupling constant
  from the inclusive jet cross section in pp collisions at $\sqrt{s} = 7$
  $\,\text {TeV}$}},  {\em Eur. Phys. J. C} \textbf{75} (2015), no.~6 288,
  [\href{http://arxiv.org/abs/1410.6765}{{\texttt{arXiv:1410.6765}}}].

\bibitem{Aaboud:2017fml}
\textbf{ATLAS} Collaboration, M.~Aaboud et~al., \textit{{Determination of the
  strong coupling constant $\alpha _\mathrm {s}$ from transverse
  energy\textendash{}energy correlations in multijet events at $\sqrt{s} =
  8~\hbox {TeV}$ using the ATLAS detector}},  {\em Eur. Phys. J. C} \textbf{77}
  (2017), no.~12 872,
  [\href{http://arxiv.org/abs/1707.02562}{{\texttt{arXiv:1707.02562}}}].

\bibitem{Chatrchyan:2013txa}
\textbf{CMS} Collaboration, S.~Chatrchyan et~al., \textit{{Measurement of the
  Ratio of the Inclusive 3-Jet Cross Section to the Inclusive 2-Jet Cross
  Section in pp Collisions at $\sqrt{s}$ = 7 TeV and First Determination of the
  Strong Coupling Constant in the TeV Range}},  {\em Eur. Phys. J. C}
  \textbf{73} (2013), no.~10 2604,
  [\href{http://arxiv.org/abs/1304.7498}{{\texttt{arXiv:1304.7498}}}].

\bibitem{Chatrchyan:2013haa}
\textbf{CMS} Collaboration, S.~Chatrchyan et~al., \textit{{Determination of the
  Top-Quark Pole Mass and Strong Coupling Constant from the $t \bar{t}$
  Production Cross Section in $pp$ Collisions at $\sqrt{s}$ = 7 TeV}},  {\em
  Phys. Lett. B} \textbf{728} (2014) 496--517,
  [\href{http://arxiv.org/abs/1307.1907}{{\texttt{arXiv:1307.1907}}}].
  [Erratum: Phys.Lett.B 738, 526--528 (2014)].

\bibitem{Ball:2018iqk}
\textbf{NNPDF} Collaboration, R.~D. Ball, S.~Carrazza, L.~Del~Debbio, S.~Forte,
  Z.~Kassabov, J.~Rojo, E.~Slade, and M.~Ubiali, \textit{{Precision
  determination of the strong coupling constant within a global PDF analysis}},
   {\em Eur. Phys. J.} \textbf{C78} (2018), no.~5 408,
  [\href{http://arxiv.org/abs/1802.03398}{{\texttt{arXiv:1802.03398}}}].

\bibitem{Forte:2020pyp}
S.~Forte and Z.~Kassabov, \textit{{Why $\alpha _s$ cannot be determined from
  hadronic processes without simultaneously determining the parton
  distributions}},  {\em Eur. Phys. J. C} \textbf{80} (2020), no.~3 182,
  [\href{http://arxiv.org/abs/2001.04986}{{\texttt{arXiv:2001.04986}}}].

\bibitem{Becciolini:2014lya}
D.~Becciolini, M.~Gillioz, M.~Nardecchia, F.~Sannino, and M.~Spannowsky,
  \textit{{Constraining new colored matter from the ratio of 3 to 2 jets cross
  sections at the LHC}},  {\em Phys. Rev. D} \textbf{91} (2015), no.~1 015010,
  [\href{http://arxiv.org/abs/1403.7411}{{\texttt{arXiv:1403.7411}}}].
  [Addendum: Phys.Rev.D 92, 079905 (2015)].

\bibitem{Dimopoulos:1981yj}
S.~Dimopoulos, S.~Raby, and F.~Wilczek, \textit{{Supersymmetry and the Scale of
  Unification}},  {\em Phys. Rev. D} \textbf{24} (1981) 1681--1683.

\bibitem{CooperSarkar:2011pa}
A.~Cooper-Sarkar, P.~Mertsch, and S.~Sarkar, \textit{{The high energy neutrino
  cross-section in the Standard Model and its uncertainty}},  {\em JHEP}
  \textbf{08} (2011) 042,
  [\href{http://arxiv.org/abs/1106.3723}{{\texttt{arXiv:1106.3723}}}].

\bibitem{Gauld:2015kvh}
R.~Gauld, J.~Rojo, L.~Rottoli, S.~Sarkar, and J.~Talbert, \textit{{The prompt
  atmospheric neutrino flux in the light of LHCb}},  {\em JHEP} \textbf{02}
  (2016) 130,
  [\href{http://arxiv.org/abs/1511.06346}{{\texttt{arXiv:1511.06346}}}].

\bibitem{Garzelli:2016xmx}
\textbf{PROSA} Collaboration, M.~V. Garzelli, S.~Moch, O.~Zenaiev,
  A.~Cooper-Sarkar, A.~Geiser, K.~Lipka, R.~Placakyte, and G.~Sigl,
  \textit{{Prompt neutrino fluxes in the atmosphere with PROSA parton
  distribution functions}},  {\em JHEP} \textbf{05} (2017) 004,
  [\href{http://arxiv.org/abs/1611.03815}{{\texttt{arXiv:1611.03815}}}].

\bibitem{Zenaiev:2015rfa}
\textbf{PROSA} Collaboration, O.~Zenaiev et~al., \textit{{Impact of
  heavy-flavour production cross sections measured by the LHCb experiment on
  parton distribution functions at low x}},  {\em Eur. Phys. J. C} \textbf{75}
  (2015), no.~8 396,
  [\href{http://arxiv.org/abs/1503.04581}{{\texttt{arXiv:1503.04581}}}].

\bibitem{Gauld:2016kpd}
R.~Gauld and J.~Rojo, \textit{{Precision determination of the small-$x$ gluon
  from charm production at LHCb}},  {\em Phys. Rev. Lett.} \textbf{118} (2017),
  no.~7 072001,
  [\href{http://arxiv.org/abs/1610.09373}{{\texttt{arXiv:1610.09373}}}].

\bibitem{Kovarik:2015cma}
K.~Kovarik et~al., \textit{{nCTEQ15 - Global analysis of nuclear parton
  distributions with uncertainties in the CTEQ framework}},  {\em Phys. Rev. D}
  \textbf{93} (2016), no.~8 085037,
  [\href{http://arxiv.org/abs/1509.00792}{{\texttt{arXiv:1509.00792}}}].

\bibitem{Eskola:2016oht}
K.~J. Eskola, P.~Paakkinen, H.~Paukkunen, and C.~A. Salgado, \textit{{EPPS16:
  Nuclear parton distributions with LHC data}},  {\em Eur. Phys. J. C}
  \textbf{77} (2017), no.~3 163,
  [\href{http://arxiv.org/abs/1612.05741}{{\texttt{arXiv:1612.05741}}}].

\bibitem{Hou:2019efy}
T.-J. Hou et~al., \textit{{New CTEQ global analysis of quantum chromodynamics
  with high-precision data from the LHC}},  {\em Phys. Rev. D} \textbf{103}
  (2021), no.~1 014013,
  [\href{http://arxiv.org/abs/1912.10053}{{\texttt{arXiv:1912.10053}}}].

\bibitem{Bailey:2020ooq}
S.~Bailey, T.~Cridge, L.~A. Harland-Lang, A.~D. Martin, and R.~S. Thorne,
  \textit{{Parton distributions from LHC, HERA, Tevatron and fixed target data:
  MSHT20 PDFs}},
  \href{http://arxiv.org/abs/2012.04684}{{\texttt{arXiv:2012.04684}}}.

\bibitem{Moffat:2021dji}
E.~Moffat, W.~Melnitchouk, T.~Rogers, and N.~Sato, \textit{{Simultaneous Monte
  Carlo analysis of parton densities and fragmentation functions}},
  \href{http://arxiv.org/abs/2101.04664}{{\texttt{arXiv:2101.04664}}}.

\bibitem{Gao:2013bia}
J.~Gao and P.~Nadolsky, \textit{{A meta-analysis of parton distribution
  functions}},  {\em JHEP} \textbf{07} (2014) 035,
  [\href{http://arxiv.org/abs/1401.0013}{{\texttt{arXiv:1401.0013}}}].

\bibitem{Carrazza:2015hva}
S.~Carrazza, J.~I. Latorre, J.~Rojo, and G.~Watt, \textit{A compression
  algorithm for the combination of pdf sets},  {\em Eur. Phys. J.} \textbf{C75}
  (2015) 474,
  [\href{http://arxiv.org/abs/1504.06469}{{\texttt{arXiv:1504.06469}}}].

\bibitem{Dulat:2015mca}
S.~Dulat, T.-J. Hou, J.~Gao, M.~Guzzi, J.~Huston, P.~Nadolsky, J.~Pumplin,
  C.~Schmidt, D.~Stump, and C.~P. Yuan, \textit{{New parton distribution
  functions from a global analysis of quantum chromodynamics}},  {\em Phys.
  Rev. D} \textbf{93} (2016), no.~3 033006,
  [\href{http://arxiv.org/abs/1506.07443}{{\texttt{arXiv:1506.07443}}}].

\bibitem{Harland-Lang:2014zoa}
L.~A. Harland-Lang, A.~D. Martin, P.~Motylinski, and R.~S. Thorne,
  \textit{{Parton distributions in the LHC era: MMHT 2014 PDFs}},  {\em Eur.
  Phys. J. C} \textbf{75} (2015), no.~5 204,
  [\href{http://arxiv.org/abs/1412.3989}{{\texttt{arXiv:1412.3989}}}].

\bibitem{Steerenberg:2645638}
by~Rende~Steerenberg and A.~Schaeffer, \textit{{LHC Report: Protons: mission
  accomplished}}, .

\bibitem{Arneodo:1996kd}
\textbf{New Muon} Collaboration, M.~Arneodo et~al., \textit{{Accurate
  measurement of $F_2^d/F_2^p$ and $R_d-R_p$}},  {\em Nucl. Phys.}
  \textbf{B487} (1997) 3--26,
  [\href{http://arxiv.org/abs/hep-ex/9611022}{{\texttt{hep-ex/9611022}}}].

\bibitem{Arneodo:1996qe}
\textbf{New Muon} Collaboration, M.~Arneodo et~al., \textit{{Measurement of the
  proton and deuteron structure functions, $F_2^p$ and $F_2^d$, and of the
  ratio $\sigma_L/\sigma_T$}},  {\em Nucl. Phys.} \textbf{B483} (1997) 3--43,
  [\href{http://arxiv.org/abs/hep-ph/9610231}{{\texttt{hep-ph/9610231}}}].

\bibitem{Whitlow:1991uw}
L.~W. Whitlow, E.~M. Riordan, S.~Dasu, S.~Rock, and A.~Bodek, \textit{{Precise
  measurements of the proton and deuteron structure functions from a global
  analysis of the SLAC deep inelastic electron scattering cross-sections}},
  {\em Phys. Lett.} \textbf{B282} (1992) 475--482.

\bibitem{Benvenuti:1989rh}
\textbf{BCDMS} Collaboration, A.~C. Benvenuti et~al., \textit{{A High
  Statistics Measurement of the Proton Structure Functions $F_2(x, Q^2)$ and
  $R$ from Deep Inelastic Muon Scattering at High $Q^2$}},  {\em Phys. Lett.}
  \textbf{B223} (1989) 485.

\bibitem{Onengut:2005kv}
\textbf{CHORUS} Collaboration, G.~Onengut et~al., \textit{{Measurement of
  nucleon structure functions in neutrino scattering}},  {\em Phys. Lett.}
  \textbf{B632} (2006) 65--75.

\bibitem{Goncharov:2001qe}
\textbf{NuTeV} Collaboration, M.~Goncharov et~al., \textit{{Precise measurement
  of dimuon production cross-sections in $\nu_{\mu}$Fe and $\bar{\nu}_{\mu}$Fe
  deep inelastic scattering at the Tevatron}},  {\em Phys. Rev.} \textbf{D64}
  (2001) 112006,
  [\href{http://arxiv.org/abs/hep-ex/0102049}{{\texttt{hep-ex/0102049}}}].

\bibitem{Mason:2006qa}
D.~A. Mason, {\em {Measurement of the strange - antistrange asymmetry at NLO in
  QCD from NuTeV dimuon data}}.
\newblock PhD thesis, Oregon U., 2006.

\bibitem{Abramowicz:2015mha}
\textbf{ZEUS, H1} Collaboration, H.~Abramowicz et~al., \textit{{Combination of
  measurements of inclusive deep inelastic ${e^{\pm }p}$ scattering cross
  sections and QCD analysis of HERA data}},  {\em Eur. Phys. J.} \textbf{C75}
  (2015), no.~12 580,
  [\href{http://arxiv.org/abs/1506.06042}{{\texttt{arXiv:1506.06042}}}].

\bibitem{Abramowicz:1900rp}
\textbf{H1 , ZEUS} Collaboration, H.~Abramowicz et~al., \textit{{Combination
  and QCD Analysis of Charm Production Cross Section Measurements in
  Deep-Inelastic ep Scattering at HERA}},  {\em Eur.Phys.J.} \textbf{C73}
  (2013) 2311,
  [\href{http://arxiv.org/abs/1211.1182}{{\texttt{arXiv:1211.1182}}}].

\bibitem{Aaron:2009af}
\textbf{H1} Collaboration, F.~D. Aaron et~al., \textit{{Measurement of the
  Charm and Beauty Structure Functions using the H1 Vertex Detector at HERA}},
  {\em Eur. Phys. J.} \textbf{C65} (2010) 89--109,
  [\href{http://arxiv.org/abs/0907.2643}{{\texttt{arXiv:0907.2643}}}].

\bibitem{Abramowicz:2014zub}
\textbf{ZEUS} Collaboration, H.~Abramowicz et~al., \textit{{Measurement of
  beauty and charm production in deep inelastic scattering at HERA and
  measurement of the beauty-quark mass}},  {\em JHEP} \textbf{09} (2014) 127,
  [\href{http://arxiv.org/abs/1405.6915}{{\texttt{arXiv:1405.6915}}}].

\bibitem{Webb:2003ps}
\textbf{NuSea} Collaboration, J.~C. Webb et~al., \textit{{Absolute Drell-Yan
  dimuon cross sections in 800-GeV/c p p and p d collisions}},
  \href{http://arxiv.org/abs/hep-ex/0302019}{{\texttt{hep-ex/0302019}}}.

\bibitem{Webb:2003bj}
J.~C. Webb, \textit{{Measurement of continuum dimuon production in 800-GeV/c
  proton nucleon collisions}},
  \href{http://arxiv.org/abs/hep-ex/0301031}{{\texttt{hep-ex/0301031}}}.

\bibitem{Towell:2001nh}
\textbf{FNAL E866/NuSea} Collaboration, R.~S. Towell et~al., \textit{{Improved
  measurement of the anti-d/anti-u asymmetry in the nucleon sea}},  {\em Phys.
  Rev.} \textbf{D64} (2001) 052002,
  [\href{http://arxiv.org/abs/hep-ex/0103030}{{\texttt{hep-ex/0103030}}}].

\bibitem{Moreno:1990sf}
G.~Moreno et~al., \textit{{Dimuon production in proton - copper collisions at
  $\sqrt{s}$ = 38.8-GeV}},  {\em Phys. Rev.} \textbf{D43} (1991) 2815--2836.

\bibitem{Aaltonen:2010zza}
\textbf{CDF} Collaboration, T.~A. Aaltonen et~al., \textit{{Measurement of
  $d\sigma/dy$ of Drell-Yan $e^+e^-$ pairs in the $Z$ Mass Region from
  $p\bar{p}$ Collisions at $\sqrt{s}=1.96$ TeV}},  {\em Phys. Lett.}
  \textbf{B692} (2010) 232--239,
  [\href{http://arxiv.org/abs/0908.3914}{{\texttt{arXiv:0908.3914}}}].

\bibitem{Abazov:2007jy}
\textbf{D0} Collaboration, V.~M. Abazov et~al., \textit{{Measurement of the
  shape of the boson rapidity distribution for $p \bar{p} \to Z/\gamma^* \to
  e^{+} e^{-}$ + $X$ events produced at $\sqrt{s}$=1.96-TeV}},  {\em Phys.
  Rev.} \textbf{D76} (2007) 012003,
  [\href{http://arxiv.org/abs/hep-ex/0702025}{{\texttt{hep-ex/0702025}}}].

\bibitem{Abazov:2013rja}
\textbf{D0} Collaboration, V.~M. Abazov et~al., \textit{{Measurement of the
  muon charge asymmetry in $p\bar{p}$ $\to$ W+X $\to$ $\mu$$\nu$ + X events at
  $\sqrt{s}$=1.96 TeV}},  {\em Phys.Rev.} \textbf{D88} (2013) 091102,
  [\href{http://arxiv.org/abs/1309.2591}{{\texttt{arXiv:1309.2591}}}].

\bibitem{D0:2014kma}
\textbf{D0} Collaboration, V.~M. Abazov et~al., \textit{{Measurement of the
  electron charge asymmetry in ${p\bar{p}\rightarrow W+X \rightarrow e\nu +X}$
  decays in ${p\bar{p}}$ collisions at ${\sqrt{s}=1.96}$ TeV}},  {\em Phys.
  Rev.} \textbf{D91} (2015), no.~3 032007,
  [\href{http://arxiv.org/abs/1412.2862}{{\texttt{arXiv:1412.2862}}}].
  [Erratum: Phys. Rev.D91,no.7,079901(2015)].

\bibitem{Aad:2013iua}
\textbf{ATLAS} Collaboration, G.~Aad et~al., \textit{{Measurement of the
  high-mass Drell--Yan differential cross-section in pp collisions at
  $\sqrt{s}$=7 TeV with the ATLAS detector}},  {\em Phys.Lett.} \textbf{B725}
  (2013) 223,
  [\href{http://arxiv.org/abs/1305.4192}{{\texttt{arXiv:1305.4192}}}].

\bibitem{Aad:2014qja}
\textbf{ATLAS} Collaboration, G.~Aad et~al., \textit{{Measurement of the
  low-mass Drell-Yan differential cross section at $\sqrt{s}$ = 7 TeV using the
  ATLAS detector}},  {\em JHEP} \textbf{06} (2014) 112,
  [\href{http://arxiv.org/abs/1404.1212}{{\texttt{arXiv:1404.1212}}}].

\bibitem{Aad:2011dm}
\textbf{ATLAS} Collaboration, G.~Aad et~al., \textit{{Measurement of the
  inclusive $W^{\pm}$ and $Z/\gamma^*$ cross sections in the electron and muon
  decay channels in pp collisions at $\sqrt{s}$= 7 TeV with the ATLAS
  detector}},  {\em Phys.Rev.} \textbf{D85} (2012) 072004,
  [\href{http://arxiv.org/abs/1109.5141}{{\texttt{arXiv:1109.5141}}}].

\bibitem{Aaboud:2016btc}
\textbf{ATLAS} Collaboration, M.~Aaboud et~al., \textit{{Precision measurement
  and interpretation of inclusive $W^+$, $W^-$ and $Z/\gamma^*$ production
  cross sections with the ATLAS detector}},
  \href{http://arxiv.org/abs/1612.03016}{{\texttt{arXiv:1612.03016}}}.

\bibitem{Aad:2015auj}
\textbf{ATLAS} Collaboration, G.~Aad et~al., \textit{{Measurement of the
  transverse momentum and $\phi ^*_{\eta }$ distributions of Drell–Yan lepton
  pairs in proton–proton collisions at $\sqrt{s}=8$ TeV with the ATLAS
  detector}},  {\em Eur. Phys. J.} \textbf{C76} (2016), no.~5 291,
  [\href{http://arxiv.org/abs/1512.02192}{{\texttt{arXiv:1512.02192}}}].

\bibitem{Aad:2014kva}
\textbf{ATLAS} Collaboration, G.~Aad et~al., \textit{{Measurement of the
  $t\bar{t}$ production cross-section using $e\mu $ events with b-tagged jets
  in pp collisions at $\sqrt{s}$ = 7 and 8 $\,\mathrm{TeV}$ with the ATLAS
  detector}},  {\em Eur. Phys. J.} \textbf{C74} (2014), no.~10 3109,
  [\href{http://arxiv.org/abs/1406.5375}{{\texttt{arXiv:1406.5375}}}].
  [Addendum: Eur. Phys. J.C76,no.11,642(2016)].

\bibitem{Aaboud:2016pbd}
\textbf{ATLAS} Collaboration, M.~Aaboud et~al., \textit{{Measurement of the
  $t\bar{t}$ production cross-section using $e\mu$ events with b-tagged jets in
  pp collisions at $\sqrt{s}$=13 TeV with the ATLAS detector}},  {\em Phys.
  Lett.} \textbf{B761} (2016) 136--157,
  [\href{http://arxiv.org/abs/1606.02699}{{\texttt{arXiv:1606.02699}}}].

\bibitem{Aad:2015mbv}
\textbf{ATLAS} Collaboration, G.~Aad et~al., \textit{{Measurements of top-quark
  pair differential cross-sections in the lepton+jets channel in $pp$
  collisions at $\sqrt{s}=8$ TeV using the ATLAS detector}},  {\em Eur. Phys.
  J.} \textbf{C76} (2016), no.~10 538,
  [\href{http://arxiv.org/abs/1511.04716}{{\texttt{arXiv:1511.04716}}}].

\bibitem{Chatrchyan:2012xt}
\textbf{CMS} Collaboration, S.~Chatrchyan et~al., \textit{{Measurement of the
  electron charge asymmetry in inclusive W production in pp collisions at
  $\sqrt{s}$ = 7 TeV}},  {\em Phys.Rev.Lett.} \textbf{109} (2012) 111806,
  [\href{http://arxiv.org/abs/1206.2598}{{\texttt{arXiv:1206.2598}}}].

\bibitem{Chatrchyan:2013mza}
\textbf{CMS} Collaboration, S.~Chatrchyan et~al., \textit{{Measurement of the
  muon charge asymmetry in inclusive pp to WX production at $\sqrt{s}$ = 7 TeV
  and an improved determination of light parton distribution functions}},  {\em
  Phys.Rev.} \textbf{D90} (2014) 032004,
  [\href{http://arxiv.org/abs/1312.6283}{{\texttt{arXiv:1312.6283}}}].

\bibitem{Chatrchyan:2013tia}
\textbf{CMS} Collaboration, S.~Chatrchyan et~al., \textit{{Measurement of the
  differential and double-differential Drell-Yan cross sections in
  proton-proton collisions at $\sqrt{s} =$ 7 TeV}},  {\em JHEP} \textbf{1312}
  (2013) 030,
  [\href{http://arxiv.org/abs/1310.7291}{{\texttt{arXiv:1310.7291}}}].

\bibitem{Khachatryan:2016pev}
\textbf{CMS} Collaboration, V.~Khachatryan et~al., \textit{{Measurement of the
  differential cross section and charge asymmetry for inclusive $\mathrm
  {p}\mathrm {p}\rightarrow \mathrm {W}^{\pm }+X$ production at ${\sqrt{s}} =
  8$ TeV}},  {\em Eur. Phys. J.} \textbf{C76} (2016), no.~8 469,
  [\href{http://arxiv.org/abs/1603.01803}{{\texttt{arXiv:1603.01803}}}].

\bibitem{Khachatryan:2015oaa}
\textbf{CMS} Collaboration, V.~Khachatryan et~al., \textit{{Measurement of the
  Z boson differential cross section in transverse momentum and rapidity in
  proton–proton collisions at 8 TeV}},  {\em Phys. Lett.} \textbf{B749}
  (2015) 187--209,
  [\href{http://arxiv.org/abs/1504.03511}{{\texttt{arXiv:1504.03511}}}].

\bibitem{Khachatryan:2016mqs}
\textbf{CMS} Collaboration, V.~Khachatryan et~al., \textit{{Measurement of the
  t-tbar production cross section in the e-mu channel in proton-proton
  collisions at sqrt(s) = 7 and 8 TeV}},  {\em JHEP} \textbf{08} (2016) 029,
  [\href{http://arxiv.org/abs/1603.02303}{{\texttt{arXiv:1603.02303}}}].

\bibitem{Khachatryan:2015uqb}
\textbf{CMS} Collaboration, V.~Khachatryan et~al., \textit{{Measurement of the
  top quark pair production cross section in proton-proton collisions at
  $\sqrt(s) =$ 13 TeV}},  {\em Phys. Rev. Lett.} \textbf{116} (2016), no.~5
  052002,
  [\href{http://arxiv.org/abs/1510.05302}{{\texttt{arXiv:1510.05302}}}].

\bibitem{Khachatryan:2015oqa}
\textbf{CMS} Collaboration, V.~Khachatryan et~al., \textit{{Measurement of the
  differential cross section for top quark pair production in pp collisions at
  $\sqrt{s} = 8\,\text {TeV} $}},  {\em Eur. Phys. J.} \textbf{C75} (2015),
  no.~11 542,
  [\href{http://arxiv.org/abs/1505.04480}{{\texttt{arXiv:1505.04480}}}].

\bibitem{Aaij:2012vn}
\textbf{LHCb} Collaboration, R.~Aaij et~al., \textit{{Inclusive $W$ and $Z$
  production in the forward region at $\sqrt{s} = 7$ TeV}},  {\em JHEP}
  \textbf{1206} (2012) 058,
  [\href{http://arxiv.org/abs/1204.1620}{{\texttt{arXiv:1204.1620}}}].

\bibitem{Aaij:2012mda}
\textbf{LHCb} Collaboration, R.~Aaij et~al., \textit{{Measurement of the
  cross-section for $Z \to e^+e^-$ production in $pp$ collisions at
  $\sqrt{s}=7$ TeV}},  {\em JHEP} \textbf{1302} (2013) 106,
  [\href{http://arxiv.org/abs/1212.4620}{{\texttt{arXiv:1212.4620}}}].

\bibitem{Aaij:2015gna}
\textbf{LHCb} Collaboration, R.~Aaij et~al., \textit{{Measurement of the
  forward $Z$ boson production cross-section in $pp$ collisions at $\sqrt{s}=7$
  TeV}},  {\em JHEP} \textbf{08} (2015) 039,
  [\href{http://arxiv.org/abs/1505.07024}{{\texttt{arXiv:1505.07024}}}].

\bibitem{Aaij:2015zlq}
\textbf{LHCb} Collaboration, R.~Aaij et~al., \textit{{Measurement of forward W
  and Z boson production in $pp$ collisions at $ \sqrt{s}=8 $ TeV}},  {\em
  JHEP} \textbf{01} (2016) 155,
  [\href{http://arxiv.org/abs/1511.08039}{{\texttt{arXiv:1511.08039}}}].

\bibitem{Aad:2011fc}
\textbf{ATLAS} Collaboration, G.~Aad et~al., \textit{{Measurement of inclusive
  jet and dijet production in pp collisions at $\sqrt{s}$ = 7 TeV using the
  ATLAS detector}},  {\em Phys. Rev.} \textbf{D86} (2012) 014022,
  [\href{http://arxiv.org/abs/1112.6297}{{\texttt{arXiv:1112.6297}}}].

\bibitem{Aad:2014vwa}
\textbf{ATLAS} Collaboration, G.~Aad et~al., \textit{{Measurement of the
  inclusive jet cross-section in proton-proton collisions at $ \sqrt{s}=7$ TeV
  using 4.5 fb$^{-1}$ of data with the ATLAS detector}},  {\em JHEP}
  \textbf{02} (2015) 153,
  [\href{http://arxiv.org/abs/1410.8857}{{\texttt{arXiv:1410.8857}}}].

\bibitem{Aad:2013lpa}
\textbf{ATLAS} Collaboration, G.~Aad et~al., \textit{{Measurement of the
  inclusive jet cross section in pp collisions at $\sqrt{s}$=2.76 TeV and
  comparison to the inclusive jet cross section at $\sqrt{s}$=7 TeV using the
  ATLAS detector}},  {\em Eur.Phys.J.} \textbf{C73} (2013) 2509,
  [\href{http://arxiv.org/abs/1304.4739}{{\texttt{arXiv:1304.4739}}}].

\bibitem{Chatrchyan:2012bja}
\textbf{CMS} Collaboration, S.~Chatrchyan et~al., \textit{{Measurements of
  differential jet cross sections in proton-proton collisions at $\sqrt{s}=7$
  TeV with the CMS detector}},  {\em Phys.Rev.} \textbf{D87} (2013) 112002,
  [\href{http://arxiv.org/abs/1212.6660}{{\texttt{arXiv:1212.6660}}}].

\bibitem{Khachatryan:2015luy}
\textbf{CMS} Collaboration, V.~Khachatryan et~al., \textit{{Measurement of the
  inclusive jet cross section in pp collisions at $\sqrt{s} = 2.76\,\text
  {TeV}$}},  {\em Eur. Phys. J.} \textbf{C76} (2016), no.~5 265,
  [\href{http://arxiv.org/abs/1512.06212}{{\texttt{arXiv:1512.06212}}}].

\bibitem{Abulencia:2007ez}
\textbf{CDF - Run II} Collaboration, A.~Abulencia et~al., \textit{{Measurement
  of the Inclusive Jet Cross Section using the $k_{\text{T}}$ algorithm in
  $p\overline{p}$ Collisions at $\sqrt{s}$=1.96 TeV with the CDF II Detector}},
   {\em Phys. Rev.} \textbf{D75} (2007) 092006,
  [\href{http://arxiv.org/abs/hep-ex/0701051}{{\texttt{hep-ex/0701051}}}].

\bibitem{Ball:2015tna}
R.~D. Ball, V.~Bertone, M.~Bonvini, S.~Forte, P.~Groth~Merrild, J.~Rojo, and
  L.~Rottoli, \textit{{Intrinsic charm in a matched general-mass scheme}},
  {\em Phys. Lett. B} \textbf{754} (2016) 49--58,
  [\href{http://arxiv.org/abs/1510.00009}{{\texttt{arXiv:1510.00009}}}].

\bibitem{Ball:2015dpa}
R.~D. Ball, M.~Bonvini, and L.~Rottoli, \textit{{Charm in Deep-Inelastic
  Scattering}},  {\em JHEP} \textbf{11} (2015) 122,
  [\href{http://arxiv.org/abs/1510.02491}{{\texttt{arXiv:1510.02491}}}].

\bibitem{Tanabashi:2018oca}
\textbf{Particle Data Group} Collaboration, M.~Tanabashi et~al.,
  \textit{{Review of Particle Physics}},  {\em Phys. Rev.} \textbf{D98} (2018),
  no.~3 030001.

\bibitem{Verbytskyi:2019zhh}
A.~Verbytskyi, A.~Banfi, A.~Kardos, P.~F. Monni, S.~Kluth, G.~Somogyi,
  Z.~Szőr, Z.~Trócsányi, Z.~Tulipánt, and G.~Zanderighi, \textit{{High
  precision determination of $\alpha_s$ from a global fit of jet rates}},
  \href{http://arxiv.org/abs/1902.08158}{{\texttt{arXiv:1902.08158}}}.

\bibitem{Bruno:2017gxd}
\textbf{ALPHA} Collaboration, M.~Bruno, M.~Dalla~Brida, P.~Fritzsch, T.~Korzec,
  A.~Ramos, S.~Schaefer, H.~Simma, S.~Sint, and R.~Sommer, \textit{{QCD
  Coupling from a Nonperturbative Determination of the Three-Flavor $\Lambda$
  Parameter}},  {\em Phys. Rev. Lett.} \textbf{119} (2017), no.~10 102001,
  [\href{http://arxiv.org/abs/1706.03821}{{\texttt{arXiv:1706.03821}}}].

\bibitem{Zafeiropoulos:2019flq}
S.~Zafeiropoulos, P.~Boucaud, F.~De~Soto, J.~Rodríguez-Quintero, and
  J.~Segovia, \textit{{The strong running coupling from the gauge sector of
  Domain Wall lattice QCD with physical quark masses}},  {\em Phys. Rev. Lett.}
  \textbf{122} (2019), no.~16 162002,
  [\href{http://arxiv.org/abs/1902.08148}{{\texttt{arXiv:1902.08148}}}].

\bibitem{Pich:2018lmu}
A.~Pich, J.~Rojo, R.~Sommer, and A.~Vairo, \textit{{Determining the strong
  coupling: status and challenges}},  in {\em {13th Conference on Quark
  Confinement and the Hadron Spectrum (Confinement XIII) Maynooth, Ireland,
  July 31-August 6, 2018}}, 2018.
\newblock \href{http://arxiv.org/abs/1811.11801}{{\texttt{arXiv:1811.11801}}}.

\bibitem{DelDebbio:2013kxa}
L.~Del~Debbio, N.~P. Hartland, and S.~Schumann, \textit{{MCgrid: projecting
  cross section calculations on grids}},  {\em Comput.Phys.Commun.}
  \textbf{185} (2014) 2115--2126,
  [\href{http://arxiv.org/abs/1312.4460}{{\texttt{arXiv:1312.4460}}}].

\bibitem{Campbell:1999ah}
J.~M. Campbell and R.~K. Ellis, \textit{{An Update on vector boson pair
  production at hadron colliders}},  {\em Phys. Rev. D} \textbf{60} (1999)
  113006,
  [\href{http://arxiv.org/abs/hep-ph/9905386}{{\texttt{hep-ph/9905386}}}].

\bibitem{Campbell:2011bn}
J.~M. Campbell, R.~K. Ellis, and C.~Williams, \textit{{Vector boson pair
  production at the LHC}},  {\em JHEP} \textbf{07} (2011) 018,
  [\href{http://arxiv.org/abs/1105.0020}{{\texttt{arXiv:1105.0020}}}].

\bibitem{Nagy:2001fj}
Z.~Nagy, \textit{{Three jet cross-sections in hadron hadron collisions at
  next-to-leading order}},  {\em Phys. Rev. Lett.} \textbf{88} (2002) 122003,
  [\href{http://arxiv.org/abs/hep-ph/0110315}{{\texttt{hep-ph/0110315}}}].

\bibitem{Rojo:2015acz}
J.~Rojo et~al., \textit{{The PDF4LHC report on PDFs and LHC data: Results from
  Run I and preparation for Run II}},  {\em J. Phys.} \textbf{G42} (2015)
  103103,
  [\href{http://arxiv.org/abs/1507.00556}{{\texttt{arXiv:1507.00556}}}].

\bibitem{Cacciari:2011ze}
M.~Cacciari and N.~Houdeau, \textit{{Meaningful characterisation of
  perturbative theoretical uncertainties}},  {\em JHEP} \textbf{1109} (2011)
  039, [\href{http://arxiv.org/abs/1105.5152}{{\texttt{arXiv:1105.5152}}}].

\bibitem{David:2013gaa}
A.~David and G.~Passarino, \textit{{How well can we guess theoretical
  uncertainties?}},  {\em Phys. Lett.} \textbf{B726} (2013) 266--272,
  [\href{http://arxiv.org/abs/1307.1843}{{\texttt{arXiv:1307.1843}}}].

\bibitem{Bagnaschi:2014wea}
E.~Bagnaschi, M.~Cacciari, A.~Guffanti, and L.~Jenniches, \textit{{An extensive
  survey of the estimation of uncertainties from missing higher orders in
  perturbative calculations}},  {\em JHEP} \textbf{02} (2015) 133,
  [\href{http://arxiv.org/abs/1409.5036}{{\texttt{arXiv:1409.5036}}}].

\bibitem{Ball:2018lag}
R.~D. Ball and A.~Deshpande, {\em {The proton spin, semi-inclusive processes,
  and measurements at a future Electron Ion Collider}}, pp.~205--226.
\newblock 2019.
\newblock \href{http://arxiv.org/abs/1801.04842}{{\texttt{arXiv:1801.04842}}}.

\bibitem{Pearson:2018tim}
R.~L. Pearson and C.~Voisey, \textit{{Towards parton distribution functions
  with theoretical uncertainties}},  {\em Nucl. Part. Phys. Proc.}
  \textbf{300-302} (2018) 24--29,
  [\href{http://arxiv.org/abs/1810.01996}{{\texttt{arXiv:1810.01996}}}].

\bibitem{zahari_kassabov_2019_2571601}
Z.~Kassabov, ``{Reportengine: A framework for declarative data analysis}.''
  https://doi.org/10.5281/zenodo.2571601, Feb., 2019.

\bibitem{Ridder:2013mf}
A.~Gehrmann-De~Ridder, T.~Gehrmann, E.~Glover, and J.~Pires, \textit{{Second
  order QCD corrections to jet production at hadron colliders: the all-gluon
  contribution}},  {\em Phys.Rev.Lett.} \textbf{110} (2013) 162003,
  [\href{http://arxiv.org/abs/1301.7310}{{\texttt{arXiv:1301.7310}}}].

\bibitem{Currie:2013dwa}
J.~Currie, A.~Gehrmann-De~Ridder, E.~Glover, and J.~Pires, \textit{{NNLO QCD
  corrections to jet production at hadron colliders from gluon scattering}},
  {\em JHEP} \textbf{1401} (2014) 110,
  [\href{http://arxiv.org/abs/1310.3993}{{\texttt{arXiv:1310.3993}}}].

\bibitem{Currie:2016bfm}
J.~Currie, E.~W.~N. Glover, and J.~Pires, \textit{{NNLO QCD predictions for
  single jet inclusive production at the LHC}},  {\em Phys. Rev. Lett.}
  \textbf{118} (2017), no.~7 072002,
  [\href{http://arxiv.org/abs/1611.01460}{{\texttt{arXiv:1611.01460}}}].

\bibitem{Czakon:2019tmo}
M.~Czakon, A.~van Hameren, A.~Mitov, and R.~Poncelet, \textit{{Single-jet
  inclusive rates with exact color at $ \mathcal{O} $ ($ {\alpha}_s^4 $)}},
  {\em JHEP} \textbf{10} (2019) 262,
  [\href{http://arxiv.org/abs/1907.12911}{{\texttt{arXiv:1907.12911}}}].

\bibitem{Martin:1987vw}
A.~D. Martin, R.~G. Roberts, and W.~J. Stirling, \textit{{Structure Function
  Analysis and psi, Jet, W, Z Production: Pinning Down the Gluon}},  {\em Phys.
  Rev.} \textbf{D37} (1988) 1161.

\bibitem{Aversa:1988fv}
F.~Aversa, P.~Chiappetta, M.~Greco, and J.~P. Guillet, \textit{{Higher Order
  Corrections to QCD Jets}},  {\em Phys. Lett.} \textbf{B210} (1988) 225.

\bibitem{Ellis:1988hv}
S.~D. Ellis, Z.~Kunszt, and D.~E. Soper, \textit{{The One Jet Inclusive
  Cross-section at Order $\alpha_s^3$: Gluons Only}},  {\em Phys. Rev. Lett.}
  \textbf{62} (1989) 726.

\bibitem{Giele:1994xd}
W.~T. Giele, E.~W.~N. Glover, and D.~A. Kosower, \textit{{The inclusive two jet
  triply differential cross-section}},  {\em Phys. Rev.} \textbf{D52} (1995)
  1486--1499,
  [\href{http://arxiv.org/abs/hep-ph/9412338}{{\texttt{hep-ph/9412338}}}].

\bibitem{Currie:2017ctp}
J.~Currie, E.~W.~N. Glover, A.~Gehrmann-De~Ridder, T.~Gehrmann, A.~Huss, and
  J.~Pires, \textit{{Single jet inclusive production for the individual jet
  $p_{T}$ scale choice at the LHC}},  in {\em {23rd Cracow Epiphany Conference
  on Particle Theory Meets the First Data from LHC Run 2 Cracow, Poland,
  January 9-12, 2017}}, 2017.
\newblock \href{http://arxiv.org/abs/1704.00923}{{\texttt{arXiv:1704.00923}}}.

\bibitem{Currie:2018xkj}
J.~Currie, A.~Gehrmann-De~Ridder, T.~Gehrmann, E.~W.~N. Glover, A.~Huss, and
  J.~Pires, \textit{{Infrared sensitivity of single jet inclusive production at
  hadron colliders}},  {\em JHEP} \textbf{10} (2018) 155,
  [\href{http://arxiv.org/abs/1807.03692}{{\texttt{arXiv:1807.03692}}}].

\bibitem{Cacciari:2019qjx}
M.~Cacciari, S.~Forte, D.~Napoletano, G.~Soyez, and G.~Stagnitto,
  \textit{{Single-jet inclusive cross section and its definition}},  {\em Phys.
  Rev.} \textbf{D100} (2019), no.~11 114015,
  [\href{http://arxiv.org/abs/1906.11850}{{\texttt{arXiv:1906.11850}}}].

\bibitem{Dasgupta:2016bnd}
M.~Dasgupta, F.~A. Dreyer, G.~P. Salam, and G.~Soyez, \textit{{Inclusive jet
  spectrum for small-radius jets}},  {\em JHEP} \textbf{06} (2016) 057,
  [\href{http://arxiv.org/abs/1602.01110}{{\texttt{arXiv:1602.01110}}}].

\bibitem{Aaboud:2017dvo}
\textbf{ATLAS} Collaboration, M.~Aaboud et~al., \textit{{Measurement of the
  inclusive jet cross-sections in proton-proton collisions at $ \sqrt{s}=8 $
  TeV with the ATLAS detector}},  {\em JHEP} \textbf{09} (2017) 020,
  [\href{http://arxiv.org/abs/1706.03192}{{\texttt{arXiv:1706.03192}}}].

\bibitem{Khachatryan:2016mlc}
\textbf{CMS} Collaboration, V.~Khachatryan et~al., \textit{{Measurement and QCD
  analysis of double-differential inclusive jet cross sections in pp collisions
  at $ \sqrt{s}=8 $ TeV and cross section ratios to 2.76 and 7 TeV}},  {\em
  JHEP} \textbf{03} (2017) 156,
  [\href{http://arxiv.org/abs/1609.05331}{{\texttt{arXiv:1609.05331}}}].

\bibitem{Aad:2013tea}
\textbf{ATLAS Collaboration} Collaboration, G.~Aad et~al., \textit{{Measurement
  of dijet cross sections in $pp$ collisions at 7 TeV centre-of-mass energy
  using the ATLAS detector}},  {\em JHEP} \textbf{1405} (2014) 059,
  [\href{http://arxiv.org/abs/1312.3524}{{\texttt{arXiv:1312.3524}}}].

\bibitem{Sirunyan:2017skj}
\textbf{CMS} Collaboration, A.~M. Sirunyan et~al., \textit{{Measurement of the
  triple-differential dijet cross section in proton-proton collisions at
  $\sqrt{s}=8\,\text {TeV} $ and constraints on parton distribution
  functions}},  {\em Eur. Phys. J.} \textbf{C77} (2017), no.~11 746,
  [\href{http://arxiv.org/abs/1705.02628}{{\texttt{arXiv:1705.02628}}}].

\bibitem{Gehrmann-DeRidder:2019ibf}
A.~Gehrmann-De~Ridder, T.~Gehrmann, E.~W.~N. Glover, A.~Huss, and J.~Pires,
  \textit{{Triple Differential Dijet Cross Section at the LHC}},  {\em Phys.
  Rev. Lett.} \textbf{123} (2019), no.~10 102001,
  [\href{http://arxiv.org/abs/1905.09047}{{\texttt{arXiv:1905.09047}}}].

\bibitem{Nocera:2017zge}
E.~R. Nocera and M.~Ubiali, \textit{{Constraining the gluon PDF at large x with
  LHC data}},  {\em PoS} \textbf{DIS2017} (2018) 008,
  [\href{http://arxiv.org/abs/1709.09690}{{\texttt{arXiv:1709.09690}}}].

\bibitem{Zurita:2018vrs}
P.~Zurita, \textit{{Recent progress in Nuclear Parton Distributions}},  in {\em
  {13th Conference on the Intersections of Particle and Nuclear Physics}}, 9,
  2018.
\newblock \href{http://arxiv.org/abs/1810.00099}{{\texttt{arXiv:1810.00099}}}.

\bibitem{Paukkunen:2018kmm}
H.~Paukkunen, \textit{{Nuclear PDFs Today}},  {\em PoS} \textbf{HardProbes2018}
  (2018) 014,
  [\href{http://arxiv.org/abs/1811.01976}{{\texttt{arXiv:1811.01976}}}].

\bibitem{Rojo:2019uip}
J.~Rojo, \textit{{The Partonic Content of Nucleons and Nuclei}},
  \href{http://arxiv.org/abs/1910.03408}{{\texttt{arXiv:1910.03408}}}.

\bibitem{Abreu:2007kv}
N.~Armesto, N.~Borghini, S.~Jeon, and U.~A. Wiedemann, eds., {\em {Proceedings,
  Workshop on Heavy Ion Collisions at the LHC: Last Call for Predictions}:
  {Geneva, Switzerland, May 14 - June 8, 2007}}, vol.~35, 2008.

\bibitem{Adams:2005dq}
\textbf{STAR} Collaboration, J.~Adams et~al., \textit{{Experimental and
  theoretical challenges in the search for the quark gluon plasma: The STAR
  Collaboration's critical assessment of the evidence from RHIC collisions}},
  {\em Nucl. Phys. A} \textbf{757} (2005) 102--183,
  [\href{http://arxiv.org/abs/nucl-ex/0501009}{{\texttt{nucl-ex/0501009}}}].

\bibitem{Alekhin:2017kpj}
S.~Alekhin, J.~Bl{\"u}mlein, S.~Moch, and R.~Placakyte, \textit{{Parton
  distribution functions, $\alpha_s$, and heavy-quark masses for LHC Run II}},
  {\em Phys. Rev.} \textbf{D96} (2017), no.~1 014011,
  [\href{http://arxiv.org/abs/1701.05838}{{\texttt{arXiv:1701.05838}}}].

\bibitem{Ball:2009mk}
\textbf{NNPDF} Collaboration, R.~D. Ball, L.~Del~Debbio, S.~Forte, A.~Guffanti,
  J.~I. Latorre, A.~Piccione, J.~Rojo, and M.~Ubiali, \textit{{Precision
  determination of electroweak parameters and the strange content of the proton
  from neutrino deep-inelastic scattering}},  {\em Nucl. Phys. B} \textbf{823}
  (2009) 195--233,
  [\href{http://arxiv.org/abs/0906.1958}{{\texttt{arXiv:0906.1958}}}].

\bibitem{Adam:2015hoa}
\textbf{ALICE} Collaboration, J.~Adam et~al., \textit{{Measurement of charged
  jet production cross sections and nuclear modification in p-Pb collisions at
  $\sqrt{s_\text{NN}} = 5.02$ TeV}},  {\em Phys. Lett. B} \textbf{749} (2015)
  68--81,
  [\href{http://arxiv.org/abs/1503.00681}{{\texttt{arXiv:1503.00681}}}].

\bibitem{Adam:2016dau}
\textbf{ALICE} Collaboration, J.~Adam et~al., \textit{{Multiplicity dependence
  of charged pion, kaon, and (anti)proton production at large transverse
  momentum in p-Pb collisions at $\mathbf{\sqrt{{\textit s}_\text{NN}}}$ = 5.02
  TeV}},  {\em Phys. Lett. B} \textbf{760} (2016) 720--735,
  [\href{http://arxiv.org/abs/1601.03658}{{\texttt{arXiv:1601.03658}}}].

\bibitem{Adam:2015xea}
\textbf{ALICE} Collaboration, J.~Adam et~al., \textit{{Measurement of dijet
  $k_T$ in p\textendash{}Pb collisions at $\sqrt{s}_{NN}$=5.02 TeV}},  {\em
  Phys. Lett. B} \textbf{746} (2015) 385--395,
  [\href{http://arxiv.org/abs/1503.03050}{{\texttt{arXiv:1503.03050}}}].

\bibitem{Aad:2016zif}
\textbf{ATLAS} Collaboration, G.~Aad et~al., \textit{{Transverse momentum,
  rapidity, and centrality dependence of inclusive charged-particle production
  in $\sqrt{s_{NN}}=5.02$ TeV $p$ + Pb collisions measured by the ATLAS
  experiment}},  {\em Phys. Lett. B} \textbf{763} (2016) 313--336,
  [\href{http://arxiv.org/abs/1605.06436}{{\texttt{arXiv:1605.06436}}}].

\bibitem{Chatrchyan:2014hqa}
\textbf{CMS} Collaboration, S.~Chatrchyan et~al., \textit{{Studies of dijet
  transverse momentum balance and pseudorapidity distributions in pPb
  collisions at $\sqrt{s_{\mathrm{NN}}} = 5.02$ $\,\text {TeV}$}},  {\em Eur.
  Phys. J. C} \textbf{74} (2014), no.~7 2951,
  [\href{http://arxiv.org/abs/1401.4433}{{\texttt{arXiv:1401.4433}}}].

\bibitem{Zhu:2015kpa}
\textbf{ALICE} Collaboration, J.~Zhu, \textit{{Measurement of W-boson
  production in p\textendash{}Pb collisions at $\sqrt{s_{NN}}$ = 5.02 TeV with
  ALICE at the LHC}},  {\em J. Phys. Conf. Ser.} \textbf{612} (2015), no.~1
  012009.

\bibitem{TheATLAScollaboration:2015lnm}
\textit{{Measurement of $W\rightarrow\mu\nu$ production in $p$+Pb collision at
  $\sqrt{s_{\text{NN}}}=5.02$ TeV with ATLAS detector at the LHC}}, .

\bibitem{Aad:2015gta}
\textbf{ATLAS} Collaboration, G.~Aad et~al., \textit{{$Z$ boson production in
  $p+$Pb collisions at $\sqrt{s_{NN}}=5.02$ TeV measured with the ATLAS
  detector}},  {\em Phys. Rev. C} \textbf{92} (2015), no.~4 044915,
  [\href{http://arxiv.org/abs/1507.06232}{{\texttt{arXiv:1507.06232}}}].

\bibitem{Khachatryan:2015pzs}
\textbf{CMS} Collaboration, V.~Khachatryan et~al., \textit{{Study of Z boson
  production in pPb collisions at $\sqrt {s_{NN}} = 5.02$ TeV}},  {\em Phys.
  Lett. B} \textbf{759} (2016) 36--57,
  [\href{http://arxiv.org/abs/1512.06461}{{\texttt{arXiv:1512.06461}}}].

\bibitem{Khachatryan:2015hha}
\textbf{CMS} Collaboration, V.~Khachatryan et~al., \textit{{Study of W boson
  production in pPb collisions at $\sqrt{s_{\mathrm{NN}}} =$ 5.02 TeV}},  {\em
  Phys. Lett. B} \textbf{750} (2015) 565--586,
  [\href{http://arxiv.org/abs/1503.05825}{{\texttt{arXiv:1503.05825}}}].

\bibitem{CMS:2015lca}
\textbf{CMS} Collaboration, \textit{{Charm-tagged jet production in pPb
  collisions at 5.02 TeV and pp collisions at 2.76 TeV}}, .

\bibitem{Adam:2016mkz}
\textbf{ALICE} Collaboration, J.~Adam et~al., \textit{{Measurement of D-meson
  production versus multiplicity in p-Pb collisions at $
  \sqrt{{\mathrm{s}}_{\mathrm{NN}}}=5.02 $ TeV}},  {\em JHEP} \textbf{08}
  (2016) 078,
  [\href{http://arxiv.org/abs/1602.07240}{{\texttt{arXiv:1602.07240}}}].

\bibitem{Adam:2015qda}
\textbf{ALICE} Collaboration, J.~Adam et~al., \textit{{Measurement of electrons
  from heavy-flavour hadron decays in p-Pb collisions at $\sqrt{s_\text{NN}} =$
  5.02 TeV}},  {\em Phys. Lett. B} \textbf{754} (2016) 81--93,
  [\href{http://arxiv.org/abs/1509.07491}{{\texttt{arXiv:1509.07491}}}].

\bibitem{Abelev:2014hha}
\textbf{ALICE} Collaboration, B.~B. Abelev et~al., \textit{{Measurement of
  prompt $D$-meson production in $p-Pb$ collisions at $\sqrt{s_{NN}}$ = 5.02
  TeV}},  {\em Phys. Rev. Lett.} \textbf{113} (2014), no.~23 232301,
  [\href{http://arxiv.org/abs/1405.3452}{{\texttt{arXiv:1405.3452}}}].

\bibitem{Khachatryan:2015sva}
\textbf{CMS} Collaboration, V.~Khachatryan et~al., \textit{{Transverse momentum
  spectra of inclusive b jets in pPb collisions at $\sqrt{s_{NN}} = $ 5.02
  TeV}},  {\em Phys. Lett. B} \textbf{754} (2016) 59,
  [\href{http://arxiv.org/abs/1510.03373}{{\texttt{arXiv:1510.03373}}}].

\bibitem{Khachatryan:2015uja}
\textbf{CMS} Collaboration, V.~Khachatryan et~al., \textit{{Study of B Meson
  Production in p$+$Pb Collisions at $\sqrt{s_{NN}}=5.02$ TeV Using Exclusive
  Hadronic Decays}},  {\em Phys. Rev. Lett.} \textbf{116} (2016), no.~3 032301,
  [\href{http://arxiv.org/abs/1508.06678}{{\texttt{arXiv:1508.06678}}}].

\bibitem{Aaij:2017gcy}
\textbf{LHCb} Collaboration, R.~Aaij et~al., \textit{{Study of prompt D$^{0}$
  meson production in $p$Pb collisions at $ \sqrt{s_{\mathrm{NN}}}=5 $ TeV}},
  {\em JHEP} \textbf{10} (2017) 090,
  [\href{http://arxiv.org/abs/1707.02750}{{\texttt{arXiv:1707.02750}}}].

\bibitem{Aaij:2019lkm}
\textbf{LHCb} Collaboration, R.~Aaij et~al., \textit{{Measurement of $B^+$,
  $B^0$ and $\Lambda_b^0$ production in $p\mkern 1mu\mathrm{Pb}$ collisions at
  $\sqrt{s_\mathrm{NN}}=8.16\,\text{TeV}$}},  {\em Phys. Rev. D} \textbf{99}
  (2019), no.~5 052011,
  [\href{http://arxiv.org/abs/1902.05599}{{\texttt{arXiv:1902.05599}}}].

\bibitem{Kusina:2016fxy}
A.~Kusina, F.~Lyonnet, D.~B. Clark, E.~Godat, T.~Jezo, K.~Kovarik, F.~I.
  Olness, I.~Schienbein, and J.~Y. Yu, \textit{{Vector boson production in pPb
  and PbPb collisions at the LHC and its impact on nCTEQ15 PDFs}},  {\em Eur.
  Phys. J. C} \textbf{77} (2017), no.~7 488,
  [\href{http://arxiv.org/abs/1610.02925}{{\texttt{arXiv:1610.02925}}}].

\bibitem{Kusina:2017gkz}
A.~Kusina, J.-P. Lansberg, I.~Schienbein, and H.-S. Shao, \textit{{Gluon
  Shadowing in Heavy-Flavor Production at the LHC}},  {\em Phys. Rev. Lett.}
  \textbf{121} (2018), no.~5 052004,
  [\href{http://arxiv.org/abs/1712.07024}{{\texttt{arXiv:1712.07024}}}].

\bibitem{Armesto:2015lrg}
N.~Armesto, H.~Paukkunen, J.~M. Pen\'\i{}n, C.~A. Salgado, and P.~Zurita,
  \textit{{An analysis of the impact of LHC Run I proton\textendash{}lead data
  on nuclear parton densities}},  {\em Eur. Phys. J. C} \textbf{76} (2016),
  no.~4 218,
  [\href{http://arxiv.org/abs/1512.01528}{{\texttt{arXiv:1512.01528}}}].

\bibitem{Eskola:2019dui}
K.~J. Eskola, P.~Paakkinen, and H.~Paukkunen, \textit{{Non-quadratic improved
  Hessian PDF reweighting and application to CMS dijet measurements at 5.02
  TeV}},  {\em Eur. Phys. J. C} \textbf{79} (2019), no.~6 511,
  [\href{http://arxiv.org/abs/1903.09832}{{\texttt{arXiv:1903.09832}}}].

\bibitem{Schienbein:2007fs}
I.~Schienbein, J.~Y. Yu, C.~Keppel, J.~G. Morfin, F.~Olness, and J.~F. Owens,
  \textit{{Nuclear parton distribution functions from neutrino deep inelastic
  scattering}},  {\em Phys. Rev. D} \textbf{77} (2008) 054013,
  [\href{http://arxiv.org/abs/0710.4897}{{\texttt{arXiv:0710.4897}}}].

\bibitem{Kovarik:2010uv}
K.~Kovarik, I.~Schienbein, F.~I. Olness, J.~Y. Yu, C.~Keppel, J.~G. Morfin,
  J.~F. Owens, and T.~Stavreva, \textit{{Nuclear Corrections in
  Neutrino-Nucleus DIS and Their Compatibility with Global NPDF Analyses}},
  {\em Phys. Rev. Lett.} \textbf{106} (2011) 122301,
  [\href{http://arxiv.org/abs/1012.0286}{{\texttt{arXiv:1012.0286}}}].

\bibitem{Alvarez-Ruso:2017oui}
\textbf{NuSTEC} Collaboration, L.~Alvarez-Ruso et~al., \textit{{NuSTEC White
  Paper: Status and challenges of neutrino\textendash{}nucleus scattering}},
  {\em Prog. Part. Nucl. Phys.} \textbf{100} (2018) 1--68,
  [\href{http://arxiv.org/abs/1706.03621}{{\texttt{arXiv:1706.03621}}}].

\bibitem{Paukkunen:2013grz}
H.~Paukkunen and C.~A. Salgado, \textit{{Agreement of Neutrino Deep Inelastic
  Scattering Data with Global Fits of Parton Distributions}},  {\em Phys. Rev.
  Lett.} \textbf{110} (2013), no.~21 212301,
  [\href{http://arxiv.org/abs/1302.2001}{{\texttt{arXiv:1302.2001}}}].

\bibitem{Paukkunen:2010hb}
H.~Paukkunen and C.~A. Salgado, \textit{{Compatibility of neutrino DIS data and
  global analyses of parton distribution functions}},  {\em JHEP} \textbf{07}
  (2010) 032,
  [\href{http://arxiv.org/abs/1004.3140}{{\texttt{arXiv:1004.3140}}}].

\bibitem{Hirai:2007sx}
M.~Hirai, S.~Kumano, and T.~H. Nagai, \textit{{Determination of nuclear parton
  distribution functions and their uncertainties in next-to-leading order}},
  {\em Phys. Rev. C} \textbf{76} (2007) 065207,
  [\href{http://arxiv.org/abs/0709.3038}{{\texttt{arXiv:0709.3038}}}].

\bibitem{Eskola:2008ca}
K.~J. Eskola, H.~Paukkunen, and C.~A. Salgado, \textit{{An Improved global
  analysis of nuclear parton distribution functions including RHIC data}},
  {\em JHEP} \textbf{07} (2008) 102,
  [\href{http://arxiv.org/abs/0802.0139}{{\texttt{arXiv:0802.0139}}}].

\bibitem{Eskola:2009uj}
K.~J. Eskola, H.~Paukkunen, and C.~A. Salgado, \textit{{EPS09: A New Generation
  of NLO and LO Nuclear Parton Distribution Functions}},  {\em JHEP}
  \textbf{04} (2009) 065,
  [\href{http://arxiv.org/abs/0902.4154}{{\texttt{arXiv:0902.4154}}}].

\bibitem{deFlorian:2003qf}
D.~de~Florian and R.~Sassot, \textit{{Nuclear parton distributions at
  next-to-leading order}},  {\em Phys. Rev. D} \textbf{69} (2004) 074028,
  [\href{http://arxiv.org/abs/hep-ph/0311227}{{\texttt{hep-ph/0311227}}}].

\bibitem{deFlorian:2011fp}
D.~de~Florian, R.~Sassot, P.~Zurita, and M.~Stratmann, \textit{{Global Analysis
  of Nuclear Parton Distributions}},  {\em Phys. Rev. D} \textbf{85} (2012)
  074028, [\href{http://arxiv.org/abs/1112.6324}{{\texttt{arXiv:1112.6324}}}].

\bibitem{Khanpour:2016pph}
H.~Khanpour and S.~Atashbar~Tehrani, \textit{{Global Analysis of Nuclear Parton
  Distribution Functions and Their Uncertainties at Next-to-Next-to-Leading
  Order}},  {\em Phys. Rev. D} \textbf{93} (2016), no.~1 014026,
  [\href{http://arxiv.org/abs/1601.00939}{{\texttt{arXiv:1601.00939}}}].

\bibitem{Walt:2019slu}
M.~Walt, I.~Helenius, and W.~Vogelsang, \textit{{Open-source QCD analysis of
  nuclear parton distribution functions at NLO and NNLO}},  {\em Phys. Rev. D}
  \textbf{100} (2019), no.~9 096015,
  [\href{http://arxiv.org/abs/1908.03355}{{\texttt{arXiv:1908.03355}}}].

\bibitem{Forte:2002fg}
S.~Forte, L.~Garrido, J.~I. Latorre, and A.~Piccione, \textit{{Neural network
  parametrization of deep inelastic structure functions}},  {\em JHEP}
  \textbf{05} (2002) 062,
  [\href{http://arxiv.org/abs/hep-ph/0204232}{{\texttt{hep-ph/0204232}}}].

\bibitem{DelDebbio:2007ee}
\textbf{NNPDF} Collaboration, L.~Del~Debbio, S.~Forte, J.~I. Latorre,
  A.~Piccione, and J.~Rojo, \textit{{Neural network determination of parton
  distributions: The Nonsinglet case}},  {\em JHEP} \textbf{03} (2007) 039,
  [\href{http://arxiv.org/abs/hep-ph/0701127}{{\texttt{hep-ph/0701127}}}].

\bibitem{Rojo:2008ke}
\textbf{NNPDF} Collaboration, J.~Rojo, R.~D. Ball, L.~Del~Debbio, S.~Forte,
  A.~Guffanti, J.~I. Latorre, A.~Piccione, and M.~Ubiali, \textit{{Update on
  Neural Network Parton Distributions: NNPDF1.1}},  in {\em {38th International
  Symposium on Multiparticle Dynamics}}, 11, 2008.
\newblock \href{http://arxiv.org/abs/0811.2288}{{\texttt{arXiv:0811.2288}}}.

\bibitem{Ball:2010de}
R.~D. Ball, L.~Del~Debbio, S.~Forte, A.~Guffanti, J.~I. Latorre, J.~Rojo, and
  M.~Ubiali, \textit{{A first unbiased global NLO determination of parton
  distributions and their uncertainties}},  {\em Nucl. Phys. B} \textbf{838}
  (2010) 136--206,
  [\href{http://arxiv.org/abs/1002.4407}{{\texttt{arXiv:1002.4407}}}].

\bibitem{Ball:2011mu}
R.~D. Ball, V.~Bertone, F.~Cerutti, L.~Del~Debbio, S.~Forte, A.~Guffanti, J.~I.
  Latorre, J.~Rojo, and M.~Ubiali, \textit{{Impact of Heavy Quark Masses on
  Parton Distributions and LHC Phenomenology}},  {\em Nucl. Phys. B}
  \textbf{849} (2011) 296--363,
  [\href{http://arxiv.org/abs/1101.1300}{{\texttt{arXiv:1101.1300}}}].

\bibitem{Ball:2011uy}
\textbf{NNPDF} Collaboration, R.~D. Ball, V.~Bertone, F.~Cerutti,
  L.~Del~Debbio, S.~Forte, A.~Guffanti, J.~I. Latorre, J.~Rojo, and M.~Ubiali,
  \textit{{Unbiased global determination of parton distributions and their
  uncertainties at NNLO and at LO}},  {\em Nucl. Phys. B} \textbf{855} (2012)
  153--221,
  [\href{http://arxiv.org/abs/1107.2652}{{\texttt{arXiv:1107.2652}}}].

\bibitem{Ball:2012cx}
R.~D. Ball et~al., \textit{{Parton distributions with LHC data}},  {\em Nucl.
  Phys. B} \textbf{867} (2013) 244--289,
  [\href{http://arxiv.org/abs/1207.1303}{{\texttt{arXiv:1207.1303}}}].

\bibitem{Aubert:1987da}
\textbf{European Muon} Collaboration, J.~J. Aubert et~al.,
  \textit{{Measurements of the nucleon structure functions $F2_n$ in deep
  inelastic muon scattering from deuterium and comparison with those from
  hydrogen and iron}},  {\em Nucl. Phys.} \textbf{B293} (1987) 740--786.

\bibitem{Ashman:1988bf}
\textbf{European Muon} Collaboration, J.~Ashman et~al., \textit{{Measurement of
  the Ratios of Deep Inelastic Muon - Nucleus Cross-Sections on Various Nuclei
  Compared to Deuterium}},  {\em Phys. Lett.} \textbf{B202} (1988) 603--610.

\bibitem{Arneodo:1989sy}
\textbf{European Muon} Collaboration, M.~Arneodo et~al., \textit{{Measurements
  of the nucleon structure function in the range $0.002-{\rm GeV}^2 < x <
  0.17-{\rm GeV}^2$ and $0.2-GeV^2 < q^2 < 8-GeV^2$ in deuterium, carbon and
  calcium}},  {\em Nucl. Phys.} \textbf{B333} (1990) 1--47.

\bibitem{Ashman:1992kv}
\textbf{European Muon} Collaboration, J.~Ashman et~al., \textit{{A Measurement
  of the ratio of the nucleon structure function in copper and deuterium}},
  {\em Z.Phys.} \textbf{C57} (1993) 211--218.

\bibitem{Amaudruz:1995tq}
\textbf{New Muon} Collaboration, P.~Amaudruz et~al., \textit{{A Reevaluation of
  the nuclear structure function ratios for D, He, Li-6, C and Ca}},  {\em
  Nucl.Phys.} \textbf{B441} (1995) 3--11,
  [\href{http://arxiv.org/abs/hep-ph/9503291}{{\texttt{hep-ph/9503291}}}].

\bibitem{Arneodo:1995cs}
\textbf{New Muon} Collaboration, M.~Arneodo et~al., \textit{{The Structure
  Function ratios F2(li) / F2(D) and F2(C) / F2(D) at small x}},  {\em
  Nucl.Phys.} \textbf{B441} (1995) 12--30,
  [\href{http://arxiv.org/abs/hep-ex/9504002}{{\texttt{hep-ex/9504002}}}].

\bibitem{Arneodo:1996rv}
\textbf{New Muon} Collaboration, M.~Arneodo et~al., \textit{{The A dependence
  of the nuclear structure function ratios}},  {\em Nucl.Phys.} \textbf{B481}
  (1996) 3--22.

\bibitem{Arneodo:1996ru}
\textbf{New Muon} Collaboration, M.~Arneodo et~al., \textit{{The $Q^2$
  dependence of the structure function ratio $F_2^{\rm Sn}/F_2^{C}$ in deep
  inelastic muon scattering}},  {\em Nucl.Phys.} \textbf{B481} (1996) 23--39.

\bibitem{Alde:1990im}
D.~M. Alde et~al., \textit{{Nuclear dependence of dimuon production at 800-GeV.
  FNAL-772 experiment}},  {\em Phys. Rev. Lett.} \textbf{64} (1990) 2479--2482.

\bibitem{Benvenuti:1987az}
\textbf{BCDMS} Collaboration, A.~C. Benvenuti et~al., \textit{{Nuclear Effects
  in Deep Inelastic Muon Scattering on Deuterium and Iron Targets}},  {\em
  Phys. Lett.} \textbf{B189} (1987) 483--487.

\bibitem{PhysRevD.49.4348}
J.~Gomez et~al., \textit{Measurement of the $a$ dependence of deep-inelastic
  electron scattering},  {\em Phys. Rev.} \textbf{D49} (1994) 4348.

\bibitem{Adams:1992vm}
\textbf{Fermilab E665} Collaboration, M.~R. Adams et~al., \textit{{Shadowing in
  the muon xenon inelastic scattering cross-section at 490-GeV}},  {\em Phys.
  Lett.} \textbf{B287} (1992) 375--380.

\bibitem{Amaudruz:1992wn}
\textbf{New Muon} Collaboration, P.~Amaudruz et~al., \textit{{Measurements of
  R(d) - R(p) and R(Ca) - R(C) in deep inelastic muon scattering}},  {\em Phys.
  Lett.} \textbf{B294} (1992) 120--126.

\bibitem{Dasu:1988ru}
S.~Dasu et~al., \textit{{Measurement of the Difference in R = $\sigma^-$l /
  $\sigma^-$t and $\sigma$(a) / $\sigma(D$) in Deep Inelastic $e D$, $e$ Fe and
  $e$ Au Scattering}},  {\em Phys. Rev. Lett.} \textbf{60} (1988) 2591.

\bibitem{Gomez:1993ri}
J.~Gomez et~al., \textit{{Measurement of the A-dependence of deep inelastic
  electron scattering}},  {\em Phys. Rev.} \textbf{D49} (1994) 4348--4372.

\bibitem{Adams:1995is}
\textbf{E665} Collaboration, M.~R. Adams et~al., \textit{{Shadowing in
  inelastic scattering of muons on carbon, calcium and lead at low x(Bj)}},
  {\em Z. Phys.} \textbf{C67} (1995) 403--410,
  [\href{http://arxiv.org/abs/hep-ex/9505006}{{\texttt{hep-ex/9505006}}}].

\bibitem{Ball:2016neh}
\textbf{NNPDF} Collaboration, R.~D. Ball, V.~Bertone, M.~Bonvini, S.~Carrazza,
  S.~Forte, A.~Guffanti, N.~P. Hartland, J.~Rojo, and L.~Rottoli, \textit{{A
  Determination of the Charm Content of the Proton}},  {\em Eur. Phys. J.}
  \textbf{C76} (2016), no.~11 647,
  [\href{http://arxiv.org/abs/1605.06515}{{\texttt{arXiv:1605.06515}}}].

\bibitem{Lin:2017snn}
H.-W. Lin et~al., \textit{{Parton distributions and lattice QCD calculations: a
  community white paper}},  {\em Prog. Part. Nucl. Phys.} \textbf{100} (2018)
  107--160,
  [\href{http://arxiv.org/abs/1711.07916}{{\texttt{arXiv:1711.07916}}}].

\bibitem{Ball:2016spl}
R.~D. Ball, E.~R. Nocera, and J.~Rojo, \textit{{The asymptotic behaviour of
  parton distributions at small and large $x$}},  {\em Eur. Phys. J.}
  \textbf{C76} (2016), no.~7 383,
  [\href{http://arxiv.org/abs/1604.00024}{{\texttt{arXiv:1604.00024}}}].

\bibitem{Gauld:2015yia}
R.~Gauld, J.~Rojo, L.~Rottoli, and J.~Talbert, \textit{{Charm production in the
  forward region: constraints on the small-x gluon and backgrounds for neutrino
  astronomy}},  {\em JHEP} \textbf{11} (2015) 009,
  [\href{http://arxiv.org/abs/1506.08025}{{\texttt{arXiv:1506.08025}}}].

\bibitem{Bertone:2018dse}
V.~Bertone, R.~Gauld, and J.~Rojo, \textit{{Neutrino Telescopes as QCD
  Microscopes}},  {\em JHEP} \textbf{01} (2019) 217,
  [\href{http://arxiv.org/abs/1808.02034}{{\texttt{arXiv:1808.02034}}}].

\bibitem{GlorotAISTATS2010}
X.~Glorot and Y.~Bengio, \textit{Understanding the difficulty of training deep
  feedforward neural networks},  in {\em JMLR W\&CP: Proceedings of the
  Thirteenth International Conference on Artificial Intelligence and Statistics
  (AISTATS 2010)}, vol.~9, pp.~249--256, May, 2010.

\bibitem{Rojo:2018qdd}
J.~Rojo, \textit{{Machine Learning tools for global PDF fits}},  in {\em {13th
  Conference on Quark Confinement and the Hadron Spectrum (Confinement XIII)
  Maynooth, Ireland, July 31-August 6, 2018}}, 2018.
\newblock \href{http://arxiv.org/abs/1809.04392}{{\texttt{arXiv:1809.04392}}}.

\bibitem{DBLP:journals/corr/KingmaB14}
D.~P. Kingma and J.~Ba, \textit{Adam: {A} method for stochastic optimization},
  {\em CoRR} \textbf{abs/1412.6980} (2014)
  [\href{http://arxiv.org/abs/1412.6980}{{\texttt{arXiv:1412.6980}}}].

\bibitem{das}
R.~D. Ball and S.~Forte, \textit{Double asymptotic scaling at hera},  {\em
  Phys. Lett.} \textbf{B335} (1994) 77--86,
  [\href{http://arxiv.org/abs/hep-ph/9405320}{{\texttt{hep-ph/9405320}}}].

\bibitem{Owens:2007kp}
J.~F. Owens et~al., \textit{{The Impact of new neutrino DIS and Drell-Yan data
  on large-x parton distributions}},  {\em Phys. Rev.} \textbf{D75} (2007)
  054030,
  [\href{http://arxiv.org/abs/hep-ph/0702159}{{\texttt{hep-ph/0702159}}}].

\bibitem{Ball:2012wy}
R.~D. Ball, S.~Carrazza, L.~Del~Debbio, S.~Forte, J.~Gao, et~al.,
  \textit{{Parton Distribution Benchmarking with LHC Data}},  {\em JHEP}
  \textbf{1304} (2013) 125,
  [\href{http://arxiv.org/abs/1211.5142}{{\texttt{arXiv:1211.5142}}}].

\bibitem{Botje:2011sn}
M.~Botje et~al., \textit{{The PDF4LHC Working Group Interim Recommendations}},
  \href{http://arxiv.org/abs/1101.0538}{{\texttt{arXiv:1101.0538}}}.

\bibitem{Alekhin:2011sk}
S.~Alekhin et~al., \textit{{The PDF4LHC Working Group Interim Report}},
  \href{http://arxiv.org/abs/1101.0536}{{\texttt{arXiv:1101.0536}}}.

\bibitem{Sirunyan:2019dox}
\textbf{CMS} Collaboration, A.~M. Sirunyan et~al., \textit{{Observation of
  nuclear modifications in W$^\pm$ boson production in pPb collisions at
  $\sqrt{s_\mathrm{NN}} =$ 8.16 TeV}},  {\em Phys. Lett. B} \textbf{800} (2020)
  135048,
  [\href{http://arxiv.org/abs/1905.01486}{{\texttt{arXiv:1905.01486}}}].

\bibitem{ATLAS-CONF-2015-056}
\textit{{Measurement of $W\rightarrow\mu\nu$ production in $p$+Pb collision at
  $\sqrt{s_{_\text{NN}}}=5.02$ TeV with ATLAS detector at the LHC}},  Tech.
  Rep. ATLAS-CONF-2015-056, CERN, Geneva, Sep, 2015.

\bibitem{Berger:2016inr}
E.~L. Berger, J.~Gao, C.~S. Li, Z.~L. Liu, and H.~X. Zhu, \textit{{Charm-Quark
  Production in Deep-Inelastic Neutrino Scattering at Next-to-Next-to-Leading
  Order in QCD}},  {\em Phys. Rev. Lett.} \textbf{116} (2016), no.~21 212002,
  [\href{http://arxiv.org/abs/1601.05430}{{\texttt{arXiv:1601.05430}}}].

\bibitem{Gao:2017kkx}
J.~Gao, \textit{{Massive charged-current coefficient functions in
  deep-inelastic scattering at NNLO and impact on strange-quark
  distributions}},  {\em JHEP} \textbf{02} (2018) 026,
  [\href{http://arxiv.org/abs/1710.04258}{{\texttt{arXiv:1710.04258}}}].

\bibitem{Candido:2020yat}
A.~Candido, S.~Forte, and F.~Hekhorn, \textit{{Can $ \overline{\mathrm{MS}} $
  parton distributions be negative?}},  {\em JHEP} \textbf{11} (2020) 129,
  [\href{http://arxiv.org/abs/2006.07377}{{\texttt{arXiv:2006.07377}}}].

\bibitem{Bazarko:1994tt}
\textbf{CCFR} Collaboration, A.~O. Bazarko et~al., \textit{{Determination of
  the strange quark content of the nucleon from a next-to-leading order QCD
  analysis of neutrino charm production}},  {\em Z. Phys.} \textbf{C65} (1995)
  189--198,
  [\href{http://arxiv.org/abs/hep-ex/9406007}{{\texttt{hep-ex/9406007}}}].

\bibitem{chorus-dimuon}
\textbf{CHORUS} Collaboration, A.~Kayis-Topaksu et~al., \textit{{Leading order
  analysis of neutrino induced dimuon events in the CHORUS experiment}},  {\em
  Nucl. Phys.} \textbf{B798} (2008) 1--16,
  [\href{http://arxiv.org/abs/0804.1869}{{\texttt{arXiv:0804.1869}}}].

\bibitem{MasonPhD}
D.~A. Mason, \textit{{Measurement of the strange - antistrange asymmetry at NLO
  in QCD from NuTeV dimuon data}}, . FERMILAB-THESIS-2006-01.

\bibitem{Mason:2007zz}
D.~Mason et~al., \textit{{Measurement of the Nucleon Strange-Antistrange
  Asymmetry at Next-to-Leading Order in QCD from NuTeV Dimuon Data}},  {\em
  Phys. Rev. Lett.} \textbf{99} (2007) 192001.

\bibitem{Samoylov:2013xoa}
\textbf{NOMAD} Collaboration, O.~Samoylov et~al., \textit{{A Precision
  Measurement of Charm Dimuon Production in Neutrino Interactions from the
  NOMAD Experiment}},  {\em Nucl.Phys.} \textbf{B876} (2013) 339,
  [\href{http://arxiv.org/abs/1308.4750}{{\texttt{arXiv:1308.4750}}}].

\bibitem{Stirling:2012vh}
W.~Stirling and E.~Vryonidou, \textit{{Charm production in association with an
  electroweak gauge boson at the LHC}},  {\em Phys.Rev.Lett.} \textbf{109}
  (2012) 082002,
  [\href{http://arxiv.org/abs/1203.6781}{{\texttt{arXiv:1203.6781}}}].

\bibitem{Chatrchyan:2013uja}
\textbf{CMS} Collaboration, S.~Chatrchyan et~al., \textit{{Measurement of
  associated W + charm production in pp collisions at $\sqrt{s}$ = 7 TeV}},
  {\em JHEP} \textbf{02} (2014) 013,
  [\href{http://arxiv.org/abs/1310.1138}{{\texttt{arXiv:1310.1138}}}].

\bibitem{CMS-PAS-SMP-18-013}
\textbf{CMS Collaboration} Collaboration, \textit{{Measurement of the
  associated production of a W boson and a charm quark at
  $\sqrt{s}=8~\mathrm{TeV}$}},  Tech. Rep. CMS-PAS-SMP-18-013, CERN, Geneva,
  2019.

\bibitem{Sirunyan:2018hde}
\textbf{CMS} Collaboration, A.~M. Sirunyan et~al., \textit{{Measurement of
  associated production of a W boson and a charm quark in proton-proton
  collisions at $\sqrt{s} =$ 13 TeV}},  {\em Eur. Phys. J.} \textbf{C79}
  (2019), no.~3 269,
  [\href{http://arxiv.org/abs/1811.10021}{{\texttt{arXiv:1811.10021}}}].

\bibitem{Aad:2014xca}
\textbf{ATLAS} Collaboration, G.~Aad et~al., \textit{{Measurement of the
  production of a $W$ boson in association with a charm quark in $pp$
  collisions at $\sqrt{s} =$ 7 TeV with the ATLAS detector}},  {\em JHEP}
  \textbf{1405} (2014) 068,
  [\href{http://arxiv.org/abs/1402.6263}{{\texttt{arXiv:1402.6263}}}].

\bibitem{Airapetian:2013zaw}
\textbf{HERMES} Collaboration, A.~Airapetian et~al., \textit{{Reevaluation of
  the parton distribution of strange quarks in the nucleon}},  {\em Phys. Rev.}
  \textbf{D89} (2014), no.~9 097101,
  [\href{http://arxiv.org/abs/1312.7028}{{\texttt{arXiv:1312.7028}}}].

\bibitem{Borsa:2017vwy}
I.~Borsa, R.~Sassot, and M.~Stratmann, \textit{{Probing the Sea Quark Content
  of the Proton with One-Particle-Inclusive Processes}},  {\em Phys. Rev.}
  \textbf{D96} (2017), no.~9 094020,
  [\href{http://arxiv.org/abs/1708.01630}{{\texttt{arXiv:1708.01630}}}].

\bibitem{Sato:2019yez}
\textbf{JAM} Collaboration, N.~Sato, C.~Andres, J.~Ethier, and W.~Melnitchouk,
  \textit{{Strange quark suppression from a simultaneous Monte Carlo analysis
  of parton distributions and fragmentation functions}},  {\em Phys. Rev. D}
  \textbf{101} (2020), no.~7 074020,
  [\href{http://arxiv.org/abs/1905.03788}{{\texttt{arXiv:1905.03788}}}].

\bibitem{Aad:2012sb}
\textbf{ATLAS} Collaboration, G.~Aad et~al., \textit{{Determination of the
  strange quark density of the proton from ATLAS measurements of the $W,Z$
  cross sections}},  {\em Phys.Rev.Lett.} (2012)
  [\href{http://arxiv.org/abs/1203.4051}{{\texttt{arXiv:1203.4051}}}].

\bibitem{Kusina:2020lyz}
A.~Kusina et~al., \textit{{Impact of LHC vector boson production in heavy ion
  collisions on strange PDFs}},
  \href{http://arxiv.org/abs/2007.09100}{{\texttt{arXiv:2007.09100}}}.

\bibitem{Brodsky:2019jla}
S.~J. Brodsky, I.~Schmidt, and S.~Liuti, \textit{{Is the Momentum Sum Rule
  Valid for Nuclear Structure Functions ?}},
  \href{http://arxiv.org/abs/1908.06317}{{\texttt{arXiv:1908.06317}}}.

\bibitem{Aaboud:2019tab}
\textbf{ATLAS} Collaboration, M.~Aaboud et~al., \textit{{Measurement of prompt
  photon production in $\sqrt{s_\mathrm{NN}} = 8.16$ TeV $p$+Pb collisions with
  ATLAS}},  {\em Phys. Lett. B} \textbf{796} (2019) 230--252,
  [\href{http://arxiv.org/abs/1903.02209}{{\texttt{arXiv:1903.02209}}}].

\bibitem{Aad_2015}
G.~Aad, B.~Abbott, J.~Abdallah, S.~Abdel~Khalek, O.~Abdinov, R.~Aben, B.~Abi,
  M.~Abolins, O.~AbouZeid, H.~Abramowicz, and et~al., \textit{Centrality and
  rapidity dependence of inclusive jet production in snn=5.02 tev proton–lead
  collisions with the atlas detector},  {\em Physics Letters B} \textbf{748}
  (Sep, 2015) 392–413.

\bibitem{Sirunyan_2018}
\textbf{CMS} Collaboration, A.~M. Sirunyan et~al., \textit{{Constraining gluon
  distributions in nuclei using dijets in proton-proton and proton-lead
  collisions at $\sqrt{s_{_\mathrm{NN}}} =$ 5.02 TeV}},  {\em Phys. Rev. Lett.}
  \textbf{121} (2018), no.~6 062002,
  [\href{http://arxiv.org/abs/1805.04736}{{\texttt{arXiv:1805.04736}}}].

\bibitem{Acharya:2020puh}
\textbf{ALICE} Collaboration, S.~Acharya et~al., \textit{{Z-boson production in
  p-Pb collisions at $\sqrt{s_{\mathrm{NN}}}=8.16$ TeV and Pb-Pb collisions at
  $\sqrt{s_{\mathrm{NN}}}=5.02$ TeV}},  {\em JHEP} \textbf{09} (2020) 076,
  [\href{http://arxiv.org/abs/2005.11126}{{\texttt{arXiv:2005.11126}}}].

\bibitem{Aaij:2017cqq}
\textbf{LHCb} Collaboration, R.~Aaij et~al., \textit{{Prompt and nonprompt
  J/$\psi$ production and nuclear modification in $p$Pb collisions at
  $\sqrt{s_{\text{NN}}}= 8.16$ TeV}},  {\em Phys. Lett. B} \textbf{774} (2017)
  159--178,
  [\href{http://arxiv.org/abs/1706.07122}{{\texttt{arXiv:1706.07122}}}].

\bibitem{Aaboud:2017cif}
\textbf{ATLAS} Collaboration, M.~Aaboud et~al., \textit{{Measurement of
  quarkonium production in proton\textendash{}lead and
  proton\textendash{}proton collisions at $5.02~\mathrm {TeV}$ with the ATLAS
  detector}},  {\em Eur. Phys. J. C} \textbf{78} (2018), no.~3 171,
  [\href{http://arxiv.org/abs/1709.03089}{{\texttt{arXiv:1709.03089}}}].

\bibitem{Adam:2016ich}
\textbf{ALICE} Collaboration, J.~Adam et~al., \textit{{$D$-meson production in
  $p$-Pb collisions at $\sqrt{s_{\rm NN}}=$5.02 TeV and in pp collisions at
  $\sqrt{s}=$7 TeV}},  {\em Phys. Rev. C} \textbf{94} (2016), no.~5 054908,
  [\href{http://arxiv.org/abs/1605.07569}{{\texttt{arXiv:1605.07569}}}].

\bibitem{Schmookler:2019nvf}
\textbf{CLAS} Collaboration, B.~Schmookler et~al., \textit{{Modified structure
  of protons and neutrons in correlated pairs}},  {\em Nature} \textbf{566}
  (2019), no.~7744 354--358,
  [\href{http://arxiv.org/abs/2004.12065}{{\texttt{arXiv:2004.12065}}}].

\bibitem{Grazzini:2017mhc}
M.~Grazzini, S.~Kallweit, and M.~Wiesemann, \textit{{Fully differential NNLO
  computations with MATRIX}},  {\em Eur. Phys. J. C} \textbf{78} (2018), no.~7
  537, [\href{http://arxiv.org/abs/1711.06631}{{\texttt{arXiv:1711.06631}}}].

\bibitem{deFlorian:2014xna}
D.~de~Florian, R.~Sassot, M.~Epele, R.~J. Hernández-Pinto, and M.~Stratmann,
  \textit{{Parton-to-Pion Fragmentation Reloaded}},  {\em Phys. Rev. D}
  \textbf{91} (2015), no.~1 014035,
  [\href{http://arxiv.org/abs/1410.6027}{{\texttt{arXiv:1410.6027}}}].

\bibitem{Anderle:2015lqa}
D.~P. Anderle, F.~Ringer, and M.~Stratmann, \textit{{Fragmentation Functions at
  Next-to-Next-to-Leading Order Accuracy}},  {\em Phys. Rev. D} \textbf{92}
  (2015), no.~11 114017,
  [\href{http://arxiv.org/abs/1510.05845}{{\texttt{arXiv:1510.05845}}}].

\bibitem{Anderle:2016czy}
D.~P. Anderle, T.~Kaufmann, M.~Stratmann, and F.~Ringer, \textit{{Fragmentation
  Functions Beyond Fixed Order Accuracy}},  {\em Phys. Rev. D} \textbf{95}
  (2017), no.~5 054003,
  [\href{http://arxiv.org/abs/1611.03371}{{\texttt{arXiv:1611.03371}}}].

\bibitem{deFlorian:2017lwf}
D.~de~Florian, M.~Epele, R.~J. Hernandez-Pinto, R.~Sassot, and M.~Stratmann,
  \textit{{Parton-to-Kaon Fragmentation Revisited}},  {\em Phys. Rev. D}
  \textbf{95} (2017), no.~9 094019,
  [\href{http://arxiv.org/abs/1702.06353}{{\texttt{arXiv:1702.06353}}}].

\bibitem{Buskulic:1994ft}
\textbf{ALEPH} Collaboration, D.~Buskulic et~al., \textit{{Inclusive pi+-, K+-
  and (p, anti-p) differential cross-sections at the Z resonance}},  {\em Z.
  Phys. C} \textbf{66} (1995) 355--366.

\bibitem{Abreu:1998vq}
\textbf{DELPHI} Collaboration, P.~Abreu et~al., \textit{{pi+-, K+-, p and
  anti-p production in Z0 ---\ensuremath{>} q anti-q, Z0 ---\ensuremath{>} b
  anti-b, Z0 ---\ensuremath{>} u anti-u, d anti-d, s anti-s}},  {\em Eur. Phys.
  J. C} \textbf{5} (1998) 585--620.

\bibitem{Akers:1994ez}
\textbf{OPAL} Collaboration, R.~Akers et~al., \textit{{Measurement of the
  production rates of charged hadrons in e+ e- annihilation at the Z0}},  {\em
  Z. Phys. C} \textbf{63} (1994) 181--196.

\bibitem{Brandelik:1980iy}
\textbf{TASSO} Collaboration, R.~Brandelik et~al., \textit{{Charged Pion, Kaon,
  Proton and anti-Proton Production in High-Energy e+ e- Annihilation}},  {\em
  Phys. Lett. B} \textbf{94} (1980) 444--449.

\bibitem{Althoff:1982dh}
\textbf{TASSO} Collaboration, M.~Althoff et~al., \textit{{Charged Hadron
  Composition of the Final State in e+ e- Annihilation at High-Energies}},
  {\em Z. Phys. C} \textbf{17} (1983) 5--15.

\bibitem{Braunschweig:1988hv}
\textbf{TASSO} Collaboration, W.~Braunschweig et~al., \textit{{Pion, Kaon and
  Proton Cross-sections in $e^+ e^-$ Annihilation at 34-{GeV} and 44-{GeV}
  Center-of-mass Energy}},  {\em Z. Phys. C} \textbf{42} (1989) 189.

\bibitem{Leitgab:2013qh}
\textbf{Belle} Collaboration, M.~Leitgab et~al., \textit{{Precision Measurement
  of Charged Pion and Kaon Differential Cross Sections in e+e- Annihilation at
  s=10.52 GeV}},  {\em Phys. Rev. Lett.} \textbf{111} (2013) 062002,
  [\href{http://arxiv.org/abs/1301.6183}{{\texttt{arXiv:1301.6183}}}].

\bibitem{Itoh:1994kb}
\textbf{TOPAZ} Collaboration, R.~Itoh et~al., \textit{{Measurement of inclusive
  particle spectra and test of MLLA prediction in e+ e- annihilation at
  s**(1/2) = 58-GeV}},  {\em Phys. Lett. B} \textbf{345} (1995) 335--342,
  [\href{http://arxiv.org/abs/hep-ex/9412015}{{\texttt{hep-ex/9412015}}}].

\bibitem{Lees:2013rqd}
\textbf{BaBar} Collaboration, J.~P. Lees et~al., \textit{{Production of charged
  pions, kaons, and protons in $e^+e^-$ annihilations into hadrons at
  $\sqrt{s}$=10.54 GeV}},  {\em Phys. Rev. D} \textbf{88} (2013) 032011,
  [\href{http://arxiv.org/abs/1306.2895}{{\texttt{arXiv:1306.2895}}}].

\bibitem{Aihara:1988su}
\textbf{TPC/Two Gamma} Collaboration, H.~Aihara et~al., \textit{{Charged hadron
  inclusive cross-sections and fractions in $e^+e^-$ annihiliation
  $\sqrt{s}=29$ GeV}},  {\em Phys. Rev. Lett.} \textbf{61} (1988) 1263.

\bibitem{Abe:2003iy}
\textbf{SLD} Collaboration, K.~Abe et~al., \textit{{Production of $\pi^+$,
  $\pi^-$, $K^+$, $K^-$, p and $\bar{\rm p}$ in Light ($uds$), $c$ and $b$ Jets
  from $Z^0$ Decays}},  {\em Phys. Rev. D} \textbf{69} (2004) 072003,
  [\href{http://arxiv.org/abs/hep-ex/0310017}{{\texttt{hep-ex/0310017}}}].

\bibitem{Adolph:2016bga}
\textbf{COMPASS} Collaboration, C.~Adolph et~al., \textit{{Multiplicities of
  charged pions and charged hadrons from deep-inelastic scattering of muons off
  an isoscalar target}},  {\em Phys. Lett. B} \textbf{764} (2017) 1--10,
  [\href{http://arxiv.org/abs/1604.02695}{{\texttt{arXiv:1604.02695}}}].

\bibitem{Airapetian:2012ki}
\textbf{HERMES} Collaboration, A.~Airapetian et~al., \textit{{Multiplicities of
  charged pions and kaons from semi-inclusive deep-inelastic scattering by the
  proton and the deuteron}},  {\em Phys. Rev. D} \textbf{87} (2013) 074029,
  [\href{http://arxiv.org/abs/1212.5407}{{\texttt{arXiv:1212.5407}}}].

\bibitem{Seidl:2020mqc}
\textbf{Belle} Collaboration, R.~Seidl et~al., \textit{{Update of inclusive
  cross sections of single and pairs of identified light charged hadrons}},
  {\em Phys. Rev. D} \textbf{101} (2020), no.~9 092004,
  [\href{http://arxiv.org/abs/2001.10194}{{\texttt{arXiv:2001.10194}}}].

\bibitem{Abbiendi:1999ry}
\textbf{OPAL} Collaboration, G.~Abbiendi et~al., \textit{{Leading particle
  production in light flavor jets}},  {\em Eur. Phys. J. C} \textbf{16} (2000)
  407--421,
  [\href{http://arxiv.org/abs/hep-ex/0001054}{{\texttt{hep-ex/0001054}}}].

\bibitem{DAgostini:2004kis}
G.~D'Agostini, \textit{{Asymmetric uncertainties: Sources, treatment and
  potential dangers}},
  \href{http://arxiv.org/abs/physics/0403086}{{\texttt{physics/0403086}}}.

\bibitem{Bertone:2015cwa}
V.~Bertone, S.~Carrazza, and E.~R. Nocera, \textit{{Reference results for
  time-like evolution up to $ \mathcal{O}\left({\alpha}_s^3\right) $}},  {\em
  JHEP} \textbf{03} (2015) 046,
  [\href{http://arxiv.org/abs/1501.00494}{{\texttt{arXiv:1501.00494}}}].

\bibitem{Altarelli:1979kv}
G.~Altarelli, R.~K. Ellis, G.~Martinelli, and S.-Y. Pi, \textit{{Processes
  Involving Fragmentation Functions Beyond the Leading Order in QCD}},  {\em
  Nucl. Phys. B} \textbf{160} (1979) 301--329.

\bibitem{Graudenz:1994dq}
D.~Graudenz, \textit{{One particle inclusive processes in deeply inelastic
  lepton - nucleon scattering}},  {\em Nucl. Phys. B} \textbf{432} (1994)
  351--376,
  [\href{http://arxiv.org/abs/hep-ph/9406274}{{\texttt{hep-ph/9406274}}}].

\bibitem{deFlorian:1997zj}
D.~de~Florian, M.~Stratmann, and W.~Vogelsang, \textit{{QCD analysis of
  unpolarized and polarized Lambda baryon production in leading and
  next-to-leading order}},  {\em Phys. Rev. D} \textbf{57} (1998) 5811--5824,
  [\href{http://arxiv.org/abs/hep-ph/9711387}{{\texttt{hep-ph/9711387}}}].

\bibitem{Anderle:2016kwa}
D.~Anderle, D.~de~Florian, and Y.~Rotstein~Habarnau, \textit{{Towards
  semi-inclusive deep inelastic scattering at next-to-next-to-leading order}},
  {\em Phys. Rev. D} \textbf{95} (2017), no.~3 034027,
  [\href{http://arxiv.org/abs/1612.01293}{{\texttt{arXiv:1612.01293}}}].

\bibitem{Anderle:2018qrw}
D.~P. Anderle, {\em {Higher order corrections to Semi-Inclusive Hadron
  Production Processes}}.
\newblock PhD thesis, U. Tubingen, 2018.

\bibitem{Stratmann:2001pb}
M.~Stratmann and W.~Vogelsang, \textit{{Towards a global analysis of polarized
  parton distributions}},  {\em Phys. Rev. D} \textbf{64} (2001) 114007,
  [\href{http://arxiv.org/abs/hep-ph/0107064}{{\texttt{hep-ph/0107064}}}].

\bibitem{Anderle:2012rq}
D.~P. Anderle, F.~Ringer, and W.~Vogelsang, \textit{{QCD resummation for
  semi-inclusive hadron production processes}},  {\em Phys. Rev. D} \textbf{87}
  (2013), no.~3 034014,
  [\href{http://arxiv.org/abs/1212.2099}{{\texttt{arXiv:1212.2099}}}].

\bibitem{Guerrero:2015wha}
J.~V. Guerrero, J.~J. Ethier, A.~Accardi, S.~W. Casper, and W.~Melnitchouk,
  \textit{{Hadron mass corrections in semi-inclusive deep-inelastic
  scattering}},  {\em JHEP} \textbf{09} (2015) 169,
  [\href{http://arxiv.org/abs/1505.02739}{{\texttt{arXiv:1505.02739}}}].

\bibitem{Hirai:2016loo}
M.~Hirai, H.~Kawamura, S.~Kumano, and K.~Saito, \textit{{Impacts of B-factory
  measurements on determination of fragmentation functions from
  electron-positron annihilation data}},  {\em PTEP} \textbf{2016} (2016),
  no.~11 113B04,
  [\href{http://arxiv.org/abs/1608.04067}{{\texttt{arXiv:1608.04067}}}].

\bibitem{Gamberg:2021lgx}
L.~Gamberg, Z.-B. Kang, D.~Pitonyak, A.~Prokudin, N.~Sato, and R.~Seidl,
  \textit{{Electron-Ion Collider impact study on the tensor charge of the
  nucleon}},
  \href{http://arxiv.org/abs/2101.06200}{{\texttt{arXiv:2101.06200}}}.

\bibitem{Pumplin:2002vw}
J.~Pumplin, D.~R. Stump, J.~Huston, H.~L. Lai, P.~M. Nadolsky, and W.~K. Tung,
  \textit{{New generation of parton distributions with uncertainties from
  global QCD analysis}},  {\em JHEP} \textbf{07} (2002) 012,
  [\href{http://arxiv.org/abs/hep-ph/0201195}{{\texttt{hep-ph/0201195}}}].

\bibitem{AbelleiraFernandez:2012cc}
\textbf{LHeC Study Group} Collaboration, J.~L. Abelleira~Fernandez et~al.,
  \textit{{A Large Hadron Electron Collider at CERN: Report on the Physics and
  Design Concepts for Machine and Detector}},  {\em J. Phys. G} \textbf{39}
  (2012) 075001,
  [\href{http://arxiv.org/abs/1206.2913}{{\texttt{arXiv:1206.2913}}}].

\bibitem{AbelleiraFernandez:2012ty}
\textbf{LHeC Study Group} Collaboration, J.~L. Abelleira~Fernandez et~al.,
  \textit{{On the Relation of the LHeC and the LHC}},
  \href{http://arxiv.org/abs/1211.5102}{{\texttt{arXiv:1211.5102}}}.

\bibitem{Klein:2018rhq}
M.~Klein, {\em {Future Deep Inelastic Scattering with the LHeC}}.
\newblock 2019.
\newblock \href{http://arxiv.org/abs/1802.04317}{{\texttt{arXiv:1802.04317}}}.

\bibitem{Aad:2016zzw}
\textbf{ATLAS} Collaboration, G.~Aad et~al., \textit{{Measurement of the
  double-differential high-mass Drell-Yan cross section in pp collisions at $
  \sqrt{s}=8 $ TeV with the ATLAS detector}},  {\em JHEP} \textbf{08} (2016)
  009, [\href{http://arxiv.org/abs/1606.01736}{{\texttt{arXiv:1606.01736}}}].

\bibitem{CMS-PAS-SMP-17-014}
\textbf{CMS} Collaboration, \textit{{Measurement of associated production of W
  bosons with charm quarks in proton-proton collisions at
  $\sqrt{s}=13~\mathrm{TeV}$ with the CMS experiment at the LHC}},  Tech. Rep.
  CMS-PAS-SMP-17-014, CERN, Geneva, 2018.

\bibitem{Aaboud:2017cbm}
\textbf{ATLAS} Collaboration, M.~Aaboud et~al., \textit{{Measurement of the
  cross section for inclusive isolated-photon production in $pp$ collisions at
  $\sqrt s=13$ TeV using the ATLAS detector}},  {\em Phys. Lett.} \textbf{B770}
  (2017) 473--493,
  [\href{http://arxiv.org/abs/1701.06882}{{\texttt{arXiv:1701.06882}}}].

\bibitem{Aaboud:2017wsi}
\textbf{ATLAS} Collaboration, M.~Aaboud et~al., \textit{{Measurement of
  inclusive jet and dijet cross-sections in proton-proton collisions at
  $\sqrt{s}=13$ TeV with the ATLAS detector}},  {\em JHEP} \textbf{05} (2018)
  195, [\href{http://arxiv.org/abs/1711.02692}{{\texttt{arXiv:1711.02692}}}].

\bibitem{Buckley:2014ana}
A.~Buckley, J.~Ferrando, S.~Lloyd, K.~Nordström, B.~Page, et~al.,
  \textit{{LHAPDF6: parton density access in the LHC precision era}},  {\em
  Eur.Phys.J.} \textbf{C75} (2015) 132,
  [\href{http://arxiv.org/abs/1412.7420}{{\texttt{arXiv:1412.7420}}}].

\bibitem{Paukkunen:2017phq}
\textbf{LHeC study Group} Collaboration, H.~Paukkunen, \textit{{An update on
  nuclear PDFs at the LHeC}},  {\em PoS} \textbf{DIS2017} (2018) 109,
  [\href{http://arxiv.org/abs/1709.08342}{{\texttt{arXiv:1709.08342}}}].

\bibitem{Cooper-Sarkar:2016udp}
\textbf{LHeC study Group} Collaboration, A.~M. Cooper-Sarkar, \textit{{Improved
  measurement of parton distribution functions and $\alpha_s(M_Z)$ with the
  LHeC}},  {\em PoS} \textbf{DIS2016} (2016) 274,
  [\href{http://arxiv.org/abs/1605.08579}{{\texttt{arXiv:1605.08579}}}].

\bibitem{LHeCtalk1}
C.~Gwenlan, ``{Talk at the DIS19, Turin, Italy,
  \url{https://indico.cern.ch/event/749003/contributions/3330763/attachments/1828248/2992940/DIS19-LHeC_Gwenlan.pdf}}.''
  2019.

\bibitem{LHeCtalk2}
A.~Cooper-Sarkar, ``{Talk at the DIS17, Birmingham, UK,
  \url{https://indico.cern.ch/event/568360/contributions/2523511/attachments/1439352/2215137/LHeC_2017.pdf}}.''
  2017.

\bibitem{Ball:2017otu}
R.~D. Ball, V.~Bertone, M.~Bonvini, S.~Marzani, J.~Rojo, and L.~Rottoli,
  \textit{{Parton distributions with small-x resummation: evidence for BFKL
  dynamics in HERA data}},  {\em Eur. Phys. J.} \textbf{C78} (2018), no.~4 321,
  [\href{http://arxiv.org/abs/1710.05935}{{\texttt{arXiv:1710.05935}}}].

\bibitem{mklein}
\url{http://hep.ph.liv.ac.uk/~mklein/lhecdata/} and
  \url{http://hep.ph.liv.ac.uk/~mklein/heavyqdata/}.

\bibitem{Klein:1564929}
M.~Klein and V.~Radescu, \textit{{Partons from the LHeC}},
  \href{http://arxiv.org/abs/CERN-LHeC-Note-2013-002
  PHY}{{\texttt{CERN-LHeC-Note-2013-002 PHY}}}.

\bibitem{Jones:2016ldq}
S.~P. Jones, A.~D. Martin, M.~G. Ryskin, and T.~Teubner, \textit{{The exclusive
  $J/\psi$ process at the LHC tamed to probe the low $x$ gluon}},  {\em Eur.
  Phys. J. C} \textbf{76} (2016), no.~11 633,
  [\href{http://arxiv.org/abs/1610.02272}{{\texttt{arXiv:1610.02272}}}].

\bibitem{Acharya:2020sxs}
\textbf{ALICE} Collaboration, S.~Acharya et~al., \textit{{Measurement of
  isolated photon-hadron correlations in $\sqrt{s_{\rm{NN}}}$ = 5.02 TeV pp and
  p-Pb collisions}},
  \href{http://arxiv.org/abs/2005.14637}{{\texttt{arXiv:2005.14637}}}.

\bibitem{Benic:2016uku}
S.~Benic, K.~Fukushima, O.~Garcia-Montero, and R.~Venugopalan, \textit{{Probing
  gluon saturation with next-to-leading order photon production at central
  rapidities in proton-nucleus collisions}},  {\em JHEP} \textbf{01} (2017)
  115, [\href{http://arxiv.org/abs/1609.09424}{{\texttt{arXiv:1609.09424}}}].

\bibitem{vanderKolk:2020fqo}
\textbf{ALICE FoCal} Collaboration, N.~van~der Kolk, \textit{{FoCal: A highly
  granular digital calorimeter}},  {\em Nucl. Instrum. Meth. A} \textbf{958}
  (2020) 162059.

\bibitem{ALICECollaboration:2719928}
C.~ALICE~Collaboration, \textit{{Letter of Intent: A Forward Calorimeter
  (FoCal) in the ALICE experiment}},  Tech. Rep. CERN-LHCC-2020-009.
  LHCC-I-036, CERN, Geneva, Jun, 2020.

\bibitem{vanLeeuwen:2019zpz}
M.~van Leeuwen, \textit{{Constraining nuclear Parton Density Functions with
  forward photon production at the LHC}},
  \href{http://arxiv.org/abs/1909.05338}{{\texttt{arXiv:1909.05338}}}.

\bibitem{Ball:2010gb}
\textbf{The NNPDF} Collaboration, R.~D. Ball et~al., \textit{{Reweighting
  NNPDFs: the W lepton asymmetry}},  {\em Nucl. Phys.} \textbf{B849} (2011)
  112--143,
  [\href{http://arxiv.org/abs/1012.0836}{{\texttt{arXiv:1012.0836}}}].

\bibitem{Accardi:2012qut}
A.~Accardi et~al., \textit{{Electron Ion Collider: The Next QCD Frontier}:
  {Understanding the glue that binds us all}},  {\em Eur. Phys. J. A}
  \textbf{52} (2016), no.~9 268,
  [\href{http://arxiv.org/abs/1212.1701}{{\texttt{arXiv:1212.1701}}}].

\bibitem{Helenius:2014qla}
I.~Helenius, K.~J. Eskola, and H.~Paukkunen, \textit{{Probing the small-$x$
  nuclear gluon distributions with isolated photons at forward rapidities in
  p+Pb collisions at the LHC}},  {\em JHEP} \textbf{09} (2014) 138,
  [\href{http://arxiv.org/abs/1406.1689}{{\texttt{arXiv:1406.1689}}}].

\bibitem{d'Enterria:2013vba}
D.~d'Enterria, K.~J. Eskola, I.~Helenius, and H.~Paukkunen,
  \textit{{Confronting current NLO parton fragmentation functions with
  inclusive charged-particle spectra at hadron colliders}},  {\em Nucl. Phys.}
  \textbf{B883} (2014) 615--628,
  [\href{http://arxiv.org/abs/1311.1415}{{\texttt{arXiv:1311.1415}}}].

\bibitem{Albino:2005me}
S.~Albino, B.~Kniehl, and G.~Kramer, \textit{{Fragmentation functions for light
  charged hadrons with complete quark flavor separation}},  {\em Nucl. Phys. B}
  \textbf{725} (2005) 181--206,
  [\href{http://arxiv.org/abs/hep-ph/0502188}{{\texttt{hep-ph/0502188}}}].

\bibitem{Hirai:2007cx}
M.~Hirai, S.~Kumano, T.-H. Nagai, and K.~Sudoh, \textit{{Determination of
  fragmentation functions and their uncertainties}},  {\em Phys. Rev. D}
  \textbf{75} (2007) 094009,
  [\href{http://arxiv.org/abs/hep-ph/0702250}{{\texttt{hep-ph/0702250}}}].

\bibitem{Albacete:2017qng}
J.~L. Albacete et~al., \textit{{Predictions for Cold Nuclear Matter Effects in
  $p+$Pb Collisions at $\sqrt{s_{_{NN}}} = 8.16$ TeV}},  {\em Nucl. Phys. A}
  \textbf{972} (2018) 18--85,
  [\href{http://arxiv.org/abs/1707.09973}{{\texttt{arXiv:1707.09973}}}].

\bibitem{Aschenauer:2017jsk}
E.~Aschenauer, S.~Fazio, J.~Lee, H.~Mantysaari, B.~Page, B.~Schenke,
  T.~Ullrich, R.~Venugopalan, and P.~Zurita, \textit{{The
  electron\textendash{}ion collider: assessing the energy dependence of key
  measurements}},  {\em Rept. Prog. Phys.} \textbf{82} (2019), no.~2 024301,
  [\href{http://arxiv.org/abs/1708.01527}{{\texttt{arXiv:1708.01527}}}].

\bibitem{Aschenauer:2017oxs}
E.~Aschenauer, S.~Fazio, M.~Lamont, H.~Paukkunen, and P.~Zurita,
  \textit{{Nuclear Structure Functions at a Future Electron-Ion Collider}},
  {\em Phys. Rev. D} \textbf{96} (2017), no.~11 114005,
  [\href{http://arxiv.org/abs/1708.05654}{{\texttt{arXiv:1708.05654}}}].

\bibitem{Aschenauer:2019kzf}
E.~C. Aschenauer, I.~Borsa, R.~Sassot, and C.~Van~Hulse,
  \textit{{Semi-inclusive Deep-Inelastic Scattering, Parton Distributions and
  Fragmentation Functions at a Future Electron-Ion Collider}},  {\em Phys. Rev.
  D} \textbf{99} (2019), no.~9 094004,
  [\href{http://arxiv.org/abs/1902.10663}{{\texttt{arXiv:1902.10663}}}].

\bibitem{Aschenauer:2012ve}
E.~C. Aschenauer, R.~Sassot, and M.~Stratmann, \textit{{Helicity Parton
  Distributions at a Future Electron-Ion Collider: A Quantitative Appraisal}},
  {\em Phys. Rev. D} \textbf{86} (2012) 054020,
  [\href{http://arxiv.org/abs/1206.6014}{{\texttt{arXiv:1206.6014}}}].

\bibitem{Aschenauer:2013iia}
E.~C. Aschenauer, T.~Burton, T.~Martini, H.~Spiesberger, and M.~Stratmann,
  \textit{{Prospects for Charged Current Deep-Inelastic Scattering off
  Polarized Nucleons at a Future Electron-Ion Collider}},  {\em Phys. Rev. D}
  \textbf{88} (2013) 114025,
  [\href{http://arxiv.org/abs/1309.5327}{{\texttt{arXiv:1309.5327}}}].

\bibitem{Aschenauer:2015ata}
E.~C. Aschenauer, R.~Sassot, and M.~Stratmann, \textit{{Unveiling the Proton
  Spin Decomposition at a Future Electron-Ion Collider}},  {\em Phys. Rev. D}
  \textbf{92} (2015), no.~9 094030,
  [\href{http://arxiv.org/abs/1509.06489}{{\texttt{arXiv:1509.06489}}}].

\bibitem{Aschenauer:2020pdk}
I.~Borsa, G.~Lucero, R.~Sassot, E.~C. Aschenauer, and A.~S. Nunes,
  \textit{{Revisiting helicity parton distributions at a future electron-ion
  collider}},  {\em Phys. Rev. D} \textbf{102} (2020), no.~9 094018,
  [\href{http://arxiv.org/abs/2007.08300}{{\texttt{arXiv:2007.08300}}}].

\bibitem{Ball:2013tyh}
\textbf{NNPDF} Collaboration, R.~D. Ball, S.~Forte, A.~Guffanti, E.~R. Nocera,
  G.~Ridolfi, and J.~Rojo, \textit{{Polarized Parton Distributions at an
  Electron-Ion Collider}},  {\em Phys. Lett. B} \textbf{728} (2014) 524--531,
  [\href{http://arxiv.org/abs/1310.0461}{{\texttt{arXiv:1310.0461}}}].

\bibitem{Faura:2020oom}
F.~Faura, S.~Iranipour, E.~R. Nocera, J.~Rojo, and M.~Ubiali, \textit{{The
  Strangest Proton?}},  {\em Eur. Phys. J. C} \textbf{80} (2020), no.~12 1168,
  [\href{http://arxiv.org/abs/2009.00014}{{\texttt{arXiv:2009.00014}}}].

\bibitem{Charchula:1994kf}
K.~Charchula, G.~Schuler, and H.~Spiesberger, \textit{{Combined QED and QCD
  radiative effects in deep inelastic lepton - proton scattering: The Monte
  Carlo generator DJANGO6}},  {\em Comput. Phys. Commun.} \textbf{81} (1994)
  381--402.

\bibitem{Kwiatkowski:1990es}
A.~Kwiatkowski, H.~Spiesberger, and H.~Mohring, \textit{{Heracles: An Event
  Generator for $e p$ Interactions at \{HERA\} Energies Including Radiative
  Processes: Version 1.0}},  {\em Comput. Phys. Commun.} \textbf{69} (1992)
  155--172.

\bibitem{Ingelman:1996mq}
G.~Ingelman, A.~Edin, and J.~Rathsman, \textit{{LEPTO 6.5: A Monte Carlo
  generator for deep inelastic lepton - nucleon scattering}},  {\em Comput.
  Phys. Commun.} \textbf{101} (1997) 108--134,
  [\href{http://arxiv.org/abs/hep-ph/9605286}{{\texttt{hep-ph/9605286}}}].

\bibitem{Sjostrand:2019zhc}
T.~Sj\"ostrand, \textit{{The PYTHIA Event Generator: Past, Present and
  Future}},  {\em Comput. Phys. Commun.} \textbf{246} (2020) 106910,
  [\href{http://arxiv.org/abs/1907.09874}{{\texttt{arXiv:1907.09874}}}].

\bibitem{Dudek:2012vr}
J.~Dudek et~al., \textit{{Physics Opportunities with the 12 GeV Upgrade at
  Jefferson Lab}},  {\em Eur. Phys. J. A} \textbf{48} (2012) 187,
  [\href{http://arxiv.org/abs/1208.1244}{{\texttt{arXiv:1208.1244}}}].

\bibitem{Garcia:2020jwr}
A.~Garcia, R.~Gauld, A.~Heijboer, and J.~Rojo, \textit{{Complete predictions
  for high-energy neutrino propagation in matter}},  {\em JCAP} \textbf{09}
  (2020) 025,
  [\href{http://arxiv.org/abs/2004.04756}{{\texttt{arXiv:2004.04756}}}].

\bibitem{Brown:1971qr}
R.~Brown, \textit{{Intermediate boson. i. theoretical production cross-sections
  in high-energy neutrino and muon experiments}},  {\em Phys. Rev. D}
  \textbf{3} (1971) 207--223.

\bibitem{Aaij:2013mga}
\textbf{LHCb} Collaboration, R.~Aaij et~al., \textit{{Prompt charm production
  in pp collisions at sqrt(s)=7 TeV}},  {\em Nucl. Phys. B} \textbf{871} (2013)
  1--20, [\href{http://arxiv.org/abs/1302.2864}{{\texttt{arXiv:1302.2864}}}].

\bibitem{Aaij:2015bpa}
\textbf{LHCb} Collaboration, R.~Aaij et~al., \textit{{Measurements of prompt
  charm production cross-sections in $pp$ collisions at $ \sqrt{s}=13 $ TeV}},
  {\em JHEP} \textbf{03} (2016) 159,
  [\href{http://arxiv.org/abs/1510.01707}{{\texttt{arXiv:1510.01707}}}].
  [Erratum: JHEP 09, 013 (2016), Erratum: JHEP 05, 074 (2017)].

\bibitem{Aaij:2016jht}
\textbf{LHCb} Collaboration, R.~Aaij et~al., \textit{{Measurements of prompt
  charm production cross-sections in pp collisions at $ \sqrt{s}=5 $ TeV}},
  {\em JHEP} \textbf{06} (2017) 147,
  [\href{http://arxiv.org/abs/1610.02230}{{\texttt{arXiv:1610.02230}}}].

\bibitem{Akhunzyanov:2018ysf}
\textbf{COMPASS} Collaboration, R.~Akhunzyanov et~al., \textit{{K$^-$ over
  K$^+$ multiplicity ratio for kaons produced in DIS with a large fraction of
  the virtual-photon energy}},  {\em Phys. Lett. B} \textbf{786} (2018)
  390--398,
  [\href{http://arxiv.org/abs/1802.00584}{{\texttt{arXiv:1802.00584}}}].

\bibitem{Alexeev:2020jia}
\textbf{COMPASS} Collaboration, G.~D. Alexeev et~al., \textit{{Antiproton over
  proton and K$^-$ over K$^+$ multiplicity ratios at high $z$ in DIS}},  {\em
  Phys. Lett. B} \textbf{807} (2020) 135600,
  [\href{http://arxiv.org/abs/2003.11791}{{\texttt{arXiv:2003.11791}}}].

\bibitem{Adolph:2013stb}
\textbf{COMPASS} Collaboration, C.~Adolph et~al., \textit{{Hadron Transverse
  Momentum Distributions in Muon Deep Inelastic Scattering at 160 GeV/$c$}},
  {\em Eur. Phys. J. C} \textbf{73} (2013), no.~8 2531,
  [\href{http://arxiv.org/abs/1305.7317}{{\texttt{arXiv:1305.7317}}}].
  [Erratum: Eur.Phys.J.C 75, 94 (2015)].

\bibitem{Bacchetta:2019sam}
A.~Bacchetta, V.~Bertone, C.~Bissolotti, G.~Bozzi, F.~Delcarro, F.~Piacenza,
  and M.~Radici, \textit{{Transverse-momentum-dependent parton distributions up
  to N$^{3}$LL from Drell-Yan data}},  {\em JHEP} \textbf{07} (2020) 117,
  [\href{http://arxiv.org/abs/1912.07550}{{\texttt{arXiv:1912.07550}}}].

\bibitem{Gottfried:1967kk}
K.~Gottfried, \textit{{Sum rule for high-energy electron - proton scattering}},
   {\em Phys. Rev. Lett.} \textbf{18} (1967) 1174.

\bibitem{Forte:1992df}
S.~Forte, \textit{{The Gottfried sum rule and the light flavor content of the
  nucleon}},  {\em Phys. Rev. D} \textbf{47} (1993) 1842--1853.

\bibitem{Abbate:2005ct}
R.~Abbate and S.~Forte, \textit{{Re-evaluation of the Gottfried sum using
  neural networks}},  {\em Phys. Rev. D} \textbf{72} (2005) 117503,
  [\href{http://arxiv.org/abs/hep-ph/0511231}{{\texttt{hep-ph/0511231}}}].

\bibitem{Brodsky:1973kr}
S.~J. Brodsky and G.~R. Farrar, \textit{{Scaling Laws at Large Transverse
  Momentum}},  {\em Phys. Rev. Lett.} \textbf{31} (1973) 1153--1156.

\bibitem{Brivio:2017vri}
I.~Brivio and M.~Trott, \textit{{The Standard Model as an Effective Field
  Theory}},  {\em Phys. Rept.} \textbf{793} (2019) 1--98,
  [\href{http://arxiv.org/abs/1706.08945}{{\texttt{arXiv:1706.08945}}}].

\bibitem{Carrazza:2019sec}
S.~Carrazza, C.~Degrande, S.~Iranipour, J.~Rojo, and M.~Ubiali, \textit{{Can
  New Physics hide inside the proton?}},  {\em Phys. Rev. Lett.} \textbf{123}
  (2019), no.~13 132001,
  [\href{http://arxiv.org/abs/1905.05215}{{\texttt{arXiv:1905.05215}}}].

\end{thebibliography}\endgroup


\end{backmatter}

\end{document}